\documentclass[aps,prd,preprint,tightenlines,superscriptaddress,nofootinbib]{revtex4}

\usepackage{amsmath,amssymb,url,mathrsfs}
\usepackage{graphicx,tabularx}
\usepackage{grffile}
\usepackage{lscape,pdflscape}

\newcommand{\lrf}[2]{\left(\frac{#1}{#2}\right)}
\newcommand{\del}{\partial}
\newcommand{\ccdot}{\!\cdot\!}

\newcommand{\etal}{\textit{et al.}}
\newcommand{\ie}{{\sl i.e. }}
\newcommand{\eg}{{\sl e.g. }}

\newcommand{\etmiss}{E_T^\text{miss}}
\newcommand{\ptmiss}{\vec{p}_T^\text{\, miss}}
\newcommand{\pmiss}{p^\text{miss}}

\renewcommand{\mtt}{M_{T2}}

\newcommand{\om}{\ensuremath\mathcal{O}}
\newcommand\one{\leavevmode\hbox{\small1\normalsize\kern-.33em1}}
\newcommand{\chk}{\checkmark}

\newcommand{\Mpl}{M_\text{Pl}}
\newcommand{\Mstar}{M_\star}
\newcommand{\mkk}{m_\text{KK}}
\newcommand{\p}{\partial}

\newcommand{\s}{\mathcal{S}}
\newcommand{\ope}{\mathcal{O}}
\newcommand{\cutoff}{\Lambda_\text{cutoff}}
\newcommand{\mtrans}{\Lambda_T}

\newcommand{\kev}{{\ensuremath\rm keV}}
\newcommand{\gev}{{\ensuremath\rm GeV}}
\newcommand{\tev}{{\ensuremath\rm TeV}}

\newcommand{\pb}{{\ensuremath\rm pb}}
\newcommand{\br}{{\ensuremath\rm BR}}
\newcommand{\tr}{{\ensuremath\rm Tr}}
\newcommand{\ns}{{\ensuremath\rm ns}}
\newcommand{\sign}{{\ensuremath\rm sign}}
\newcommand{\ifb}{{\ensuremath\rm fb^{-1}}}

\def\slashchar#1{\setbox0=\hbox{$#1$}           
   \dimen0=\wd0                                 
   \setbox1=\hbox{/} \dimen1=\wd1               
   \ifdim\dimen0>\dimen1                        
      \rlap{\hbox to \dimen0{\hfil/\hfil}}      
      #1                                        
   \else                                        
      \rlap{\hbox to \dimen1{\hfil$#1$\hfil}}   
      /                                         
   \fi}

\begin{document}

\preprint{ANL-HEP-PR-09-114\\}
\preprint{UCI-TR-2009-17\\}
\preprint{NUHEP-TH/09-19\\}


\title{Physics searches at the LHC}

\author{David E. Morrissey}
\affiliation{TRIUMF, Vancouver, Canada\\
Jefferson Physical Laboratory, Harvard University, Cambridge, USA}

\author{Tilman Plehn}
\affiliation{Institut f\"ur Theoretische Physik, Universit\"at Heidelberg, Germany}

\author{Tim M.P. Tait}
\affiliation{University of California, Irvine, USA\\
Northwestern University, Evanston, USA\\
Argonne National Laboratory, Argonne, USA}

\begin{abstract}
With the LHC up and running, the focus of experimental and
theoretical high energy physics will soon turn to an interpretation of
LHC data in terms of the physics of electroweak symmetry breaking and
the TeV scale. We present here a broad review of models for new TeV-scale
physics and their LHC signatures.  
In addition, we discuss possible new physics signatures and describe
how they can be linked to specific models of physics beyond the 
Standard Model.  Finally, we illustrate how the LHC era could culminate 
in a detailed understanding of the underlying principles of 
TeV-scale physics.
\end{abstract}

\maketitle

\newpage
\tableofcontents
\newpage

\section{Why new physics at the TeV scale?}
\label{sec:why}

A large number of experimental tests performed over many years
have given us confidence that the Standard Model~(SM) is the correct 
effective theory of elementary particles at energies up to the weak scale.  
These tests include a wide range of direct particle searches as well 
as high-precision tests of quantum effects.  
Since the Standard Model, including the yet to be observed Higgs boson,
is based on a renormalizable gauge field theory,
the model can also be consistently extrapolated to energies
many orders of magnitude above what we have probed directly. 

Despite these successes, the central ingredient of electroweak 
symmetry breaking is not fully understood --- we still have not yet 
discovered the Higgs boson responsible for the breakdown of the electroweak 
$SU(2)_L\times U(1)_Y$ symmetry to its smaller electromagnetic $U(1)_\text{em}$
subgroup within the Standard Model.  The search for the source of electroweak
symmetry breaking has been the major motivation for experimental searches 
as well as theoretical model building for the past 25 years.\bigskip

  Beyond our ignorance of the cause of electroweak symmetry breaking,
there are many reasons to hope and expect that new particles and interactions
beyond the Standard Model will be discovered in the near future.
These include:
\begin{enumerate}
\item A fundamental scalar Higgs boson is not the only way to induce 
\emph{electroweak symmetry breaking}.  New strong interactions can also play 
this role.  All we know to date is that a light Higgs boson is
consistent with precision electroweak data,
but this does
not generally preclude other possibilities.
\item If the Higgs boson is a fundamental scalar, its mass parameter --- which
is closely tied to the scale of electroweak symmetry breaking --- 
is extremely sensitive to quantum corrections.  As a result, attempts 
to extrapolate the Standard Model to energies much above the electroweak 
scale lead to the \emph{gauge hierarchy problem}, where an extreme
fine-tuning of the underlying model parameters is required to
maintain the electroweak scale at its observed value.  
This is not inconsistent theoretically, but it is at the very least
extremely puzzling.
\item The Standard Model is unable to account for the dark matter in 
the universe.
On the other hand, dark matter can be explained by a new stable 
weak-scale particle with weak couplings.
Such a state can naturally obtain the correct relic density to make up
the dark matter through the process of \emph{thermal freeze-out}
in the early universe.  Stable new weakly-interacting states
also arise in many theories that attempt to protect the scale
of electroweak symmetry breaking.
\item The Standard Model cannot explain the asymmetry of visible matter 
over antimatter.  New physics near the electroweak scale can potentially 
give rise to this baryon asymmetry. 
\end{enumerate}
This is only a partial list of shortcomings of the Standard Model.
Other challenges include the three almost unifying gauge couplings,
a quantum description of gravity,
an explanation for the cosmological constant,
and an understanding of the sources of flavor mixing and the 
masses of neutrinos.
The items on our list, however, seem the most likely to
be solved by new physics near the electroweak scale.\bigskip

  Many theories have been proposed for new physics beyond the Standard Model 
that address one or more of the challenges on our list.  
Based on current data, it is not clear which of them, 
if any, are correct.  However, this is about to change.
With the recent start of the Large Hadron Collider~(LHC) and a wide
range of other collider experiments and cosmological probes already
underway or about to begin, in the next few years
we may finally be able to uncover the origin of electroweak symmetry
breaking and new physics related to it.\bigskip

In this work we review some of the most popular proposals for new
physics beyond the Standard Model and we describe how they might 
be discovered at the LHC.  Our goal is to present a broad overview of both 
models and collider signatures, which we hope will be useful to theorists
and experimentalists alike. There is no true target audience for this
review, and many readers might find those parts
dealing with their area of expertise rather short and somewhat
incomplete.  At the same time, we hope that only very few
people would say that about this entire volume. Certainly the three
authors do not feel that way -- all of us are still trying to learn all
the physics which are included in our own work.

The outline of this review is as follows. In the remainder of this
section we discuss in more detail the motivation for physics beyond
the Standard Model. The main body of the review has two orthogonal organization
principles. First, in Section~\ref{sec:models} we review some of the
many proposals for this new physics, including supersymmetry, extra
dimensions, and strong interactions.  The focus in this section will
be on the structures of TeV-scale models and their generic
features at colliders.  Secondly, in Section~\ref{sec:sig} we collect
and describe collider signals these and other models predict. In
addition, some of the signals will not actually be linked to a
complete model, but are included because they pose interesting
challenges to the LHC experiments.  Finally, in Section~\ref{sec:para}
we describe how these signals can be mapped back to identify and
extract information about particular models, reflecting our hopes 
for how the physics landscape will look towards the end of the LHC era.

The matrix structure linking new physics models and new physics
signatures at the LHC we also illustrate in the following
table. Single check marks indicate possible and established links,
double check marks mean generic signatures whose observation would be
taken as a strong hint concerning underlying models.
\bigskip

\emph{
  Of course, the enormous number of existing models and analyses, as
  well as our limitations of space and time, make it impossible to be
  truly complete.  We apologize in advance if your preferred model was
  not included, and we welcome suggestions for additions to the text.
}


\begin{turnpage}
\begin{table}
\renewcommand{\arraystretch}{1.1}
{\small
\begin{tabular}{|c|@{}c@{}|@{}c@{}|@{}c@{}|@{}c@{}|@{}c@{}|@{}c@{}|@{}c@{}|@{}c@{}|@{}c@{}|@{}c@{}|@{}c@{}|@{}c@{}|@{}c@{}|@{}c@{}|@{}c@{}|@{}c@{}|}
\hline
    \begin{picture}(50,15)(0,0)
      \put(20,8){\bf signature}
      \put(-9,15){\line(3,-1){72}}
      \put(3,-7){\bf model}
    \end{picture}
&\begin{tabular}{c}missing\\energy\\(p.89)\end{tabular}
&\begin{tabular}{c}cascade\\decays\\(p.91)\end{tabular}
&\begin{tabular}{c}mono-\\jets/$\gamma$\\(p.15)\end{tabular}
&\begin{tabular}{c}lepton\\resnce\\(p.109)\end{tabular}
&\begin{tabular}{c}di-jet\\resnce\\(p.109)\end{tabular}
&\begin{tabular}{c}$t\bar{t}$\\resnce\\(p.120)\end{tabular}
&\begin{tabular}{c}WW/ZZ\\resnce\\(p.15)\end{tabular}
&\begin{tabular}{c}$\ell\nu_{\ell}$\\resnce\\(p.93)\end{tabular}
&\begin{tabular}{c}top\\partner\\(p.116)\end{tabular}
&\begin{tabular}{c}charged\\tracks\\(p.123)\end{tabular}
&\begin{tabular}{c}displ.\\vertex\\(p.123)\end{tabular}
&\begin{tabular}{c}multi-\\photons\\(p.29)\end{tabular}
&\begin{tabular}{c}spherical\\events\\(p.47,76)\end{tabular}
\\
\hline
\begin{tabular}{c}SUSY (heavy $m_{3/2}$)\\(p.17,26)\end{tabular}&\chk\chk&\chk\chk&&&&&&&\chk&&&&\\
\hline
\begin{tabular}{c}SUSY (light $m_{3/2}$)\\(p.17,27)\end{tabular}&\chk&\chk&\chk&&&&&&\chk&\chk&\chk&&\\
\hline
\begin{tabular}{c}large extra dim\\(p.39)\end{tabular}&\chk\chk&&\chk\chk&&&&&&&&&&\chk\\
\hline
\begin{tabular}{c}universal extra dim\\(p.47)\end{tabular}&\chk\chk&\chk\chk&&\chk&\chk&\chk&\chk&\chk&\chk&&&&\\
\hline
\begin{tabular}{c}technicolor (vanilla)\\(p.51)\end{tabular}&&&&\chk&\chk&\chk&\chk&\chk\chk&&&&&\\
\hline
\begin{tabular}{c}topcolor/top seesaw\\(p.53,54)\end{tabular}&&&&&\chk&\chk\chk&\chk&&&&&&\\
\hline
\begin{tabular}{c}little Higgs (w/o $T$)\\(p.55,58)\end{tabular}&&&&\chk&\chk&\chk&\chk&\chk&&&&&\\
\hline
\begin{tabular}{c}little Higgs (w $T$)\\(p.55,58)\end{tabular}&\chk\chk&\chk\chk&\chk&\chk&\chk&\chk&\chk&\chk&\chk&&&&\\
\hline
\begin{tabular}{c}warped extra dim (IR SM)\\(p.61,63)\end{tabular}&&&&\chk&\chk&\chk&\chk&&&&&&\\
\hline
\begin{tabular}{c}warped extra dim (bulk SM)\\(p.61,64)\end{tabular}&&&&\chk&\chk&\chk\chk&\chk&\chk&&&&&\\
\hline
\begin{tabular}{c}Higgsless/comp. Higgs\\(p.69,73)\end{tabular}&&&&\chk&\chk&\chk\chk&\chk\chk&&&&&&\\
\hline
\begin{tabular}{c}hidden valleys\\(p.75)\end{tabular}&\chk&\chk&\chk&\chk&\chk&\chk&\chk&\chk&\chk&\chk&\chk&\chk&\chk\\
\hline
\end{tabular}
}
\end{table}
\end{turnpage}

\newpage

In contrast to other reviews this work does not attempt to be self
contained. There are several reasons for this: first, the field of
physics searches at the LHC itself is so wide that it would be
impossible to cover it in any kind of satisfactory manner in less that
1000 pages. Writing a text book on physics searches at the LHC, on the
other hand, will most likely have to wait until we know what kind of
new physics we will have discovered. Following the comments from the
introduction it is clearly debatable if we really have an interest in
documenting all wrong TeV-scale model building directions taken in the
run up to LHC.\bigskip

Even worse, physics searches at the LHC extensively build on other
fields of physics, like QCD for the description of the incoming (and
outgoing) particles, particle physics and kinematics to interpret our
measurements, electroweak symmetry breaking to judge their relevance,
and astrophysics to see if some of our greatest hopes on unifying
particle physics and cosmology have really come true.\bigskip

In a first step, we can recommend a few standard text books which we
find useful to advanced students with some knowledge of field theory
and particle physics. Focussing on QCD there are a few text books we
can recommend: the classical work for LHC physics is the book by Keith
Ellis, James Stirling and Bryan Webber~\cite{book_bryan}, in
particular the chapters based on hadron collider physics. A very
useful addition is the monograph by G\"unther Dissertori, Ian Knowles
and Michael Schmelling~\cite{book_dissertori}. Finally, for more
technical questions concerning next-to-leading order calculations for
the LHC Rick Field's book has been of great use to
us~\cite{book_field}. For a little broader scope on the entire
Standard Model of particle physics and less focus on QCD we can
recommend the work by Otto Nachtmann~\cite{book_nachtmann} or
the hands-on introduction to all of collider physics by Vernon Barger,
Roger Phillips, and Tao Han~\cite{book_tao}. Slightly deviating from
the main topic of LHC physics Wolfgang Kilian's notes on electroweak
symmetry breaking offer precisely the perspective we will adopt when
discussing strongly interacting physics
models~\cite{review_wolfgang}.\bigskip

Unfortunately, even starting from those text books it is not unlikely
that there will still be a small gap to bridge before really enjoying
our condensed review article, written from the clear perspective of
LHC new physics searchers. Luckily, there exists an invaluable source
of reviews targeted at the level of slightly advanced graduate
students: the lecture notes triggered by the TASI school series in
Boulder, Colorado. For those of us who have been involved the
perseverance with which these notes are collected might at times have
appeared as a big pain. On the other hand, it leaves us with an
excellent series of pedagogical writeups for example on collider
physics~\cite{tasi_lhc}, QCD~\cite{tasi_qcd,Plehn:2009nd}, Higgs
physics~\cite{Plehn:2009nd,tasi_higgs}, electroweak precision
data~\cite{tasi_ew}, dark matter~\cite{tasi_dm},
supersymmetry~\cite{tasi_susy}, extra dimensions~\cite{tasi_extrad},
or random topics of our interest~\cite{tasi_stuff}.  When historians
of science will at some time in the future look back at our era of
unlimited creativity at the TeV scale, which was so abruptly ended by
LHC observations, this collection of notes will serve as proof in the
most painful detail.


\subsection{Electroweak symmetry breaking}
\label{sec:why_ewsb}

Electroweak symmetry breaking lies at the core of the Standard
Model~\cite{Glashow:1961tr,qcd,Fox:2005yp}.
The structure of the Standard Model is built around the gauge group
$SU(3)_c\times SU(2)_L\times U(1)_Y$.  Of this, only the smaller
$SU(3)_c\times U(1)_\text{em}$ subgroup is manifest at low energies.
\emph{Electroweak symmetry breaking} is the process of reducing
$SU(2)_L\times U(1)_Y$ to $U(1)_\text{em}$.  It gives rise to the
underlying masses of the Standard Model fermions and is the reason why
the weak interactions appear to be so much weaker than
electromagnetism.  In the Standard Model, electroweak symmetry
breaking is induced by the scalar Higgs field, and the associated
\emph{Higgs boson} excitation is the only particle in the model that
has not yet been discovered~\cite{higgs}.  A wide range of precision electroweak
measurements have given us confidence that the underlying gauge
structure of the Standard Model is correct.  However, until we
discover the Higgs boson (or convince ourselves that it does not
exist), we will not fully understand electroweak symmetry
breaking.\bigskip

The Higgs field $H = (H^+,H^0)^t$ is the only fundamental scalar in
the Standard Model, and transforms as $({1},{2},1/2)$
under $SU(3)_c\times SU(2)_L\times
U(1)_Y$~\cite{Gunion:1989we,Carena:2002es,tasi_higgs,
Plehn:2009nd}.
Gauge invariance and renormalizability allow the Higgs potential 
\begin{equation}
V_\text{Higgs} = m_{H}^2|H|^2 + \frac{\lambda}{2}|H|^4 \; .  
\end{equation} 
If $m_{H}^2$ is negative, the minimum of the potential defining the
vacuum of the theory can be taken to be
\begin{equation}
\langle H\rangle = \left(
\begin{array}{c}
0\\v
\end{array}
\right) \qquad \qquad
\text{with} \qquad v= \sqrt{\frac{-m_{H}^2}{\lambda}} \; .
\label{hvev}
\end{equation}
Since the Higgs field transforms non-trivially under $SU(2)_L\times U(1)_Y$,
this vacuum spontaneously
`breaks' this gauge symmetry to the smaller $U(1)_\text{em}$ subgroup
with charges $Q = (t^3_L+Y)$ under which the vacuum is neutral.

  Expanding about this \emph{vacuum expectation value}~(VEV) of 
the Higgs field, there is a physical Higgs boson scalar $h^0$ with 
mass~\cite{Carena:2002es,tasi_higgs}
\begin{equation}
m_h = \sqrt{2\lambda}\,v.
\end{equation}
The Higgs VEV also allows the $SU(2)_L\times U(1)_Y$
gauge bosons to mix yielding a massless photon $\gamma$ corresponding
to the unbroken $U(1)_\text{em}$ gauge group, along with 
massive $W^{\pm}$ and $Z^0$ vector bosons with tree-level masses
\begin{equation}
m_W^2 = \frac{g^2}{2}\,v^2, \qquad \qquad \qquad
m_Z^2 = \frac{g^2+g'^2}{2}\,v^2,
\end{equation}
where $g$ and $g'$ are the $SU(2)_L$ and $U(1)_Y$ gauge couplings.
Electroweak measurements fix the vacuum expectation value
of the Higgs to be $v \simeq 174\,\gev$.  Counting degrees of freedom,
three of the four real components of $H$ have gone to generate the
longitudinal components of the $W^{\pm}$ and $Z^0$ gauge bosons,
while the fourth component shows up as the physical Higgs boson $h^0$.\bigskip

  A wide range collider and low-energy measurements have tested 
the electroweak structure of the Standard Model to an extremely high 
level of precision.
They also provide indirect information about 
the Higgs~\cite{Peskin:1990zt,tasi_ew}.
Indeed, these probes are so precise that it is necessary to include 
quantum  corrections when making theoretical predictions for the 
Standard Model.  Once such corrections are incorporated, 
the Standard Model provides a good fit to the 
data giving us confidence that we are on the right track~\cite{Alcaraz:2007ri}.

\begin{figure}[t]
\begin{center}
  \includegraphics[width=0.4\textwidth]{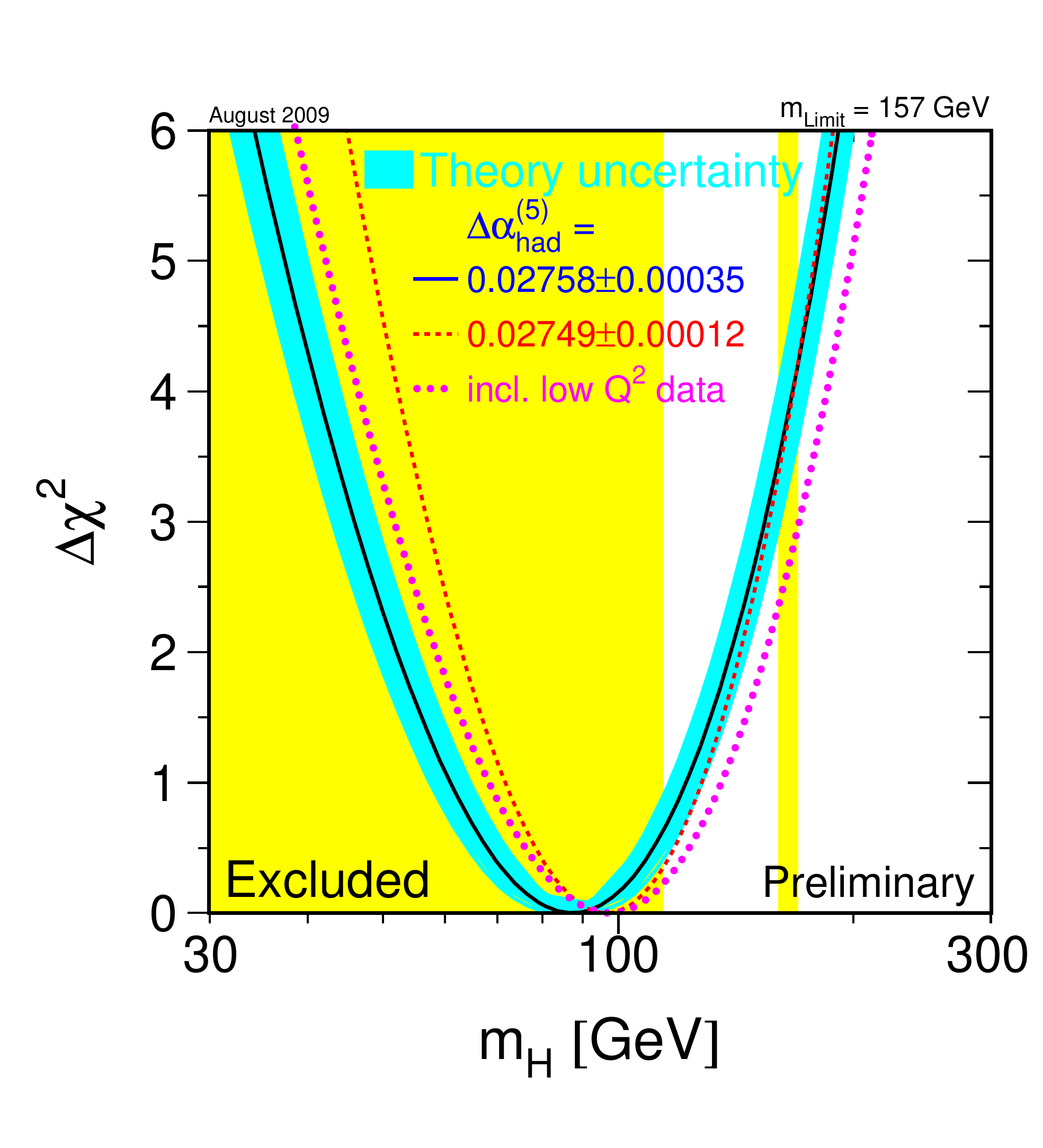}
\end{center}
\vspace{-0.5cm}
\caption{Limits on the Higgs mass within the Standard Model from precision
  electroweak constraints, and direct Higgs searches by the LEP and
  Tevatron experiments.  Figure from
  Refs.~\cite{Alcaraz:2007ri,lepewwg}}
\label{fig:blueband}
\end{figure}

However, among the quantum corrections that must be included
are loop diagrams involving the Higgs.
These corrections depend logarithmically on the Higgs mass, 
and an agreement between the Standard Model and data requires that 
the Higgs not 
be too much heavier than $m_h \simeq 90\,\gev$.  We show the 
precise constraint on the Higgs mass from Ref.~\cite{Alcaraz:2007ri,lepewwg}
in Fig.~\ref{fig:blueband} in terms of the $\Delta \chi^2$ relative 
to the best-fit point.  This figure also shows the lower bound on the 
Standard Model Higgs mass obtained by LEP~\cite{Schael:2006cr},
\begin{equation}
m_h > 114.4\,\gev~~~~(95\,\% \text{C.L.}),
\label{lephiggs}
\end{equation}
as well as the Higgs mass exclusion $160 < m_h < 170\,\gev$ obtained
by the Tevatron collaborations~\cite{tevatron:higgs}.  
Note that all of these experimental result are very sensitive to 
the underlying model, which means that they only hold for the 
Standard Model with its minimal Higgs sector. Extended models could,
for example, prefer a heavy Higgs boson based on the same data.
Both the Tevatron and the LHC are expected to probe the remaining 
electroweak-compatible range of Standard Model Higgs masses 
in the coming years.
\bigskip

  Despite these successes, the fundamental scalar SM Higgs field 
also presents a severe theoretical puzzle that we will discuss 
in the next section.  In light of this, it worth looking at 
what would happen if the Standard Model did not contain a Higgs.  
To be more precise, 
consider a theory with the same particle content and interactions 
as the Standard Model, but with massive $W^{\pm}$ and $Z^0$ vector bosons 
and no Higgs.  This theory is not renormalizable, but it can be 
understood as an effective theory with a high-energy cutoff.  
Computing the amplitude for $WW\to WW$ scattering at tree-level, 
we obtain~\cite{Lee:1977yc,Lee:1977eg}
\begin{equation}
\mathcal{A}_{WW} = \mathcal{A}^{(2)}(\cos\theta)\frac{s}{m_W^2}
+ \mathcal{A}^{(0')}(\cos\theta)\ln\lrf{s}{m_W^2} 
+ \mathcal{A}^{(0)}(\cos\theta),
\label{wwamp}
\end{equation}
where $\sqrt{s}$ is the center-of-mass energy and $\cos\theta$ is
the scattering angle.  We can always decompose this amplitude
into partial waves according to
\begin{equation}
\mathcal{A} = 16\pi\,\sum_{\ell=0}^{\infty}(2\ell+1) \; 
              P_\ell(\cos\theta)\,a_\ell.
\end{equation}
Perturbative unitarity of the scattering process requires that $|a_\ell| \leq 1/2$
for all $\ell$.
Comparing to Eq.~\eqref{wwamp}, we see that this condition
is violated by the tree-level amplitude for $s\gg m_W^2$.  
Careful analyses~\cite{Lee:1977yc,Lee:1977eg}
find that this breakdown occurs for $\sqrt{s} = (4\pi\sqrt{2}/G_F)^{1/2}
\simeq 1.2\,\tev\sim 2\pi\,v$.

At this point let us emphasize strongly that quantum mechanical
unitarity \emph{is not} violated --- even without a Higgs the
underlying quantum field theory is manifestly unitary.  What happens
is that the tree-level amplitude given in Eq.~\eqref{wwamp} does not
give the whole story.  For $\sqrt{s} \gg m_W$, higher-order
corrections are expected to become large and correct the amplitude to
maintain unitarity.  This implies that the electroweak theory without
a Higgs becomes strongly coupled at energies on the order of
$1\,\tev$.  It is not known precisely how the theory with massive
vector bosons behaves at higher energies, but deviations from the
Standard Model (with a Higgs) are expected to be large enough to be
visible at the LHC~\cite{Chanowitz:1985hj,Bagger:1992vu,
  Bagger:1993zf}.
Putting the Higgs back in, 
the dangerous $\mathcal{A}^{(2)}$ and $\mathcal{A}^{(0')}$ 
terms in Eq.~\eqref{wwamp} are cancelled by tree-level diagrams involving 
the Higgs~\cite{'tHooft:1971rn,'tHooft:1972fi,
Cornwall:1973tb,Cornwall:1974km}.  As long as the
mass of the Higgs boson is less than $m_h \lesssim 1\,\tev$, 
$WW$ scattering remains weakly-coupled to up arbitrarily high energies.\bigskip 

In the coming years the Tevatron and the LHC will probe the
interesting range of masses for a Higgs boson within the Standard
Model.  If there is new physics beyond the Standard Model allowing the
Higgs to be heavier or not present at all, these colliders also stand
an excellent chance of discovering it.  Given the success of the
Standard Model with a Higgs boson, both in terms of precision
electroweak constraints and calculability of scattering amplitudes,
our primary focus is on scenarios with a fundamental 
Higgs boson throughout much of this review.

\subsection{Gauge hierarchy problem}
\label{sec:why_hierarchy}

The Higgs is a fundamental scalar field in the Standard Model, and
this fact leads to a severe theoretical puzzle.  Electroweak symmetry
breaking is induced by the Higgs at the energy scale $v \simeq
174\,\gev$.  In contrast, the characteristic energy scale of gravity
appears to be the much larger Planck mass, $M_{\rm Pl} \simeq
2.4\times 10^{18}\,\gev$.  This enormous difference of energy scales
is unstable against quantum corrections or loop correction, and it is
a mystery why the electroweak scale is not much closer to $M_{\rm
  Pl}$~\cite{Weinberg:1975gm,Gildener:1976ai,
'tHooft:1980xb}.
Attempts to solve this \emph{gauge hierarchy problem} require new
physics beyond the Standard Model near the electroweak scale.
Discovering this new physics is a major motivation for the LHC.

To see where the problem arises, suppose we try to extrapolate the
Standard Model up to much higher energies near the Planck scale,
assuming there is no new physics along the way.  Since the Standard
Model is renormalizable and weakly-coupled this is a sensible thing to
do, at least up to near the Planck scale, where we expect that
gravitational interactions will become relevant.  Approaching the
Planck scale, gravitational interactions might become relevant, and we
will have to expand our description of elementary particles to include
gravity. Unfortunately, we have no clear guiding principle to decide
how to write down an ultraviolet completion to gravity, as we will
discuss in Section~\ref{sec:models_add}.  But until then, it should be
clear sailing.\bigskip

\begin{figure}[t]
\begin{center}
  \includegraphics[width=0.35\textwidth]{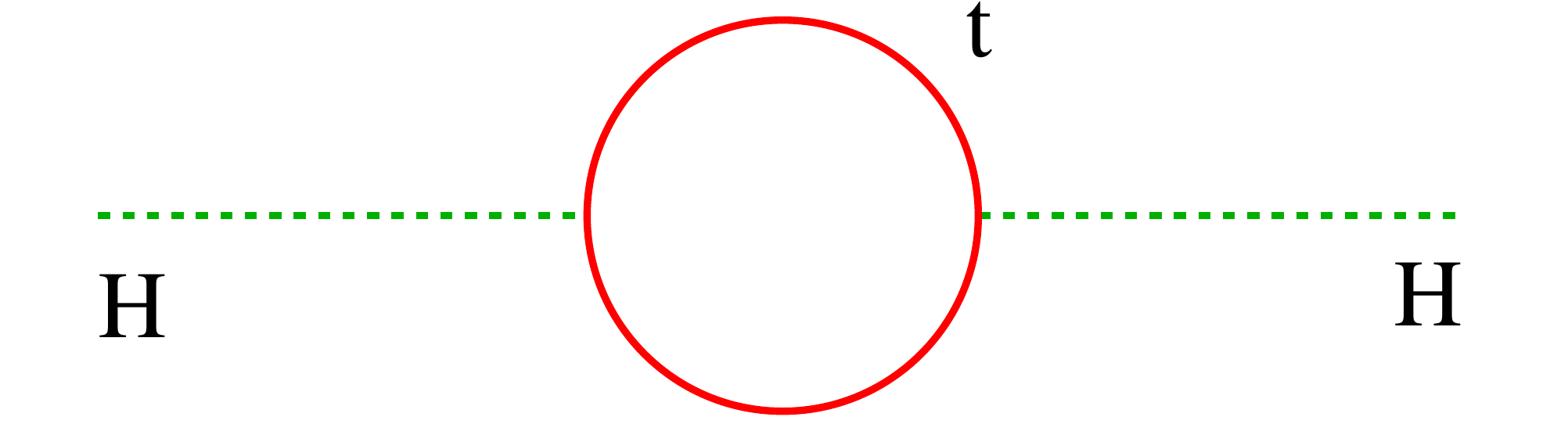}
\end{center}
\caption{Correction to $m_H^2$ in the Standard Model from the top quark. 
 Similar contributions arise from the weak gauge boson and from the Higgs
 boson itself.}
\label{fig:toploop}
\end{figure}

The problem with such an extrapolation is that quantum loop effects 
generate large corrections to the Higgs mass parameter $m_H^2$.  
The most dangerous correction comes from the top quark loop shown in 
Fig.~\ref{fig:toploop}, and is quadratically divergent in the 
ultraviolet~(UV).  Cutting it off at scale $\Lambda$, the leading
correction to the Higgs mass parameter is~\cite{Martin:1997ns} 
\begin{equation}
\Delta m_H^2 = -\frac{3y_t^2}{(4\pi)^2}\,\Lambda^2.
\label{quadratic}
\end{equation}
We see that for $\Lambda \gg v$, the correction to $m_H^2$ from the top 
loop is much larger than $v^2$.  Since the electroweak scale $v$ is 
determined by $m_H^2$ through Eq.~\eqref{hvev}, maintaining $v\ll \Lambda/4\pi$
would seem to require a severe fine-tuning of the underlying parameters
of the theory.

In writing Eq.~\eqref{quadratic} we have implicitly treated the
\emph{UV-regulated} loop diagram as a physical correction.  A reader
experienced in quantum field might object to this since upon
renormalizing the theory the dependence on $\Lambda$ disappears, and
the remainder is an acceptably small finite correction proportional to
the top quark mass.  This renormalization procedure, however, is
simply the act of parameterizing the unknown UV physics in terms of a
finite number of model parameters.  One candidate for the correct UV
theory (and a quantum theory of gravity) is superstring
theory~\cite{Polchinski:1998rq}.  Within string
theory, the would-be divergences are regulated by the appearance of
stringy states whose masses begin at a scale $M_S$ that is usually
close to $M_{\rm Pl}$.  From this point of view, we should take the UV
cutoff as a physical quantity with $\Lambda \sim M_S$, and the act of
renormalizing away the top quark quadratic divergence is completely
equivalent to a fine-tuning of the underlying parameters.  A similar
conclusion holds in other UV completions with a very large
characteristic energy scale.\bigskip

Independent of the details of renormalization and the correct UV
completion of the Standard Model, there are also large \emph{finite}
corrections to the Higgs mass from any heavy new particles that couple
to the Standard Model.  For example, a new fermion with mass $m_f$ and
Higgs Yukawa coupling $y_f$ produces a finite one-loop threshold
correction on the order of~\cite{Martin:1997ns}
\begin{equation}
\Delta m_H^2 \sim -\frac{y_f^2}{(4\pi)^2} \; m_f^2.
\end{equation}
Corrections of this form arise in a number of models of new physics
such as heavy right-handed neutrinos, or new states associated with gauge 
unification or quantum gravity.  Even if the new states do not couple 
directly to the Higgs, they can still contribute dangerously 
to $m_H^2$ at higher loop order.\bigskip

Solutions to the gauge hierarchy problem almost always involve new
particles with masses below $4\pi\,v \sim \tev$.  These solutions fall
into three general classes.  In the first class, new \emph{partner}
particles in the theory cancel off the dangerous corrections to
$m_H^2$ from quadratically divergent loops of Standard Model
particles.  Examples include supersymmetry~\cite{Martin:1997ns} where
loops of superpartners cancel against loops of Standard Model states,
and little Higgs theories~\cite{Perelstein:2005ka} in which new
same-spin partners cancel against their Standard Model
counterparts.
 
The second class of solutions involves lowering 
the fundamental ultraviolet cutoff of the theory to near the electroweak scale.
Examples in this class typically have additional spacetime dimensions
that effectively reduce the underlying scale of gravity
to near the electroweak scale~\cite{tasi_extrad}.  

In the third class, 
the Higgs boson is either removed altogether or emerges as a composite 
bound state of fermions.  Examples include technicolor and composite 
Higgs models~\cite{Hill:2002ap}.
The really exciting part about all these solutions to the hierarchy
problem is that they lead to new signals at the LHC.

\subsection{Dark matter}

\begin{figure}[t]
\begin{center}
\vspace*{-1.5cm}
\includegraphics[width=0.45\textwidth]{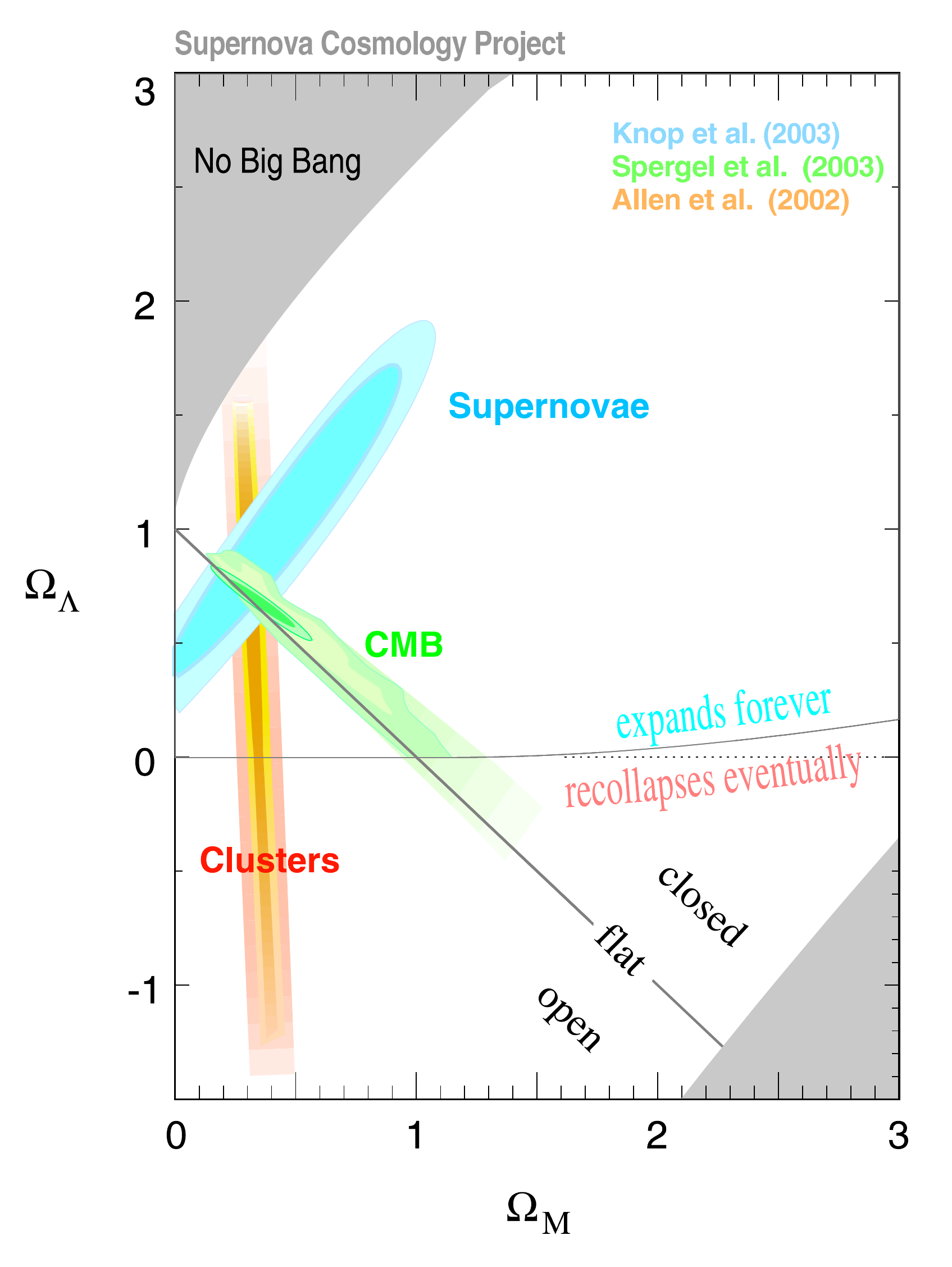}
\end{center}
\vspace*{-0.75cm}
\caption{Regions in the plane of the energy density of the Universe in the form
of dark matter ($x$-axis) and dark energy ($y$-axis) expressed as a fraction
of the critical density, which are consistent with observations of the
CMB, supernova, and structure formation.  From Ref.~\cite{Knop:2003iy}.}
\label{fig:dmdefit}
\end{figure}

  There is overwhelming evidence that most of the matter in our 
Universe is composed of a non-relativistic particle species which 
interacts only very feebly with the Standard 
Model~\cite{Jungman:1995df,Olive:2003iq,Bertone:2004pz,Feng:2003zu}.  
As can be seen in Figure~\ref{fig:dmdefit}, measurements
of the cosmic microwave background, supernova luminosities 
versus distance, and structure formation all favor a Universe with 
a large fraction of dark matter.  The Standard Model does not
contain a suitable candidate: its weakly-interacting particles
are either too light (photons and neutrinos) or have too short lifetimes
(as in the case of the Higgs or the $Z$) to be dark matter.
The existence of dark matter therefore represents concrete experimental
evidence for physics beyond the Standard Model.  

A new massive stable neutral particle can give rise to dark matter
through the mechanism of \emph{thermal freeze out}~\cite{Kolb:1990vq}.
At the very high temperatures expected in the early universe, such a
state will reach thermodynamic equilibrium provided it has even very
small interactions with the Standard Model.  As the universe cools
below the particle mass, it becomes much more probable for the stable
particles to annihilate than to be created, and the equilibrium
particle density starts to become exponentially small.  The particle
density eventually becomes so small that its annihilation effectively
stops.  The remaining \emph{relic} stable particle density
subsequently remains constant up to a dilution by the expansion of the
Universe, and can make up the required dark matter density.

The relic particle energy density due to thermal freeze out 
is given approximately by~\cite{Jungman:1995df}
\begin{equation}
\Omega_{DM}h^2 \equiv \frac{\rho_{DM}}{\rho_\text{tot}}h^2  
\simeq 
\frac{3\times 10^{-27}\text{cm}^3\text{s}^{-1}}{\langle \sigma v\rangle},
\label{eq:dmdens1}
\end{equation}
where $\langle \sigma v\rangle$ is the thermal average of the annihilation
cross section during freeze-out, and $h = H_0/100\,\text{km}\,\text{s}^{-1}\text{Mpc}^{-1} 
\simeq 0.72$ corresponds to the Hubble constant today.  
By comparison, the measured value is 
$\Omega_{DM}h^2 = 0.1131\pm 0.0034$~\cite{Komatsu:2008hk}.
In the case of a neutral particle
with a mass $M$ annihilating to the Standard Model through a dimensionless
coupling of strength $g$, the characteristic size of the annihilation 
cross-section is expected to be roughly
\begin{equation}
\langle \sigma v\rangle \sim \frac{g^4}{16\pi}\frac{1}{M^2}
\simeq (6\times 10^{-26}\text{cm}^3\,\text{s}^{-1})
\lrf{g}{0.5}^4\lrf{500\,\gev}{M}^2
\label{eq:dmdens2}
\end{equation}

From Eqs.~\eqref{eq:dmdens1} and \eqref{eq:dmdens2} we see that one 
obtains the correct DM density if the new stable particle has mass 
$M\sim 500\,\gev$ and a coupling to the Standard Model $g\sim 0.5$.  

  In light of the discussion in the previous subsections, this result
is highly suggestive.  Because the hierarchy problem is intrinsically
a question about the electroweak interaction, solutions to it naturally 
contain weakly-interacting massive particles (WIMPs) -- 
new particles with electroweak couplings ($g\sim 0.5$) and masses close 
to the electroweak scale ($v=174\,\gev$).  These two features
are precisely what is needed to generate the observed dark matter density,
provided one of the new states is stable.  This seeming coincidence
is often called the ``WIMP miracle'', and has guided most of the past 
theoretical work on thermal relics~\cite{Jungman:1995df,Olive:2003iq,
Bertone:2004pz,Feng:2003zu,cdm}

  While a thermal WIMP as dark matter makes for a compelling picture, 
it could also simply be a coincidence of scales~\cite{Feng:2008ya}. 
Our estimation of the WIMP relic density relies on extrapolating the 
thermal history of the Universe backwards from the time of nucleosynthesis 
to temperatures of order 10-100~GeV.  The history of the Universe could
well deviate from our extrapolation if there are unexpected light
degrees of freedom, or if there was an injection of entropy after
freeze-out from a phase transition or the decay of a heavy
particle no longer in thermodynamic equilibrium\cite{Kamionkowski:1990ni}.
Alternatively,
the production of WIMPs in the early Universe can arise in a natural
way from a number of \emph{non-thermal} mechanisms, such as the late
decay of a heavy state into dark matter~\cite{Moroi:1999zb}, 
or from a connection to a primordial charge asymmetry between particles 
and anti-particles~\cite{Kaplan:1991ah}
(such as the asymmetry between baryons and anti-baryons(.
There are also many plausible candidates for dark matter that aren't
WIMPs at all such as the Peccei-Quinn axion~\cite{Turner:1989vc}, 
or super-heavy relics~\cite{Chung:1998ua}.

  The subtle component in a theory of dark matter is explaining why 
the dark matter particle is massive, but 
(at least to a very good approximation) stable.  
In order for a state to have survived long enough to be the dark matter,
it must have a lifetime longer than the age of the Universe itself.  
This is strongly suggestive of a symmetry, either exact or broken only
at very high ($\gtrsim M_\text{GUT} \sim 10^{16}\,\gev$) scales.
In fact, such a symmetry may also explain other features (not) seem in high 
energy data.  The fact that no obvious deviations from the SM predictions 
are present in precision electroweak observables~\cite{Collaboration:2009jr} 
suggests that the scale of new physics should be larger than a few TeV, 
at odds with a natural solution to the hierarchy problem which should live 
at scales of a few hundred GeV.  A simple way to reconcile these two scales 
(the so-called the ``Little Hierarchy Problem'') is to 
require that the new physics only contribute to precision measurements 
at the loop level
by inventing a symmetry under which new states are odd and Standard Model 
states are even.  This symmetry may arise naturally from the structure 
of the theory, for example, as a remnant of higher-dimensional Lorentz 
invariance (KK-parity) as in models with 
Universal Extra Dimensions~\cite{Appelquist:2000nn} or as a technibaryon 
number in models of Technicolor~\cite{Nussinov:1985xr}.
Curing the little hierarchy problem by combatting large corrections to
precision measurements it acts similarly as $T$-parity in Little Higgs
Models~\cite{Cheng:2003ju}.  Like $R$-parity in supersymmetric models
or GUT models of warped extra dimensions~\cite{Agashe:2004ci} such a
symmetry can also prevent proton decay.  In generally, it is important
to remember that whatever the origin of such a protective symmetry is,
it may (if sufficiently exact) also serve to stabilize dark matter.\bigskip

  Given the strong potential for a deep connection between dark matter 
and electroweak symmetry breaking, we will focus much of our discussion 
in this review on models that address both in a unified framework.  
If the dark matter particle is related to electroweak symmetry breaking,
there is every reason to believe that it will be produced at the LHC.
The controlled initial state and relatively well-understood backgrounds 
in LHC collisions make this collider a powerful tool to extract information
about the nature of dark matter from a wide range of measurements, as 
discussed in Section~\ref{sec:sig_met}.
This stands in contrast to other programs
to probe dark matter directly in the laboratory or indirectly from 
astrophysics, which usually include only few distinct measurements and
are sensitive to uncertainties in dark matter 
distributions and astrophysical backgrounds.  
Even so, these techniques could inform us of the most promising search
channels for dark matter at the LHC.  In any case, LHC will provide 
an interesting and orthogonal take on dark matter, and has the potential 
to greatly enhance our understanding of the results of searches for 
dark matter in the heavens.

\subsection{Baryon asymmetry}
\label{sec:why_baryon}

  A second major cosmological shortcoming of the Standard Model
is that it is unable to explain the dominance of matter over antimatter. 
The only quantity in the Standard Model that distinguishes between matter and 
anti-matter is the CP-violating phase in the Cabbibo-Kobayashi-Maskawa~(CKM)
matrix, and it does not appear to be enough to explain the observed 
asymmetry.  On the other hand, many models of physics beyond the Standard Model 
contain new sources of CP violation, and in some cases these 
can be enough to account for the matter asymmetry.  The discovery of 
new physics at colliders and experiments searching for CP violation
may therefore provide clues to the cosmological matter mystery.
Reviews of baryogenesis can be found in Refs.~\cite{Riotto:1998bt}.
 
  Nearly all the non-exotic (\emph{i.e.} non-dark) matter in the universe
consists of baryons in the form of protons and neutrons.  The net baryonic
energy density extracted from measurements of the cosmic microwave
background radiation is~\cite{Komatsu:2008hk}
\begin{equation}
\rho_b/\rho_{tot} = 0.0456 \pm 0.0015 \; .
\end{equation}
This value agrees with the nucleosynthesis estimates for 
light element abundances~\cite{Olive:1999ij}.
The net density of antimatter in the universe is consistent with
pair production in collisions of matter accelerated by high-energy 
astrophysical processes.

  Baryogenesis is the process of creating an excess of baryons 
over anti-baryons in the early universe, starting from a Universe
with equal numbers of both.  Any mechanism of baryogenesis 
must satisfy three \emph{Sakharov
conditions}~\cite{Sakharov:1967dj}: 
\begin{enumerate}
\item violation of baryon number $B$;
\item violation of C and CP; 
\item departure from thermodynamic equilibrium.  
\end{enumerate}
The need for $B$-violation and a departure from thermodynamic
equilibrium are obvious, since we want to go from a state with ${B}=0$ to 
one with ${B}\neq 0$.  C and CP violation are also necessary because 
without them, a baryon-number violating process would produce just 
as many baryons as anti-baryons on the average.

  All three of these conditions are met by the Standard Model in the 
early universe.  Departure from thermodynamic equilibrium can occur due 
to the expansion of spacetime, and we know that C and CP violation are 
already present.  Baryon number violation also occurs in the Standard Model 
due to non-perturbative quantum effects.  Both baryon ($B$) and lepton ($L$) 
number are symmetries of the Standard Model at the classical level, 
but the combination $({B+L})$ is explicitly broken 
at the quantum level through nonperturbative $SU(2)_L$ interactions.
The only processes in the Standard Model that actually violate $({B+L})$
are tunnelling transitions in the $SU(2)_L$ gauge theory.  
The rate for these so-called \emph{instanton} transitions is proportional to 
$e^{-16\pi^2/g^2} \sim 10^{-{400}}$, and thus they are completely 
unobservable today~\cite{'tHooft:1976fv}.  At finite temperatures, 
however, it is possible for the system to pass over the tunnelling 
barrier by classical thermal fluctuations called 
\emph{sphaleron transitions}~\cite{Klinkhamer:1984di}.  
These transitions are active in the early universe
for temperatures below $T < 10^{12}\,\gev$~\cite{Moore:2000mx}, 
but shut off quickly once the electroweak symmetry is broken at 
temperatures near $T \sim 100\,\gev$~\cite{Khlebnikov:1988sr}.\bigskip

  Despite satisfying the necessary Sakharov conditions, there is no
known mechanism to generate the baryon asymmetry within the Standard Model.
The greatest obstacle is that the Standard Model does not have large enough
CP violation~\cite{Gavela:1993ts}.
However, extensions of the Standard Model 
proposed to solve the gauge hierarchy problem or to generate the 
dark matter frequently contain new sources of CP 
violation~\cite{Grossman:1997pa}. 
Within many of these theories, there exist viable mechanisms through 
which the baryon asymmetry can be created~\cite{Riotto:1998bt}.

  Two of the most popular baryogenesis mechanisms are 
leptogenesis~\cite{Fukugita:1986hr}
and electroweak baryogenesis~\cite{Kuzmin:1985mm}
In (standard) leptogenesis, an asymmetry 
in the lepton number $L$ is created by the out-of-equilibrium decays of 
a very heavy neutrino.  Such heavy neutrinos are well-motivated since
they can give rise to very small masses for the neutrinos in the 
Standard Model through
the \emph{seesaw} mechanism.  The asymmetry in $L$ created by heavy neutrino
decays is reprocessed by sphaleron transitions into a net baryon
number.  Unfortunately, while leptogenesis can generate the observed
matter asymmetry, all the new physics associated with it is frequently
much heavier than the electroweak scale making it inaccessible at
upcoming particle colliders.

  In contrast to leptogenesis, electroweak baryogenesis
makes testable predictions for the Tevatron and the 
LHC.
Baryon production in electroweak baryogenesis occurs during 
the \emph{electroweak phase transition}.  At temperatures well above 
the electroweak scale in the 
early universe the full $SU(2)_L\times U(1)_Y$ electroweak gauge symmetry 
is unbroken.  As the universe cools, this symmetry is broken spontaneously
down to the $U(1)_\text{em}$ gauge symmetry of electromagnetism.  
If this electroweak phase transition is strongly first order
it proceeds through the nucleation of bubbles of broken $U(1)_\text{em}$ 
phase within the surrounding plasma of symmetric $SU(2)_L\times U(1)_Y$ phase.
CP-violating interactions in the bubble walls generate a
chiral asymmetry which is reprocessed by the sphalerons into
a net baryon number density.  The Standard Model does not have a 
strongly first-order phase transition for Higgs masses above 
$\mathcal{O}(60~\gev)$~\cite{Buchmuller:1995sf},
nor does it have enough CP violation~\cite{Gavela:1993ts} for electroweak 
baryogenesis to be viable.
However, with new physics such as an extended Higgs sector or supersymmetry,
electroweak baryogenesis can generate the matter asymmetry.  
This requires new particles that can potentially be observed at the LHC, 
as well as new sources
of CP violation that might be found in upcoming searches for
permanent electric dipole moments~\cite{Murayama:2002xk}.

\newpage

\section{Models}
\label{sec:models}

  In view of the experimental and theoretical shortcomings of the
Standard Model, there has been a steady flow of models for
new TeV-scale physics over the last two decades. As we will see in this
section, not all of these models address all of the problems of the
Standard Model.  Some of them actually address none of them, and are
simply motivated by experimental limitations or field-theoretic structures. 
Others aim to connect physics at the electroweak scale to a more
fundamental description of Nature, such as a theory of quantum gravity. 

  Despite these differences, the vast majority of models of new
physics proposed to date focus on two primary issues.
First and foremost, they seek to explain electroweak symmetry breaking
and the associated puzzle of fundamental scalars.
Secondly, any model which claims to describe physics near the 
electroweak scale also has the potential to account for the
observed dark matter.  The former is comparably hard and can
involve many different approaches, from protecting a light fundamental
scalar with some symmetry, to denying the existence of fundamental high
mass scales, to generating the weak scale through dimensional
transmutation.  The latter can usually be solved by introducing some
kind of discrete $\mathbb{Z}_2$ symmetry into a model of electroweak 
symmetry breaking containing new stable particles.\bigskip

  In passing, let us mention that we do not describe extended
Higgs boson sectors in this section except as they relate to the
structure of the models we discuss.  There is an extensive literature
on the Higgs boson and its extensions, including many excellent
reviews~\cite{Gunion:1989we,Carena:2002es,tasi_higgs}.
Suffice it to say there are general ways to parameterize Higgs sectors 
involving nearly any variety of light Higgs resonance arising from different 
underlying models~\cite{silh}.  In addition, we know how to combine all 
LHC signals channels from a light Higgs boson and compare them to an effective
theory of the Higgs sector, including a general set of
couplings~\cite{Duhrssen:2004cv,Lafaye:2009vr}.

\subsection{Supersymmetry} 
\label{sec:models_susy}

  Supersymmetry~(SUSY) has long been a favorite candidate for
new physics beyond the Standard Model.
It naturally accounts for the large hierarchy between the Planck
and the electroweak scales, it provides an explanation for the underlying
cause of electroweak symmetry breaking, it strongly suggests the unification
of the $SU(3)_c\times SU(2)_L\times U(1)_Y$ gauge couplings into
a simple group at high energies,
and it can lead to viable candidates for the dark matter and the source 
of the baryon asymmetry.  At the same time, SUSY maintains the greatest 
strength of the Standard Model -- as a renormalizable perturbative 
field theory, it allows for accurate theoretical predictions.  

  As such, many different models of SUSY have been constructed,
and several of them have been used as benchmark theories by the
ATLAS and CMS collaborations~\cite{cms_tdr,atlas_tdr,atlas_csc}.
Even within the minimal supersymmetric extension of the Standard Model~(MSSM),
a broad palette of final state signatures are possible, allowing for a
thorough test of the capabilities of the LHC detectors for discovering
new physics at the $\tev$ scale.  In this section we will describe
the MSSM and some of its simple extensions, 
as well as some generic features of their phenomenology.
For more detailed phenomenological discussions of supersymmetry,
we refer the reader to the many reviews~\cite{Martin:1997ns,Nilles:1983ge,tasi_susy}
and textbooks~\cite{Wess:1992cp}
on the topic.
 
\subsubsection{MSSM}
\label{sec:models_mssm}

  A primary field-theoretic motivation for supersymmetry
is that it can stabilize the scale of electroweak symmetry breaking
against large quantum corrections from new physics at higher energies.
Supersymmetry is an extension of the Poincar\'e symmetries
of spacetime that relates bosons to fermions~\cite{Golfand:1971iw}.
In doing so, it extends
the generic chiral protection of fermion masses to the Higgs 
allowing for a fundamental scalar with a naturally small mass parameter.

  The most simple realization of supersymmetry that is also the most
straightforwardly consistent with the Standard Model
has a single anti-commuting chiral fermionic generator 
$Q$ such that~\cite{Wess:1992cp}
\begin{equation}
\{Q,Q\}=0,~~~\{Q,Q^{\dagger}\} = 2\sigma_{}^{\mu}P_{\mu},
\end{equation}
where $P_{\mu}$ is the usual momentum generator.  Acting on single-particle
states, 
\begin{equation}
Q\left|\text{boson}\right> \sim \left|\text{fermion}\right>,~~~
Q\left|\text{fermion}\right> \sim \left|\text{boson}\right>,
\end{equation}
we see that supersymmetry relates bosons and fermions.
Theories with a single chiral fermion generator $Q$ are usually 
referred to as $\mathcal{N}=1$, reflecting that there is only a one
irreducible generator.
If Nature is supersymmetric, particles must fill out complete
representations of the supersymmetry group, in the same way
that Poincar\'e invariance allows us to classify states
into particles with definite mass and spin (or helicity).  
Within each representation, all component particles must have
the same mass and quantum numbers, and the numbers of real fermionic 
and scalar degrees of freedom are equal to each 
other~\cite{Wess:1992cp,Salam:1974yz}.  

  For phenomenological applications, the most useful 
representations (in four spacetime dimensions) are the chiral 
and the massless vector supermultiplets~\cite{Martin:1997ns,Wess:1992cp}.
A chiral multiplet $\Phi$ contains a complex scalar $\phi$ and 
a two-component chiral Weyl fermion $\psi$~\cite{Wess:1992cp} 
\begin{equation}
\Phi = \left(\phi,\,\psi\right).
\end{equation}
Here, we have used \emph{superfield} notation, where $\Phi$ represents
all of its components at once~\cite{Wess:1992cp,Salam:1974yz}.  
Note that the chiral multiplet has an equal number of fermion and boson
real degrees of freedom: the 
complex scalar $\phi$ has 
two real components and the chiral fermion $\psi$ has two 
independent helicities.
The massless vector multiplet $V$ contains a vector field $A_{\mu}$ 
and a single chiral Weyl fermion $\lambda$~\cite{Wess:1992cp},
\begin{equation}
V = \left(\lambda,\,A_{\mu}\right).
\end{equation} 
As for the chiral supermultiplet, the numbers of bosonic and fermionic
degrees of freedom match within the vector multiplet: the massless vector 
$A_{\mu}$ on its mass shell has two independent physical polarization 
states while the Weyl fermion has two helicities. This Weyl or Majorana
nature of the fermions involved will have important phenomenological 
consequences which we will discuss later.\bigskip

  To extend the Standard Model to a supersymmetric theory, 
all the Standard Model states must be embedded in 
supermultiplets~\cite{Fayet:1977yc}.  
Each of the $SU(3)_c\times SU(2)_L\times U(1)_Y$ gauge fields is completed 
to form its own massless vector multiplet.  This implies the existence of 
a color octet Weyl fermion \emph{gluino} partner to the gluon, 
a $SU(2)_L$ triplet \emph{wino} partner to the $W^{\pm,3}$ gauge bosons, 
and a singlet \emph{bino} partner to the $U(1)_Y$ $B^0$ gauge boson. 
All these \emph{gaugino} states are listed in Table~\ref{tab:mssmvec}.  
The wino and bino states mix to form
electromagnetically charged and neutral fermions after
electroweak symmetry breaking and including degrees of freedom 
from the supersymmetric Higgs sector.

  The Standard Model fermions are placed in chiral multiplets.  For each generation,
the set of chiral multiplets is $Q$, $U^c$, $D^c$, $L$, $E^c$,
where we list the corresponding component states and quantum numbers 
in Table~\ref{tab:mssmchi}.  Each Standard Model fermion therefore has a 
complex scalar \emph{sfermion} superpartner.
The matched number of degrees of freedom -- two real scalar degrees of
freedom for each of the helicities of a chiral Standard Model fermion --
implies that one complex scalar superpartner field can be mapped onto
each of the Weyl spinors forming a Standard Model Dirac fermion. 
We will see below that after accounting for electroweak and
supersymmetry breaking, the sfermion partners of the left- and
right-handed components of a Standard Model fermion $f$ are not mass-degenerate,
and mix with each other to form mass eigenstates $\tilde{f}_{1,2}$.\bigskip

\begin{table}[t]
\centering
\begin{tabular}{|c|c|c|c|c|c|c|}
\hline
&$SU(3)_c\times SU(2)_L\times U(1)_Y$&Fermions&Bosons&${}$~$B$~${}$&${}$~$L$~${}$&${}$~$R$~${}$\\
\hline\hline
\phantom{i}$G^a$\phantom{\Huge{I}}&(8, 1, 0)&
$\tilde{g}^a$&$g^a_{\mu}$&0&0&-1\\
\hline
\phantom{i}$W^d$\phantom{\Huge{I}}&(1, 3, 0)&
$\tilde{W}^d$&$W^d_{\mu}$&0&0&-1\\
\hline
\phantom{i}$B$\phantom{\Huge{I}}&(1, 1, 0)&
$\tilde{B}^0$&$B_{\mu}$&0&0&-1\\
\hline
\end{tabular}
\caption{Vector supermultiplets in the MSSM.
\label{tab:mssmvec}}
\end{table}

  The Standard Model Higgs boson is also embedded in a pair of chiral multiplets.  
Two multiplets are needed for this, $H_u$ and $H_d$, for a couple of 
reasons.  First, each multiplet has a chiral fermion \emph{higgsino} 
component that would induce a gauge anomaly if not for having a partner 
with opposite charge.  Second, the structure of supersymmetric
interactions (that we will describe below) forces a chiral superfield
and its conjugate to have the same interactions up to complex conjugation.
As a result, we cannot use the Higgs field and its conjugate to give mass to up-type
and down-type fermions, as we do in the Standard Model. Instead,
two distinct Higgs doublets are needed to give masses 
to both the up-type and down-type fermions of the Standard Model 
after electroweak symmetry breaking in a supersymmetric theory. \bigskip

 Supersymmetry also constrains the possible interactions in the theory.
The general rule of thumb is that for each Standard Model vertex, supersymmetry
leads to the corresponding vertices where any two of the legs on the 
Standard Model vertex are transformed into superpartners~\cite{Martin:1997ns}.  
The two classes of interactions in the Standard Model are gauge and Yukawa.  
Incorporating them into SUSY leads to new testable predictions.

  Gauge interactions in SUSY are obtained by promoting
ordinary derivatives to gauge covariant derivatives, 
$\del_{\mu} \to D_{\mu} \equiv (\del_{\mu} + ig\,t^aA_{\mu}^a)$, where $t^a$
is the gauge generator for the representation of the target field
and $A^a$ is the gauge field.  This gauge field originates from
a massless vector supermultiplet which describes the on-shell
degrees of freedom of the component.  To formulate a quantum-mechanical
gauge theory, we should also include off-shell degrees of freedom,
and doing so in supersymmetry forces us to add a real scalar \emph{auxiliary}
$D^a$ as part of the vector multiplet~\cite{Wess:1992cp}.  
It is auxiliary in the sense that it does not have a kinetic term, 
and will disappear from physical observables.
Ensuring that the gauge couplings are supersymmetric also requires additional
couplings involving a matter fermion $\psi_i$, 
its scalar superpartner $\phi_i$, the gaugino $\lambda^a$, 
and the auxiliary $D^a$ field~\cite{Martin:1997ns,Wess:1992cp}:
\begin{equation}
-\mathscr{L}_\text{gauge} = 
\sqrt{2}g\, \left( \sum_i\phi_i^{\dagger}\,t^a\,\lambda^a\psi_i 
+ \text{h.c.} \right) 
- g\,\sum_i\phi_i^\dagger\,t^a\phi_i\,D^a - \frac{1}{2}D^a\!D^a.
\label{supergauge}
\end{equation}
The first term can be thought of as a supersymmetrization of the usual
fermion gauge coupling, with one fermion replaced by its scalar
superpartner and the gauge field replaced by its gaugino.  

  The second and third terms are the only Lagrangian couplings involving
the auxiliary $D^a$ field.  Since it does not have a kinetic term,
its equation of motion is algebraic, $D^a = -g\,\sum_i\phi_i^\dagger\,t^a\phi_i$,
and we can integrate it out of the theory.  Doing so leads to the
so-called $D$-term scalar potential~\cite{Martin:1997ns,Wess:1992cp} 
\begin{equation}
-\mathscr{L}_\text{gauge} \supset V_{D} =
\frac{1}{2}D^aD^a = 
\frac{g^2}{2}\left(\sum_i\phi_i^*t^a\phi_i\right)^2.
\label{dterms}
\end{equation}

 In the specific case of supersymmetric QCD, the three gauge couplings are 
the Standard Model vertex factor $g_{q q g}$ and its two SUSY-mirrored 
partners $g_{\tilde{q} \tilde{q} g}$ and $g_{\tilde{q} q \tilde{g}}$.  
All three of them have the same value to all orders in perturbation 
theory due to supersymmetry.  However, computing higher orders using 
dimensional regularization can break supersymmetry explicitly if it is 
not applied carefully: 
while the Weyl gluino has two degrees of freedom in any number of dimensions, 
the gluon has $d\!-\!2$ degrees of freedom in $d$ dimensions. 
This spurious SUSY-breaking effect can be compensated by a finite shift 
in $g_{\tilde{q} q \tilde{g}}$~\cite{martin_vaughn,
prospino}.
\bigskip

\begin{table}[t]
\centering
\begin{tabular}{|c|c|c|c|c|c|c|}
\hline
&$SU(3)_c\times SU(2)_L\times U(1)_Y$&Fermions&Bosons&$B$&$L$&$R$\\
\hline\hline
$Q= \left( \begin{array}{c} U_L \\ D_L \end{array} \right)$ 
&(3, 2, 1/6)&
$u_L, d_L$&$\tilde{u}_L, \tilde{d}_L$&1/3&\phantom{3}0\phantom{3}&-1\\
\hline
$U^{c}$&($\bar{3}$, 1, -2/3)&
${u}_R^{\dagger}$&$\tilde{u}_R^*$&-1/3&0&-1\\
\hline
$D^{c}$&($\bar{3}$, 1, 1/3)&
${d}_R^{\dagger}$&$\tilde{d}_R^*$&-1/3&0&-1\\
\hline
$L= \left( \begin{array}{c} \nu_L \\ e_L \end{array} \right)$ 
&(1, 2, -1/2)&
${\nu}_L, e_L$&$\tilde{\nu}_L, \tilde{e}_L$&0&1&-1\\
\hline
$E^c$&(1, 1, 1)&
$e_R^{\dagger}$&$\tilde{e}_R^*$&0&-1&-1\\
\hline\hline
$H_u= \left( \begin{array}{c} H_u^+ \\ H_u^0 \end{array} \right)$ 
&(1, 2, 1/2)&
$\tilde{h}_u^+,\tilde{h}_u^0$&$H_u^+, H_u^0$&0&0&+1\\
\hline
$H_d= \left( \begin{array}{c} H_d^0 \\ H_d^- \end{array} \right)$ 
&(1, 2, -1/2)&
$\tilde{h}_d^0,\tilde{h}_d^-$&$H_d^0, H_d^-$&0&0&+1\\
\hline
\end{tabular}
\caption{Chiral supermultiplets in the MSSM.
\label{tab:mssmchi}}
\end{table}
 
  All other renormalizable non-gauge interactions in a supersymmetric
theory can be derived from a \emph{superpotential} $W(\Phi_i)$ 
that is a function of the chiral supermultiplets $\Phi_i = (\phi_i,\psi_i)$ 
in the theory~\cite{Martin:1997ns,Wess:1992cp}.  An important property 
of the superpotential is that it is \emph{holomorphic}, in that
it depends only on the supermultiplet fields $\Phi_i$ and not on
their complex conjugates.  For example, the most general
possible cubic superpotential is given by
$W(\Phi) = f_i\Phi_i + m_{ij}\Phi_i\Phi_j + \lambda_{ijk}\Phi_i\Phi_j\Phi_k$.
The superpotential implies the Lagrangian 
terms~\cite{Martin:1997ns,Wess:1992cp}
\begin{equation}
-\mathscr{L}_W = 
\frac{1}{2} \, \left( \sum_{ij}
   \left.\frac{\partial^2 W}{\partial \Phi_i \partial \Phi_j}\right|_0\;
   \psi_i \psi_j + \text{h.c.} \right)
- \left(\sum_i\left.\frac{\del W}{\del\Phi_i}\right|_0F_i + \text{h.c.}\right) 
- \sum_iF_i^{\dagger}F_i.
\label{lagf}
\end{equation}
Here, the derivatives of $W$ are evaluated on the scalar components
of the chiral superfields, 
$\Phi_i\to \left.\Phi_i\right|_0 = \phi_i$.  The $F_i$ in Eq.~\eqref{lagf} 
are complex scalar auxiliary fields related to $\Phi_i$.  They are analogous to 
the $D^a$ fields in that they account for the off-shell degrees of freedom
of the chiral multiplets.  They also do not have kinetic terms,
so their algebraic equations of motion are simply 
$F_i^{\dagger} = -\left.\del W/\del{\Phi_i}\right|_0$.
Plugging these back into Eq.~\eqref{lagf} produces the scalar 
$F$-term potential~\cite{Martin:1997ns,Wess:1992cp},
\begin{equation}
-\mathscr{L}_W \supset V_F = 
\sum_i|F_i|^2 =
\sum_i\left|\frac{\del W}{\del\Phi_i}\right|_0^2.
\end{equation}  
The structure given in Eq.~\eqref{lagf}
links the scalar masses and the fermion masses to each other.
This ensures that, in the absence of supersymmetry breaking, 
the scalar masses will vanish in the chiral limit of the fermion
masses going to zero.\bigskip

  The superpotential of the MSSM is given by
\begin{equation}
W_\text{MSSM} = y_u\,Q\ccdot H_u U^c - y_d\,Q\ccdot H_d D^c 
- y_e\,L\ccdot H_dE^c
+ \mu\,H_u\ccdot H_d,
\label{wmssm}
\end{equation}
where the Yukawa couplings $y_u$, $y_d$ and $y_e$ are $3\times 3$ matrices
in flavor space, and $SU(2)_L$ doublet indices are connected anti-symmetrically
such that $A\ccdot B = a_1b_2-a_2b_1$ for $A = (a_1,a_2)^t$ 
and $B=(b_1,b_2)^t$.  The bilinear fermion couplings derived from this
superpotential reproduce the Yukawa interactions of the Standard Model 
along with
a Dirac mass term for the Higgsinos.  Note that as claimed above, 
the holomorphicity of the superpotential requires two Higgs doublets
to give mass to all the Standard Model fermions.  Scalar masses and self-couplings
are generated by the $F$-term potential.  Since the MSSM
superpotential has terms only up to cubic order, the Lagrangian derived
from it using the formula of Eq.~\eqref{lagf} is renormalizable.\bigskip

  Unlike the Standard Model Lagrangian, the MSSM superpotential of 
Eq.\eqref{wmssm} does not contain every possible renormalizable 
term that is consistent with the gauge symmetries of the theory.  
Several perfectly gauge-invariant superpotential terms such as 
$L\ccdot Q\,U^c$, $U^cD^cD^c$, and $L\ccdot H_u$ are not included in the MSSM 
because they violate baryon~($B$) or lepton ($L$) number conservation 
and would lead to severe problems like the rapid decay of the proton
or lepton flavor mixing.  
The standard way to justify the absence of these dangerous operators 
in the MSSM is to impose an additional discrete symmetry on the 
Lagrangian called \emph{R-parity}~\cite{Martin:1997ns,Farrar:1978xj}.
This is a $\mathbb{Z}_2$ symmetry under which each field has charge 
\begin{equation}
R = (-1)^{(3B-L) + 2s},
\end{equation}
where $s$ is the spin of the state.  Thus, the fermions of the
Standard Model as well as the gauge and Higgs bosons all have $R = +1$, 
while the sfermions, higgsinos, and gauginos have $R=-1$.
Note that the components within a given supermultiplet have
opposite $R$-parities.  A very important phenomenological 
implication of $R$-parity is that the $R = -1$ \emph{superpartner} 
particles can only be created or destroyed in pairs.  
As a result, the lightest superpartner, the LSP,
is absolutely stable if $R$-parity is an exact symmetry.  
If the LSP is also neutral under color and electromagnetism,
it can provide a promising candidate for the dark matter
in our universe~\cite{cdm,Jungman:1995df}.\bigskip

\begin{figure}[t]
\begin{center}
\vspace{1cm}
  \includegraphics[width=0.45\textwidth]{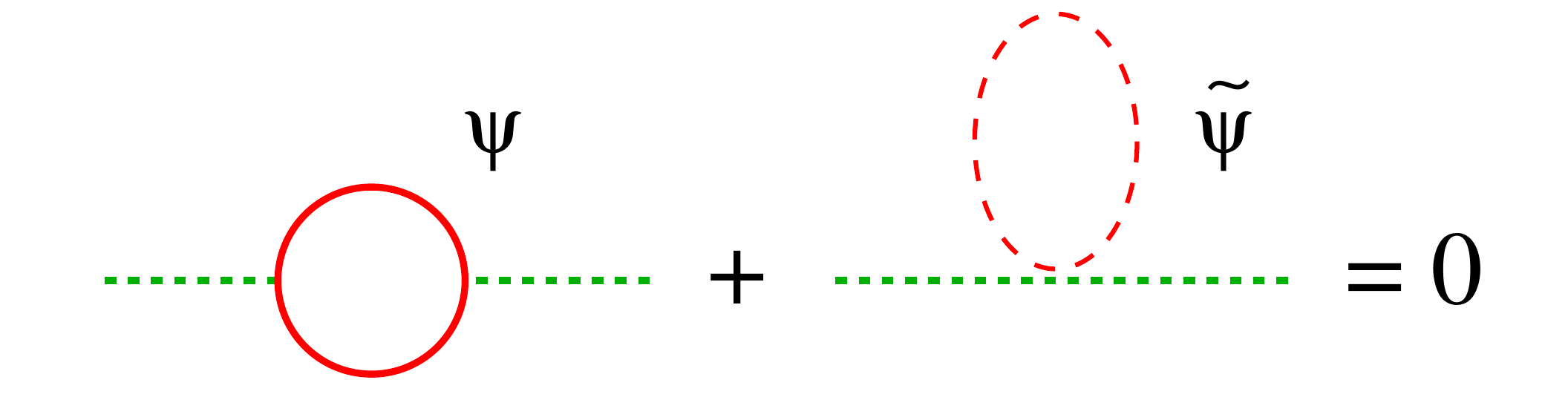}
\end{center}
\vspace*{0cm}
\caption{Cancellation of quadratic corrections in SUSY}
\label{fig:susymagic}
\end{figure}

  Having extended the Standard Model to include supersymmetry, 
let us now go back and examine our motivation for going to all 
this trouble in the first place.  As discussed in Section~\ref{sec:why}, 
the electroweak scale in the Standard Model is unstable under 
quantum corrections 
from new physics at high energies~\cite{Weinberg:1975gm,Gildener:1976ai}
The source of 
the instability are quadratic corrections to the mass term for the 
scalar Higgs field.  In supersymmetry, the Higgs is embedded into 
a pair of chiral supermultiplets, and supersymmetry relates its mass 
to that of the fermionic higgsino superpartners.  Since the masses of
fermions enjoy a chiral protection and are only logarithmically sensitive 
to new physics at higher energies, the corrections to the Higgs mass parameters
in supersymmetry can only be logarithmic as well.  These remaining
corrections are numerically small enough to not be dangerous.  In terms of
Feynman diagrams, the quadratic corrections to the Higgs are
found to cancel between loops of superpartners.  For example,
the quadratic divergence in the top quark loop correction to the
Higgs mass parameter is cancelled off by a loop containing
the scalar top superpartner, which we illustrate in Fig.~\ref{fig:susymagic}.
Beyond stabilizing the gauge hierarchy, supersymmetry with $R$-parity 
can also produce a viable dark matter candidate~\cite{Jungman:1995df}, 
and can lead to the grand unification of the 
$SU(3)_c\times SU(2)_L\times U(1)_Y$ gauge couplings
at very high energies near $M_\text{GUT} \simeq 10^{16}\,\gev$~\cite{
Dimopoulos:1981zb}.

  This great blessing of supersymmetry, that it relates the masses 
of bosons to fermions, is also its curse.  If supersymmetry were 
a manifest symmetry of Nature, each Standard Model fermion would have
a scalar superpartner 
with exactly the same mass and charge and we would have discovered 
them already~\cite{Martin:1997ns,Wess:1992cp,no_sbottom}.
Instead, supersymmetry must be broken.  As long as the breaking is 
\emph{soft}, with only non-supersymmetric interactions that have 
dimensionful couplings 
of size $m_\text{soft}$, the theory will be approximately supersymmetric 
at energies above $m_\text{soft}$~\cite{Girardello:1981wz}.  
In particular, the quadratic corrections to
the Higgs mass terms will be cut off near $m_\text{soft}$ and on
the order of $\Delta m_H \sim m_\text{soft}/4\pi$~\cite{Girardello:1981wz}.  
Therefore softly broken supersymmetry can stabilize the scale of 
electroweak symmetry breaking provided 
$m_\text{soft} \lesssim 4\pi\,m_W \sim 1\,\tev$.  
This is the main motivation to search for $\tev$-scale supersymmetry
in particle colliders.\bigskip

  The complete set of soft supersymmetry breaking terms 
in the MSSM is~\cite{Martin:1997ns}
\begin{alignat}{5}
-\mathscr{L}_\text{soft} 
&= m_{H_d}^2 |H_d|^2 +  m_{H_u}^2\,|H_u|^2
  -(B\mu\,H_u\ccdot H_d + \text{h.c.})\notag \\
&  + m_{Q}^2\,|\tilde{Q}|^2
   + m_{U^c}^2\,|\tilde{U}^{c}|^2
   + m_{D}^2\,|\tilde{D}^{c}|^2 
   + m_{L}^2\,|\tilde{L}|^2
   + m_{E}^2\,|\tilde{E}^{c}|^2 \notag \\
&  + \left(y_uA_{u}\,\tilde{Q}\ccdot H_u\,\tilde{U}^c \ccdot\tilde{Q}_i\,\tilde{U}^c_j 
   -  y_dA_d\,\tilde{Q}\ccdot H_d \tilde{U}^c
   - y_eA_e\,\tilde{L}\ccdot H_d\tilde{E^c} + \text{h.c.} \right) \notag \\
&  +\frac{1}{2}\,\left(M_3\,\tilde{g}^a\tilde{g}^a 
   + M_2\,\tilde{W}^d\tilde{W}^d
   + M_1\,\tilde{B}^0\tilde{B}^0 + \text{h.c.} \right) \; .
\label{softterms}
\end{alignat}
The terms in the first line of this equation contribute to 
the potential for the Higgs bosons.  The squared masses in the 
second line and the trilinear couplings on the third and line
generate masses and mixings for the sfermions. 
Couplings of the form $c_{ijk}\phi_i^*\phi_j\phi_k$ are also soft 
within the MSSM, but they tend to be tiny in most models 
of supersymmetry breaking~\cite{Girardello:1981wz}.
Note that we have suppressed flavor indices in these terms.
The final line contains mass terms for the gauginos.  These masses 
are Majorana, which are possible here since the gauginos transform
in the adjoint of their respective gauge group,
and a pair of adjoints can be contracted to form a singlet.

  The presence of Majorana gluino masses is an important reflection of
minimal supersymmetry.  Matching the degrees of freedom of the gluon, 
the gluino must have two on-shell degrees of freedom --- 
not enough to form a Dirac spinor. Two degrees of freedom,
on the other hand, are sufficient to construct a Majorana spinor out
of one Weyl spinor.
In this sense, the Majorana nature of the gluino is a
prediction of  minimal embedding of $SU(3)_c$ in
supersymmetry. 
The situation is slightly more subtle for the electroweak gauginos, 
because electroweak symmetry breaking 
marries at least the charged gauginos and higgsinos into Dirac states.
Combined with the soft supersymmetry-breaking masses for the winos and bino, 
the result is four Majorana neutralinos and two Dirac charginos, 
where the latter despite being Dirac reflect their Majorana origins through 
the presence of fermion number violating couplings to fermion/sfermion pairs.\bigskip

  Superpartner masses, or more precisely the mass splitting between 
Standard Model and superpartner masses, are determined 
by the soft terms in Eq.~\eqref{softterms}.  The precise values of these 
soft terms depend on the origin of supersymmetry breaking.  
This is a messy business~\cite{Witten:1981nf},
and nearly all viable theories for its origin have the underlying dynamics
in a separate sector that is \emph{hidden} from the MSSM 
fields~\cite{Martin:1997ns,Nilles:1983ge}.
Supersymmetry breaking is then communicated to the MSSM by 
\emph{messenger} interactions. 
In \emph{gravity mediation}, these interactions consist
of higher-dimensional operators suppressed by the Planck mass 
$M_\text{Pl}$~\cite{Ibanez:1981yh,Chamseddine:1982jx,Barbieri:1982eh,Ohta:1982wn,Hall:1983iz}.
If $\sqrt{F}$ characterizes the mass scale of supersymmetry breaking 
in the hidden sector, the MSSM soft terms are then on the order of
\begin{equation}
m_\text{soft} \sim \frac{F}{M_\text{Pl}},
\end{equation}
implying that $F \sim (10^{11}\,\gev)^2$ is needed for $m_\text{soft}\sim \tev$.  
In \emph{gauge mediation}, supersymmetry breaking is communicated
by a set of messenger fields with masses $M_\text{mess}$ that couple to
the source of supersymmetry breaking and are charged under
$SU(3)_c\times SU(2)_L\times U(1)_Y$~\cite{Dine:1981za,
Dine:1993yw,Giudice:1998bp}.  
They lead to MSSM soft terms
on the order of
\begin{equation}
m_\text{soft} \sim \frac{g^2}{(4\pi)^2}\frac{F}{M_\text{mess}},
\end{equation}
where $g$ represents a Standard Model gauge coupling.
Relative to gravity mediation, gauge mediation allows for a much
lower scale of supersymmetry breaking provided 
$M_\text{mess}\ll M_\text{Pl}/(4\pi)^2$.
Other popular forms of supersymmetry breaking include 
\emph{anomaly mediation}~\cite{Randall:1998uk,Giudice:1998xp,Pomarol:1999ie,Gherghetta:1999sw}
and \emph{gaugino mediation}~\cite{Kaplan:1999ac,Chacko:1999mi,Schmaltz:2000ei}.

  While supersymmetry breaking is needed to make the superpartners
heavy enough to account for why we have not discovered them yet,
it can also be a source of new phenomenological problems.
The scalar soft terms in Eq.~\eqref{softterms} can be a new source
of flavor mixing~\cite{Gabbiani:1996hi}.  For example, $m_Q^2|\tilde{Q}|^2$ 
represents $(m_Q^2)^{ij}\tilde{Q}^{\dagger}_i\tilde{Q}_j$, and 
$y_uA_u\,\tilde{Q}\ccdot H_u\,\tilde{U}^c$ stands for 
$(y_uA_u)^{ij}\,\tilde{Q}_i\ccdot H_u\,\tilde{U}^c_j$.
Both $(m_{\tilde{f}}^2)_{ij}$ and $(y_fA_{f})^{ij}$ can have off-diagonal
entries that mix the different generations.

  New flavor mixing beyond the CKM matrix of the Standard Model is strongly
constrained, and the soft terms must obey some kind of flavor symmetry 
similar to the Standard model;
for example, they could be approximately universal
and diagonal or aligned~\cite{Nir:1993mx}.  Moreover, the gaugino 
masses and the $A$- and $B$-type scalar soft terms in Eq.~\eqref{softterms} 
are generically complex, and could lead to unacceptably large
CP violating effects~\cite{Gabbiani:1996hi} such as
permanent electric dipole moments.  Consistency with observations 
typically requires such phases to be fairly 
small~\cite{with_toby}.
It is a challenge for models of supersymmetry breaking to generate
soft terms that respect these phenomenological constraints on flavor
and CP.  Gravity mediation often runs into difficulties on both fronts,
while gauge mediation produces acceptably flavor-universal soft terms
but can still induce too much CP violation~\cite{Giudice:1998bp}.\bigskip

  On top of determining the masses of the superpartners,
the soft supersymmetry breaking couplings are also responsible for
electroweak symmetry breaking within the MSSM.  The scalar
potential for the Higgs scalars at tree-level is given by
\begin{alignat}{5}
V_H = V_\text{soft}+V_F+V_D
=& \; m_{H_u}^2|H_u|^2+m_{H_d}^2|H_d|^2 
-(B\mu\,H_u\ccdot H_d + \text{h.c.}) 
\notag \\ 
&+ |\mu|^2(|H_u|^2+|H_d|^2)\label{higgspot} 
+ \frac{g^2+g'^2}{8}\left(|H_u|^2-|H_d|^2\right)^2.
\end{alignat}
Here, the first line comes from soft SUSY breaking terms, the second
line from the $F$-term and $D$-term potentials.  
This potential can induce a vacuum expectation value for the Higgs
fields if one or both of their scalar soft masses is negative at low energies.  

Indeed, this occurs very naturally in many models of supersymmetry 
breaking due to the effect of the large top quark Yukawa coupling on 
the renormalization group running of $m_{H_u}^2$ below the messenger mass 
scale~\cite{Ellis:1983bp}.
After being destabilized at
the origin by a negative soft mass, the potential will be stabilized
further out in the field space by the quartic terms in the potential,
provided $|H_u|\neq |H_d|$.  Note that the MSSM's quartic terms 
in Eq.~\eqref{higgspot} come entirely from the $D$-term contributions 
to the potential.  This is much more restrictive than the Standard Model where
the Higgs quartic coupling is an independent and potentially 
troublesome parameter in the theory.

  By making an electroweak gauge transformation, any Higgs vacuum expectation
can be rotated into the neutral components with no loss of generality.
A further change of field variables allows us to assume that $B\mu$ is
real and positive~\cite{Martin:1997ns}.  The minimum of the potential
then has real and positive expectation values for the neutral
components of $H_u$ and $H_d$:
\begin{equation}
\langle H_u^0\rangle = {v} \,\sin\beta, \qquad \qquad \qquad
\langle H_d^0\rangle = {v}\,\cos\beta,
\end{equation}
where $v = 174\,\gev$ and $\sin\beta,\,\cos\beta > 0$. 
After electroweak symmetry breaking, one neutral and one charged 
pseudoscalar combination of the Higgs multiplets are eaten by the
electroweak gauge bosons~\cite{Carena:2002es,tasi_higgs}.  
The remaining five physical scalar states are
two CP-even neutral states $h^0$ and $H^0$ with $m_{h^0}<m_{H^0}$, 
one CP-odd neutral state $A^0$, and one charged Higgs pair $H^{\pm}$.
We list these states in Table~\ref{tab:mssmstates}.

  Electroweak symmetry breaking leads to mixing among the gauginos
and Higgsinos.  The neutral components of these states mix to form
four Majorana \emph{neutralinos}.  In the basis 
$\tilde{\psi}_i^0 = (\tilde{B}^0,\tilde{W}^3, \tilde{H}_d^0,\tilde{H}_u^0)$, 
the neutralino mass matrix is given by~\cite{Martin:1997ns}
\begin{equation}
\mathcal{M}^{(0)} = \left(
\begin{array}{cccc}
M_1&0&-c_{\beta}s_W\,m_Z&s_{\beta}s_W\,m_Z\\
0&M_2&c_{\beta}c_W\,m_Z&-s_{\beta}c_W\,m_Z\\
-c_{\beta}s_W\,m_Z&c_{\beta}c_W\,m_Z&0&-\mu\\
s_{\beta}s_W\,m_Z&-s_{\beta}c_W\,m_Z&-\mu&0
\end{array}
\right) \; .
\label{eq:neutmass}
\end{equation}
This can be diagonalized by a unitary matrix $N$ such
that $N^*\mathcal{M}^{(0)}N^{\dagger}$, producing the neutralino
mass eigenstates $\tilde{\chi}_i^0 = N_{ij}\tilde{\psi}_i^0$, ordered so that 
$m_i \leq m_{i+1}$~\cite{Martin:1997ns,Dreiner:2008tw}.
Since the neutralino mass matrix contains Majorana gaugino soft masses,
all mass eigenstates will be Majorana fermions as well.
The off-diagonal terms in Eq.~\eqref{eq:neutmass} mixing gauginos
with Higgsinos originate from supersymmetrized gauge couplings of the form
given in the first term of Eq.~\eqref{supergauge}, where the scalar
field is a Higgs replaced by its expectation value. To measure these 
off-diagonal entries we cannot rely only on determining the four 
physical neutralino masses, but have to include an actual 
neutralino-Higgs coupling measurement~\cite{Kilian:2004uj}. 

  The neutral color octet gluino is also a Majorana fermion with 
tree-level mass $M_3$.  On account of its quantum numbers, it is unable
to mixing with the Higgsinos or anything else. 

  The charged components of the gauginos and Higgsinos mix to form
\emph{chargino} fermions.  Their mass matrix in the basis
$(\tilde{\psi_i}^+,\tilde{\psi_i}^-) = 
(\tilde{W}^+,\tilde{H}_u^+,\tilde{W}^-,\tilde{H}_d^-)$
according to
\begin{equation}
\mathscr{M}^{(\pm)} = \left(
\begin{array}{cc}
0&X^t\\
X&0
\end{array}
\right)
\qquad \qquad \text{where} \qquad
X =  \left(
\begin{array}{cc}
M_2&\sqrt{2}s_{\beta}\,m_W\\
\sqrt{2}c_{\beta}\,m_W&\mu
\end{array}
\right).
\label{eq:charmass}
\end{equation}
This matrix can be bi-diagonalized by a pair of unitary matrices
$U$ and $V$ according to 
$U^*\mathcal{M}^{(\pm)}V^{\dagger}$~\cite{Martin:1997ns,Dreiner:2008tw}.
The mass eigenstates are then given by 
$\tilde{\chi}_i^+ = V_{ij}\tilde{\psi}_j^+$ 
and $\tilde{\chi}_i^- = U_{ij}\tilde{\psi}_j^{-}$, and are ordered so that
$m_{\tilde{\chi}_1^{\pm}} \leq m_{{\tilde{\chi}_2}^{\pm}}$.  
In contrast to the neutralinos,
the charginos are Dirac fermions, as required for electrically charged states. 
As for the neutralinos, gaugino-Higgsino mixing among these charged states 
still originates from terms of the form of Eq.~\eqref{supergauge}.\bigskip

  Electroweak and supersymmetry breaking also causes the scalar 
superpartners of the left- and right-handed fermions to mix with each other.
The mass-squared matrix for the up-type sfermion $\tilde{f}$ 
corresponding to the up-type SM fermion $f$, is given in the basis 
$(\tilde{f}_L,\tilde{f}_R)$, by~\cite{Martin:1997ns}
\begin{equation}
\mathcal{M}^2_{\tilde{f}} = \left(
\begin{array}{cc}
m_{\tilde{f}_L}^2+ m_f^2+D_{f_L}&m_f(A_f^*-\mu/\tan\beta)\\
m_f(A_f-\mu^*/\tan\beta)&m_{\tilde{f}_R}^2+m_f^2+D_{f_R}
\end{array}
\right),
\label{upsferm}
\end{equation}
where $m_f$ is the Standard Model fermion mass and 
\begin{equation}
D_{f_i} = (t^3_{f_i} - Q_{f_i}s^2_W)\cos^2(2\beta)\,m_Z^2.
\end{equation}
As the notation suggests, this contribution to the sfermion masses
comes from the $D$-term potential of Eq.~\eqref{dterms}.
The mass eigenstates are labelled $\tilde{f}_1$ and $\tilde{f}_2$
with $m_{\tilde{f}_1} < m_{\tilde{f}_2}$.
For the down-type sfermions, including the sleptons, their mass-squared
matrices have the similar form
\begin{equation}
\mathcal{M}^2_{\tilde{f}} = \left(
\begin{array}{cc}
m_{\tilde{f}_L}^2+ m_f^2+D_{f_L}&m_f(A_f^*-\mu\tan\beta)\\
m_f(A_f-\mu^*\tan\beta)&m_{\tilde{f}_R}^2+m_f^2+D_{f_R}
\end{array}
\right).
\label{downsferm}
\end{equation}
Note that the left-right sfermion mixing for both types of sfermions
is proportional to the corresponding fermion mass $m_f$
(subject to our definition of the trilinear $A$ terms). 
Thus, as long as the trilinear soft terms $A_f$ are 
of similar size for all generations, this mixing will be the most pronounced in 
the third generation.  In many phenomenological analyses, the mixing among
the first two generation is neglected and the mass eigenstates are identified
with the chiral eigenstates.  We list the complete set of Higgs 
and superpartner mass eigenstates in the MSSM in 
Table~\ref{tab:mssmstates}.\bigskip

\begin{table}[t]
\centering
\begin{tabular}{|c|c|c|c|c|}
\hline
State&Symbol&$SU(3)_c$&$\phantom{3}Q_{em}\phantom{3}$&$\phantom{i3}R\phantom{i3}$\\
\hline\hline
\phantom{\Huge{i}}\!\!
CP-even Higgs &$h^0,H^0$&1&0&+1\\
CP-odd Higgs&$A^0$&1&0&+1\\
charged Higgs&$H^{\pm}$&1&$\pm1$&+1\\
\hline\hline
\phantom{\Huge{i}}\!\!
gluino&$\tilde{g}$&8&0&-1\\
neutralinos&$\tilde{\chi}_1^0$, $\tilde{\chi}_2^0$, $\tilde{\chi}_3^0$, $\tilde{\chi}_4^0$&1&0&-1\\
charginos&$\chi^{\pm}_1$, $\chi^{\pm}_2$&1&$\pm 1$&-1\\
\hline
\phantom{\Huge{i}}\!\!
up Squarks&$\tilde{u}_{L,R}$, $\tilde{c}_{L,R}$, $\tilde{t}_{1,2}$&3&2/3&-1\\
down Squarks&$\tilde{d}_{L,R}$, $\tilde{s}_{L,R}$, $\tilde{b}_{1,2}$&3&-1/3&-1\\
sleptons&$\tilde{e}_{L,R}$, $\tilde{\mu}_{L,R}$, $\tilde{\tau}_{1,2}$&
1&-1&-1\\
sneutrinos&$\tilde{\nu}_{e}$, $\tilde{\nu}_{\mu}$, $\tilde{\nu}_{\tau}$&1&0&-1\\
\hline
\phantom{\Huge{i}}\!\!
gravitino&$\tilde{G}$&1&0&-1\\
\hline
\end{tabular}
\caption{Higgs and superpartner mass eigenstates after electroweak 
symmetry breaking.  In writing these out we assume that left-right 
scalar mixing is only significant for the third generation.  
States are conventionally labelled in order of 
increasing mass: $m_i\leq m_{i+1}$.
\label{tab:mssmstates}}
\end{table}

  Beyond the Standard Model superpartners, the MSSM contains one additional
superpartner related to gravity.  Even though the MSSM is a model 
of particle physics at energies well below the Planck scale, 
if supersymmetry and supersymmetry breaking are present we expect that 
they will apply to the gravitational interactions as well.  
The extension of supersymmetry
to include gravity requires that this global symmetry is extended
to a local symmetry, unsurprisingly called \emph{supergravity}~\cite{
Freedman:1976xh}.
This is the supersymmetric analog of extending the global Poincar\'e symmetry 
of special relativity to the local coordinate invariance symmetry 
of general relativity.
In doing so, the spin-$2$ graviton field is embedded within a supermultiplet
containing a spin-$3/2$ \emph{gravitino} field $\tilde{G}$.  
The gravitino is massless when supersymmetry is exact, 
but it acquires a mass through 
the \emph{super-Higgs} mechanism by eating the spin-$1/2$ 
\emph{goldstino} fermion generated when supersymmetry 
is broken~\cite{Deser:1977uq,Fayet:1977vd}.  
The resulting gravitino mass is 
\begin{equation}
m_{3/2} = \frac{F}{\sqrt{3}M_\text{Pl}}.
\label{mgrav}
\end{equation}
Thus, the gravitino is about as heavy as the MSSM superpartners
in gravity mediation, but can be much lighter in gauge mediation.
Since the gravitino has odd $R$-parity, the lightest MSSM superpartner 
in scenarios with gauge mediation can be unstable against decaying
to its Standard Model counterpart and the gravitino.  This has important
phenomenological implications that we will address below.\bigskip

Collider signals of the MSSM depend very much on the superpartner mass
spectrum as well as the relative mass of the gravitino.  The case most
commonly studied is when the lightest superpartner particle~(LSP) is
one of the MSSM states, and not the gravitino.  This occurs in
theories where supersymmetry is broken at a high scale, such as
gravity mediation or anomaly mediation.  Cosmological bounds together
with searches for exotic stable charged particles typically require
that the LSP be neutral under color and electromagnetism if it is to
make up the dark matter~\cite{DeRujula:1989fe}.
This leaves the neutralinos and the sneutrinos as possible 
weakly interacting candidates.
Of these two, the lightest neutralino makes for a much better dark
matter candidate in the presence of experimental limits from direct
detection experiments~\cite{Falk:1994es}, and most investigations of
MSSM collider signals assume that it is the LSP as we will do here.

To study supersymmetric signatures and develop strategies for LHC
searches between theory and experiment it is useful to have simple
benchmark models~\cite{Allanach:2002nj}.  Historically, a standard
framework including a neutralino LSP was \emph{minimal supergravity}
(mSUGRA), also called the \emph{constrained MSSM}~(cMSSM).  Here, a
set of universal soft terms are assumed at a very high input scale on
the order of $M_\text{GUT} = 2\times 10^{16}\,\gev$.  Specifically,
all scalar soft masses are set to a common flavor-diagonal value
$m_0$, all trilinear $A$ terms are set to $A_0$, and all gaugino soft
masses are taken to be $m_{1/2}$ at the scale $M_\text{GUT}$.  The
low-energy value of $\tan\beta$ and the sign of the superpotential
$\mu$ term are specified at the weak scale, in addition to assuming
that electroweak symmetry is broken.  With these five inputs the
low-energy Higgs and superpartner spectrum is fully determined after
extrapolating the soft terms down to near the electroweak scale by
renormalization group running.  Let us emphasize, however, that while
the mSUGRA framework is very useful for defining benchmark models, it
has very little theoretical motivation and should not be taken too
seriously. In particular, we should not mistake experimental
constraints on such simplified benchmark models for constraints on
TeV-scale supersymmetry~\cite{with_dan,how_light_1,Berger:2008cq}. 
For example, the assumption of gaugino mass unification places a
very strong constraint on the spectrum that the bino
will be roughly half as heavy as the wino.  This limits the bino mass
since the wino leads to a chargino whose mass limit is roughly half
the LEP2 energy.
Instead, we can derive the lower bound on the LSP mass either
including the measured relic density (yielding values in the ~GeV
range)~\cite{how_light_1}, or without this requirement in which case
the lightest neutralino can be practically
massless~\cite{how_light_2}.

  With a neutralino LSP and a superpartner mass spectrum that is not
too spread out,  the generic picture of MSSM phenomenology at a hadron
collider is as follows.  Most of the superpartner production
is in the form of squarks and gluinos on account of their QCD charges.
They must be created in pairs due to $R$-parity.
Once produced, each colored superpartner decays promptly to a 
lighter superpartner and one or more Standard Model states.  
These decays continue 
until the LSP is created, at which point the decay cascade ends.
Being stable and uncharged, the LSP leaves the detector carrying
some of the energy and momentum of the event with it.
This process gives rise to events with energetic jets,
possibly a few leptons, and missing transverse energy. 
Such signatures and their LHC-specific issues will be discussed in detail in 
Section~\ref{sec:sig_met}, while the extraction of underlying parameters 
including detailed experimental simulations are presented 
in Section~\ref{sec:para}.\bigskip

  The collider phenomenology of the MSSM can be very different
when a light gravitino is the LSP.  This can occur when supersymmetry
breaking is transferred to the MSSM by messengers that are much lighter
than the Planck scale, as occurs in gauge mediation.  
The massive spin-$3/2$ gravitino $\tilde{G}$ 
has \emph{transverse} components coming from the massless 
spin-$3/2$ superpartner of the graviton as well as 
\emph{longitudinal} components from the massless 
spin-$1/2$ goldstino~\cite{Fayet:1977vd}.  At energies much larger than 
the gravitino mass, the goldstino components couple to the MSSM fields 
with strength $1/F$, which is much stronger than the transverse 
coupling proportional to $1/M_\text{Pl}^2$~\cite{Fayet:1977vd,Fayet:1979qi,
Fayet:1979yb}. This feature is sometimes called the
\emph{goldstino equivalence principle}, in analogy with the
similar situation for Higgsed gauge bosons.
The gravitino-matter coupling connects a particle $X$ with its
superpartner $\tilde{X}$, and therefore allows the superpartner to
decay to the gravitino $\tilde{G}$ at the rate~\cite{Fayet:1979qi}
\begin{equation}
\Gamma(\tilde{X}\to \tilde{G}X) = \frac{1}{16\pi}\frac{m_{\tilde{X}}^5}{F^2}\,
\left(1-\frac{m_X^2}{m_{\tilde{X}^2}}\right)^4
= \frac{1}{48\pi}\frac{m_{\tilde{X}}^5}{m_{3/2}^2M_\text{Pl}^2}\,
\left(1-\frac{m_X^2}{m_{\tilde{X}^2}}\right)^4.
\label{nlspwidth}
\end{equation}
For low supersymmetry breaking scales $\sqrt{F} \lesssim 10^6\,\gev$, 
or equivalently $m_{3/2} \lesssim 100\,\kev$, these superpartner decays
can be fast enough to be seen in particle colliders.

  When the gravitino is the LSP, the typical collider picture that emerges
is that superpartners are produced in pairs, and decay in a 
cascade down to the next-to-lightest superpartner~(NLSP) in much the 
same way as when the LSP is a Standard Model 
superpartner~\cite{Dimopoulos:1996vz}.  
At the LHC we generally expect that the QCD-charged superpartners
will be produced most often, so these cascades will generally be
accompanied by hard jets.
If the NLSP is long-lived on collider time scales, it will simply leave 
the detector.  However, since it will later decay to the neutral 
gravitino LSP, it can carry an electromagnetic or color
charge and still be consistent with cosmological constraints~\cite{
Jedamzik:2004er}.
A charged NLSP can leave highly-ionizing charged tracks in 
the detector, while a colored NLSP will hadronize 
and may also produce visible tracks~\cite{Nisati:1997gb,Allanach:2001sd,Fairbairn:2006gg}.  
When the NLSP decays to its superpartner and a gravitino within the detector, 
there will be an additional energetic particle and a reduced amount 
of missing energy carried by the gravitino.

  The collider signatures of supersymmetry with a gravitino LSP depend 
strongly on the lifetime and the identity of the NLSP.  The NLSP lifetime 
is the inverse of the decay width given in Eq.~\eqref{nlspwidth}. 
Details of supersymmetry breaking determine the NLSP identity.  
In minimal models of gauge mediation, the NLSP is either a mostly-bino 
neutralino or a mostly right-handed stau 
(scalar tau)~\cite{Dimopoulos:1996vz,Ambrosanio:1996zr}
More general models of gauge mediation~\cite{Cheung:2007es}
can lead to other
varieties of NLSPs such as a mostly-wino or higgsino neutralino~\cite{
Matchev:1999ft,
Meade:2009qv},
a chargino~\cite{Kribs:2008hq}, a sneutrino~\cite{Katz:2009qx},
or a squark~\cite{DiazCruz:2007fc} or 
gluino~\cite{Raby:1998xr,Rajaraman:2009ga}.

\begin{figure}[t]
\begin{center}
  \includegraphics[width=0.5\textwidth]{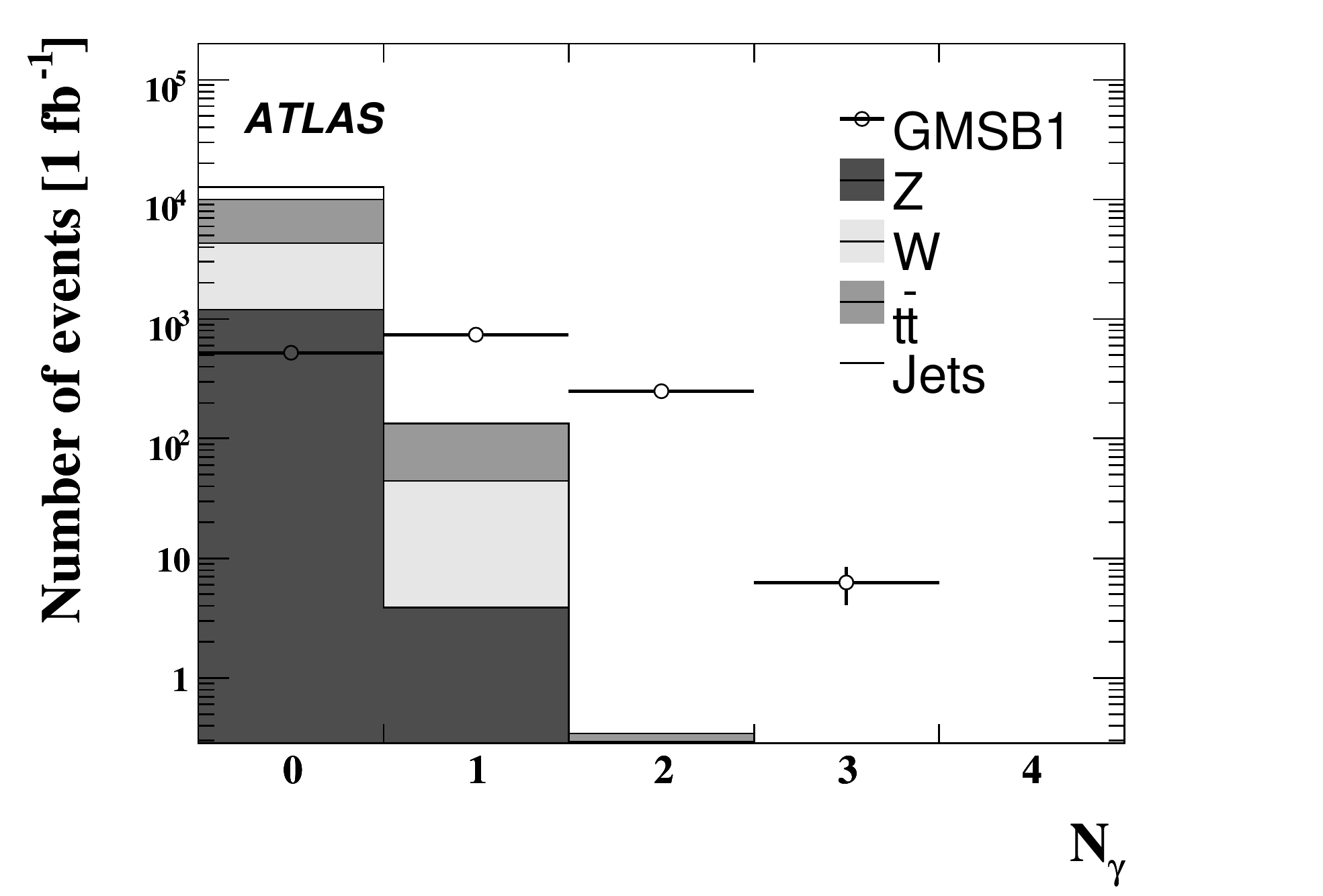}
\end{center}
\vspace*{0cm}
\caption{Estimated number of GMSB signal events at the LHC containing
photons with $p_T^{\gamma}> 20\,\gev$ and $|\eta|<2.5$ from a GMSB
scenario with a mostly bino NLSP that decays promptly to the gravitino.
Also shown are the estimated Standard Model backgrounds.  There is a clear
excess above background in the multi-photon channels.  
The diagram is from Ref.~\cite{atlas_csc}.}
\label{fig:lhcgmsb}
\end{figure}

  A mostly-bino neutralino decays primarily through 
$\tilde{\chi}_1^0\to\gamma\tilde{G}$ with a subleading decay 
$\tilde{\chi}_1^0 \to Z^0\tilde{G}$.
For prompt decays of a bino NLSP, the resulting pair of photons
in each supersymmetric event provides a handle to efficiently reduce the
Standard Model backgrounds~\cite{Dimopoulos:1996vz,Ambrosanio:1996zr,
Hinchliffe:1998ys}.  
We illustrate this feature in Fig.~\ref{fig:lhcgmsb}, which shows
the results from a study by the ATLAS collaboration of a GMSB scenario
containing a mostly-bino NLSP that decays promptly.  There is a clear
excess of multi-photon signal events over the estimated background.
The kinematic distribution of the outgoing photons can also be
used to extract the mass of the neutralino NLSP~\cite{Shirai:2009kn}.
In the intermediate regime where the the neutralino decay is delayed
but still occurs inside the detector, the resulting photons need 
not point back to the interaction vertex.  This makes it harder to
identify the photons, but may also allow for a determination of
the NLSP lifetime from the amount of vertical 
displacement~\cite{Kawagoe:2003jv,atlas_csc}.
Very slow decays of a bino NLSP simply lead to missing energy
with no visible photons.\bigskip
 
  When a mostly right-handed stau $\tilde{\tau}_1$ is the NLSP
it will decay through $\tilde{\tau}_1 \to \tau\tilde{G}$, and a new 
set of signals emerges~\cite{Dimopoulos:1996vz,Ambrosanio:1996zr,
Dicus:1997yp}.
Prompt decays imply that most supersymmetric
events will be accompanied by a pair of taus and missing energy.
Moreover, for lower values of $\tan\beta$ and small $A$-terms
there is very little left-right mixing, and the stau NLSP is nearly 
degenerate with the right-handed selectron $\tilde{e}_R$ and the 
right-handed smuon $\tilde{\mu}_R$.  In this case the very slightly 
heavier selectrons and smuons can behave as co-NLSPs, 
decaying primarily to their respective superpartners and the gravitino 
instead of first going to the stau.  This leads to muons or electrons 
and missing energy in some of the supersymmetric cascades, 
in addition to taus.
Longer-lived $\tilde{\tau}_1$ NLSPs can produce displaced vertices 
from their decays as well as charged tracks.
When the $\tilde{\tau}_1$ is stable on collider scales, it will 
produce a highly-ionizing charged track that extends all the way out to 
the muon chamber~\cite{Allanach:2001sd,Fairbairn:2006gg}.
Signatures of such long-lived charged exotics will be discussed
in Section~\ref{sec:sig_stable}.
\bigskip 

  Signatures of other NLSP candidates have not been studied in as 
much detail.  A mostly-wino neutralino NLSP can decay through
$\tilde{\chi}_1^0\to \gamma\tilde{G}$ and $\tilde{\chi}_1^0\to Z^0\tilde{G}$, 
while a mostly higgsino neutralino NLSP can also decay significantly via 
$\tilde{\chi}_1^0\to h^0\tilde{G}$~\cite{Meade:2009qv}.  These neutralino 
varieties can also have a chargino co-NLSP that decays through 
$\tilde{\chi}_1^+\to W^+\tilde{G}$~\cite{Feng:1999fu}.
Promptly decaying wino or higgsino NLSPs are typically more difficult 
to distinguish from Standard Model backgrounds than a mostly-bino 
state decaying 
to a photon~\cite{Meade:2009qv}.  Chargino~\cite{Kribs:2008hq} 
or sneutrino~\cite{Katz:2009qx} NLSPs can also lead to interesting
signatures.  A squark or gluino NLSP would produce additional hard jets
in every supersymmetric event if it decays quickly.
\bigskip

  In certain regions of the MSSM parameter space there can arise 
squarks and gluinos that are quasi-stable on collider scales.  
This includes long-lived squark and gluino NLSPs in gauge mediation. 
A further possibility is \emph{split supersymmetry}, which is simply
the MSSM in the limit where all scalar superpartners are pushed to be 
much heavier than the superpartner 
fermions~\cite{Wells:2003tf,Giudice:2004tc}.
The low energy
spectrum then consists of the Standard Model with a single light Higgs boson
together with the gluino, chargino, and neutralino superpartners.  
While the MSSM in this limit no longer stabilizes the electroweak scale, 
it still preserves the nice properties of gauge unification and a dark matter
candidate~\cite{Pierce:2004mk}.  Due to the decoupled scalars, 
split supersymmetry generically avoids constraints from flavor mixing 
and CP violation.

  The only new accessible colored particle in split supersymmetry 
is the gluino.  To decay, it has to go through a heavy squark, 
either off-shell in a three body $\tilde{g}\to q\bar{q}'\chi^{0,\pm}$ mode, 
or through a loop in the radiative $\tilde{g} \to g\chi^0$ 
channel~\cite{Toharia:2005gm}.
These channels lead to a gluino lifetime on the order of
\begin{equation}
\tau \sim (10\,s)\lrf{\tilde{m}}{10^9\,\gev}^4\lrf{1000\,\gev}{m_{\tilde{g}}}^5,
\end{equation}
where $\tilde{m}$ is the typical mass of the squarks. 
Cosmological bounds on late decays of a gluino limit 
${\tilde{m}} \lesssim 10^9\,\gev$ for $m_{\tilde{g}} > 300\,\gev$,
assuming an otherwise standard cosmological history.
Under these circumstances, the gluino will hadronize long before it
decays by combining with two or three quarks or a gluon to
form a heavy color-singlet 
$R$-hadron~\cite{Farrar:1978xj,Baer:1998pg,Kilian:2004uj,Fairbairn:2006gg}.  
Stable squarks will also form $R$-hadrons with similar collider properties.
We will discuss some of the novel LHC signals of $R$-hadrons
in more detail in Section~\ref{sec:sig_stable}.\bigskip

  The very heavy scalars in split supersymmetry make it more difficult
to establish the presence of supersymmetry using collider measurements.  
The existence these scalars can only be deduced indirectly through their
effects on mediating three-body (or loop-level two-body) gluino decays.
To gain information about these effects it is crucial to measure the
gluino production and decay rates very carefully, taking into account 
QCD effects described in Section~\ref{sec:sig_qcd}~\cite{Turlay:2008dm}.
A second important test of the underlying supersymmetric structure
is to measure the off-diagonal terms within the neutralino and 
chargino mass matrices given in Eq.~(\ref{eq:neutmass}) and 
Eq.~(\ref{eq:charmass}).  These couplings are fixed by the 
supersymmetrization of gauge interactions described before 
Eq.~(\ref{supergauge}), with a Higgs field coupling to a higgsino and a gaugino.
When supersymmetry is broken, however, these couplings evolve differently
under the renormalization group below the scale of the sfermion masses
than the couplings of the gauge bosons to fermions.   
The size of these effects are in the range of 
10-20\%~\cite{Giudice:2004tc,Arvanitaki:2005fa},
but it is unclear whether the LHC will be able to measure this
even in the most optimistic scenario.

\subsubsection{Extended supersymmetric models}
\label{sec:models_xssm} 

  The minimal supersymmetric extension of the Standard Model
is able to stabilize the gauge hierarchy and can give rise to
a viable dark matter candidate and gauge 
unification~\cite{Martin:1997ns}.
While these results are impressive, the model also leads to some new puzzles,
and the most pressing of them have to do with the minimal structure of the 
MSSM Higgs sector.  As a result, many of the proposed modifications of 
the MSSM extend the minimal two-Higgs doublet format.  The simplest
possibility is to add a singlet field $S$, but other well-motivated 
alternatives include extended gauge structures and additional 
higher-dimensional Higgs operators.  Going beyond the Higgs sector
there is a wide range of additional interesting possibilities,
such as a fourth generation of quark and lepton supermultiplets, 
scenarios with an exact $U(1)_R$ global symmetry, and extensions 
containing a new hidden sector.
In this section we discuss some of these proposed MSSM extensions, 
their motivation and the structure of their collider signals.\bigskip

  Before discussing MSSM extensions, let us first examine where the
MSSM itself runs into difficulty.  There are two main puzzles,
both related the MSSM Higgs sector: 
the \emph{fine-tuning} problem, and the $\mu$-problem.
The fine-tuning problem arises from the fact that the structure of 
the MSSM Higgs sector makes it difficult for all the Higgs states to 
avoid the direct and indirect constraints on them.
At tree-level, the mass of the lighter CP-even Higgs boson is bounded
by~\cite{Martin:1997ns,Carena:2002es,tasi_higgs}: 
\begin{equation} m_{h}^2
\leq m_{Z}^2\cos^22\beta < m_Z^2 \; .
\label{mssmhiggs}
\end{equation}
The maximum value is reached in the decoupling limit, where all heavy
Higgs states are governed by a high mass scale $m_{A^0} \sim m_{H^0}
\sim m_{H^\pm} \gg m_Z$. In this limit all heavy Higgs states decouple
from the theory, and the $h^0$ state behaves exactly like the Standard Model Higgs
boson whose mass is bounded by the LEP experiments $m_{h_\text{SM}} >
114.4$~GeV at 95\%~CL~\cite{Schael:2006cr}. Note that this decoupling
limit of the supersymmetric Higgs sector generally holds, but it
requires an appropriate treatment of $\mu$ and the sparticle masses,
to avoid artificially introducing non-decoupling effects~\cite{delta_b}.

  The LEP limit does not rule out the MSSM since loop corrections increase
the physical Higgs mass above the tree-level value.  
The leading contribution generally comes from the scalar tops, 
and in the absence of left-right stop mixing (and in the decoupling limit)
is given by~\cite{mh_feyn,mh_eff,Martin:1997ns,Carena:2002es,tasi_higgs}
\begin{equation}
\Delta m_h^2 = 
\frac{12}{(4\pi)^2}\sin^2\!\beta\,y_t^2m_t^2\,
\log \frac{m_{\tilde{t}_L}m_{\tilde{t}_R}}{m_t^2} 
\end{equation}
This mass shift can be large due to the top Yukawa
coupling and an enhancement by the large logarithm of the
stop masses relative to the top mass.  For heavy stops
$m_{\tilde{t}_{L,R}} \gtrsim 1\,\tev$ and relatively large values of
$\tan\beta \gtrsim 10$, this is enough to push the physical $h^0$ mass
above the LEP limit.

  Unfortunately heavy stops also imply a certain degree of fine-tuning in the model.
One of the conditions for electroweak symmetry breaking in the MSSM 
is~\cite{Martin:1997ns,Carena:2002es,tasi_higgs}
\begin{equation}
m_Z^2 = -2\,|\mu|^2 - \frac{2\,m_{H_u}^2\tan^2\beta - 2\,m_{H_d}^2}
{\tan^2\beta-1}
\simeq -2\,|\mu|^2 -2\,m_{H_u}^2,
\label{mssmmin}
\end{equation} 
where the second approximate equality holds for $\tan\beta\gg 1$.
The large top Yukawa does not only have a beneficial effect on the 
light Higgs mass. 
When the MSSM stops are heavy, the value of $m_{H_u}^2$ is also driven
to large negative values in the course of renormalization 
group~(RG) running~\cite{Martin:1997ns}
\begin{equation}
\Delta m_{H_u}^2 \simeq -\frac{12\,y_t^2}{(4\pi)^2}\,m_{\tilde{t}}^2\,
\log \frac{\Lambda}{m_{\tilde{t}}},
\end{equation}
where $m_{\tilde{t}}$ represents a common stop mass and $\Lambda$ is a
large energy scale, possibly motivated by unification, where we start
the RG running.  If $|m_{H_u}^2|$ is much larger than $m_{Z}^2$, a
very precise cancellation between $|\mu|^2$ and $m_{H_u}^2$ is then
required in Eq.~\eqref{mssmmin}.  Since the values of $|\mu|^2$ and
$m_{H_u}^2$ need not be related to each other in any particular way,
this represents a \emph{fine-tuning} problem of the MSSM.  It can be
slightly alleviated splitting the two stop masses and introducing a
large stop mixing parameter ($A_t - \mu/\tan \beta$)~\cite{mh_eff,feynhiggs}.
\bigskip

  The MSSM fine-tuning problem leads in directly to the MSSM 
\emph{$\mu$-problem}.
To induce electroweak symmetry breaking at the observed scale,
the supersymmetric $\mu$-term must cancel against $m_{H_u}^2$ which
arises only from supersymmetry breaking.  However, from the point of view 
of the low-energy effective theory, there is no reason why the value of $\mu$ 
should even be near the weak scale, instead of having a much larger 
value on the order of $M_\text{GUT}$ or $M_\text{Pl}$.  
This puzzle is called the $\mu$-problem.  
The only good news is that since $\mu$ is supersymmetric, once it
is set to a small value it will stay there after including 
quantum corrections.

  In some cases it is possible to connect the $\mu$-term to
supersymmetry breaking, such as through the Giudice-Masiero
mechanism~\cite{Giudice:1988yz}.  This mechanism involves imposing
additional symmetries on the theory at high energy that forbid a bare
$\mu$-term, but allow a coupling between the source of supersymmetry
breaking and the operator $H_u\ccdot H_d$.  An effective $\mu$-term is
then generated when supersymmetry is broken.  However, in many cases
(usually when $m_\text{soft} \nsim m_{3/2}$), this mechanism also
leads to a $B\mu$ soft term that is much too large.  The value of
$\mu$ induced in this way also need not have any special relationship
to $m_{H_u}^2$, so it does not fix the fine-tuning problem in complete
generality. On the other hand, for gravity mediation, \ie one-scale
SUSY breaking without any knowledge about Standard Model charges, the
Giudice-Masiero mechanism works fine.\bigskip

  Another way to simultaneously address both the fine-tuning problem 
and the $\mu$-problem is to add a gauge singlet chiral superfield $S$ to the 
MSSM~\cite{Ellwanger:2009dp}. 
With a singlet, the MSSM $\mu$-term superpotential operator  
can be replaced by 
\begin{equation}
W \supset \lambda\,S\,H_u\ccdot H_d.
\label{lambdamu}
\end{equation}
This leads to an effective $\mu$-term when the scalar
component of $S$ develops a vacuum expectation value
$\mu_\text{eff} = \lambda\,\langle S\rangle$.
Such a VEV is induced naturally by supersymmetry breaking, thereby linking
the value of the effective $\mu$-term to the soft breaking mass scale
and giving a simple solution to the MSSM $\mu$-problem.  Even better, 
there is a new superpotential contribution to the
tree-level CP-even Higgs mass, and the lower bound of Eq.~\eqref{mssmhiggs}
is increased to~\cite{Ellwanger:2009dp} 
\begin{equation}
m_h^2 \geq M_Z^2\cos^22\beta + \lambda^2\,v^2\sin^22\beta
\end{equation}
where $v \simeq 174\,\gev$.
For $\tan\beta \sim 2$ and $\lambda \sim 1$,
this easily pushes the tree-level Higgs mass above 
the LEP limit of $114.4\,\gev$ without any need for large stop 
masses and their consequent fine-tuning.

  The most popular singlet extension 
is the next-to-MSSM~(NMSSM)~\cite{Ellwanger:2009dp,nmssm,Maniatis:2009re},
with the Higgs-sector superpotential and the related soft terms
\begin{alignat}{5}
W_\text{NMSSM} &\supset \lambda\,S\,H_u\ccdot H_d + \frac{\kappa}{3}S^3 
\notag \\
-\mathscr{L}_\text{soft} &\supset m_S^2|S|^2 
+ (\lambda\,A_{\lambda}\,S\,H_u\ccdot H_d 
- \frac{\kappa}{3}A_{\kappa}S^3 + \text{h.c.}).
\label{wnmssm}
\end{alignat}
This theory has a global $\mathbb{Z}_3$ symmetry, 
under which $S$ and all the MSSM chiral superfields have the same charge.
It forbids the usual bare $\mu$-term as well as any possible dimensionful
couplings involving the singlet $S$ in the supersymmetric Lagrangian. 
However, if this $\mathbb{Z}_3$ is exact it will lead to dangerous 
domain walls in the early universe~\cite{Abel:1995wk}.
These can be avoided with a very small explicit breaking of the symmetry
by a singlet tadpole operator~\cite{n2mssm}.
The cubic $S^3$ interaction in Eq.~\eqref{wnmssm} is needed to
stabilize the $S$ field at a non-zero VEV.
With one additional singlet superfield, the Higgs boson sector of the
NMSSM has one extra CP-even Higgs and one extra CP-odd Higgs compared
to the MSSM, while the \emph{singlino} fermion component of the singlet  
gives rise to a fifth neutralino state.

  Collider signals of the NMSSM depend primarily on the 
amount of mixing between the MSSM states and the singlets~\cite{nmssm}. 
They become identical to the usual MSSM signals in the limit 
of $\lambda \to 0$ and $\langle S\rangle \to \infty$ 
with $\mu_\text{eff} = \lambda \langle S\rangle$ held fixed.  
In this limit the singlet states decouple completely from the rest 
of the theory.  On the other hand, when $\lambda$ is of order unity 
and $\mu_\text{eff}$ is not too large, there can be significant 
mixing between the singlet and charged states.\bigskip

   An especially interesting example of this mixing arises when the 
singlet sector has an approximate global symmetry that produces 
a light pseudoscalar.  
Two promising candidates are a $U(1)_R$ symmetry when 
$A_{\lambda},\,A_{\kappa}\to 0$~\cite{Dobrescu:2000yn}, 
and a $U(1)_\text{PQ}$ symmetry for 
$\kappa,\,A_{\kappa}\to 0$~\cite{Miller:2003ay}.
If either of these holds approximately, they will be broken spontaneously 
when $S$, $H_u$, and $H_d$ develop VEVs giving rise to an approximate 
Nambu-Goldstone boson~\cite{Dobrescu:2000yn,Miller:2003ay}.
This state takes the form of a light, mostly singlet 
CP-odd Higgs $a^0$ in the spectrum,
and can easily be lighter than half the mass of the lightest 
SM-like CP-even Higgs $h^0$. This way
the Higgs decay mode $h^0 \to a^0a^0$ can become dominant, 
beating out the usual $h^0\to b\bar{b}$ channel 
for $m_h \lesssim 130\,\gev$~\cite{Dermisek:2005ar,Dermisek:2005gg}.
The light pseudoscalar $a^0$ inherits couplings to the Standard Model from its mixing with
the heavier non-singlet CP-odd Higgs, and decays primarily through
$a^0\to b\bar{b}$ for $m_{a} \gtrsim 10\,\gev$ and $a^0\to \tau\bar{\tau}$
for $3\,\gev \lesssim m_{a} \lesssim 10\,\gev$. Generically, this leads
to four-body Higgs decays producing  4$b$, 4$\tau$, or 2$b$2$\tau$ 
and even lighter final states.  Even more unusual decay modes can 
arise going beyond the NMSSM~\cite{Chang:2005ht}.

  These non-standard decays of the otherwise SM-like $h^0$ Higgs 
can significantly weaken the LEP limits on this 
state~\cite{Dermisek:2005ar,Dermisek:2005gg,Chang:2005ht,Chang:2008cw}.  
When the $h^0 \to a^0a^0$ decay mode is dominant, the LEP bound can be 
as low as $m_h \gtrsim 107\,\gev$ for $a^0\to b\bar{b}$ primarily, 
and $m_h\gtrsim 86\,\gev$ when $a^0\to \tau\bar{\tau}$ 
is dominant~\cite{Abbiendi:2002in}. 
The weaker bounds reflect that such decays are 
generically harder to extract due to the higher multiplicity and softer
energy spectrum of the decay products, but they are also due in part
to the fact that the original analyses were optimized for other signatures.
A good example is the 4$b$ Higgs final state that actually appears in 
the usual $h^0 \to b\bar{b}$ search, 
just with lower efficiency~\cite{Chang:2008cw}. 
For the 4$\tau$ channel a preliminary reanalysis of ALEPH data 
suggests that the actual LEP bound might be closer to 
$m_{h^0} \gtrsim 105\,\gev$~\cite{kyle-aleph}.   This scenario is
also constrained by $B$-physics~\cite{Dermisek:2006py}. 

  For all such exotic Higgs decays there is a model-independent 
LEP limit of $m_h\gtrsim 82\,\gev$ that relies only on the recoiling 
$Z$ boson in $e^+ e^- \to Zh$, and therefore
applies no matter how the $h^0$ decays~\cite{Abbiendi:2002qp}.
Lowering the LEP bound on the $h^0$ Higgs to this value can reduce the 
fine-tuning in these models to some extent~\cite{Dermisek:2005ar,
Dermisek:2005gg,Schuster:2005py,Dermisek:2006wr}. 
Most importantly, Higgs searches at the LHC are much less sensitive
to soft decay products, which means that non-standard multi-body
decay modes can seriously complicate the LHC Higgs searches if too many 
of the standard discovery channels are 
suppressed~\cite{Chang:2008cw,Forshaw:2007ra}.
A promising new channel is the $4b$ or $4\tau$ final state with 
the $h^0$ Higgs produced in association with a 
$W$ or $Z$~\cite{Carena:2007jk}.  
When the $4\tau$ mode from $a^0\to \tau\bar{\tau}$ is dominant,
the $h^0$ Higgs can also be discovered at the Tevatron or the LHC
through the subleading but clean $h^0\to a^0a^0\to 2\tau 2\mu$
decay channel~\cite{Lisanti:2009uy}.\bigskip

  The NMSSM is only one of many singlet extensions of the MSSM.
In the nMSSM (not-quite-MSSM or MNSSM), the cubic coupling is 
replaced by a singlet tadpole~\cite{n2mssm},
\begin{equation}
W_\text{nMSSM}\supset \lambda S\,H_u\ccdot H_d + t_F\,S \; ,
\end{equation}
along with corresponding soft terms.  This superpotential, with
$t_F \sim m_\text{soft}^2$, can be motivated in supergravity with 
discrete $R$-symmetries~\cite{n2mssm,Panagiotakopoulos:2000wp}.
It also arises as the low-energy effective superpotential in 
\emph{Fat-Higgs} scenarios where the singlet and Higgs fields
arise as composite operators~\cite{Harnik:2003rs,Chang:2004db,Delgado:2005fq}.
The nMSSM is similar to the NMSSM in its phenomenology,
but differs in that it predicts a light mostly-singlet neutralino
state whose mass is bounded above by~\cite{Menon:2004wv,Hesselbach:2007te}
\begin{equation}
m_{\tilde{\chi}_1^0}^2 \lesssim \frac{\lambda^2\,v^2
(m_Z^2c_{2\beta}^2+\lambda^2v^2s_{2\beta}^2)}{|\mu_\text{eff}|^2 + \lambda^2v^2(1+s_{2\beta}^2)+m_Z^2c_{2\beta}^2}
\end{equation}
For larger $\lambda$, the SM-like $h^0$ Higgs can decay predominantly
into pairs of these neutralinos producing invisible Higgs 
decays~\cite{Menon:2004wv,Cao:2009ad}.

Besides the nMSSM, there are many possible multi-singlet 
MSSM extensions~\cite{Barger:2006dh}.
In nearly all such extensions, the effect of the coupling of 
Eq.~\eqref{lambdamu} on raising the $h^0$ Higgs mass is strongly limited 
if we require perturbative gauge unification.  To maintain $\lambda(Q)$ 
perturbatively small up to the unification scale, its low-energy value 
must lie below $\lambda(m_W) \lesssim 0.7$~\cite{Kane:1992kq}.
In the NMSSM, the corresponding limit is roughly 
$\sqrt{\lambda^2+\kappa^2} \lesssim 0.7$~\cite{Ellwanger:2009dp}.  
The nMSSM thus allows a maximal value of $\lambda$ since effectively
$\kappa \to 0$.  Fat Higgs models allow for even larger values by interpreting
the Landau pole in $\lambda$ as the compositeness scale.
This bound can also be relaxed by extending the gauge structure of
the theory to include new non-Abelian gauge 
symmetries that modify the running of 
$\lambda$~\cite{Batra:2003nj},
allowing the SM-like Higgs boson to be heavier than 200\,$\gev$
without the need for heavy stops.

  The new singlet $S$ introduced in the NMSSM and its brethren
can be problematic unless it is charged under some symmetry: global or
local, discrete or continuous~\cite{Bagger:1993ji}.
Instead of the approximate $\mathbb{Z}_3$
of the NMSSM, an attractive alternative is to introduce a new $U(1)_x$
gauge symmetry under which $S$ is charged~\cite{Barger:2006dh}.  
To forbid a bare $\mu$ term and at the same time allow for the  
VEV-based superpotential term 
of Eq.~\eqref{lambdamu}, the Standard Model fields must also be charged under $U(1)_x$.  
Having done this, there is no need for other superpotential terms 
involving $S$ since the $U(1)_x$ $D$-term potential generates 
a stabilizing quartic $|S|^4$ coupling:
\begin{equation}
V_{D_x} = \frac{g_x^2}{2}\left(x_S|S|^2 + x_{H_u}|H_u|^2+x_{H_d}|H_d|^2 
+ \ldots\right)^2,
\end{equation}
where $x_i$ are $U(1)_x$ charges.
This $D$-term potential also increases the theoretical tree-level 
upper bound on the $h^0$  mass to
\begin{equation}
m_h^2 \leq M_Z^2\cos^22\beta + \lambda^2\,v^2\sin^22\beta
+ 2\,g_x^2\,v^2\left(x_{H_U}\cos^2\beta+x_{H_d}\sin^2\beta\right)^2
\end{equation}
As in the NMSSM, the Higgs sector of these theories contains
an additional CP-even and CP-odd Higgs relative to the MSSM,
as well as at least one new neutralino derived from the fermion part of $S$.
Such new states can modify the Higgs and supersymmetry signals at the LHC.
In addition, there can be new supersymmetric signals from the 
massive $Z_x$ gauge boson~\cite{Baumgart:2006pa}.

  For these $U(1)_x$ extensions to be self-consistent and phenomenologically
acceptable, the $U(1)_x$ charges $x_i$ of $S$ and the MSSM fields must 
typically satisfy anomaly cancellation conditions, they should be
the same for all three generations to avoid flavor violation, 
and they should allow the usual MSSM superpotential interactions.
(However, see Refs.~\cite{Langacker:2000ju}
for specific counterexamples.) 
These requirements, together with the assumption that $x_S \neq 0$
can be shown to imply that the theory must also contain new exotic
color-charged states~\cite{Erler:2000wu}.
Such states can be vector-like in their Standard Model charges but must be chiral 
under $U(1)_x$ to avoid re-introducing a $\mu$-problem.  
For example, these states could include 
$D^c = (\bar{3},1,1/3,x_{D^c})$ and 
$D = (3,1,-1/3,x_D)$ under $SU(3)_c\times SU(2)_L\times U(1)_Y\times U(1)_x$.
If $(x_S+x_D+x_{D^c}) = 0$, the exotics states can get a mass
from the superpotential coupling
\begin{equation}
W \supset \xi_D\,S\,D\,D^c \; .
\end{equation}
Such states are frequently long-lived, often decaying to the MSSM 
only through higher-dimensional operators, and can produce massive 
charged tracks at the LHC~\cite{Kang:2007ib}.

The case of $U(1)_x$ is a representative example of $D$-term contributions to the Higgs
mass.  More complicated structures involving non-Abelian gauge groups are also
possible, and can result in much heavier Higgs bosons while remaining consistent with
perturbative unification provided the new gauge interactions are
asymptotically free~\cite{Batra:2003nj,Maloney:2004rc}.
\bigskip

  The specific extensions of the MSSM Higgs sector 
described above are only a small subset of a vast range of possibilities.  
If the new states modifying the Higgs sector are somewhat heavier than
their MSSM counterparts, they can can be integrated out leaving behind 
an effective theory consisting of the minimal MSSM Higgs sector augmented by 
higher-dimensional operators.  
The strategy of Refs.~\cite{Dine:2007xi}
is to study 
the effects of such operators without specifying where they come from.
For example, the superpotential operator
\begin{equation}
W \supset \int d^2\theta\,\frac{\zeta}{M}(H_u\ccdot H_d)^2
\label{dst-op}
\end{equation}
modifies the Higgs potential according to~\cite{Dine:2007xi} 
\begin{equation}
\Delta V = \frac{2 \zeta\mu^*}{M}H_u\ccdot H_d\,
\left(H_u^{\dagger}H_u + H_d^{\dagger}H_d\right) + \text{h.c.}
\end{equation}
where $\mu$ is the usual MSSM $\mu$-term.
With supersymmetry breaking, there can arise an additional related
correction to the Higgs potential of the form~\cite{Dine:2007xi}
\begin{equation}
\Delta V = \frac{\lambda\,m_\text{soft}}{M}(H_u\ccdot H_d)^2 + \text{h.c.}
\end{equation}
where $m_\text{soft}$ is on the order of the soft SUSY breaking terms.
These corrections can significantly increase the mass of the lightest
SM-like Higgs boson provided the ratio of $H_u$ and $H_d$ VEVs $\tan\beta$ 
is not too large and $M/\zeta$ is less than a few TeV.  
Again, this eliminates the need for heavy stops and their associated fine-tuning
to push up the Higgs mass.  The operator of Eq.~\eqref{dst-op} also
alters the Higgs-higgsino couplings
and can modify the branching fractions of the heavy Higgses $H^0$, $A^0$,
and $H^\pm$.  It can also be a source of CP 
violation~\cite{Blum:2008ym}, as well as induce a new electroweak
symmetry breaking minimum that persists in the supersymmetric limit
$m_\text{soft}\to 0$~\cite{Batra:2008rc}.\bigskip

  Extensions of the MSSM Higgs sector can also contain new
states that are not singlets under the Standard Model.
In general, we can add any number of doublets and singlets 
or higher representations to the Higgs sector provided
they do not generate gauge anomalies or ruin the prediction
for electroweak observables.  While such models do not necessarily
have a strong theory motivation, they can receive experimental
motivation from the LEP2 results in the four-jet channel or simply 
to challenge the LHC analysis setup such as the so-called
\emph{stealth models}~\cite{stealth_jochum,stealth_jack}
and beyond~\cite{Schabinger:2005ei}.
These models can contain multiple
Higgs resonances or even a near continuum.

  One strategy to search for such objects is to 
choose the LHC Higgs discovery channel best suited for a Higgs mass
reconstruction and simply try resolve the resonant structures.
For example, regular weak-boson-fusion production with a 
collinear $m_{\tau \tau}$ reconstruction~\cite{boosted_taus} 
can resolve two Higgs resonances separated by 
$\mathcal{O}(15~\gev)$~\cite{wbf_h_susy}.  Much better resolution
is available from the $h\to ZZ \to 4\ell$ provided it has a
reasonable rate.
An orthogonal approach starts from a Higgs channel which only very
approximately reconstructs the Higgs resonance, for example as a wide
$WW$ transverse mass~\cite{Dittmar:1996ss,wbf_h_ww}, as described in
Section~\ref{sec:sig_met}. As a production process we can either
choose gluon fusion or weak boson fusion, but depending on the models
considered weak boson fusion might be more stable with respect to
model variations, it has a wider reach in Higgs masses, and it has a
generically better signal-to-background ratio of order one. Changing
the shape of a Higgs `resonance' from a single mass peak to many mass
peaks or even a continuum will now affect the shape of the transverse
$WW$ mass reconstruction~\cite{Alves:2002tj} and allows us to extract
the structure of the Higgs sector by an appropriate numerical fit to
the $m_{T,WW}$ shape.\bigskip

  Besides modifying the Higgs sector of the MSSM, there are many
other well-motivated extensions of other parts of the theory.  
A simple example that also has important implications for the Higgs sector 
is the inclusion of a (supersymmetric) fourth generation of quarks 
and leptons.  This possibility can be consistent with precision electroweak
bounds provided there is a moderate mass splitting between the new
up- and down-type quarks and leptons~\cite{Kribs:2007nz}.  
Current Tevatron bounds imply that the new quarks must be heavier 
than about $300\,\gev$, forcing their Yukawa couplings to the Higgs 
field to be considerably larger than unity~\cite{teva-tprime}.  
This helps to push up the mass of the lightest CP-even Higgs boson, 
allowing $m_{h^0} > 114.4\,\gev$ with superpartner masses near $300\,\gev$ 
and fixing the Higgs fine-tuning problem of the MSSM~\cite{Fok:2008yg,Litsey:2009rp}.  
A fourth supersymmetric generation will give rise to highly visible signals 
at the LHC, although the details depend on the mass 
spectrum~\cite{Fok:2008yg}.  The fourth-generation
quarks will also induce significant changes to the production and decay
properties of the $h^0$ Higgs boson~\cite{Kribs:2007nz,Fok:2008yg}.

  Despite these attractive features, the large fourth-generation Yukawa 
couplings lead to a breakdown of perturbation theory near $10\,\tev$ 
unless new states enter below that scale to moderate their 
running~\cite{Fok:2008yg,Murdock:2008rx}.  
A simple alternative is to augment the MSSM with new supersymmetric 
vector-like quarks and leptons that have direct masses as well as 
Yukawa couplings to the Higgses~\cite{Martin:2009bg}.
Such extensions can also repair the Higgs fine-tuning problem of the MSSM
while still allowing perturbative unification.\bigskip

  A more radical modification of the MSSM is to elevate $R$-parity
to a continuous global $U(1)_R$ symmetry~\cite{Hall:1990hq,Kribs:2007ac}.  
Such a symmetry arises naturally when supersymmetry is unbroken.
The $R$-charge of a supermultiplet is usually specified by that
of its lowest-spin component, with the charges of the higher components
given by $R = [R_0 - 2(s-s_0)]$, where $R_0$ is the charge of the lowest-spin
component and $s_0$ is its spin.  Using the rules for deriving couplings
from the superpotential given in Section~\ref{sec:models_mssm}, it is not hard
to check that the total $R$-charge of any term in 
the superpotential must be equal to $+2$ if the Lagrangian of the
theory is to be invariant under $U(1)_R$ transformations~\cite{Martin:1997ns}. 
The gaugino in a vector multiplet must also have $R = 1$ to allow 
a standard vector boson kinetic term~\cite{Martin:1997ns}.  
Without soft supersymmetry breaking, the MSSM is symmetric under a $U(1)_R$
where $H_u$ and $H_d$ have $R= 1$ and the rest of the chiral superfields
all have $R=1/2$.  This symmetry is explicitly broken by the soft Majorana 
masses for the gauginos and the soft trilinear $A$-terms, and spontaneously 
broken by the Higgs VEVs.

  To extend the MSSM to include an unbroken $R$-symmetry, a new source
of gaugino soft masses is required.  These can arise in an $R$-symmetric way
when adjoint chiral multiplets $\Phi_i$ with $R = 0$ are introduced 
for each gauge group.  Together with supersymmetry breaking, 
this allows soft Dirac gaugino masses~\cite{Fox:2002bu,Choi:2008pi}
\begin{equation}
-\mathscr{L} \supset m_{i}\,\lambda_i\psi_i,
\end{equation}
where $\lambda_i$ is the gaugino of the $i$-th gauge group and $\psi_i$
is the fermion component of the adjoint $\Phi_i$.
This form of gaugino masses arises naturally in theories with an
enhanced number of supersymmetries ($\mathcal{N}=2$)~\cite{Fox:2002bu}.
To avoid spontaneously breaking the $U(1)_R$, the Higgs multiplets
$H_u$ and $H_d$ must both have $R=0$.  All the usual Yukawa couplings
will then be allowed provided the matter fields all have $R=1$.
This charge assignment forbids the superpotential $\mu$-term, 
leaving the Higgsinos unacceptably light.  To fix this, it is necessary
to add two new $R=2$ chiral multiplets $R_u$ and $R_d$ with the same
quantum numbers as $H_d$ and $H_u$ respectively.  With these fields,
new $R$-symmetric contributions to the higgsino masses arise from the 
superpotential couplings~\cite{Kribs:2007ac}
\begin{equation}
W \supset \mu_u\,R_u\ccdot H_u + \mu_d R_d\ccdot H_d
 + \sum_{i=1,2}\left(\xi_u^iH_u\Phi_iR_u+\xi_d^iH_d\Phi_iR_d\right).
\end{equation}
Together, these fields make up the minimal $R$-symmetric supersymmetric
extension of the Standard Model~(MRSSM)~\cite{Kribs:2007ac}.

  The collider signals of the MRSSM are similar to those of the MSSM,
but with a few important differences.  Relative to the MSSM, the MRSSM
can accommodate a much larger amount of flavor mixing while remaining
consistent with experimental bounds~\cite{Kribs:2007ac}.  This arises 
primarily for two reasons:
the absence of Majorana gaugino masses and trilinear $A$-terms forbids 
most dimension-five flavor mixing operators; and the Dirac gaugino masses
can be naturally much larger than the sfermion masses leading to an
additional suppression by factors of 
$m_{\tilde{f}}/m_{\tilde{g}} \ll 1$~\cite{Kribs:2007ac}. 
Large amounts of flavor mixing can generate new collider signals,
such as single-top plus missing energy through the flavor-violating decays
$\tilde{u} \to \tilde{\chi}_1^0\,t$ and $\tilde{d} \to \tilde{\chi}_1^-\,t$~\cite{Kribs:2009zy}.
The MRSSM also accommodates a chargino NLSP and a gravitino LSP
in the context of gauge mediation much more readily than the MSSM,
allowing displaced charged tracks~\cite{Kribs:2008hq}.

\begin{figure}[t]
\begin{center}
  \includegraphics[width=0.55\textwidth]{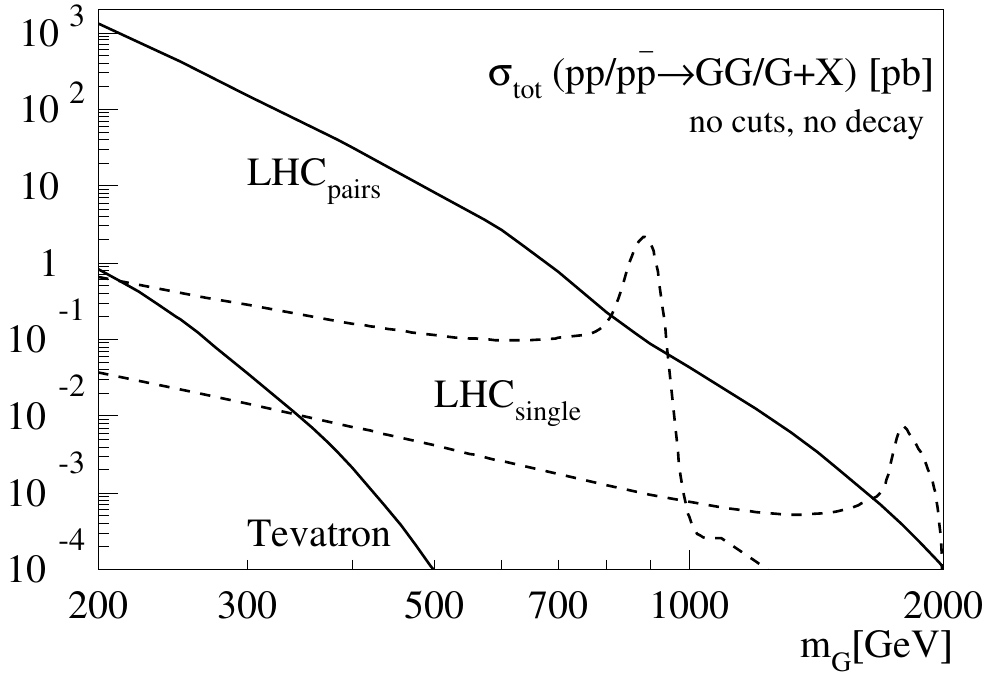}
\end{center}
\vspace*{0cm}
\caption{Production rates for sgluons at the Tevatron and at the
  LHC. For the LHC we show pair production (solid) and single
  production (dashed) as a function of the sgluon mass. The two curves
  for single sgluon production assume a gluino mass of 1~TeV and
  squark masses of 500~GeV (upper curve) and 1~TeV (lower curve).
  Figure from Ref.~\cite{Plehn:2008ae}}
\label{fig:models_slguon}
\end{figure}

Independent of the details of the comparably messy weak sector, an
enhanced ($\mathcal{N}=2$) supersymmetry or a Dirac gaugino mass requires
the introduction of an additional strongly interacting complex scalar
field $\Phi_3$, or \emph{sgluon} $G$.  At the LHC, this new particle
will dominate the phenomenology, owing to its QCD-sized $gGG$
coupling. The primary question is whether sgluons are dominantly
pair-produced through their tree-level color charge or singly produced
via a loop-induced coupling to quarks.  With a flavor diagonal squark
sector, the light sgluon decays to either two gluons, two light
quarks, or two tops, all at one-loop
order~\cite{Plehn:2008ae,Choi:2008ub}.  Because of the scaling of this
effective operator, a heavy sgluon will prefer the gluon decay
channel. On the other hand, once the main flavor constraints in the
MSSM have vanished, there is no reason to expect the squark sector to
be flavor diagonal. In that case the sgluon will decay to a single-top
final state $tq$. From sgluon pair production the most striking
signature would then be like-sign top--top
production~\cite{Plehn:2008ae}.\bigskip

A further possibility is that the MSSM couples to a new gauged
\emph{hidden sector} with a characteristic mass scale below a
$\gev$~\cite{ArkaniHamed:2008qp}.  Models of this type have received
attention recently in relation to potential hints of dark matter, but
they are also interesting in their own right since they can produce
unusual signals at particle colliders~\cite{ArkaniHamed:2008qp}.  The
simplest supersymmetric example consists of an Abelian $U(1)_x$ gauge
multiplet together with hidden Higgs fields $H$ and $H'$ and the
superpotential~\cite{ArkaniHamed:2008qn}
\begin{equation}
W_\text{hid} = \mu'\,H H'.
\end{equation} 
These hidden Higgses are assumed to develop vacuum expectation values on the order of a $\gev$.
The hidden sector can connect very weakly to the MSSM through 
\emph{gauge-kinetic mixing} with hypercharge of the 
form~\cite{Polchinski:1982an}
\begin{equation}
\mathscr{L} \supset \epsilon\left(-\frac{1}{2}X_{\mu\nu}B^{\mu\nu}
+ i\tilde{X}^{\dagger}\bar{\sigma}^{\mu}\del_{\mu}\tilde{B} + \text{h.c.} \right),
\label{kinmix}
\end{equation}
where $X_{\mu\nu}$ is the hidden $U(1)_x$ gauge field strength and
$\tilde{X}$ is the hidden gaugino.  A value of $\epsilon \sim
10^{-4}\!-\!10^{-2}$ is generated naturally by integrating out heavy
states charged under both $U(1)_x$ and $U(1)_Y$~\cite{Holdom:1985ag}.
The first mixing term in Eq.~\eqref{kinmix} induces a coupling between
the hidden massive $Z_x$ gauge boson and the MSSM states.  If the
$Z_x$ boson is unable to decay to other states in the hidden sector
due to kinematics, it will decay back to the Standard Model with width
proportional to $\epsilon^2$.  Existing searches for light physics
limit $\epsilon \lesssim 10^{-3}$ for $m_{Z_x} \sim 1\,\gev$ when such
decays dominate.  The second term in Eq.~\eqref{kinmix} leads to
superpartner decays from the MSSM to the hidden sector of the form
$\tilde{B} \to H\tilde{H}$~\cite{Baumgart:2009tn}.

A light hidden sector of this form can produce striking signals at the
LHC~\cite{ArkaniHamed:2008qp}.  These originate primarily from
supersymmetric MSSM cascade decays.  Instead of terminating at the
lightest MSSM superpartner, these cascades will continue into the
hidden sector.  For example, $\tilde{\chi}_1^0\to
\tilde{\chi}_\text{hid}\,h_\text{hid}$, where
$\tilde{\chi}_\text{hid}$ is a hidden sector neutralino and
$h_\text{hid}$ a hidden Higgs.  The cascade will continue in the
hidden sector until the lightest superpartner is reached.  If these
produce one or more $Z_x$ gauge bosons, a distinctive signature will
arise from decays of the $Z_x$ to Standard Model leptons.  Such lepton pairs will
typically have a small invariant mass on account of their origin, but
they will also be highly collimated since their energy scale is set by
the mass of the much heavier MSSM
states~\cite{Baumgart:2009tn,Cheung:2009su}.

\subsection{Extra dimensions}
\label{sec:models_extra}

  Having discussed supersymmetric models, we turn next to models with
extra spacetime dimensions~\cite{tasi_extrad,earlyxd}. 
There exists a wide range of such models, 
and we can categorize them in a broad way. 
To have avoided detection, the extra dimension(s) should typically be
\emph{compact}, having a finite extent.  Very small extra dimensions
arise frequently in string theory but are difficult to probe experimentally.
In most cases only extra dimensions with a characteristic scale around 
the TeV scale can be observed at the LHC, and we will concentrate primarily
on these. Next, extra dimensions can be flat and separable,  
as discussed in this section, or involve a warped
extra-dimensional metric like the RS models discussed in
Section~\ref{sec:models_rs}.  Last, the extra dimensions can be accessible 
to gravity only, like the ADD models discussed in Section~\ref{sec:models_add},
or incorporate Standard Model fields within the additional dimensions, 
like the universal extra dimensions introduced in 
Section~\ref{sec:models_ued}.
All of these models offer distinct solutions to various challenges facing 
the Standard Model, and they predict a broad class of signatures at
colliders.  Some of the more specific LHC aspects will be discussed 
in Section~\ref{sec:sig}.

\subsubsection{Large extra dimensions}
\label{sec:models_add}

  Large extra dimensions, also known as ADD models, aim to solve
the hierarchy problem of the Standard Model with a minimal change
to its particle content~\cite{add}.  They consist of $(4+n)$ spacetime
dimensions, with the $n$ additional dimensions all being compact
but with some of them having a relatively large (but microscopic)
characteristic radius.  The Standard Model is confined to a 
four-dimensional subsurface of this space called a \emph{brane}.  
Gravity propagates in all dimensions, and has a characteristic
higher-dimensional scale $M_* \sim \tev$ marking the onset of
quantum gravitational effects and the end of the Standard Model description.  
However, the effective strength of gravity acting on the Standard Model 
is reduced to $M_\text{Pl}$ due to a dilution by the volume of the 
extra dimensions.  This setup explains the observed hierarchy between 
the electroweak scale and the perceived Planck~\cite{add} 
(or GUT~\cite{unification}) scale to the extent that the size of 
the extra dimensions is stabilized.

  At the classical level, extra-dimensional gravity can be described
by the Einstein-Hilbert action.  This presents a number of basic features,
and the propagating degrees of freedom arise through the
symmetric and dimensionless metric $g_{MN}$. The mass
dimension of the Ricci scalar is independent of the underlying
space-time dimension since this is determined by derivatives.
ADD models also make a number of further specific assumptions about
the background metric of the $(4+n)$ dimensional space:
\begin{itemize}	
\item[--] \emph{spatial}: the $n$ extra dimensions are all spatial with
signature $(-1,-1, \cdots)$.
\item[--] \emph{separable}: the metric decomposes as a product space
  $g^{(4+n)} = g^{(4)} \otimes g^{(n)}$.
\item[--] \emph{flat}: matter is restricted to a four-dimensional brane
with the energy-momentum tensor
\begin{equation}
T_{AB}(x; y) = \eta^\mu_A \, \eta^\nu_B \, T_{\mu\nu}(x) \, \delta^{(n)}(y)
             = \left( \begin{array}{cc}
                      T_{\mu\nu}(x) \, \delta^{(n)}(y) & 0 \\ 0 & 0 
                      \end{array} \right).
\label{eq:conEMT}  			
\end{equation}
The assumption of an infinitely thin brane for our 4-dimensional world
might have to be weakened to generate realistic higher-dimensional
operators for flavor physics or proton
decay~\cite{fat_brane}. Alternatively, we can avoid those experimental 
constraints using a discrete symmetry~\cite{discrete_gauge}. 
\item[--] \emph{compact/periodic}: the extra dimensions are compact,
and we will assume here that they are simply periodic.
\end{itemize}
We can then generalize the Einstein-Hilbert action to $(4+n)$
dimensions and match it to what is seen by a four-dimensional 
(brane-localized) observer:
\begin{equation}
 - \frac{1}{2} \Mstar^{n+2} 
                   \int d^{4+n}z \; \sqrt{|g^{(4+n)}|} \; R^{(4+n)} 
               \equiv - \frac{1}{2} \Mpl^2 \int d^4 x
                    \sqrt{-g^{(4)}} \; R^{(4)},
\end{equation}
where the observed Planck scale is given in terms of the fundamental
gravity scale $M_*$ and the radius $r$ of the $n$ extra dimensions by
\begin{equation} 
\Mpl = \Mstar \; (2\pi r\Mstar)^{n/2} = M_*\;\sqrt{V_nM_*^n}, 
\end{equation}
where $V_n$ is the volume of the compact dimensions.
(In general, the extra dimensions can have different sizes
in which case the second expression should be used.)
Assuming $\Mstar= 1\;\tev$ to solve the gauge hierarchy problem,
we can transfer the problem into spacetime geometry: 
$r = 10^{-3}$--$10^{-11}\text{m}$ for $n=2$--$6$.  
These dimensions are large in the sense that $r \gg M_*^{-1}$.

  In the minimal model where the extra dimensions are all periodic 
with radius $r$, the $n=1$ case is ruled out by classical bounds 
on gravity~\cite{gravity_tests} as well as astrophysical data, 
unless there is and enhanced mass gap between massless 
and massive excitations~\cite{gps}.  This argument is purely
classical, and to describe the physical degrees of freedom we resort
to the ideas of Kaluza and Klein and map the higher-dimensional
gravitational theory onto an effective d=4
theory~\cite{kaluza_klein}.\bigskip

  The phenomenology of ADD models rests on the properties of
gravitons propagating in all dimensions~\cite{grw,coll_tao}.
Generically, a massless graviton in higher dimensions can be described
by an effective theory of massive gravitons and gauge fields in four
dimensions. For such a field the Pauli--Fierz mass
term~\cite{fierz_pauli} and the coupling to matter fields is highly
restricted.  In particular, it is inconsistent to introduce massive
spin-2 fields not originating from some type of Kaluza--Klein
decomposition~\cite{kk_sugra}.  In terms of the graviton field $g_{AB}
= \eta_{AB} + 2 h_{AB}/\Mstar^{1+n/2}$ the linearized
Einstein--Hilbert action leads to
\begin{equation}
\mathscr{L} = -\frac{1}{2}h^{MN}\Box h_{MN}+\frac{1}{2}h \Box h
-h^{MN}\p_M\p_N h +h^{MN}\p_M\p_Lh_L^N
-\Mstar^{-(1+n/2)}h^{MN}T_{MN},
\label{eq:KKLAG} 
\end{equation}
Periodic boundary conditions allow us to Fourier decompose the 
$(4+n)$-dimensional graviton field into a tower of 
four-dimensional real bosonic Kaluza-Klein modes
$h^{(\vec{n})}_{AB}(x)$ with mass dimension one.  Since these modes 
do not transform irreducibly under the Lorentz group in four
dimensions, we further decompose them into a set
irreducible representations that includes the four-dimensional massive
graviton~\cite{grw}.  In the original
Kaluza--Klein~\cite{kaluza_klein} decomposition a 5-dimensional metric
$g_{AB}$ is decomposed into a four-dimensional metric $g_{\mu\nu}$, a
vector $A_\mu$ and a scalar $\phi$.  Including KK masses, the vector and
scalar fields are eaten by the graviton to build a massive
4-dimensional graviton with five degrees of freedom.  Similarly, we
define $G_{\mu\nu}^{\vec{n}}$ as a massive graviton field.  However,
this does not exhaust the degrees of freedom; there is an additional
massive $(n-1)$-multiplet of vectors, $(n^2-n-2)/2$ scalars and an
additional singlet scalar. In terms of these fields the Einstein
equations are
\begin{alignat}{5}
(\Box+\mkk^2)& G_{\mu\nu}^{(\vec{n})}
&&= \frac{1}{\Mpl} \left[ -T_{\mu\nu}
                         +\left(\frac{\p_\mu \p_\nu}{\mkk^2}
                               +\eta_{\mu\nu}\right)
                          \frac{T_\lambda ^\lambda}{3}
                  \right] 
      = - \frac{T^{\mu\nu}}{\Mpl} \notag \\ 	
(\Box+\mkk^2)& V_{\mu j}^{(\vec{n})} &&= 0 \notag \\
(\Box+\mkk^2)& S_{jk}^{(\vec{n})}    &&= 0 \notag \\
(\Box+\mkk^2)& H^{(\vec{n})}         &&= \frac{\sqrt{3(n-1)/(n+2)}}{3\Mpl} T_\mu^\mu \; .
\label{eq:ee_new}
\end{alignat}
At the LHC the most relevant couplings are to effectively massless
degrees of freedom.
The physical graviton couplings are then degenerate
and individually suppressed by the four-dimensional Planck mass
$1/\Mpl$.  The fields $V_{\mu j}$ and $S_{jk}$ do not couple to the
energy momentum tensor, \ie to the Standard Model. The massless radion
$H$ corresponds to a volume fluctuation of the compactified extra
dimension. We assume that the compactification radius $r$ is
stabilized in some way~\cite{stabilize}, giving it mass~\cite{radion}.
Such a radion couples proportionally to masses, so as a scalar with no
Standard Model charge it will mix with a Higgs boson without overly
drastic effects.  

  The mass splitting between the KK states inside the
tower is given by $1/r$, which translates into ($\Mstar=1$~TeV):
\begin{alignat}{5}
\delta
\mkk\sim\frac{1}{r}=2\pi\Mstar\left(\frac{\Mstar}{\Mpl}\right)^{2/n}&=
\left\{
\begin{array}{lll}
\displaystyle{0.003~\text{eV}}
\qquad &&(n=2) \\
\displaystyle{0.1~\text{MeV}}
\qquad &&(n=4) \\
\displaystyle{0.05~\text{GeV}}
\qquad &&(n=6)
\end{array} \right.
\end{alignat}
These mass splittings are too small for the LHC to resolve.

  ADD models are constrained by a number of different ways
including electroweak precision data~\cite{tao_danny} and 
astrophysical observations~\cite{add_astro}, in part
orthogonal to their collider effects~\cite{gps}.  Current limits
strongly constrain ADD models with $n \leq 2$.  
For two to seven extra
dimensions, strong direct constraints on $\Mstar$ come from the
Tevatron experiments~\cite{tevatron_real,tevatron_virt}.\bigskip

  We can probe ADD models at colliders by searching for virtual graviton 
exchange as well as real graviton emission in high-energy reactions.  
An emitted tower of KK gravitons appears as missing transverse momentum, 
for example in association with single jet
production~\cite{grw,lhc_real,coll_maxim}.  Similarly, hard single
photon events with missing energy (\emph{monophotons}) 
would constitute a striking
signature for physics beyond the Standard Model.  The Drell--Yan
process $q \bar{q} \to \gamma^*,Z \to \ell^+ \ell^-$ is arguably the
best known hadron collider process. A large amount of missing energy
in this channel would be a particularly clean signal for physics
beyond the Standard Model~\cite{dave}.  Depending on the detailed
analysis, both of these electroweak signatures do have smaller rates
than a jet+graviton final state, but the lack of QCD backgrounds and
QCD-sized experimental and theory uncertainties result in a similar
discovery potential~\cite{grw,lhc_real}.\bigskip

  In computing the collider signals of ADD models it is essential to
sum over the production or exchange of all the graviton modes.
Since the modes are closely spaced, these sums can be approximated
reliably by an integral over an $n$-dimensional sphere in KK density
space with measure
\begin{equation}
d \sigma^\text{tower}  
= d\sigma^\text{graviton} \;
  \frac{S_{n-1} \, \mkk^{n-1} \, d \mkk}{(2\pi\Mstar)^n}
  \; \left( \frac{\Mpl}{\Mstar} \right)^2 \; .
\label{eq:KKTxsec}
\end{equation}
with $S_{n-1}=2\pi^{n/2}/\Gamma(n/2)$.  The factor $M^2_\text{Pl}$
from the KK tower summation defines a KK tower effective coupling
$E/\Mstar$ instead of $E/\Mpl$, roughly of the same size as the
Standard Model gauge couplings.  For larger $n$, 
the integral over $\mkk$ is ultraviolet-dominated.  This results from
the KK modes becoming more tightly spaced as we move to higher masses, 
and even more so for an increasing number of extra dimensions $n$.  If we
assign a cutoff $\cutoff$ to the $\mkk$ integration in
Eq.~(\ref{eq:KKTxsec}),  the integral scales at the partonic level like
$\cutoff^n$.  In other words, our effective theory of KK gravity
crucially requires a cutoff to regularize an ultraviolet divergence,
reflecting the fact that gravity is not perturbatively
renormalizable. 
The major question is then whether LHC observables are sensitive 
to $\cutoff$.  Luckily, for real graviton
production the kinematic constraint $\mkk < \sqrt{s}$
provides a natural ultraviolet cutoff, implying that the cross
section prediction is insensitive to physics far above the LHC energy
scale, which might or might not cover the fundamental Planck scale.\bigskip

  As a second collider signature, virtual gravitons exchanged in the 
$s$-channel generate the dimension-8 operator
\begin{equation}
              \frac{1}{\Mpl^2}
              \sum \frac{1}{s-\mkk^2} \; T_{\mu\nu} T^{\mu\nu} 
            \equiv \s(s) \; \mathcal{T} \; .
\label{eq:d8}
\end{equation}
Similarly, a loop-induced dimension-6 operator will couple two
axial-vector currents.  and can be estimated by naive dimensional
analysis~\cite{gs}.

 In the Standard Model, leptons and weak gauge bosons can only be
produced by a $q\bar{q}$ initial state. Because at LHC energies the
protons mostly consist of gluons, indirect graviton channels with
these final states get a head start.  The Tevatron mostly looks for new
physics of this type in two-photon or two-electron final 
states~\cite{tevatron_virt}.  
At the LHC, the cleanest signal taking into account backgrounds as well as
experimental complications is a pair of muons~\cite{coll_joanne}.
Aside from the squared amplitude for graviton production, 
Standard Model Drell-Yan amplitudes interfere with the graviton amplitude,
affecting the total rate as well as kinematic distributions. This mix
of squared amplitudes and interference effects make it hard to apply
any kind of golden cut to cleanly separate signal and background. One
useful property of the $s$-channel process is that the final state
particles decay from a pure $d$-wave (spin-2) state.  This results in
a distinctive angular separation of the final state
muons~\cite{coll_joanne}.

\begin{figure}[t]
\begin{center}
\includegraphics[width=8cm]{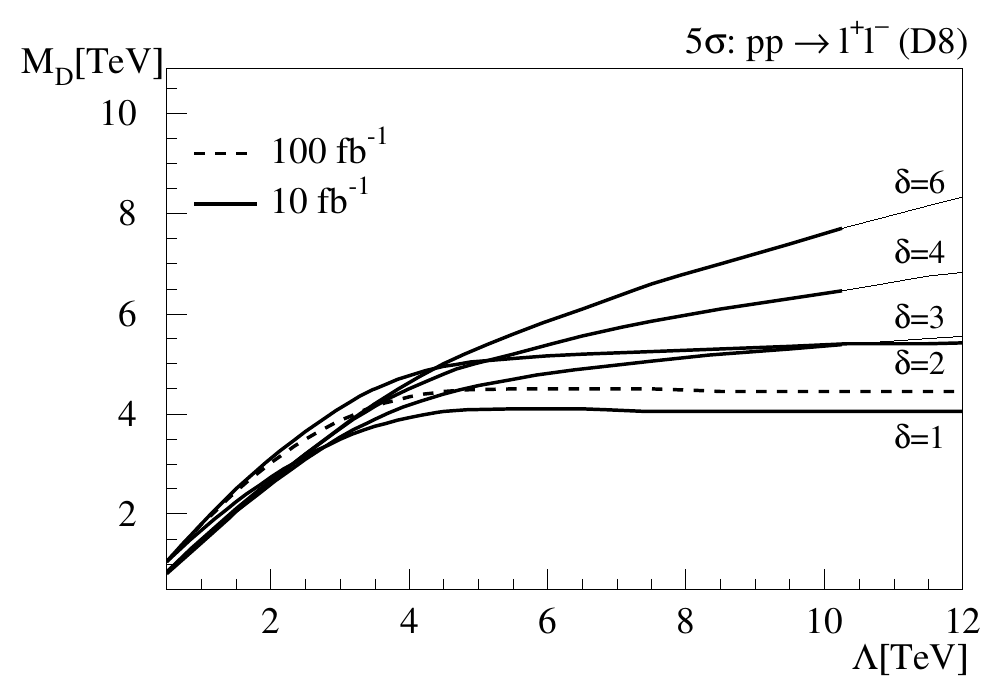}
\end{center}
\vspace*{0cm}
\caption{$5\sigma$ discovery contours for discovery of extra
  dimensions via virtual graviton contributions to the Drell--Yan
  process. Figure from Ref.\cite{gps}.}
\label{fig:um23}	
\end{figure}

At this stage we are most interested in the derivation of the
dimension-8 operator $\s$ from the KK effective theory.  Its dimension
$1/m^4$ coefficient arises partly from the coupling and partly from
the propagator structure.  In a similar vein to the real emission case,
we can replace $\Mpl$ with its definition in terms of $\Mstar$.  The
factor of $r^n$ appearing in the denominator is precisely $(\Delta
\mkk)^2$, giving us a (radial) integral over KK masses. Again, the
result is UV divergent for $n>1$.  In the spirit of an effective
theory we study the leading terms in $s/\cutoff^2$
\begin{equation}
\s(s)
= \frac{S_{n-1}}{\Mstar^4 (n-2)}
   \left(\frac{\cutoff}{\Mstar}\right)^{n-2}
   \left[1+\ope \left( \frac{s}{\cutoff^2} \right) \right]
\approx \frac{S_{n-1}}{n-2} \; \frac{1}{\Mstar^4} 
\label{eq:conabsa}
\end{equation}
where in the final line we identify $\Mstar \equiv \cutoff$, lacking
other reasonable options. Obviously, this relation should be
considered an order-of-magnitude estimate rather than an exact
relation valid.  For simplicity this term can
further be approximated in terms of a generic mass scale
$\s=4\pi/M^4$~\cite{grw,gs}, assuming (or hoping) that the effective
KK theory captures the dominant effects.  The function $\s(s)$
including $\cutoff$ can be integrated analytically for a Wick rotated
graviton propagator $1/(s+\mkk^2)$, but the interpretation of
particles in an effective field theory will be lost beyond a leading
approximation $s \gg \mkk^2$ or $s \ll \mkk^2$.

The LHC reach in $\Mstar \equiv M_D$ is given in Fig.\ref{fig:um23},
again in the presence of a variable cutoff $\cutoff$. As opposed to
the real emission case shown in Fig.~\ref{fig:ReEmm}, the result is
clearly sensitive to $\cutoff$.  Only for $n=1$ is this not true,
reflecting that $\int_0^{\infty} dm\; \mkk^{-2}$ in fact
converges.\bigskip

As explicitly seen above, the effective field theory description of
extra-dimensional gravity breaks down once the LHC energy approaches
the fundamental Planck scale, no matter if we study real or virtual
gravitons.  Once this ratio of the typical energy over the Planck
scale becomes large, gravity appears to become strongly interacting and
we expect to encounter ultraviolet physics.  There are a number of
effective-theory proposals which side-step this complication by
defining the integral in a cutoff-independent scheme.  One such
treatment relates to the eikonal approximation to the $2\rightarrow 2$
process~\cite{eikonal}, another involves introducing a finite brane
thickness~\cite{thick_brane}. In the latter, the gravitational
coupling is exponentially suppressed above $\Mstar$ by a brane rebound
effect.  This is applied to the case of high energy cosmic rays
interacting via KK graviton exchange~\cite{pluemi}.  However, none of
these models offer a compelling UV completion for the KK effective
theory.

At a conceptual level this potential breakdown of the KK effective theory
at LHC energies insinuates the immediate need for a more
complete description of gravity.  We know several possible
modifications of this dangerous ultraviolet
behavior~\cite{weinberg_new}: most well known, string theory includes
its own fundamental scale $M_S$ which is related to a finite size of
its underlying objects.  Such a minimum length acts as a an
ultraviolet cutoff in the energy, which regularizes all observables
described by the gravitational interaction~\cite{veneziano,lhc_strings}.  
An alternative approach
which avoids any ad-hoc introduction of radically different physics
above some energy scale is based on the ultraviolet behavior of
gravity itself~\cite{weinberg}: the asymptotic safety scenario is
based on the observation that the gravitational coupling develops an
ultraviolet fixed point which avoids the ultraviolet divergences
naively derived from power
counting~\cite{fp_early,ergf,trunc_misc,fp_extrad,review_misc}.
This idea can be applied to LHC phenomenology in ADD~\cite{lp} or
Randall-Sundrum~\cite{tom_joanne} models.\bigskip

  For example, the scattering process $q \bar{q} \rightarrow \mu^+
\mu^-$ can be computed using open string perturbation theory.  The
helicity amplitudes for $2\to 2$ scattering are simply analytic
functions in $s$, $t$ and $u$ together with the common Veneziano
amplitude~\cite{veneziano}
\begin{equation} 
\s(s,t)=
\frac{\Gamma(1-s/M_S^2) \; \Gamma(1-t/M_S^2)}{\Gamma(1-(s+t)/M_S^2)} 
\end{equation}
in terms of the inverse string scale $M_S$.  While we do not exactly
know the size of this scale it should lie between the weak scale and
the fundamental Planck scale $\Mstar$. We consider three distinct
limits: first, the leading correction below the string scale is $\s =
(1 - \pi^2/6 \; st/M_S^4)$.  This form corresponds to our KK effective
field theory, modulo a normalization factor which relates the $M_S$
and $\Mpl$.  Going back to Eq.~(\ref{eq:conabsa}) we find that the
dimension-8 operator in terms of the string scale is given by
$\pi^2/32 \times g^4/M_S^4$.  While this series in $M_S$ is not what
we are interested in as the UV completion of our theory, it is the reason
why we expect string excitations to be the dominant process for high
energy particle scattering.  Next, in the hard scattering limit
$s\to\infty$ and for a fixed scattering angle the amplitude behaves as
$\s \sim \exp -\alpha'(s\,\log s+t\,\log t)$.  Due to the finite
string scale all scattering amplitudes becomes weak in the
UV. Unfortunately, this limit is not very useful for LHC
phenomenology.  Last but not least, in the Regge limit for small angle
high energy scattering $s \gg t$ the poles in the $\Gamma$ functions
determine the structure.  They describe string resonances with masses
$\sqrt{n} M_S$.  The physical behavior for scattering amplitudes above
the string scale is a combination of Regge and hard scattering
behavior.  In other words, equally spaced string resonances together
alongside exponential suppression, but at colliders the resonances
should be the most visible effects.

Aside from these gauge boson string resonances, the string theory
equivalent of our process generating the effective dimension-8
operator is the scattering of four open strings via the exchange of a
closed string.  This amplitude is insignificant compared to the string
excitations in the vicinity of the string scale.  The KK mass
integration is finite for all $n$ due to an exponential suppression of
similar origin as the hard scattering behavior noted above.  For
fields confined to a D3 brane this integral is
\begin{equation}
\s(s) \sim \int d^6 m \; 
      \frac{e^{\alpha'(s-m^2)/2}}{s-m^2} \; .
\end{equation}
A similar exponential factor $\exp (-\alpha' m^2)/(s-m^2)$
regularizing the $m$ integral appears in the modification using a
finite brane thickness~\cite{thick_brane}.  However, this approach
still violates unitarity for large $s$. 

The main issue with the naive phenomenology of RS gravitons described
in Section~\ref{sec:models_rs} as well as string excitations is that it is
no experimental or phenomenological challenge to the Tevatron or LHC
communities. For example the Tevatron experiments have been searching
for (and ruling out) heavy gauge bosons for a long time. The
discovery of a $Z'$ resonance at the LHC would simply trigger a
discussion of its origin, including KK gravitons, string resonances,
KK gauge bosons, or simple heavy $Z'$ gauge bosons from an additional
gauged $U(1)$ symmetry~\cite{Carena:2004xs}.\bigskip

  In contrast to the common lore there might be no need at all to alter
the structure of gravity at high energies --- gravity can simply be
its own ultraviolet completion~\cite{weinberg,fp_early,ergf}.
Although asymptotically safe gravity is developed in four
dimensions~\cite{fp_review,trunc_misc,review_misc}, the
results generalize in a straightforward manner~\cite{fp_extrad}.
Evidence comes in many forms: the concept of asymptotic safety or
non-perturbative renormalizability was proposed originally in
1980~\cite{weinberg}. The first hints that gravity might have a UV
fixed point were uncovered using a $2+\epsilon$ expansion for the
space-time dimensionality~\cite{two_plus_eps}.  Further evidence was
collected in the $1/N$ expansion where $N$ is the number of matter
fields coupled to gravity~\cite{fp_early}.  More modern results use
exact renormalization group methods~\cite{ergf}.  There are a number
of reviews of the subject~\cite{fp_review}, and on the necessary
non-perturbative techniques, namely the exact flow equation for the
effective average action or Wetterich equation~\cite{wetterich_eq}.
Gravitational invariants including $R^8$ and minimal coupling to
matter are consistently included in flow equations without destroying
the fixed point~\cite{fp_review}.  More recently, it has been shown
that including invariants proportional to divergences in perturbation
theory do not give divergent results
non-perturbatively~\cite{include_divergences}.  Furthermore, there is
independent evidence for asymptotic safety coming from recent lattice
simulations, causal dynamical triangulations~\cite{cdt}.  Universal
quantities, like \eg critical exponents, computed using this method
agree non-trivially with results derived using renormalization group
methods.

The key point for our application is that in the UV the coupling
exhibits a finite fixed point behavior.  This ensures that the UV
behavior of the complete theory is dominated by fixed point scaling,
rendering all our computed transition rates weakly interacting at all
energy scales.\bigskip

Again, we start from the Einstein--Hilbert action to calculate the
scaling behavior for the gravitational coupling. Applying $\mu \,
d/d\mu = d/\log \mu$ we find
\begin{equation}
\beta_g = \frac{d}{d \log \mu} \; g(\mu) = 
\left[ 2+n+\eta(g(\mu)) \right] \; g(\mu) \; ,
\label{eq:canbeta}
\end{equation}
The parameter $\eta(g)$ will in general contain contributions from all
couplings in the Lagrangian, not only the dimensionless Newton's
constant. One important property is immediately apparent: for $g=0$ we
have a perturbative gaussian fixed point, \ie an IR fixed point which
corresponds to classical general relativity where we have not observed
a running gravitational coupling. Secondly, depending on the
functional form of $\eta(g)$ the prefactor $2+n+\eta(g)$ can vanish,
giving rise to a a non-gaussian fixed point $g_\star \ne 0$.  Using
the exact renormalization group flow equation we can compute the
anomalous dimensions~\cite{ergf}, which depend on the shape of the UV
regulator. The beta function of the gravitational coupling becomes
\begin{equation}
\beta_g(g) =\frac{(1-4 (4+n) g)(2+n)g}{1-(4+2n)g}\,
\qquad \qquad 
\eta(g) =\frac{2(2+n)(6+n)\,g}{2(2+n)\,g-1}
\end{equation}
One way of interpreting the physical effects of the gravitational UV
fixed point is to modify the original calculations by defining a
running Newton's coupling and evaluate it at the energy scale given by
the respective process~\cite{bonanno_reuter,tom_joanne}.  The leading
effects from the renormalization group running of the gravitational
coupling we can now include into a form factor
\begin{equation}
\frac{1}{\Mstar^{2+n}} \; h^{MN}T_{MN}
\rightarrow 
\frac{1}{\Mstar^{2+n}} \;
\left[1+\frac{1}{8\pi g_*}\left(\frac{\mu^2}{\Mstar^2}\right)^{1+n/2}
\right]^{-1} \; h^{MN}T_{MN}
\equiv \frac{F(\mu^2)}{\Mstar^{2+n}} \; h^{MN}T_{MN}
\end{equation}
which carries through in the decomposition of $h_{MN}$ to the
4-dimensional field $G_{\mu\nu}^{n}$. At high energies the form factor
scales like $F(\mu^2) \propto (\Mstar/\mu)^{2+n}$.  The factor $1/(8
\pi g_*)$ is an $\ope(1)$ parameter controlling the transition to
fixed point scaling.  For collider processes, the simplest choice is
$\mu=\sqrt{s}$.  If we treat $\mkk$ and $\sqrt{s}$ as separate scales
in the KK mass kernel Eq.~(\ref{eq:d8}) and evaluate the form factor in
terms of $\sqrt{s}$, the $\mkk$ integration still requires a
cutoff. On the other hand, as far as the $s$ integration is concerned,
the form factor solves the unitarity problem associated with graviton
scattering amplitudes at the LHC~\cite{tom_joanne}.\bigskip

\begin{figure}[t]
\begin{center}
\includegraphics[width=8cm]{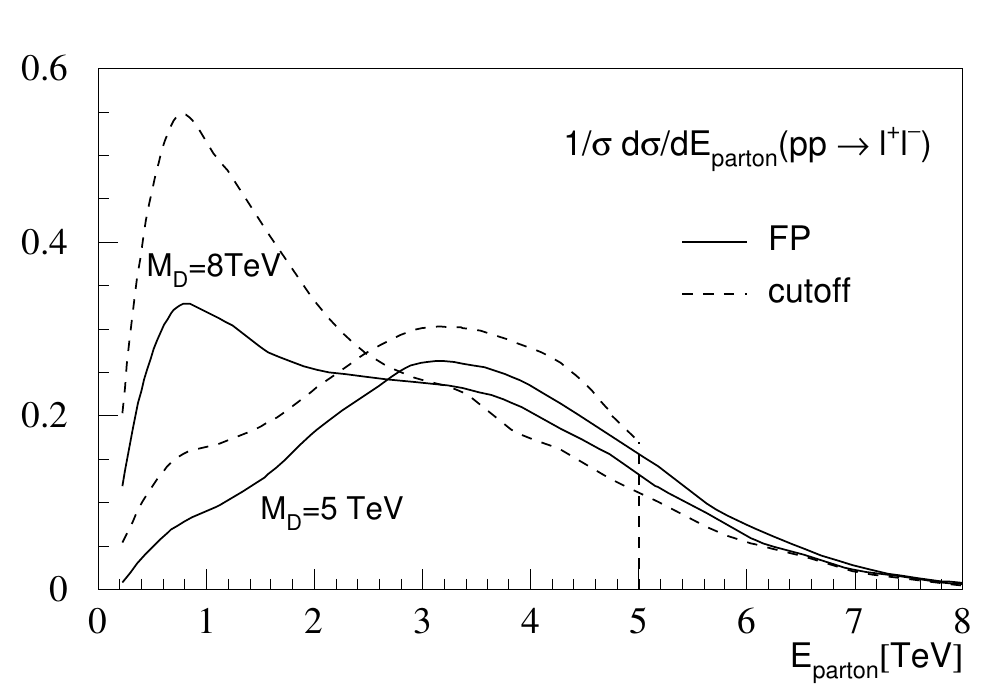}
\end{center}
\vspace*{0cm}
\caption{Normalized distribution of the partonic energy (or $m_{\ell
    \ell}$) in the Drell--Yan channel for $n=3$.  The non-trivial
  shape difference between the $\Mstar=5\;\tev$ and $\Mstar=8\;\tev$
  is the result of interference effects. Figure from Ref.~\cite{lp}}
\label{fig:ERG2}	
\end{figure}

An alternative method which is better suited for the case of virtual
gravitons is based directly on the form of the anomalous dimension of
the graviton Eq.~(\ref{eq:canbeta})~\cite{jan}.  In this picture, there
is a phase transition occurring at a critical point with anomalous
dimension $\eta_\star =-2-n$ at or beyond the Planck scale.  The
momentum-space two point function generically has the form $\Delta(p)
\sim 1/(p^2)^{1-\eta/2}$, which reproduces the classical result for
small $\eta$.  In the vicinity of the non-gaussian fixed point, this
becomes $1/(p^2)^{-(4+n)/2}$~\cite{daniel_optimized}.  The massive
graviton propagator in the fixed point region is then modified
as~\cite{lp}
\begin{equation}
\frac{1}{s-\mkk^2}\rightarrow \frac{\mtrans^{2+n}}{(s-\mkk^2)^{(4+n)/2}}
\end{equation}
where the transition scale $\mtrans \sim \Mstar$ in the numerator
maintains the canonical dimensions for the propagator.  The graviton
kernel $\s(s)$ integrated over the entire $\mkk$ range is then finite
to for all $n$, and the transition scale $\mtrans$ parameterizes the
crossover to fixed point scaling.  The low-energy and high-energy
contributions we can calculate to leading order in $s/\mtrans^2$
\begin{equation}
\s  = \frac{S_{n-1}}{4\Mstar^4}
\left(\frac{\mtrans}{\Mstar}\right)^{n-2}
     + \frac{S_{n-1}}{(n-2)\Mstar^4 }
\left(\frac{\mtrans}{\Mstar} \right)^{n-2}
     =  \frac{S_{n-1}}{(n-2) \Mstar^4} \;
 \left(\frac{\mtrans}{\Mstar} \right)^{n-2} \left(1+\frac{n-2}
 {4}\right)
\label{eq:irkernel}
\end{equation}
With this result we do not need any artificial cutoff scale for the
$\mkk$ integration.  The result only has a small sensitivity to the
transition scale $\mtrans \sim\Mstar$, but including a more elaborate
description of the transition region will remove this.  For hadronic
cross sections there is still the $s$ integral from the convolution
with the parton distribution functions, and only to leading-order $\s$
is independent of $s$.  For the full $\s(s)$ perturbative unitarity is
guaranteed by its large-$s$ scaling behavior $1/s^2$.  Most
importantly, we do not expect any resonance structure, which makes
asymptotic safety clearly distinguishable from a strong theory
completion.

The UV completion of the KK prediction helps an LHC study in two ways:
first of all, the signal is enhanced by including the UV portion of
the integral.  In addition, the graviton kernel has a distinctive
shape which depends on the number of extra dimensions $n$.  At lower
partonic energies, the dominant interference term between gravitons
and the Standard Model diagrams implies a scaling with $\s \sim
(n-1)$.  For higher energies the pure graviton amplitude is dominant
and the rate scales as $(n-1)^2$.  This fact is demonstrated in
Fig.\ref{fig:ERG2}.\bigskip

Independent of the UV completion of gravity all of the descriptions
above fails once the spatial separation of the two partons colliding
at the LHC becomes smaller than the Schwarzschild radius of TeV-scale
gravity. In this configuration a black hole will form~\cite{bh_first},
more or less thermalize, and then decay again. Their production cross
section is approximately given in terms of the Schwarzschild radius
$\pi r_S^2$~\cite{bh_semiclassical}.  Highly spherical decays to many
particles we can search for and unambiguously identify at the
LHC~\cite{charybdis}, as long as the black hole involves a large
number of particles and does not simply appear as a higher-dimensional
operator~\cite{bh_higherdim}. The main question that the UV completion
of gravity has to answer is if and how the black hole actually decays,
after radiating a sufficient number of particles. Both, string theory
and fixed-point gravity indicate that there may be a stable remnant
delaying the final decay to infinity, which again can be tested in
experiment~\cite{bh_remnant}.

\subsubsection{Universal extra dimensions}
\label{sec:models_ued}

  Theories with Universal Extra Dimensions~(UED) also posit additional
compact spatial dimensions.  The distinguishing feature is that all
fields live equally in all of the
dimensions~\cite{Appelquist:2000nn,unification}, unlike ADD or
RS-like constructions, which confine some or all of the Standard Model
to a brane.  This simple feature has the consequence that no point is
singled out prior to compactification, and thus the extra-dimensional 
Lorentz invariance remains.  Some of these symmetries are partially 
broken by the compactification, but remnant symmetries often persist.
These can ameliorate constraints from precision measurements 
and predict a stable particle which can play the role of dark matter.

  In contrast to other models involving large extra dimensions, UED
theories do not make a serious attempt to solve the hierarchy problem,
and a compactification radius on the order of a TeV is typically 
put in by hand.  (However, see Ref.~\cite{ArkaniHamed:2000hv} for a 
topcolor-like construction.)  
Instead, they are motivated for other reasons.
Six-dimensional implementations of UED offer anomaly cancellation as an
explanation for why there are three generations in the Standard Model,
or more precisely $3n$ where $n$ is an integer $\geq 0$~\cite{Dobrescu:2001ae},
and can suppress dangerous operators leading to rapid proton 
decay~\cite{Appelquist:2001mj}.  UED theories also have intrinsic worth 
as a serious model of dark matter~\cite{Servant:2002aq}.
Their SUSY-like spectra and phenomenology of $\tev$-scale KK modes 
can also serve as a straw-man theory with which we can test our ability 
to distinguish genuine SUSY signals from similar phenomena arising in other 
theories~\cite{Cheng:2002ab}.\bigskip

  The primary characteristics defining a model of UED are the number of
compact dimensions, their topology, and their sizes.  A five-dimensional
theory with one compact dimension with chiral fermions must be
compactified into a line segment (orbifold), which is characterized by
a single quantity: its length $L$.  Six-dimensional theories with
chiral fermions have been considered with the topology of an
orbifolded torus, characterized by two lengths $L_1$ and $L_2$, as
well as the chiral square, defined as with $L_1 = L_2 = L$ and its
adjacent sides identified~\cite{Dobrescu:2004zi,Burdman:2006gy}.  
The KK spectrum depends very sensitively on the sizes of the extra
dimensions according to
\begin{equation}
M_j^2 = \frac{\pi^2}{L^2} j^2 + m_0^2 \quad (\text{5d}) \qquad \qquad
M_{j,k}^2 = \pi^2 \left( \frac{j^2}{L_1^2} + \frac{k^2}{L_2^2} \right)+ m_0^2 \quad (\text{6d}) \; ,
\label{eq:kkmass}
\end{equation}
where $j$ and $k$ label the mode number(s) and $m_0$ is the
corresponding zero mode or Standard Model particle mass.

\begin{figure}[t]
\begin{center}
\includegraphics[width=0.45\textwidth]{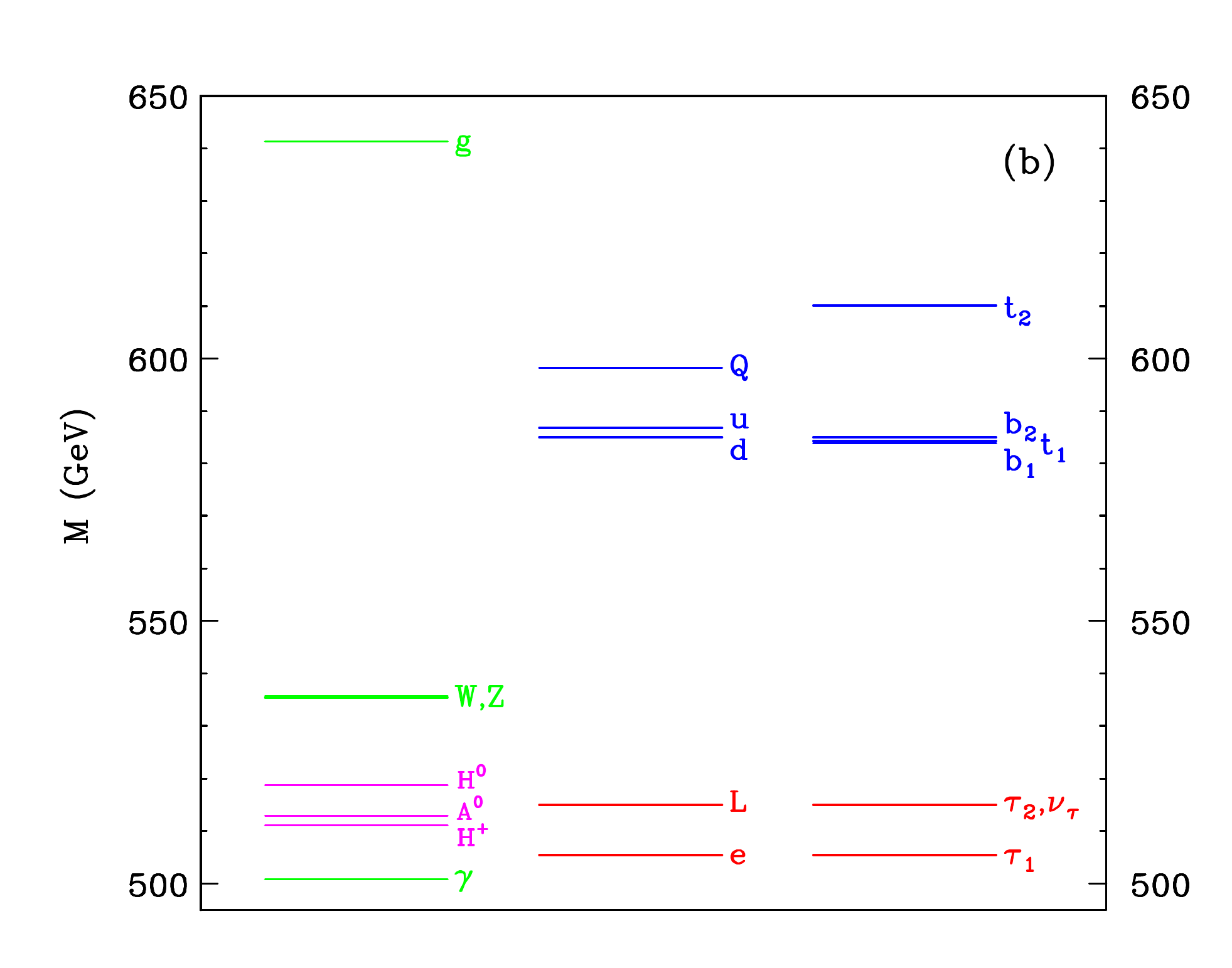}
\hspace*{10mm}
\includegraphics[width=0.45\textwidth]{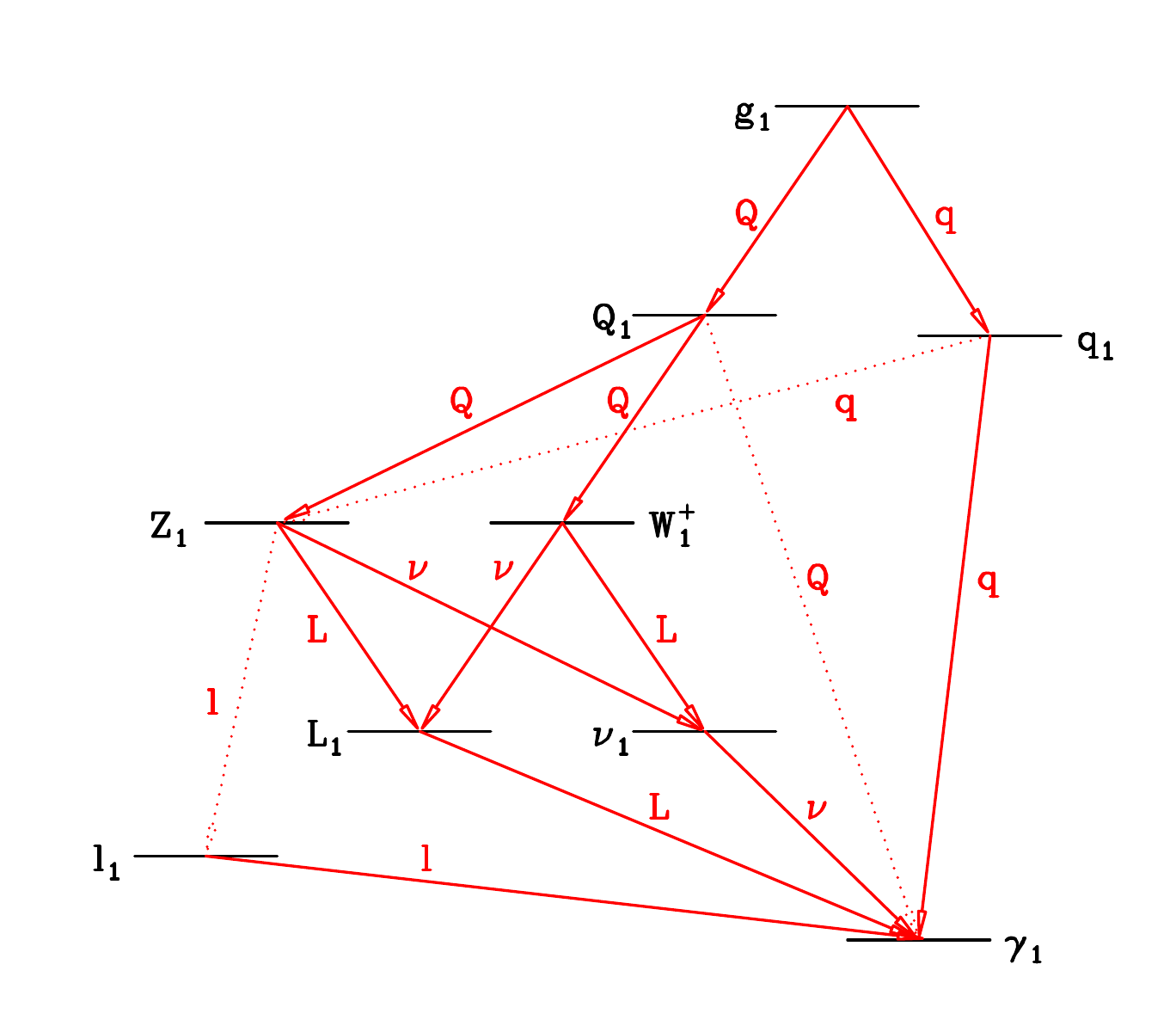}
\end{center}
\vspace*{-0.25cm}
\caption{Sample spectra in a 5d model (left, from Ref.~\cite{Cheng:2002iz})
under the assumption that boundary terms vanish at the cut-off
scale. Also shown (right, from Ref.~\cite{Cheng:2002ab})
are typical transitions between KK levels, showing the
Standard Model particle emitted in a given 2-body decay, for the dominant decays (solid lines)
and subdominant decays (dashed lines).}
\label{fig:5duedspectra} 
\end{figure}

  In addition to the bulk characteristics, UED theories are
characterized by terms which live on the boundaries of the extra
dimensions.  The subtle breaking of extra-dimensional Lorentz
invariance by the compactification renders such terms inevitable, 
and they are typically divergent, depending on the unknown UV
completion~\cite{Georgi:2000ks}.
A popular
prescription, termed minimal UED, is to choose a scale for the UV
completion consistent with a perturbative extra-dimensional
description up to some scale $\Lambda$, and then assume that in the
UV theory the boundary terms vanish at $\Lambda$~\cite{Cheng:2002iz}.  
Under this assumption, the boundary terms will be generated at the 
one-loop level proportional to $\log \Lambda L$.  These terms modify the mass
spectrum of the KK modes and induce mixing among what would have been
the KK states of the theory with zero boundary
terms~\cite{Carena:2002me}. In analogy to soft SUSY breaking, 
modifications of the UED mass spectrum should be expected to be 
required by experimental data~\cite{Chen:2009gz}.

  If these states are induced purely through loops of bulk physics, they
will break KK number conservation, but preserve a discrete subgroup
called KK parity.  The five-dimensional model has a $\mathbb{Z}_2$ symmetry
under which $(j)$ states transform as $(-1)^j$.  The orbifolded torus
compactification has a $\mathbb{Z}_2 \otimes \mathbb{Z}_2$ symmetry
under which $(j,k)$ states transform as $(-1)^j \otimes (-1)^k$.  The
chiral square has a $\mathbb{Z}_2$ symmetry under which $(j,k)$ states
transform as $(-1)^{j+k}$.  These KK parities are radiatively stable
in the sense that provided the initial boundary terms respect them,
loops of bulk physics will not destroy them. On the other hand,
arbitrary boundary terms will break them explicitly, so their existence
amounts to an assumption about the UV physics.  As in supersymmetric
or little Higgs models, such a KK parity guarantees that the lightest
odd-level KK particle (LKP) is stable, so it can serve as a dark
matter agent.  At colliders, odd-level KK particles can
only be produced in pairs.\bigskip

\begin{figure}[t]
\begin{center}
\includegraphics[width=0.45\textwidth]{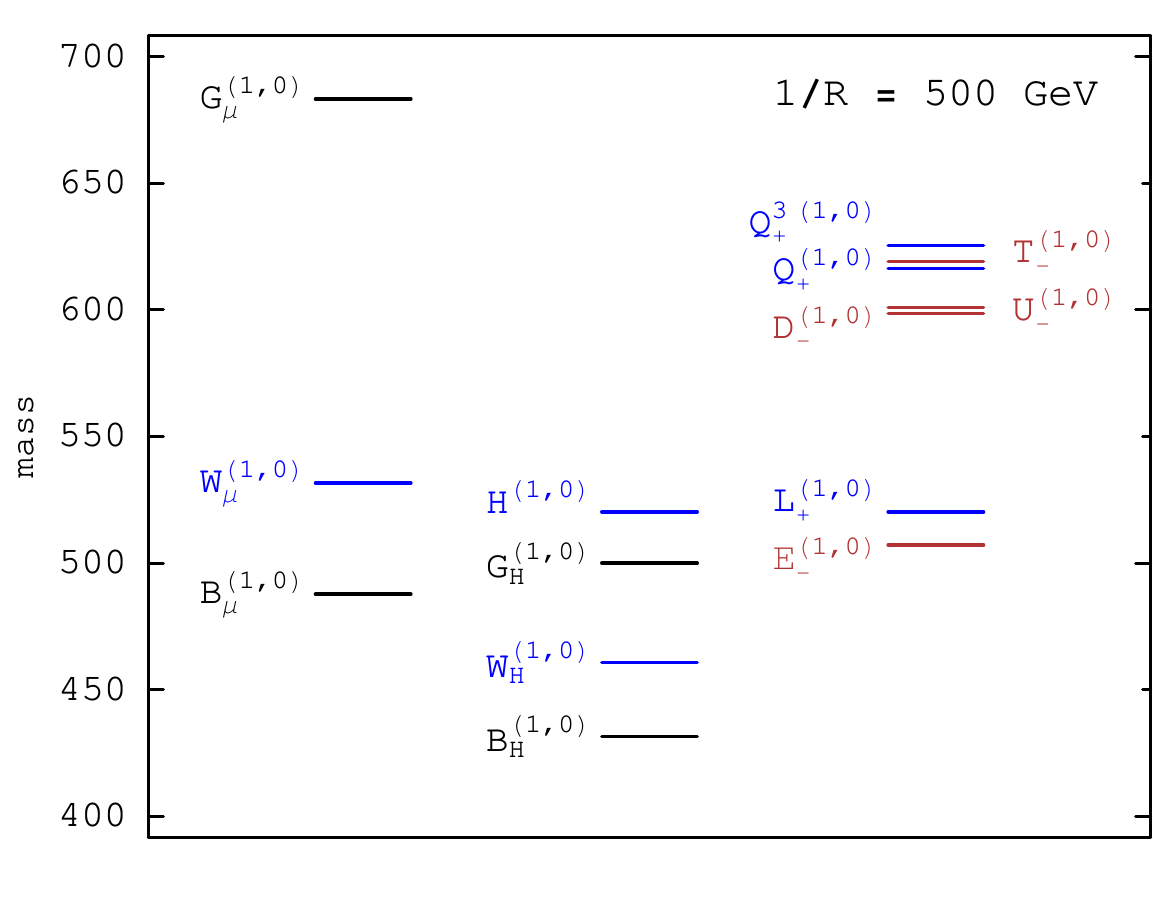}
\hspace*{10mm}
\includegraphics[width=0.45\textwidth]{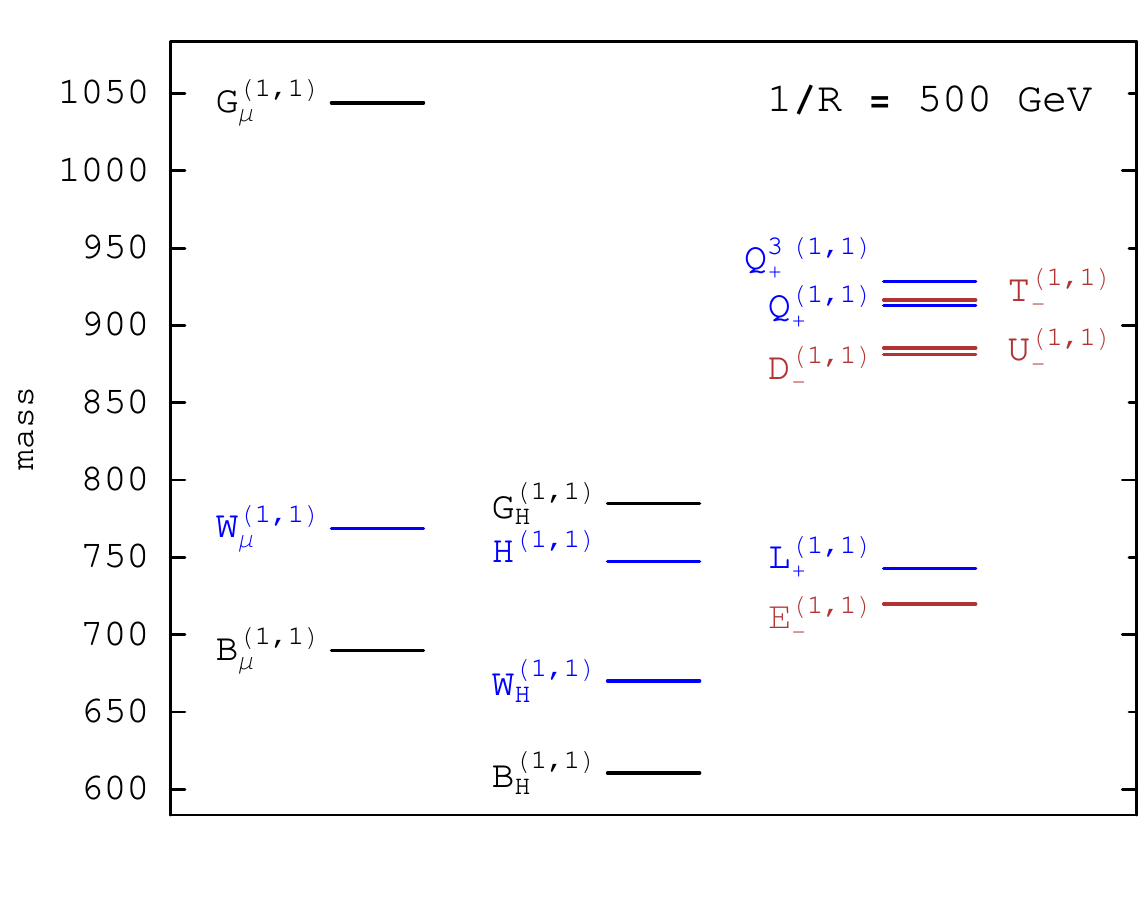}
\end{center}
\vspace*{-0.25cm}
\caption{Sample spectra of the KK-odd $(1,0)$ modes (left) and
KK-even $(1,1)$ modes (right) in 
the chiral square 6d model (right, from Ref.~\cite{Burdman:2006gy})
for $L= 1 / 500$~GeV,
under the assumption that boundary terms vanish at the cut-off
scale.}
\label{fig:6duedspectra} 
\end{figure}

In Figs.~\ref{fig:5duedspectra} and \ref{fig:6duedspectra} we show the
resulting spectra for the 5d minimal UED model and a 6d chiral square
when the compactification radius is set to $L = 1/500\,\gev$.  
The spectrum of the lightest tier of KK modes is highly reminiscent 
of a typical supersymmetric theory, with the colored KK particles ending up
heaviest, followed by $SU(2)_L$-charged electroweak states in the
middle, and only $U(1)_Y$-charged states as the lightest.  The identity
of the LKP determines the features of possible cascade decays.  It is
remarkable that in both the 5d model and the chiral square, minimal
boundary terms result in a neutral LKP suitable as dark matter.  In
the 5d case the LKP is a vector particle, the first KK mode of the
$U(1)_Y$ gauge boson.  On the chiral square it is a combination of the
$5$ and $6$ components of the $U(1)_Y$ boson, which appear as a
Lorentz scalar to an observer in four dimensions.  Non-minimal boundary
terms can lead to different LKPs and spectra, as explored for the 5d
theory in~\cite{Flacke:2008ne,Park:2009cs}.

  The lightest tier of KK particles is a set of KK-odd states which have
cascade decays down to the LKP, resulting in missing energy signatures
at the LHC. Figure~\ref{fig:5duedspectra} shows the dominant cascade
decays in the 5d UED model.  At the LHC we expect the dominant
production modes to be through KK quarks and gluons, because despite
being slightly heavier than the uncolored states they have large
couplings to the initial-state quarks and gluons. Thus, just as in
supersymmetric theories, a typical UED event produces multiple final
state jets and leptons~\cite{Appelquist:2000nn}.  While this 
is positive in the sense that it
means the standard SUSY searches will automatically also apply to
large regions of UED parameter space, it raises the question as to how
well we can distinguish SUSY from UED at the 
LHC~\cite{Cheng:2002ab,Appelquist:2000nn,ued}.\bigskip

  The most obvious distinguishing feature between SUSY and UED is that
UED theories have an entire tower of KK excitations, in contrast to
the single set of superpartners in SUSY.  The discovery of elements of 
the next higher level ($2$-modes for 5d theories or $(1,1)$-modes for 6d
theories), which we will collectively refer to as level-2 modes from
here on, would suggest the extra-dimensional scenario.  The level-2
modes are again even under KK parity, meaning that they can be singly
produced and decay into pairs of ordinary Standard Model fermions --- call them
yet another reincarnation of a $Z'$ boson.  The most promising targets
for an LHC discovery are indeed the level-2 modes of the charged and
neutral electroweak bosons.  At the parton level we estimate that with
$100~\ifb$ of data, the LHC can discover the $\gamma^{(2)}$ and
$Z^{(2)}$ neutral bosons decaying into leptons, and reconstruct them
as separate resonances, provided their masses are less than about
2~TeV~\cite{Datta:2005zs}. While discovery of the two level-2 modes
would be suggestive, it is not definitive.  One can imagine
supersymmetric theories with additional $U(1)$ gauge groups with
similar phenomenology.\bigskip

A second approach is to reconstruct the underlying spins of the first
level particles.  For a SUSY theory, these spins will differ from their
Standard Model counterparts by $1/2$, whereas in UED the spins will be
the same as the corresponding Standard Model particle.  Measuring 
spins is a challenge at a hadron collider, particularly because the
initial state for a given event is unknown, and the LKPs or LSPs escape
detection and carry spin angular momentum which is thus lost.  
We will describe techniques to distinguish between different spin
hypotheses for new physics states in Section~\ref{sec:sig_cascade}.

\subsection{Compositeness}
\label{sec:models_tc}

  An alternative approach to the hierarchy problem relies on the fact
that fermions, unlike scalars, have mass terms protected by chiral
symmetry.  As a result, quantum corrections to fermion masses are 
at most logarithmically sensitive to physics in the UV, 
and a theory with only fermions and no fundamental scalars is natural.  
This suggests a another class of models of electroweak symmetry breaking
that avoid the gauge hierarchy problem by replacing the fundamental
Higgs scalar with the vacuum expectation value~(VEV) of a composite
scalar operator made up of fundamental fermion fields.
\bigskip

  A simple example of this already occurs within the Standard Model.
Even without a Higgs VEV, electroweak symmetry breaking would still
be induced by QCD, albeit at an unacceptably low scale near $\Lambda_{QCD}$.
To see this, consider the fermion bilinear operator of two quarks
$\overline{\Psi}_L \Psi_R$, where the $SU(3)_c$ indices are contracted,
$\Psi_L$ is an $SU(2)_L$ doublet, and $\Psi_R$ is an $SU(2)_L$ singlet.
This operator has the quantum numbers of a Higgs boson: $(1,2,\pm 1/2)$.
QCD confinement generates a non-zero expectation value for this
bilinear operator on the order of $\Lambda_\text{QCD}^3 \sim (\gev)^3$,
thereby breaking the electroweak symmetry.  In composite (or dynamical)
models of electroweak symmetry breaking, this structure is extended to a 
new strongly-coupled gauge group in place of $SU(3)_c$ 
with a scaled-up characteristic energy.

  The interpretation of the composite operator 
$\overline{\Psi}_L\Psi_R$ depends on the nature of the underlying 
dynamics which produces its VEV.  If there is a scalar bound state whose Fock
decomposition is mostly $\overline{\Psi}_L \Psi_R$, we have a theory
of a composite Higgs particle. It may be approximated as a theory
containing a Higgs doublet up to the energy scale $\Lambda$ 
characterizing the binding energy of the $\overline{\Psi}_L \Psi_R$
system, but breaks down at higher scales.  As such, naturalness
requires $\Lambda \sim v$ so that the UV physics characterized by
$\Lambda$ does not destabilize the weak scale.  The Little Higgs
theories to be discussed in Section~\ref{sec:models_lhiggs} 
are a framework which extend this idea by providing an explanation
for a gap of $\Lambda \sim 4 \pi v$ while retaining naturalness.
Other theories have no light particles corresponding to
$\overline{\Psi}_L \Psi_R$, and are truly theories without a Higgs
boson.\bigskip

  Whatever the details of the composite particle spectrum, strong
coupling is required.  At tree level, one expects the coefficient of
the quadratic term in the effective potential for $\langle
\overline{\Psi}_L \Psi_R \rangle$ will be related to the masses of the
fermions. The corresponding coefficient is given by $(m_L + m_R)^2$
at tree level and is at best zero when the fermions in question 
are massless.  A VEV requires the quadratic term in the effective 
potential be less than zero.  In order to overcome the tree level 
expectation, large corrections from higher-order processes must 
drive the quadratic term negative; something that can only occur 
in the limit of strongly-coupled dynamics.

\subsubsection{Technicolor}

  Electroweak symmetry breaking in technicolor models follows
closely what we know about QCD and the breakdown of the approximate chiral 
$SU(2)_L\times SU(2)_R$ flavor symmetry of the theory with two flavors of
light quarks~\cite{Weinberg:1975gm,
Hill:2002ap}.  Because
QCD is asymptotically free, at energies below $\Lambda_\text{QCD}$ the
quarks form condensates, \ie two--quark operators will develop a
vacuum expectation value.  This operator spontaneously breaks the
chiral flavor symmetry into its diagonal isospin $SU(2)$ subgroup. 
Below $\Lambda_\text{QCD}$, composite color-singlet mesons and baryons 
become the relevant physical degrees of freedom, with masses are of the 
order of the nucleon masses.  The only light particles are the pions,
corresponding to Nambu Goldstone Bosons~(NGBs) from the breakdown of 
$SU(2)_L \times SU(2)_R$.  Their coupling strength (or \emph{decay rate}) 
is governed by $f_\pi \sim 100$~MeV, defined by 
$\left< 0 | j_\mu^5 | \pi \right> = i f_\pi p_\mu$.
In terms of $N_c$ and $\Lambda_\text{QCD}$ there
are scaling rules in QCD which are based on for example $\beta \propto
N_c$ (and which strictly speaking do not hold arbitrarily well):
\begin{equation}
f_\pi \sim \sqrt{N_c} \; \Lambda_\text{QCD} 
  \qquad \qquad
\left< \overline{Q} Q \right> \sim N_c \; \Lambda_\text{QCD}^3
  \qquad \qquad
m_\text{fermion} \sim \; \Lambda_\text{QCD}
\end{equation}

The chiral symmetry breaking of QCD results in $W$ and $Z$ masses on the
order of MeV.  The simplest technicolor models simply scale up 
$\Lambda_\text{QCD}$ to the TeV scale by postulating a new strong 
sector, typically with an $SU(N)$ gauge group with $N_f$ flavors 
of technifermions $Q_i$ which are fundamentals of $SU(N)$. 
Each flavor contains two Weyl-fermion components, reflecting a global 
$SU(N_f) \times SU(N_f)$ symmetry. Under the electroweak $SU(2)_L$ gauge 
group they have the usual left-right charge structure characteristic
of the quarks in the Standard Model.  The new $SU(N)$ interaction confines,
$\langle \bar{Q}_i Q_j \rangle \sim \Lambda_\text{TC}^3 \delta_{ij}$,
and breaks the electroweak symmetry.  A subset of the would-be
Goldstone bosons of the broken $SU(N_f) \times SU(N_f)$ flavor
symmetry serve as the longitudinal degrees of freedom for the $W$ and
$Z$ bosons.  All we have to ensure is that $f_\text{TC}$, the
scaled-up version of $f_\pi$, gives us the correct value for the
electroweak scale. The scaling rules read
\begin{equation}
f_\text{TC} \sim \sqrt{\frac{N}{N_c}} \, \frac{\Lambda_\text{TC}}{\Lambda_\text{QCD}} \; f_\pi
  \qquad \qquad
\sqrt{2} v = \sqrt{N_f} f_\text{TC} \sim  \sqrt{\frac{N_f N}{N_c}} \, f_\pi
\end{equation}
We can solve these relations for $f_\text{TC}$ and
$\Lambda_\text{TC}$.  
For example for $N=N_f=4$ and $\alpha_s(M_\text{GUT}) \sim 1/30$ we
find $\Lambda_\text{TC} \sim 800~\Lambda_\text{QCD} \sim
165$~GeV. This gives a reasonable $v = 190$~GeV and generates the
required hierarchy between $v$ and $M_\text{GUT}$ via dimensional
transmutation.\bigskip

  Unfortunately for technicolor, electroweak symmetry has a second
function: it also generates the fermion masses. To explain all fermion
masses with a composite condensate requires higher dimensional
operators that link the technicolor condensate to the Standard Model fermions, with a
specific operator for each of the vastly different fermion masses in
the Standard Model.  The most popular means to accomplish this task is
through extended technicolor
(ETC)~\cite{Dimopoulos:1979es}
which introduces a set
of gauge bosons with masses of the order of $\Lambda_{ETC}$ which
couple to both the technifermions and the Standard Model fermions.  At low
energies, integrating out the ETC bosons and Fierzing the four-fermion
operators results in operators such as
\begin{equation}
\frac{1}{\Lambda^2_\text{ETC}} \overline{Q} Q \overline{\psi}_L \psi_R + \text{h.c.}
\rightarrow m_\psi \overline{\psi}_L \psi_R + \text{h.c.} \; ,
\end{equation}
with $m_\psi \sim N \Lambda^3_\text{TC} / \Lambda^2_\text{ETC}$.  For
quark masses in the GeV range this means a reasonable
$\Lambda_\text{ETC} \sim 2$~TeV.  But to generate the top mass, 
the ETC scale cannot be much larger than the technicolor scale;
$\Lambda_\text{ETC} \sim \Lambda_\text{TC} \sim v$.  This is a problem,
since in order to generate the observed mixings in the quark and
lepton sectors of the Standard Model, the ETC bosons must encode 
nontrivial flavor structure in their coupling to Standard Model fermions
and can therefore induce dangerous new flavor-mixing effects.

  ETC gauge boson exchange also leads to four-fermion operators
involving just the Standard Model fermions, leading to flavor-violating effects
which are strongly bounded, naively requiring $\Lambda_\text{ETC}
\gtrsim 1000$~TeV~\cite{utfit}.  The most popular remedy to this
problem with FCNCs is to adjust the matter content charged under the
technicolor group such that it becomes approximately conformal, or
\emph{walks}~\cite{Holdom:1981rm}. This allows for large anomalous dimensions
to rescale the ETC interactions which involve technifermion bilinears,
enhancing them and allowing for larger choices of
$\Lambda_\text{ETC}$.  The dangerous SM-only FCNC interactions do not
feel the effects of technicolor, and are thus relatively suppressed.
The large top mass remains a challenge, even for modern theories of
walking technicolor, requiring multiple ETC scales, non-commuting family
structure, larger technifermion representations, and/or fundamental
scalars~\cite{Appelquist:1993sg,Appelquist:2007hu}.

  Extended technicolor models are further challenged by measurements
of the $Z$-$b$-$\bar{b}$ interaction~\cite{Chivukula:1992ap}.
The same ETC interactions which produce the top quark mass
operator also result in the vector operator 
\begin{equation}
\frac{1}{\Lambda_\text{ETC}^2}
\left( \overline{Q}_L \gamma_\mu \tau^a \psi_{3L}   \right) \;
\left( \overline{\psi}_{3L}   \gamma^\mu \tau^a Q_L \right) \;
\rightarrow 
\frac{1}{\Lambda_\text{ETC}^2}
\left( \overline{Q}_L \gamma_\mu \tau^a Q_{L}   \right) \;
\left( \overline{\psi}_{3L}   \gamma^\mu \tau^a \psi_{3L} \right) \;
\end{equation}
including techni-quarks $Q$ and the third generation Standard Model quark
doublet $\psi_{3L}$.  The second expression is
obtained by a Fierz transformation.
Since this operator is also responsible for the top mass,
even in a multi-scale ETC model, the relevant scale is 
$\Lambda_\text{ETC} \sim v$.
Replacing the vector operator of techniquarks with the corresponding 
techni-pion operator,
\begin{equation}
\overline{Q}_L \gamma_\mu \tau^a Q_{L} 
\rightarrow \frac{v^2}{2} \text{Tr}~\Sigma^\dagger \tau^a i D_\mu \Sigma
\end{equation}
where $\Sigma = \text{Exp}~ 2 i \pi_T^a \tau^a / v$, and moving to the
unitary gauge, $\Sigma \rightarrow 1$, results in a
correction to the $Z$-$b$-$\bar{b}$ interaction of
$v^2 / (2 \Lambda^2_\text{ETC}) \sim 1$, generically in conflict
with the experimental measurement of
$R_b = \Gamma_Z(b\bar{b})/\Gamma_Z(\text{hadrons})$.

Non-perturbative symmetry breaking dynamics at the weak scale are also a
challenge to other precision electroweak measurements.  In terms of the
oblique parameters~\cite{Peskin:1990zt},
the generic
expectation is that $\Delta T$ will be of order one, as expected in
the limit of a heavy Higgs boson, and $\Delta S$ will depend on the
number of technifermion doublets as $\Delta S = N_f/(6
\pi)$~\cite{Peskin:1990zt,Holdom:1990tc}, in clear conflict with
electroweak measurements~\cite{Alcaraz:2006mx}.  Corrections to
$\Delta T$ can be controlled with an $SU(2)$ custodial symmetry, and
their sign depends on the relative contribution of the new up-type and
down-type fermions.  In contrast, the new particles' corrections to
$\Delta S$ tend to be positive, unless we apply drastic changes to
the Standard Model setup.  Non-standard effects on $\Delta S$ are
notoriously difficult to compute for near-conformal theories and may
be controlled by a variety of means~\cite{Golden:1990ig}. An example
of computable negative corrections to $\Delta S$ arises in
the framework of Higgsless theories to be discussed 
in Section~\ref{sec:models_hless}.

  The detailed spectrum of a technicolor model depends on the
technifermion sector, and will typically include massive vector and
scalar particles, some of which may be colored if some of the
technifermions are charged under $SU(3)$~\cite{Eichten:1986eq,Eichten:2007sx,Hirn:2007we}.  
The most generic collider signals are related to heavy 
vector particles (the techni-rhos) which are largely responsible 
for unitarizing high
energy $W$-$W$ scattering
.  As a result, they must have large coupling
to the longitudinal $W$ and $Z$ bosons, but may have small or
non-universal coupling to fermions.

\subsubsection{Topcolor-assisted technicolor}

Top-color-assisted technicolor models provide an alternate means to
generate the large top mass, addressing the single largest defect of a
typical technicolor model~\cite{Hill:1991at,Hill:1994hp}.  These theories
build on the idea of top condensation~\cite{Bardeen:1989ds},
dynamically generating a Higgs-like scalar with a vacuum expectation 
value that explains part of electroweak symmetry breaking and has large 
coupling to the top.  The theory doubles the color gauge sector of the 
Standard Model to $SU(3)_1 \times SU(3)_2$.  The top quark 
(both $Q_3$ and $t_R$) are
triplets under $SU(3)_1$ and the light quarks are triplets under
$SU(3)_2$.  At the scale of a few TeV, this structure breaks down
through some usually unspecified dynamics, resulting in an unbroken
Standard Model color $SU(3)_c$ and a massive color octet of topgluons.
The couplings are usually chosen such that the topgluons are mostly
massive $SU(3)_1$ bosons, and the ordinary gluons are mostly the
$SU(3)_2$ gluons.  This choice results in the topgluons coupling
strongly to top quarks and weakly to the light quarks.  The unbroken
$SU(3)_c$ symmetry guarantees that the massless bosons couple
universally to all of the quarks.\bigskip

  The effects of the topgluon below its mass 
are  encapsulated by contact interactions, the most important of which is
\begin{equation}
\label{eq:models_4top}
-\frac{g^2}{M^2} \left[ \bar{t}_R \gamma^\mu T^a t_R \right] \left[\bar{Q}_L \gamma_\mu T^a Q_L \right]
= -\frac{g^2}{M^2} \left[ \bar{t}_R Q_L \right] \left[ \bar{Q}_L t_R \right]
\end{equation}
where $g$ is the topgluon coupling to top quarks and $M$ is its mass.
$Q_L$ is the third family quark doublet 
and the last equality is easily understood as a Fierz transformation.  

Four-fermion interactions of this kind, but involving the light quarks,
have been extensively studied as a low energy model of QCD 
and chiral symmetry breaking, the Nambu-Jona Lasinio model~\cite{Nambu:1961tp}.
If $g$ is large enough, this interaction results in scalar bound states. 
To study these we can write down 
an effective theory for the four-fermion interaction
in terms of an auxiliary field $\Phi$,
\begin{equation}
\mathscr{L}(\Lambda ) = \frac{1}{2} g_t \bar{Q}_L t_R \Phi + g_t^* \bar{t}_R Q_L \Phi^* - \Lambda^2 |\Phi|^2
\label{eq:effphiL}
\end{equation}
where $\Phi$ transforms as a Standard Model Higgs, but has no kinetic terms
at the scale $\Lambda$.  We require $g^2 / M^2 = g_t^2 / \Lambda^2$ to match
to the topgluon operator.\bigskip

Below the scale $\Lambda$, kinetic terms for $\Phi$ as well as a 
quartic interaction are induced at one loop, and the parameters 
$\Lambda^2$ and $g_t$ are renormalized. 
The form of the effective theory at the scale $\Lambda$ determines
the renormalization constants
$Z (\Lambda ) = 0$ and 
$m^2 (\Lambda ) = \Lambda^2$. The Lagrangian then reads
\begin{equation}
\mathscr{L}(\mu ) =  Z(\mu ) \left( D^\mu \Phi \right)^2
- m^2 (\mu ) | \Phi |^2 - \lambda (\mu ) |\Phi|^4 
+ \frac{1}{2} g_t(\mu ) \left( \bar{Q}_L t_R \Phi + \text{h.c.} \right)
\end{equation}
where
\begin{alignat}{5}
Z (\mu ) & = N_c \frac{g_t^2}{16\pi^2} \log \left( \frac{\Lambda}{\mu} \right) + ... \notag \\
m^2 (\mu )& = \Lambda^2 - N_c \frac{g_t^2}{8\pi^2} \Lambda^2 + ...
\label{eq:models_topcol}
\end{alignat}
with similar expressions for $g_t (\mu )$ and $\lambda ( \mu )$.  
Note that beyond a critical coupling value 
$g_t^2 > 8\pi^2/N_c \equiv g_\text{crit}^2$ the scalar mass squared
$m^2 (\mu )$ becomes negative, and the scalar field develops a 
vacuum expectation value, $\langle \Phi \rangle = v_t$,
breaking the electroweak symmetry.

  Its kinetic terms are canonically normalized by
scaling $\Phi \rightarrow Z^{-1/2}\Phi$.  In terms of the canonically 
normalized field the prefactors in the Lagrangian 
Eq.~(\ref{eq:models_topcol}) become
$m^2 \rightarrow Z^{-1} m^2$, $\lambda \rightarrow Z^{-2} \lambda$, and
$g_t \rightarrow Z^{-1/2} g_t$.  A useful approximation to determine 
the physics at the electroweak scale is to fix these parameters such that
$m^2 (\Lambda ) \sim \infty$ and $g_t (\Lambda ) \sim \lambda (\Lambda) 
\sim \infty$, since all of these are finite in the unnormalized theory 
at $\Lambda$ and going to the canonical normalization divides each by 
a positive power of $Z (\Lambda ) \simeq 0$ at that scale. 
We then use the renormalization group to determine the
parameters at scales $\mu \leq \Lambda$.\bigskip

Far below $\Lambda$, the theory looks like the Standard Model with the
Higgs potential parameters $m^2$ and $\lambda$ and the top Yukawa
coupling $g_t$ all predicted in terms of the original parameters
$\Lambda$ and $g_t$, provided $g_t > g_\text{crit}$ so the electroweak
symmetry is broken.  The top mass is then given by $m_t = g_t
v_t$. Since part of electroweak symmetry breaking also arises from
technicolor, the topcolor vacuum expectation value cannot account for the entire amount of
symmetry breaking, \ie $v_t < v$.  Therefore, $g_t$ is larger than the
SM top Yukawa coupling, implying enhanced rates \eg for single Higgs
production $gg \rightarrow h^0$~\cite{Hashimoto:2002cy}.  In addition,
the longitudinal $W$ and $Z$ are now mixtures predominantly composed
of the technifermion Goldstone bosons, but with some components from
$\Phi$.  The orthogonal mixtures are physical scalar states with large
couplings to top, the \emph{top pions}~\cite{Balazs:1998nt}.

  At the LHC, a topgluon can be produced by $q \bar{q}$ annihilation.
While the coupling of the topgluon to light quarks is not huge, the
fact that the light quark parton densities are sizeable at large $x$
values usually renders this the dominant production mechanism.  Once
produced, the large coupling to top quarks dictates a dominant topgluon decay
into a $t \bar{t}$ pair.  Thus, the signature is a resonant structure in
the invariant mass of $t \bar{t}$ pairs.  While not as clean as a
decay into leptons, decays into tops still have a lot of potential
compared to backgrounds, because top decays produce jets enriched with
bottom quarks which can be tagged and also a fair fraction of leptonic
$W$ decays, see also Section~\ref{sec:sig_top}.

\subsubsection{Top seesaw}

 One might be tempted to abandon the technicolor part of
topcolor-assisted technicolor, and try to use topcolor as the entire
mechanism for electroweak symmetry breaking.  That is, in fact, what
the original top condensation models set out to
do~\cite{Bardeen:1989ds}.  However, in practice, low scale topcolor
generally has difficulty getting the right amount of electroweak
symmetry breaking and the correct top mass.  The reason is that full
electroweak symmetry breaking with $v_t = 174$~GeV requires very large
$g_t$ in Eq.~(\ref{eq:effphiL}), which in turn predicts a top mass in
excess of 700~GeV.  A simple fix is provided by the topcolor
top-seesaw models~\cite{Dobrescu:1997nm}.\bigskip

 The top-seesaw introduces an additional vector-like quark $\chi$. Its
right-handed component is charged under $SU(3)_1$ while its
left-handed component and $t_R$ are charged under $SU(3)_2$, along
with the rest of the light Standard Model quarks.  There is a
gauge-invariant mass term linking $\chi_L$ and $t_R$, and a term
induced by $SU(3)_1 \times SU(3)_2$ breaking which links $\chi_L$ and
$\chi_R$.  The four-fermion operators induced by the topgluon now
contain $Q_{3L}$ and $\chi_R$, and the Higgs is largely a composite of
those two fields. Its vacuum expectation value induces a mass term 
linking $Q_{3L}$ and $\chi_R$ of size $g_t v \sim 700$~GeV.  
The top and $\chi$ masses may then be written in matrix form
\begin{equation}
\left( \bar{t}_L ~~ \bar{\chi}_L \right) \; 
\left( \begin{array}{cc}
     0 & g_t v \\ M_{t \chi} & M_{\chi \chi}
       \end{array}
\right) \; 
\left( \begin{array}{c}
t_R \\ \chi_R
       \end{array}
\right) + \text{h.c.}
\end{equation}
By adjusting the values of $M_{t \chi}$ and $M_{\chi \chi}$, one can
fix the lighter eigenstate to have mass $m_t$, which means we can
identify it with the top. There is also a heavier eigenstate with a
mass in the TeV range.  At scales below $\Lambda$, the theory thus
contains the Standard Model including a heavy composite Higgs doublet
whose would-be Goldstones are now completely eaten, plus the TeV-scale
quark.  This heavy quark provides a relatively large contribution to
$\Delta T$ and only a small contribution to $\Delta S$, and can
balance the heavy Higgs, resulting in a reasonably good fit to
precision data provided $\Lambda \gtrsim
1$~TeV~\cite{Dobrescu:1997nm}.\bigskip

The ingredients of a top-seesaw theory occur naturally in extra 
dimensional models, with the role of the topgluon played by the KK gluons, 
and $\chi$ by the KK modes of $t_R$~\cite{Dobrescu:1998dg}.
The vector-like quark can be pair-produced at the LHC, 
or singly produced through its mixing with top, as in $t$-channel 
single top production.  Decays into $W b$, $Z t$, 
are kinematically allowed for typical regions of parameter space, 
and the decay to a Higgs $h t$ is present when the Higgs is light enough.  
In the limit of $m_{\chi} \gg m_t, m_h$, these occur in the
ratio $2:1:1$, much as in the case of the vector-like top partner quark 
found in the Little Higgs theories to be discussed next.

\subsection{Little Higgs models}
\label{sec:models_lhiggs}

  As we have seen in the preceding sections, when the Standard Model is
extended to address the gauge hierarchy problem new experimental conflicts 
frequently appear.
Without dedicated symmetries, new physics near the $\tev$ scale will 
typically generate flavor and CP violation 
well beyond what we observe, and it can alter the values 
of precision electroweak observables in conflict with experiment.  
To avoid these constraints, the mass scale of new physics typically 
has to be greater than $10\,\tev$ or more.  However, new physics
this heavy doesn't do a very good job at stabilizing the electroweak
scale near $200\,\gev$, and some degree of fine-tuning is needed.  
While this is not necessarily disastrous, and the fine-tuning involved 
is much smaller than in the standard gauge hierarchy problem, 
this \emph{little hierarchy problem} is a puzzling feature 
when naturalness was the motivation for introducing new physics 
in the first place.  

    Little Higgs~(LH) models are extensions of the Standard Model 
designed specifically to address the little hierarchy 
problem~\cite{ArkaniHamed:2001nc}.
In other words, we do not require them to complete as a weakly interacting 
renormalizable field theory up to some kind of GUT scale.
The key idea is to keep the Higgs very light
relative to the new physics by making it a Nambu-Goldstone 
boson~\cite{Georgi:1974yw}
of an approximate global symmetry. This is the standard way to protect 
small masses in field theories.  Note that if the symmetry 
is not exact, the would-be NGB does not have to be massless.
Unfortunately, this general idea turns out to be problematic,
and an additional property of \emph{collective symmetry breaking} 
is needed to keep the Higgs significantly lighter than the new states 
beyond the Standard Model~\cite{ArkaniHamed:2001nc}.
Similar to supersymmetry, this hierarchy is maintained by a 
cancellation of one-loop quadratic divergences between loops of 
Standard Model particles with loops involving new particle states.  
In contrast to SUSY, however, these new states have the same spin 
as their Standard Model 
partners~\cite{Schmaltz:2005ky,Perelstein:2005ka}.\bigskip

  LH models are usually written in the language of non-linear 
sigma models\,(NLSM)~\cite{Coleman:1969sm}.
These are an
efficient way to describe the interactions of Nambu-Goldstone bosons
that arise from the spontaneous breaking of a continuous global symmetry $G$
down to a subgroup $H$.  A NLSM is specified by the scale of symmetry 
breaking $f$ and the pattern of symmetry breaking $G\to H$.  
We then obtain a $G/H$ model as an effective field theory 
of the NGBs encoding the full symmetry structure of the underlying 
high scale theory.  Given a symmetry breaking scale $f$, 
the NLSM effective field theory is valid up to
a UV cutoff on the order of $\Lambda \simeq 4\pi f$.  
The physics above the cutoff $\Lambda$ is not specified other than
it is assumed to respect the symmetry group $G$ and to spontaneously 
break to $H$ at the energy scale $f$. It may even involve strong 
interactions and technicolor-like dynamics or 
supersymmetry~\cite{Katz:2003sn}.
\bigskip

  Early attempts to realize the Higgs boson as an approximate NGB
encountered serious problems~\cite{Georgi:1974yw}.  The underlying global 
symmetry $G$ must be explicitly broken to generate an acceptable 
potential for the Higgs and induce electroweak symmetry breaking.  
Since the Higgs should arise from an electroweak doublet, this can 
be achieved by gauging the $SU(2)_L\times U(1)_Y$ subgroup 
of a larger group $G$. This setup explicitly breaks $G$ because some components 
of $G$ multiplets couple to $SU(2)_L\times U(1)_Y$ gauge bosons 
while others do not.  Radiative corrections, including 
quadratically divergent gauge boson corrections to the Higgs mass 
cut off at $\Lambda = 4 \pi f$, then induce a potential for the 
Higgs leading to an electroweak breaking scale at 
$v \sim f$.  On the other hand, precision electroweak 
constraints push $f\gtrsim 1\,\tev$, so the minimal models is killed 
by the little hierarchy problem.

  The source of difficulties in these early NGB Higgs models are 
quadratically divergent one-loop corrections to the Higgs mass parameter 
from the electroweak gauge bosons and the top quark.  LH models avoid them
through the mechanisms of
collective symmetry breaking~\cite{ArkaniHamed:2001nc,ArkaniHamed:2002pa}.  
In product group LH models, a $G_1\times G_2$ subgroup of the 
global symmetry group $G$ is gauged, where each $G_i$ factor contains 
a copy of the electroweak group.  
Upon breaking $G\to H$, the gauged subgroup is broken to the 
$SU(2)_L\times U(1)_Y$ electroweak group of the Standard Model.  
Note that by gauging a subgroup of $G$, the global symmetry is 
explicitly broken and the would-be NGBs from the spontaneous symmetry 
breaking at $f$ acquire masses.  The Higgs is embedded in this symmetry 
structure as an exact NGB if one or 
the other of the gauge couplings $g_1$ or $g_2$ of $G_1\times G_2$ 
vanishes~\cite{ArkaniHamed:2001nc,ArkaniHamed:2002pa}.
Put the other way around, both $g_1$ and $g_2$ must be non-zero for the Higgs
boson to develop a potential, and they collectively break the global
symmetry group $G$. As an alternative we can also build
simple group LH models that realize collective symmetry
breaking in a slightly different way~\cite{Schmaltz:2005ky}.\bigskip

  It is straightforward to check that quadratically divergent one-loop
gauge boson contributions to the Higgs mass parameter must be proportional to
either $g_1^2$ or $g_2^2$. In contrast, collective symmetry breaking ensures 
that electroweak loop corrections to the Higgs potential depend only 
on the product $g_1g_2$.  Therefore the dangerous quadratic gauge boson 
corrections are absent at one-loop order. In terms of the gauge boson 
mass eigenstates, the lack of quadratic divergences can be interpreted
as a cancellation between Standard Model gauge boson loops and new gauge bosons 
from the extended gauge group, which during symmetry breaking acquire 
masses of order $g\,f$.  The global symmetries ensure that the 
couplings of these states indeed lead to a cancellation. This is analogous
to SUSY, where the cancellation occurs between loops of bosons and fermions.
The new feature of LH models is that this cancellation
occurs between fields having the same spin.\bigskip

\begin{figure}[t]
\begin{center}
  \includegraphics[width=0.6\textwidth]{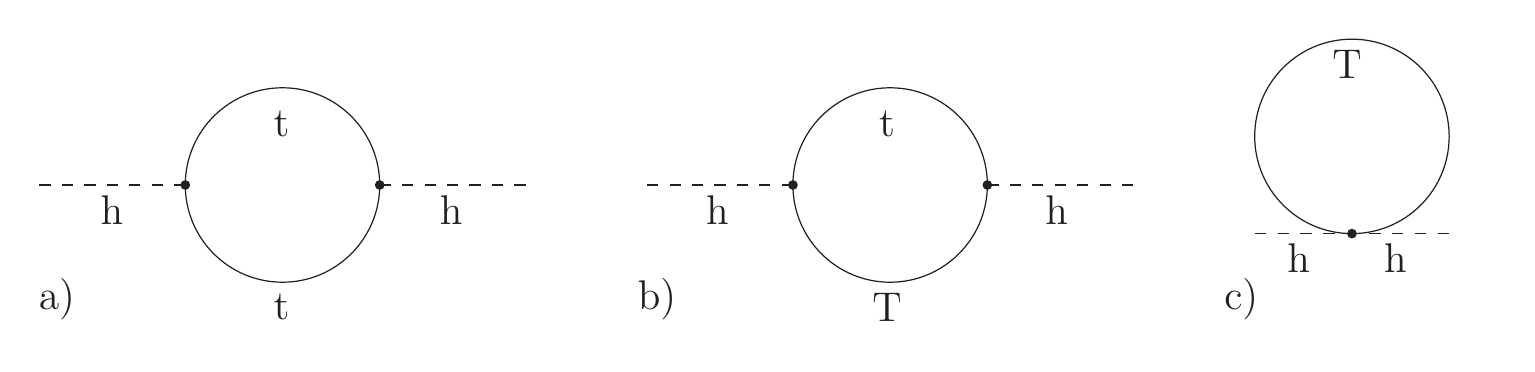}
\end{center}
\vspace*{0cm}
\caption{Cancellation of the top-quark quadratic correction to the 
Higgs mass by loops of the top-partner $T$ fermion in LH theories.
Figure from Ref.~\cite{Perelstein:2003wd}.}
\label{fig:lh-tprime}
\end{figure}

Dealing with gauge boson loops through the symmetry structure unfortunately 
does not cure all problems of a NGB Higgs boson.
Coupling the top to the Higgs also requires an
explicit breaking of the underlying global symmetry. We
introduce this coupling in a way that respects the collective 
symmetry breaking structure by adding new
vectorial top-partner $T$ fermions. 
This ensures that quadratically divergent top-sector corrections 
to the Higgs mass are absent at one-loop order. In terms of the fermion 
mass eigenstates, this arises 
as a cancellation between the top quark loop with loops of the heavy 
top partner $T$ states which again has a mass of the order 
$f$.  We illustrate the corresponding diagrams 
in Fig.~\ref{fig:lh-tprime}.

  While collective symmetry breaking forces the cancellation of quadratic
divergences at one-loop order, it does not remove logarithmic 
one-loop divergences or quadratic divergences at higher loop orders
This is not a problem since these corrections are numerically small 
enough that they do not destabilize the electroweak scale within the 
NLSM, which is valid only up the scale $\Lambda = 4\pi f$.  
In fact, the one-loop logarithmic corrections are essential to
generate a potential for the Higgs field.  This potential can induce 
electroweak symmetry breaking for certain ranges of model parameters.
The leading contribution in this case typically comes from loops
of the top quark and its heavy partners~\cite{ArkaniHamed:2002pa}.\bigskip

Going beyond these generic structures, the precise particle content and 
phenomenology of LH models depends on the pattern of symmetry breaking.  
Popular examples include the \emph{minimal moose} model based on  
$[SU(3)_L\times SU(3)_R/SU(3)_V]^4$
with a gauged $SU(3)\times SU(2)\times U(1)$ 
subgroup~\cite{ArkaniHamed:2002qx},
the \emph{littlest Higgs} model with $SU(5)/SO(5)$ global symmetry and
a gauged $[SU(2)\times U(1)]^2$ subgroup~\cite{ArkaniHamed:2002qy},
and the \emph{simplest LH} model based on $[SU(3)\times U(1)/SU(2)]^2$ 
with a gauged diagonal $SU(3)\times U(1)$ 
subgroup~\cite{Kaplan:2003uc}.
For some other examples of LH models, 
see Refs.~\cite{Gregoire:2002ra}.
Despite a wide variety of constructions, all models share 
some common features:
to ensure the cancellation of one-loop quadratic corrections to 
the Higgs mass, new gauge boson and top quark partners with masses 
around $f$ must appear in the spectrum. Embedding the 
Higgs boson into a multiplet of NGBs also gives rise to additional 
scalar states, often carrying electroweak charges.\bigskip

  The strongest current bounds on LH models come from precision 
electroweak constraints. The new heavy gauge bosons 
mix with the Standard Model electroweak gauge bosons and modify their masses
and couplings to matter fields, and condensing electroweak triplet scalars
modify the pattern of electroweak symmetry 
breaking~\cite{Csaki:2002qg,Hewett:2002px,Han:2003wu}.  
To satisfy the precision electroweak data,
the symmetry breaking scale $f$ 
must typically be larger than several $\tev$, although values as low as
$1\,\tev$ are possible within restricted parameter regions of 
some models~\cite{Csaki:2002qg}. 
When the scale $f$ grows larger than a $\tev$, we need a fine tuning 
to stabilize the value of electroweak symmetry breaking, so 
we yet again run into a little hierarchy problem.
This tension has motivated several extensions of the LH models
described above. For example, the minimal moose model can be extended 
to include an explicit $SU(2)_C$ custodial symmetry~\cite{Chang:2003un},
as can the littlest Higgs model~\cite{Chang:2003zn}. 

  A more radical approach to the precision electroweak tension in
LH models is to introduce a new discrete symmetry called
$T$ parity~\cite{Cheng:2003ju,Cheng:2004yc}
Under this $\mathbb{Z}_2$ symmetry, which acts by exchanging the product 
gauge groups $G_1$ and $G_2$, 
the Standard Model gauge bosons are even while the new gauge bosons and scalars 
are (generally) odd. 
This symmetry forbids tree-level corrections to the electroweak observables 
from gauge boson mixing.  Instead, the leading electroweak corrections 
only arise at loop level, and values $f < 1$~TeV 
are be consistent with current 
data~\cite{Hubisz:2005tx}.  
Adding fermions while preserving $T$-parity requires a doubling of 
the spectrum of $SU(2)_L$ doublet states.  The particle spectrum 
below the scale $f$ consists of $T$-even Standard Model fermions, 
along with a set of heavy $T$-odd partner states.
In the case of the top quark and its partners,
additional vector $SU(2)_L$ singlet states must also be included
to preserve the collective symmetry breaking structure.
Depending on how this is done, the lighter top quark partners
can consist exclusively of $T$-odd states, or additional massive
$T$-even states may be present as well~\cite{Hubisz:2004ft,Cheng:2005as}. 
Interestingly, the additional contribution of the heavy top partners
to the precision electroweak fit allows the Higgs to be heavier than
several hundred $\gev$ in some regions of the parameter 
space~\cite{Hubisz:2005tx}.

  Among the many new states in LH models with $T$ parity, 
the lightest $T$-odd particle~(LTP) is stable to the extent that 
$T$-parity is conserved.   In the littlest Higgs model with $T$-parity,
the LTP is typically the $T$-odd hypercharge gauge boson partner $B_H$,
in turn providing a viable dark matter 
candidate~\cite{Hubisz:2004ft,Birkedal:2006fz}.
We should point out, however, that in some UV completions of LH 
theories with $T$ parity, this new discrete symmetry is 
anomalous~\cite{Hill:2007zv}.  The operator induced by the anomaly
leads to a prompt decay of the LTP to Standard Model particles ruling it out as
dark matter candidate in these models.  Even so, the anomalous $T$-parity
remains effective at loosening the electroweak constraints.
It is also possible to extend the model structure to remove 
$T$-parity anomalies~\cite{Krohn:2008ye}.
Unfortunately, it is not possible to implement
$T$-parity in simple group LH models~\cite{Schmaltz:2005ky}.\bigskip

  The collider signatures of LH theories depend crucially 
on whether the model has a $T$ parity or not.  Without $T$ parity
the new partner states
can be singly produced, and can decay directly to Standard Model particles. 
At the same time, electroweak constraints force these states to be
very heavy, with masses often well above a $\tev$.

In the littlest Higgs model there are exotic heavy charged 
$W_H^{\pm}$ vectors along with $W_H^3$ and $B_H$ 
neutral gauge bosons.  Mixing among the neutral states will produce
mass eigenstates $Z_H$ and $A_H\,(=V_H)$.  However, the mixing angle
is typically very small, going like $v^2/f^2$, so it is usually a
good approximation to identify $Z_H \simeq W_H^3$ and $A_H \simeq B_H$.
These exotic vector bosons are produced at the LHC primarily through
Drell-Yan ($q\bar{q}$ annihilation), and decay to Standard Model fermions or 
electroweak boson states~\cite{Han:2003wu,Burdman:2002ns}.  
For example, the $W_H^3$ state has the decay 
widths~\cite{Perelstein:2003wd,Han:2003wu,Burdman:2002ns}
\begin{alignat}{5}
\Gamma(W_H^3\to \ell^+\ell^-) &= \frac{g^2\cot^2\theta}{96\pi}M_{W_H^3}
&\Gamma(W_H^3\to q\bar{q}) &= \frac{g^2\cot^2\theta}{32\pi}M_{W_H^3}
\notag \\
\Gamma(W_H^3\to Zh) &= \frac{g^2\cot^22\theta}{192\pi}M_{W_H^3} \qquad
&\Gamma(W_H^3\to W^+W^-) &= \frac{g^2\cot^22\theta}{192\pi}M_{W_H^3}
\label{lhbrw}
\end{alignat}
while the charged exotic vector decays through
$W_H^{-}\to \ell\bar{\nu},\;q\bar{q}',\;W^{-}h$.
The LHC reach for these states as a function of their masses and the model 
mixing parameter $\theta$ 
is shown in Fig.~\ref{fig:lh-lhc-gauge}.  
This reach is large enough that the LHC will cover regions of LH model
parameter space that have not yet been probed by the precision 
electroweak analyses.  By measuring cross sections and branching ratios
it may also be possible to study the 
underlying LH global symmetry structure~\cite{Perelstein:2003wd,Burdman:2002ns,
Azuelos:2004dm,Han:2005ru}.  

\begin{figure}[t]
\begin{center}
  \includegraphics[width=0.45\textwidth]{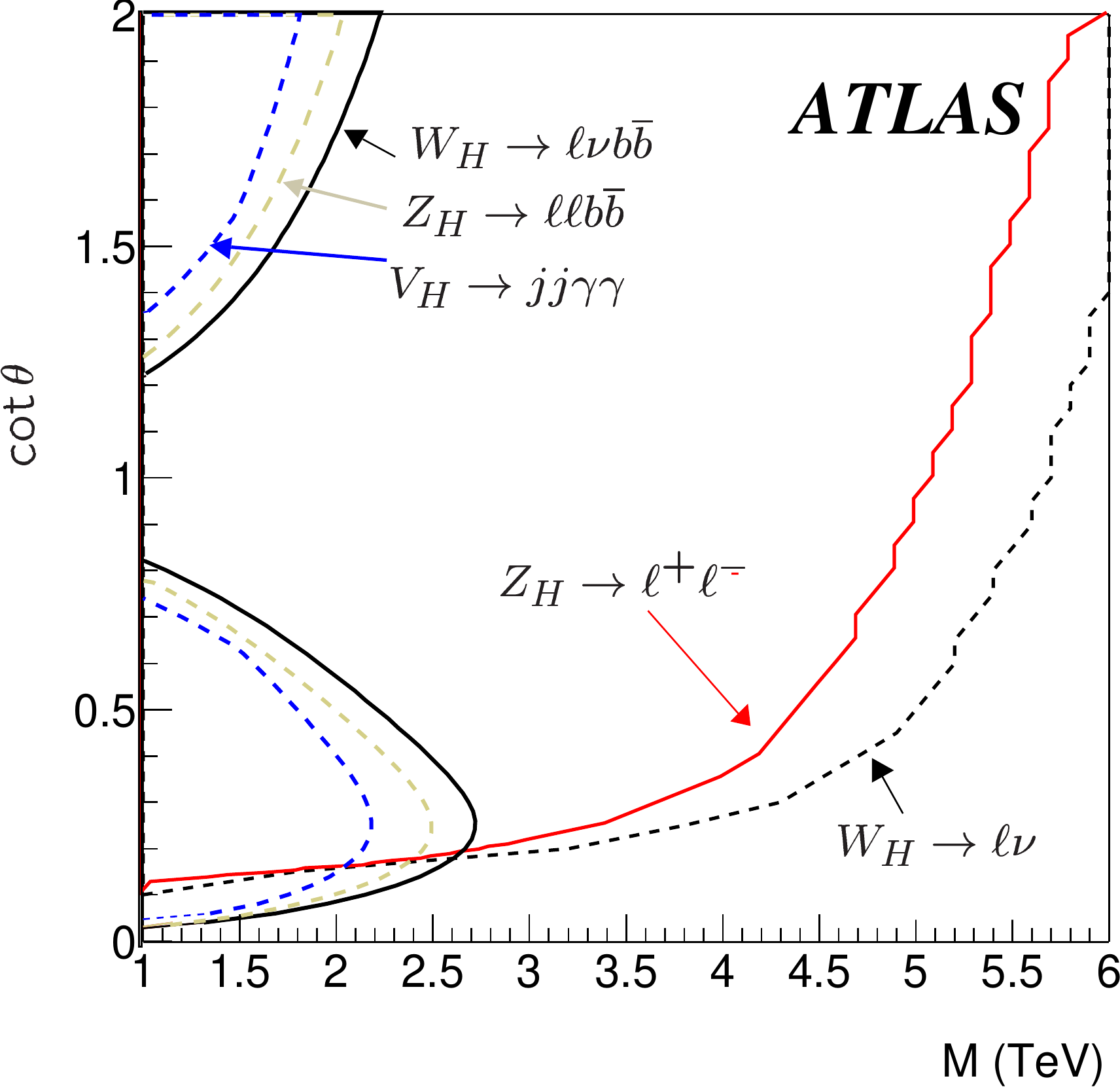}
\end{center}
\vspace*{-0.5cm}
\caption{ATLAS reach for the massive electroweak vector bosons in the
littlest Higgs model with $300\,\ifb$ of data.  
Figure from Ref.~\cite{Azuelos:2004dm}.}
\label{fig:lh-lhc-gauge}
\end{figure}

   The LHC discovery reach for the heavy top partner $T$ quark 
is more limited, extending to about $2.5\,\tev$
in the littlest Higgs model~\cite{Azuelos:2004dm}. 
Naturalness of the electroweak scale
strongly favors $T$ masses that are not too heavy.
Even so, it often turns out that the $T$ quark is heavy enough that 
single production through the $t$-channel exchange of a $W$ boson 
beats the QCD pair 
production rate~\cite{Han:2003wu,Perelstein:2003wd}.  
Once created, the $T$ quark can decay back to Standard Model states
through the modes $T\to th$, $T\to Z^0t$, and $t\to W^{+}b$,
but may also decay to lighter LH exotics.
The branching ratios are related 
to the underlying global symmetry structure~\cite{Perelstein:2003wd}.
For the littlest Higgs model we find
\begin{equation}
\Gamma(T\to th) = \Gamma(T\to tZ) = \frac{1}{2}
\qquad \qquad
\Gamma(T\to bW^+) = \frac{\lambda_T^2}{64\pi^2}M_T \; ,
\label{lhbrt}
\end{equation}
where $\lambda_T$ is the effective $T$-Higgs Yukawa coupling.
As for the gauge bosons, the underlying LH structure of 
the theory can be probed by measuring these $T$-quark
branching ratios~\cite{Perelstein:2003wd,Han:2005ru}.

Finally, in the Higgs sector, a distinctive doubly charged $\phi^{++}$ scalar
is present in many models including the littlest Higgs.  This state can
be produced at the LHC by in weak boson fusion.
Its most spectacular decay goes to like-sign leptonic $W$ 
bosons~\cite{Han:2003wu,Azuelos:2004dm}.  
As in any extended Higgs or scalar model, the light SM-like Higgs boson 
can mix with new heavy 
neutral scalar states modifying its couplings by corrections 
on the order of $v^2/f^2$~\cite{Han:2003gf}.
\bigskip

  The collider signatures of LH models change drastically if there
is a conserved $T$ parity.  All $T$-odd partners must 
be produced in pairs, and the LTP is stable on collider time scales.  
When a pair of $T$-odd states is produced, they will each decay
through a cascade down to the LTP giving rise to missing energy signatures
analogous to the those occurring in supersymmetry with $R$-parity
or UED with KK parity and discussed in detail in Section~\ref{sec:sig_met}.
$T$-parity also allows the new partner states to be much lighter
(than without the symmetry) while remaining consistent with precision
electroweak constraints.

  The $T$-odd top partner state $T_-$ is created primarily through
QCD pair production.  These will cascade down to the LTP, 
typically emitting top quarks along the way and giving rise to
top-rich signatures with missing energy~\cite{Hubisz:2004ft,Cheng:2005as,Matsumoto:2008fq}.  
In the littlest Higgs model with
$T$-parity, the LTP is the hypercharge gauge boson partner $B_H$,
and the odd top partner $T_-$ generally decays through 
$T_-\to t\,B_H$~\cite{Hubisz:2004ft}, although the $T\to W_H^+\,b$ and
$T_-\to W_H^3 t$ channels can open up for certain implementations of
the top partners~\cite{Cheng:2005as}.  Some models also contain
a $T$-even top partner $T_+$ whose behavior is similar to the top
partner without $T$-parity except that $T_+\to T_-B_H$ can sometimes
occur as well~\cite{Hubisz:2004ft,Cheng:2005as}.

  In addition to the top partners, electroweak processes can produce pairs 
of the $T$-odd vector boson partners $W_H^{\pm}$, $W_H^3$, and $B_H$, 
as well as $T$-odd scalar partners at the LHC.  With $B_H$ being the LTP, 
the dominant decay modes of the heavier vector partners are 
$W_H^{\pm} \to W^{\pm}\,B_H$ and $W_H^3 \to h\,B_H$~\cite{Hubisz:2004ft}.  
These decays can mimic chargino and neutralino decays in the MSSM.  
The decay modes of the scalar partners are more model-dependent.  
Additional signals can arise from the heavy $T$-odd $SU(2)_L$ 
doublet partners of the other quarks and leptons~\cite{Carena:2006jx}.
When $T$-parity is anomalous, the LTP will generally decay promptly 
to Standard Model states. For instance, in the littlest Higgs model 
the $B_H$ vector LTP then decays quickly to $W^+W^-$ 
or $Z^0Z^0$~\cite{Barger:2007df}.\bigskip

Beyond LH models, a number of related scenarios have been proposed
that also realize the Higgs as an approximate NGB.  In \emph{Twin Higgs}
models the particle content of the Standard Model is doubled by including
an independent mirror sector with the same field content as the Standard Model.
Both sectors are related by a softly broken
$\mathbb{Z}_2$ exchange symmetry. 

The Higgs sectors of both copies are also embedded in a global
symmetry structure with a gauged product
subgroup~\cite{Chacko:2005pe,Barbieri:2005ri,Chacko:2005vw}.  For
example, the original model consists of a global $SU(6)\times
SU(4)\times U(1)$ group in the Higgs sector, with a gauged
$[SU(3)\times SU(2)\times U(1)]_A\times [SU(3)\times SU(2)\times
  U(1)]_B$ subgroup~\cite{Chacko:2005pe,
  Barbieri:2005ri,Chacko:2005vw}.  These two factors make up the gauge
groups of each of the two sectors, and their coupling strengths are
(approximately) equal to each other due to the exchange symmetry.  The
first $A$ copy is identified with the Standard Model and the second
$B$ copy with the mirror sector.  A portion of the $B$-sector
symmetries are assumed to be spontaneously broken at the scale $f$,
and a light Higgs boson in the $A$-sector then arises as a pseudo-NGB.
The approximate global symmetry together with the exchange symmetry
ensure that it does not receive any quadratically divergent loop
corrections to its mass at one-loop order.

  In terms of Feynman diagrams, the cancellation of quadratic Higgs mass 
corrections occurs between contributions from particles having the 
same spins as their Standard Model counterparts, but charged under 
a different set of gauge symmetries.
A similar structure arises in \emph{folded supersymmetry} scenarios,
except that here the quadratic Standard Model corrections to the Higgs
mass are cancelled by loops of superpartners in the mirror
sector~\cite{Burdman:2006tz}.
\bigskip

  The phenomenology of twin and folded Higgs scenarios depends very 
much on the underlying global and gauge symmetry structure of the model.
Much of the new mirror sector can be largely decoupled and difficult
to probe at the LHC~\cite{Chacko:2005pe}.  Other constructions, 
however, contain new vector-like quarks and additional vector bosons
and scalars~\cite{Chacko:2005vw}, which can provide interesting
new signals~\cite{Goh:2006wj}.  A more unusual feature of some of these
constructions are massive vector-like fermions that are charged under
both the visible electroweak group as well as a confining gauge group
in the mirror sector~\cite{Chacko:2005pe,Burdman:2006tz}.
When the fermion mass is much larger than the mirror confinement scale,
these states are called \emph{quirks}, leading to highly 
unusual LHC signals at the LHC~\cite{Kang:2008ea,Burdman:2008ek,Harnik:2008ax}.
We will discuss them further in Section~\ref{sec:models_hidden}.

\subsection{Warped Extra Dimensions}
\label{sec:models_warped}

  Warped or Randall-Sundrum~(RS) models are a wide
class of extra-dimensional scenarios in which the extra dimension has
an intrinsic curvature~\cite{Randall:1999ee,Randall:1999vf}.  This
\emph{warping} of spacetime is their defining feature, and can
provide a natural explanation for the large hierarchy between the
electroweak and Planck scales~\cite{Randall:1999ee,Goldberger:1999wh}.
For a bulk curvature $k$, the natural energy scales at any two points
separated by a distance $\Delta y$ in the extra dimension will differ
by a \emph{warp factor} $e^{-2k\Delta y}$.  We will show how this 
exponential dependence allows the electroweak scale to be derived
from the Planck scale according to $\tev \sim e^{k\Delta y}M_\text{Pl}$.

  RS models, like other theories of extra dimensions, can also provide 
a geometric interpretation for the wide range of Yukawa coupling
strengths of the Standard Model fermions.  This feature relies on the 
Standard Model fermions' wave functions in the extra dimension having
very different overlaps with the Higgs field.  
Since the relative overlap depends exponentially on the underlying
model parameters, a very large coupling hierarchy can arise from 
inputs of order unity.

  Warped extra dimensions are also thought to be related in a very deep
way to four-dimensional approximately conformal field theories
through something called the AdS/CFT correspondence.  
As such, RS models may be either 
interpreted as theories with a genuine extra spatial dimension, 
or as a powerful tool to perform calculations in classes of theories
similar to walking technicolor that cannot be treated perturbatively.
Since the infrared physics is thought to be the same in both cases, 
the distinction is not very important from a phenomenological
point of view, even though the language used to describe
the two classes theories is often quite different.  

  In this subsection we discuss several phenomenological implementations
of warped extra dimensions.  This includes minimal and bulk RS models,
Higgsless models of electroweak symmetry breaking, and holographic
composite Higgs models.  For more detailed
discussions of RS scenarios, we refer to one of the many reviews of
these models~\cite{tasi_extrad,Csaki:2005vy,
  Gherghetta:2006ha,Davoudiasl:2009cd}.

\subsubsection{Minimal warped models}
\label{sec:models_rs}

  The standard RS scenario consists of a 5-dimensional bulk spacetime
bounded on either end by two 4-dimensional surfaces called \emph{branes}.
For technical convenience, the extra dimension is often written in terms 
of an $S^1/\mathbb{Z}_2$ orbifold, which just means that the coordinate
of the fifth dimension lives in the periodic interval 
$y\in [-\pi\,R,\,\pi\,R]$, and that the points $y\leftrightarrow -y$
are identified with each other. As far as the low-energy
theory is concerned, this is completely equivalent to working on
an interval between $y=0$ and $y=\pi\,R$~\cite{Csaki:2005vy}.
In terms of $y$ and the usual four-dimensional coordinates  $x^{\mu}$,
the spacetime between the branes is described by 
the metric~\cite{Randall:1999ee}
\begin{equation}
ds^2 = g_{MN}\,dX^MdX^N = e^{-2k|y|}\eta_{\mu\nu}dx^{\mu}dx^{\nu} - dy^2
\label{rs-metric}
\end{equation}
where $k$ is a constant describing the curvature of the extra dimension
and $\eta_{\mu\nu}$ is the usual Minkowski metric in four dimensions.  
We illustrate this spacetime in Fig.~\ref{fig:rs-diag}.
This metric is a solution to the five-dimensional classical Einstein 
equations in the presence of a bulk cosmological constant.  To obtain
an effective Minkowski metric in the four-dimensional low-energy
effective theory,
it is necessary to tune the tensions of the two branes to cancel
off the total cosmological constant in the low-energy four-dimensional
effective theory.  

\begin{figure}[t]
\begin{center}
  \includegraphics[width=0.35\textwidth]{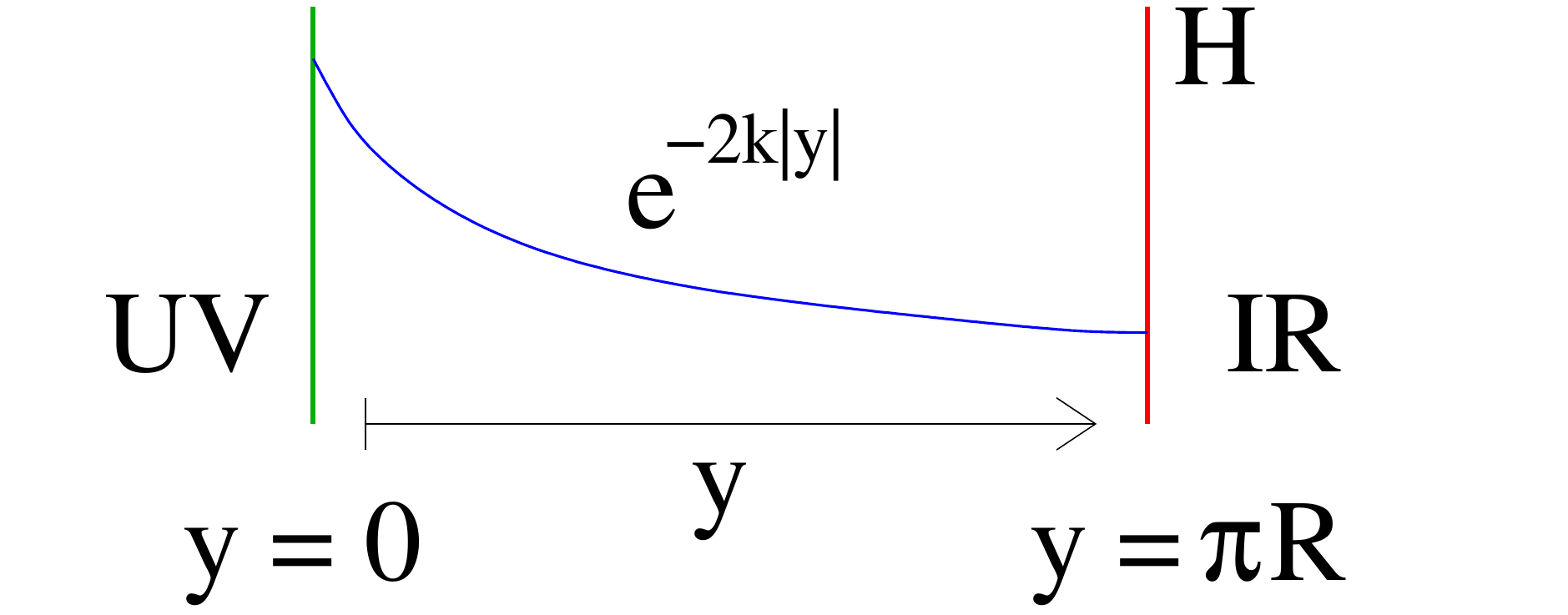}
\end{center}
\vspace*{0cm}
\caption{Diagram of the RS spacetime.  The natural mass scale on the
UV brane at $y=0$ is on the order of $M_*\sim M_\text{Pl}$, while the natural
mass scale on the IR brane at $y=\pi R$ is $e^{-k\pi R}M_* \sim \tev$.
In the minimal RS model, the Higgs is localized on the IR brane.}
\label{fig:rs-diag}
\end{figure}

  This warped metric corresponds to a finite slice of the symmetric
spacetime $AdS_5$ with curvature $k$.  A natural value of $k$ is of the 
same order as the fundamental five-dimensional Planck scale $M_*$.  
In terms of these underlying fundamental parameters, the effective 
four-dimensional Planck scale is given by~\cite{Randall:1999ee}
\begin{equation}
M_\text{Pl}^2 = \frac{M_*^3}{k}(1-e^{-2\pi\,kR}).
\label{rs-mpl}
\end{equation}
Thus, for $k\,R \gtrsim 1$ we expect $k \sim M_* \sim M_\text{Pl}$. In
contrast to the case of flat large extra dimensions, this means that
the compactification of warped extra dimensions does not affect the
hierarchy problem.

Instead, RS models provide a nice explanation for the gauge hierarchy
when the Higgs field $H$ is confined to the 
brane at $y = \pi\,R$.  The Higgs Lagrangian is then given by
\begin{alignat}{5}
\mathscr{L}_\text{Higgs} 
&= \int d^4x\; \sqrt{|g_\text{ind}|}\,
\left[ g_\text{ind}^{\mu\nu}(D_{\mu}H)^{\dagger}D_{\nu}H
      - \lambda\,(|H|^2-V^2)\right] \notag \\
&= \int d^4x\;  
\left[\eta^{\mu\nu}(D_{\mu}H')^{\dagger}D_{\nu}H'
- \lambda\,(|H'|^2-e^{-2\pi kR}\,V^2)\right].
\label{rs-lhiggs1}
\end{alignat}
where $g_{ind}^{\mu\nu} = G^{\mu\nu}(y=\pi\,R)$ is the 
\emph{induced} metric on the corresponding brane.
In the second line we rescale 
the Higgs field variable to $H' = e^{\pi\,kR}H$ to get
a canonical kinetic term.
From this expression we see that the expectation value
of the Higgs boson, \ie the electroweak scale,
is given by 
\begin{equation}
\left<H'\right> = v = e^{-\pi kR}\,V.
\end{equation}
Without another mass scale in the game, the natural value of 
$V$ in the full extra-dimensional theory
is on the order of the fundamental scale $M_* \sim M_\text{Pl} \sim k$.
However, the four-dimensional Higgs expectation value
is warped down from $V$ by an exponential factor.  This exponential
warping therefore provides a solution to the gauge hierarchy 
problem provided we postulate $k R \sim 11$:
\begin{equation}
kR \sim 11 \qquad \Rightarrow \qquad
\Lambda_{\pi} \equiv e^{-\pi kR}M_\text{Pl} \sim \tev.
\end{equation}
Such values of $kR$ can arise naturally if there is
a scalar \emph{radion} field in the bulk that stabilizes
the size of the fifth dimension~\cite{Goldberger:1999wh}.
In the spirit described above, the natural value of any dimensionful coupling  
in the underlying theory is on the order of $M_*$.  When the corresponding 
term in a dimension-four Lagrangian is related to fields localized 
near the brane at $y=\pi\,kR$,
its value in the low-energy four-dimensional effective theory is
warped down to near the electroweak scale.  For this 
reason, the brane at $y=\pi\,R$ is frequently called the IR or TeV brane
while the brane at $y=0$ is often called the UV or Planck brane.\bigskip

  In the minimal RS model (RS1) all Standard Model fields are localized on the 
IR brane~\cite{Randall:1999ee}.  In this case, the primary source 
of new physics signals comes from the graviton.  This field describes 
fluctuations of the spacetime structure and necessarily propagates 
in all the dimensions.  Just like in ADD models, we can expand the graviton 
field $h_{MN}$ into components $h_{\mu\nu}$ and $\phi$.  
Of these, the $h_{\mu\nu}$ field is symmetric and traceless, 
couples to the energy-momentum tensor, and contains
the usual four-dimensional massless graviton.  The $\phi$ field is 
called the \emph{radion} and describes fluctuations in the size of the fifth 
dimension.  It couples to the trace of the energy momentum tensor.
There are also other components in the expansion of $h_{MN}$, 
but as discussed in Section~\ref{sec:models_add}, they
do not couple to the IR brane and can be neglected.
Concentrating on $h_{\mu\nu}$, we can expand it into a tower
of spin-2 Kaluza-Klein~(KK) excitations,
\begin{equation}
h_{\mu\nu}(x,y) = \sum_{n=0}^{\infty}h_{\mu\nu}^{(n)}(x)\,g_n(y).
\label{rskkgrav}
\end{equation}
The $n=0$ \emph{zero mode} in this expansion is massless and corresponds 
to the four-dimensional graviton.  In contrast to flat extra dimensions,
however, the higher KK modes have widely spaced masses on the order 
of a $\tev$~\cite{Randall:1999ee,Davoudiasl:1999jd}, 
\begin{equation}
m_n^\text{KK} = x^G_n\, k e^{-k \pi R}, \qquad n=1,2,\ldots
\end{equation}
The constants $x_n^G$ are given to an excellent approximation 
by Bessel function zeroes $J_1(x_n^G)= 0$.  The first few values are 
$x_{1,2,3}^G \simeq 3.83,7.02,10.17$, and $x^G_n\to n\pi$ for $n\gg 1$.

  The graviton zero and KK modes couple to the Standard Model fields 
on the IR brane
through the energy-momentum tensor according to~\cite{Davoudiasl:1999jd}
\begin{equation}
\mathscr{L} \supset -\frac{1}{M_\text{Pl}}\,T^{\mu\nu}h_{\mu\nu}^{(0)}
-\frac{1}{\Lambda_{\pi}}\,T^{\mu\nu}\sum_{n=1}^{\infty}h_{\mu\nu}^{(n)},
\end{equation}
with $\Lambda_{\pi} = e^{-k\pi R}M_\text{Pl}$.  Note that 
the massive KK modes couple with an inverse TeV-scale couplings to matter,
while the massless zero mode couples like the usual four-dimensional 
graviton.  This arises because the massive KK modes have 
wave functions in the extra dimension, $g_n(y)$ in eq.~\eqref{rskkgrav}, 
that are peaked (or \emph{localized}) at the $\tev$ brane.  
In contrast, the massless zero mode is spread throughout the 
extra dimension and the fundamental scale associated with its coupling 
is set by the Planck mass resulting in the weakness of gravity 
at long-distances.

  New graviton KK-modes give rise to heavy dilepton and dijet resonances 
at colliders, similar to old-fashioned $Z'$ searches.  The existing bounds 
and future reach for these resonances can be expressed in terms of 
$k/M_\text{Pl}=k e^{- k \pi R}/\Lambda_{\pi}$ 
and the mass of the first KK resonance 
$m_1^{KK}$~~\cite{Davoudiasl:1999jd}.  
Perturbative calculability in the RS scenario requires 
$k/M_\text{Pl} \lesssim 0.1$, while $\Lambda_{\pi} \lesssim 10\,\tev$
is needed to avoid re-introducing a fine-tuning problem~\cite{Randall:1999ee}.
In Fig.~\ref{fig:rs-cmsgrav} we show the projected reach for the first 
RS KK graviton in dimuon final states at CMS~\cite{cms_tdr}.  
For $k/M_\text{Pl}$ between $0.01$ and $0.1$, the reach varies between
 about $m_1^\text{KK} = 1.7$~TeV and 3.5~TeV.  The spin-2 nature 
of the graviton KK mode can also be distinguished from a spin-0 or spin-1 
resonance by examining angular correlations in the production and decay 
processes~\cite{Davoudiasl:1999jd,cms_tdr,Cousins:2005pq,Osland:2008sy,Boudjema:2009fz}.
\bigskip

\begin{figure}[t]
\begin{center}
  \includegraphics[width=0.60\textwidth]{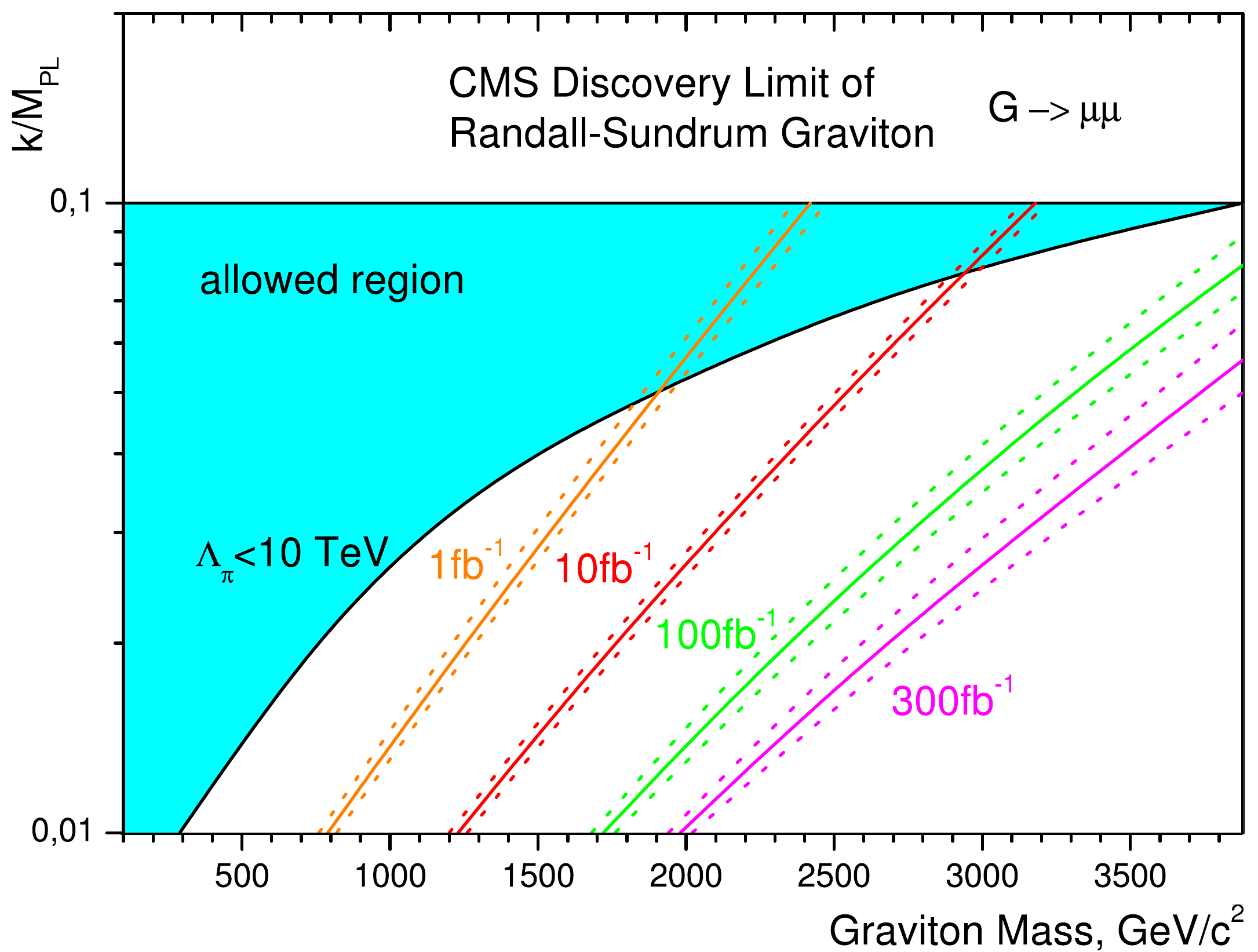}
\end{center}
\vspace*{0cm}
\caption{LHC reach in the $\mu^+\mu^-$ final state for the 
lightest RS graviton KK mode when all Standard Model fields are confined
to the IR brane.  The theoretically understood region with
$k/M_\text{Pl} < 0.1$ and $\Lambda_{\pi} < 10\,\tev$ is shaded blue.
Figure from Ref.~\cite{cms_tdr}.}
\label{fig:rs-cmsgrav}
\end{figure}

  The minimal RS model described above can be generalized by allowing 
some or all of the Standard Model gauge and fermion fields to propagate 
in the five-dimensional bulk~\cite{Gherghetta:2000qt,Davoudiasl:1999tf,
Pomarol:1999ad,Grossman:1999ra,Chang:1999nh,Davoudiasl:2000wi,Huber:2000ie}.  
Putting a gauge field in the bulk,
we must take into account the fact that it now has five components,
$A_{M}$ with $M=0,1,2,3,5$.  Just like a gauge field in four dimensions,
not all these components lead to physical degrees of freedom.
To separate the physical modes it is convenient to apply the gauge condition
$\del_5A_5 = 0$.  We must also specify the boundary values of this bulk field 
on the two branes, and the natural choice that does not explicitly break 
any of the four-dimensional gauge symmetries
we would like to preserve is the \emph{Neumann condition} 
$\left.\partial_5\,A_{\mu}\right|_{y=0,\pi R} = 0$ for the four-dimensional
components and the \emph{Dirichlet condition} $\left.A_5\right|_{y=0,\pi R} = 0$
for the fifth~\cite{Csaki:2005vy,Davoudiasl:1999tf,Pomarol:1999ad}. 
Together with our gauge choice, we see that $A_5=0$ vanishes everywhere. 
Decomposing the remaining $A_{\mu}$ vector field into KK modes, just like the
graviton field in the RS1 realization,
we find a massless four-dimensional zero mode gauge field 
with a constant wave function in the extra dimension, 
as well as a tower of $\tev$-scale massive four-dimensional vector 
KK excitations localized towards the IR brane. Their  
masses are similar to the RS1 graviton case~\cite{Davoudiasl:1999tf,Pomarol:1999ad}
\begin{equation}
m_n^\text{KK} = x_n^A\, k e^{-\pi k R}\; , 
\end{equation}
where $x_{1,2,3}^A = 2.45,5.57,8.02$, and $x_n^A \to n\pi$ for $n\gg 1$.
In this picture the fifth component $A_5$ of the bulk gauge field
does not appear explicitly in the spectrum. 
At each KK level it provides the extra degrees of freedom for 
the massive KK vector bosons.  This is analogous to the Higgs mechanism, 
where the would-be Goldstone boson is eaten by the gauge field to produce 
a massive vector with an extra longitudinal degree of freedom.

  A phenomenological problem arises if only the Standard Model gauge fields 
propagate in the bulk while the Standard Model fermions remain confined to the 
IR brane~\cite{Davoudiasl:1999tf}.
The effective gauge-matter couplings are proportional to the integrated 
overlap of their respective wave functions. When we fix the couplings
of the zero-mode gauge fields to their Standard Model counterparts,
the couplings of the massive vector KK-excitations to
the Standard Model fermions are automatically enhanced by a 
factor of $\sqrt{2\pi\,kR}$
due to the localization of the gauge KK-modes towards the $\tev$ brane.  
This is the same effect which enhances the KK graviton couplings starting 
from the first excitation.
As a result, the KK-modes generate dangerously large corrections 
to precision electroweak observables, and the
scenario is ruled out unless $k\,e^{-\pi kR} \gtrsim 11\,\tev$~\cite{Davoudiasl:1999tf, Csaki:2002gy}.
The corresponding KK masses, both graviton and gauge, would then be
too large to probe directly at the LHC, and this scenario would
no longer provide a convincing solution to the gauge 
hierarchy problem. This negative conclusion can be avoided 
to some extent by adding large brane-localized kinetic terms on 
the $\tev$ brane which modify the boundary conditions and have the 
effect of pushing the KK modes away from the 
brane~\cite{Carena:2002me,Carena:2002dz,Davoudiasl:2002ua}.\bigskip

  The situation is much better if both the Standard Model fermions
and gauge fields are allowed to propagate in the fifth 
dimension~\cite{Gherghetta:2000qt,Grossman:1999ra,Chang:1999nh,
Davoudiasl:2000wi}.
The challenge is that bulk fermions in five 
dimensions are inherently non-chiral.  The minimal possible fermion 
representation in this case has two opposite chiral components,
corresponding to left- and right-handed spinors in four dimensions.
To regain the chiral fermion structure of the Standard Model, we can make use of
the explicit breaking of the five-dimensional Lorentz symmetry by the branes.
For each Standard Model fermion in the bulk, we must add a partner with the
opposite chirality.  However, it is not necessary for both chiral
components to have the same boundary conditions on the branes, 
and only those chiral components with Neumann boundary conditions 
at \emph{both} branes, $\left.\del_{y} \psi_{L,R}\right|_{y=0,\pi R}=0$ 
will have a massless zero mode (up to coupling to the Higgs boson).  
By specifying Neumann boundary conditions for some of the chiral 
fermion components, and mixed or Dirichlet boundary conditions for the rest, 
we can regain a chiral set of fermion zero modes that coincides with 
the Standard Model fermions.  On the other hand, the massive fermion KK modes
will consist of fermions with both chiralities.

  In general, the Lagrangian for a five-dimensional fermion contains 
a bulk mass term which is conventionally expressed in terms of the curvature 
$c\,k$ with a scaling factor $c$.  Despite the bulk mass, 
there can still arise a massless
chiral zero mode for the left-handed component with the natural boundary
conditions $\left.(\del_{y} +ck)\psi_L\right|_{y=0,\pi R}=0$ 
and $\left.\psi_R\right|_{y=0,\pi R}=0$.
The unpaired left-handed zero mode then has a profile in the fifth 
dimension proportional to~\cite{Gherghetta:2000qt,Grossman:1999ra}
\begin{equation}
\psi^{(0)}_L \propto e^{-(c-1/2)k|y|}.
\label{rsfermion}
\end{equation}
Thus, the left-handed zero mode is exponentially localized towards 
the UV brane for $c> 1/2$, and exponentially localized towards the 
IR for $c < 1/2$.  For a right-handed fermion zero-mode,
the natural boundary condition is 
$\left.(\del_{y} -ck)\psi_R\right|_{y=0,\pi R}=0$ 
and $\left.\psi_L\right|_{y=0,\pi R}=0$, which yields the profile
\begin{equation}
\psi^{(0)}_R \propto e^{(c+1/2)k|y|}.
\end{equation}
Now, the right-handed zero mode is UV-localized for $c<-1/2$
and IR-localized when $c>-1/2$.  
By comparison, the lowest fermion KK modes are all localized near
the IR brane with approximate masses 
\begin{equation}
m_n^{(L,R)} \simeq 
\pi\left[n+\frac{1}{2} \left( |c\mp 1/2|-1 \right)-(-1)^n\frac{1}{4}\right]\,
k\,e^{-\pi kR}
\end{equation}
with $n=1,2,\ldots$ Note that since the gauge boson zero modes have
flat profiles, they will couple to each bulk fermion with equal strength
(up to the charge of the state)
no matter where in the bulk the fermion is localized.\bigskip

  The exponential localization of bulk fermion zero modes can provide 
a nice explanation for the large variation among the observed Yukawa couplings,
corresponding to Standard Model fermion masses~\cite{Gherghetta:2000qt}.
With the Higgs confined to the to IR brane, the effective Yukawa
coupling of a given Standard Model fermion will be determined by how much the
corresponding zero-mode wave function overlaps with the IR brane.  
The heavy top quarks should therefore be localized near the IR brane,
while the very light electrons should lie closer to the UV brane.
We illustrate this localization in 
Fig.~\ref{fig:rsflav}.
Since the overlap on the IR brane is determined by an exponential
of the bulk mass parameter $c$ as in eq.~\eqref{rsfermion}, order unity
variations in the bulk mass can produce Yukawa couplings that range
over several orders of magnitude.  Thus, the large difference $m_e \ll m_t$ can
be obtained for similar and natural parameter values in the underlying
extra-dimensional theory.  

Localizing fermions in different parts of the 
fifth dimension has the additional benefit of suppressing many of the 
potential new and dangerous sources of flavor violation coming from 
higher-dimensional operators relative to what they would be if all 
the Standard Model fermions were confined to the IR 
brane~\cite{Huber:2000ie,Agashe:2004ay,Agashe:2004cp}.
Even so, some further flavor alignment appears to be required
to obtain a KK scale below 
$5\,\tev$~\cite{Csaki:2008zd,Blanke:2008zb,Blanke:2008yr,Albrecht:2009xr}.  

\begin{figure}[t]
\begin{center}
  \includegraphics[width=0.3\textwidth]{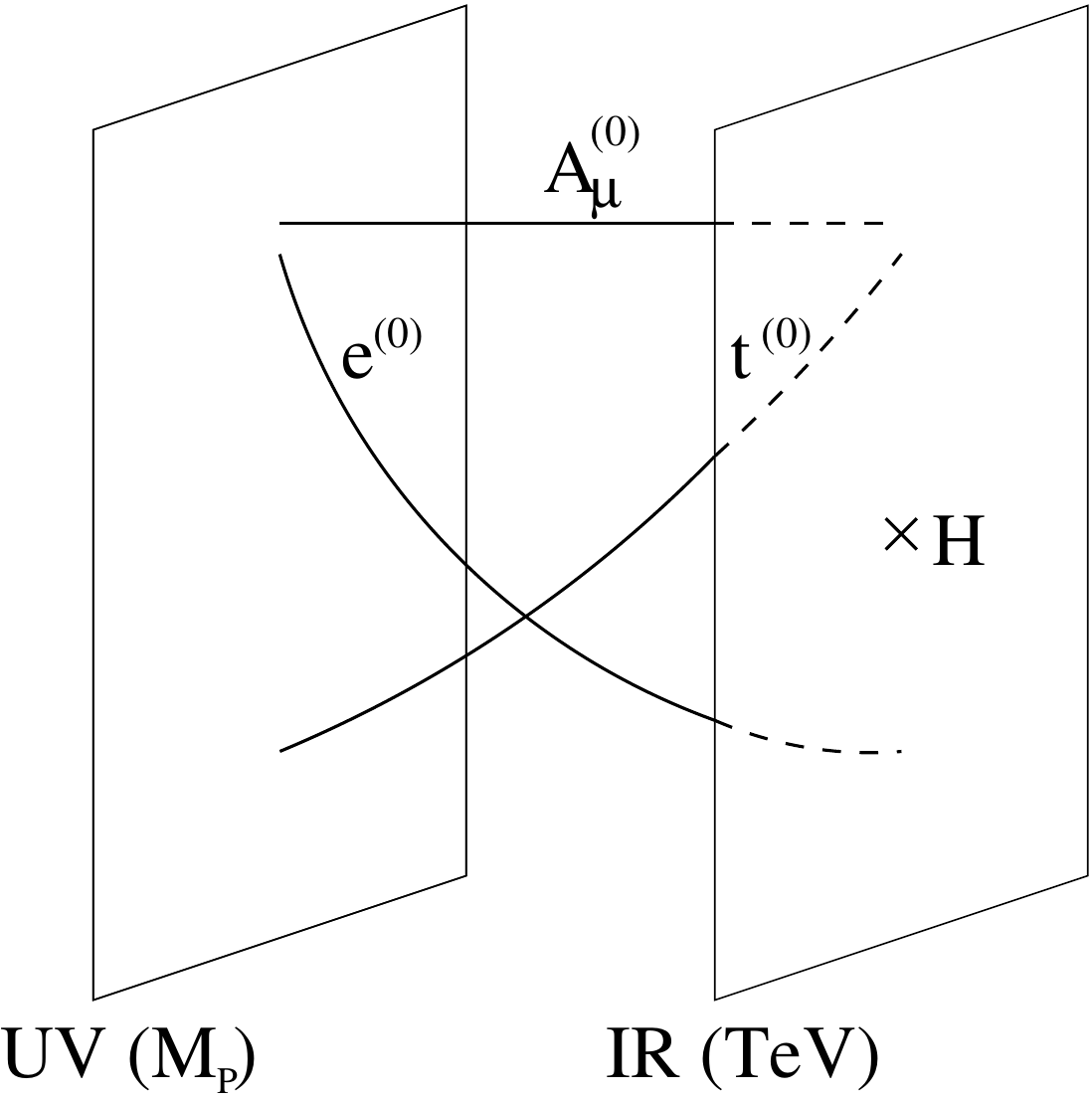}
\end{center}
\vspace*{0cm}
\caption{Localization of fermions and origin of the flavor hierarchy
in bulk RS models.  The diagram is from Ref.~\cite{Gherghetta:2006ha}.}
\label{fig:rsflav}
\end{figure}

More importantly, this class of RS scenarios with Standard Model fermions 
and the gauge bosons residing in the bulk and the Higgs confined
to the IR brane
are strongly constrained by precision electroweak
measurements.  With no additional structure, these bounds imply a KK scale
$k\,e^{-\pi kR} \gtrsim 5\,\tev$~\cite{Hewett:2002fe,Burdman:2002gr,Carena:2003fx}.  
The electroweak constraints can be weakened significantly by extending 
the bulk gauge symmetry to 
$SU(3)_c\times SU(2)_L\times SU(2)_R\times U(1)_{X}$~\cite{Agashe:2003zs},
including the approximate custodial $SU(2)_R$ global symmetry
of the Standard Model~\cite{Sikivie:1980hm}.
The extended gauge structure is explicitly broken down to 
the Standard Model subgroup $SU(3)_c\times SU(2)_L\times U(1)_Y$, 
with $Y = (t_R^3+X)$, by a choice of (non-Neumann) boundary conditions 
for the gauge fields.  This leaves KK modes for all the gauge bosons 
in the theory, but only allows massless zero modes for the Standard Model gauge bosons.  

Even with an extended custodial bulk gauge symmetry, 
there can remain dangerously large corrections to the 
$Z$-$b_L$-$\bar{b}_L$ coupling, in contradiction to the measured value of
$R_b = \Gamma_Z(b\bar{b})/\Gamma_Z(\text{hadrons})$~\cite{Alcaraz:2007ri}.
These can be further reduced to an acceptable level 
by including a discrete $P_{L,R}$ parity that enforces equal 
$SU(2)_L$ and $SU(2)_R$ bulk gauge couplings, and by placing $b_L$ 
into a non-minimal representation of the bulk group with 
$t_L^3(b_L) = t_R^3(b_L)$~\cite{Agashe:2006at}.  With this extended 
custodial gauge and fermion structure, precision electroweak bounds 
can be consistent with a KK scale as low as 
$k\,e^{-\pi kR} \simeq 1\,\tev$~\cite{Carena:2006bn,Carena:2007ua}
and $KK$ gauge boson excitations with masses below $3\,\tev$.  
The non-minimal fermion representations required in this framework 
lead to exotic top-like custodial KK fermions with masses often 
below $1\,\tev$.\bigskip

  The bulk RS scenario can lead to interesting signatures at the LHC.
Given the relative masses and couplings of the various low-lying KK 
excitations, the most accessible mode is typically the first KK excitation 
of the gluon~\cite{Agashe:2006hk,Lillie:2007yh}.  The KK gluon is a 
heavy color octet, and can be produced 
resonantly in the $s$-channel.
Its decay modes are related to the geography of the warped extra dimension.
As discussed above, the low-lying KK excitations of the Standard Model 
gauge bosons are localized towards the IR brane.  The fermion zero-modes 
corresponding to the top quark are also localized near the IR brane while 
the lighter Standard Model fermions are localized in the UV.  This has the consequence
that the KK gauge bosons couple much more strongly to the heavier
SM fermions than to the lighter ones.  The massive KK gluon resonances,
as well as the KK resonances of the other gauge bosons, therefore decay
preferentially into top quarks, providing a way to differentiate them
from Standard Model backgrounds and other types of heavy resonances.  

  Detecting a massive KK gluon at the LHC presents a number of experimental
challenges.  Unless the $t\bar{t}$ final state can be distinguished
from ordinary dijets, the signal will be swamped by Standard Model 
backgrounds~\cite{Agashe:2006hk,Lillie:2007yh}.
For relatively slow-moving top quarks this is straightforward
to do by reconstructing the final state tops by identifying 
the bottom quarks, jets, and leptons from $t\to Wb$.
However, the top quarks emitted by a KK gauge boson will typically
have a very large boost, and as a result the decay products can be 
highly collimated.  Traditional jet identification
algorithms will often combine all these products into a single jet,
possibly mistaking it for a more common light jet. The phenomenology of 
massive boosted particles we discuss in Section~\ref{sec:sig_top}, 
since they naturally appear in many models, even including the Standard Model.

  Another slightly less general source of collider signals for RS models 
are the exotic top-like custodial KK fermions motivated by precision 
electroweak bounds~\cite{Agashe:2006at,Carena:2006bn,Carena:2007ua}.  These
states can have non-standard electromagnetic charges such as 
$Q=5/3,\,-1/3$ in addition to $Q=2/3$, and may be lighter than $1\,\tev$.
This contrast with other scenarios such as little Higgs models,
which typically on have top-partner states with $Q=2/3$.  
The lightest of the exotic KK states tend to be localized in the IR, 
and therefore have enhanced couplings to the KK gauge 
bosons~\cite{Agashe:2006at,Carena:2006bn}. 
Depending on their masses and charges, single and pair production 
can both be significant.  These exotic fermions can also arise as 
the decay products of a KK gauge boson resonance.  Once produced,
the dominant decay channels are to top and bottom quarks along
with a longitudinal $W$, $Z$, or Higgs bosons.  For example,
$T_{5/3} \to t\,W^{+}\to b\,W^{+}W^{+}$ can give rise to a distinctive
same-sign dilepton signal\cite{Dennis:2007tv,Contino:2008hi}.

Finally, potential RS signals at the LHC also arise from the KK excitations
of the electroweak gauge bosons~\cite{Djouadi:2007eg,
Agashe:2007ki,Agashe:2008jb}, KK 
fermions~\cite{Carena:2006bn,Carena:2007ua}, mixing between the Higgs and 
the radion field~\cite{Csaki:2000zn},
and black holes~\cite{Giddings:2001bu,Dimopoulos:2001hw}.\bigskip

  Before moving on, let us mention that the AdS/CFT 
correspondence~\cite{Maldacena:1997re,Gubser:1998bc,Witten:1998qj}
suggests that the RS models discussed in this section have a 
fundamental connection to strongly-coupled scenarios in four dimensions.  
This correspondence was first seen in the many remarkable coincidences 
between type IIB supergravity in the ten-dimensional space $AdS_5\times S_5$ 
with $\mathcal{N}=4$ super-conformal $SU(N)$ Yang-Mills in four dimensions
with $N \gg 1$.  
As a result, these two seemingly very different theories are thought 
to be \emph{dual} to each other in that they describe the same physics
at low energies.  More generally, the AdS/CFT correspondence relates
certain conformal field theories in $d$ dimensions to theories with
gravity in $(d+1)$ dimensions.  It is useful because 
when one description is strongly coupled, the other is often 
weakly coupled and easier to numerically analyze.  The correspondence has been applied
all the way from superstring theory~\cite{Aharony:1999ti}, 
to phenomenological model building, to QCD~\cite{Erlich:2009me}, 
to condensed matter physics~\cite{Hartnoll:2009sz}.

  The application of the AdS/CFT correspondence to phenomenology,
and RS models in particular, was initially studied in 
Refs.~\cite{ArkaniHamed:2000ds,Rattazzi:2000hs}.  
The location in the fifth dimension of the RS spacetime is related to 
the energy scale in a nearly conformal four-dimensional gauge theory.
The branes on either side of the slice of $AdS_5$ bulk
are interpreted as a breaking of conformal invariance
in the gauge theory at high and low energies.  Starting at a very high 
energy in the dual gauge
theory description, the theory enters a strongly-coupled conformal 
regime at $M_\text{Pl}$ corresponding to the Planck brane.  The gauge
coupling remains nearly constant over many decades during 
the renormalization group running down to lower energies.  This continues 
until the energy approaches $\Lambda_{\pi}$, corresponding to the IR brane,
which is interpreted as a spontaneous breaking of conformality.
Here, the gauge coupling grows large and the theory confines
producing massive bound states. This behavior of the gauge 
theory coincides with that of \emph{walking technicolor} discussed above, 
and thus RS models are thought to be special (dual) examples of these theories.
In the limit of $k/M_\text{Pl} \ll 1$ where the RS solution can be trusted,
the gauge theory dual is strongly coupled.

  Beyond connecting energy scales with the location in the extra dimension,
the AdS/CFT correspondence also provides a matching between the particles
and symmetries in the two theories.  The KK excitations in RS 
correspond to bound-state resonances in the dual gauge theory.
Bulk RS fields are interpreted as mixtures of fundamental
and composite fields in the gauge theory, with fields localized closer 
to the IR brane being mostly composite and fields near the Planck 
brane mostly fundamental.  The Higgs boson confined to the IR brane is
entirely composite.  Weakly-coupled gauge symmetries in the bulk
(such as the Standard Model gauge forces) represent genuine symmetries 
of the conformal sector.  However, they mostly act as spectators to the 
much stronger conformal dynamics, and do not have to be conformal themselves.
Global symmetries on the CFT side are also interpreted as 
gauge symmetries in the bulk.  This accounts for why an expanded 
bulk gauge symmetry is used to impose a custodial $SU(2)_R$
global symmetry on RS theories to protect them from large electroweak
corrections.  The distinguishing feature is that global symmetries
in the dual gauge theory correspond to bulk gauge symmetries in RS that 
are broken explicitly on the UV brane.  Weak spectator gauge symmetries 
in the gauge theory are only broken spontaneously on the IR brane, if at all.

\subsubsection{Higgless models}
\label{sec:models_hless}

  Higgless and holographic composite Higgs models are variations 
of the Randall-Sundrum scenario discussed above.  
Both classes of models are motivated by a little hierarchy problem 
in RS models.  Indirect experimental constraints, like electroweak 
precision data, force the natural mass scale of KK resonances to be 
greater than a $\tev$, 
while the mass of the Higgs boson and its corresponding expectation value
generally cannot be much larger than a few hundred GeV.
Some degree of fine-tuning is needed to maintain this hierarchy.  

  In Higgsless models the Higgs is removed altogether.  
In holographic composite Higgs models, the Higgs comes from a bulk gauge field 
and is a Nambu-Goldstone Boson of an approximate global symmetry.  
These approaches are reminiscent of technicolor and little Higgs models 
respectively.  Indeed, we will describe below how the AdS/CFT 
correspondence relates Higgsless models to technicolor, 
and shows that composite Higgs models share features with little Higgs 
scenarios at low energy.\bigskip

Higgless models of electroweak symmetry breaking are built around
the RS spacetime described above, consisting 
of a warped five-dimensional bulk space bounded on either end by two 
four-dimensional 
branes~\cite{Csaki:2003dt,Csaki:2003zu,Nomura:2003du,Barbieri:2003pr,
Cui:2009dv}.
The Standard Model vector fields propagate in all five dimensions, 
and the $W$ and $Z$ vector bosons arise as massive 
Kaluza-Klein excitations.  This is qualitatively different from 
bulk RS models, where the $W$ and $Z$ states are massless 
zero modes before electroweak symmetry breaking induced by an 
explicit Higgs field on the IR brane.  

Standard model fermions in Higgless 
models also extend into all dimensions, and they have explicit mass terms 
that are allowed by the non-chiral structure of the theory in the bulk 
and on the IR brane.  Higgless models are reviewed in 
Ref.~\cite{Csaki:2005vy,Grojean:2006wr}.\bigskip

  For our purposes, it is most convenient to write the bulk RS metric as
\begin{equation}
ds^2 = \lrf{R}{z}^2\left(\eta_{\mu\nu}dx^{\mu}dx^{\nu} - dz^2\right),
\end{equation}
with the branes located at $z=R=k^{-1}$ and $z=R'>R$.  This metric is related 
to the one we used previously by the change of variables $z = R\ e^{y/R}$.  
For a ratio of $R'/R \sim 10^{16}\,\gev$ this framework can account for
the hierarchy between the Planck mass and the electroweak scale.
Gauge fields are assumed to propagate within all five 
dimensions, and the electroweak gauge structure of the minimal viable model 
is $SU(2)_L\times SU(2)_R\times U(1)_{X}$. 
Boundary conditions for the bulk gauge fields are chosen so that 
this gauge group is broken down to $SU(2)_L\times U(1)_Y$ on the UV 
brane at $z=R$ with $Y = (t_R^3+X)$, and the diagonal $SU(2)_D$ 
$SU(2)_L\times SU(2)_R$ on the 
IR brane at $z=R'$~\cite{Csaki:2003zu,Nomura:2003du,Barbieri:2003pr},
as sketched in Fig.~\ref{fig:hlessgauge}.
The $SU(3)_c$ QCD group is unbroken everywhere in the bulk and
on the branes.\bigskip

\begin{figure}[t]
\begin{center}
  \includegraphics[width=0.4\textwidth]{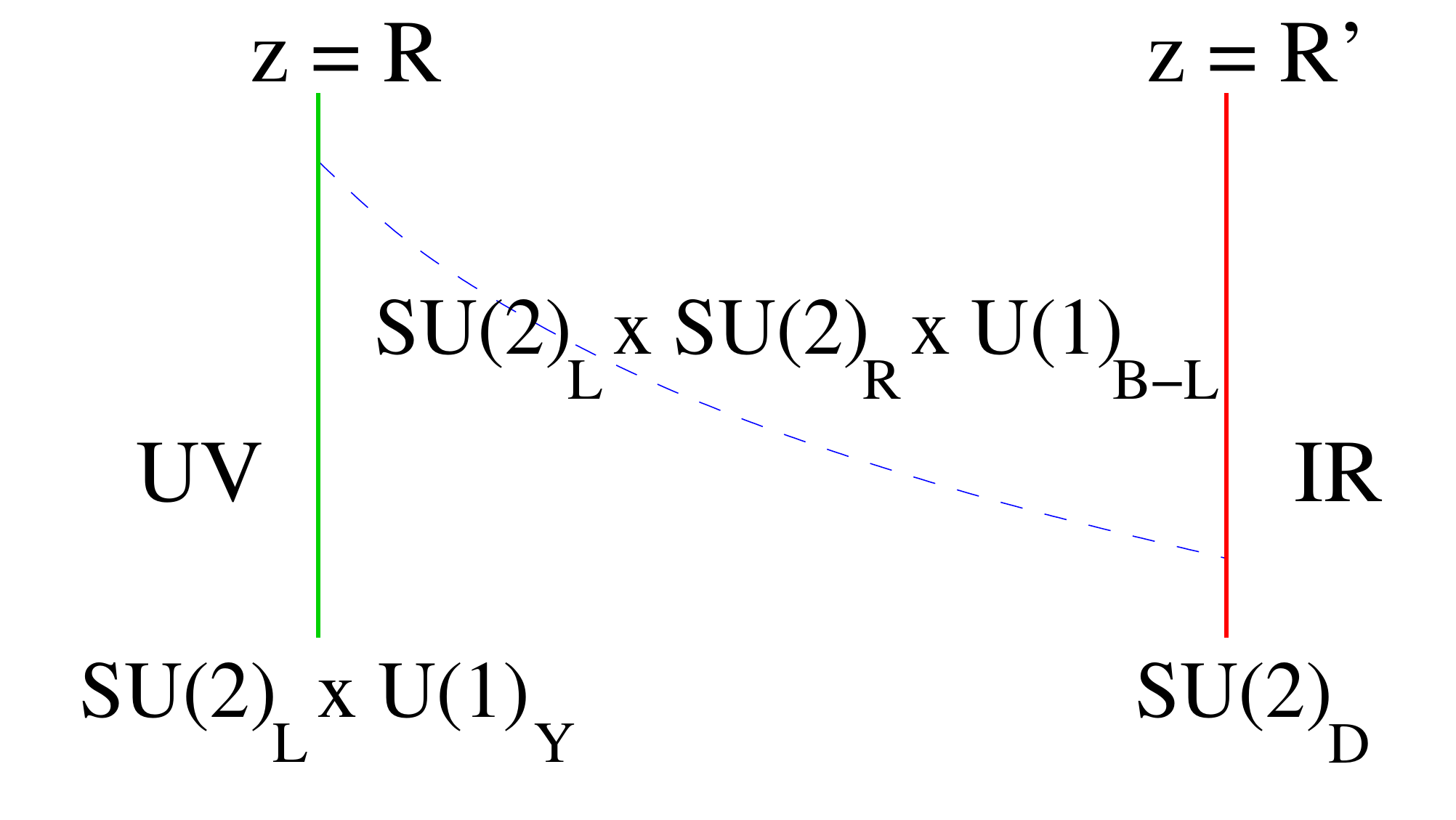}
\end{center}
\vspace*{0cm}
\caption{Bulk electroweak gauge symmetry structure of the 
warped-space Higgsless model.  The $SU(3)_c$ QCD group is also present
and unbroken everywhere in the bulk and on the branes.}
\label{fig:hlessgauge}
\end{figure}

  The spectrum of electroweak vector bosons in this scenario consists of
a single massless photon along with KK towers of charged $W_n$
and neutral $Z_n$ states~\cite{Csaki:2003zu}.  
The massive $W$ and $Z$ vectors of the Standard Model are identified with the 
lowest KK modes of the $W_n$ and $Z_n$ towers.  
This is very different from the RS scenarios discussed 
previously where the $W$ and $Z$ arise as massless zero 
modes with flat profiles and get their masses entirely from 
a Higgs VEV on the IR brane.  
Instead, in Higgless models there is no explicit Higgs boson, 
and the $W$ and $Z$ get KK masses from the 
$SU(2)_L\times U(1)_Y$-violating boundary condition on the IR brane.  
The masses and effective gauge couplings to fermions on the UV brane
of these states are 
(to leading order in $R/R'$)~\cite{Csaki:2003zu}
\begin{alignat}{5}
M_W^2  &\simeq \frac{1}{R'^2} \; \log \frac{R'}{R} &
\frac{1}{g^2} &\simeq R \; \log \frac{R'}{R}\, \frac{1}{g_5^{2}} ,\notag \\
M_Z^2  &\simeq \frac{g_5^2+2\tilde{g}_5^2}{g_5^2+\tilde{g}_5^2} \;
               \frac{1}{R'^2} \; \log \frac{R'}{R} \qquad \qquad &
\frac{1}{g'^{2}} &\simeq R \log \frac{R'}{R}\,
        \left( \frac{1}{g_5^{2}} + \frac{1}{\tilde{g}_5^{2}} \right) \; ,
\end{alignat}
where $g_5$ is the bulk gauge coupling of $SU(2)_L$ and $SU(2)_R$,
assumed to be equal here for simplicity. The 
bulk gauge coupling of $U(1)_{X}$ is $\tilde{g}_5$.  The masses of the massive
KK modes are on the order of $n\pi/R'$, $n=1,2,\ldots$. 

These expression also show that the inverse radius must be less 
than about ${R'}^{-1}\lesssim \tev$ to reproduce the observed 
$W$ and $Z$ masses.  Higher KK modes will enter 
near a $\tev$.\bigskip  

  From the point of view of a low-energy four-dimensional observer,
this scenario looks like a theory with explicit masses for the $W$
and $Z$ vectors.  On the face of it, this seems to present a puzzle.
Recall from Section~\ref{sec:why_ewsb} that massive vector theories
 have problems at high energies.  Without a light Higgs boson underlying 
the vector mass, these theories typically become strongly-coupled 
at energies near $4\pi\,M_W \sim 2\,\tev$, where $M_W$ is the vector mass.

On the other hand, the Higgless theory was derived from
a weakly-coupled extra-dimensional theory that should be valid up to energies
near $4\pi/R' \gg \tev$.  The resolution to this apparent puzzle is
that compared to a generic theory of massive vectors, Higgsless models
also contain a tower of heavier KK modes~\cite{Csaki:2003dt}.  
These should be included when computing the self-scattering 
of the $W$ and $Z$ vectors at energies near and above a $\tev$.  
After summing over all the KK modes contributing to the scattering, 
many cancellations occur that postpone the onset of strong coupling to 
energies well above $4\pi\,M_W$.
A precise calculation of the perturbative unitarity behavior shows
that vector boson self-scattering becomes strongly coupled at energies 
on the order of the cutoff of the extra-dimensional gauge theory, 
as one might expect~\cite{Papucci:2004ip}.\bigskip  

The Standard Model fermions must now get masses, but without 
the help of a Higgs.  This can be achieved by putting them
in the bulk and writing down mass terms which mix the $SU(2)$ doublet
and singlet fields on the IR 
brane~\cite{Nomura:2003du,Barbieri:2003pr,Csaki:2003sh}.
Without such terms, the theory would have chiral zero modes, and KK towers
which, as discussed in Section~\ref{sec:models_rs}, are non-chiral.
Turning on the boundary masses mixes the would-be zero modes with the
original basis KK modes, resulting in a massive spectrum which (together with
bulk masses as usual and brane kinetic terms) allow one to generate the 
correct mass spectrum for the fermions~\cite{Csaki:2003sh}.  
As usual, there will be towers of both left- and right-chiral fermion 
KK modes.\bigskip

  The most serious challenge to building viable models of Higgsless
electroweak symmetry breaking is satisfying the constraints from
precision electroweak measurements~\cite{Barbieri:2003pr,Davoudiasl:2003me,
Davoudiasl:2004pw,Hewett:2004dv,Burdman:2003ya,Cacciapaglia:2004jz}.  
Mixing of the $W$ and $Z$
with higher KK modes changes their couplings to fermions relative
to the Standard Model.  Heavier KK modes are preferred to reduce these 
deviations to an acceptable level, but the KK modes can't be too 
heavy if they are to unitarize vector boson scattering.
Both requirements can be satisfied simultaneously if there are localized
kinetic terms on each of the branes~\cite{Cacciapaglia:2004jz},
and if the Standard Model fermions have nearly flat profiles in the 
bulk~\cite{Cacciapaglia:2004rb}.
In this case the first vector boson KK modes
above the Standard Model $Z$ and $W$ typically have masses in the
$500$-$1500\,\gev$ range~\cite{Cacciapaglia:2004rb}.  

Further model structure is needed to generate a sufficiently large top
quark mass while not overly disrupting the measured $Zb_L\bar{b}_L$
coupling.  Some examples include new top-like custodial bulk
fermions~\cite{Agashe:2006at,Cacciapaglia:2006gp}, or a second warped
bulk space on the other side of the UV brane with its own IR
brane~\cite{Cacciapaglia:2005pa}.\bigskip

  The collider signatures of Higgless models are closely related 
to the constraints imposed on them by the unitarization of gauge boson 
scattering and precision electroweak bounds.  For the scattering of
the $W$ and $Z$ to remain perturbative at energies 
above a $\tev$, these states must couple in a specific way to the heavier KK
excitations~\cite{Csaki:2003dt}.  This places a lower bound 
on the decay width of a given electroweak vector KK mode to the light 
Standard Model vector bosons.  Precision electroweak constraints and 
flavor bounds can be satisfied if the light Standard Model fermions have nearly 
flat but slightly UV-localized bulk profiles so that their couplings 
to the IR-localized 
KK vector bosons are somewhat suppressed.  The couplings of the KK modes 
to the third generation are typically larger.

When the vector KK modes have small couplings to the light generations
but significant couplings with the third, the Higgless collider
signals are similar to those of the bulk RS scenarios discussed
previously, with the KK modes decaying frequently to top quark
pairs~\cite{Schwinn:2005qa}.  The main difference is that the new
Higgsless KK resonances are considerably lighter than what is
typically considered for bulk RS models.  If the KK couplings to the
third generation are not too large, additional new signals can arise
in weak boson fusion.  This is the dominant production channel for the
electroweak KK modes when their couplings to light fermions are
suppressed.  In this context, new electroweak KK modes appear as
resonances electroweak gauge boson scattering.  For example, $WZ\to
WZ$ can be modified by an $s$-channel $W_1$ under the assumption that
$W_1$ does not couple significantly to
fermions~\cite{Birkedal:2004au}.  Reconstructing a purely leptonic
$WZ$ invariant mass yields a distinctive transverse mass peak, as
discussed in Section~\ref{sec:sig_met}.  In
Fig.~\ref{fig:hlessmatchev} we show the estimated signal compared to
the Standard Model background, as well as a pair of phenomenological
schemes designed to describe the unitarization of massive vector boson
scattering. This is only one example of electroweak scattering
channels which will serve as discovery channels for Higgsless KK
excitation~\cite{He:2007ge,Ohl:2008ri,Alves:2008up,
  Alves:2009aa,Cata:2009iy,Han:2009qr}.  Alternative signals appear in
Higgless models with an extended top quark sector, including exotic
vector-like custodial
fermions~\cite{Cacciapaglia:2006gp,Cacciapaglia:2006mz,Martin:2009gi}.\bigskip

\begin{figure}[t]
\begin{center}
  \includegraphics[width=0.50\textwidth]{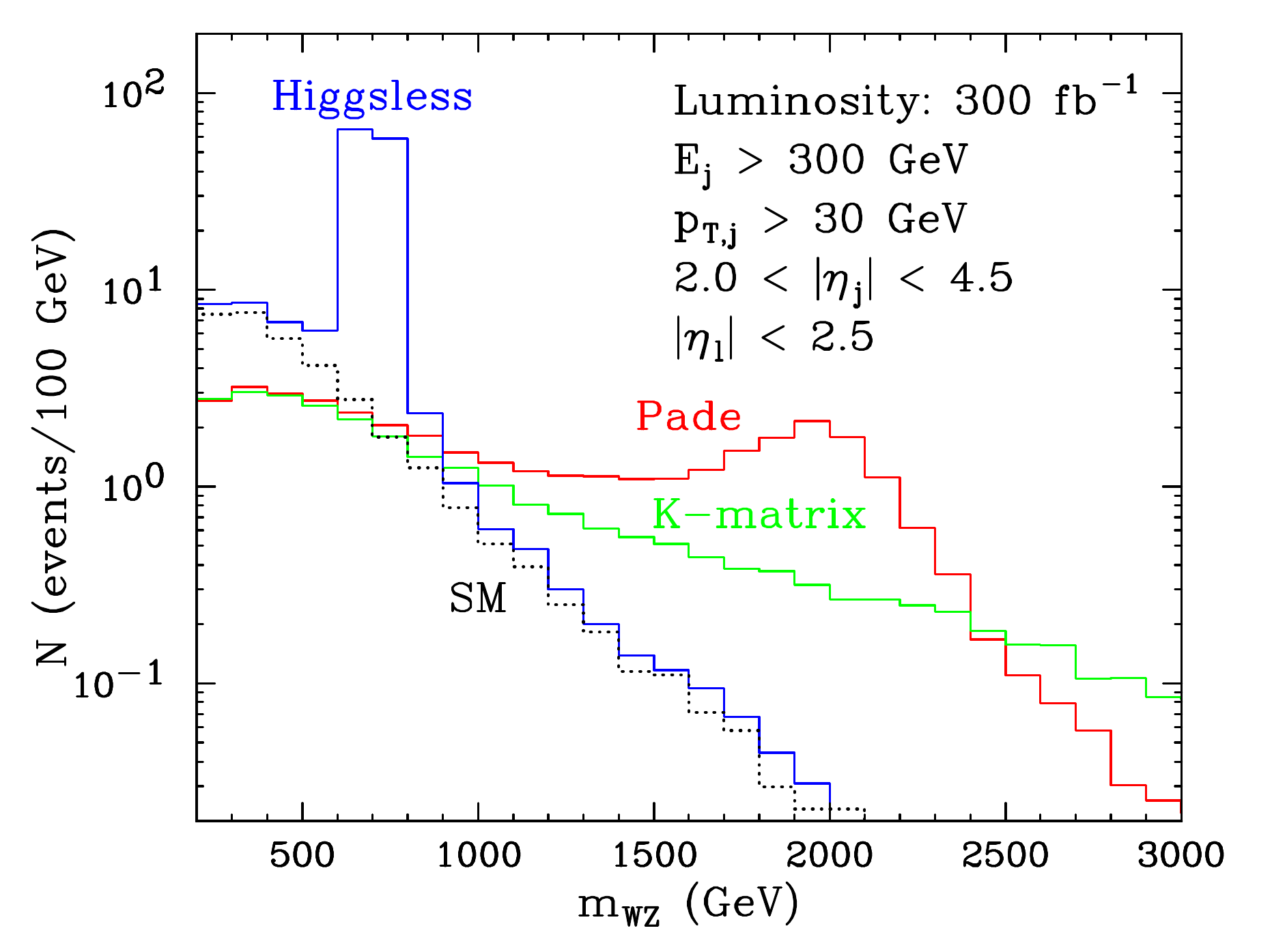}
\end{center}
\vspace*{0cm}
\caption{Higgless signals in $WZ\to WZ$ scattering at the LHC. The
  resonance comes from a $W_1$ KK mode without any couplings to
  fermions. Also shown are two parameterizations (Pad\'e and
  $K$-matrix) unitarizing vector boson scattering.  Figure from
  Ref.~\cite{Birkedal:2004au}.}
\label{fig:hlessmatchev}
\end{figure}

  Many of the properties of Higgsless theories are also found to
arise in models of technicolor.  In both cases it is challenging
to satisfy the precision electroweak constraints and to generate
a large top quark mass.  It is perhaps not so surprising then that 
the AdS/CFT correspondence discussed in Section~\ref{sec:models_rs} suggests 
that these five-dimensional warped models are dual to 
certain classes four-dimensional conformal 
technicolor theories~\cite{Csaki:2003zu,Burdman:2003ya}.  
Under this correspondence, the $z=R$ brane
matches onto a high energy scale where the technicolor gauge theory enters into
a nearly conformal regime, while the $z=R'$ brane represents the 
energy scale where the conformal running ends spontaneously and 
the gauge theory confines.  The $SU(2)_L\times U(1)_Y$ gauge symmetry
present in the bulk and on the IR brane coincides with a weakly-coupled 
spectator gauge group in the conformal technicolor theory that is a subgroup 
of a larger global $SU(2)_L\times SU(2)_{R}\times U(1)_{X}$ symmetry.
Another subgroup of this global symmetry is a custodial $SU(2)$ that 
protects the $W$ and $Z$ mass ratio.
The explicit breaking of $SU(2)_L\times U(1)_Y$ on the IR brane matches onto
a spontaneous breaking of this group by the confining transition of the 
technicolor dynamics.  In this context, the heavier KK modes are 
bound-state resonances in the technicolor theory.  The advantage of the 
extra-dimensional formulation of these theories is that precision
electroweak (and flavor) constraints can be computed explicitly.\bigskip
  
  Higgless models can also be related to gauge theories in four dimensions
through \emph{dimensional 
deconstruction}~\cite{ArkaniHamed:2001ca,Hill:2000mu,ArkaniHamed:2001nc}.  
There we 
replace an extra dimension with a one-dimensional lattice of 
four-dimensional gauge theories connected by bifundamental link fields 
charged under nearest neighbor pairs.  It is motivated by the analogy 
that a gauge symmetry in five dimensions is similar to having an independent 
four-dimensional gauge symmetry at each point along the extra dimension.  
Each of the bi-charged link fields has a non-vanishing expectation value
$f_i$ that breaks the adjacent groups down to a smaller subgroup.
With varying $f_i$ along the lattice, this construction describes a 
latticized warped extra dimension with an overall gauge group
given by the subgroup of the lattice left unbroken by the
link VEVs.  Not surprisingly, deconstructed four-dimensional 
Higgsless models~\cite{Foadi:2003xa,Chivukula:2004pk,Georgi:2004iy,Casalbuoni:2005rs} 
face the same challenges from precision electroweak 
constraints~\cite{Chivukula:2004pk,Foadi:2004ps,Georgi:2005dm} 
and the top quark mass~\cite{Foadi:2005hz} as their 
five-dimensional counterparts.  The collider phenomenology in these 
scenarios is also similar to the five-dimensional RS-based 
constructions~\cite{He:2007ge}.

\subsubsection{Holographic composite models}

  In the standard RS models described previously, the Higgs
is added to the theory by hand and is assumed to be confined
to the IR brane.  A more holistic approach within the RS framework 
is to derive the Higgs from the fifth ($A_5$) component of 
a bulk gauge field~\cite{Fairlie:1979at,Manton:1979kb,Hosotani:1983xw,
Hosotani:1983vn,Csaki:2002ur,Burdman:2002se,Scrucca:2003ra}.  
This occurs in 
holographic composite Higgs scenarios~\cite{Contino:2003ve,
Agashe:2004rs}.  
In the dual interpretation of these theories via AdS/CFT, the Higgs 
field emerges as a composite state that is an approximate Nambu-Goldstone 
boson.  For obvious reasons, these models are also sometimes 
called holographic NGB Higgs or gauge-unified Higgs scenarios.\bigskip

  The minimal viable holographic composite Higgs model is based on
the RS spacetime with a bulk electroweak gauge group 
$SO(5)\times U(1)_X$~\cite{Agashe:2004rs}.  
The fifth dimension is bounded by a UV brane at $z=R$ and an 
IR brane at $z=R'>R$.  Boundary conditions for the bulk gauge
fields are chosen so that on the UV brane the bulk gauge group 
is broken down to the familiar $SU(2)_L\times U(1)_Y$ of the Standard Model
with $Y=(t^3_R + X)$, while on the IR brane the remaining electroweak 
gauge symmetry is $SO(4)\times U(1)_X$.  This gauge structure is similar to 
Higgless models.  The important difference, however, is that the
larger group $SO(4)=SU(2)_L\times SU(2)_R$ is left unbroken 
on the IR brane, so that the electroweak $SU(2)_L\times U(1)_Y$ gauge 
symmetry is unbroken everywhere.  

  The low-energy particle spectrum from this gauge structure
consists of massless $SU(2)_L\times U(1)_Y$ gauge bosons 
prior to electroweak symmetry breaking, a collection of charged and
neutral massive vector KK modes, and a complex scalar $SU(2)_L$ 
doublet localized towards the IR brane.  This doublet comes from 
the $A_5$ components of the gauge fields corresponding to $SO(5)/SO(4)$.
Because the adjoint of $SO(N)$ has $N(N-1)/2$
elements, there are four real broken generators in $SO(5)/SO(4)$;
just enough to make up a complex doublet.
Recall that in standard bulk RS and Higgless scenarios
the $A_5$ components of the gauge fields do not lead to physical excitations.
Here, those degrees of freedom from $A_5$ which are not in $SO(5)/SO(4)$ do not
lead to physical excitations either. 

However, in general the presence or absence of a physical
$A_5$ mode depends on its boundary conditions.  
The boundary conditions for all the gauge fields 
in standard bulk RS and Higgsless models, as well as the non-$SO(5)/SO(4)$
components in the holographic case, are either 
Dirichlet ($\left.A_5\right.=0$) on both branes 
or Dirichlet on one brane and Neumann on the other.  The only solution 
in this case is $A_5 = 0$.
In the holographic case, for an appropriate 
choice of gauge the bulk equation of motion for $A^5$ 
is $\del_z^2(A_5/z) = 0$~\cite{Csaki:2005vy,Cacciapaglia:2005pa}.
Both boundary conditions for the $A_5$ generators of $SO(5)/SO(4)$ 
in this gauge are Neumann conditions 
$\left.\del_z(A_5/z)\right|_{z=R,R'} = 0$. These to 
allow for a profile $A_5 \propto z \propto e^{k|y|}$.\bigskip

  Electroweak symmetry breaking in holographic composite Higgs models is
driven by the complex scalar doublet, which is identified with 
the Higgs field~\cite{Agashe:2004rs}.
At tree level this field has no potential at all,
but loop corrections predominantly from the
top quarks and electroweak gauge bosons generated it. The absence of 
a classical 
potential is due to the five-dimensional $SO(5)$ gauge symmetry.  
To obtain a non-vanishing potential and
induce electroweak symmetry breaking, simultaneous $SO(5)$-breaking effects 
from both branes are required, 
because symmetry breaking at only one of the two branes does not lead to a 
physical $A_5$ mode.  This non-local simultaneous  symmetry breaking
does not allow quadratic corrections to the loop-generated
potential, so the scale of electroweak symmetry is set
by the natural mass scale $M_{KK} = k\,e^{-\pi\,kR}$ on the IR brane.

  The origin of this scalar doublet can also be understood in terms of
the four-dimensional dual theory suggested by the AdS/CFT 
correspondence~\cite{Contino:2003ve,Agashe:2004rs}.
Under this correspondence, the bulk $SO(5)$ gauge symmetry is interpreted 
as an approximate global symmetry of a strongly-coupled conformal field theory,
with a weakly-coupled gauged $SU(2)_L\times U(1)_{Y}$ subgroup.  The explicit 
breaking of $SO(5)$ down to $SO(4)\times U(1)_X$ on the IR brane 
corresponds to a spontaneous breaking of this symmetry by confining dynamics 
at the $\tev$ scale (corresponding to the truncation of $AdS_5$ by 
the IR brane).  In this context, the Higgs doublet arises as a composite state
that is an approximate NGB of the $SO(5)/SO(4)$ global symmetry breaking.  
%
The advantage of the five-dimensional formulation is that the theory
is weakly coupled, allowing for a reliable calculation of the Higgs
potential and the phenomenological constraints on the model.\bigskip

  Fermions in holographic composite Higgs models propagate in the bulk, 
and therefore have to be embedded in full representations of
the $SO(5)$ bulk gauge symmetry~\cite{Agashe:2004rs,Agashe:2005dk}.  
One of the simplest viable options for the Standard Model quarks is to put 
each left-handed quark doublet into a ${5}$, 
each right-handed up singlet into a separate
${5}$, and each right-handed down singlet into a 
${10}$~\cite{Contino:2006qr,Carena:2007ua}, where 
the ${5}$ representation of $SO(5)$ 
is the fundamental while the ${10}$ is the adjoint.
Since the Standard Model fermions only make up a fraction of 
these larger representations,
the model also contains exotic quark states, which in this case
have electric charges $5/3,\,2/3,\,-1/3$.  Boundary conditions for
these fermions can be chosen so that only the Standard Model states have  
massless zero modes prior to electroweak symmetry breaking.  

  Just like in the standard bulk RS scenario, the masses of the zero mode 
fermions are generated by the expectation value of the Higgs field.  
Their precise values depend on the overlap of the fermion wave functions 
with the IR-localized Higgs profile.  In turn, these wave functions 
are determined by explicit bulk and IR brane mass terms along with 
brane-localized kinetic terms~\cite{Agashe:2005dk,Contino:2006qr}.  
The first and second 
generation fermions can be made light by arranging their wave functions 
to peak in the UV, while the third generation states should be localized 
nearer to the IR brane.  

Beyond simple localization, the Higgs coupling 
to fermions is also restricted by the bulk gauge
symmetry: since the Higgs consists of $A_5$ components of
a bulk gauge field, the underlying Yukawa couplings are set by
the gauge coupling.  Gauge couplings only connect states within
the same representation of the gauge group, while the Standard Model fermions reside
in different representations in the 
${5}\oplus{5}\oplus{10}$ embedding described above.  
However, a mixing between different 
multiplets is induced by the IR brane mass and kinetic terms, 
which need only respect $SO(4)\times U(1)_X$ instead of the full $SO(5)$ 
bulk gauge symmetry.  Together, the bulk masses and brane terms can give 
rise to an acceptable pattern of Standard Model fermion masses, provided we are willing 
to tune the model parameters accordingly.\bigskip

As usual, the most stringent current bounds on holographic Higgs scenarios
come from precision electroweak and flavor measurements.  Both sets 
of constraints are very sensitive to the localization of 
third generation bulk fermions.  Most importantly, the one-loop potential
for the $A_5$ Higgs doublet must spontaneously break the electroweak symmetry.
This potential depends primarily on the bulk electroweak 
gauge coupling, the IR brane scale, and the profiles of the left-
and right-handed top quarks~\cite{Agashe:2004rs}.  Detailed analyses 
of the potential find a range of parameters that lead to acceptable 
symmetry breaking~\cite{Contino:2006qr,Carena:2007ua,Medina:2007hz}.  

But even with a reasonable electroweak vacuum, precision electroweak 
bounds further limit the allowed model parameters.  The most dangerous 
effect is is typically a too-large shift in the $Zb_L\bar{b}_L$ coupling.  
This can be fixed up by embedding the Standard Model quarks into 
the ${5}\oplus{5}\oplus{10}$ representations described
above which leads to an enhanced custodial protection~\cite{Agashe:2006at}, 
and by carefully arranging the $Q_{L3}$ and $u_{R3}$ profiles.
IR brane scales as low as $M_\text{KK}\simeq 1\,\tev$ then turn out to be 
acceptable~\cite{Contino:2006qr,Carena:2007ua,Medina:2007hz,
Carena:2006bn,Carena:2007tn}.

Of course, flavor is another problem.  With generic flavor mixing
parameters, an IR scale above $20\,\tev$ is found to be needed,
even when localization is used to set the light quark masses
and reproduce the CKM mixing matrix~\cite{Csaki:2008zd,Agashe:2009di}.
Lower IR scales are possible in restricted parameter regions, 
or when we postulate an appropriate flavor symmetry, as we observe 
it in the Standard Model.\bigskip

Collider signals in holographic composite Higgs models are often similar
to the bulk RS scenario (as for example in~\cite{Agashe:2009bb}), though  
the Higgs itself may show significant deviations from SM behavior
which may challenge conventional Higgs searches
\cite{silh,Bellazzini:2009xt}.
The IR scale can be low enough that KK modes
of the gluon and electroweak gauge bosons are potentially observable
at the LHC.  These states couple most strongly to the top quark states
localized near the IR brane, but only very weakly to the light fermions.  
The extended fermion representations required to fill out
$SO(5)$ multiplets and satisfy electroweak constraints lead to exotic
top-like quark states with charges $5/3,\,2/3,\,-1/3$~\cite{
Contino:2006qr,Carena:2007ua}.  In some parameter regions 
one or more of these heavy quarks can be lighter than 
$1\,\tev$~\cite{Contino:2006qr,Carena:2007ua,Carena:2007tn}.
These states can be produced singly or in pairs, and they tend to
decay primarily to top quarks and longitudinal $W$ or $Z$ gauge bosons,
such as $T_{5/3} \to t\,W^+$.  Final states with multiple gauge bosons
can arise from the decay of one exotic to another when they are split
in mass, $T_{2/3} \to T_{5/3}W^- \to t\,W^{+}W^{-}$~\cite{Contino:2006qr}. 
Heavy KK gluon states can have significant decays to pairs of relatively light 
exotic quarks as well~\cite{Carena:2007tn}.

\subsection{Hidden sector models}
\label{sec:models_hidden}

  All the models we have described up to now are motivated by specific 
shortcomings of the Standard Model, such as the instability of the 
electroweak scale or the failure to account for the observed cosmology.  
These models lead to a very diverse set of collider signatures.  
However, it is entirely possible that new physics will turn up at the 
LHC that is different from what we have already thought of, or unrelated 
to the hierarchy or cosmological problems of the Standard Model.  
If we focus our search techniques too narrowly on specific models, 
there is always a danger of missing the unexpected.  
For this reason it is useful to consider possibilities for phenomena 
that give rise to particularly unusual collider signatures
or that might be challenging to discover using standard search techniques.

  A very broad class new physics of this type falls into the paradigm
of a \emph{hidden valley}~\cite{Strassler:2006im,Strassler:2006qa}.
As we will see, some of the models we have already discussed can be 
classified as examples of hidden valleys.
A generic hidden valley model consists of a visible sector containing
the Standard Model, a hidden sector with suppressed couplings to the 
visible sector, and a set of heavy mediators that couple directly to 
both sectors.  

This framework gives rise to a novel source of new signals
at high energy colliders, which we illustrate schematically 
in the left panel of Fig.~\ref{fig:hiddenvalley}.
Collisions of Standard Model particles create mediator particles, 
which then decay some part of the time to the hidden sector.  
If the hidden sector states are heavier than some of the Standard Model 
particles 
(for which there is strong motivation from cosmology~\cite{Strumia:2006db}), 
they can potentially decay back to the Standard Model.  
The decay products may be prompt or delayed and could have 
a large boost if the mediators are much heavier than the hidden states.  
Moreover, at energies well below the masses of the mediators, 
the mediators can be integrated out giving rise to new suppressed 
and often higher-dimensional direct couplings between the visible 
and hidden sectors.\bigskip

  There is a great variety of hidden valley models and we do not have
room to list all their signatures,
so we will restrict ourselves to a few simple examples.
As a first example, consider a situation where the Standard Model is 
the visible sector, the hidden sector consists of a $SU(N_v)$ ($N_v>2$)
non-Abelian gauge theory with two flavors of $v$-quarks
and both sectors are charged under an exotic $U(1)_x$ gauge symmetry
that is spontaneously broken near the $\tev$ scale~\cite{Strassler:2006im}.
The massive $U(1)_x$ gauge boson $Z_x$ will act as a mediator
between the visible and hidden sectors.  Even after applying
the constraints from the LEP experiments, the LHC production cross-section 
for $v$-quarks through an $s$-channel $Z_x$ can be significant.
In the hidden sector, the $SU(N_v)$ gauge theory with two flavors
will confine at the energy scale $\Lambda_v$.  If both flavors of $v$-quarks 
are lighter than $\Lambda_v$, direct $v$-quark production will lead 
to $v$-hadrons which decay in hidden-sector cascades down to 
the lightest $v$-pion.  This pion will in turn decay back to the Standard Model
through an off-shell $Z_x$ with a strong preference for heavier 
flavors (due to chirality) such as $b\bar{b}$ 
and $t\bar{t}$~\cite{Strassler:2006im,Strassler:2008fv}.  
If only one of the two $v$-quark flavors is lighter than $\Lambda_v$, 
there can also be decays from the hidden sector to the Standard Model 
that go to both light and heavy flavors~\cite{Strassler:2006im}.  
In both cases the decays can be prompt or displaced, and showering
in the hidden sector can generate relatively spherical
events with a high particle multiplicity~\cite{Han:2007ae}.\bigskip

\begin{figure}[t]
\begin{center}
  \includegraphics[width = 0.4\textwidth]{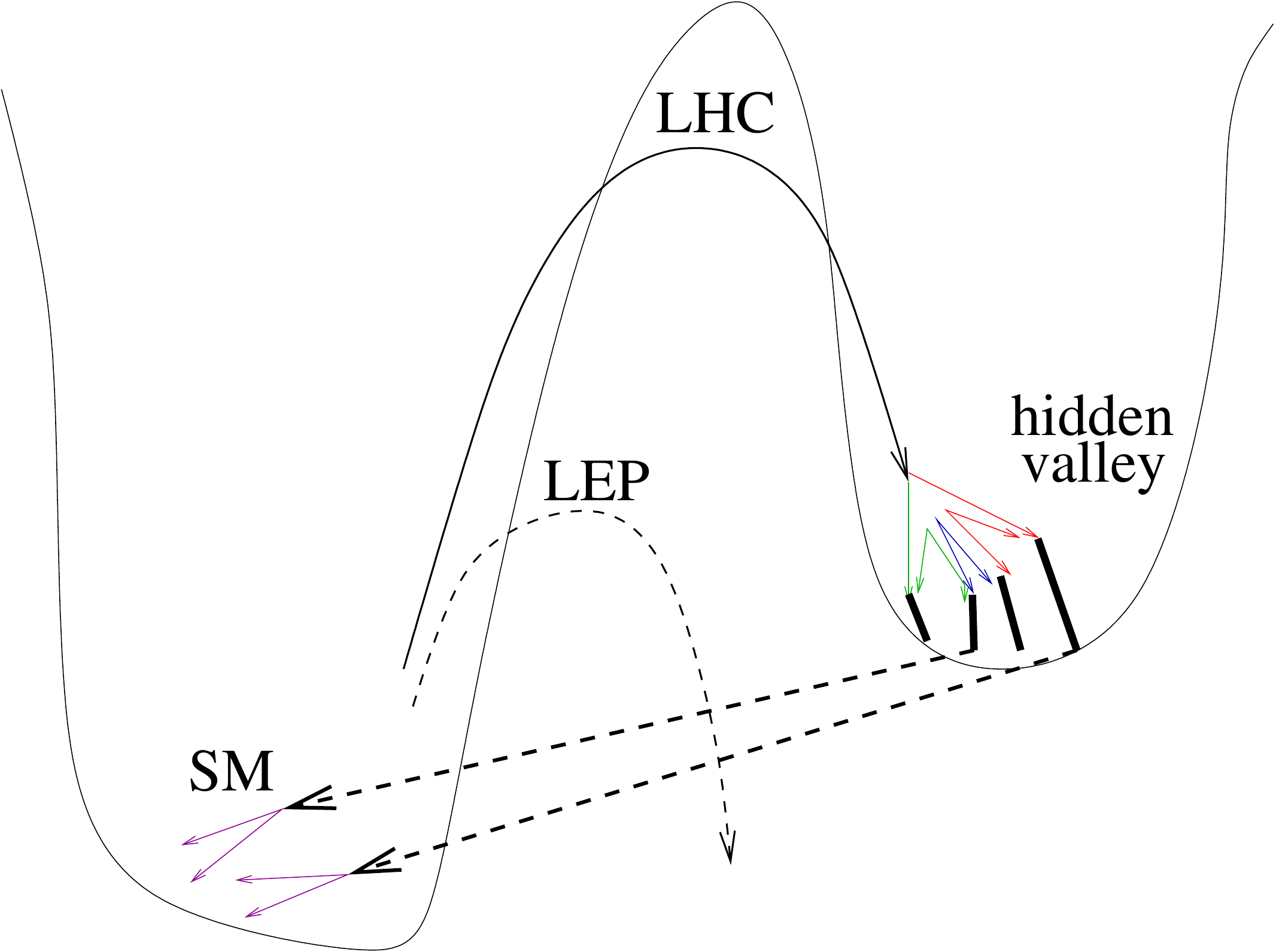}
  \hspace*{0.05\textwidth}
  \includegraphics[width = 0.4\textwidth]{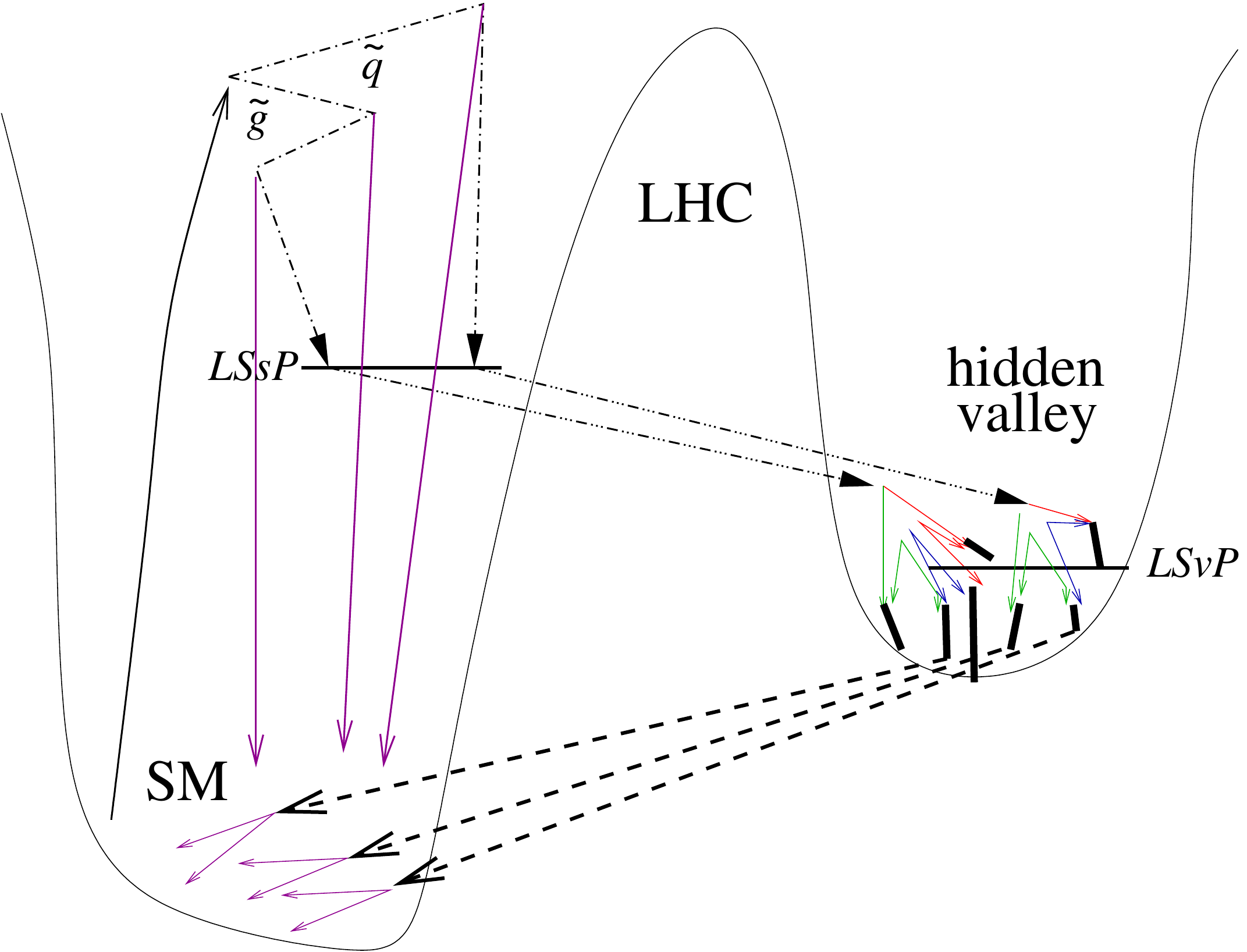}
\end{center}
\vspace*{0cm}
\caption{Schematic representation of hidden valley scenarios,
from Ref.~\cite{Strassler:2006qa}. Left:
standard situation, where a massive mediator is produced
and decays to the hidden valley, with the hidden valley
states decaying back to the Standard Model. Right:
the hidden and visible sectors share a discrete
symmetry, forcing some heavy visible states to decay to the hidden sector.}
\label{fig:hiddenvalley}
\end{figure}

  A simple supersymmetric example of an hidden valley is
described in Section~\ref{sec:models_xssm}, where the visible sector
is the MSSM and the hidden sector consists of a $\gev$-scale 
supersymmetric $U(1)_x$ Abelian Higgs model.  In this case the mediator 
is not a specific particle, but rather the kinetic mixing between the
$U(1)_x$ and $U(1)_Y$ gauge groups. Such kinetic mixing can
be induced by massive states charged under both gauge groups, so this
scenario can be thought of as a very heavy  mediator sector.
It is qualitatively different from the previous example because
both the hidden and visible sectors share an $R$-parity symmetry.
Supersymmetric production in the visible sector will lead to a
cascade decay down to the lightest MSSM $R$-odd state.  This state
will subsequently decay to the lighter hidden sector since $R$-parity
forbids it from decaying further within the visible 
sector~\cite{Strassler:2006qa}.  We illustrate
the decay process schematically in the right panel of 
Fig.~\ref{fig:hiddenvalley}.  
The same qualitative behavior can also arise in any kind of scenario
where the visible and hidden sectors share a discrete 
symmetry~\cite{Strassler:2006qa}.\bigskip

  As a third and particularly unusual example of a hidden valley, 
we take the visible sector to be the Standard Model and the hidden sector
 to be a pure $SU(N_v)$ gauge theory. The mediator sector is a set of 
heavy fermions $Q$ with mass $m_Q\gtrsim 300\,\gev$ that are charged 
under both the Standard Model gauge groups and $SU(N_v)$.  Let us assume 
further that the confinement scale $\Lambda_v$ of the $SU(N_v)$ gauge theory 
is significantly less than its fermion masses, 
$\Lambda_v \ll m_Q$~\cite{Strassler:2006qa,Kang:2008ea,Burdman:2008ek}.  
The mediator fermions in this case are sometimes called \emph{quirks},
and the hidden sector consists of a spectrum of $v$-glueball 
states~\cite{Kang:2008ea}. 
Loops of quirks generate an effective coupling between pairs of
$v$-gluons and pairs of photons or gluons~\cite{Strassler:2006qa,Kang:2008ea}
\begin{equation}
-\mathscr{L} \supset \frac{g_\text{SM}g_v}{16\pi^2m_Q^2}
(F_{\mu\nu}^\text{SM})^2
(F_{\alpha\beta}^v)^2.
\end{equation} 
This operator allows for a very slow decay of $v$-glueballs to the 
Standard Model with a decay length on the order of
\begin{equation}
c\tau \sim (10\,\text{m})\,\lrf{\Lambda}{50\,\gev}^9\lrf{m_Q}{\tev}^8.
\end{equation}
Thus, the $v$-glueballs tend to be stable on collider scales.

At the LHC quirk-antiquirk pairs will be produced directly 
through their couplings to the Standard Model gauge bosons.  Once created,
the quirk and antiquirk will be connected by a tube of
confining $SU(N_v)$ flux, much like the color connection 
between a quark and an antiquark in QCD~\cite{Kang:2008ea}.  
In contrast to QCD, however, it is nearly impossible for this flux tube 
to break through the nucleation of a $Q\bar{Q}$ pair.  
Creating a pair of quirks from the vacuum takes an energy of 
about $2m_Q$, which is much larger than the typical reduction of 
the flux tube energy of $\Lambda_v^2/m_Q$.
The flux tube connecting the quirks can be modelled as a string.
Quirks produced at the LHC will have a typical kinetic energy 
on the order of $m_Q$.  They will initially fly apart, but the string
will pull them back together after stretching to a maximal length on 
the order of
\begin{equation}
L \sim \frac{m_Q}{\Lambda_v^2} \sim (10\,\text{m})\lrf{m_Q}{\tev}
\lrf{100\,\text{eV}}{\Lambda_v}^2 = (1\,\text{fm})\lrf{m_Q}{\tev}
\lrf{10\,\text{GeV}}{\Lambda_v}^2.
\end{equation}
Each time the quirks are pulled back together there is a small probability
for them to annihilate, but they will usually go through many oscillations
before doing so~\cite{Kang:2008ea}.\bigskip 

The qualitative collider signatures of quirks depend primarily on the
length of the string connecting them and their relative couplings to
the $SU(N_v)$ and Standard Model gauge bosons~\cite{Kang:2008ea}.
Macroscopic strings ($L \gtrsim 10^{-3}$~m) can lead to a pair of
quirk tracks with a distinctive curvature in the $r\!-\!z$ plane
(where $z$ is the beam axis).  Quirks connected by mesoscopic strings
with $10^{-10}~\text{m} \lesssim L \lesssim 10^{-3}~\text{m}$ cannot
be individually resolved, but produce a single track.  Interactions of
the quirk system with matter in the detector in this length regime can
potentially prevent them from annihilating.  Microscopic quirks ($L
\lesssim 10^{-10}$~m) will oscillate rapidly, and emit soft pions,
photons, and $v$-glueballs.  This emission can reduce the energy of
the quirk-string system, allowing the string to shrink enough for the
quirks to annihilate with close to zero kinetic energy. These decays
can produce leptons or photons that can be extracted from the
backgrounds.  Reconstructing the invariant mass of these decay
products will yield a peak at $2m_Q$.  This peak can be potentially be
correlated with a distinctive pattern of QCD or electromagnetic
radiation from the soft emission during the string
oscillation~\cite{Kang:2008ea,Harnik:2008ax}.\bigskip

Yet another class of hidden valley scenarios are
\emph{unparticles}~\cite{Georgi:2007ek,Georgi:2007si}.  The hidden
sector is assumed to be an approximately conformal field theory,
without a characteristic mass scale.  As such, the fields in the
hidden sector do not have an obvious particle interpretation.
Coupling the unparticle sector to the Standard Model through
irrelevant higher dimensional operators can produce interesting
phenomenological effects.  For example, consider coupling an
unparticle operator $\mathcal{O}^{(d_U)}$ of mass dimension $d_U$ to
fermions $f$ and $f'$ according to
\begin{equation}
-\mathscr{L} \supset \frac{1}{M^{d_U-1}}\bar{f}'f\,\mathcal{O}^{(d_U)},
\end{equation}
where $M$ is a large mass scale.
This operator allows the decay $f \to f'+$~unparticles, provided
$f$ is heavier than $f'$.  The characteristic feature of the invisible
unparticle final state is that the kinematic distribution of the
outgoing $f'$ fermion will correspond to the decay of $f$ to $f'$
and $d_U$ invisible particles~\cite{Georgi:2007ek}.  What makes this 
slightly unusual is that $d_U$ need not be an integer.  

Although such an apparent non-integral number of final states might
strange, a similar behavior arises in QCD and
QED~\cite{Cacciapaglia:2007jq,Neubert:2007kh}.  To understand the
unparticle behavior, suppose we add a very small mass term (or
relevant operator) to the unparticle theory that breaks the conformal
symmetry at a very low energy~\cite{Fox:2007sy}.  
As explained below, such a term is inevitable once the unparticles have
relevant interactions with the SM.
This breaking allows
us to interpret the theory in terms of light particles.  In a process
whose characteristic energy is much larger than the unparticle sector
mass, any amplitude involving an unparticle leg will be significantly
corrected at higher orders by the multiple emission of soft or
collinear unparticle radiation.  Similar corrections are present both
in QCD and QED and are cut off by the characteristic mass scales,
$\Lambda_\text{QCD}$ or
$m_e$~\cite{Cacciapaglia:2007jq,Neubert:2007kh}.  Unparticles differ
primarily in that the mass cutoff is taken to zero, so that the soft
emissions never stop.  Put another way, for an unparticle process,
such as the decay to unparticles described above, the funny kinematic
distribution can be understood in terms of the repeated showering of
unparticles.  
 
Interactions between the Standard Model and the unparticle sector 
(which are inevitable if unparticles are relevant at colliders) will 
lead to a breaking of conformality.  
For instance, unparticles
can couple to the Higgs through the operator~\cite{Fox:2007sy}
\begin{equation}
-\mathscr{L} \supset \frac{\lambda}{M^{d_U-2}} \; |H|^2 \; \mathcal{O}^{(d_U)}
\end{equation}
This coupling generates a potentially relevant operator when the Higgs
develops an expectation value (and may have an interesting back-reaction
on the Higgs VEV~\cite{Delgado:2007dx} as well).  As a result, we generally
expect a
mass gap in unparticle models, and this can open up decays back to the
Standard Model as in hidden valley models~\cite{Strassler:2008bv}.
The mediator sector responsible for coupling unparticles to the
Standard Model will also generate new higher-dimensional operators
involving only Standard Model fields.  Constraints on such operators often imply
that unparticle effects will be very difficult to observe, unless they
acquire a mass gap and decay back to the Standard
Model~\cite{Grinstein:2008qk}.


\newpage

\section{Signatures}
\label{sec:sig}

A wide and diverse collection of models have been put forward as
candidates for new physics beyond the Standard Model.  These  
models give rise to an equally vast multitude of new signals at the
LHC.  Both the ATLAS and CMS detectors have been designed as
multi-purpose instruments capable of measuring nearly the entire range
of possibilities.  They reconstruct particles coming from the
proton-proton interaction point, determine their four-momenta, and
search for kinematic patterns which cannot be explained by the
Standard Model.  The great advantage of the LHC stems from its highly
energetic proton beams. They provide an enormous luminosity, a large
density of gluons inside the protons at the relevant energy scales,
and a large coupling $\alpha_s$ which governs the dominant
production cross sections.\bigskip
 
The greatest challenges for the LHC are closely connected with these
advantages: any high-energy, high-luminosity proton--proton collider
will produce huge numbers of massless quark and gluon jets yielding
too many events to even write to tape.  Therefore we need to
determine early on in the recording chain which kinds of events we
would like to store and analyze in detail.  Among the subsets of
events that are kept to be studied, signal events must be extracted
from samples often containing many more background events.  For these
reasons it is enormously useful to have concrete models of new physics
to inform us of the most promising directions for a new discovery. On
the other hand, we have to ensure that we do not weaken our
experimental program due to a theoretical bias, for example by
focussing only on specific models.

  The first step towards separating interesting signals from uninteresting 
background events is the trigger, which favors photons,
leptons, heavy quarks, and missing energy -- features which we expect
from new-physics events.  Even so, some models are particularly challenging for
the LHC triggers such as new strongly-interacting stable particles 
or the hidden--valleys described in Section~\ref{sec:models_hidden}.
Fortunately, the ATLAS and CMS triggers are flexible, and can be adjusted 
if we find interesting new features in the data.  
Note that triggers are much less of an issue at electron--positron colliders
which have a much cleaner environment, and are also much less rigid
at the Tevatron simply because of the comparatively smaller number
of events.

  The second challenge is to distinguish new signals from background 
within the experimental resolution. To be able to say something about
a final-state particle, it has to pass basic acceptance cuts reflecting
the geometry of the detector components.  For the particles that are
accepted, we will likely have to study specific kinematic features
to extract to extract an often relatively small number of new-physics
events from the Standard Model, and QCD in particular, as we understand it.  
As an example, the LHC will produce roughly one million times as many two-jet
events as top quarks, and still a factor $100$ fewer light Higgs
bosons.  Even when an unexpected kinematic feature is seen,
we will need to carefully rule out the possibility that we are
simply seeing very rare manifestations of Standard Model physics 
or unexpected detector effects before we can claim a discovery of 
new physics.

  If a discovery is made, we will have to develop a strategy
to identify the new physics that is observed.  The difficulty in
doing so depends on what is seen: 
resonances decaying to pairs of leptons, photons, or jets are 
relatively easy to study;
models involving a dark matter candidate which leads 
to missing transverse energy in the detector are considerably harder; 
particles that decay to many light jets without an obvious resonance 
structure are a serious challenge.\bigskip

These aspects of LHC detectors, Standard Model backgrounds, and the
specific signatures predicted by theory will all have to be carefully
considered in our quest for new physics.  An enormous asset for this
is our expertise developed in Tevatron analyses, which have very
successfully circumvented most of these challenges, albeit in a more
friendly QCD environment.  In this section we will discuss LHC signals
of new physics beyond the Standard Model, and we will describe how to
go about measuring them.  As much as possible, we will present these
signals in a general way without concentrating too much on any
specific model.  However, since studies of these signals often involve
very specific questions, it is impossible to answer them in a
completely model-independent way.  The frequent appearance of
supersymmetry reflects a long history of physics sociology, but
essentially all of the results can be easily applied to other
consistent TeV-scale completions of the Standard Model.

\subsection{QCD features of new particle production}
\label{sec:sig_qcd}

If new physics at the TeV scale consists of a strongly and a weakly
interacting sector, the strongly-interacting particles will likely be
produced at the LHC in much larger numbers.  In supersymmetry, for
example, all the superpartner production rates have been calculated at
next-to-leading order in QCD~\cite{prospino} and the weak
coupling~\cite{susy_weak}, and are shown in
Fig.~\ref{fig:sig_prospino}.  Strongly interacting particles evidently
have much larger production rates in most cases.  Particles with
masses up to several hundred GeV will be produced at rates around
$10$--$100$~pb, corresponding to $10^4$--$10^5$ events in the first
$1~\ifb$ of integrated luminosity. This number is the main feature of
the LHC compared to the Tevatron --- if such particles indeed exist we
can do more than just produce them and observe a signal for physics
beyond the Standard Model; we can also study their properties in their
decays.  To do so, it essential that we understand the production
processes involved, as well as the QCD backgrounds from the Standard
Model. For pedagogical reviews of QCD processes at hadron colliders we
refer the reader to Refs.~\cite{Plehn:2009nd,qcd_reviews}.\bigskip

\begin{figure}[t]
\begin{center}
  \includegraphics[width=0.60\textwidth]{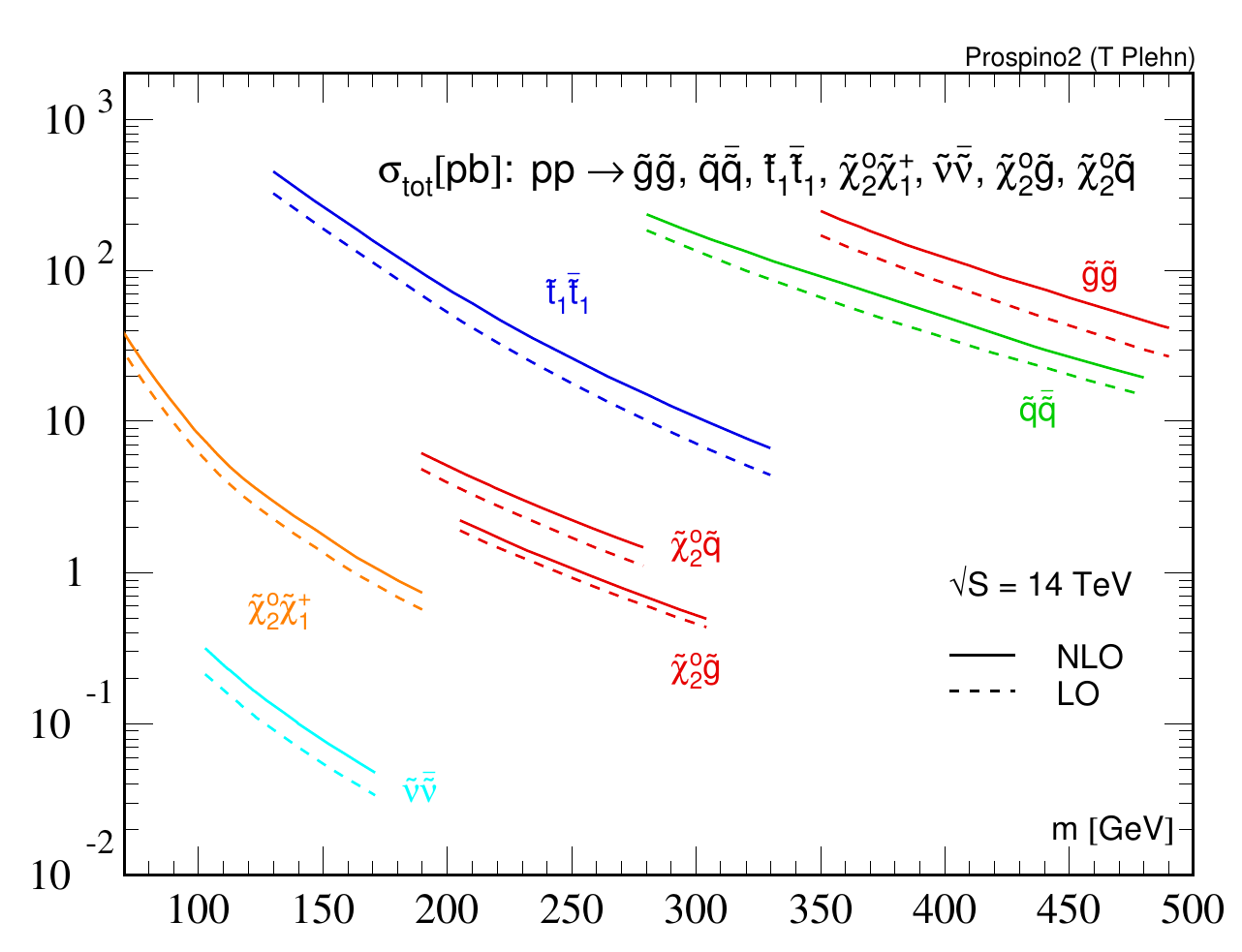}
\end{center}
\caption{ LHC production cross sections for supersymmetric particles
with $\sqrt{s}=14\,\tev$.
All rates are shown as a function of the average final-state mass.
Figure from Ref.~\cite{prospino}.}
\label{fig:sig_prospino}
\end{figure}

The main tree--level LHC production processes for strongly interacting
states are determined almost entirely by their spins and color
charges.  New colored scalar particles most often arise as fundamentals
(scalar quarks~\cite{Martin:1997ns}) or adjoints (scalar
octets~\cite{mrssm,mrssm_early,Plehn:2008ae,Choi:2008ub,color_octets}) 
under $SU(3)_c$, and couple to gluons correspondingly. The couplings 
to one or more gluons are related through gauge invariance.  Their 
couplings to quarks need not be mediated by QCD at tree level, 
and can therefore be suppressed.  Leptoquarks, as discussed in
Section~\ref{sec:sig_lepto}, are another example in this class, and
from a production point of view are equivalent to squarks even
though their decay channels can be much different.

New strongly-interacting fermions also arise most often as fundamental
(heavy quarks~\cite{Appelquist:2000nn}) or adjoint 
(gluinos~\cite{Martin:1997ns} color
representations.  A color adjoint fermion can be either Dirac or
Majorana, since the adjoint of $SU(N)$ is a real representation.  In
the case of minimal ($\mathcal{N}=1$) supersymmetry, there is only a
single color adjoint fermion gluino with two on-shell degrees of
freedom to match those of the gluon, so the only possible mass term 
is Majorana.  A supersymmetric Dirac gluino requires the introduction of
a second color adjoint fermion and its scalar gluon superpartner (or
sgluon~\cite{Plehn:2008ae}), either by extending the amount of
supersymmetry in the gauge sector
($\mathcal{N}=2$)~\cite{Choi:2008ub} or as an ad-hoc modification of
the MSSM~\cite{mrssm}. As discussed in Section~\ref{sec:models_mssm},
squarks and gluinos in the MSSM share a new
$q$-$\tilde{q}$-$\tilde{g}$ interaction which is proportional to the
quark's color charge, and must be present for the gauge interactions
to be supersymmetric~\cite{martin_vaughn}.

Last but not least, strongly-interacting massive vector bosons with
adjoint color charge (heavy gluons) arise in many models.  Such new
states are generically subject to strong experimental constraints from
flavor-changing neutral currents. These can be avoided by requiring
large heavy gluon masses or a discrete $\mathbb{Z}_2$ charge of new
particles. This setup then determines whether a single heavy gluon can
couple to Standard Model quarks and gluons or if there is only a 
three-particle vertex involving two heavy adjoint vectors
and one Standard Model gluon.\bigskip

  QCD couplings can also enter into loop-level or higher-dimensional
operators that contribute to the production rates for new particles.
This feature plays a very important role for the rate of Higgs boson
production at the LHC.  In contrast, for states that are charged under
the color group such operators are typically much less important for production
than the direct coupling to gluons.  Nevertheless, these operators can
dominate the decay signatures when direct QCD-mediated decays are forbidden
by phase space.

  Theories of new physics can sometimes produce gluon couplings to 
new strongly-interacting states that are not flavor-diagonal.
For example, supersymmetric squarks can mix between generations 
to produce non-diagonal $q$-$\tilde{q}$-$\tilde{g}$ coupling.
An obvious question is whether we can look for ``rare production processes''
at the LHC that involve flavor mixing, in analogy to rare decays in $B$ physics.
In most cases the answer appears to be negative due to the very strong
constraints on new sources of flavor mixing associated with the gluon.
Even considering modes enhanced by factors of $10$--$100$, 
such as Higgs or direct top production, it is not clear whether rare
flavor-violating production channels are a promising
path for discovery~\cite{rare_prod,direct_top,single_top_us}.\bigskip

  When computing the production rates for new strongly-interacting
heavy states at the LHC, QCD effects tend to spoil our theory predictions. 
First of all, even at high energy scales well above the $W/Z$ mass
the strong coupling $\alpha_s$ is not small enough. 
This is not immediately obvious, since our fixed-order perturbation
series can be organized in powers of $\alpha_s \sim 1/10$ 
would suggest that unknown higher orders should lead to uncertainties 
in the few--percent range.  Unfortunately, there can arise large
infrared logarithmic enhancements to higher-order terms in the fixed order
expansion that ruin its apparently rapid convergence.\bigskip

A historically well-known (but not quite devastating at the LHC)
example of this commonly occurs if we produce two strongly 
interacting heavy states at once.  One or many gluons can be exchanged
between these two particles if they are moving slowly relative
to each other.  Since the steep energy dependence of the
gluon density inside the proton pushes the phase space for heavy
particles towards the threshold region, this effect will be relevant at
the LHC. In the limit of small relative velocity between the new states, 
the perturbative series in $\alpha_s$ gets dressed with logarithms 
of the form $\log (1-4 m^2/s)$ which diverge at threshold. 
In processes where these logarithms appear they need to be summed 
or re-summed, depending on the philosophy and prior knowledge of 
the theorist involved.  Such threshold corrections are particularly 
dangerous if the leading-order amplitude vanishes at threshold while the
next-to-leading-order amplitude has a finite limit for 
$\sqrt{s} \to 2m$.  In that case the relative next-to-leading order 
corrections are formally infinite but integrable and are correctly 
described by a Sommerfeld form factor~\cite{sommerfeld}.\bigskip

  A numerically more important effect for the LHC is the soft and collinear logarithms 
that arise from gluons radiated off the incoming partons. 
When we integrate a scattering amplitude (squared) over the phase space of a
radiated gluon, the integrand diverges like $1/p_T$ of the massless
final-state parton. Given a $p_T$ range, this behavior
integrates to a logarithm $\log(p_T^\text{max}/p_T^\text{min})$, where
the upper end is regularized by the finite energy of the incoming
partons. The lower integration limit can be arbitrarily small, implying
that our perturbative series in $\alpha_s$ gets spoiled by
large logarithms of the kind $\log(m/p_T^\text{min})$ with
$p_T^\text{min} \to 0$. These logarithms can be (re-)summed, 
such that radiated jets below $p_T^\text{max}$ are no longer part of the 
specific hard process and instead get absorbed into
the parton densities of the inclusive process.
For example, an $n+1$ jet process with one low-$p_T$ gluon in the final
state is replaced by a hard process with $n$ jets where the soft
gluon is incorporated into the parton shower.  The limiting
value of $p_T$ below which a soft jet is incorporated into
the parton densities is called the factorization scale and
evolves according to the DGLAP equation~\cite{Plehn:2009nd,qcd_reviews}. 
Turning around this argument, parton showers should not be used to 
radiate jets harder than the hard (factorization) scale of the process, 
even though in practice this used to be done by the standard Monte Carlos, 
and their results can often be tuned to agree with data.

  Due to this large logarithmic enhancement, we can deduce that 
jets with a transverse momentum considerably smaller than the masses 
of the particles produced (\ie the hard scale of the process) 
will be ubiquitous at the LHC.  To understand QCD jets at the LHC 
it is therefore crucial that we describe properly the radiation of 
quarks and gluons in phase space regions where the logarithm $\log(m/p_T)$ 
is small, as well as in regions where it becomes large and gets 
(re-)summed.  Appropriate CKKW and MLM schemes 
that \emph{match} between these two regimes have been developed only
recently and are now available as part of several Monte Carlo
generators~\cite{ckkw,mlm,torbjorn_merging}.
\bigskip

\begin{figure}[t]
\begin{center}
  \includegraphics[width=0.99\textwidth]{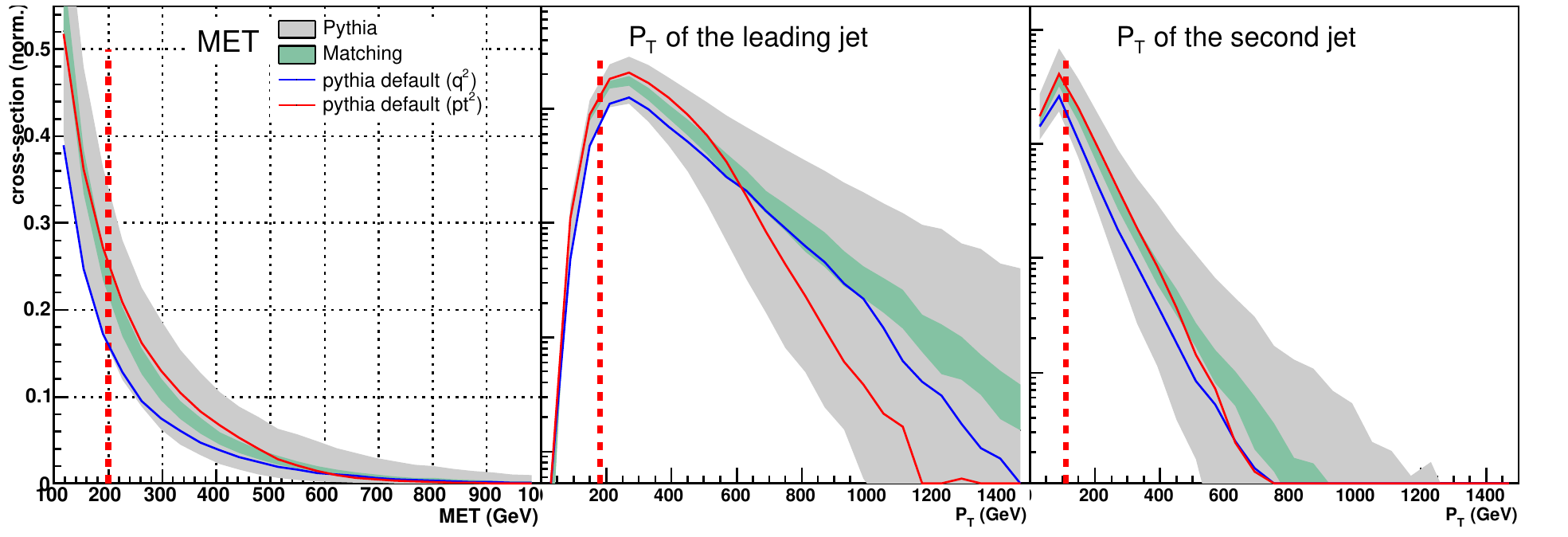}
\end{center}
\caption{Missing energy and jet transverse momenta for gluino pair
  production at the LHC with a gluino mass of 600~GeV, an LSP mass of
  500~GeV and decoupled squarks. The matching scheme applied to the
  green shaded range is MLM~\cite{mlm}. Figure from
  Ref.~\cite{heavy_mg}.
\label{fig:sig_susy_jets}}
\end{figure} 

The behavior of QCD jets in association with heavy new particles has
been computed for supersymmetric squarks and
gluinos~\cite{heavy_skands,heavy_mg} as well as for
sgluons~\cite{Plehn:2008ae}.  In Fig.~\ref{fig:sig_susy_jets} we show the
missing energy spectrum for gluino pair production at the LHC, as well
as the transverse momentum spectra for the two leading QCD jets.  The
different curves correspond to Monte Carlo simulations using
Pythia~\cite{pythia_63}, and an improved estimate using MLM matching.
There are two features we observe.  First, the two Pythia showers that
use virtuality and transverse momentum ordering of collinear jet
radiation give consistent results within an error band which increases
from $10\%$ for soft jets to at best an order-of-magnitude for
$p_{T,j} \gtrsim m_{\tilde g}$.  This widening is expected from the
collinear approximation applied in the parton shower approach used 
in Pythia.  The predictive power of an untuned shower vanishes 
once the jet transverse momenta exceeds the hard scale of the
process. The only reason why the jet radiation from the parton shower
does not simply drop to zero is that they are tuned to explain gauge
boson and top pair production at the Tevatron, roughly probing the
same partonic energy fraction as heavy gluinos at the LHC.

When we instead consistently add all different leading-order 
$n$-jet rates, the result agrees with the parton shower
for small transverse momenta, where the results from such a matching
procedure actually rely on the parton shower. In addition, by definition 
it extrapolates
smoothly and reliably to very hard jets, described correctly by the 
matrix element. In between these two regimes the matching procedure has 
to be set up such that it produces smooth
jet distribution. The corresponding error band,
however, should be taken as a lower limit on the theory error --- any
leading-order prediction for LHC processes at the LHC should not be
expected to have a theory error below about $50\%$
based on the typical size of perturbative QCD corrections.

  From the point of view of LHC phenomenology, 
Fig.~\ref{fig:sig_susy_jets} has a reassuring aspect: in the typical 
range $p_T \lesssim 300$~GeV where phase space still allows for an 
appreciable jet radiation probability, the merged or matched results 
lie right in between the two Pythia curves~\cite{heavy_skands,heavy_mg}.
We generally expect other showering codes such as Herwig to produce similar
results, possibly with more relatively hard jets~\cite{Marchesini:1991ch}.
This means that for the production of heavy particles we do not expect 
hard QCD activity to completely the spoil results obtained with 
(more recent) parton shower tunes. This simply reflects the fact that 
jets in the range of $p_{T,j} \lesssim m_\text{heavy}/2$ accompanying 
the production of heavy states with mass $m_\text{heavy}$ are well 
covered by the collinear approximation used in the parton showers. 
For other processes, such as the $W$+jets background, the answer 
could be very different.\bigskip

\begin{figure}[t]
\begin{center}
  \includegraphics[width=0.45\textwidth]{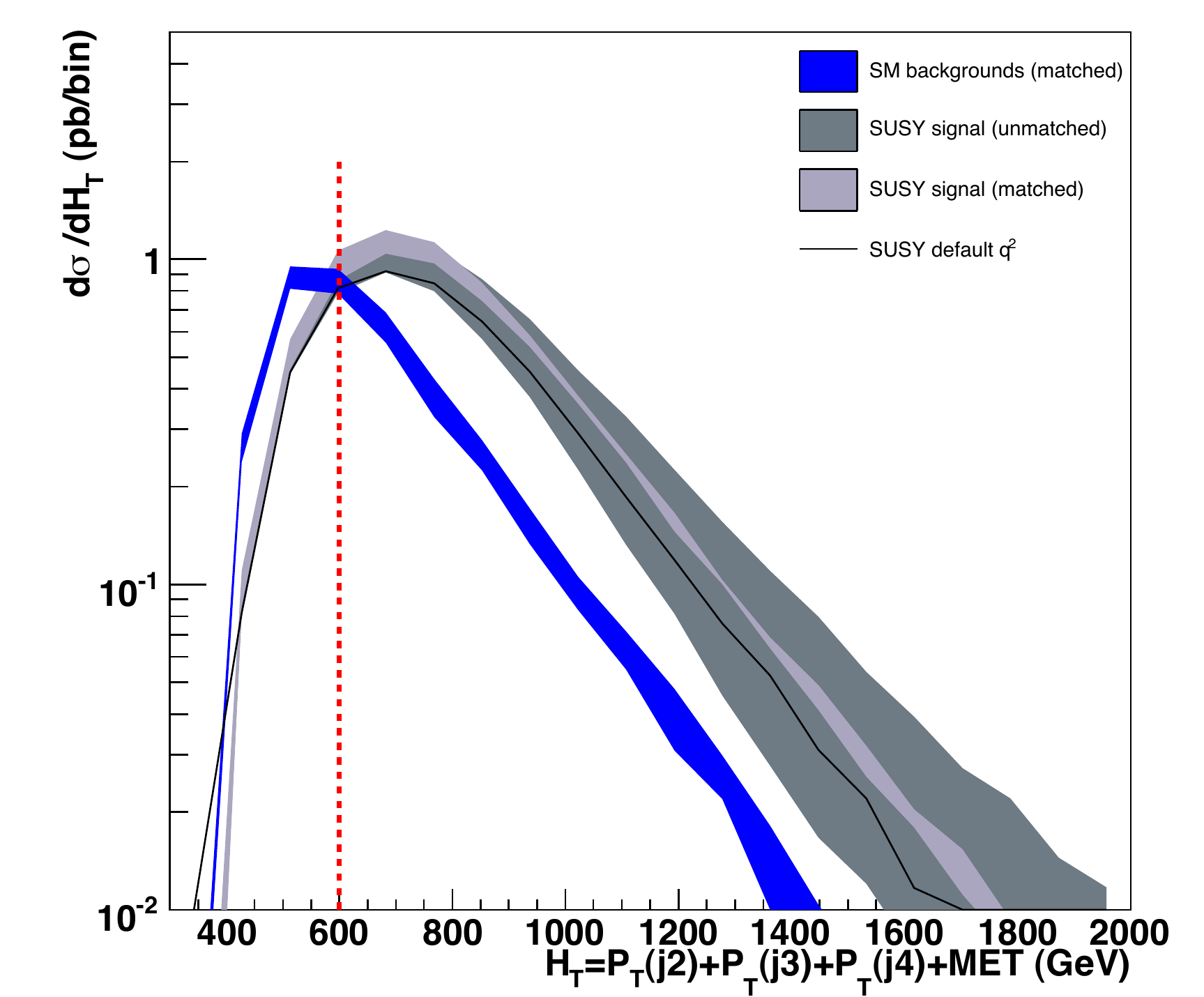}
\end{center}
\caption{$m_\text{eff} = \sum p_{T,j} + \etmiss$ for squark and gluino
  production with jet matching~\cite{mlm}. We also show the
  appropriately matched background prediction. Note that $H_T$ in the
  label does not correspond to our $H_T$ as defined in
  Eq.(\ref{eq:sig_def_ht}), but to our $m_\text{eff}$ defined in
  Eq.(\ref{eq:sig_def_meff}). Figure from Ref.~\cite{heavy_mg}.
\label{fig:sig_ht_jets}}
\end{figure} 

  A second example of the importance of jet matching can be
seen by looking at the distribution of the variable $m_\text{eff}$, 
which is defined as the sum of the
transverse jet momenta plus the missing energy. 
This quantity acts as a measure of the mass of the heavy
particles produced, and we will discuss it in more detail in 
Section~\ref{sec:sig_cascade}.  In Fig.~\ref{fig:sig_ht_jets} we 
show the distribution of $m_\text{eff}$ for a particular MSSM spectrum.
We see from this figure that consistently adding any number of hard jets 
to the final state slightly increases the over-all transverse energy 
of the event. On the other hand, as suggested by the discussion above, 
the result is not (far) outside the error band from the tuned parton shower 
simulation and just slightly higher than the default virtuality-ordered 
Pythia shower.  The signal distribution is clearly different for the 
Standard Model background from $W$+jets background with
its comparably low hard scale. 

As advertised many times, but not explicitly shown in 
Fig.~\ref{fig:sig_ht_jets}, the shape of
the $H_T$ or $m_\text{eff}$ distributions for Standard Model
backgrounds to new heavy particles shifts significantly~\cite{alpgen}.
The reason is that kinematic cuts only leave us with phase space regions where
jet radiation off gauge boson production is not at all described by a
collinear parton shower.\bigskip

\begin{figure}[t]
\begin{center}
\includegraphics[width=0.32\textwidth]{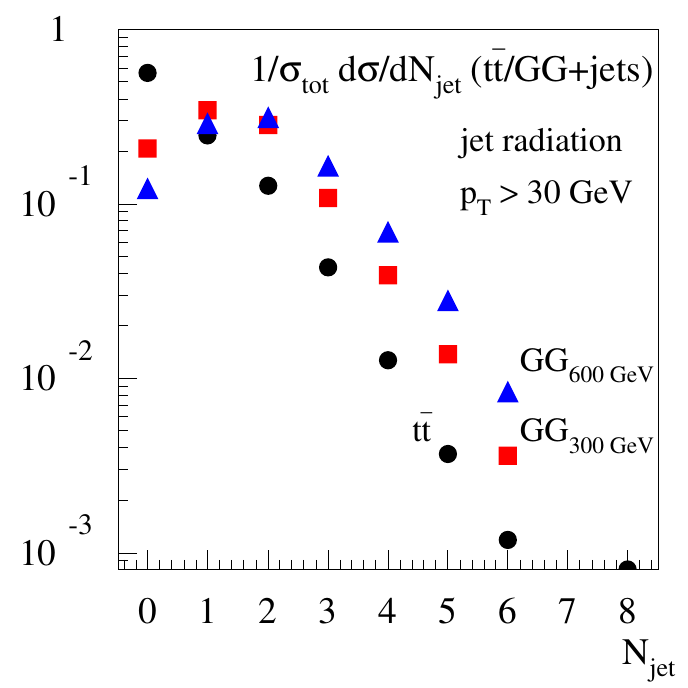} \hspace*{0.0cm}
\includegraphics[width=0.32\textwidth]{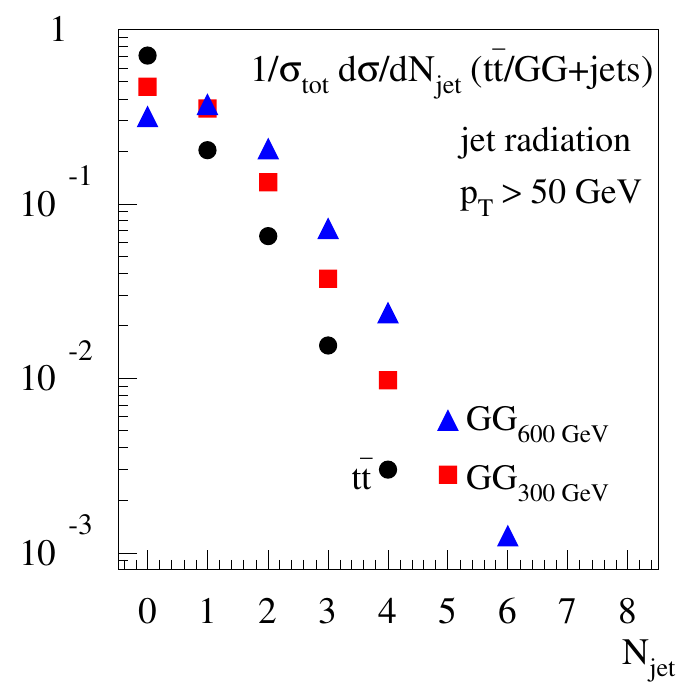} \hspace*{0.0cm}
\includegraphics[width=0.32\textwidth]{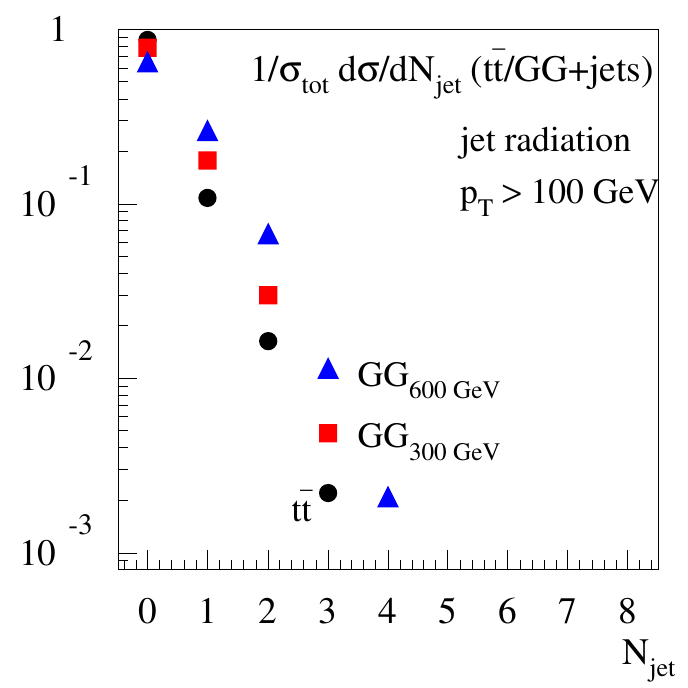}
\end{center}
\vspace*{-5mm}
\caption{Number of jets produced in QCD jet radiation in addition to
  the sgluon or top decay jets. We use MLM merging as implemented in
  MadEvent.  Figure from Ref.~\cite{Plehn:2008ae}.
\label{fig:sig_sgluon_jets}}
\end{figure} 

  Our discussion of heavy new particles and QCD is largely independent
of the identity of the new states, whether they are supersymmetric
squarks and gluinos, scalars with an adjoint color
charge, or massive vector gluon partners.  In Fig.~\ref{fig:sig_sgluon_jets} 
we show the number 
of jets radiated in association with Standard Model top quarks and 
heavy scalar gluons (sgluons) with masses of 300 and 600~GeV.  
With increasing minimum transverse momentum, most events with two 
600~GeV sgluons also involve two, one, or zero jets from 
initial-state radiation. If we want to use jets
from the decay of new particles for background rejection or for the
construction of useful observables, it would probably be a good idea to
limit ourselves to jets with transverse momenta of at least
100~GeV. In contrast, jets from $W/Z$ decays or top quark decays are
most likely in the 50~GeV range and will be subject to serious
combinatorial errors.

 Another example of a potentially useful but very challenging observable at
the LHC is the momentum of a hadronically decaying top quark or a $W$
boson coming from the resonant production of a new state: each of
these decay jets will be in the $40$--$70$~GeV
range. Figs.~\ref{fig:sig_susy_jets} and \ref{fig:sig_sgluon_jets}
indicate that this is precisely where we should expect several QCD
jets to accompany the hard production process, which will lead to a
challenging combinatorial background. For heavier gluino pair
production, Ref.~\cite{mihoko_johan} shows how to get rid of this 
combinatorial background by extracting the (relevant) decay jets from 
the complete jet activity based on the $\mtt$ variable defined in
Eq.~\eqref{eq:sig_def_mt2}.  Fig.~\ref{fig:sig_mihoko_jets}
shows that a QCD jet from initial state radiation and its rapidity is
strongly correlated with a jet extracted by a $\mtt$ criterion
described below. Note, however, that only jets above $p_T = 100$~GeV
are included in this analysis, which limits the number of additional
QCD jets to at most one. From Fig.~\ref{fig:sig_sgluon_jets} it is
obvious that this procedure would be at the least considerably more
challenging for jets from $W$ decays with $p_T = 30$--$50$~GeV.
\bigskip

\begin{figure}[t]
\begin{center}
  \includegraphics[width=0.4\textwidth]{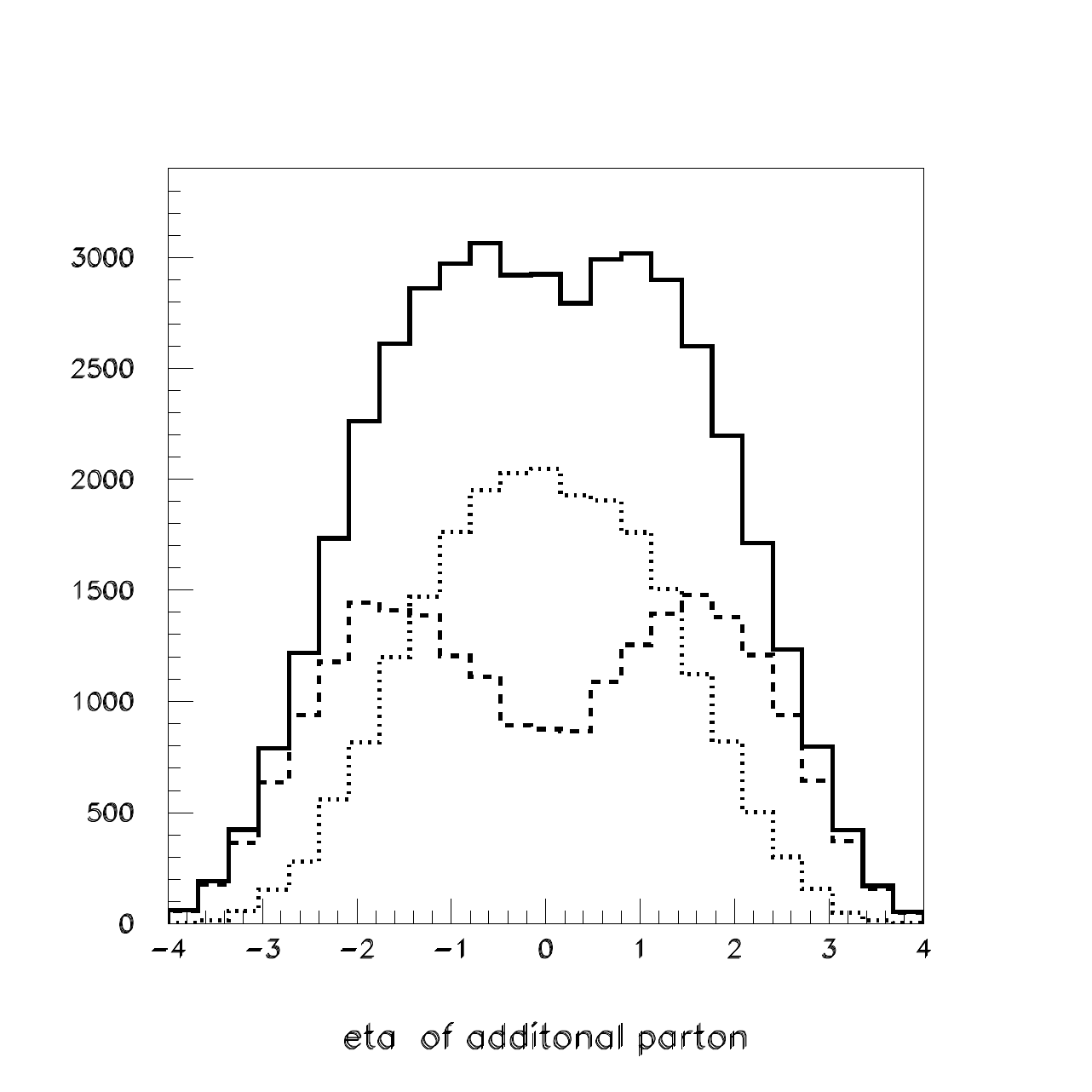}
  \hspace*{0.05\textwidth}
  \includegraphics[width=0.4\textwidth]{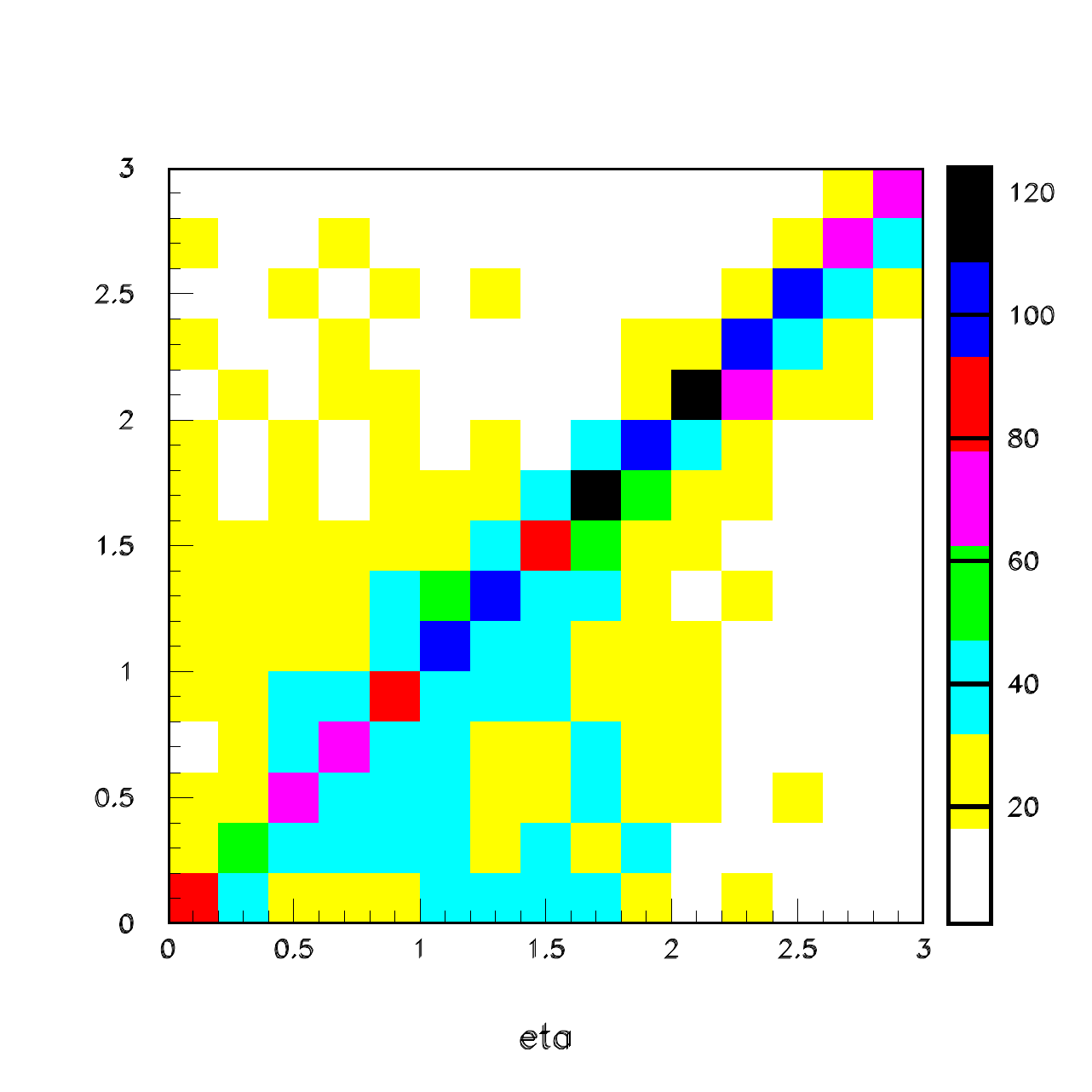}
\end{center}
\vspace*{-5mm}
\caption{Left: rapidity distribution for quark (dashed) and gluon
  (dotted) radiation off gluino pair production ($m_{\tilde g} =
  685$~GeV) and the sum of the two (solid). Only jets above 100~GeV
  are included. Right: correlation of the rapidity of the radiated QCD
  jet ($x$ axis) with the rapidity of the jet extracted using $\mtt$
  ($y$ axis), as described in Section~\ref{sec:sig_cascade}.  Figure
  from Ref.~\cite{mihoko_johan}.
\label{fig:sig_mihoko_jets}}
\end{figure} 

  Up to this point in our discussion we have presented additional
QCD jets as a problem to be overcome.  They can also be very useful.
A prime example is in searching for direct QCD top production
through a flavor-violating vertex generated by new 
physics~\cite{direct_top,single_top_jets,single_top_us}
\begin{alignat}{5}
ug, cg \to tg \qquad \qquad \text{(effective $q$-$t$-$g$ vertex)}.
\end{alignat}
This process can be induced by the flavor structure of 
the general MSSM described in Section~\ref{sec:models_susy}.
Electroweak single top production in the Standard Model
is an important background to this new physics mode:
\begin{alignat}{5}
\bar{q} b &\to \bar{q}' t 
&& \text{($t$-channel $W$ exchange)}
\notag \\
q \bar{q}' &\to \bar{b} t 
&& \text{(off-shell $s$-channel $W$ production)}
\notag \\
b g &\to t W \to t \; q \bar{q}' \qquad \qquad
&& \text{(associated $t$-$W$ production)}
\end{alignat}
All processes listed above differ only by the different kinds of jets 
they contain in addition to the top quark, which is assumed to decay 
leptonically to trigger the event.  The background processes arise 
in part from bottom quarks
in the initial state.  Even though incoming bottom partons are not
present in the proton as valence quarks, they can be generated by
gluon splitting for far off-shell gluons~\cite{bottom_partons}. In
this case we expect an additional far-forward bottom jet at large
rapidities where flavor tags cannot be applied.  While these bottom
tags would certainly help to disentangle these different top
production mechanisms, we can try to do the same thing based only on
QCD arguments without such flavor tags.

  The variable of choice is the angle $\cos \theta^*(P_1,P_2)$. It
parameterizes the angle between $\vec{p}_1$ in the rest frame of the
$P_1 + P_2$ system and $(\vec{p}_1 + \vec{p}_2)$ in the lab frame. It
is not symmetric in the arguments: if the two particles are back to
back and $|\vec{p}_1| > |\vec{p}_2|$ it approaches $\cos \theta^* =1$,
whereas for $|\vec{p}_1| < |\vec{p}_2|$ it becomes $-1$. In between, it
vanishes in the case where $\vec{p}_1$ in the center of mass frame is
orthogonal to the lab-frame momentum of this center-of-mass system. In
Fig.~\ref{fig:sig_single_top} we see that $s$-channel and $t$-channel
single top production and direct BSM top production have distinctively
different jet distributions. While each distribution individually is 
not sufficient to apply a kinematic cut, we should be able to 
train a neural network to analyze these QCD features of jet radiation. 
Note that we have left out associated $Wt$ production at this stage since 
two of its jets should distinctively reconstruct a $W$ mass and the third jet
should be a forward bottom quark.
\bigskip

\begin{figure}[t]
\begin{center}
  \includegraphics[width=0.4\textwidth]{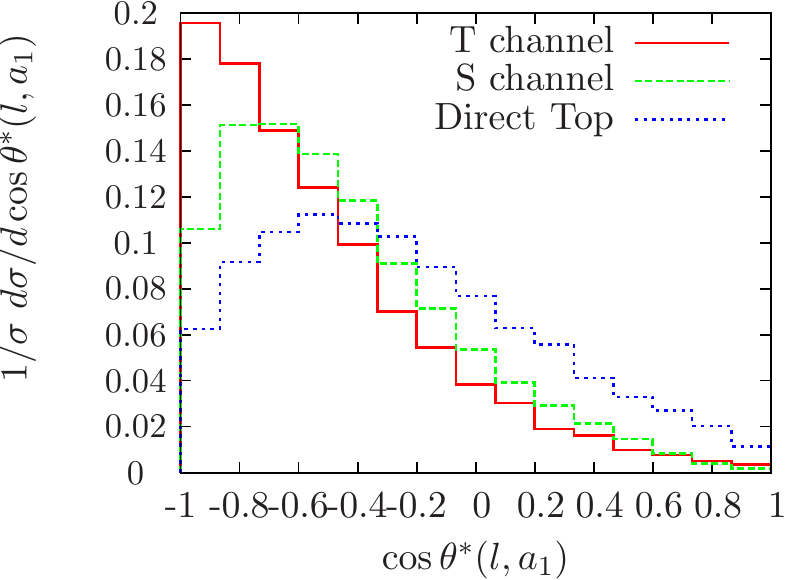}
  \hspace*{0.05\textwidth}
  \includegraphics[width=0.4\textwidth]{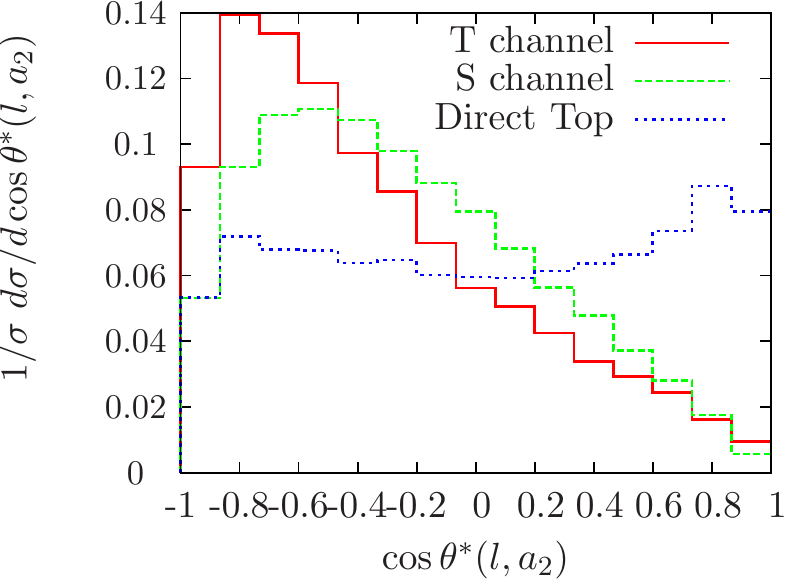}\\[3mm]
  \includegraphics[width=0.4\textwidth]{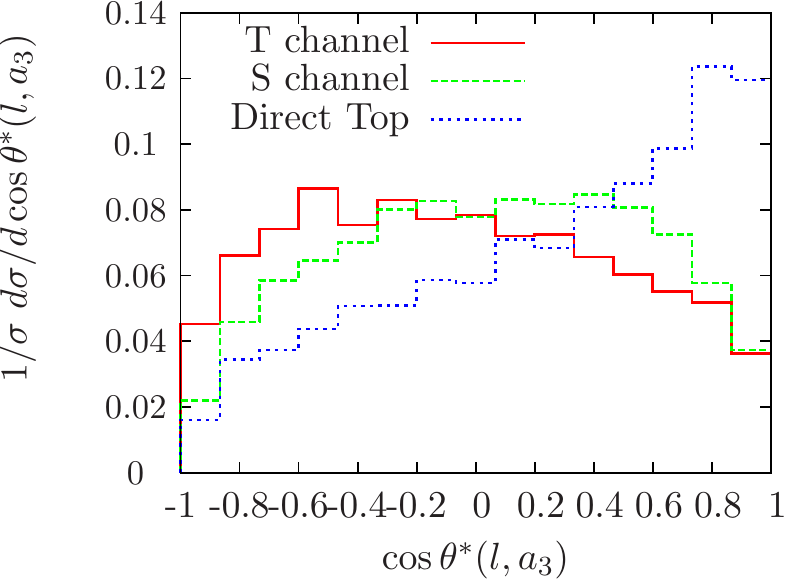}
  \hspace*{0.05\textwidth}
  \includegraphics[width=0.4\textwidth]{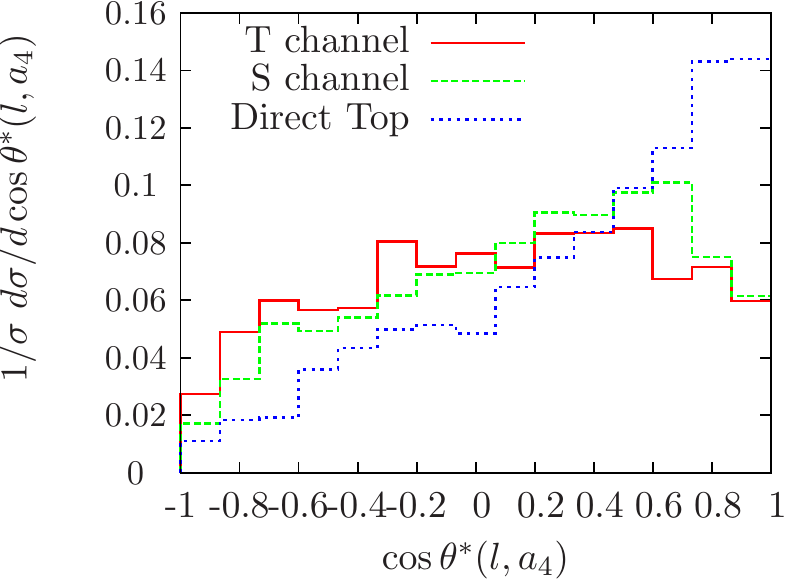}
\end{center}
\vspace*{-5mm}
\caption{Angular separation $\cos \theta^*$ between the lepton from
  the top decay and the four leading jets $a_j$ for three
  single/direct top production mechanisms. Figure from
  Ref.~\cite{single_top_us}.
\label{fig:sig_single_top}}
\end{figure} 

After this long discussion of QCD features let us next turn to the real
question for the LHC: what can we learn about new particles produced
in hadronic $pp$ collisions?  Our first guess might be that if 
the LHC production cross section of new particles only depends on
their masses, spins, and color representations, this
production rate should be a very good observable which could allow us
to measure properties of new physics at the LHC early on and without
detailed simulations of the particle decays.

  Unfortunately, the total production rate is not an easy observable at
all, both theoretically as well as experimentally.  First, we hardly
ever measure the actual production cross section.  While we can of
course formulate an exclusion limit for a new particle based on its
production rate under the explicit assumption that it only decays to a
single final state, this is not very likely in models as complex 
as the MSSM. Therefore, we have to analyze the products of cross sections
and branching ratios assuming we see only few allowed final states.  
One example for a feasible analysis is the squark cascade
and its relative branching ratios to $e+\mu$ and $\tau$ leptons, 
which contain useful information about the stau sector~\cite{lhc_dm}.
One notable exception is the top squark in the MSSM with minimal flavor
violation which, with increasing mass, first decays dominantly through
loop-induced amplitudes to charm-neutralino, then goes on-shell to
bottom-chargino through its weak charge, and finally to top-gluino
via its color charge~\cite{stop_decays}.

  Secondly, we know from the discussion above that higher-order
corrections to QCD cross sections are by no means of the order
$\alpha_s \sim 1/10$, possibly suppressed by additional factors of $1/\pi$.
Collinear logarithms lead to next-to-leading-order corrections 
of typically $30$--$80\%$, depending on the color charges of the 
initial state and the produced particle~\cite{prospino,nnlo_susy}. 
Even for weakly interacting Drell-Yan-type processes like 
neutralino-chargino production, we observe corrections of the order of 
$30 \%$.  For production processes which are at the leading order 
proportional to $\alpha_s^2$, the theoretical uncertainty due to higher-order 
corrections is roughly covered by the variation of the renormalization 
and factorization scales. The argument for this error measure is simply 
that these scales are artifacts of perturbative QCD, meaning that 
to all orders in perturbation theory the scale dependence should
vanish. However, for leading-order processes going like $\alpha^2$
this error estimate undershoots the known higher-order corrections by
large factors. In this situation it is not only crucial to know the
theory error on a cross section prediction, but to actually use the
higher-order results~\cite{prospino,nnlo_susy}.
\bigskip

  For convoluted spectra like the squarks and gluinos in the MSSM or
similarly heavy quarks and gluons in universal extra dimensions we
face another challenge: the list of production processes includes
$\tilde{q}\tilde{q}$ and $\tilde{q} \tilde{q}^*$ production, 
gluino pairs $\tilde{g}\tilde{g}$ and last but not least the 
associated $\tilde{q} \tilde{g}$ process. These processes can only 
be distinguished by the number of jets in the final state. 
The mixed associated production has a
quark-gluon initial state which is particularly well-suited for the
production of heavy states at the LHC: while valence quarks dominate
the large-$x$ region of the available partonic energy, for smaller $x
< 0.01$ the gluon densities practically explode. In general, for heavy
states the mixed $qg$ initial state turns out to be the best
compromise for average $x$ values around 0.1, which means that heavy
associated production processes typically dominate over heavy pair
production processes.

\subsection{Missing energy}
\label{sec:sig_met}

  Many theories of physics beyond the Standard Model predict the 
existence of quasi-stable neutral particles.  When created
in particle colliders, they will leave the detector without generating
a direct signal.  Instead, they can make their presence felt indirectly
by producing an imbalance in the vector sum of the visible transverse 
particle momenta in the event,
\begin{equation}
\ptmiss = \sum_\text{invisible} \vec{p}_{T_i} 
= -\sum_\text{visible} \vec{p}_{T_j}.
\end{equation}
This \emph{missing energy} is a useful tool for separating signals
from Standard Model backgrounds, but it also makes life more difficult
when it comes to measuring the detailed particle properties underlying
these signals. Note that this formula can be used at the theory level,
but taking into account detector effects it is a bad approximation to
the missing energy resolution at the LHC.\bigskip

  There are two primary reasons why quasi-stable particles are nearly
ubiquitous in models of new physics.  First, an exactly stable neutral 
particle with a weak-scale mass and electroweak couplings can potentially 
make up the dark matter in our universe.  The stability of such a state 
simply requires some sort of $\mathbb{Z}_2$ parity symmetry under which 
it is the lightest odd state.  Secondly, by extending such a parity 
symmetry to the entire theory, the precision electroweak and flavor 
constraints on new physics that is parity-odd are significantly 
weakened, even when the parity is only approximate.  Specific examples
of theories with a global parity symmetry include supersymmetry with
$R$-parity, universal extra dimensions with KK-parity, 
and little Higgs models with $T$-parity.  It is precisely on account
of their parity symmetries that the new particles in these theories
can be as light as a few hundred GeV while remaining consistent
with current experimental constraints.  

In theories with a parity symmetry that extends to the entire theory,
there are typically heavy new QCD-interacting states with odd
parity.  These will be produced in pairs in large quantities at the
LHC.  Each of them will decay in a cascade down to the lightest
parity-odd particle, which will escape the detector if it is neutral.
The decay process will therefore involve two possibly distinct decay
chains -- one for each parity-odd heavy QCD state created in the
initial collision.  Along the way, the color charge of the
initially-produced states must be shed by emitting Standard Model
quarks and gluons, so the resulting events are characterized by a high
jet multiplicity and large missing transverse energy.  Moreover, these
heavy particle decays are expected to be more spherical on average
than the QCD backgrounds.  A simple example of such a cascade in
supersymmetry is illustrated in Fig.~\ref{fig:sig_cascsusy}, where a
pair of gluinos is produced that decay down to the stable neutralino
LSP by emitting quarks.  Such a process will
produce a net signal consisting of jets and missing energy.\bigskip

\begin{figure}[t]
\begin{center}
  \raisebox{5mm}{\includegraphics[width=0.45\textwidth]{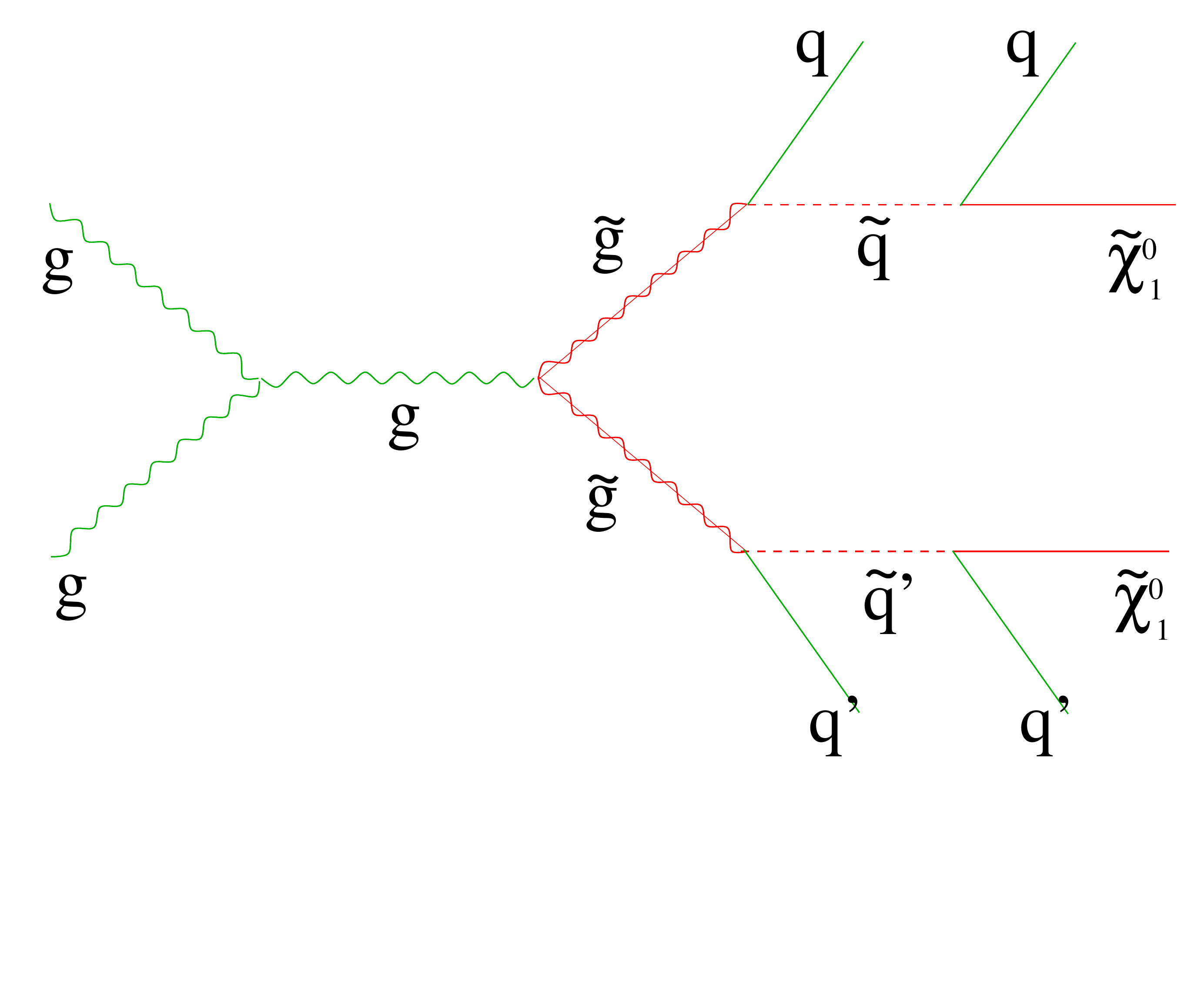}}
\end{center}
\vspace{-2.5cm}
\caption{Supersymmetric cascade decay with a neutralino LSP.  Two
  gluinos are produced in the initial interaction, and each
  subsequently decays down to the neutralino LSP by emitting quarks.}
\label{fig:sig_cascsusy}
\end{figure}

  Signals of missing energy can also appear in models of new physics
that do not have any sort of parity, but that contain neutral states that
are long-lived due to very weak couplings or kinematics.
This is the case for the light Kaluza-Klein graviton excitations
in theories with large extra dimensions.  

  Missing energy is an extremely powerful tool to extract new signals
from Standard Model backgrounds.  There are, however, detector and 
Standard Model sources of missing energy as well.  On the detector
side, a momentum imbalance can be generated by a mis-measurement
of energy in the detector, typically from QCD events.  
Controlling such backgrounds will require careful studies of
the detector performance based on collision data and a minimum of
several months of work.  The second source of backgrounds
are events with genuine missing energy from neutrinos produced
by Standard Model processes involving $W$ and $Z$ bosons, such as
($Z\to \nu\nu$)+jets or $\bar tt$ production where the latter even 
has a high jet multiplicity.  Again, collider data will be essential to 
understanding these backgrounds: $Z\to \ell^+\ell^-$ can be used to
determine the rate for $Z\to \bar{\nu}\nu$, and fully reconstructed
$t\bar{t}$ events provide the top production rate.  Since we will 
need sufficient statistics to accurately predict missing energy background
rates from the Standard Model and detector effects, the amount of time 
required to claim a discovery of new physics through missing energy 
could well be dominated by the collection of these control samples,
and not by signal statistics. \bigskip

  In the remainder of this section we will discuss signals of missing
energy from cascade decays in models with a parity symmetry, as well as 
the direct production of long-lived neutral states.  We will also
describe techniques to measure the underlying physics giving rise
to the missing energy.  To do so, it is necessary to use kinematic
distributions sensitive to new heavy particles that are perturbatively
well-behaved.  The major question is whether we can use
decay kinematics in the presence of an invisible final state particle, 
without reconstructing either the (partonic) initial state or the
final state. For example, in supersymmetry or universal extra
dimensions, any extraction of model parameters (to be discussed in
Section~\ref{sec:para}) hinges on the identification and isolation of
complete new-physics decay chains.  Only then can we hope to measure
the kinematic properties of the decay products and extract masses and
couplings of the new states, which is the main task of the LHC 
once new physics is discovered at the TeV scale.

\subsubsection{Cascade decays}
\label{sec:sig_cascade}

  In this subsection we will focus on the detection and measurement 
of new physics scenarios with a parity symmetry.  
We will assume further that a fraction of the
new particles are odd under the parity, including some carrying QCD charges,
and that the lightest parity-odd state is neutral and stable on collider
time scales.  New parity-odd states will be therefore be produced in pairs
and decay through a cascade down to the lightest parity-odd state, 
generating a signal of missing energy.
While these assumptions are specific, they cover some of
the most popular theories for physics beyond the Standard Model
including supersymmetry with $R$-parity and some little Higgs and
extra-dimensional scenarios.  

  As discussed in Section~\ref{sec:sig_qcd} and illustrated 
in Fig.~\ref{fig:sig_prospino},
the production of strongly-interacting states is dominant at hadron colliders.
Thus, the great majority of decay chains will begin with pairs of
QCD-charged states.  This is especially true at the LHC, since the
$x$ values of the incoming partons dominating the production
are below $10^{-2}$ where protons consist mostly of gluons. 
Since production through electroweak interactions requires quarks 
in the initial state, it is not clear whether this will be relevant
for LHC searches.  Note that the situation can be different at the 
Tevatron with its lower collision energy where incoming gluons are 
less likely to dominate the parton luminosities.
  
  Decay cascades further illustrate the primary advantages
and challenges of the LHC.  Starting from a pair of parity-odd 
strongly-interacting particles, produced for example in gluon fusion, 
they will decay down to the weakly-interacting lightest parity-odd state. 
The different decay steps are ideally two-body decays from on-shell 
particle to on-shell particle, but they may also involve off-shell
intermediate states.  The huge luminosity of the LHC can yield
very high production rates:
strongly interacting particles with masses around 500~GeV have 
production cross sections close to $\mathcal{O}(100~\pb)$ and the LHC 
will create $10^5$ of them within the first $1~\ifb$ of data.
The price for this luminosity is that these events will be accompanied by
a large number of jets, both as part of the new physics process 
or from the underlying event.  Decay products containing leptons
or $b$-jets can be separated more reliably.\bigskip

\begin{figure}[t]
\begin{center}
  \includegraphics[width=0.6\textwidth]{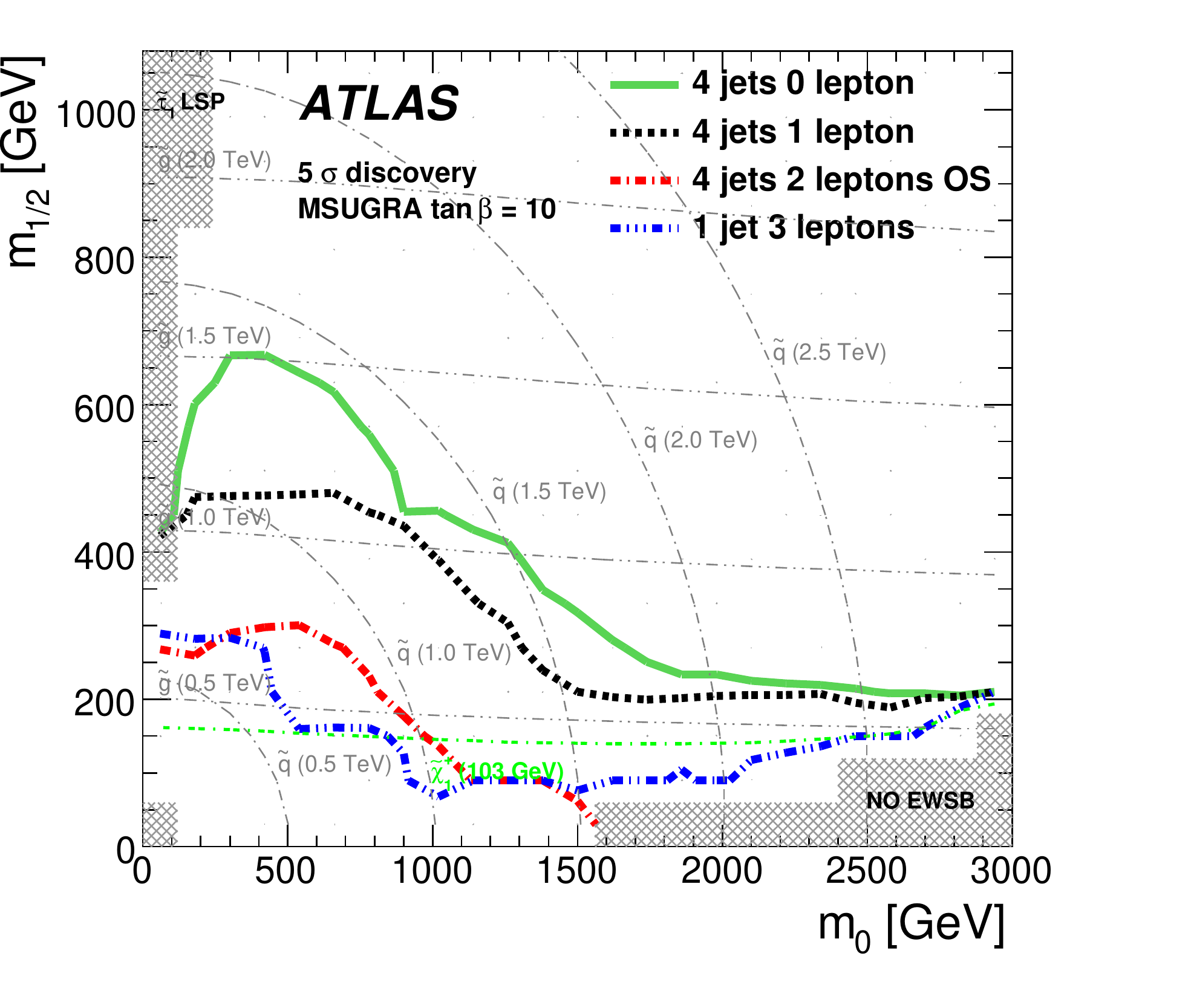}
\end{center}
\vspace{-1.0cm}
\caption{ATLAS $5\sigma$ reach in various inclusive channels
  presented in the mSUGRA (toy) mass plane. We assume a low integrated
  luminosity of $1~\ifb$~\cite{atlas_csc}.}
\label{fig:sig_reach}
\end{figure}

  It is useful to organize inclusive missing energy signatures
at the LHC and the Tevatron according to the number of hard
leptons in the event.  The standard classification is~\cite{atlas_tdr,
Baer:1995nq,ian_frank}:
\begin{itemize}
\item[] $\etmiss+\text{jets}+X$: inclusive jets plus missing energy 
($X$ = anything)
\item[] $\etmiss+\text{jets} +0\ell$: lepton veto ($0\ell$)
\item[] $\etmiss+\text{jets} +1\ell$: one lepton ($1\ell$)
\item[] $\etmiss+\text{jets}+2\ell$ with $Q(\ell_1)=-Q(\ell_2)$: 
opposite-sign dileptons ($2\ell$\,OS)
\item[] $\etmiss+\text{jets}+2\ell$ with $Q(\ell_1)=+Q(\ell_2)$:
same-sign dileptons ($2\ell$\,SS)
\item[] $\etmiss+\text{jets}+3\ell$: trileptons ($3\ell$) 
\end{itemize}
Signatures can be further classified according to additional features,
such as the number $b$-tags or the flavor of the outgoing 
leptons~\cite{Belyaev:2008pk}.

  The discovery reach of the LHC and the Tevatron is maximized by
combining all these signature channels.  In typical supersymmetric models,
the optimal strategy involves a cut on $\etmiss$ of a few hundred $\gev$.
In this case the reach is dominated by the inclusive $\etmiss$+jets
search, but the other channels can be reasonably sensitive as well 
and would help to confirm a discovery based on the fully inclusive 
signature. 
The reach of some of these channels we illustrate in Fig.~\ref{fig:sig_reach}.
For example, the approximate LHC discovery reach at $\sqrt{s}=14\,\tev$ 
is about 1.3~TeV in squark/gluino masses for an assumed luminosity of $1~\ifb$.
This number includes systematic errors on the background 
evaluation for the same luminosity. One year of design luminosity,
or $100~\ifb$, would increase the mass reach to 2.5--3~TeV and is
limited by the signal cross section dropping sharply for high masses. 
While these results were obtained for ``typical'' supersymmetric models
with squarks and gluinos of comparable mass, similar results should
apply to any model with strongly-interacting parity--odd states decaying
to a stable lightest parity-odd state~\cite{johan_gluinos}.  
Let us emphasize, however,
that these signals will often be buried in a significant standard
model background, and any discovery will be limited by our
theoretical and experimental understanding of these backgrounds.
\bigskip

  Beyond just making a discovery, we would also like to learn about the
physics giving rise to any anomalous signals that are found.
A first step in this program is to compare the relative event rates 
from different signal channels.  While this doesn't provide enough
information to deduce the source of the signals, it is very useful
in narrowing down the possibilities.  To do significantly better,
it is necessary to study kinematic distributions of the signal
decay products, and for this we must identify specific kinematic
properties that are not ruined by the fact that we are not able
to directly reconstruct the individual momenta of the pair of 
outgoing invisible particles.

  To get an idea of which kinematic distributions might be useful
for the identification and measurement of new physics models with
missing energy signatures, it is helpful to look at Standard Model 
processes that also contain missing energy.  Examples that have
already been measured or studied in detail include $W$+jets,
Higgs boson decays to $W$ pairs, and single top production.
These channels all include genuine missing energy from neutrinos
when the relevant $W$s decay leptonically: $W^- \to \ell\bar{\nu}_{\ell}$.

  In $W+$~jets production we know how to reconstruct the mass of
a leptonically decaying $W$, decaying into one observed particle (lepton) 
and one missing particle (neutrino).  A useful quantity for this 
is the transverse mass, defined in analogy to the invariant mass 
of two particles $m_{ab}^2 \equiv (p_a+p_b)^2$, 
but neglecting the longitudinal components of the final state momenta~\cite{kinematic_reviews}:  
\begin{alignat}{5}
m_{T}^2 &= \left( \etmiss + E_{T, \ell} \right)^2
       - \left( \ptmiss + \vec{p}_{T,\ell} \right)^2 
\notag \\
      &= m_\text{miss}^2 + m_{\ell}^2 
       + 2 \left( E_{T,\ell} \etmiss
       - \vec{p}_{T,\ell} \cdot \ptmiss \right), 
\label{eq:sig_mt}
\end{alignat}
where $m_\text{miss}\simeq 0$ is the mass of the invisible neutrino and
the transverse energy is defined to be $E_T^2 = \vec{p}_T^2 + m^2$.  
The transverse mass is always less than or equal to the actual $W$ mass, 
so we can extract $m_W$ from the upper edge of the $m_{T,W}$ distribution.
Obviously, we can define the transverse mass in many different 
reference frames.  However, its value is invariant under (independent of) 
longitudinal boosts, and given that we construct it as the transverse 
projection of an invariant mass it is also invariant under transverse boosts.  
Note that the transverse mass does give the $W$ mass from any single event,
but requires a large number of events to populate the upper end of the
distribution.

From studies of Higgs searches we know how to extend the transverse
mass to a pair of leptonically decaying $W$ bosons, $W^+W^-\to
\ell^+\ell'^-\nu_{\ell}\bar{\nu}_{\ell'}$, which contains two
invisible neutrinos in the final state.  In this case, a transverse
mass can again be defined, but now using the transverse components of
the sum of the visible lepton momenta, $p_{\ell\ell} =
p_{\ell}+p_{\ell'}\equiv
(E_{\ell\ell},\vec{p}_{\ell\ell})$~\cite{dave_thesis,wbf_exis}:
\begin{alignat}{5}
 m^2_{T,WW} &= \left( \etmiss + E_{T, \ell \ell} \right)^2
            - \left( \ptmiss + \vec{p}_{T, \ell \ell} \right)^2 
\notag \\
      &= m_\text{miss}^2 + m_{\ell \ell}^2 
       + 2 \left( E_{T,\ell \ell} \etmiss
       - \vec{p}_{T,\ell \ell} \cdot \ptmiss \right). 
\end{alignat}
This definition is not unique since it is not clear how to define
$m_\text{miss}$, which also implicitly enters through $\etmiss$.  From
Monte-Carlo analyses it is found that identifying $m_\text{miss}
\equiv m_{\ell \ell}$, which is correct at threshold, is the most
strongly peaked~\cite{dave_thesis}. On the other hand, setting
$m_\text{miss}=0$ for a proper bounded-from-above transverse mass
variable seems to improve the Higgs mass
extraction~\cite{barr_lester}.

  Investigations of single top production have led to another method for
conditionally reconstructing masses and momenta in the presence
of a single invisible particle.  When $t\to bW^+$ with 
$W^+\to\bar{\ell}\nu_{\ell}$, the longitudinal momentum of the neutrino
is not directly measurable.  On the other hand, we know that for
signal events the neutrino and the lepton combine to an on-shell $W$ boson,
$(p_{\ell}+p_{\nu})^2=m_W^2$.  This constraint allows us to deduce
the longitudinal momentum of the invisible particle (the neutrino)
if make an assumption for its mass; $m_{\nu}\simeq 0$ in this case.  
The main difference between such a conditional reconstruction and a
direct momentum reconstruction is that we must use a global measurement 
like $\ptmiss$, which uses the entire detector, instead of simply 
reconstructing the momentum of a single track.

  A third and more global class of observables can
be useful in the presence of several leptonic $W$ decays.
They are based on the assumption that we are looking for the decay of 
two heavy non-relativistic new states that are produced close to threshold,
which is likely due to falling parton densities. 
We can then approximate the partonic collision energy as
$\sqrt{\hat{s}} \sim (m_1 + m_2)$ and relate it to an observable based on
the total amount of visible energy in the event.
Since we assume that the heavy states are produced with little
energy, boost invariance is not required for these constructions.
Without taking into account sources of missing energy, such an
observable could take the form~\cite{higgs_pair}
\begin{equation}
m^2_\text{visible} = \left[ \sum_{\ell,j} \; E \right]^2 
                   - \left[ \sum_{\ell,j} \; \vec{p} \right]^2
\end{equation}
The Tevatron experiments have frequently used another variable in this
class: the transverse mass scale $H_T$,
\begin{equation}
 H_T = \sum_{i} \; E_{T_i} 
     = \sum_{i} \; p_{T_i} 
\label{eq:sig_def_ht}
\end{equation}
where the last step assumes that all final state particles are
massless.  This has traditionally been evaluated using jet momenta only, 
but it can trivially be extended to leptons.  If the signal originates
from a pair of non-relativistic heavy states and there is not too
much missing energy in the event, we expect that $m_\text{visible} 
\sim H_T \sim (m_1+m_2)$. 

  More generally, we would like to include missing transverse momentum
into such a measure.  This leads to the question of whether it is
better to combine the missing transverse momentum with the visible 
transverse momenta, or with the complete visible momenta. 
For example, we can use the scalar sum of all transverse
momenta in the event, now including the missing transverse momentum,
to stay longitudinally boost invariant:
\begin{equation}
m_\text{eff} = \etmiss + H_T
\label{eq:sig_def_meff}
\end{equation}
This effective mass is known to trace well the mass of the heavy new
particles decaying to jets and missing energy~\cite{effective_mass}.
We should add that the definition of global variables is not entirely
unique. For example, in many references $H_T$ is defined as $H_T =
E_{T_2} + E_{T_3} + E_{T_4} + \etmiss$.

In practice, there are two potential problems with the $m_\text{eff}$ variable.
The first is that it only tracks the masses of the initially produced states
to the extent that these are produced non-relativistically.
Second, some of the jets included in forming $H_T$ may not actually
come from the signal decay chains, but originate instead as hard
associated jets as discussed in Section~\ref{sec:sig_qcd}.  
There is also a danger of picking up jets from the underlying event,
and for this the jets used in $H_T$ and in the other methods
discussed in this section should be narrow and well-defined~\cite{bryan_ue}.
Ways to deal with QCD activity like the underlying event will be 
discussed in detail in Section~\ref{sec:sig_direct}. 
\bigskip

  These techniques to reconstruct invariant masses can also be enhanced 
by focussing on certain kinematic configurations rather than on
different events.
A classic example is the  approximate collinear
reconstruction of a heavy particle such as a Higgs decaying to boosted tau
pairs~\cite{boosted_taus,wbf_tau}, where all the decay products
of each $\tau$ are assumed to travel in the same direction. 
Another set of useful phase space regions is the vicinity of 
kinematic endpoints, where individual particles are
produced at rest~\cite{atlas_tdr}.
\bigskip

The methods described above that were developed for Standard Model
processes can be adapted to new physics searches once we specify the
relevant signatures.  Most importantly, the approach to missing energy
needs to be modified to account for the unknown finite mass of the
carrier of the missing energy, and for the possibility of various
lengths and varieties of decay chains.  This represents a problem of
relativistic kinematics and at least at leading order does not require
a detailed knowledge of QCD or new-physics models~\cite{ian_frank}.
Proposals to solve this problem can be broadly classified into three
classes.
\begin{enumerate}
\item 
\underline{Endpoint methods} extract masses from lower (threshold)
and upper (edge) kinematic endpoints of invariant mass distributions
of visible decay
products~\cite{ian_frank,edges,atlas_tdr,sps1a_per,mgl_per,mll_chris}.
These endpoints correspond to functions of the particle masses involved
in the cascades.
This method is best suited to long decay chains, where the number of
independent endpoint measurements in one of the legs of the cascade
is at least as large as the number of unknown masses in that leg.

  For this method to work we need to either clearly separate the decays
of two heavy new states, or combine a short decay chain on
one side with a long chain on the other side. In supersymmetry, this
is often naturally the case for associated $\tilde{q}_R \tilde{g}$
production (with $\tilde{q}_R \to q\chi_1^0$ and $\tilde{g}$ decaying in
a longer chain), but such a feature is not generic. Moreover, when looking
at long cascade decays we usually cannot tell where along the chain
a given final state was emitted.  For example, in a cascade chain
where two leptons are emitted, it can be difficult to determine 
which lepton was emitted earlier in the cascade.  Endpoint techniques 
will therefore always be plagued by combinatoric uncertainties from 
the mapping of particle momenta onto the decay topology. 
In this situation it is useful to consider the correlation
of different invariant masses and their endpoints~\cite{costanzo_tovey}.  
The endpoint method can be extended to use invariant mass distributions 
from both sides of the event (hidden threshold techniques), 
and correlations between the distributions from each leg 
(wedgebox techniques)~\cite{hidden_threshold}.

  A general caveat for these endpoint techniques is that they can only
be applied to a distribution of many events, just like for the transverse
mass reconstruction discussed above.   Evidently, a sufficiently large number of
events is required to fill out the region near the endpoint.  This can be
problematic when the matrix element describing the decay becomes suppressed
in the kinematic region near the endpoint.  Endpoints vanishing smoothly
in the Standard Model background noise are unlikely to be useful
for mass determination.
\bigskip

\item
\underline{Mass relation methods} generalize the single-top example
to completely reconstruct the kinematics with only a few events, assuming
that all the decaying particles are on-shell. This assumption is
also implicit in the endpoint method, but off-shell effects are not
expected to have a huge effect on the results.

  For each event with the topology $Z \to Y \to X \to N$,
as illustrated in the left panel of Fig.~\ref{fig:sig_massrel}, 
and where $N$ crosses the detector invisibly, 
this method uses the kinematic constraints~\cite{massrelation_bryan}
\begin{alignat}{5}
(p_1+p_2+p_3+\pmiss)^2 =& m_Z^2 \notag \\
(p_2+p_3+\pmiss)^2 =& m_Y^2 \notag \\
(p_3+\pmiss)^2 =& m_X^2 \notag \\
(\pmiss)^2 =& m_N^2
\label{eq:sig_massrel}
\end{alignat}
While for any one event the number of free parameters can be 
larger than the number of measurements, using several events
increases the number of measurements while keeping the number
of unknowns constant.  Eventually, the system of equations for the
masses will be solvable, provided all events are really signal events.
This technique can be generalized to either identical or different
legs in the pair of cascade chains or to to different decay chain lengths.

\begin{figure}[t]
\begin{center}
  \raisebox{5mm}{\includegraphics[width=0.35\textwidth]{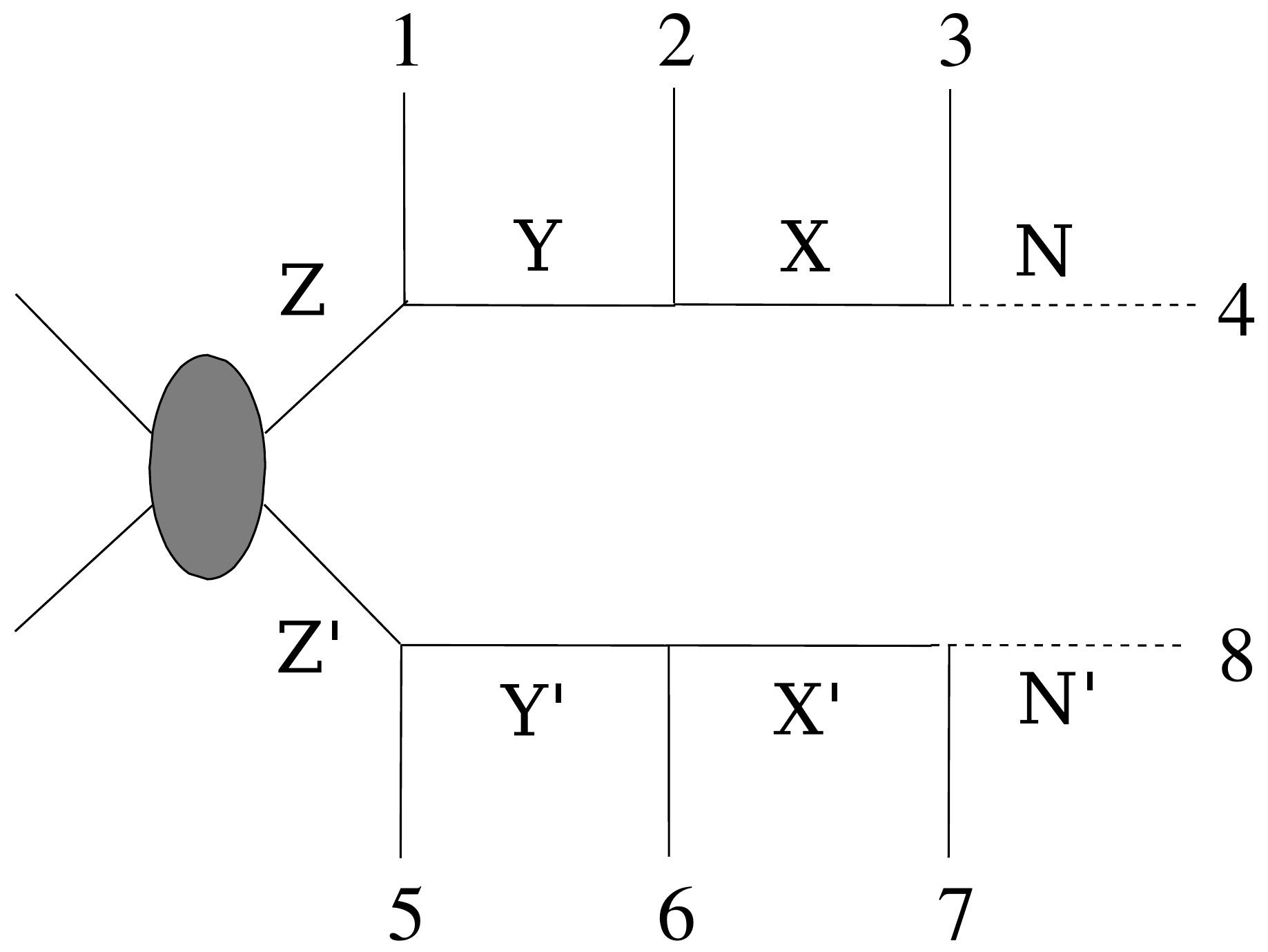}}
  \hspace*{0.05\textwidth}
  \includegraphics[width=0.45\textwidth]{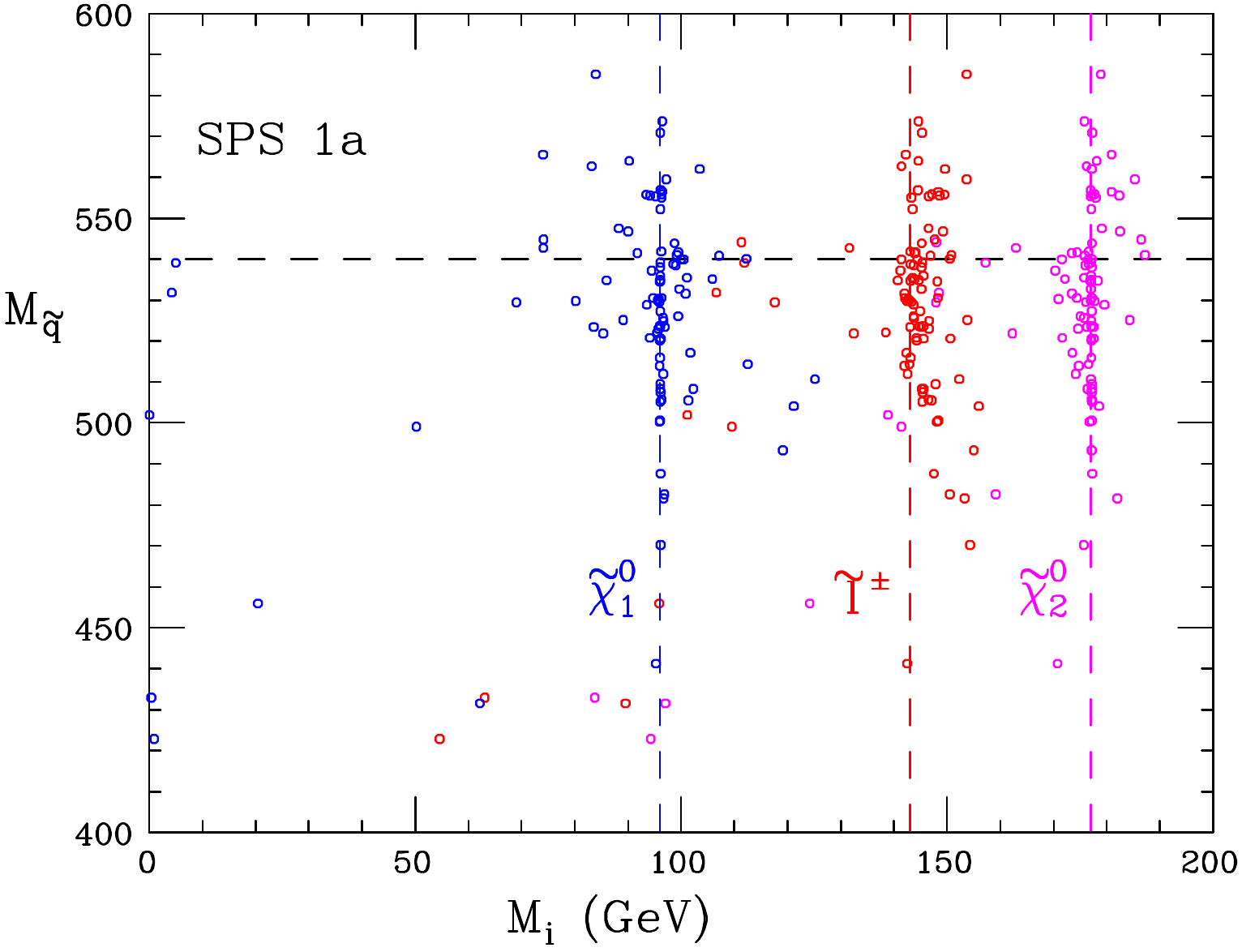}
\end{center}
\caption{Best-fit values for all masses involved in squark decays in
  the supersymmetric parameter point SPS1a. Each of the 100 points is
  extracted from a sample of 25 events. The momenta include a $10\%$
  simulated detector smearing. Figures from
  Ref.~\cite{massrelation_bryan}}
\label{fig:sig_massrel}
\end{figure}

The only unknown in Eq.(\ref{eq:sig_massrel}) aside from the four
masses is the missing momentum vector $\pmiss$. Under a given mass
hypothesis $\left\{m_i\right\} = \left\{m_Z,m_Y,m_X,m_N\right\}$ 
we can compute $\pmiss$ from all other measured momenta; 
this matrix relation for the missing momentum is a sparse matrix 
and can be inverted. The goodness of the mass hypothesis
can then be checked by testing the self-consistency condition
$(\pmiss)^2 = m_N^2$. The structure of Eq.(\ref{eq:sig_massrel})
implies that this condition does not only test the last mass condition
but all of them iteratively.  More precisely, for a given sample of 
events (or sets of measured momenta) we can minimize over the space of
hypothetical particle masses $\left\{m_i\right\}$ a quantity of 
the form
\begin{alignat}{5}
\xi^2\left( \{ m_j \} \right) = \sum_\text{events $n$} 
 \left[ (\pmiss)^2_n - m_N^2 \right]^2.
\end{alignat}
In the right panel of Fig.~\ref{fig:sig_massrel} we illustrate a
mass extraction with this technique based on a sample of 25 events
for the MSSM SPS1a parameter point.  Even after including the
leading detector smearing effects, all the relevant masses in
the decay chain can be determined to within about $10\%$.

  This mass relation method has been developed to address single
legs~\cite{mihoko_giacomo}. Using both legs allows us to consider two
decays at a time and to include the missing transverse energy as
another measurement~\cite{mihoko_giacomo,bob}.  It has also been
successfully applied as a hybrid, combining mass relations and
kinematic endpoints~\cite{cotransverse}.

\item
\underline{$\mtt$ methods} used for two one-step decay chains per event
are based on a global variable
$\mtt$~\cite{mt2}. It generalizes the transverse
mass developed for $W$ decays to the case of two massive invisible
particles, one from each leg of the event. The observed missing energy
in the event can be divided into two scalar fractions, $\ptmiss =
\slashchar{q}_1 + \slashchar{q}_2$. Given the two fractions
$\slashchar{q}_j$ we can construct a transverse mass for each side
of the event assuming we know the invisible particle's mass,
$m_{T,j}(\slashchar{q}_j;m_\text{miss})$, where the second argument
represents our external assumption.  Inspired by the usual
transverse mass, we would like to incorporate this into a mass variable 
with a well-defined upper edge, so we need to use some kind of minimum of
the $m_{T,j}$ as a function of the splitting of $\ptmiss$. 
Without any additional constraints, this minimum will occur with 
zero transverse momentum on one leg, which is not very interesting. 
However, in this limit the transverse mass from the other leg reaches 
a maximum, so we can instead define the quantity
\begin{equation}
\mtt(m_\text{miss}) 
     = \min_{\ptmiss = \slashchar{q}_1 + \slashchar{q}_2}
       \left[ \max_j \; m_{T,j}(\slashchar{q}_j;m_\text{miss}) \right]
\label{eq:sig_def_mt2}
\end{equation}
as a function of the unknown missing particle mass. By construction,
each of the $m_{T,j}$ lie between the sum of the masses of the two decay 
products and the mass of the decaying particle, so for massless Standard
Model decay products there will be a global $\mtt$ threshold at the
missing particle's mass. 
Moreover, for the correct value of $m_\text{miss}$, 
the $\mtt$ distribution has a sharp edge at the mass
of the parent particle. This means that in favorable cases $\mtt$ may
allow the measurement of the masses of both the parent particle and 
the lighter stable particle based on a single-step decay
chain~\cite{stransverse,kinks,mtgen,cotransverse,mihoko_mt2}. 

  An interesting aspect of $\mtt$ is that it is trivially boost invariant
if $m_\text{miss}$ is indeed the mass of the missing particle. In
contrast, for a wrong assignment of $m_\text{miss}$ it has nothing to
do with the actual kinematics and hence with any kind of invariants,
which means it will not be boost invariant. This aspect can be
exploited by scanning over $m_\text{miss}$ and looking for so-called
kinks, \ie points where different events kinematics all return the
same value for $\mtt$.~\cite{kinks}

  Similar to the more global $m_\text{eff}$ variable, we can generalize
$\mtt$ to the case where we do not have a clear assignment of the two
decay chains involved. This modification $M_{T 
text{Gen}}$~\cite{mtgen}
again has an upper edge, which unfortunately is not quite as sharp as
the one in $\mtt$.  This procedure can be further generalized to any one-step
decay, for example a three-body decay with either one or two missing
particles on each side of the event. Such $M_{TX}$ distributions are
useful as long as they have a sharp enough edge.
This can become a problem in some cases, such as when looking at 
the $M_{T4}$ distribution for decays with two heavy invisible particles
and two neutrinos in the final state.

  Beyond just mass reconstruction, $\mtt$ can also be used to
disentangle central jets from heavy particle decays from forward
initial-state jet radiation~\cite{mihoko_johan}, a problem 
discussed in Section~\ref{sec:sig_qcd}. The situation for
hard jets with $p_{T,j}> 100$~GeV is illustrated by the left panel in
Fig.~\ref{fig:sig_mihoko_jets}, where we see that a simple cut on the
jet rapidity is not enough to extract a sample with purely
decays jets. Instead, an unwanted QCD jet can be eliminated by constructing
a set of reduced $\mtt$ variables,
\begin{equation}
\mtt^\text{min} = \min_j \mtt^{(j)} 
\equiv \min_j \mtt(p_1,...,p_{j-1},p_{j+1},...),
\end{equation}
where the definition of $\mtt$ might have to be adapted to the number
of decay jets expected.  We then search for the value $j$ with the
smallest $\mtt^{(j)}$. By construction, combinations of decay jets
will tend to produce a non-zero value in the minimization procedure
constructing $\mtt$. In contrast, QCD jets are uncorrelated and will
drive $\mtt$ to zero. This correlation is shown in the right panel
of Fig.~\ref{fig:sig_mihoko_jets} where we find that for sufficiently
forward jets $\mtt^\text{min}$ strongly correlates with the QCD jet in
the sample.
\bigskip

A slight variation to the usual definition of the transverse mass is
the co-transverse mass, which is designed to cancel any smearing due
to the transverse momentum of the heavy decaying particle. For two
identical decaying particles with short decay chains like in the $WW$ or
$\mtt$ scenario, we can instead use~\cite{cotransverse}
\begin{alignat}{5}
m_\text{CT}^2 
&= \left( E_{T, \ell 1} + E_{T, \ell 2}
   \right)^2 
 - \left( \vec{p}_{T, \ell 1} - \vec{p}_{T, \ell 1} 
   \right)^2 
\notag \\
&= m_{\ell 1}^2 + m_{\ell 2}^2 
 + 2 \left( E_{T, \ell 1} E_{T, \ell 2} 
          + \vec{p}_{T, \ell 1} \cdot \vec{p}_{T, \ell 2} 
      \right)                    
 < \frac{m_\text{heavy}^2 - m_\text{miss}^2}{m_\text{heavy}^2}
\label{eq:sig_def_mct}
\end{alignat}
While it is not invariant under transverse boosts anymore it still
retains the endpoint and sharpens its drop by cancelling part of the
transverse momentum dependence on both sides.
\end{enumerate}
\bigskip

A generic feature of all three of these methods based on decay
kinematics is that it is easier to constrain the differences of masses
than to determine the absolute mass scale.  This is particularly true
for endpoint methods, since the formulas for the endpoints involve
depend on squared mass differences.  Experimentally, correlated jet
and lepton energy scale uncertainties do not make life easier
either. All these techniques are sensitive to a lesser extent to
absolute mass scale than to mass differences, but the common lore that
kinematics only constrain mass differences does not hold up to a
detailed analysis. Just as a reminder we emphasize that all of these
methods only work if we are dealing with one well-defined massive
particle responsible for the missing energy in new-physics
events. \bigskip
 
While all three strategies should in principle work and would then
differ mostly by statistics, it is not clear how QCD effects affect
them.  To illustrate them in action, let us now describe how they have
been applied to the supersymmetric MSSM parameter point
SPS1a~\cite{Allanach:2002nj}, which is the only parameter point studied at the
necessary level of experimental sophistication~\cite{lhc_ilc,def_masses}.  
This popular sample
point has $m_{\tilde{g}} \simeq 600\,\gev$, $m_{\tilde{q}}\simeq 550\,\gev$,
$m_{\tilde{t}_1} \simeq 400\,\gev$, $m_{\tilde{\ell}} \simeq
150\,\gev$, $m_{\tilde{\chi}_1^0}\simeq 100\,\gev$,
$m_{\tilde{\chi}_2^0}\simeq m_{\tilde{\chi}_1^{\pm}} \simeq
200\,\gev$, and $m_{\tilde{\chi}_{3,4}^0} \sim
m_{\tilde{\chi}_2^{\pm}} \simeq 400\,\gev$.  These known
results will also be the basis of the model reconstruction described
in Section~\ref{sec:para}.

\begin{figure}[t]
\vspace*{-5cm}
  \includegraphics[width=0.38\textwidth]{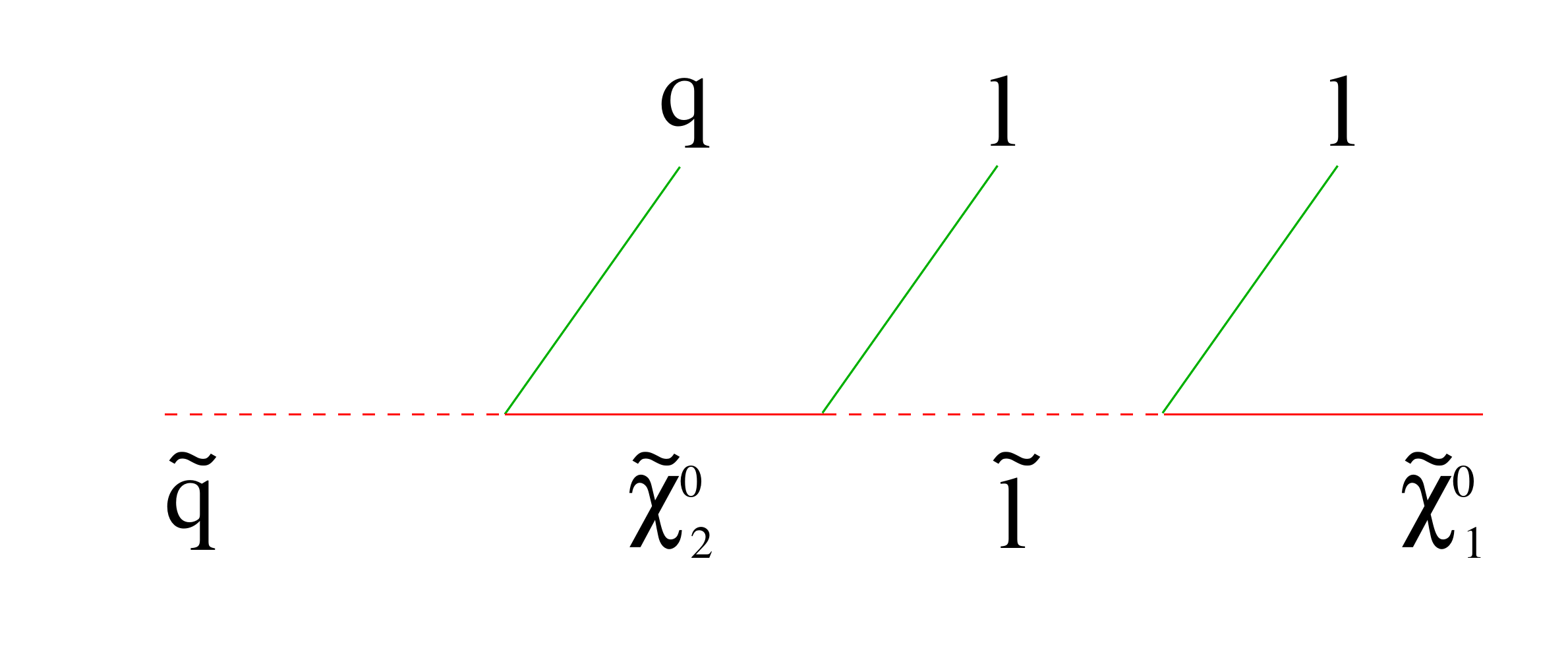}
  \hspace*{0.2\textwidth}
  \includegraphics[width=0.38\textwidth]{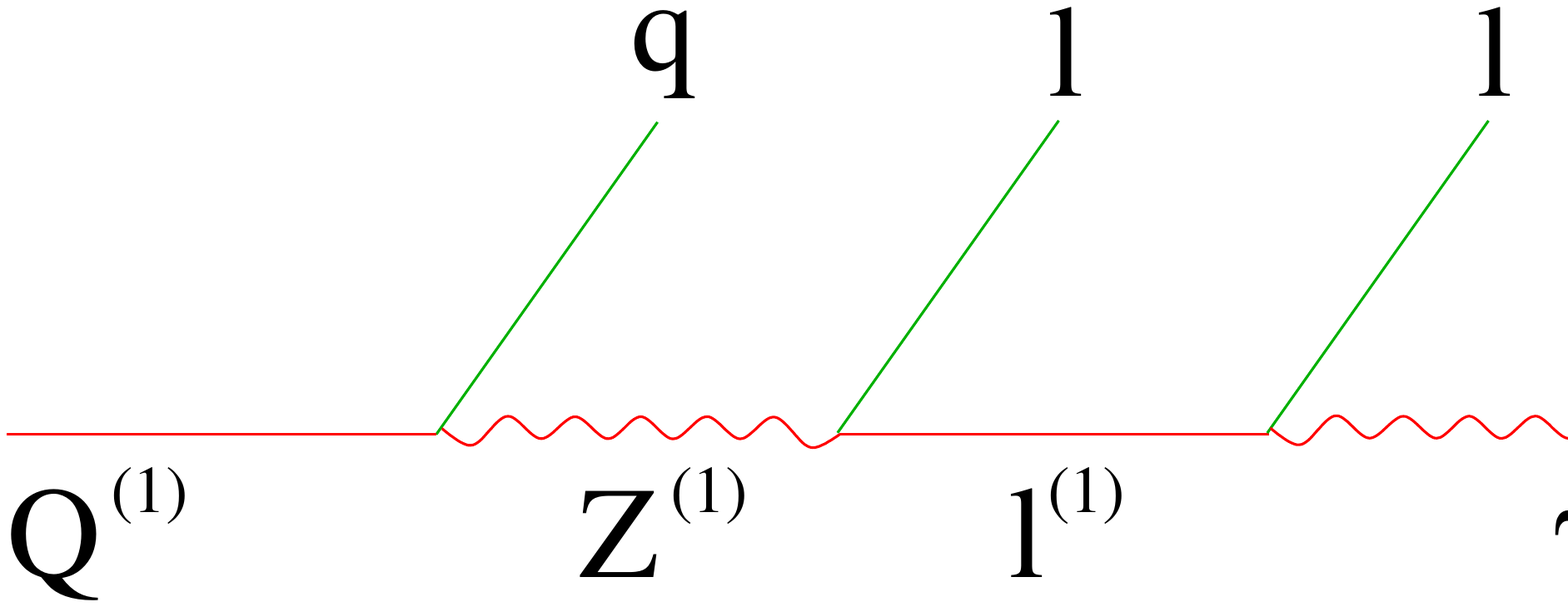}
\caption{The long squark decay chain in supersymmetry (left) and the long the KK
quark decay chain in UED (right) described in the text around 
Eq.~\eqref{eq:sig_squarkchain}.}
\label{fig:longchain}
\end{figure}

  The classic application of the endpoint method is to the long 
left-handed squark decay chain built out of successive two-body
decays~\cite{edges}.  This decay chain can also arise from the 
first KK excitations of Standard Model states in UED.  
The relevant decay patterns are:
\begin{equation}
  \tilde{q}_L \to \tilde{\chi}_2^0 q \to \tilde{\ell}^\pm
  \ell^\mp q \to \tilde{\chi}_1^0\ell^+\ell^-q
 \qquad \qquad
  Q^{(1)}_L \to Z^{(1)} q \to {\ell^{(1)}}^\pm
  \ell^\mp q \to \gamma^{(1)} \ell^+\ell^-q
\label{eq:sig_squarkchain}
\end{equation}
We illustrate these decay chains in the left and right panels 
of Fig.~\ref{fig:longchain}.
While the branching ratio for this decay chain might not be
particularly large, it is relatively easy to identify given 
the pair of leptons and missing energy.  For the SPS1a parameter point 
the long squark decay chain of Eq.~\eqref{eq:sig_squarkchain} 
has a branching ratio around $4\%$.

  When we use kinematic endpoints to extract mass parameters, it is
crucial to determine them from a signal-rich sample without a deep sea
of background events. Whenever we require leptons and missing energy a
major background will be top pairs. The key observation is that in
long cascade decays the leptons are flavor-locked, which means the
combination $e^+e^- + \mu^+\mu^- - e^-\mu^+- e^+\mu^-$ becomes roughly
twice $\mu^+\mu^-$ for the signal, while it cancels for top
pairs. Using such a combination for the endpoint analysis means the
top background is subtracted purely from data, as illustrated in
Fig.~\ref{fig:sig_mll}. Note that for rate measurements this method is
not quite as powerful as it looks at first sight -- subtracting
backgrounds using data still leaves us with twice the error from the
original background events $2 \sqrt{B}$.  For the extraction of
endpoints, the situation is obviously different.

\begin{figure}[t]
\vspace*{-3cm}
  \includegraphics[width=0.45\textwidth]{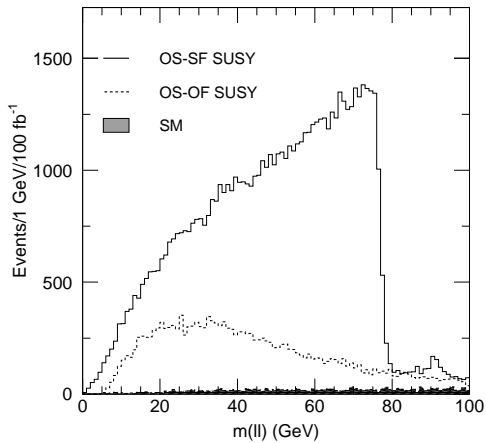}
\caption{Invariant mass distribution of two leptons after selection cuts 
  for the SPS1a model: SUSY signal Opposite Sign Same Flavor (OS-SF): full
  line; SUSY signal Opposite Sign Opposite Flavor (OS-OF): dotted
  line; Standard Model background: grey. Figure from Ref.~\cite{atlas_tdr}}
\label{fig:sig_mll}
\end{figure}

  The long squark decay chain for SPS1a-like parameter points has two important
advantages.  First, for a large enough mass hierarchy we might be able
to isolate the one decay jet just based on its
energy~\cite{heavy_skands,heavy_mg,Plehn:2008ae}. More importantly, the
invariant mass of the two leptons has a distinctive triangular shape
shown in Fig.~\ref{fig:sig_mll}. 
While this shape does not have to
survive for other underlying new-physics models~\cite{Smillie:2005ar}, the
$m_{\ell \ell}$ range is always given by
\begin{equation}
0 < m^2_{\ell\ell} 
  < \frac{(m_{\tilde{\chi}^0_2}^2-m_{\tilde{\ell}}^2)
          (m_{\tilde{\ell}}^2-m_{\tilde{\chi}_1^0}^2)}{m_{\tilde{\ell}}^2}
\qquad 
0 < m^2_{\ell\ell} 
  < \frac{(m_{Z^{(1)}}^2-m_{\ell^{(1)}}^2)
          (m_{\ell^{(1)}}^2-m_{\gamma^{(1)}}^2)}{m_{\ell^{(1)}}^2}
\label{eq:sig_mll}
\end{equation}
In complete analogy to the di-lepton endpoint, but with somewhat
reduced elegance we can measure the threshold and edge of the
$\ell^+\ell^- q$ distribution and the edges of the two $\ell^\pm q$
combinations~\cite{atlas_tdr,edges}. Then, we solve the
system for the intermediate without any model assumption, which allows
us to even measure the dark matter mass to about $10\%$. The limiting
factors will likely be our ability to observe enough endpoints in
addition to $m_{\ell \ell}^\text{max}$, and the jet energy scale
uncertainty. An interesting question is how well we will do with tau
leptons, where the edge is softened by neutrinos from tau
decays~\cite{pmz_spin}.\bigskip

  If the gluino (or heavy KK gluon) is heavier than the squarks (or KK quarks),
its mass can be measured by simply extending the
squark chain by one step: $\tilde{g}\to q\tilde{q} $~\cite{mgl_per}.
This measurement becomes more complicated, however, if one of the two 
jets from the gluino
decay is not very hard, because its information will be buried by the
combinatorial error due to QCD jet radiation. The way around is to ask
for two bottom jets from the strongly interacting decay: $\tilde{g}
\to b \tilde{b}$. In Fig.~\ref{fig:sig_gluinochain} we see that for
example the gluino mass can be extracted at the per-cent level, a point
at which we might have to start thinking about off-shell propagators
and at some point even define what exactly we mean by `masses
appearing in cascade decays'.

\begin{figure}[t]
\begin{center}
  \includegraphics[width=0.9\textwidth]{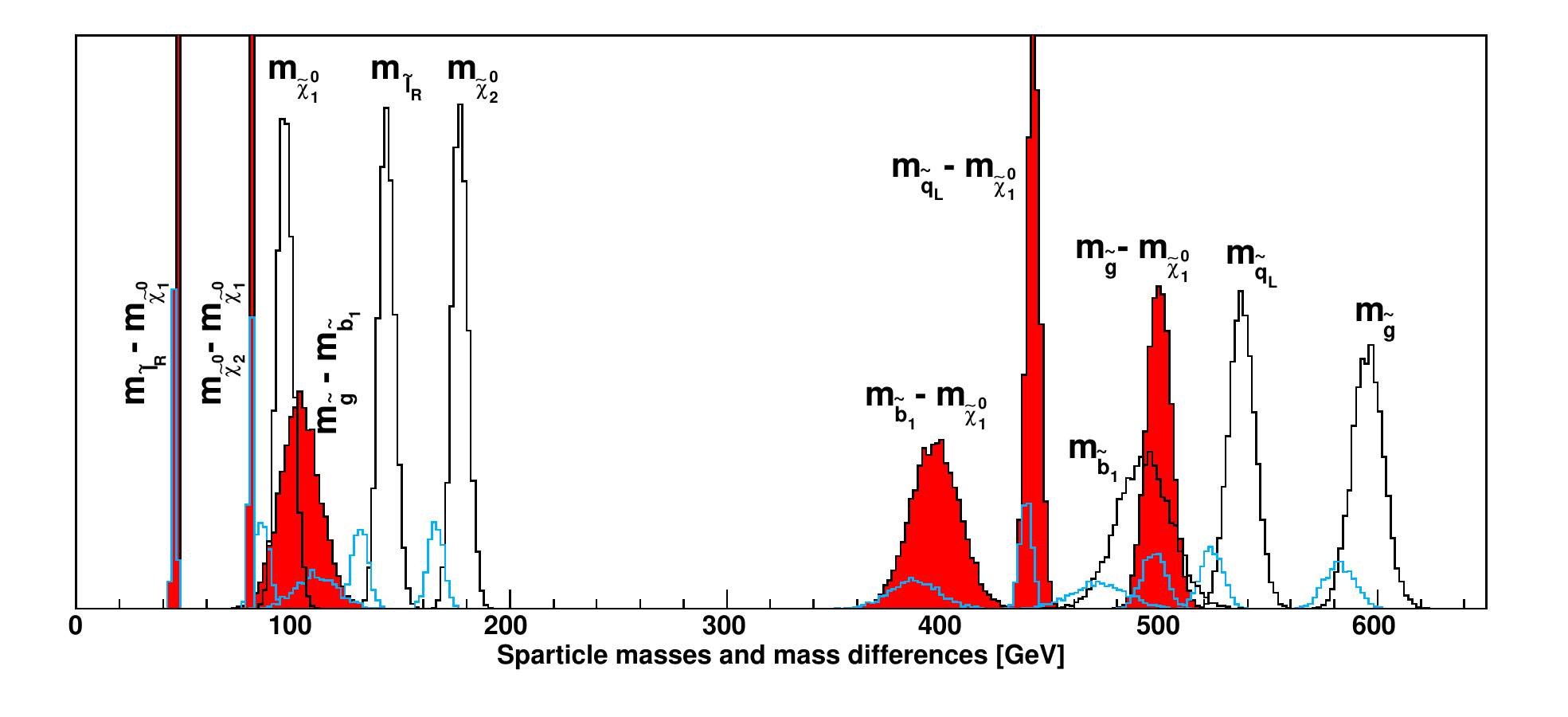}
\end{center}
\caption{Masses extracted from the gluino-sbottom decay chain,
  including estimated errors. The faint blue lines indicate wrong
  solutions when inverting the endpoint-mass relations. Figure from
  Ref.~\cite{mgl_per}.}
\label{fig:sig_gluinochain}
\end{figure}

Alternatively, we can use the same gluino decay to first reconstruct
the intermediate neutralino or KK $Z$ momentum for lepton pairs near
the $m_{\ell \ell}$ edge. In that case the invisible heavy state is
produced approximately at rest, and the momenta are correlated
as~\cite{atlas_tdr}
\begin{equation}
  \vec{p}_{\tilde{\chi}^0_2} = \left( 1 - \frac{m_{\tilde{\chi}_1^0}}{m_{\ell\ell}}
                      \right) \; \vec{p}_{\ell\ell},
\qquad \qquad \qquad
  \vec{p}_{Z^{(1)}} = \left( 1 - \frac{m_{\gamma^{(1)}}}{m_{\ell\ell}}
                      \right) \; \vec{p}_{\ell\ell}.
\end{equation}
If both neutralino masses (or the KK photon and $Z$ masses) are known
we can extract the sbottom (KK bottom) and gluino (KK gluon) masses by
simply adding the measured bottom momenta.  Again, for the SPS1a
parameter point we can measure the gluino mass to few per-cent,
depending on the systematic errors. This second method of
reconstructing an invariant mass of the intermediate state could
ideally distinguish different intermediate states, like left or right
handed sbottoms.\bigskip

  For short decay chains we can use $\mtt$ to measure the masses of
particles decaying to the missing energy state directly. In
supersymmetry, this could be a right-handed slepton or squark decaying
to a mostly bino LSP. The challenge short decay chains is that extracting 
them relies on some kind of veto, which from Section~\ref{sec:sig_qcd} 
we know is problematic for low-$p_T$ jets.  For the analyses included in
Ref.~\cite{lhc_ilc} it is assumed that the missing energy particle mass is 
known from other measurements, so we only the mass of the decaying states
are extracted from $\mtt$.  As mentioned above, it might well be possible 
to measure both masses in one step, using kinks in the $\mtt$
distribution as a function of the hypothesized dark-matter
mass~\cite{kinks}.\bigskip

  Given that endpoint analyses use only a small fraction of the set of
signal events, namely those with extreme kinematics, an obvious way to
improve the precision of mass measurements is to investigate the entire
shape of the invariant mass distributions~\cite{sps1a_per}. 
One might also hope to use the shape of a kinematic distribution
to infer the location of a kinematic endpoint that is not directly
visible.  This can occur when the decay matrix element
has a smooth falloff approaching the endpoint such that the 
distribution near the endpoint is softened and disappears into
the background noise.  Using the shape of the full distribution
could allow us to circumvent this challenge, but will typically
be much more sensitive to backgrounds than a sharp kinematic endpoint.

While with the appropriate care we can claim that a clearly visible
endpoint depends only on the masses of the particles involved, and not
on their interaction matrix elements (as illustrated for example by
Eq.(\ref{eq:sig_def_mct}) or Eq.(\ref{eq:sig_mll})) invariant mass
distributions are just an invariant way of writing angular
correlations between the outgoing particles. Those will definitely
depend on the spin assignments of all particles involved, which means
that our mass measurement will always involve a hypothesis on the
nature of the particles inside the decay chain.\bigskip

Putting this handicap to work
the same technique can be applied to measure the quantum numbers
of the the particles involved in a decay cascade. 
The determination of the spins of new particles is particularly
important, since many of the inclusive signatures of supersymmetry
can also arise from other models where the Standard Model states
have new odd partners such as UED, and there is a possibility of
confusing these very different scenarios.  In general,
the determination of discrete quantum numbers like the spin of new particles 
is difficult in the absence of fully reconstructed events. 
The usual threshold behavior is not visible in hadron collider 
production processes, in particular if the final state includes missing 
transverse energy.  Instead, we have to rely on angular correlations 
in decays (or other clever techniques).  

  A well-studied example of spin determination is the squark or KK quark
decay chains given in Eq.(\ref{eq:sig_squarkchain})~\cite{Barr:2004ze,Smillie:2005ar}.
This analysis can be treated in three steps: 
\\(1) We notice that cascade decays radiate Standard Model particles 
with known spins. For the squark decay chain of Eq.~\eqref{eq:sig_squarkchain}
the particles emitted are all fermions.  Hence, the spins
inside the decay chain alternate between fermions and bosons. 
This is also true in the corresponding UED process listed 
in Eq.~\eqref{eq:sig_squarkchain}, but in contrast all the spins of
the exotic states in the decay chain are reversed.
\\(2) Both the thresholds and the edges of all invariant masses of the
radiated fermions are completely determined by the masses inside the
decay chains.  On the other hand, the shape of the distribution between 
the endpoints is nothing but an angular correlation (in some rest
frame). This means that the well-established $m_{j \ell}$ jet-lepton 
invariant mass distribution allows us to analyze spin correlations 
in squark/KK~quark decays in a Lorentz invariant way.
\\(3) Proton--proton collisions produce considerably more squarks than
antisquarks in the squark--gluino associated channel, and the same for
heavy KK quarks and gluons. A decaying squark or KK quark radiates a
quark while an antisquark or KK antiquark radiates an antiquark, which
means that we can define a non-zero production asymmetry between $m_{q
  \ell^+}$ and $m_{q \ell^-}$
\begin{equation}
A = 
\frac{\dfrac{dN(\ell^+ q)}{dm_{\ell q}} - \dfrac{dN(\ell^- q)}{dm_{\ell q}}}
     {\dfrac{dN(\ell^+ q)}{dm_{\ell q}} + \dfrac{dN(\ell^- q)}{dm_{\ell q}}} \; .
\end{equation}
The symbol $q$ refers to either a quark or an antiquarks, since on the
jet level they are indistinguishable.  This asymmetry is shown in
Fig.~\ref{fig:sig_bryan}, generated for hierarchical SUSY-like and
degenerate UED-like spectra. For each spectrum we compare the SUSY and
UED hypotheses. We see that provided the radiated Standard Model
particles are hard enough we can clearly distinguish the two spin
hypotheses.  The difference between UED and SUSY becomes more
prominent with larger mass splittings between the parent and daughter
particles, but is a subtle enough effect so that we need to worry
about detector effects and combinatorial backgrounds.  Considering
other observables in the decay chain may further improve the spin
reconstruction in squark
decays~\cite{Athanasiou:2006ef,Athanasiou:2006hv}.\bigskip

\begin{figure}[t]
\begin{center}
  \includegraphics[width=0.42\textwidth]{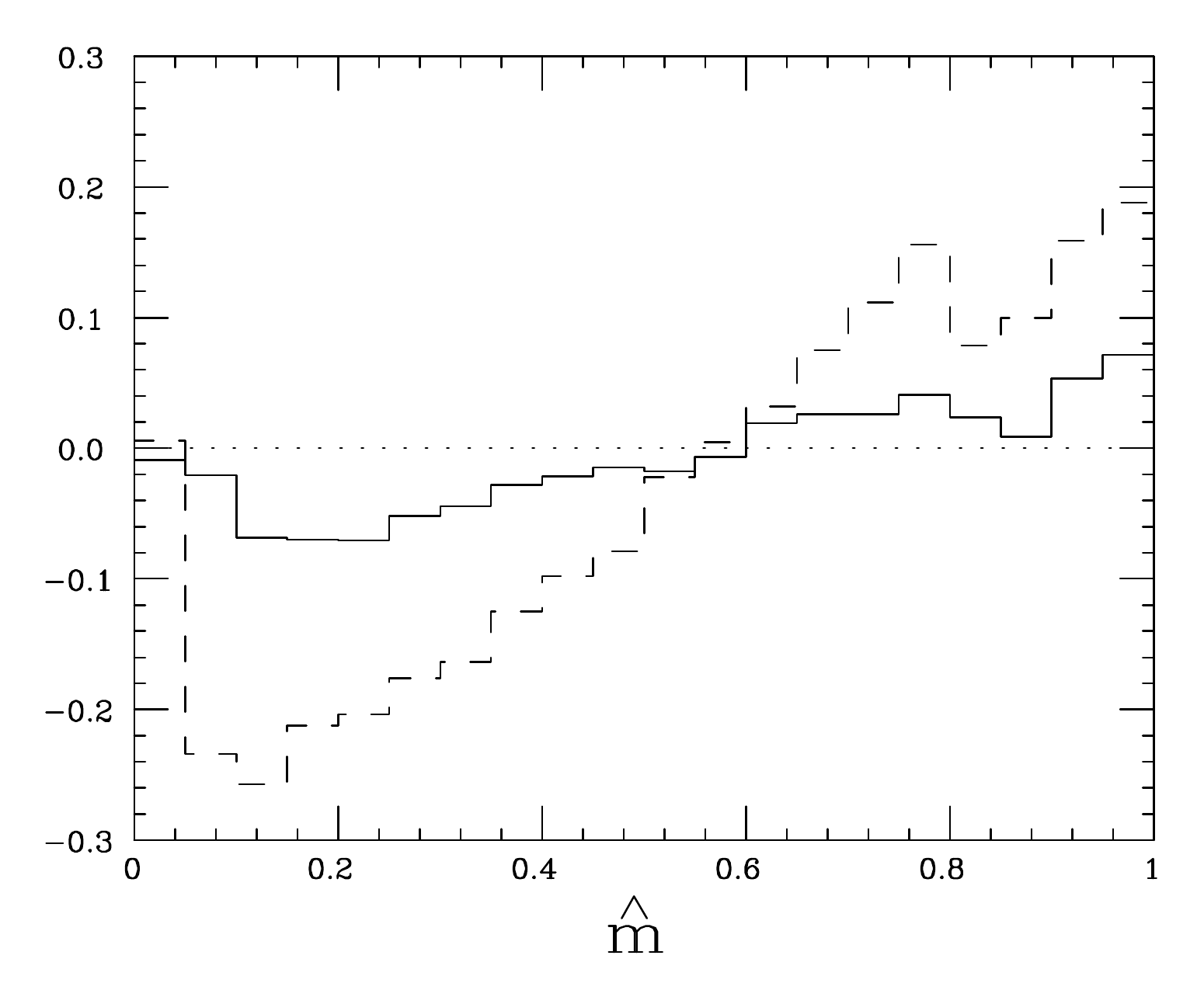}
  \hspace*{0.10\textwidth}
  \includegraphics[width=0.45\textwidth]{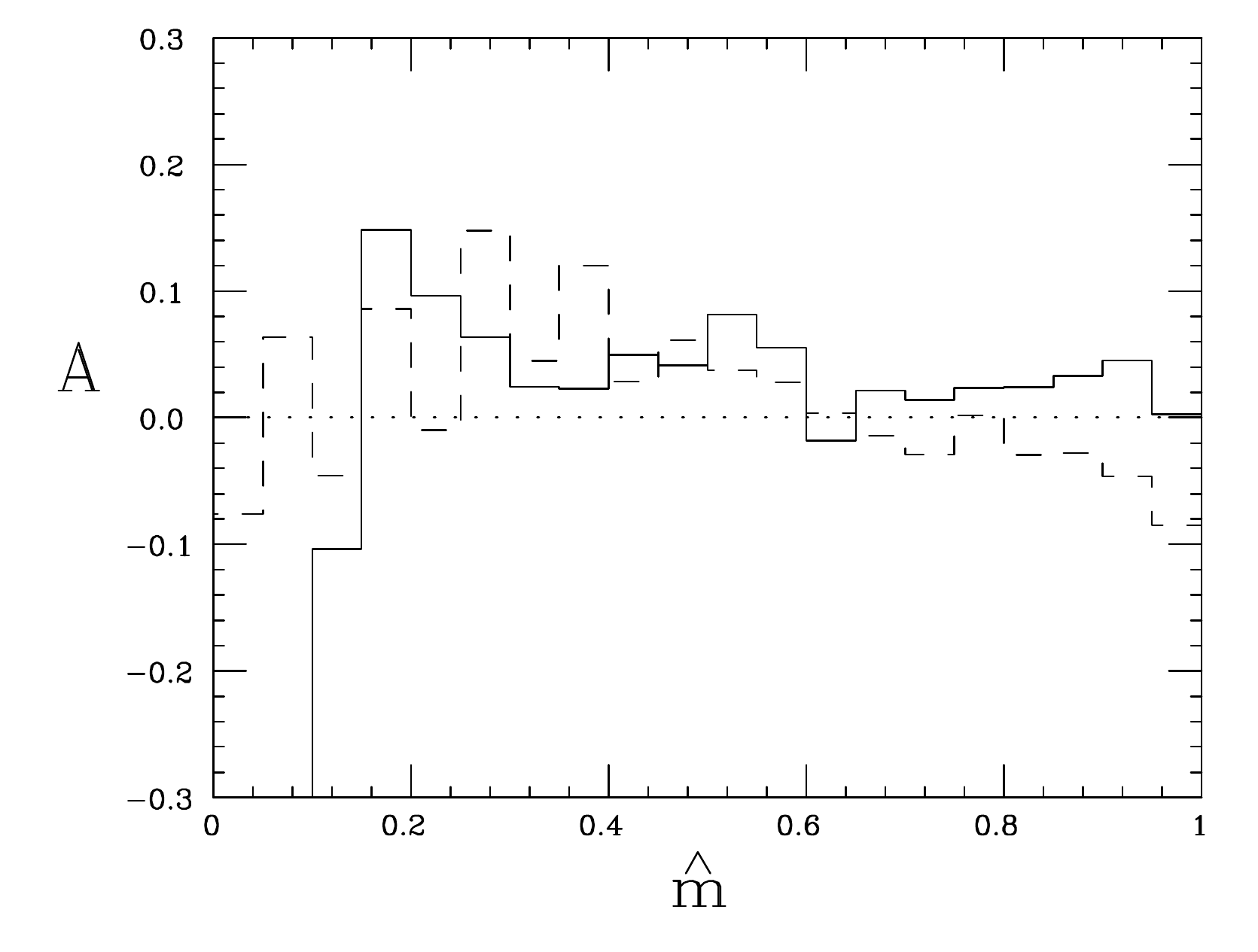}
\end{center}
\caption{Asymmetry in $m_{j \ell}$ assuming underlying supersymmetry
  (dashed) or universal extra dimensions (solid). The mass spectra for the simulation
  are either the hierarchical SPS1a supersymmetric spectrum (left) or
  the degenerate UED spectrum (right). Figures taken from
  Ref.~\cite{Smillie:2005ar} }
\label{fig:sig_bryan}
\end{figure}

  These basic ideas have also been applied to many other 
situations, like decays including gauge bosons~\cite{jennie_w},
three-body decays~\cite{three_body_gluino}, gluino decays with
decay-side asymmetries~\cite{gluino_spin}, cascades including
charginos~\cite{liantao_spin_1}, weak-boson-fusion
signatures~\cite{wbf_susy_2}, and more~\cite{liantao_spin_2,cambridge_spin}. 
The spin of the gluino is a particularly important LHC measurement.
A characteristic signal of the Majorana nature of the gluino are
like-sign dileptons in supersymmetric decay cascades originating
from QCD production~\cite{like_sign}.  
This can arise either from squark pair production 
$qq \to \tilde{q} \tilde{q}$ via a $t$-channel gluino
exchange, or the symmetric decays
\begin{alignat}{5}
\tilde{g} \to q \tilde{q}^* \to q \bar{q}' \, \ell^- \bar{\nu} \tilde{\chi}_1^0,
\qquad \qquad
\tilde{g} \to \bar{q} \tilde{q} \to \bar{q} q' \, \ell^+ \nu \tilde{\chi}_1^0.
\label{eq:likesign}
\end{alignat}
In both cases, the Majorana nature of the gluino is necessary for these
processes to occur.   On the other hand a vector KK gluon in UED
does not have an associated fermion flow direction, and can also
produce same-sign dileptons via
\begin{alignat}{5}
G^{(1)} \to q \bar{Q}^{(1)} \to q \bar{q}' \, \ell^- \bar{\nu} \gamma^{(1)},
\qquad \qquad
G^{(1)} \to \bar{q} Q^{(1)} \to \bar{q} q' \, \ell^+ {\nu} \gamma^{(1)}.
\end{alignat}
Thus, if we can show that the gluino is indeed a fermion, we will also
be able to deduce that it is a Majorana state, which is an immediate
effect of minimal supersymmetry which relates the physical gluino
degrees of freedom to the two polarizations of the gluon.  If we can
now show that the gluino is indeed a fermion we also know it is a
Majorana fermion, where this Majorana nature is an immediate effect of
the supersymmetry which conserves the on-shell number of degrees of
freedom of the gluon and the gluino, as discussed in
Section~\ref{sec:models_mssm}.

\subsubsection{Direct production of missing energy}
\label{sec:sig_direct}

  Direct production of new particles giving rise to missing energy
at the LHC has been studied in considerably less detail than missing 
energy from cascade decays.  The primary reason for this is
simple.  For a particle to produce missing energy it must be quasi-stable,
either because it is prevented from decaying by a symmetry or due to
a combination of weak couplings and kinematics.  In the first case,
the LHC production rates tend to be small unless there are new QCD-charged
states also charged under the symmetry that decay to the invisible state,
and this brings us back to the cascade decay scenario.
The second possibility is much less common in models of new physics, 
and also often implies that the direct production rate for the missing 
energy particle at the LHC will be very small.
In the present subsection we will concentrate on two specific scenarios
of direct missing energy: models of large extra dimensions, and invisible
Higgs boson decays.  We will also comment on ways to probe the direct
production of invisible stable particles protected by a symmetry,
such as a neutralino LSP in supersymmetry.

One situation in which the direct production of missing energy can be
significant occurs when there are very many new particles, each of
which couples very weakly to the Standard Model and is stable on the
time scale of particle colliders.  In this case, the small coupling
and tiny production rate of each individual particle can overcome by
the large number of individual production channels.  This is precisely
what occurs in models of large extra dimensions (ADD) discussed in
Section~\ref{sec:models_add}.  These models contain a tower of very
closely spaced KK graviton resonances, each of which according to
Eq.~\eqref{eq:ee_new} couples to the Standard Model with strength
$1/M_\text{Pl}$ and is long-lived.  However, upon summing over the
production of a large number of KK gravitons, the total inclusive
graviton production rate becomes an observable $E/\tev$ effect.
Gravitons created this way will leave the detector and generate a
signal of direct missing energy.\bigskip

  Potentially observable effects in ADD models arise collectively, 
involving the contributions of many independent gravitons.  
This means that there will not be any kind of well-defined 
resonance or similar kinematic feature.  Instead, the radiation 
of graviton towers will add missing energy to Standard Model processes.
This will alter the kinematic distributions of transverse momenta
or invariant masses much differently than radiating off
a $Z$ boson or neutrinos.  The primary search strategy for such
effects is thus to start from well-measured standard candles at the LHC,
such as Drell-Yan processes or the production of gauge bosons in
association with jets, and to look for anomalous deviations.

\begin{figure}[t]
\begin{center}
  \includegraphics[width=0.40\textwidth]{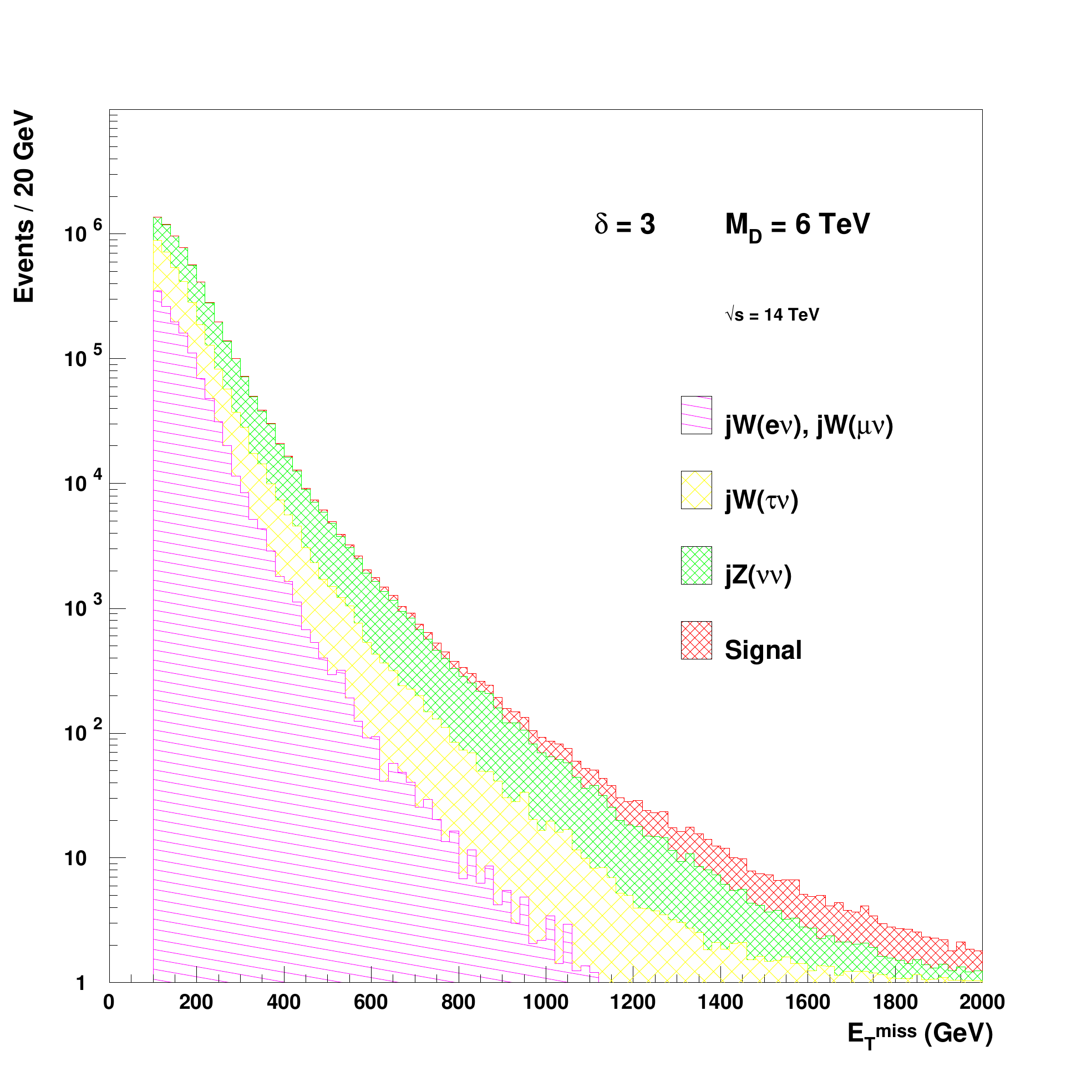}
  \hspace*{0.15\textwidth}
  \includegraphics[width=0.40\textwidth]{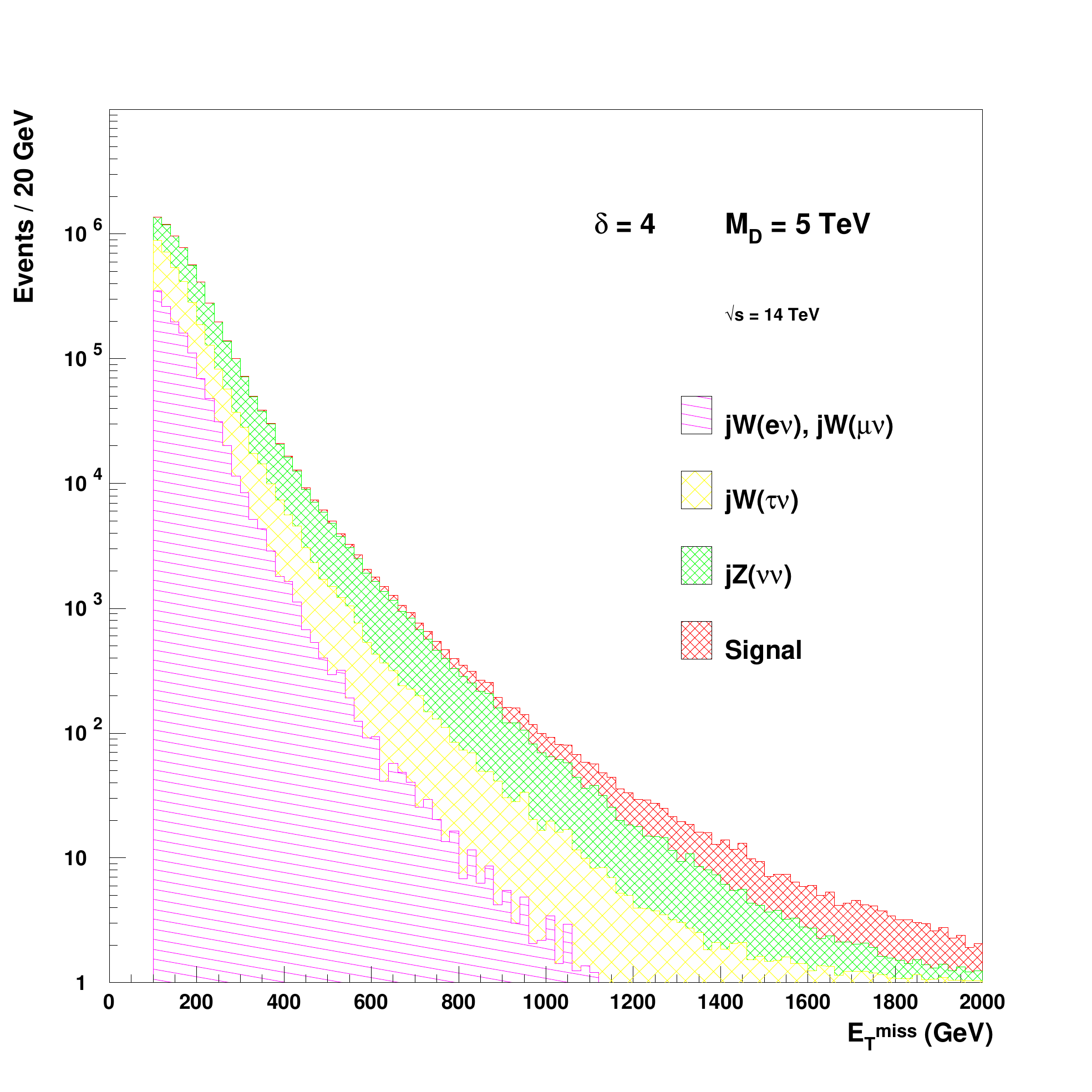}
\end{center}
\vspace*{-5mm}
\caption{Missing transverse energy from real KK graviton emission in
  models with large flat extra dimensions (ADD). We show the single
  jet plus gravitons signal and backgrounds after selection cuts and
  for an integrated luminosity of $100~\ifb$. Figure from
  Ref.~\cite{lhc_real}.}
\label{fig:sig_add}
\end{figure}

  Possibly the best way to probe ADD models is to search for
a KK graviton recoiling off a hard jet by looking at the 
transverse momentum distribution of the jet.  The primary
background to this comes from invisible $W$ or $Z$ decays with
neutrinos or an undetected lepton.  This background can be
computed directly, or simply extrapolated from measurements of 
leptonic $Z$ decays.  When computing the backgrounds it is
essential to include Sudakov logarithms of the form $\log E_j/m_Z$
arising from a mismatch or massive virtual and
real $Z$ bosons. 
Usually, LHC signatures
are not sensitive to such effects, because new physics particles tend to
be heavy, and heavy states are preferably produced
non-relativistically. However, the situation for backgrounds can be different.
In particular the search for KK gravitons relies on a very
precise understanding of the kinematic distributions at large energy
scales, where experimental measurements might be statistically
limited.  A sample distribution of missing energy
produced in association with one or more hard jets is shown in 
Fig.~\ref{fig:sig_add}.  For the leading contributions the
missing energy will be balanced by the transverse momentum of one jet,
but including additional jet radiation as well as unseen leptons it is
not clear which of the two transverse momenta will turn out to be more useful.\bigskip

\begin{figure}[t]
\includegraphics[width=8cm]{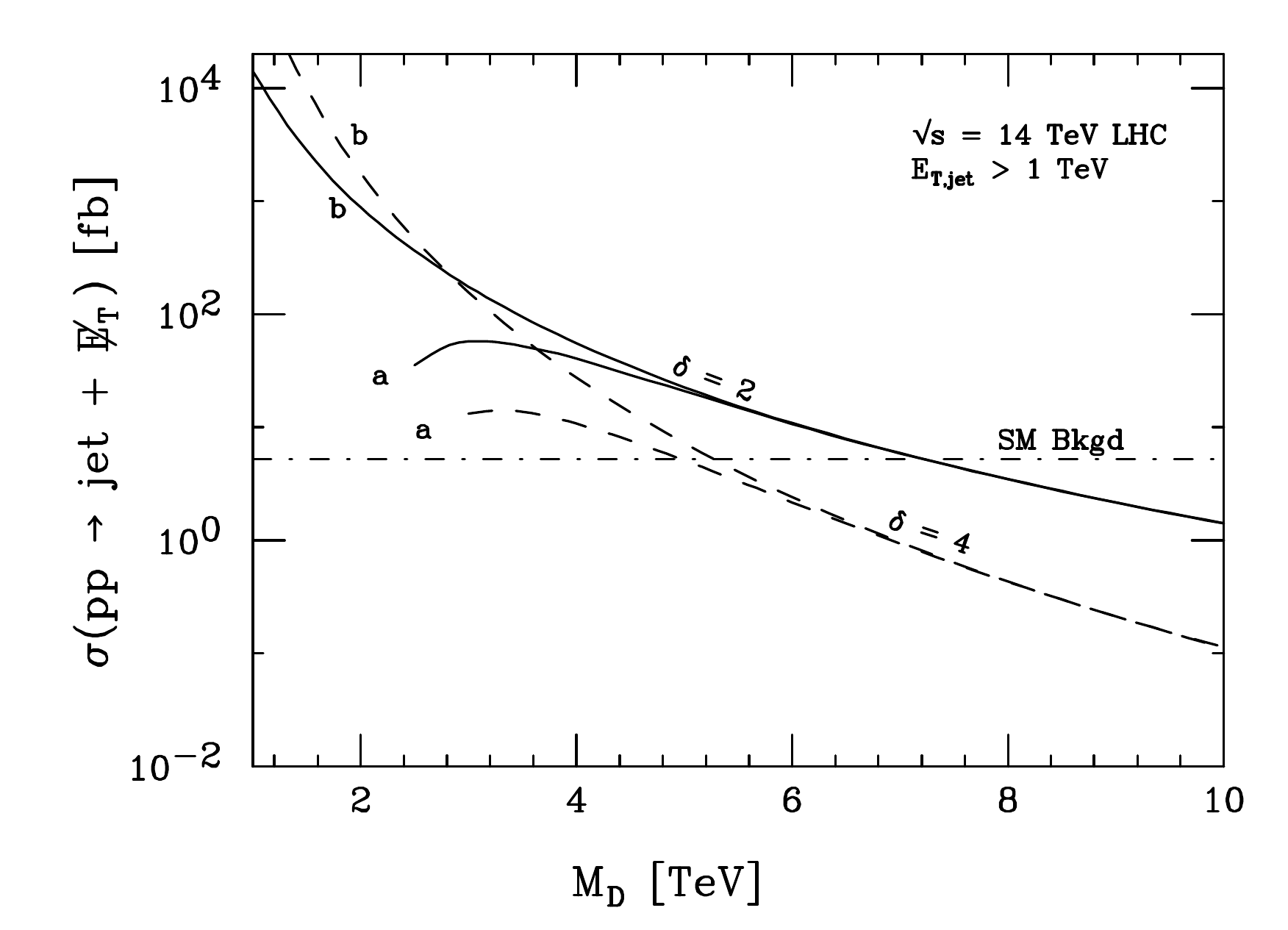}
\hspace*{15mm}
\raisebox{4mm}{\includegraphics[width=5.6cm]{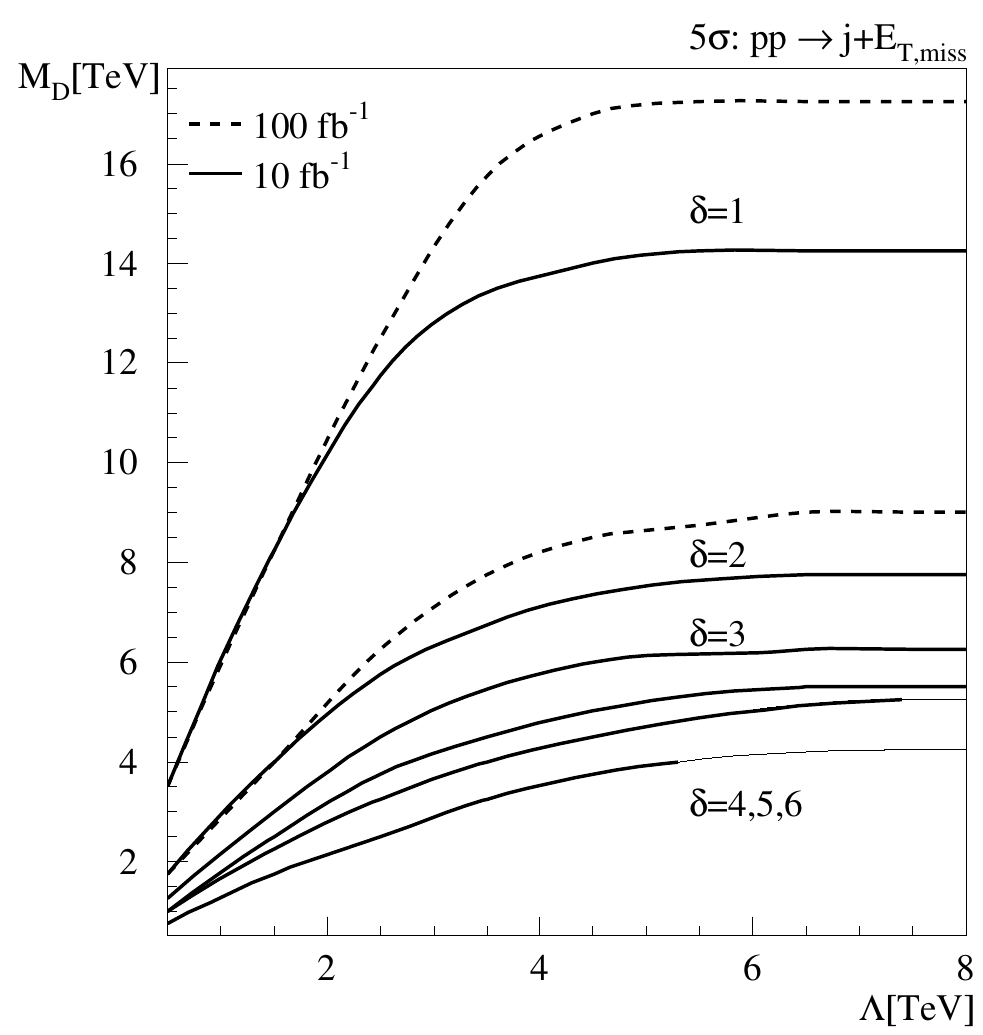}}
\caption{Left: production rates for graviton--jet production at the
  LHC.  The cut on the jet's transverse energy as a cut on the
  transverse momentum from the graviton tower, without the additional
  experimental smearing.  For the curve (a) we set $\sigma(s)=0$
  whenever $\sqrt{s}>\Mstar$. For curve (b) the $\mkk$ integration in
  Eq.(\ref{eq:KKTxsec}) includes $\sqrt{s}>\Mstar$. Figure from
  Ref.~\cite{grw}. Right: $5 \sigma$ discovery contours for real
  emission of KK gravitons in the plane of $\Mstar$ and the UV cutoff
  on $\sigma(s)$.  The transition to thin lines indicates a cutoff
  $\cutoff$ above $\Mstar$.  Figure from Ref.~\cite{gps}.  Note that
  $M_D$ in both figures corresponds to $\Mstar$ in the text.}
\label{fig:ReEmm}	
\end{figure}

In Section~\ref{sec:models_add} we discuss the impact of a UV
completion of extra dimensional gravity on relevant LHC predictions.
In the left panel of Fig.\ref{fig:ReEmm} we see that the jet +
$\ptmiss$ cross section becomes seriously dependent on physics above
$\Mstar$ the moment the Planck scale enters the range of available
energies at the LHC $\sqrt{s} \lesssim 3$~TeV. Above this threshold
the difference between curves (a) and (b) is small.

In the region where the curves differ significantly, UV effects of our
modelling of the KK spectrum become dominant and any analysis based on
the Kaluza--Klein effective theory will fail. Luckily, the parton
distributions, in particular the gluon density, drop rapidly towards
larger parton energies.  This effects effectively constrains the
impact of the UV region which includes $\sqrt{s}>\Mstar$.

In the right panel of Fig.\ref{fig:ReEmm} we see the $5\sigma$
discovery reach at the LHC with a variable ultraviolet cutoff
$\cutoff$ on the partonic collider energy. For each of the lines there
are two distinct regimes: for $\cutoff < \Mstar$ the reach in $\Mstar$
increases with the cutoff. Once the cutoff crosses a universal
threshold around 4~TeV the discovery contours reach a plateau and
become cutoff independent. This universal feature demonstrates and
quantifies the effect of the rapidly falling parton densities.  The
fact that the signal decreases with increasing dimension shows that
the additional volume element from the $n$-sphere integration is less
than the $1/\Mstar$ suppression from each additional dimension.\bigskip

  Instead of looking for real graviton emission we can also search for
their virtual effects in Drell-Yan production of dilepton pairs,
where the $s$-channel gauge bosons linking incoming quarks and 
outgoing leptons can be replaced by a graviton tower.  
There will also be a gluon fusion contribution to the graviton 
Drell-Yan process that is not present for (on-shell) $s$-channel gauge bosons. 
The observable best suited to extract this signal from the 
background is the invariant mass of the dileptons $m_{\ell \ell}$,
which is expected to show a deviation at high energies. 
As in the case of real graviton emission, the search
for new physics in tails of distributions is challenging and hinges on
our understanding of Standard Model processes and their QCD effects.

  Virtual graviton effects are strongly dependent on the detail of
the ultraviolet completion of the model, as discussed in detail in
Section~\ref{sec:models_add}.  Such an ultraviolet completion can 
include stringy resonance structures or fixed-point regularized 
operator kernels.  Following the logic for
example of the cascade decay measurements discussed in
Sections~\ref{sec:sig_cascade} and \ref{sec:para} we can study the
$m_{\ell \ell}$ behavior in the Drell-Yan process to probe the
underlying model of higher-dimensional gravity, discussed in detail in
Section~\ref{sec:models_add}.\bigskip

\begin{figure}[t]
\begin{center}
  \includegraphics[width=0.45\textwidth]{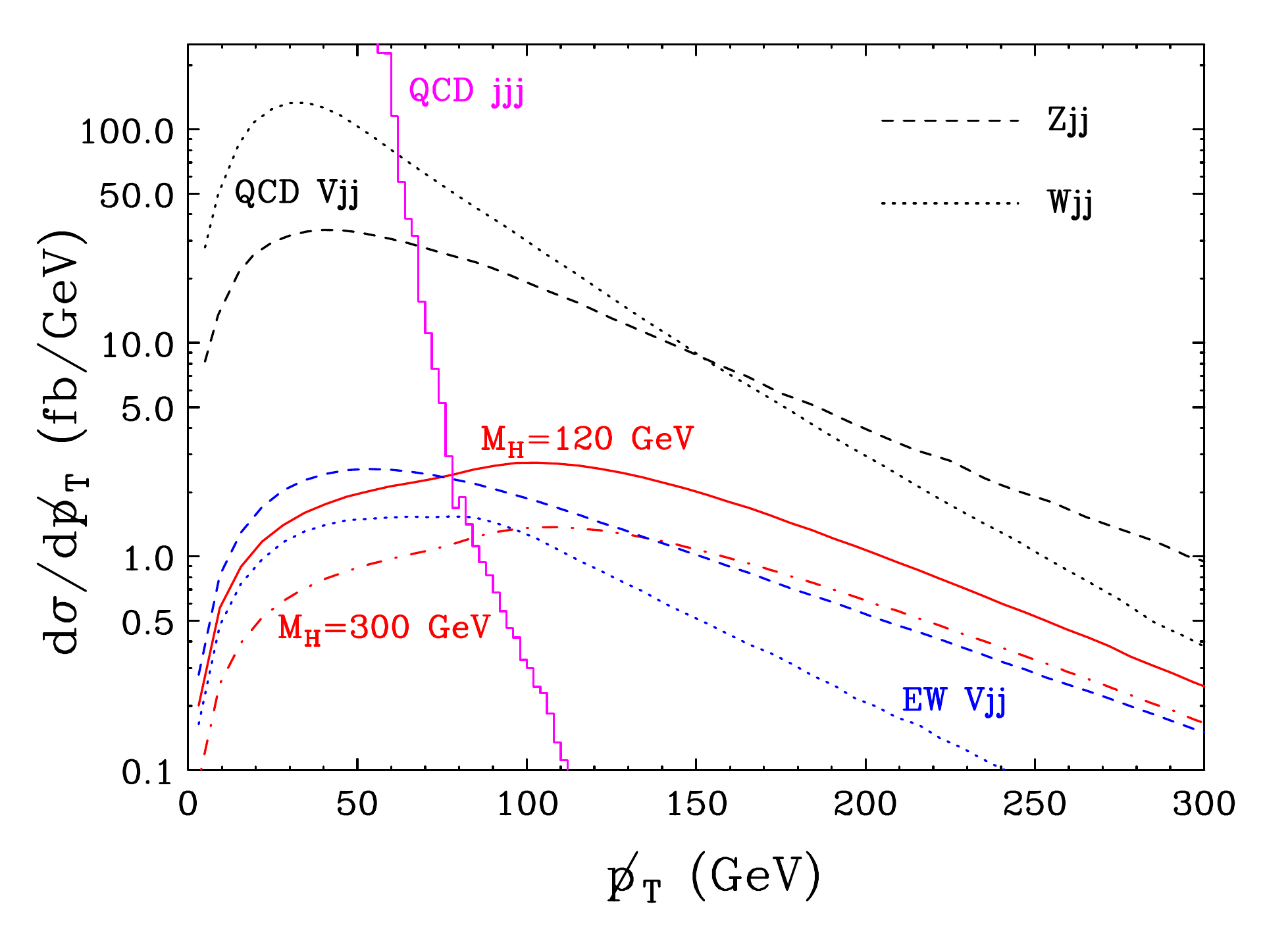}
  \hspace*{10mm}
  \includegraphics[width=0.47\textwidth]{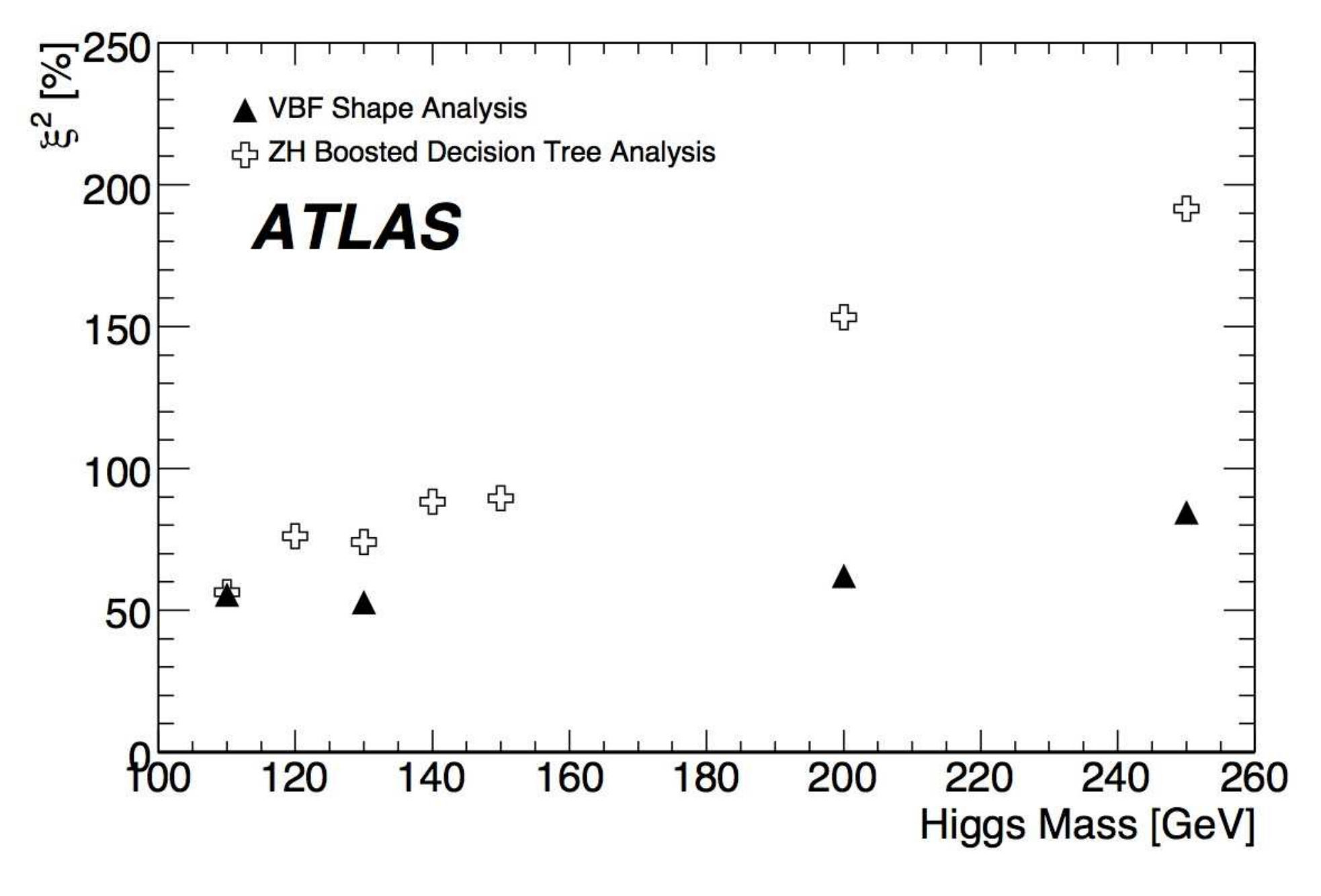}
\end{center}
\caption{ Left: distributions of $\ptmiss$ distribution for
  weak-boson-fusion Higgs at the LHC production and its backgrounds.
  Figure from Ref.~\cite{oscar_dieter}. Right: expected LHC reach for
  an invisible Higgs signature. Figure from
  Ref.~\cite{DeRoeck:2009id}.}
\label{fig:sig_invhiggs}
\end{figure}

  Another way for missing energy to be produced directly at the LHC 
is through a coupling to the Higgs $H$.  For any new scalar $\varphi$ 
the \emph{Higgs Portal} coupling is always gauge invariant~\cite{Patt:2006fw}:
\begin{equation}
-\mathscr{L} \supset \zeta\,|\varphi|^2|H|^2.
\end{equation}
As long as $\varphi$ does not develop of VEV it will not mix with the Higgs.
If $\varphi$ is light enough, this coupling will also allow the Higgs boson
decay mode $h\to \varphi^*\varphi$ which can easily dominate the Higgs
branching fraction.  When $\varphi$ is neutral under the Standard Model
and long-lived these Higgs decays will be invisible, 
and will be a source of direct missing energy.\bigskip


Missing energy from the Higgs sector leads us to the question if
more complex production channels can be useful if they
provide us with more information.  Weak boson fusion is known to offer 
such an opportunity in a few key observables without even considering 
the details of the new particles produced 
in the central event~\cite{wbf_tau}.  The prime example for this
strategy is the search for an invisibly decaying Higgs boson produced
through gluon fusion~\cite{oscar_dieter}, whose signature
consists of missing transverse energy and a pair of hard forward jets. 
These additional jets from the weak boson fusion process can provide a
trigger and a way to reduce the Standard Model backgrounds.
These forward tagging jets are required to a rapidity gap of 
$\Delta \eta \gtrsim 4.5$ and a very large invariant mass, 
$m_{jj} \gtrsim 1.2$~TeV.  These cuts are stiffer than in the usual
Higgs searches, reflecting the fact that aside from a cut on missing
energy $\ptmiss \gtrsim 100$~GeV, there is not much else to use
to reduce backgrounds.  The $\ptmiss$ distributions for both the
Higgs signal and the primary backgrounds are shown in 
Fig.~\ref{fig:sig_invhiggs}. 
If we know as well that the missing particles have a scalar coupling to the
longitudinal components of the weak gauge boson, we can apply an
additional cut on the azimuthal angle between the two jets, 
$\Delta \phi_{jj} < 1$, to suppress weak-boson-fusion 
$Z \to \nu \nu$ backgrounds~\cite{wbf_tau,wbf_coupling}.

  A second channel which can be exploited to find an invisible
Higgs is inspired by the decay-channel independent Higgs
searches at LEP. If we reconstruct a leptonic $Z$ boson
accompanying a large amount of missing energy, neither triggering nor
physics backgrounds are huge problems.  The main issue for this
associated $Zh$ production mode will likely be controlling the 
jet activity and fake missing energy
from jets missed in the detector~\cite{invhiggs_zh}. The result for
invisible Higgs searches are usually quantified in terms of
\begin{equation}
\xi^2(m_H) = \frac{\sigma_\text{BSM}}{\sigma_\text{SM}} \; 
        \br(H \to \text{inv}) 
\equiv \frac{(\sigma \cdot \br)_{H \to \text{inv}}}
            {(\sigma \cdot 1)_\text{SM}} 
\end{equation}
for a given Higgs mass. In the right panel of
Fig.~\ref{fig:sig_invhiggs} we see the experimental results for these
two search channels. For a light Higgs boson a $50\%$ branching ratio
into an invisible particle will be observable already at low
luminosities.  It may also be possible to extract the Higgs mass from
the ratio of the weak-boson-fusion and associated $Zh$ Higgs
production rates.\bigskip

\begin{figure}[t]
\begin{center}
  \includegraphics[width=0.45\textwidth]{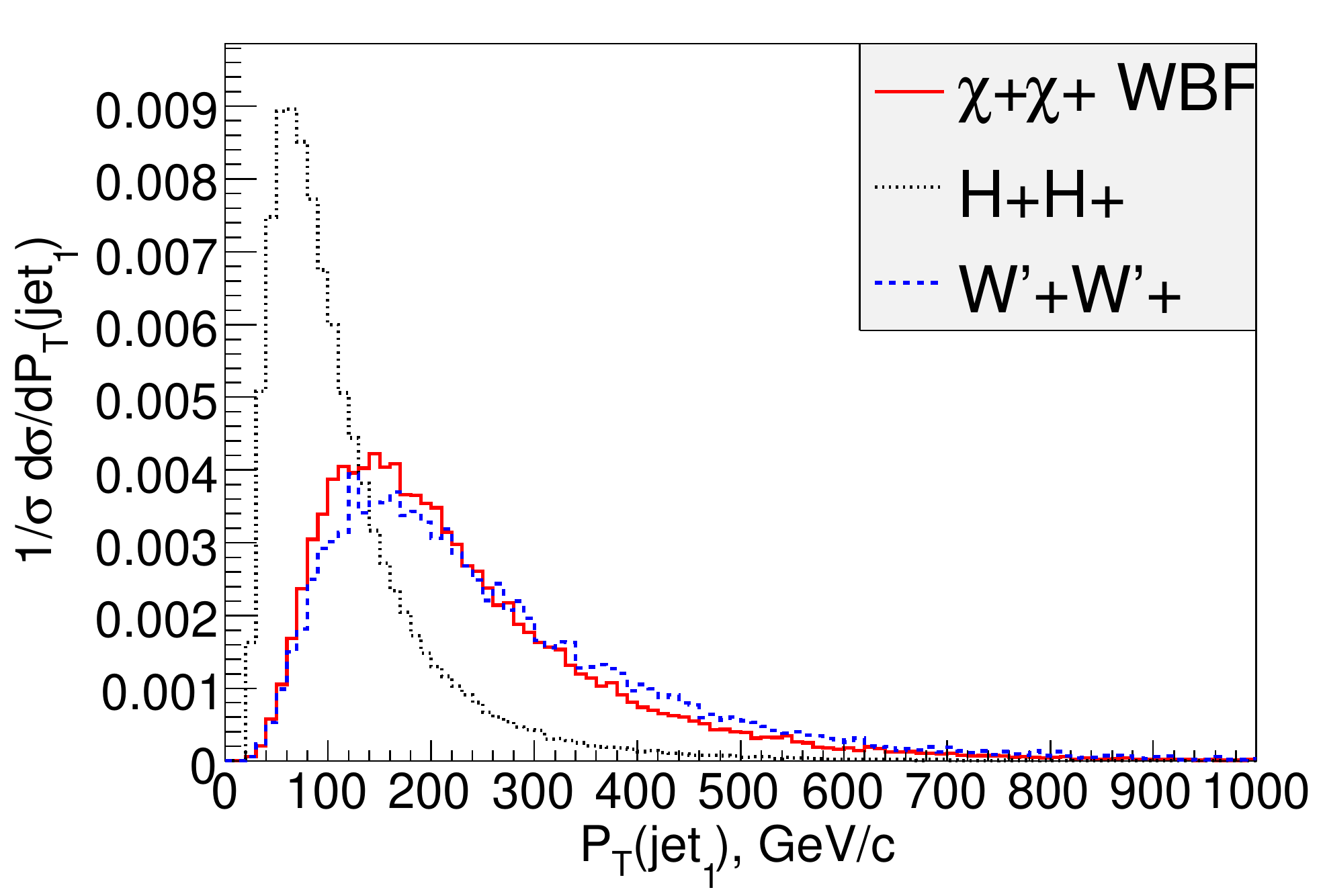}
  \hspace*{0.05\textwidth}
  \includegraphics[width=0.45\textwidth]{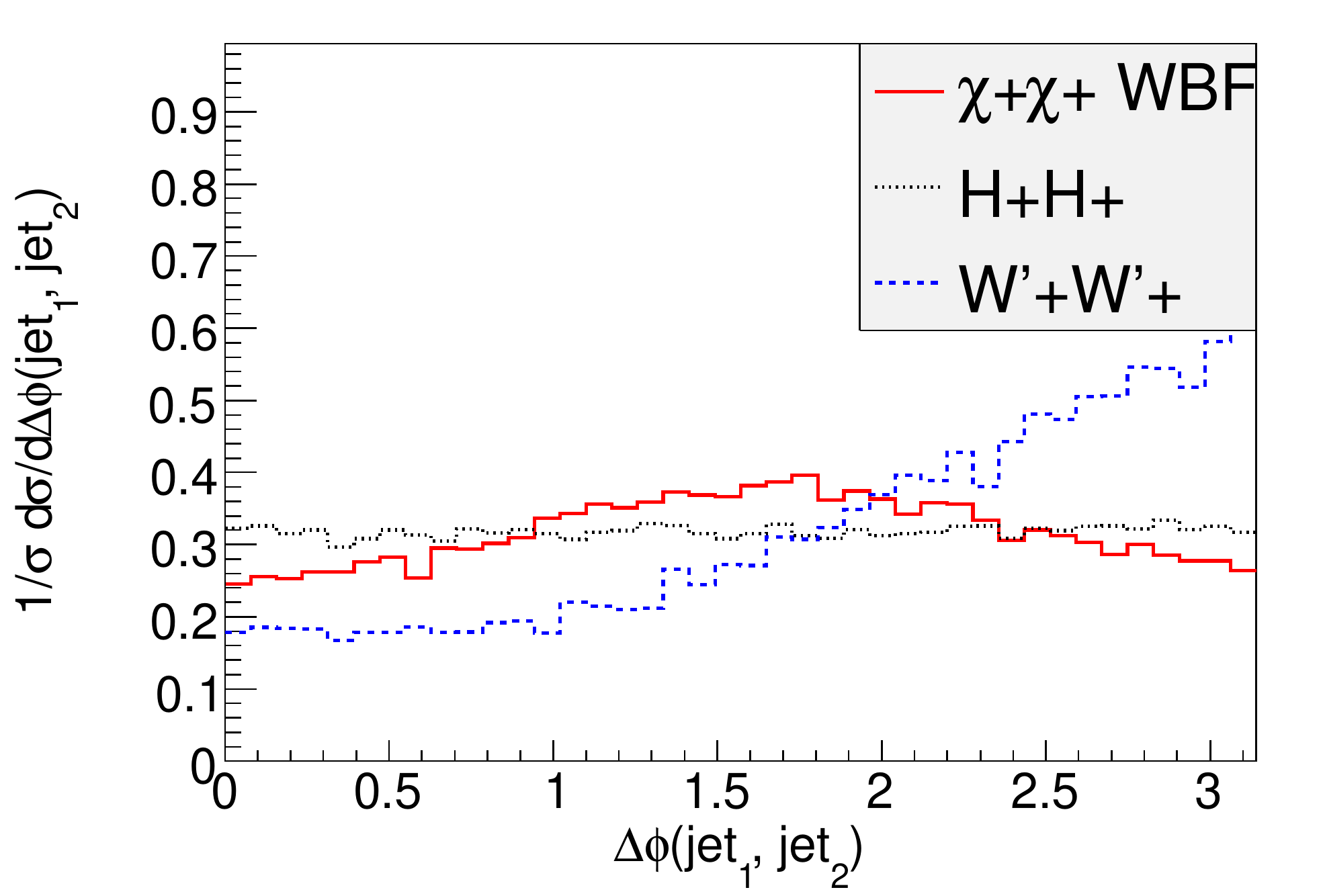}
\end{center}
\caption{Jet distributions in weak-boson-fusion. The production
  processes considered are like-sign supersymmetric charginos,
  like-sign Higgs-like scalars, and like-sign heavy gauge bosons with
  masses around 250~GeV. Figure from Ref.~\cite{wbf_coupling}.}
\label{fig:sig_wbf}
\end{figure}

The strategy applied for invisible Higgs searches can be generalized
to weak-boson-fusion production of any kind of system. Since the
number of possibilities to produce just one particle in $WW$ fusion is
limited by the $H$ and $Z$ production rates, we illustrate this for
pair production of new weakly interacting particles
Fig.~\ref{fig:sig_wbf}.  One physics motivation would be look for
like-sign particle production: in supersymmetric models this could be
like-sign charginos which are produced through a $t$-channel
neutralino exchange. Seeing such like-sign chargino pairs would imply
that the neutralinos in the $t$-channel (and possibly the dark matter
agent) is a Majorana fermion, one of the core predictions of minimal
supersymmetry.  On the other hand, such like-sign dilepton signatures
can come as well from the pair production of exotic heavy charged
gauge bosons $W'^+W'^+$ (that arise in UED and little Higgs models),
or from a pair of charged Higgs bosons $H^+H^-$ (as in supersymmetry
or little Higgs).  Note, however, that in many concrete models the
rates for these processes will be too small to be seen. From our
experience with the invisible Higgs search we can try to ask: can we
make any statements about the nature of particles produced in weak boson
fusion, but only based on the kinematics of the tagging jets, \ie
treating any kind of central particle production as `invisible'.

First of all, whatever particle(s) we produce in weak boson fusion
needs to couple to $W$ and/or $Z$ bosons. This coupling can involve
the transverse modes from the actual gauge sector or the longitudinal
Goldstone modes effectively from the Higgs sector. Radiating
transverse or longitudinal gauge bosons off quarks leads to different
energy and transverse-momentum spectra. In weak boson fusion, we can
measure this spectrum in the forward jets recoiling against the gauge
bosons; if the new physics produced couples to longitudinal $W$
bosons, like charged Higgs bosons will typically do, the transverse
momentum spectrum of the outgoing jets will be much stronger peaked,
as shown in Fig.~\ref{fig:sig_wbf}
\begin{alignat}{5}
P_T(x,p_T) &\sim \frac{1+(1-x)^2}{2 x} \;
                 \frac{p_T^2}{(p_T^2 + (1-x) \, m_W^2)^2} 
          &&\longrightarrow  
                 \frac{1+(1-x)^2}{2x} \;
                 \frac{1}{p_T^2} 
                 \notag \\
P_L(x,p_T) &\sim \frac{(1-x)^2}{x} \;
                 \frac{m_W^2}{(p_T^2 + (1-x) \, m_W^2)^2} 
          &&\longrightarrow  
                 \frac{(1-x)^2}{x} \;
                 \frac{m_W^2}{p_T^4} 
\end{alignat}

A second observable which is sensitive to the Lorentz structure of the
centrally produced new physics is the azimuthal angle between the
forward jets $\Delta \phi_{jj}$; as we know for example from
weak-boson-production of an invisibly decaying Higgs boson it has a
flat distribution for the production of a scalar particle coupling to
the two gauge bosons proportional to $g^{\mu \nu}$. For a
pseudo-scalar with a dimension-five coupling to gauge bosons instead,
it has a maximum at $\Delta \phi = \pi/2$, while for a scalar coupling
to the transverse tensor at higher dimensions it has a minimum at this
point~\cite{wbf_coupling,Hagiwara:2008iv}. Modulo a slight tilt due to
acceptance cuts we see precisely this behavior in
Fig.~\ref{fig:sig_wbf}, where the gauge bosons and fermions with their
non-trivial coupling to gauge bosons can be distinguished from the
pair of scalars.

The combination of these two jet distributions would therefore allow
us to distinguish between scalar vs fermion/gauge boson production
($p_{T,j}$) and between gauge boson vs scalar/fermion production
($\Delta \phi_{jj}$) without even looking at these new states.\bigskip

As a side remark, the angular observable can be linked to the `golden'
Higgs decay to two $Z$ bosons: the angle between the two $Z$ decay
planes is well known to include information on the spin of the Higgs
resonance~\cite{higgs_spin}. From this decay correlation we get to a
production jet correlation by switching two of the final-state
fermions to the initial state and the new (Higgs) particle into the
final state. The Feynman diagrams are identical to the Higgs decay,
and the correlation between the decay planes is replaced by an angular
correlation between the forward jets.\bigskip

The same analysis becomes even more interesting when we include the
case of KK gravitons, \ie spin-two particles produced in weak boson
fusion~\cite{mg_gravitons,Hagiwara:2008iv}. In the case of spin-zero
particles the production and decay of the heavy states are
uncorrelated which means that the only observable is the difference of
the azimuthal angles $\Delta \phi = \phi_1 - \phi_2$, and for a
$g^{\mu \nu}$ coupling its distribution has to be flat. Once we
introduce a dimension-five coupling proportional to the two allowed
$W$-$W$-scalar tensor structures~\cite{dieter_operators}
\begin{equation}
g^{\mu \nu} - \frac{q_1^\nu q_2^\mu}{(q_1 q_2)} \qquad \qquad \qquad
\epsilon^{\mu \nu \rho \sigma} \, q_{1 \rho} q_{2 \sigma}
\end{equation}
the $\Delta \phi$ distribution develops a shape, but it is still the
only combination of $\phi_1$ and $\phi_2$ we can define. The spin-one
hypothesis we usually skip because of the Landau-Yang theorem which
forbids at least an on-shell coupling of a heavy gauge boson. At the
spin-two level production and decay are not uncorrelated anymore,
which means we can probe the ($\phi_1 - \phi_2$) distribution of the
two jets as well as ($\phi_1 + \phi_2$). Ordered by the polarization
of the two incoming gauge bosons the actual spin-two production
dominates the amplitude and depends only on ($\phi_1 + \phi_2$), while
all other combinations including ($\phi_1 - \phi_2$) from the
spin-zero production at subleading~\cite{Hagiwara:2008iv}. In contrast to the
argument above this implies that to study gravitons in weak boson
fusion we cannot rely on jet-jet correlations anymore but have to
include the decay products in our analysis.

\subsection{Resonances}
\label{sec:sig_resonance}

  A standard method to find new physics at particle colliders is to
look for new resonances in the production rates of Standard Model
particle.  Such resonances arise in a number of models extending
the Standard Model and can produce a wide but related range of
signals at the LHC.  We discuss here some of the potential new resonance
signals that might be seen at the LHC.

\subsubsection{Massive neutral vector bosons}

\begin{figure}[t]
\begin{center}
  \includegraphics[width=0.50\textwidth]{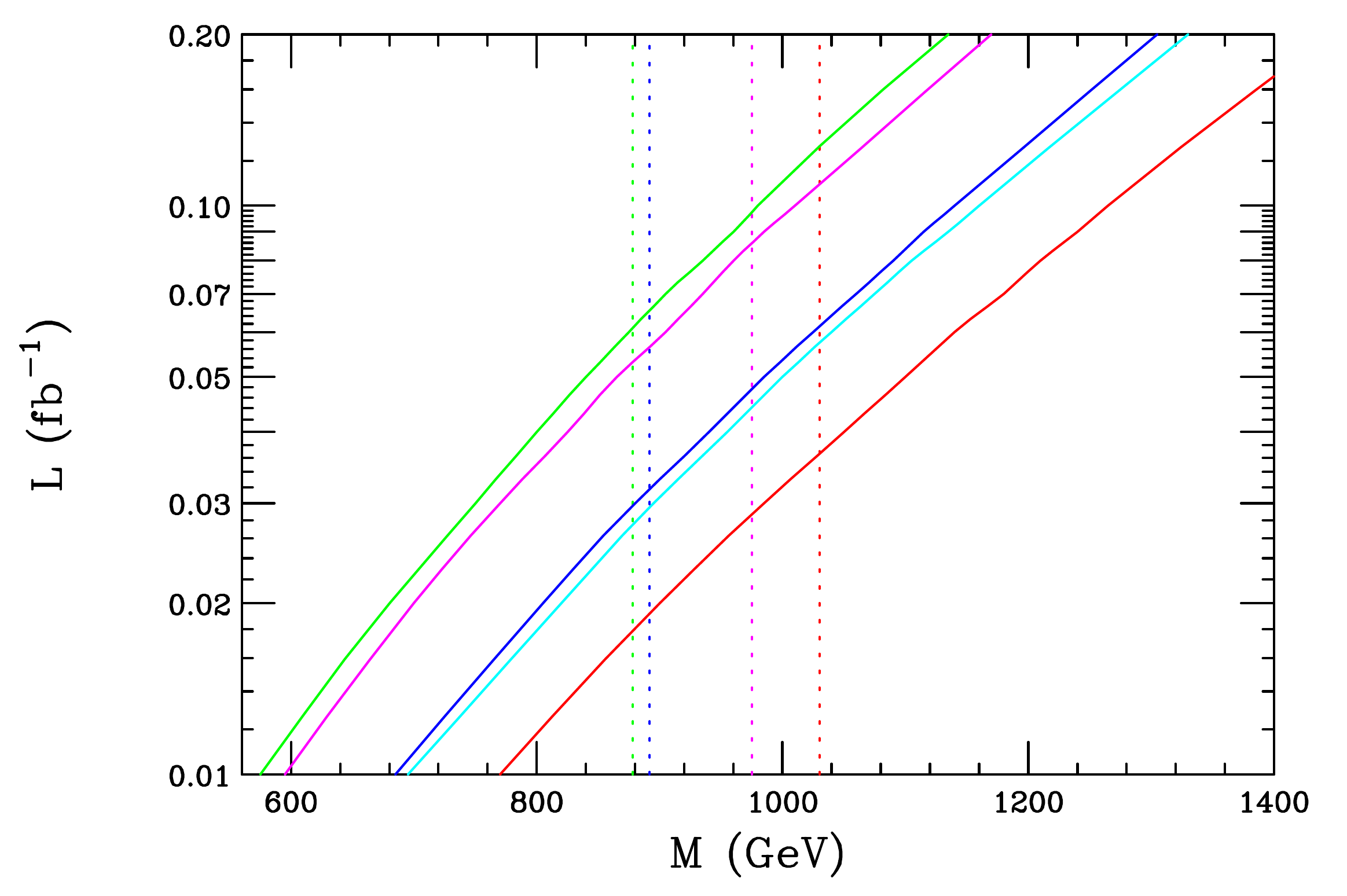}
\end{center}
\vspace*{-5mm}
\caption{Luminosity required for a $5\sigma$ LHC discovery of
  $E_6$-inspired $Z^\prime$ models (top to bottom: $\psi$, $\chi$,
  $\eta$, LRM model) as a function of $M_{Z^\prime}$.  From
  Ref.~\cite{Rizzo:2008cs}.}
\label{fig:zprdisc}
\end{figure}

Additional $U(1)$ gauge forces are a simple extension of the Standard
Model that occur in many theories of physics beyond the Standard Model.  
Sometimes a general resonance (particularly one decaying to leptons) 
is referred to as a $Z^\prime$; we reserve this designation for 
a massive neutral vector particle, with couplings both to quarks and 
to leptons. 

  The most common examples of $Z^\prime$s arise in extensions of the 
gauge sector of the Standard Model, such as $SU(3)_c \times SU(2)_W
\times U(1)_Y \times U(1)^\prime$.  More complicated constructions 
can have one or more of the Standard Model gauge factors enlarged 
to a bigger group containing an extra neutral generator.  This includes 
the case where the entire Standard Model gauge structure
is unified into a higher rank GUT group such as $SO(10)$.  In any of these
options, extra massive gauge fields arise from breaking these larger
symmetries down to the Standard Model itself.

 In this section we will take the $U(1)^\prime$ extension of the 
Standard Model as a prototype $Z^\prime$ model, with a well defined structure 
and parameters.  Larger gauge group cases fit comfortably within the
$U(1)^\prime$ framework, but with additional relations among
parameters and more new particles to search for.  Because of its
universality and the connection to larger and potentially unified
gauge groups, a $Z^\prime$ is one of the most extensively studied
extensions of the Standard Model in the
literature~\cite{Eichten:1984eu,Rizzo:2008cs,Carena:2004xs,Petriello:2008zr,Rizzo:2009pu}.
If it is not excessively massive and has a large coupling to quarks
and leptons, a $Z^\prime$ should lead to a relatively early discovery at the
LHC~\cite{Bauer:2009cc}.\bigskip

  The classic search strategy for a $Z^\prime$ is a resonance in the
$e^+ e^-$ or $\mu^+ \mu^-$ invariant mass spectra above the Drell-Yan
background.  The production rate for such a resonance therefore
depends on the coupling of the $Z^\prime$ to both quarks and leptons.
If we assume that the $Z^\prime$ has
family-independent couplings, motivated in order to evade strong
bounds from FCNCs, and no exotic states into which it can decay, the
physics at the LHC may be expressed in terms of its mass
$M_{Z^\prime}$, its width $\Gamma_{Z^\prime}$, and its couplings to
quark doublets, up-type singlets, down-type singlets, lepton doublets,
and charged lepton singlets $z_{Q}, z_u, z_d, z_{L}, z_e$.
LHC search prospects for a set of benchmark $E_6$ $Z^\prime$ models 
are shown in Fig.~\ref{fig:zprdisc}.  

\begin{figure}[t]
\begin{center}
 \includegraphics[width=0.35\textwidth,angle=90]{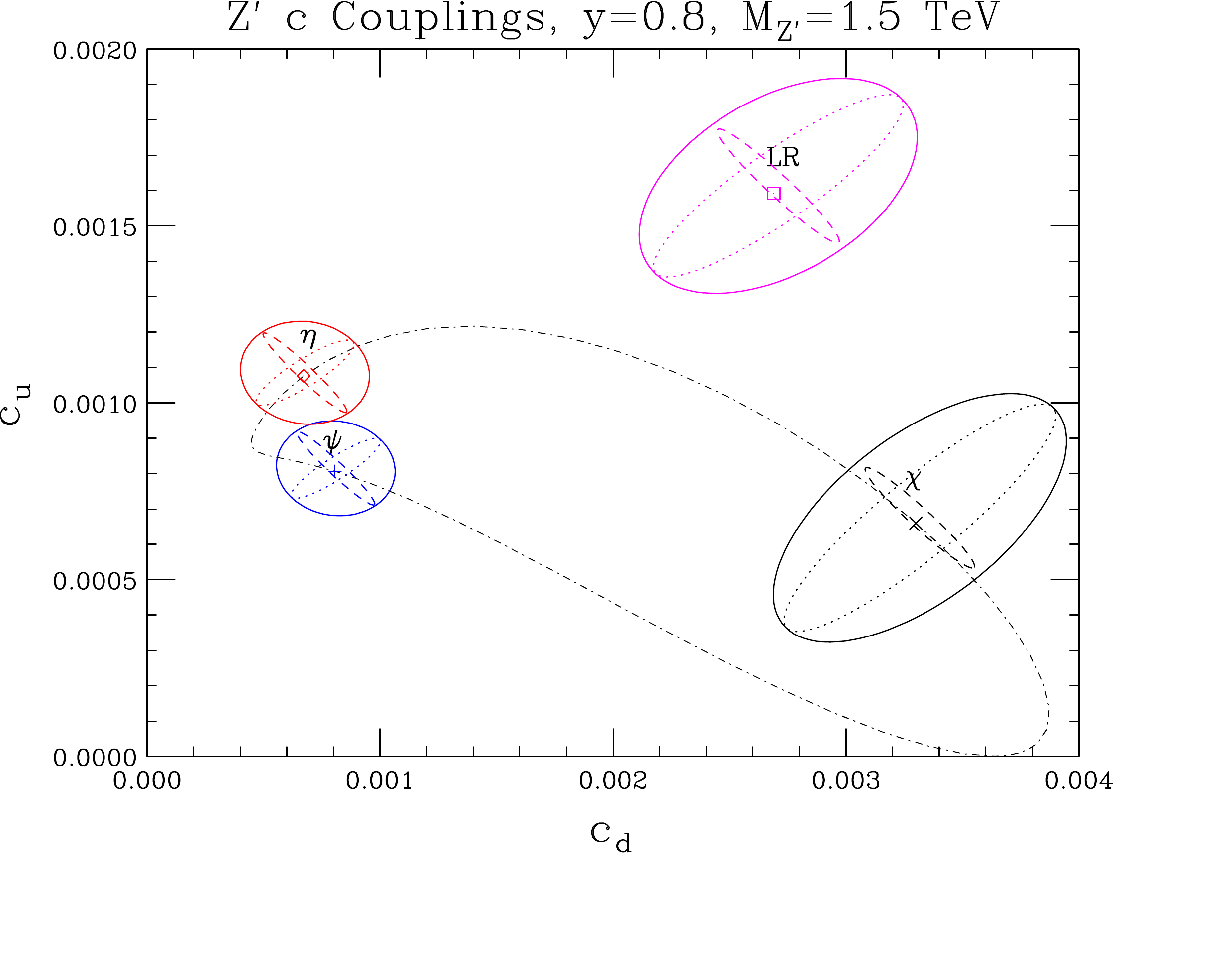}
 \hspace*{10mm}
 \includegraphics[width=0.35\textwidth,angle=90]{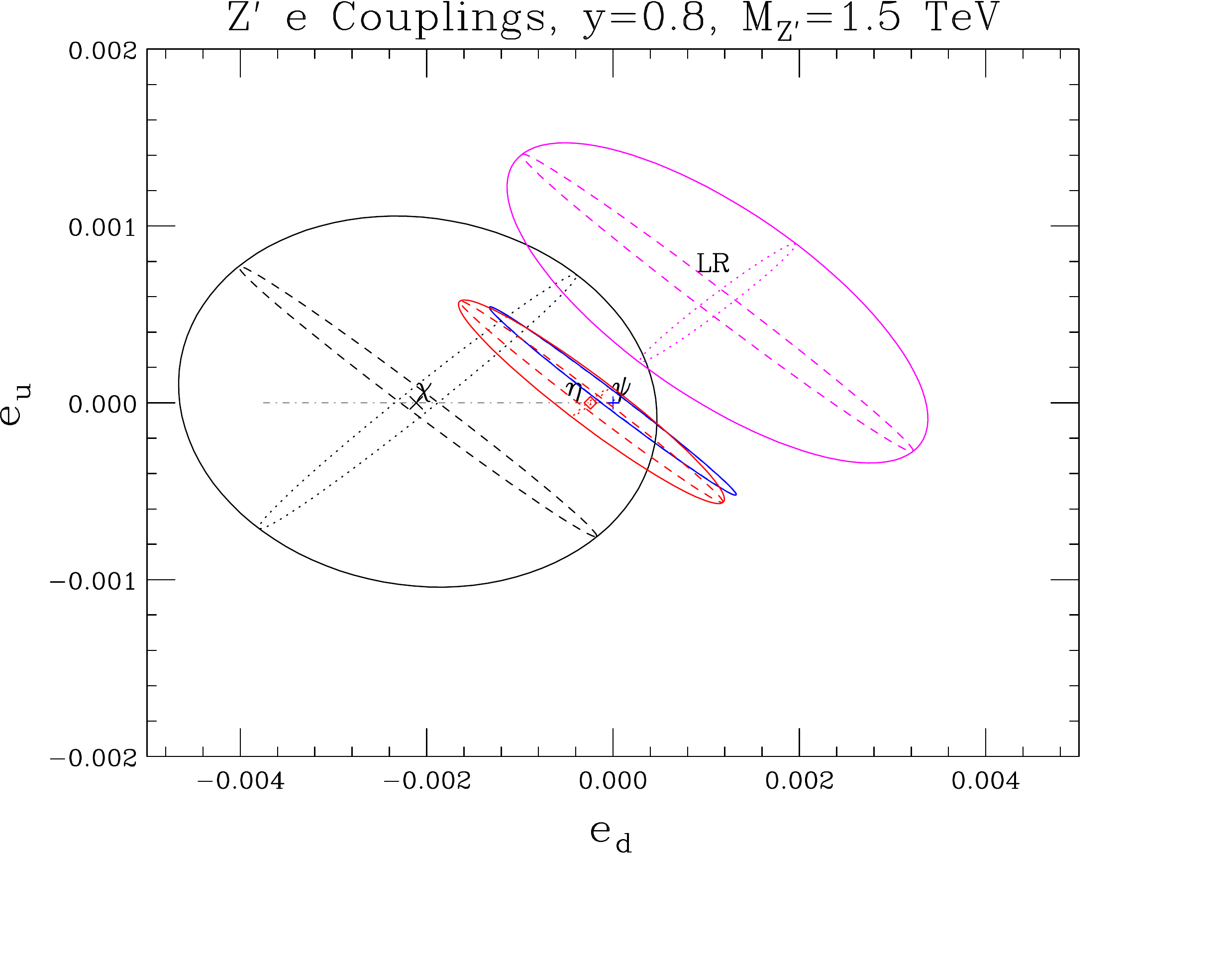}
\vspace*{-10mm}
\end{center}
\caption{$Z^\prime$ coupling measurements at the LHC with $100~\ifb$,
  for four different $E_6$-inspired $Z^\prime$ models with mass
  1.5~TeV.  Dashed lines are statistical errors and dotted lines are
  PDF uncertainties.  The solid lines are combined errors.  From
  Ref.~\cite{Petriello:2008zr}.}
\label{fig:zpcoup}
\end{figure}

  If a $Z^\prime$ signal is discovered at the LHC, further measurements
can help us to determine its fundamental properties.
A $Z^\prime$ dilepton resonance mass peak can be reconstructed with 
high precision.  The width of the peak may be also observable if it 
is larger than the lepton energy resolution, expected to be sub-percent 
at large invariant masses.  If the $Z^\prime$ does have a substantial width,
the mass estimate tends to be on the low side, because the PDFs distort
the resonance in favor of low invariant masses.
\bigskip

  The $Z^\prime$ production cross section provides further information,
although subject to the uncertainties discussed in Section~\ref{sec:sig_qcd}.
Under our assumption that the $Z^\prime$ couples universally to all fermions 
of a given charge and chirality, the cross-section is a PDF-weighted 
sum of two terms~\cite{Carena:2004xs},
\begin{equation}
c_{u,d}  \equiv  \frac{M_{Z^\prime}}{24 \pi \Gamma_{Z^\prime}}
\left( z_{Q}^2 + z_{u,d}^2 \right) \left( z_{L}^2 + z_{e}^2 \right)
= \left( z_{Q}^2 + z_{u,d}^2 \right) \text{BR}\left( Z^\prime \rightarrow \ell^+ \ell^- \right) .
\end{equation}
A cross section measurement then constrains one linear combination of
$c_u$ and $c_d$.  

  With larger and experimentally understood data samples, one can also 
consider the ratio of $Z^\prime$ production at large rapidities 
relative to central production.  At large rapidities, the
dependence on $c_u$ is enhanced due to the greater valence content 
of up-quarks in the proton, allowing us to determine $c_u$ and $c_d$
independently~\cite{Petriello:2008zr}.  A further parity-violating
combination~\cite{Petriello:2008zr},
\begin{equation}
e_{u,d} \equiv  \frac{M_{Z^\prime}}{24 \pi \Gamma_{Z^\prime}}
\left( z_{u,d}^2 - z_{Q}^2 \right) \left(  z_{e}^2 - z_{L}^2 \right),
\end{equation}
reflects itself in the forward-backward asymmetry of the produced
leptons.  Measurements of the forward-backward asymmetry at the LHC
are complicated by the fact that in principle one cannot tell the
direction of initial state quark.  However, we can use the fact that
the valence quarks have more support at higher partonic energy
fractions than the sea anti-quarks to reconstruct the quark direction
on a statistical basis~\cite{Dittmar:1996my}.  Figure~\ref{fig:zpcoup}
shows possible outcomes for the $E_6$-inspired $Z^\prime$ models with
mass 1.5 TeV, assuming $100~\ifb$ of integrated luminosity.
Further improvements are possible by combining LHC resonance data with
high luminosity low energy scattering measurements, and off-shell
$Z^\prime$ effects at the LHC~\cite{Petriello:2008zr,Rizzo:2009pu}.

\subsubsection{$SU(2)$ Gauge Extensions}

Larger gauge extensions of the SM than $U(1)$ are possible, for example $SU(2)$.  When
the extended gauge symmetry does not commute with the $SU(2)_L$ of the SM, some of the
resulting additional gauge bosons carry electroweak charge, leading to heavy 
electrically charged states such as e.g. $W^\prime$.  The detailed phenomenology of the
$W^\prime$s depends rather sensitively as to how the SM fermions are charged under the
additional $SU(2)$.

The most prevalent $SU(2)$ extension is $SU(2)_R$~\cite{Mohapatra:1974hk}.  In this 
construction, the right-handed quarks and leptons are combined to form $SU(2)_R$
doublets.  Such a gauge group is a fairly common stage from a single grand unified symmetry
to the Standard Model.  In a supersymmetric context, they are popular for
automatically enforcing $R$-parity conservation
and as constructions which lead naturally to a see-saw theory
of neutrino masses ~\cite{Mohapatra:1986su}.  Left-right symmetric models have an extremely
rich phenomenology (particularly supersymmetric constructions).  The most immediate
ingredients of interest at hadron colliders are the $W^\prime$ bosons, which have large
(electroweak size) couplings to SM quarks and leptons, and thus are 
readily produced as a resonance with large branching ratio into leptons, $\ell \nu_\ell$
or single top quarks, $t b$.
Tevatron direct searches and precision electroweak bounds require masses greater than about
800 GeV -- 1 TeV~\cite{Aaltonen:2009qu} and LHC searches are expected to probe masses
up to roughly 5 TeV at the LHC~\cite{Sullivan:2002jt}.

More general $SU(2)$ extensions may act differently on the various SM fermions.  
The un-unified model~\cite{Georgi:1989xz} 
replaces the $SU(2)$ of the SM with $SU(2)_1 \times SU(2)_2$,
where $SU(2)_1$ couples to leptons, and $SU(2)_2$ to quarks.  In the mass basis, this
results in the usual three light gauge bosons as well as an approximately degenerate set
of $W^{\prime \pm}$ and $Z^\prime$ which couple either preferentially to quarks or to leptons,
depending on the parameter choices.  Such a $W^\prime$ thus has both of the commonly
searched-for decay modes ($\ell \nu$ and $t b$)
of the left-right models, but with the branching ratio between them controlled by a free parameter.
Its couplings will also be left- rather than right-chiral, which will reflect itself, for example,
in the polarization of top quarks from the decay mode into $t b$.

A final class of $SU(2)$ extensions differentiate the various SM generations. 
Top-flavor~\cite{Li:1981nk} also replaces $SU(2)_{SM}$ with $SU(2)_1 \times SU(2)_2$,
where now $SU(2)_1$ couples to the 3rd family, and $SU(2)_2$ to
the first and second generations.  The result is a set of $W^\prime$ and $Z^\prime$ with
enhanced coupling to the third generation fermions, resulting in larger branching ratios
for $W^\prime \rightarrow \tau \nu_\tau$ and $W^\prime \rightarrow t b$.  Such models
contain non-perturbative effects which
lead to novel (but difficult to see experimentally against background) baryon-number
violating interactions at the LHC~\cite{Morrissey:2005uza}
which may further explain the baryon asymmetry of the Universe~\cite{jing}.

\subsubsection{Techni-hadrons}

\begin{figure}[t]
\begin{center}
 \includegraphics[width=0.51\textwidth]{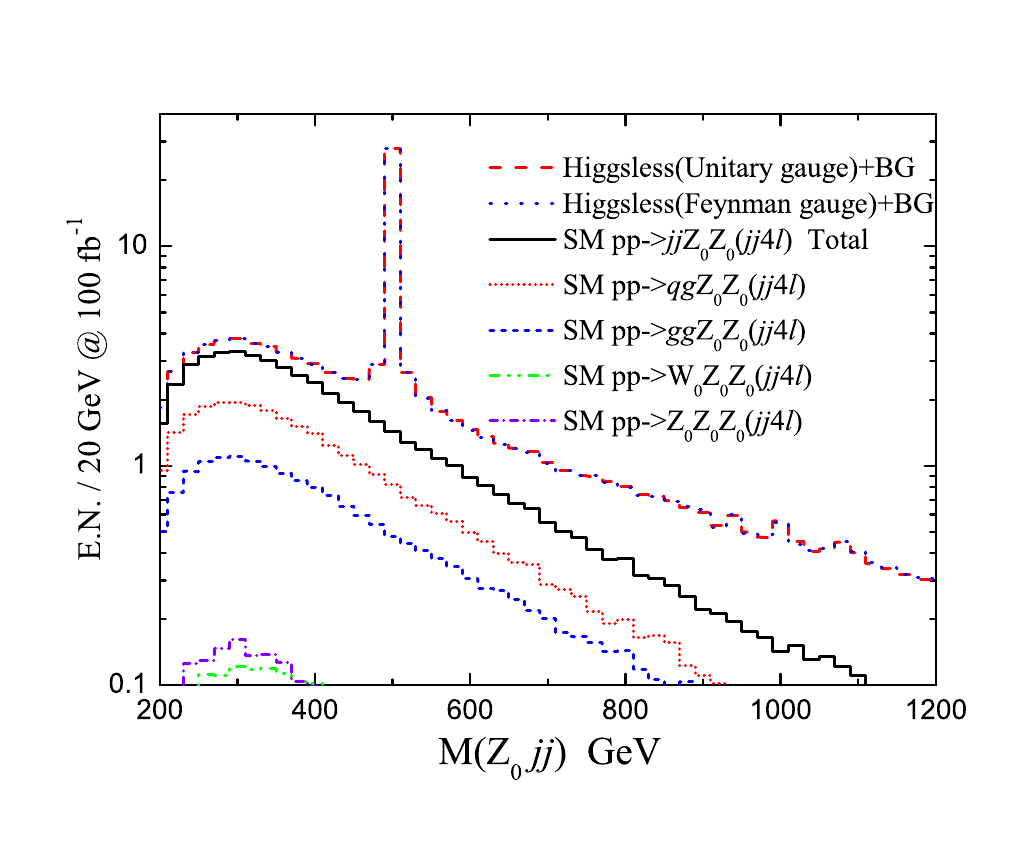}
 \hspace*{-0.90cm}
 \includegraphics[width=0.53\textwidth]{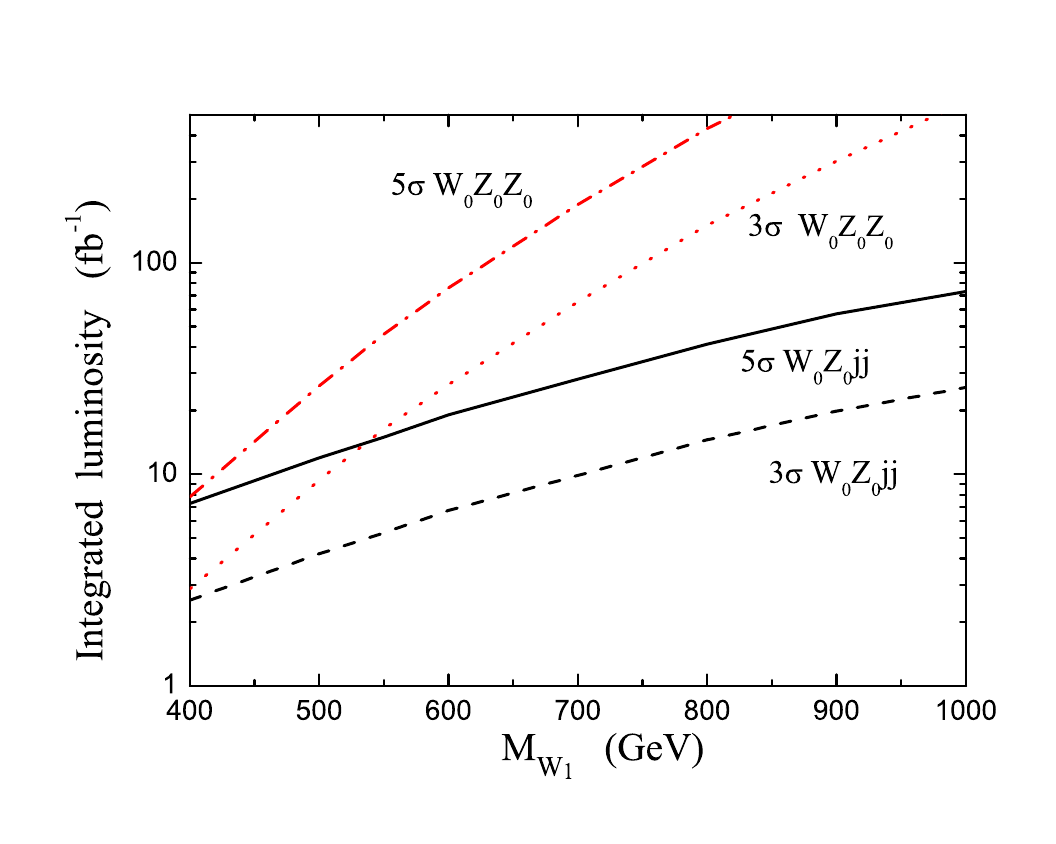}
\end{center}
\vspace*{-10mm}
\caption{Left: the invariant mass of $jj \ell^+ \ell^-$, showing how a
  peak from $\rho_T^\pm$ is visible above Standard Model backgrounds in the
  Drell-Yan-like process $pp \rightarrow W^* \rightarrow \rho_T^\pm
  Z$. Right: the luminosities required for $5\sigma$ and $3\sigma$
  signals as a function of the $\rho_T$ mass, in the Drell-Yan process
  (dotted and dot-dashed curves) and weak boson fusion production
  modes (solid and dashed curves).  From Ref.~\cite{He:2007ge}.}
\label{fig:technirho}
\end{figure}

Technicolor models lead to a rich spectrum of techni-hadrons, whose
properties reflect the details of the underlying model.  The most
general signatures arise from a weak isotriplet of vector particles
called techni-rhos, whose presence is required to at least
partially unitarize $WW$ scattering.  Most collider studies of the
$\rho_T$s~\cite{Eichten:1986eq} are in the context of the `technicolor
straw man model' implemented in Pythia~\cite{pythia_63}.  They may
have reasonably large couplings to Standard Model fermions, 
in which case they can
be caught by standard $Z^\prime$ or $W^\prime$ searches.  The dominant
decays are expected to be into a techni-pion $\pi_T$, which itself is
expected to decay largely into heavy fermion pairs plus an electroweak
boson, $W$, $Z$, or $\gamma$~\cite{Eichten:2007sx}.  ATLAS studies of
the channels 
\begin{alignat}{5}
p p &\rightarrow \rho_T^\pm \rightarrow \pi_T^\pm Z^0
     \rightarrow b q \ell^+ \ell^- \notag \\
p p &\rightarrow \rho_T^\pm \rightarrow \pi_T^0 W^\pm 
     \rightarrow b \bar{b} \ell^\pm \nu
\end{alignat}
indicate that $5\sigma$ discoveries are possible for masses in excess
of 800 GeV with $30~\ifb$~\cite{atlas_tdr}.\bigskip

Additional generic signatures of the $\rho_T$s consider no couplings
to light fermions, and perhaps no decay modes into techni-pions. They
result in objects whose relevant couplings are entirely to massive
vector bosons~\cite{He:2007ge,Bagger:1993zf,Hirn:2007we}.  In this
case, the techni-rho decays through gauge bosons, for example
$\rho_T^\pm \rightarrow W^\pm Z^0$. 
Such a $\rho_T^\pm$ can first of all be produced through a
Drell-Yan-like channel, $Z^0 \rho_T^\pm$, resulting in a $W^\pm Z^0
Z^0$ final state. Alternatively, we can study the weak boson fusion
process $pp \rightarrow \rho_T^\pm jj$, resulting in a $W^\pm Z^0$
plus forward jets final state.  In Fig.~\ref{fig:technirho}, we show a
simulation of the distribution of the invariant mass of $\rho_T^\pm
\rightarrow W^\pm Z^0 \rightarrow jj \ell^+ \ell^-$ events at the
parton level, with the $\rho_T$ produced in the Drell-Yan process and
the discovery luminosity as a function of
the $\rho_T$ mass in the Drell-Yan and weak boson fusion
channels~\cite{He:2007ge}.
If heavy enough, such resonances may also have large decays into top quarks, leading
to visible resonances in the invariant mass of $t \bar{t}$~\cite{Evans:2009ga}.

\subsubsection{Heavy Kaluza-Klein modes}

\begin{figure}[t]
\begin{center}
 \includegraphics[width=0.45\textwidth]{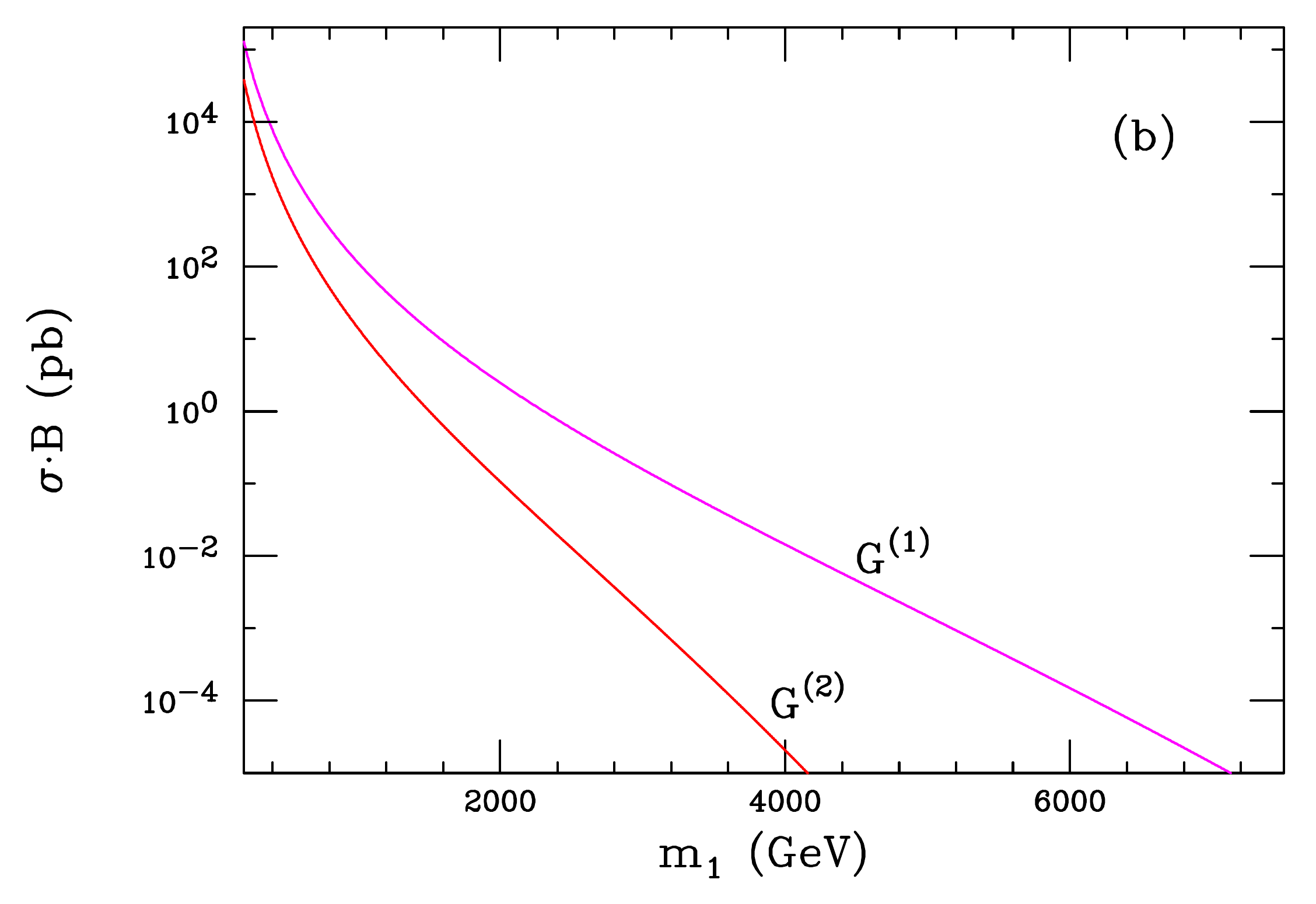}
 \hspace*{0.05\textwidth}
 \includegraphics[width=0.45\textwidth]{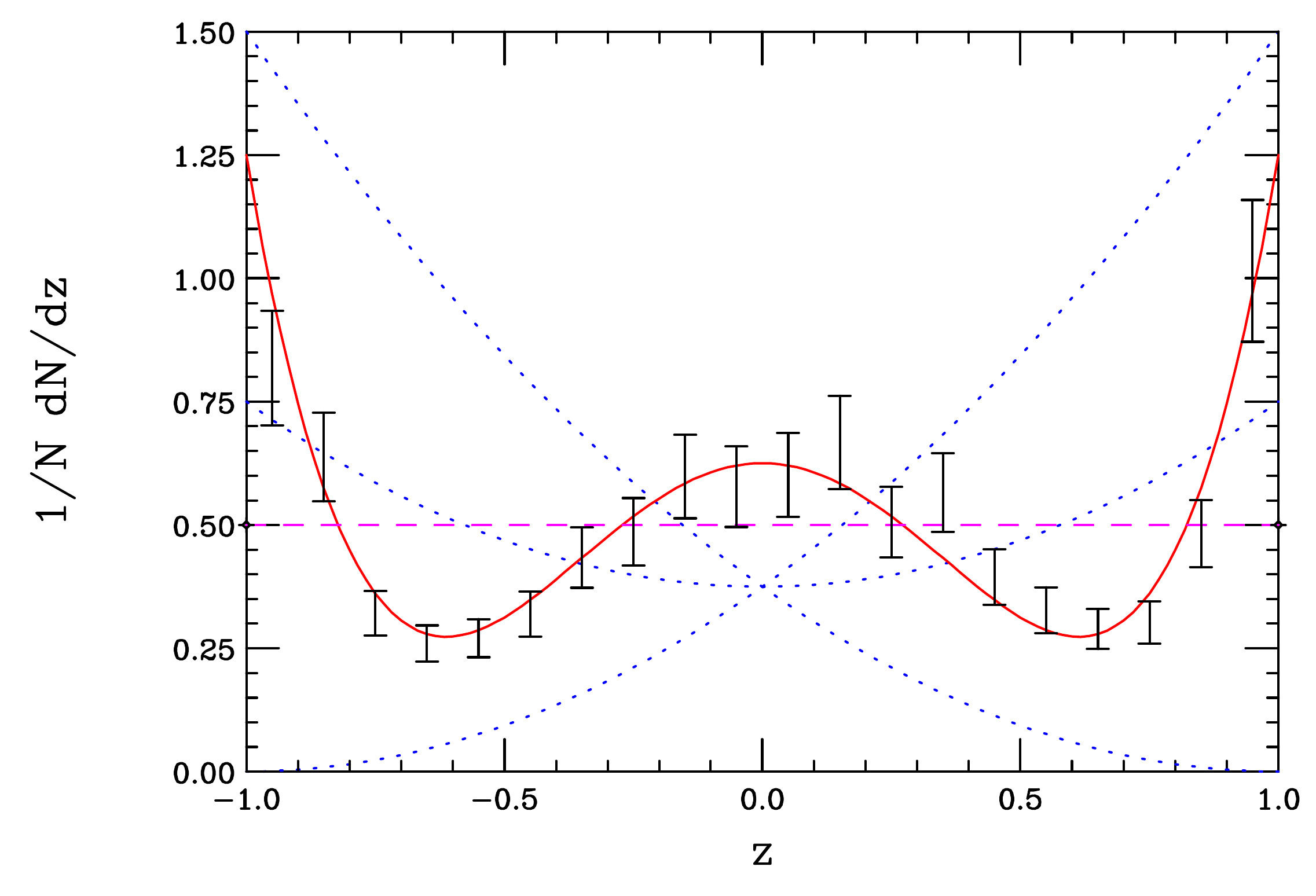}
\end{center}
\vspace*{-5mm}
\caption{Left: LHC cross sections times branching ratios into $\ell^+
  \ell^-$ for the first and second graviton KK modes in an RS model
  with $k / \bar{M}_\text{Pl} = 0.1$. Right: angular distribution of
  $\ell^+ \ell^-$ for a spin-2 KK graviton (solid), spin~0 boson
  (dashed), and three different chiral couplings of spin~1 vectors
  (dotted).  Also shown are simulated data points with error bars
  corresponding to a 4.2~TeV graviton with $100~\ifb$. From
  Ref.~\cite{Davoudiasl:2000wi}.}
\label{fig:rsgrav}
\end{figure}

  In Randall-Sundrum models, KK graviton excitations arise at 
the LHC as spin-2 resonances produced by $gg$ or $q \bar{q}$ fusion.
They decay into pairs of jets, leptons, and Higgs and weak bosons 
including photons. The relative branching ratios depend on where the 
Standard Model zero-mode wave functions have support in the extra dimension.
When the Standard Model is confined to the IR brane, the couplings
are democratic.  On the other hand, when the Standard Model
fields propagate in the bulk, the KK gravitons couple most strongly
to fields localized toward the IR brane.  In this case, there
also arise spin-1 KK excitations of the Standard Model gauge bosons
which we will discuss in Section~\ref{sec:sig_top}.

  When the branching ratios into leptons are substantial enough, 
RS KK gravitons may be visible through a standard $Z^\prime$ dilepton search.  
If the KK scale is low enough, more than one KK level may be visible.  
LHC cross sections and branching ratios into leptons for the first two
KK modes are shown in the left panel of Fig.~\ref{fig:rsgrav}, 
where it is assumed that the Standard Model fields are confined to the
IR brane and that the curvature is given by
$k / \bar{M}_\text{Pl} = 0.1$.

  Spin-2 resonances lead to a distinctive angular distribution of their
fermionic decay products~\cite{Davoudiasl:2000wi,Cousins:2005pq,Osland:2008sy,Boudjema:2009fz,Hagiwara:2008iv}.
The right panel of Fig.~\ref{fig:rsgrav} illustrates the
distributions for a spin-2 resonance, a scalar resonance, and three
different vector resonances with left-chiral, right-chiral, and
vector-like coupling to the fermions, and shows simulated data
corresponding to a 1000 events of a massive spin-2 particle's decay
into leptons.\bigskip

\begin{figure}[t]
\begin{center}
 \includegraphics[width=0.47\textwidth]{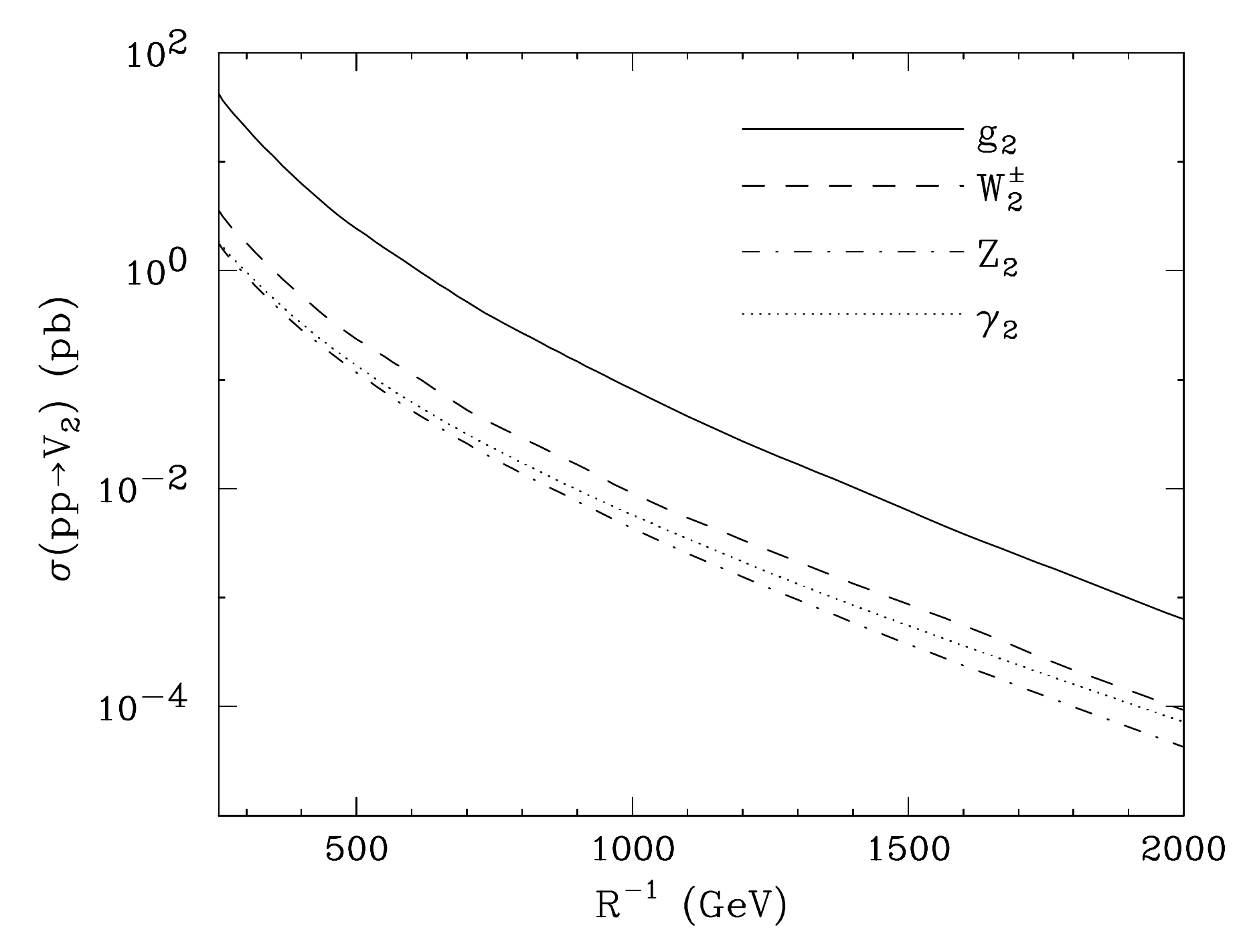}
 \includegraphics[width=0.435\textwidth]{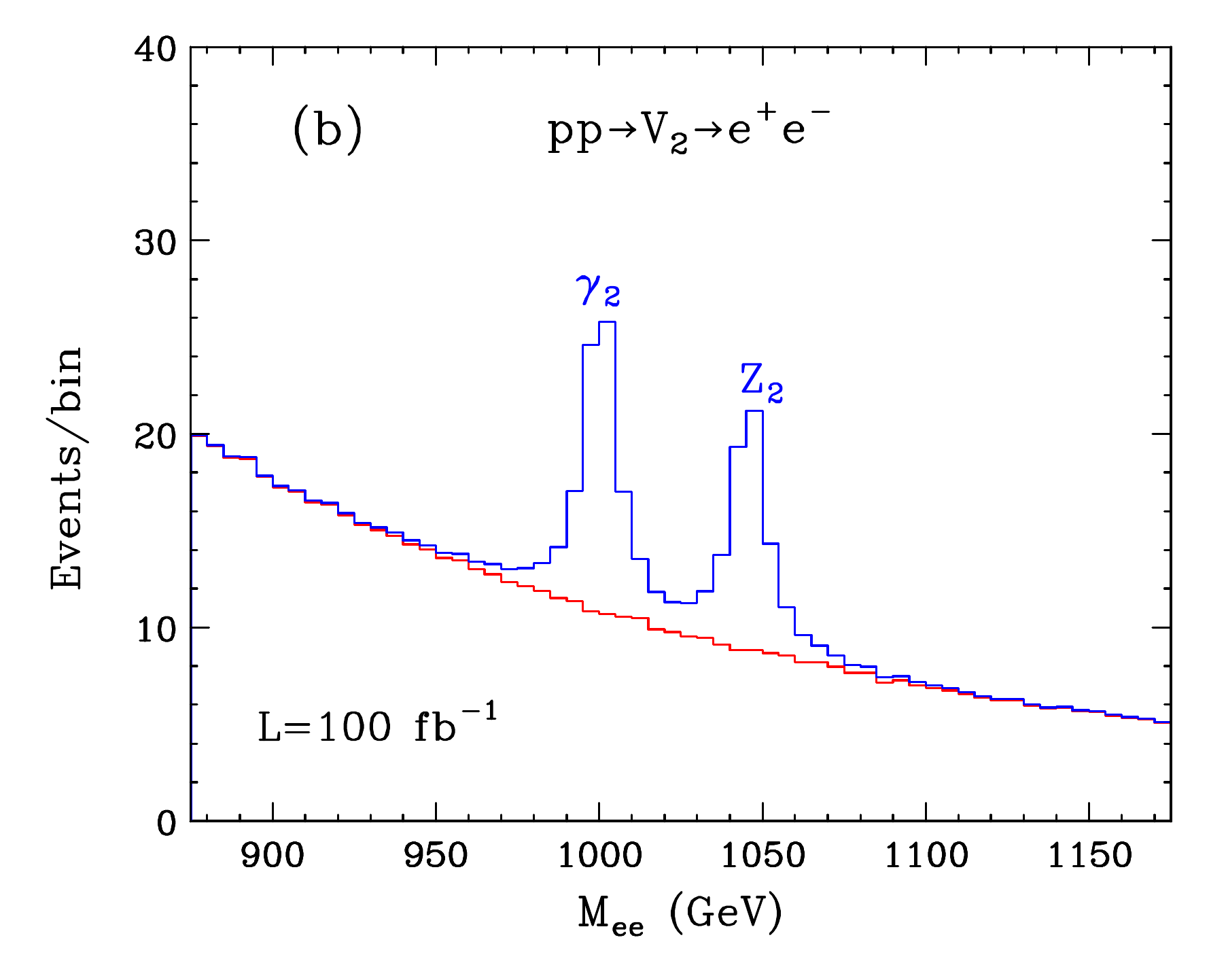}
\end{center}
\caption{Left: cross sections for single production of level-2 KK
  modes of the gauge bosons in the minimal UED 5D model as a function
  of the size of the extra dimension.  Right: invariant mass of $e^+
  e^-$ pairs for $R^{-1} = 500$~GeV with $100~\ifb$ of collected
  luminosity, showing clear separation between the $\gamma^{(2)}$ and
  $Z^{(2)}$ KK modes.  From Ref.~\cite{Datta:2005zs}.}
\label{fig:uedlvl2}
\end{figure}

In models with universal extra dimensions~(UED), the higher KK modes 
of Standard Model particles are even
under KK parity, and couple to zero-mode fields through boundary
operators.  As a result, they can be singly produced at the LHC
through processes such as $q \bar{q} \rightarrow G^{(2)},
\gamma^{(2)}, Z^{(2)}$, where the index $(2)$ can be replaced by any
KK-even state index whose mass is accessible at the LHC.  For the
chiral square, the lightest of these are the $(1,1)$ modes.

  While the $G^{(2)}$ decays into jets, and is challenged by the large
dijet backgrounds, the $\gamma^{(2)}$ and $Z^{(2)}$ states have
significant branching ratios into leptons, and would be detected by a
standard $Z^\prime$ search.  In some cases, decays into pairs of the
lightest level of KK-odd modes will also be open, and if so, tend to
dominate because the $(2)$-$(1)$-$(1)$ coupling is through bulk
physics as opposed to boundary terms.  In Fig.~\ref{fig:uedlvl2} we
show the cross sections for single production of the level-2 KK modes
of the gauge bosons in the 5D UED theory, assuming minimal boundary
terms~\cite{Datta:2005zs}. That study concluded that both resonances
can be discovered with $300~\ifb$ up to masses of about~2 TeV, or
$R^{-1}$ of about 1~TeV.\bigskip

\subsubsection{$R$-parity violation and leptoquarks}
\label{sec:sig_lepto}

Leptoquarks are fundamental particles which carry baryon and lepton
numbers~\cite{Buchmuller:1986zs,lq_tc}. They come as scalars or vector
particles, reflecting the Lorentz structure of their
coupling to one quark and one lepton. Their baryon and lepton numbers
can either be aligned, such they decay to $q \ell^+$ or
anti-aligned, which means they decay to $q \ell^-$. 

Most importantly, the couplings to quarks and leptons do not have to
be family-universal. This classification can best be seen in a
supersymmetric realization of scalar leptoquarks as $R$-parity 
violating squarks.  Their lepton couplings arise from the
superpotential terms
\begin{equation}
 - W \supset \lambda'_{ijk} L_i\ccdot Q_j {D}^c_{k} \; ,
\end{equation}
where ($i,j,k$) are generation indices~\cite{Barger:1989rk}. 
For scalar leptoquarks decaying to $q\ell^+$, the coupling matrix 
$\lambda'$ includes all possible cases. Including a further $R$-parity
and baryon number violating superpotential coupling of the form 
$\lambda'' {U}^c{D}^c{D}^c$ would generate rapid proton decay through
the combination $\lambda' \ccdot \lambda''$, 
which is why in the MSSM we require $R$ parity
conservation.\bigskip

  However, the bounds exclusively on $\lambda'$ are considerably weaker,
which means that such an operator does not have to be forbidden by an
exact symmetry.  The main constraints come from generic limits on
four-fermion operators, for example from flavor-changing neutral
currents. Including the mass scale $M$ of the leptoquark these limits
scale like $(\lambda'\ccdot\lambda')/M$, where the combination of
coupling matrices is really a matrix multiplication. Limits from rare
decays as well as from HERA scale like $(\lambda' \ccdot
\lambda')/M^2$~\cite{Plehn:1997az}, as do limits from atomic parity
violation~\cite{lq_limits,rpv_limits}. For the pair production of
leptoquarks at hadron colliders only their mass is relevant --- the
coupling to leptons comes in when we consider decay channels which
should easily be detected at the LHC~\cite{lq_lhc}.

While HERA with its $e p$ initial state was designed to search for
leptoquarks in the $s$ channel, its reach is roughly matched by the
Tevatron, where both proton beams have similar energies. Modern
multi-purpose hadron colliders are indeed perfectly suited to search
for new states as long as they carry color charge.  As far as
leptoquark signatures are concerned, two leptons plus two hard jets,
reconstructing two identical leptoquark mass peaks are a perfect
hadron collider signature, both from a triggering and a reconstruction
point of view. Only $Z'$ searches require less of an understanding of
QCD at the LHC.

\subsubsection{Multi-jet resonances}

\begin{figure}[t]
\begin{center}
  \includegraphics[width=0.48\textwidth]{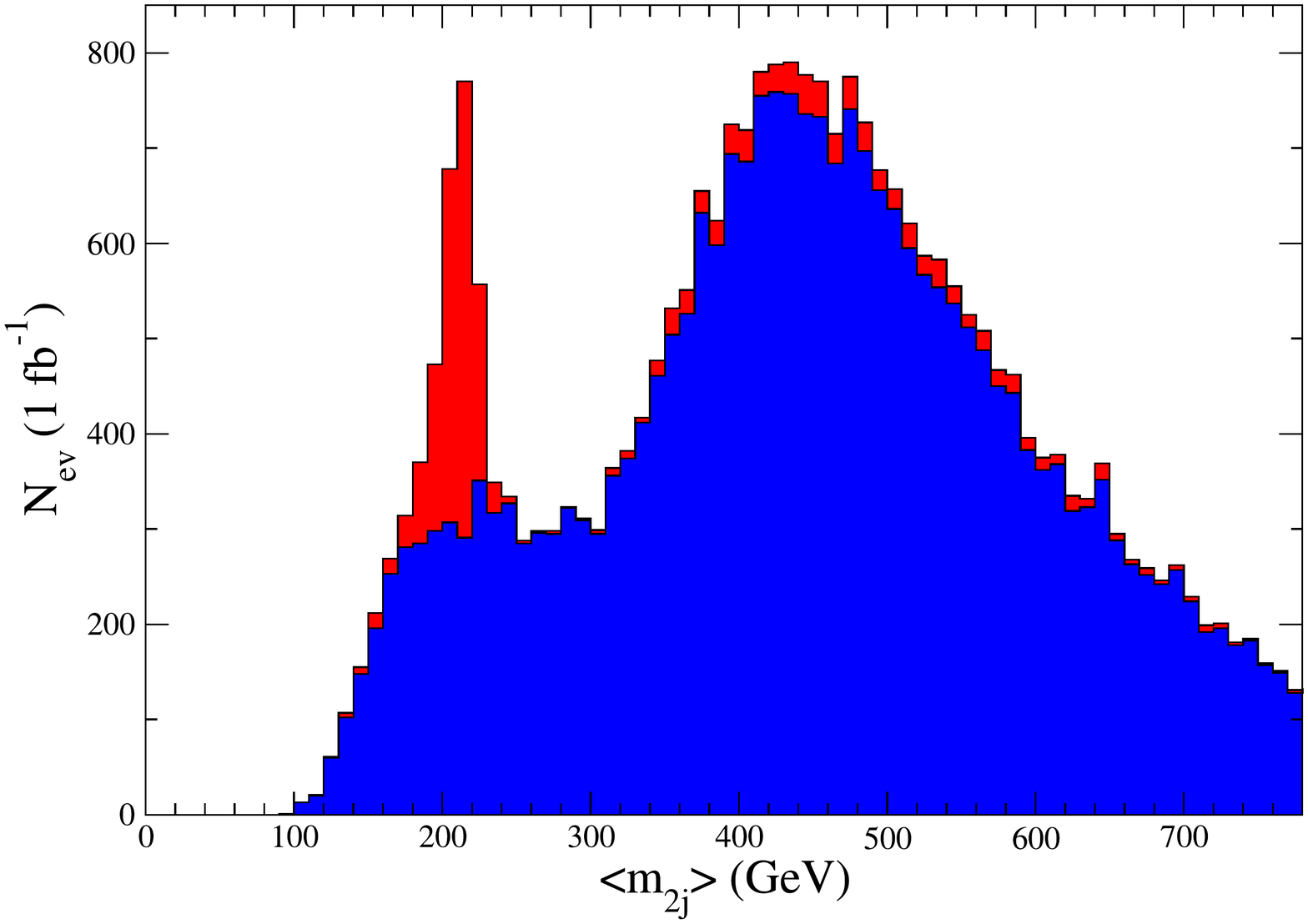}
  \hspace*{0\textwidth}
  \includegraphics[width=0.48\textwidth]{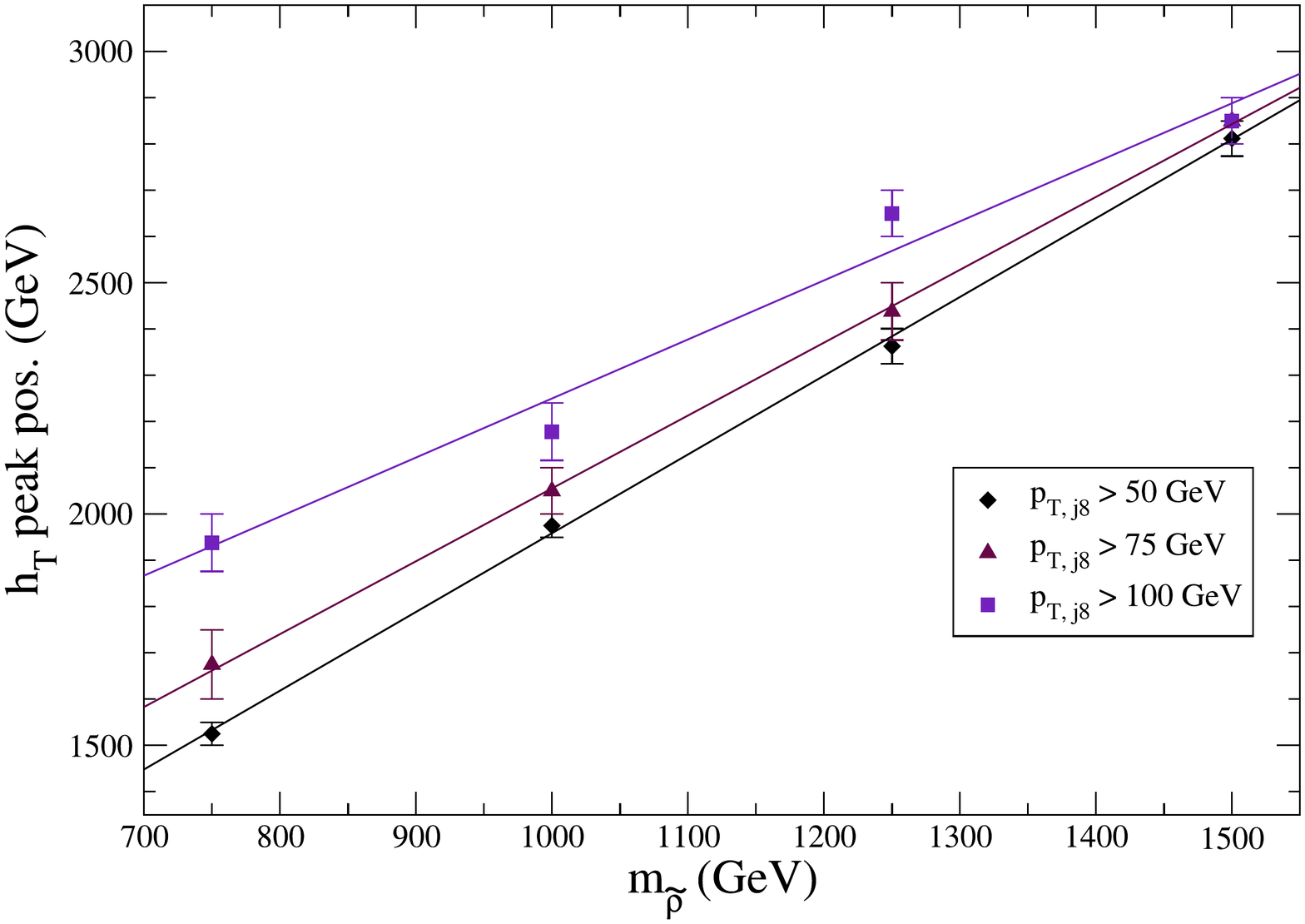}
\end{center}
\caption{Reconstruction of the hyperpion and coloron masses in the
  four-jet and eight-jet channels. The masses in the left panel are
  225~GeV for the hyperpion and 750~GeV for the coloron. Figures from
  Ref.~\cite{steffen_hyper}.}
\label{fig:sig_hyper}
\end{figure}

One of the most important questions concerning resonances at the LHC
is if and how well we can determine the masses of heavy states
decaying exclusively to jets. In supersymmetry or other models with dark matter candidates, large missing
energy with its usual advantages and disadvantages will usually accompany such multi-jet events.
Instead, we now introduce a set of
fermions which are charged under $SU(3)_c$ as well as a `hypercolor'
gauge group with a confinement scale
$\Lambda_\text{HC}$~\cite{raman_hyper}. At low energies we would see a
$SU(3)_c$ adjoint scalar hyperpion as well as a $SU(3)_c$ adjoint
vector-like coloron. Just like in QCD the hyperpion is expected to be
the lighter of the two states. The corresponding Lagrangian
\begin{alignat}{5}
 \mathcal{L}^\text{HC}
  \sim& - g_3 \; \bar{q} \gamma^{\mu} \varepsilon \rho_{\mu} q \notag \\
  &+i\chi g_3 \; \tr \left( G_{\mu\nu} \left[ \rho^\mu,\rho^\nu \right] \right)
   +i g_3^2 \xi \; \frac{\sqrt{N_\text{HC}}}{2 \pi m^2_\rho} \;
    \tr \left( \rho^\mu_\nu \left[G^\nu_\sigma,G^\sigma_\mu \right]
        \right) \notag \\
  &-g_{\rho\pi\pi} f^{abc} \rho^a_\mu \pi^b \partial^\mu \pi^c
   -\frac{3g_3^2}{16\pi^2 f_\pi}
     \tr \left[ \pi G_{\mu\nu} \widetilde{G}^{\mu\nu} \right]
   + \cdots
\end{alignat}
includes a color coupling to quarks ($\varepsilon \sim 0.2$) and to
gluons, assuming $\chi=1$ and $\xi=0$. It also includes a decay
coupling of the heavier coloron to two hyperpions and the hyperpion
coupling to gluons. The Signatures relevant at the LHC in this toy model
are~\cite{steffen_hyper}
\begin{alignat}{5}
pp \to \pi \pi \to 4~\text{jets} 
\qquad \qquad \qquad
pp \to \rho \rho \to 4~\pi \to 8~\text{jets} \; .
\end{alignat}
In Fig.~\ref{fig:sig_hyper} we see that extracting the hyperpions with
their fairly low mass poses no problem to the LHC. The analysis is
based on the four hardest jets and a pairing criterion which forces
two combinations of two jets each to reconstruct the same invariant
mass within 50~GeV.

The problem becomes harder once we try to extract the coloron as
well. Extending the left panel of Fig.~\ref{fig:sig_hyper} to include
also the four-jet invariant mass we can check that there is no peak at
the coloron mass, because most hyperpions are generated as a continuum
in gluon fusion. Instead, we have to search for coloron pair
production in the eight-jet channel, with eight staggered $p_{T,j}$
cuts between 320 and 40~GeV. From Section~\ref{sec:sig_qcd} we know
that we should expect this softest jet to already be affected by QCD
jet combinatorics. The coloron signal we again extract through pairing
criteria, assuming we know the hyperpion's mass. In the right panel of
Fig.~\ref{fig:sig_hyper} we see that the peak of the inclusive $H_T$
with optimized $p_{T,j}$ cuts indeed traces the coloron masses.

\subsection{Top signatures}
\label{sec:sig_top}

The top quark in the Standard Model plays a special role in several
ways. From an experimental point of view it is the only quark which
decays before it can hadronize~\cite{bigi_zerwas}. Its decays produce both
leptonic and hadronic structures, which can be separately reconstructed
as jets, leptons, and missing energy. These are precisely the
structures used to find the Higgs boson or new physics at the
LHC. This means that the top quark is a laboratory for many
experimental or simulational aspects which must be controlled
in searches for new physics

As discussed in Section~\ref{sec:why_hierarchy}, from a theory point
of view the top quark, as the Standard Model particle with the strongest Higgs
coupling, gives the largest contribution to the quadratic divergence in
the Higgs mass parameter.  As a result, naturalness suggests that
there exists a top partner sector whose masses must be close to
1~TeV. Thus, taking naturalness seriously,  
light top partners are one of the most robust
predictions of solutions for the hierarchy problem. The determination
of their properties is one of the primary means by which 
the different proposed solutions can be discriminated.

\subsubsection{Top Quarks from Cascade Decays}

\begin{figure}[t]
\begin{center}
 \includegraphics[width = 0.43\textwidth]{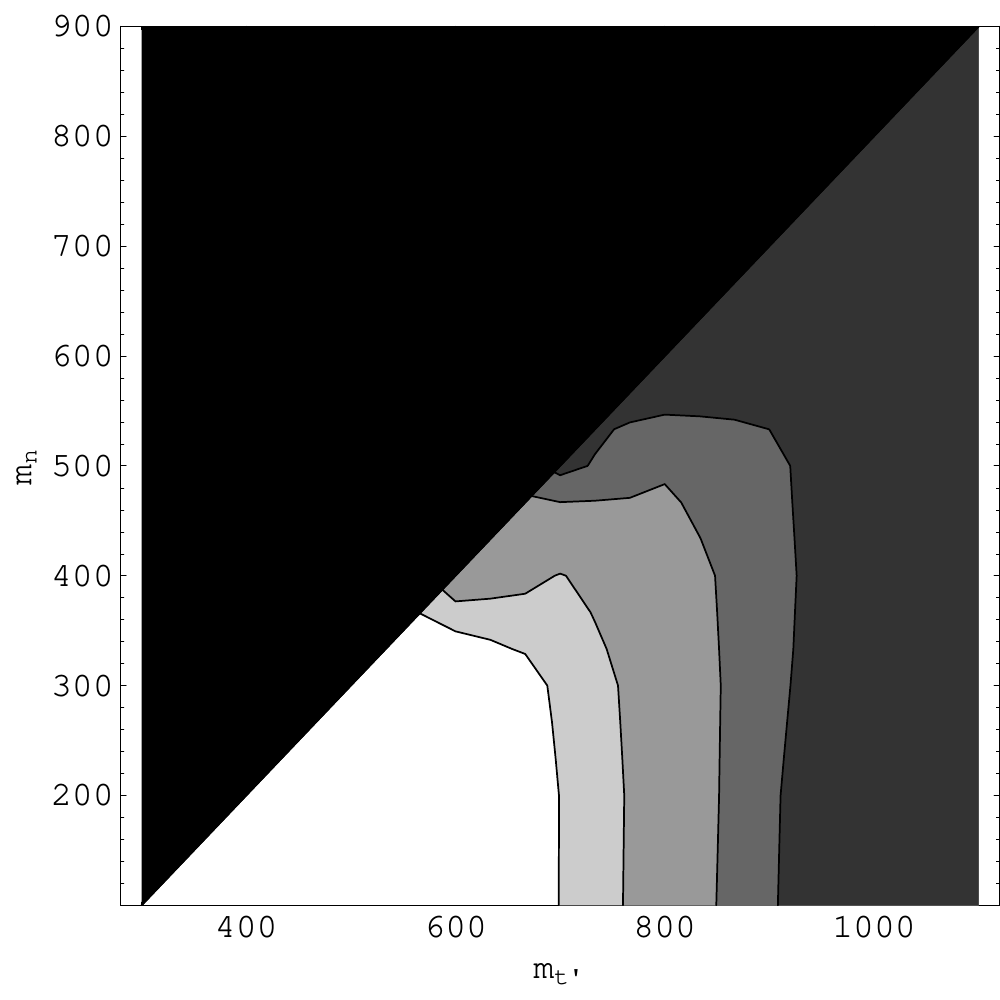}
 \hspace*{12mm}
 \includegraphics[width = 0.43\textwidth]{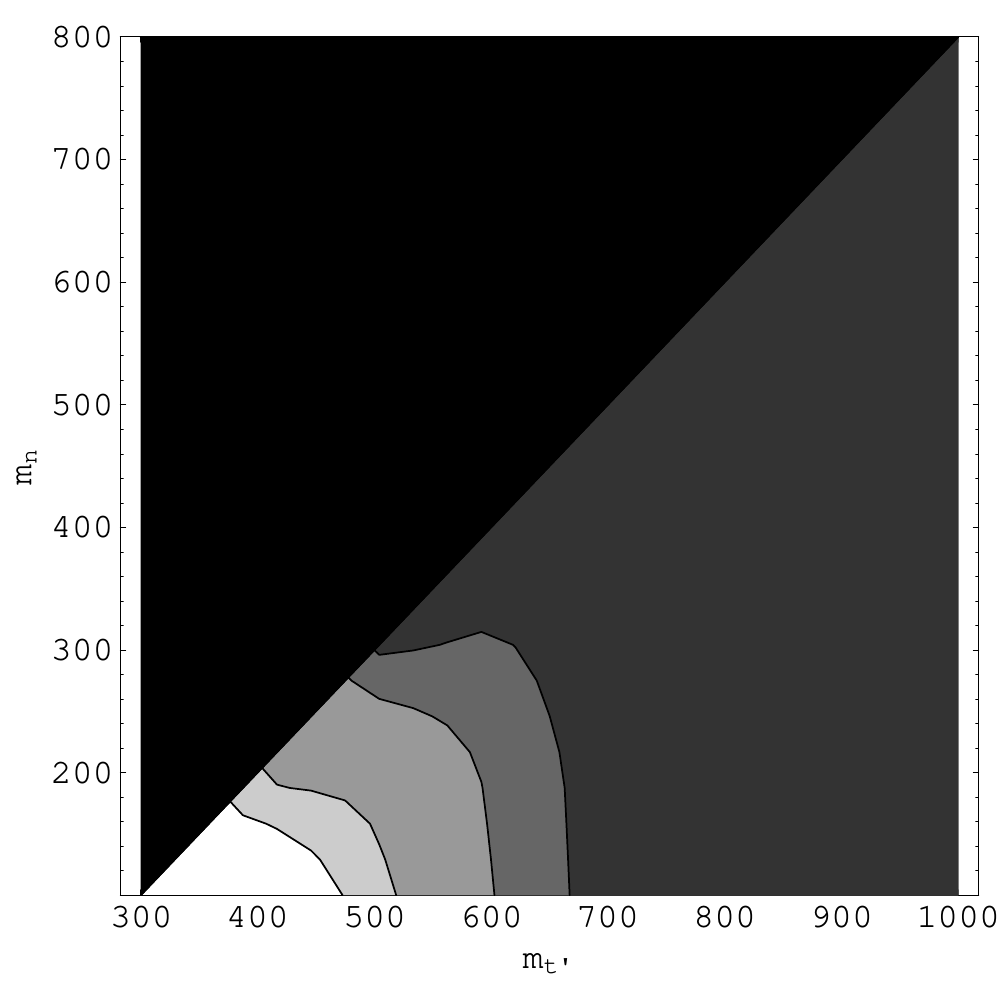}
\end{center}
\caption{Discovery prospects in the $M_{t^\prime}$-$M_N$ plane for
  pair production of fermion (left) and scalar (right) top partners
  $t^\prime$ decaying into $t \etmiss$.  The light-shaded contours are
  (left to right) signal significances of $15\sigma$, $10\sigma$,
  $5\sigma$, $>3\sigma$, $<3\sigma$.  Figures from
  Ref.~\cite{Meade:2006dw}.}
\label{fig:tpartner}
\end{figure}

At the LHC we can study effective theories containing both fermionic
(as expected in little Higgs theories) and scalar (as expected in
supersymmetric theories) top partners $t^\prime$ which, motivated by
dark matter, decay into an on-shell top quark and a dark matter particle
$N$.  The top partner spin is then either $s=0$ or $s=1/2$ so that the 
decay $t^\prime \rightarrow t N$ is allowed~\cite{Meade:2006dw,heavy_tops}. 
The discovery mode for such decays rely on hadronic top decays, 
leading to a final state containing two $b$-initiated jets, 
four light-quark jets, two pairs of which come from a $W$ decays, 
and missing energy from the pair of dark matter particles.  
To suppress QCD and $W$+jets backgrounds a standard requirement
of $\etmiss > 100$~GeV and $H_T > 500$~GeV is usually applied.
In addition, two $b$-tags and loose cuts to reconstruct pairs of 
light quarks into $W$s help to further reduce backgrounds.
The reconstructed $W$ bosons and $b$-tagged jets can then
be reconstructed into top quarks.
A plethora of relevant Standard Model backgrounds have to be 
considered in addition to the generic signal~\cite{Meade:2006dw}.

  Simulations of these signals and the relevant backgrounds suggest
that with $10~\ifb$, the LHC can discover a fermion
top partner up to about 850 GeV or a scalar top partner up to about
600 GeV, provided $M_{t^\prime} - M_N > 200$~GeV.  Detailed contours
in the $M_{t^\prime}$-$M_N$ plane are shown in
Fig.~\ref{fig:tpartner}.  In the region where a discovery is possible,
kinematic variables such as $m_\text{eff}$ allow us to measure the
mass difference $M_{t^\prime} - M_N$, but provide limited information
about the absolute scale of the masses themselves.  If one is willing
to assume that the pair production of the colored $t^\prime$ pair is
through ordinary $SU(3)_c$ interactions, for a fixed $t^\prime$ spin
the leading-order cross section is only a function of $M_{t^\prime}$
and $\alpha_s (M_{t^\prime})$. This is similar to stop pair production
in supersymmetry which does not allow for a $t$-channel gluino
exchange because there is no top flavor content inside the proton at
the LHC~\cite{prospino}.

This way we can use the production rate times branching ratio to
determine $M_{t^\prime}$. Section~\ref{sec:sig_qcd} lists many reasons
for why this is a very hard measurement, but it may be the only handle
available lacking better observables. Generic leading-order errors on
the cross section alone ranging around $50-100\%$, plus errors due to
detector efficiencies and coverage imply that we would only rely on
such ($\sigma \cdot \br$)-type information in the absence of other
useful measurements~\cite{focuspoint,ayres_lhc,Turlay:2008dm}.
Next-to-leading order predictions for new physics processes are then
mandatory, to reduce the theory uncertainty to the $20\%$
level~\cite{prospino}.  The same can be true for the branching
ratios~\cite{sdecay}.

In contrast, we remember that QCD corrections to kinematics are known
to be under control: additional jet radiation is well described by
modern matching schemes and will not lead to unexpected QCD
effects. Off-shell effects in cascade decays can of course be large
once particles become almost mass degenerate~\cite{off_shell}, but in
the standard SPS1a cascades these effects are expected to be small.
This confirms that the key to distinguishing between solutions to the
hierarchy problem is to directly construct the spin of the $t^\prime$,
for example by making use of the techniques discussed in
Section~\ref{sec:sig_met}.\bigskip

Longer decay chains are possible. For example, in supersymmetric
theories with a sufficiently heavy gluino, the decay $\tilde{g}
\rightarrow t \tilde{t}^*$ may be important.  In particular, the
stop may be among the lightest of the squarks because of the larger
mixing induced by the top mass.
If light enough, the stop may dominantly decay into $c
\tilde{\chi}^0$.  Since each gluino decays $50\%$ of the time into
$t \tilde{t}^*$ or $\bar{t} \tilde{t}$, half of the gluino
pairs produce $tt$ or $\bar{t} \bar{t}$, like-sign top quarks whose
leptonic decays produce like-sign
leptons~\cite{Kraml:2005kb,Martin:2008aw}. This analysis is completely
analogous to the usual like-sign dileptons from gluino decays,
Eq.~\eqref{eq:likesign}, with the additional complication of a large
final state multiplicity leading to combinatorial effects. 
A possible way out would be to employ the top-tagging we will
discuss below.
Searches based on like-sign leptons, $b$-tagged jets, and missing
energy are expected to yield a $5\sigma$ discovery of the gluino pair
production above the Standard Model and other SUSY backgrounds for gluino masses
up to about 900~GeV provided $m_{\tilde{g}} > M_{\tilde{q}}$~\cite{Kraml:2005kb}.  
There are also good prospects to
reconstruct the spectrum by studying kinematic
endpoints~\cite{Kraml:2005kb,Martin:2008aw}.\bigskip

  Higher mass stops may decay via $\tilde{t} \rightarrow \chi^+ b$,
$\tilde{t} \rightarrow \chi^0_2 t$, or $\tilde{t} \rightarrow
t \tilde{\chi}^0$~\cite{Hisano:2002xq,Hisano:2003qu,
Kim:2009nq,Acharya:2009gb}. Top decays to a tagged bottom jet and a 
chargino are one of the standard search channels at the Tevatron, 
even though the fact that they are not separable from top pair 
production makes them hard to extract from the $t \bar{t}$ background. 
At the LHC, angular correlations and a different $\etmiss$ spectrum 
are unlikely to be sufficient to extract stop pair production out 
of the hugely increased top pair background. Nevertheless, 
for benchmark points where the gluino mass
varies between roughly 500~GeV and 900~GeV an early LHC discovery is
limited only by our understanding of the detector, and information
about masses can be reconstructed using kinematic quantities such as
$m_\text{eff}$ and $m_{tb}$.\bigskip

\begin{figure}[t]
\begin{center}
 \includegraphics[width = 0.65\textwidth]{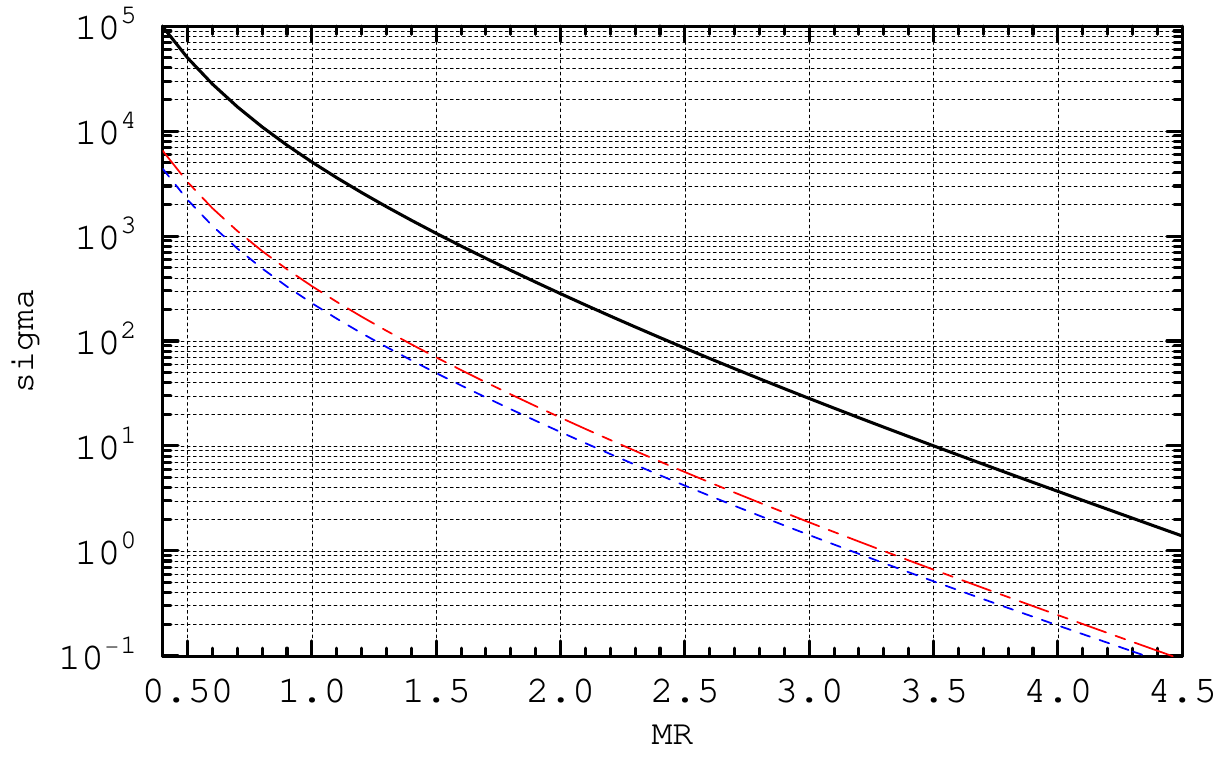}
 \end{center}
\caption{Cross sections for production of $(1,1)$ modes of (top to
  bottom) $g^{(1,1)}_\mu$, $W^{(1,1)}_{3\mu}$, and $B^{(1,1)}_\mu$ at
  the LHC. Figure from Ref.\cite{Burdman:2006gy}.}
\label{fig:11vecxsec}
\end{figure}

Top signatures also arise in cascade decays from the 6D UED chiral
square model~\cite{Burdman:2006gy} with minimal boundary terms.  As
explained in Sec.~\ref{sec:models_ued}, the spectrum contains $(1,1)$
KK modes with boundary term coupling to Standard Model quarks, 
and so the LHC can produce the vector states resonantly 
through $q \bar{q} \rightarrow V_{\mu}^{(1,1)}$ with the largest 
cross section for the KK gluon, $g^{(1,1)}_\mu$, see Fig.~\ref{fig:11vecxsec}.  
Once produced, the dominant decay is through the KK-number preserving 
bulk coupling into a $(1,1)$ fermion and its zero mode counterpart.  
The $(1,1)$ fermions decay mainly into a zero-mode fermion and 
one of the $(1,1)$ scalar adjoint particles, KK modes of the 
physical $5$ and $6$ components of the KK gauge fields.  
With minimal boundary terms, the scalar adjoints are the lightest states 
in a given gauge representation at that KK level, 
and the $(1,1)$ states tend to decay into a pair of zero-mode fermions.  
However, the boundary operator by which they couple is such that for 
on-shell fermion the coupling is proportional to $m_f$, meaning in 
practice they always decay into a pair of top quarks. The net result 
is thus a pair of light jets and a pair of top quarks, with the top 
quark invariant mass reflecting the
mass of the spinless adjoint and the four body invariant mass
reconstructing the vector $(1,1)$ mode resonance.\bigskip

Top quarks can also be produced through the decays of new 
colored fermions.  For example, a model of a chiral fourth generation 
can have the decays $b^\prime \rightarrow t W^-$
\cite{Kribs:2007nz,Holdom:2009rf}.  Vector-like quarks also have 
charged current decays into top quarks, as well as FCNC decays 
$t^\prime \rightarrow t Z$ and $t^\prime \rightarrow t h$
\cite{Choudhury:2001hs,Han:2003wu,Dobrescu:2009vz}, including
KK modes of top and its partners~\cite{Contino:2008hi}.

\subsubsection{A Composite Top Quark}

In the dual interpretation of RS warped extra dimensions, the top is
largely composite.  RS is a compelling vision of compositeness, which
manages to solve the hierarchy problem, and the related phenomenology
is discussed separately in Sections~\ref{sec:models_warped} and
\ref{sec:sig_resonance}.  However, other constructions are possible in
which the top is a among the lightest bound states of some new
confined force.  Such theories may be explored model-independently
using effective field
theory~\cite{Georgi:1994ha,Suzuki:1991kh,Lillie:2007hd,Pomarol:2008bh,Kumar:2009vs}.
Following Ref.~\cite{Georgi:1994ha}, we consider the case where the
right-handed top is composite.  A composite $t_R$ is less bounded by
precision electroweak data, and allows for a lower compositeness
scale.

If the top is composite at scale $\Lambda$, the leading deviation at
energies below that scale is in the form of higher dimensional
operators characterized by that scale. Naive dimensional analysis
would suggest that the most important operator is the dimension six
coupling of four top quarks~\cite{Georgi:1994ha},
\begin{equation}
\frac{g^2}{\Lambda^2} \left[ \bar{t}^i \gamma^\mu P_R t_j \right] 
\left[ \bar{t}^k \gamma_\mu P_R t_l \right]
\label{eq:sig_4top}
\end{equation}
where the color indices $i$, $j$, $k$, and $l$ may be contracted
either into pairs of singlets or octets and the dimensionless coupling
$g$ is expected to be $\sim 4 \pi$ by naive dimensional analysis.

Existing bounds from the Tevatron are rather weak.  Since Tevatron has
insufficient energy to produce four top quarks at any appreciable
rate, the operator of Eq.~(\ref{eq:sig_4top}) is basically impossible to
probe.  One can look for subleading operators, which may influence the
production of $t \bar{t}$ or single top production by modifying the
top coupling to the gluons and weak
bosons~\cite{Tait:2000sh,AguilarSaavedra:2008zc,Atwood:1994vm}.  Since
these deviations are the effects of higher dimension operators, they
lead to differences in kinematics, and are more pronounced for more
energetic tops.  Nonetheless, such constraints are relatively mild,
allowing compositeness scales as low as a few hundred
GeV~\cite{Kumar:2009vs}.  Such a low scale opens the door to the
possibility for huge effects at the LHC.

\begin{figure}[t]
\begin{center}
 \includegraphics[width = 0.45\textwidth]{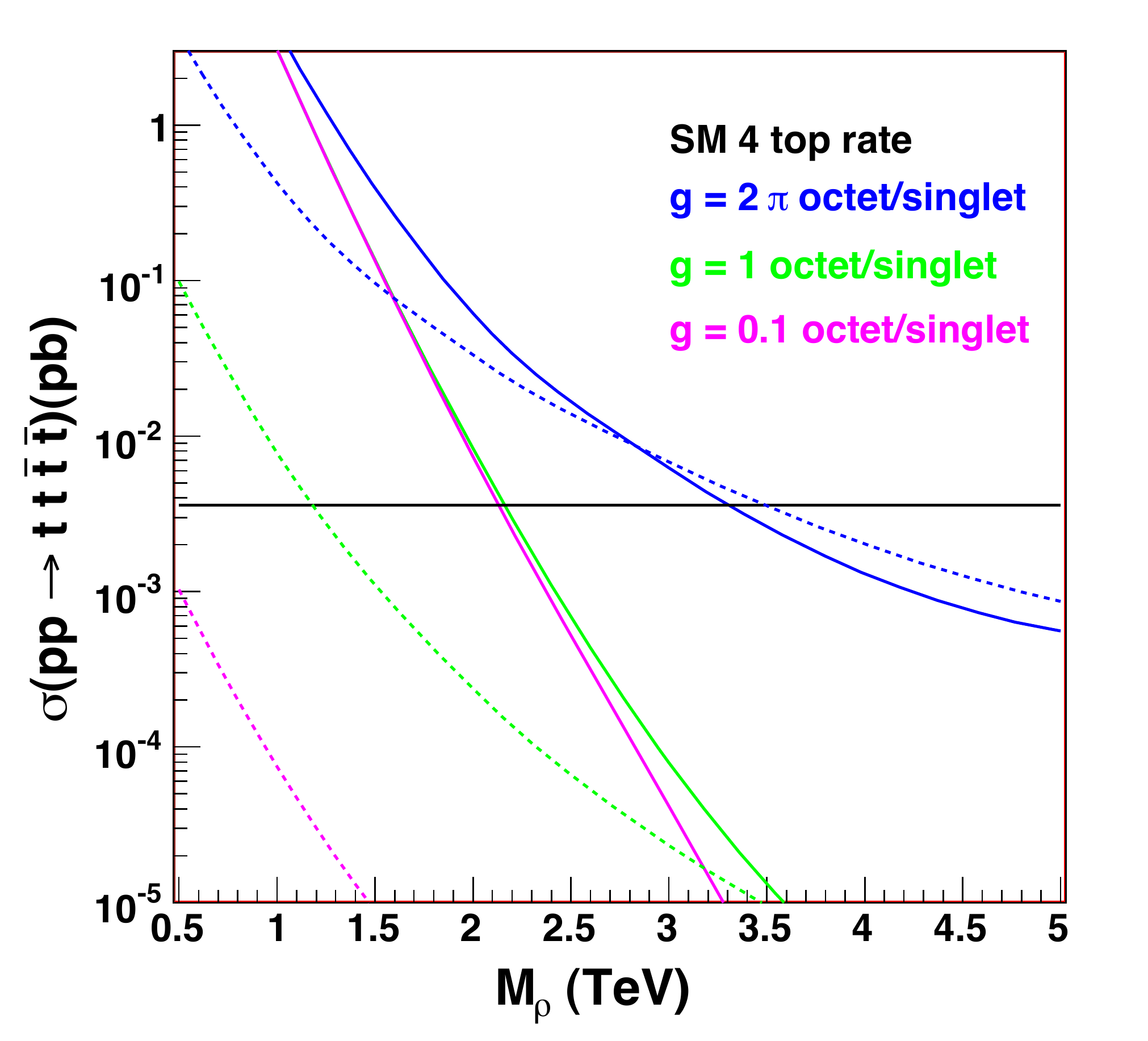}
\end{center}
\vspace{-0.5cm}
\caption{LHC signal rates for $ pp \rightarrow \rho \rho$ 
(color octet $\rho$) and $pp \rightarrow t \bar{t} \rho$ 
(color singlet or octet $\rho$ (both of which lead to a 4 top final state)
as a function of the $\rho$ mass, and for
different values of the $\rho$-$t$-$\bar{t}$ coupling.
The SM four top rate is indicated by the black line.
From Ref.~\cite{Lillie:2007hd}.}
\label{fig:sig_4top}
\end{figure}

In fact, it may be that the scale of compositeness is so low that the
LHC need not be content with subtle deviations in top couplings, but
can directly access higher resonances.  One can imagine an analogue of
the $\rho$ meson of the new force, one of the higher excited
resonances of the top itself.  Since some of the top's constituents
must be colored, one can easily imagine color octet or singlet vector
particles, with large coupling to the top but perhaps negligibly weak
coupling to light fermions.  In Fig.~\ref{fig:sig_4top} we show the
production rates for the processes $ pp \rightarrow \rho \rho$ (color
octet $\rho$) and $pp \rightarrow t \bar{t} \rho$ (color singlet or
octet $\rho$) as computed in Ref.~\cite{Lillie:2007hd}.  Since the
$\rho$ has essentially 100\%  branching fraction into $t \bar{t}$, both of these
processes lead to a four top quark final state.

The Standard Model rate for four top production is of order a few fb, whereas the
$\rho$ signals may increase it by more than a factor of $10^{3}$ if
the $\rho$ is sufficiently light.  Four top signals may be extracted
from backgrounds making use of the decay mode leading to like-sign
charged leptons, the dominant backgrounds to which are weak boson
production and $t \bar{t}$.
Studies~\cite{Lillie:2007hd,Pomarol:2008bh} indicate
that with 100 fb$^{-1}$, the LHC can extract such signals with
5$\sigma$ significance provided they are of order 10s of fb,
corresponding to $\rho$ masses of a few TeV.

\subsubsection{Massive top resonances}
\label{sec:sig_boosted}

\begin{figure}[t]
  \includegraphics[width = 0.43\textwidth]{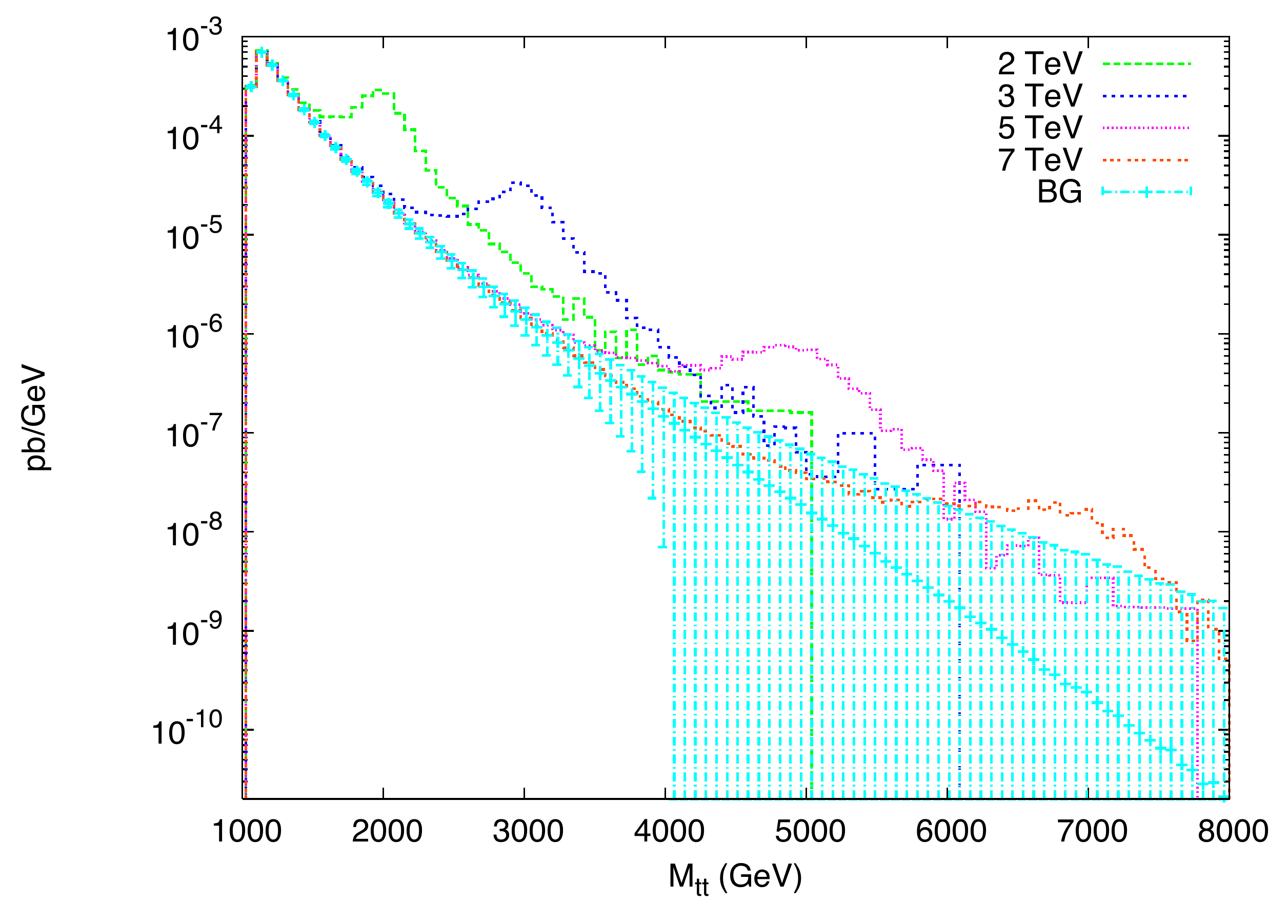}
  \hspace*{0.1\textwidth}
  \includegraphics[width = 0.43\textwidth]{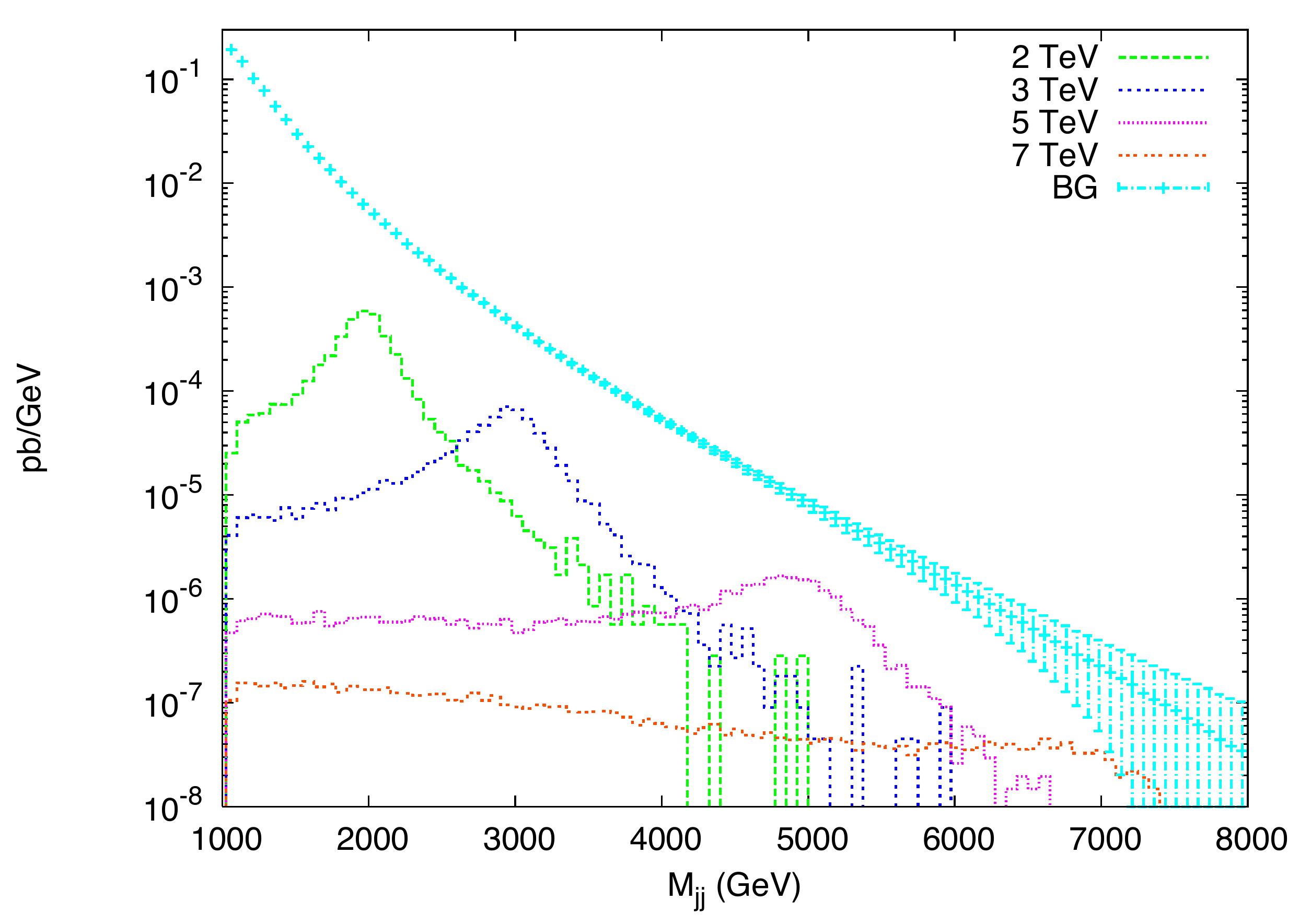}
\caption{LHC signals from KK gluons of mass $m_1^V = 2,3,5,7\,\tev$
decaying to top quark pairs.
In the left panel we show the signal relative to the irreducible 
Standard Model $t\bar{t}$ background, while in the right panel we show the signal 
relative to the rate for hard central di-jets with $p_T>500\,\gev$ and
$|\eta| < 0.5$.  Both figures come from Ref.~\cite{Lillie:2007yh}.}
\label{fig:sig_gluontt}
\end{figure}

As discussed in Sections~\ref{sec:models_ued},\ref{sec:models_lhiggs} ,
\ref{sec:models_tc}, and \ref{sec:models_rs} the LHC might face the 
situation where a very heavy resonance --- some kind of `heavy gluon' --- 
is produced and then decays into two Standard Model particles. 
From $Z^\prime$ searches we know how to look for decays to a pair of 
leptons or a pair of jets, simply using a side-bin analysis. 
A problem arises if these new very heavy states decay preferably 
to a pair of top quarks~\cite{tt_resonance,Lillie:2007ve}.  The top quarks produced 
will be highly boosted, implying that their decays products will be 
highly collimated.  As a result, the standard top quark identification
techniques that rely on identifying well-separated decay objects
may no longer work.  

  Resolving collimated top decay products is 
limited by the granularity of the ATLAS and CMS calorimeters, 
typically of the order of $R \sim 0.1$. As long as the top decay 
products can be resolved on this scale, we can attempt to reconstruct 
semi-leptonic top pairs in analogy to the current Tevatron 
analyses~\cite{top_rec_semilep}.  For very boosted tops, 
the jet algorithms will combine all decay products into 
a single jet, possibly faking a more common light jet.

  The prospects for discovering a very heavy gluon decaying to tops
at the LHC depend on how well we can distinguish collimated top jets 
from QCD jets.  If no separation is possible relative to QCD dijets,
the signal will almost certainly be swamped by the Standard Model background.
We illustrate this feature in Fig.~\ref{fig:sig_gluontt}, 
where we show the predicted $t\bar{t}$ KK-gluon signal along with 
projected Standard Model backgrounds.  In
the left panel we show only the Standard Model $t\bar{t}$ background.  Evidently,
the signal is visible above this optimistic background estimate.  In
the right panel we show the potential background from hard central
dijets with $p_T>500\,\gev$ and $|\eta|< 0.5$.  If this background is
not distinguishable at all from the $t\bar{t}$ events from a KK gluon,
the signal will be extremely difficult to find.\bigskip

  Recent investigations of the decays of boosted objects have found that,
both in the Standard Model~\cite{subjet_wh,Plehn:2009rk} as well
as in searches for heavy resonances~\cite{subjet_ww,subjet_susy,our_tagger,fast_tops,
jetfct_tops,david_e}, they can be treated in analogy with
the usual tau and bottom reconstruction methods. 
A heavy state decaying to two jets will produce a 
``fat jet'' of a size around
\begin{equation}
R_{jj} = \frac{1}{\sqrt{z (1-z)}} \; \frac{m_\text{heavy}}{p_T}
       > \frac{2 m_\text{heavy}}{p_T},
\end{equation}
depending on its transverse momentum and on the momentum balance $z$
vs ($1-z$) of the two decay products. If we form such a fat jet out of
the two massless decay products, this fat jet automatically acquires a
substructure --- keeping track of all jet masses in the clustering or
de-clustering of the fat jet we will find one splitting where the jet
mass drops from $m_\text{heavy}$ to much smaller values. If we manage
to analyze such fat jets and distinguish them from QCD backgrounds we
can identify $W$ bosons~\cite{subjet_ww}, top
quarks~\cite{fast_tops,david_e}, Higgs
bosons~\cite{subjet_wh,Plehn:2009rk}, as well as supersymmetric
particles~\cite{subjet_susy,our_tagger} in their purely hadronic decay modes and
without significant combinatorial backgrounds.\bigskip

\begin{figure}[t]
\begin{center}
  \raisebox{2mm}{\includegraphics[width=0.48\textwidth]{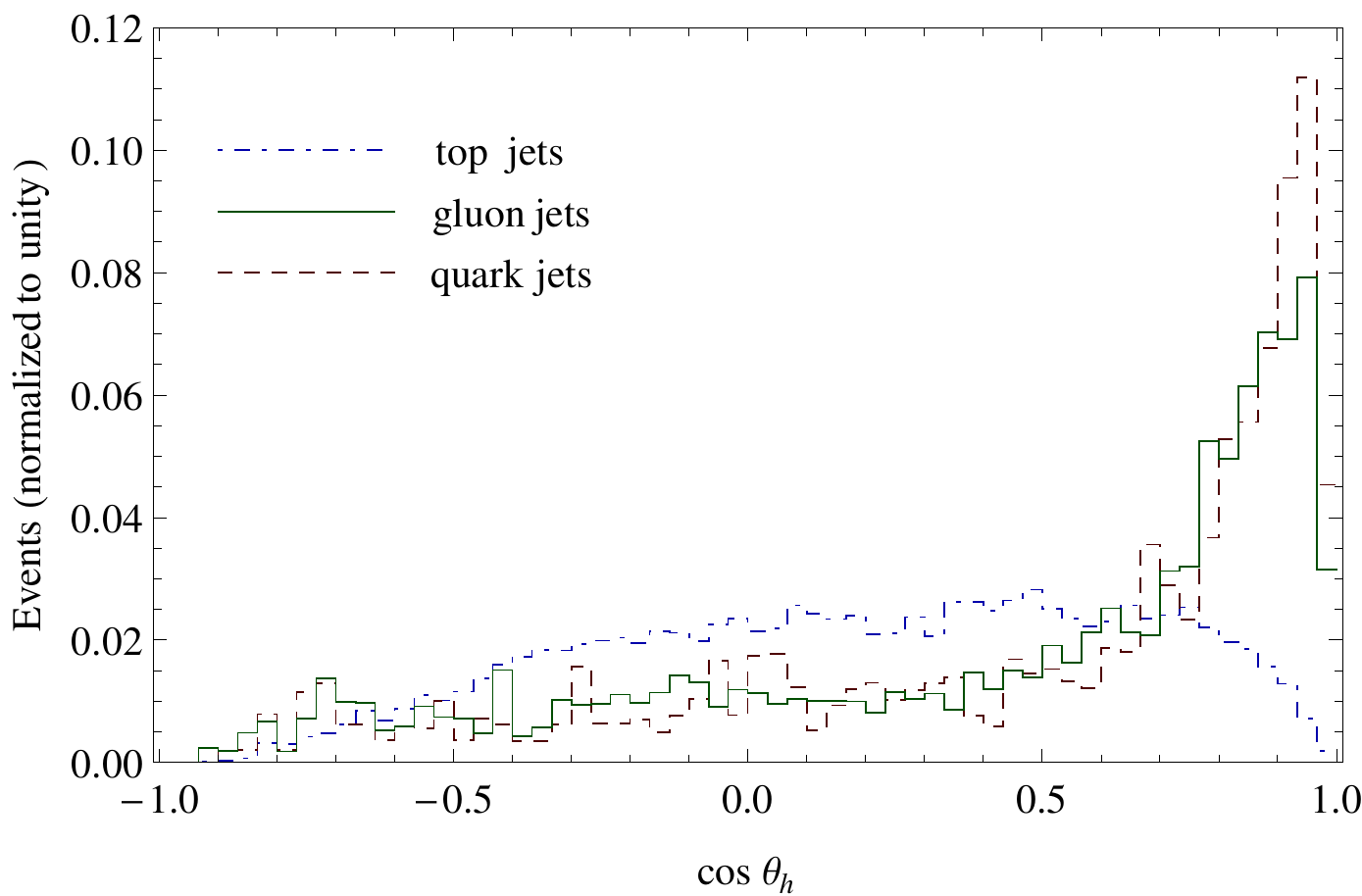}}
  \hspace*{0.05\textwidth}
  \includegraphics[width=0.42\textwidth]{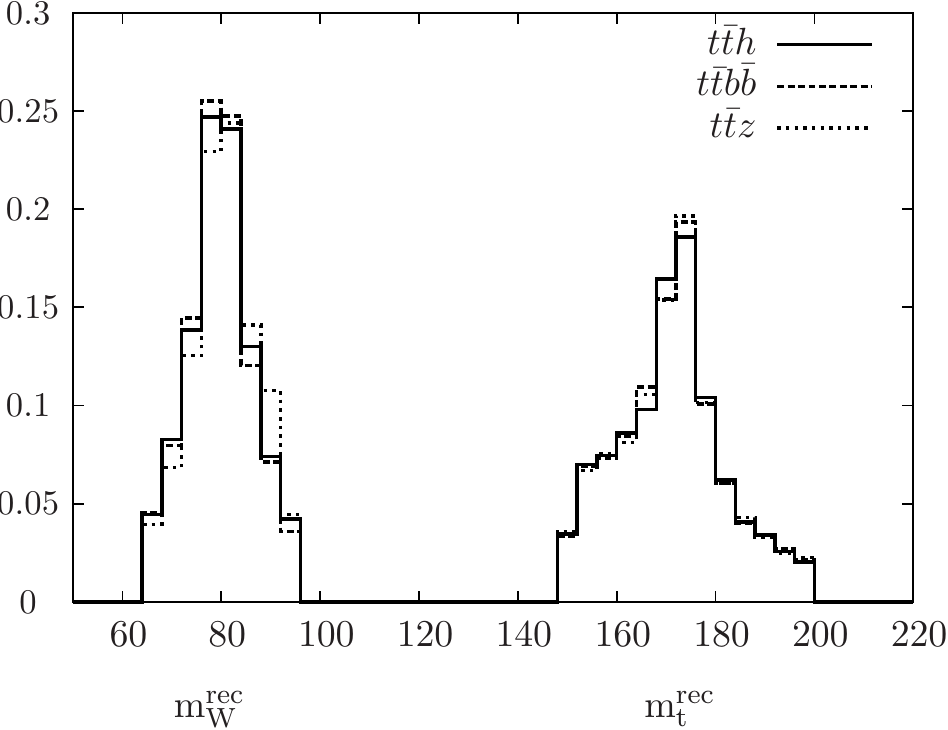}
\end{center}
\vspace{-0.5cm}
\caption{Left: helicity angle distribution for top and QCD jets with
  $p_T > 700$~GeV, after applying the $W$ and top mass windows.
  Figure from Ref.\cite{david_e}.  Right: reconstructed $W$ and top
  masses for a Standard Model top tagger with initial jet size of
  $R=1.5$ in the $t\bar{t}H$ process. Filtering is applied to reduce
  the contamination by the underlying event. Figure from
  Ref.~\cite{Plehn:2009rk}.}
\label{fig:sig_toptag}
\end{figure}

  In particular, top tagging~\cite{david_e,Plehn:2009rk} should be a
reliable method at the LHC because we can apply three different
criteria to the fat jet and its subjets, and hence minimize its
mis-tagging probability.  First, we know all particle masses involved, 
which means that we do not have to avoid an
implicit mass criterion which sculpts the background and destroys a
possible side bin analysis in the jet mass. Typical mass windows for
the reconstructed masses are $20\%$. In addition, we can exploit the
left-handed coupling for the $W$ boson to fermions. In the rest frame
of the reconstructed $W$ we can measure the angle $\theta_h$ between
the top direction and the softer of the two $W$ decay jets. In
Fig.~\ref{fig:sig_toptag} we see that for top jets this distribution
is essentially flat, while for QCD jets it diverges as a soft
singularity proportional to $1/(1- \cos\theta_h)$. With all three
criteria the top tagging efficiency, depending on the $p_T$ of the top
jet, reaches up to $50\%$ with a mis-tagging probability of $1 - 5\%$. 
An alternative to the search for massive splittings using a
Cambridge-Aachen jet algorithm~\cite{ca_algo} is to look for jet
functions inside the fat jet~\cite{jetfct_tops}. A detailed comparison
of these different approaches might only be possible once we have LHC
data.\bigskip

One final aspect we need to study before we can claim that top tagging
works at the LHC is the underlying event~\cite{bryan_ue}: the larger
the size of the initial jet becomes the more dangerous additional QCD
activity from the underlying event, proportional to $R_{jj}^4$,
becomes. At some stage of the analysis we have to remove soft QCD
activity from the constituents of the fat jet~\cite{pruning}, for
example to ensure that for example the $W$ and top mass windows in the
top tagger cover the correct distributions. The most effective method
is filtering the constituents at a finer resolution than the size of
the fat jet~\cite{subjet_wh}. For example, a fat Higgs jet in $WH$
production has a maximum size of $R_{bb} = 1.2$; in the filtering
procedure we increase the resolution to $R_\text{filt} = \min(0.3,
R_{bb}/2)$ and identify the three hardest subjets which appear at this
finer scale. Only these three subjets enter the calculation of the jet
mass, \ie the mass of the heavy decaying object. The same technique we
can apply to fat top jets which involve to mass drops inside the fat
jet, one from the top mass to the $W$ mass and another one from the
$W$ mass to massless quark jets. For the $W$ candidate we take the
three hardest subjets within $R_\text{filt} = \min(0.3, R_{12}/2)$,
based on the separation of the two hardest subjet. For the top
candidate we add two more subjets with a resolution $R_\text{filt} =
\min(0.3,R_W)$. For the most challenging case of an initial top jet
size of $R = 1.5$ for $t \bar{t}H$ production in the Standard Model we
show the reconstructed $W$ and top mass distributions including
underlying event in Fig.~\ref{fig:sig_toptag}.

\subsection{Metastable charged states}
\label{sec:sig_stable}

  Many of the models of new physics discussed in Section~\ref{sec:models}
contain heavy exotic particles charged under electromagnetism or QCD
that are quasi-stable on collider time scales.  In contrast to heavy particles 
that decay promptly to the Standard Model, such metastable charged states 
interact directly with the components of the LHC detectors.
A qualitatively different set of collider search strategies are 
therefore needed to detect and identify them.\bigskip

  Long-lived particles carrying an electric charge can arise in 
a number of ways.  In supersymmetry with gauge-mediated supersymmetry
breaking it is possible to have a slepton or chargino NLSP that decays very
slowly to a stable gravitino LSP~\cite{Ambrosanio:2000ik}.  
These possible NLSP candidates would be produced at the LHC primarily 
at the end of cascade decay chains originating with a pair of 
strongly-interacting superpartners.  
Thus, their net production rate is dominated by QCD, and a pair of 
charged massive particles will arise in every supersymmetric event.
Another example in this is a heavy lepton that only decays to 
the Standard Mode through higher-dimensional operators.  
Their production proceeds primarily through electroweak processes.
For both of these examples and also more generally, the direct collider 
effects of long-lived charged states depend primarily on their masses, 
electric charges, and the energies at which they are created.

Once produced, a heavy electromagnetically charged particle will 
propagate through the detector.  Along the way, it will leave charged
tracks in the tracking chambers, lose energy to ionizing interactions
with the detector components, and finally pass through the muon chamber,
which is sufficient to trigger on~\cite{cms_tdr,atlas_csc,Nisati:1997gb,Allanach:2001sd,Kilian:2004uj,Fairbairn:2006gg,Raklev:2009mg}.
These features are similar to those exhibited by a muon in the LHC detectors.
In contrast to hard muons, which are ultra-relativistic at the LHC, 
charged particles with masses on the order of 100---1000\,GeV are expected 
to be produced semi-relativistically with $\beta\gamma \sim 1$. 
This leads to a higher rate of energy loss $dE/dx$ due to ionization 
and a slower propagation time.  By measuring $dE/dx$ in the tracker 
or the time of flight in the tracking and muon chambers, it is possible
to determine the velocity $\beta$ of the heavy 
state.
The momentum of a charged state can be obtained from the curvature of 
its track.  Putting these measurements together allows for a precise 
determination of the mass of the heavy charged 
particle.

  The detection of a heavy stable charged particle becomes more complicated
if it is produced non-relativistically.  For $\beta \lesssim 0.7$ the
tracking efficiency falls quickly.
There is also a danger of timing
confusion between different events, since the LHC bunch crossing time of
25\,ns corresponds to the time taken for a particle with $\beta \sim 0.7$
to reach the ATLAS muon 
chambers~\cite{cms_tdr,atlas_csc,Nisati:1997gb,Allanach:2001sd,Kilian:2004uj,Fairbairn:2006gg}.
On the other hand, very slowly moving
charged states can stop in the detector or the surrounding
region~\cite{Buchmuller:2004rq,Feng:2004yi}.  If they decay at some point, 
it may be possible to determine their decay lifetime~\cite{Hamaguchi:2004df,
Asai:2009ka}.  
For the case of a metastable charged NLSP
in gauge mediated supersymmetry breaking, such a measurement would
also give the mass of the light gravitino and the energy scale of
supersymmetry breaking~\cite{Ibe:2007km}.\bigskip  

QCD-charged quasi-stable particles have more complicated interactions
in the detector. From Section~\ref{sec:models_mssm} we know that such
states can arise in the form of gluinos in split supersymmetry, as a
squark of gluino NLSP in scenarios with gauge mediated supersymmetry
breaking, or from massive exotic quarks and other colored fermions or
leptons in more general scenarios.  In all these cases, the new
strongly interacting states will be singly or pair produced abundantly
at hadron colliders.  After creation, they will combine with quarks
and gluons to form a heavy color-neutral
$R$-hadron~\cite{Farrar:1978xj,Farrar:1978rk,Baer:1998pg}.  For
example, a long-lived gluino will form $\tilde{g}q\bar{q}'$ and
$\tilde{g}g$ $R$-mesons, as well as $\tilde{g}qqq$ $R$-baryons.  The
precise spectrum of these states is not fully
understood~\cite{Chanowitz:1983ci,Foster:1998wu}.  For their collider
phenomenology, the most important question is whether the lowest-lying
particles carry an electric
charge~\cite{Kraan:2004tz,Kilian:2004uj,Fairbairn:2006gg}.  The answer
is not known, and it is therefore important to study both charged and
neutral $R$-hadrons. From these studies we know that both LHC
detectors can discover charged massive $R$ hadrons well into the TeV
range and measure their masses at least at the per-cent level.

  A neutral $R$-hadron will pass through the detector relatively easily,
depositing only a small amount of energy along the way.  They
can produce a signal of jets plus missing energy when produced in 
association with hard jets~\cite{Kilian:2004uj,Hewett:2004nw}.
Charged $R$-hadrons produce visible tracks in the tracking and muon
chambers.  These can be distinguished from genuine muons at the LHC
using the same techniques discussed above for uncolored massive charged
particles~\cite{Kilian:2004uj,Hewett:2004nw,Kraan:2005te,Kraan:2005ji,
Fairbairn:2006gg}. 
An interesting further possibility is that neutral and charged 
$R$-hadrons can interchange when they interact with material in the 
detector producing `dashed' tracks~\cite{Hewett:2004nw}.  
Slow-moving $R$-hadrons can potentially stop in the detector 
and become bound to nuclei~\cite{Arvanitaki:2005nq}.  
These will lead to displaced jets that are out of time with beam 
collisions when they decay.

\newpage

\section{Measuring the underlying theory}
\label{sec:para}
 
  Most modern scenarios of new physics designed to address
the challenges listed in Section~\ref{sec:why} propose a comprehensive
and detailed extension of the Standard Model above the electroweak scale.
Such models impact a wealth of measurements at the LHC and other 
experiments probing similar energy scales.  The LHC will 
give us a great many information about these new physics scenarios, and it will
certainly rule out large classes of these theories, but it will
definitely not give a one-to-one map between the limited number 
of hadron collider observables and a the full set of model parameters. 
Seriously testing any UV extension of the Standard Model therefore 
requires combining different experiments and investigating the features 
of the new physics parameter space from many angles.\bigskip

Some of the new-physics models discussed in Section~\ref{sec:models}
allow us to more easily determine the underlying parameters
than others.  Some models give rise to singly-produced resonances
which can be reconstructed to give precise mass measurements.
Other theories produce cascade decays with missing energy and
present a greater challenge to determining the overall mass scale
of new physics.  From an experimental point of view, some theories are
simply more suited to measurement by the LHC than others.

On the theory side, some models, such as
the MSSM, represent a perturbative and renormalizable
UV completion of the Standard Model up to very high energy scales
where Grand Unification or quantum gravity are expected to 
distort the picture.  Such theories have 
the advantage that the higher dimensional operators which
represent physics at still higher scales have a negligible effect
on LHC and related measurements.  Other theories break down
at lower energy scales, and consequently are more difficult to pin down,
because the UV physics may be relevant enough to subtly affect
observations at the LHC.\bigskip

   To illustrate the process of measuring the underlying theory of new physics, 
we will concentrate on a supersymmetric particle spectrum.
However, the SUSY-focussed analysis we present here rests on only few 
assumptions about the measurement process, and can be applied
to a number of other promising scenarios as well.
Our first assumption is that the new physics mass spectrum is measured 
through the cascade decays of heavy strongly-interacting particles into a 
weakly-interacting dark matter agent. Such mass measurements are heavily 
correlated because they rely on measuring kinematic endpoints. 
Secondly, we expect the new physics parameter space to separate into 
a fraction that is well determined at the LHC, and a fraction that we 
will not see at the LHC, involving for example particles that are not 
observed directly.  And last but not least, we assume that the TeV-scale
parameters can be extrapolated reliably to higher energies
using bottom-up renormalization group running.  The major question
we address in this section is to what extent is it possible to uniquely 
determine the parameter values in a model of new physics using data
from the LHC alone and including additional experiments.\bigskip

  The MSSM mass spectrum we study is the parameter point SPS1a~\cite{Allanach:2002nj},
based on the mSUGRA framework described 
in Section~\ref{sec:models_mssm}, and
which has been studied in detail by the experimental community. It
is somewhat on the easier side in terms of LHC parameter determination. 
This parameter point is characterized by moderately heavy squarks 
and gluinos, which decay through different long cascades including 
the neutralinos and sleptons down to the lightest neutralino. 
The gauginos are lighter than the higgsinos, and the lightest Higgs 
mass is close to the LEP limit. The mass spectrum as well
as projections for how well these mass can be measured at the LHC and
a $500\,\gev$ $e^+e^-$ linear collider (LC) are 
listed in Table~\ref{tab:para_mass_errors}. While for this parameter
point the dark matter relic density comes out slightly too large, it can 
be modified to include some amount of stau coannihilation to then
agree with all available experimental data today (SPS1a').

\begin{table}[t]
\begin{small} \begin{center}
\begin{tabular}{|l|cccc||l|cccc|}
\hline
 & $m_\text{SPS1a}$ & LHC & LC & LHC+LC &
 & $m_\text{SPS1a}$ & LHC & LC & LHC+LC\\
\hline
\hline
$h$  & 108.99& 0.25 & 0.05 & 0.05 &
$H$  & 393.69&      & 1.5  & 1.5  \\
$A$  & 393.26&      & 1.5  & 1.5  &
$H^\pm$& 401.88&      & 1.5  & 1.5  \\
\hline
$\chi_1^0$ &  97.21& 4.8 & 0.05  & 0.05 &
$\chi_2^0$ & 180.50& 4.7 & 1.2   & 0.08 \\
$\chi_3^0$ & 356.01&     & 4.0   & 4.0  &
$\chi_4^0$ & 375.59& 5.1 & 4.0   & 2.3 \\
$\chi^\pm_1$ & 179.85 & & 0.55 & 0.55 &
$\chi^\pm_2$ & 375.72 & & 3.0  & 3.0 \\
\hline
$\tilde{g}$ &  607.81& 8.0 &  & 6.5 & & & & & \\
\hline
$\tilde{t}_1$ & 399.10&     &  2.0  & 2.0 & & & & & \\
$\tilde{b}_1$ & 518.87& 7.5 &       & 5.7 &
$\tilde{b}_2$ & 544.85& 7.9 &       & 6.2 \\
\hline
$\tilde{q}_L$ &  562.98&  8.7 & &  4.9 &
$\tilde{q}_R$ &  543.82&  9.5 & &  8.0 \\
\hline
$\tilde{e}_L$    & 199.66   & 5.0 & 0.2  & 0.2  &
$\tilde{e}_R$    & 142.65   & 4.8 & 0.05 & 0.05 \\
$\tilde{\mu}_L$  & 199.66   & 5.0 & 0.5  & 0.5  &
$\tilde{\mu}_R$  & 142.65   & 4.8 & 0.2  & 0.2  \\
$\tilde{\tau}_1$ & 133.35   & 6.5 & 0.3  & 0.3  &
$\tilde{\tau}_2$ & 203.69   &     & 1.1  & 1.1  \\
$\tilde{\nu}_e$  & 183.79   &     & 1.2  & 1.2  & & & & & \\
\hline
\end{tabular}
\end{center} \end{small} 
\caption[]{Errors for the mass determination in SPS1a. Shown are the
  nominal parameter values (from SuSpect~\cite{suspect}), the errors
  for the LHC alone, for the LC alone, and for a combined LHC+LC
  analysis. Empty boxes indicate that the particle cannot, to current
  knowledge, be observed or is too heavy to be produced. All values
  are given in GeV. Table taken from Ref.~\cite{lhc_ilc}.}
\label{tab:para_mass_errors}
\end{table}

 The LHC mass measurements in Table~\ref{tab:para_mass_errors} come
from the kinematic endpoints and mass differences listed in
Table~\ref{tab:para_edges}. The systematic error is essentially due to
the uncertainty in the lepton and jet energy scales, expected to be
0.1\% and 1\%, respectively.  As discussed in
Section~\ref{sec:sig_qcd}, precision mass measurements at the LHC are
not possible based production rates (\ie $\sigma \ccdot\br$). 
The reason for this are sizeable QCD
uncertainties~\cite{susy_prod,prospino,nnlo_susy}, largely due to gluon
radiation from the initial state, but by no means restricted to this
one aspect of higher-order corrections.
As argued in Section~\ref{sec:sig_qcd} serious LHC results on new
physics can only be extracted with an appropriate understanding and
simulation of QCD, which suggests in turn that statements in the
literature about LHC results should be interpreted with care.

  For a linear collider the rule of thumb is that if particles are light enough to
be produced in pairs their masses can be determined with impressive
accuracy. This rule is only violated by states without electroweak
charges, like gluinos or heavy gluons. The mass determination either
comes from direct reconstruction or from the cross section curve at
threshold, with comparable accuracy but different
systematics. Precision measurements of branching ratios are also 
possible. We see from Table~\ref{tab:para_mass_errors} that indeed 
the LHC has a better coverage of the strongly interacting sector, 
whereas the better coverage and precision in the weakly interacting 
sector is achieved at a LC.\bigskip

\begin{table}[t]
\begin{small} \begin{center}
\begin{tabular}{|ll|r|rrrr|}
\hline
\multicolumn{2}{|c|}{ type of } & 
 \multicolumn{1}{c|}{ nominal } & 
 \multicolumn{1}{c|}{ stat. } & 
 \multicolumn{1}{c|}{ LES } & 
 \multicolumn{1}{c|}{ JES } & 
 \multicolumn{1}{c|}{ theo. } \\
\multicolumn{2}{|c|}{ measurement } & 
 \multicolumn{1}{c|}{ value } & 
 \multicolumn{4}{c|}{ error } \\
\hline
\hline
$m_h$ & 
 & 108.99& 0.01 & 0.25 &      & 2.0 \\
$m_t$ & 
 & 171.40& 0.01 &      & 1.0  &     \\
$m_{\tilde{l}_L}-m_{\chi_1^0}$ & 
 & 102.45& 2.3  & 0.1  &      & 2.2 \\
$m_{\tilde{g}}-m_{\chi_1^0}$ & 
 & 511.57& 2.3  &      & 6.0  & 18.3 \\
$m_{\tilde{q}_R}-m_{\chi_1^0}$ & 
 & 446.62& 10.0 &      & 4.3  & 16.3 \\
$m_{\tilde{g}}-m_{\tilde{b}_1}$ & 
 & 88.94 & 1.5  &      & 1.0  & 24.0 \\
$m_{\tilde{g}}-m_{\tilde{b}_2}$ & 
 & 62.96 & 2.5  &      & 0.7  & 24.5 \\
$m_{ll}^\mathrm{max}$: & three-particle edge($\chi_2^0$,$\tilde{l}_R$,$\chi_1^0$)  
 & 80.94 & 0.042& 0.08 &      & 2.4 \\
$m_{llq}^\mathrm{max}$: & three-particle edge($\tilde{q}_L$,$\chi_2^0$,$\chi_1^0$)  
 & 449.32& 1.4  &      & 4.3  & 15.2 \\
$m_{lq}^\mathrm{low}$: & three-particle edge($\tilde{q}_L$,$\chi_2^0$,$\tilde{l}_R$)
 & 326.72& 1.3  &      & 3.0  & 13.2 \\
$m_{ll}^\mathrm{max}(\chi_4^0)$: & three-particle edge($\chi_4^0$,$\tilde{l}_R$,$\chi_1^0$)
 & 254.29& 3.3  & 0.3  &      & 4.1 \\
$m_{\tau\tau}^\mathrm{max}$: & three-particle edge($\chi_2^0$,$\tilde{\tau}_1$,$\chi_1^0$)
 & 83.27 & 5.0  &      & 0.8  & 2.1 \\
$m_{lq}^\mathrm{high}$: & four-particle edge($\tilde{q}_L$,$\chi_2^0$,$\tilde{l}_R$,$\chi_1^0$)
 & 390.28& 1.4  &      & 3.8  & 13.9 \\
$m_{llq}^\mathrm{thres}$: & threshold($\tilde{q}_L$,$\chi_2^0$,$\tilde{l}_R$,$\chi_1^0$)
 & 216.22& 2.3  &      & 2.0  & 8.7 \\
$m_{llb}^\mathrm{thres}$: & threshold($\tilde{b}_1$,$\chi_2^0$,$\tilde{l}_R$,$\chi_1^0$)
 & 198.63& 5.1  &      & 1.8  & 8.0 \\
\hline
\end{tabular}
\end{center} \end{small}
\caption[]{ LHC measurements in SPS1a. Shown are the nominal values
  (from SuSpect~\cite{suspect}) and statistical errors, systematic
  errors from the lepton (LES) and jet energy scale (JES) and
  theoretical errors.  All values are given in GeV. Table taken from
  Refs.~\cite{lhc_ilc,sfitter}.}
\label{tab:para_edges}
\end{table}

  For the favorable SPS1a parameter point we can illustrate the ideal
analysis scenario of new physics in the LHC era.  If data allows,
we will start from a fully exclusive likelihood map of the parameter 
space and a ranked list of the best-fitting points.  Under the 
highly restrictive assumption that the underlying theory is mSUGRA,
the LHC measurements will provide strong correlations among the
small number of underlying model parameters.  Even so, we will see
that it is realistic to expect several distinct likelihood maxima.
This will only get worse for any proper TeV-scale ansatz.

  Starting from this exclusive likelihood map, we can apply 
frequentist and Bayesian constructions to define lower-dimensional 
probability distributions. Bayesian probabilities and profile likelihoods 
are two ways to study an imperfectly measured parameter space, where some
model parameters might be very well determined, others heavily
correlated, and even others basically unconstrained. This is different
for example from $B$ physics, where theoretical degeneracies and
symmetries have become a major challenge~\cite{ckmfitter, utfit}. 
A careful comparison of the benefits and traps of both approaches in the
light of new-physics searches is therefore helpful.  Finally, we can
attempt to reconstruct the TeV-scale Lagrangian with proper errors. Such
TeV-scale results will then serve as starting points to a bottom-up probe
of the underlying source of supersymmetry breaking or unification
without theoretical bias.

  Our hope is that different experiments probing the same energy scales,
like the LHC, electric dipole moments, the anomalous magnetic moment
of the muon, flavor observables like rare decays, and finally a future
LC can be combined to measure as large a fraction of the TeV-scale
Lagrangian as possible. 

Technically, our setup starts from the LHC and then adds more
measurements. This reflects the time arrow as well as the feature that
collider experiments are true multi-purpose detectors with broader
analyses strategies. We largely rely on SFitter~\cite{sfitter} and
Fittino~\cite{fittino} results for TeV-scale parameter
studies. However, in particular for the conceptual bases and for
mSUGRA toy analyses many other results are
available~\cite{ben,leszek,ellis_olive,Akrami:2009hp}. Unfortunately, there is no
similar analysis available for any of the other TeV-scale models
discussed in Section~\ref{sec:models}, which does not say much about
the features of these models but is due to the level of experimental
involvement.

\subsection{Parameter extraction}

  We all know from our first-year labs that it is impossible to compare
interpretations with data based purely on central values. No error bar
means no measurement. For reliable errors on new physics parameters we
need a proper treatment of all errors.  The complete set includes
statistical experimental errors, systematic experimental errors, and
theory errors. The statistical experimental errors can be treated as
uncorrelated in the measured observables. In contrast, the systematic
experimental errors, such as from the jet and lepton energy
scales~\cite{lhc_ilc}, are strongly correlated. Hence, both are
non-trivially correlated if we extract masses from the measured
endpoints. While at first sight it might be surprising,  both
experimental errors can be approximated as gaussian.
We can do this for the systematic errors since they describe the
detector or analysis response which is measured from Standard Model physics.
This is why we bother looking at top quarks or $W$ and $Z$
bosons at the LHC -- we do not expect to learn much new about them, 
but they allow us to understand the detector.\bigskip

Theory errors arise when we compute observables like kinematic
endpoints or rates from TeV-scale parameters or when we compute
TeV-scale parameters from either other TeV-scale parameters or from
high-scale parameters. One problem with theory errors is that it is
difficult to reliably estimate the uncertainties due to unknown higher
orders in perturbation theory.  A definite lower limit on a theory
error is given by the dependence of observables on unphysical
renormalization and factorization scales.  These scales arise at
finite orders of perturbation theory, and the dependence on the scale
choices has to vanish once we sum all orders in the perturbative
series.  Therefore we can vary them within a `reasonable range' to get
an estimate of the theory error~\cite{Plehn:2009nd,qcd_reviews}.

  One thing we do know about perturbative series is that there is no
reason why the unknown higher-order corrections should be centered around
a given value. Hence, theory errors are taken to be box-shaped rather
than gaussian:  the probability assigned to any measurement does not
depend on its actual value as long as it is within the interval
covered by the theory error. The size of this box is of course
negotiable.  Tails could be attached to these theory-error
distributions, but higher-order corrections are not expected
to become arbitrarily large if the perturbative expansion is to be useful.  
Taking this interval approach seriously impacts not only the distribution 
of the theory error, but also its combination with the combined (gaussian)
experimental error. Just integrating over the theory prediction
nuisance parameter within the error band means a convolution of a
box-shaped theory error with a gaussian experimental error. It gives
us the difference of two one-sided error functions. This function has
a maximum, so the convolution still knows about the central value of
theoretical prediction. Moreover, it requires a (Bayesian) measure in
the cross section, giving different results if we integrate over the
number of events or over its logarithm.

  A better solution is the Rfit distribution used by CKMfitter~\cite{ckmfitter},
which is flat as long as the measured value is within the theoretically 
acceptable interval and drops off outside like the experimental gaussian. 
It corresponds to a profile likelihood construction marginalizing over
the theory error bar. The log-likelihood given a set of measurements
$\vec d$ and in the presence of a general correlation matrix $C$
reads~\cite{ckmfitter}
\begin{alignat}{7}
\chi^2 &= -2 \log \mathcal{L}
        = {\vec{\chi}_d}^T \; C^{-1} \; \vec{\chi}_d  \notag \\ 
\chi_{d,i} &=
  \begin{cases}
  0  
          &|d_i-\bar{d}_i | <   \sigma^{\text{(theo)}}_i \\
  \dfrac{|d_i-\bar{d}_i | - \sigma^{\text{(theo)}}_i}{\sigma^{\text{(exp)}}_i}
  \qquad  &|d_i-\bar{d}_i | >   \sigma^{\text{(theo)}}_i \; ,
  \end{cases}
\label{eq:flat_errors}
\end{alignat}
where $\bar{d}_i$ is the $i$-th data point predicted by the model
parameters and $d_i$ is the actual measurement. Both experimental errors
are summed quadratically. The statistical error is assumed to be
uncorrelated between different measurements. 

\subsubsection{Setup and mSUGRA toy analysis}
\label{sec:para_sugra}

It would be scientific cheating to assume a specific model for
supersymmetry breaking for actual LHC parameter extraction analyses
unless the goal is to test only such a particular model.  Instead, the
underlying structure of masses and couplings should be inferred from
data.  However, attempting to fit the spectrum using a simple mSUGRA
Ansatz for the underlying structure of the soft parameters allows us
to identify many key features of a real TeV-scale new physics
analysis.  A given fit in this framework determines the most likely
values of the input parameters $m_0, m_{1/2}, A_0$, and
$\tan\beta$~\cite{Allanach:2002nj} defined at the input scale $M_\text{GUT} \sim
10^{16}\,\gev$ from mass or endpoint measurements at the LHC.  The
correct sign of $\mu$ is determined by the quality of the fit.  This
small number of theory parameters greatly simplifies the analysis
compared to a proper MSSM analysis, and serves as a convenient way to
begin investigating the statistical method of parameter scans.

  Strictly speaking, the mSUGRA toy map covers the parameters 
$m_0,\,m_{1/2},\,A_0,\, B\mu$, and $m_t$, where $B\mu$ is traded for 
the TeV-scale $\tan\beta$, as described in
Section~\ref{sec:para_highscale}. Similarly, the $\mu$ parameter is
traded for the electroweak VEV $v=174\,\gev$.
Usually, we show our results in terms of $\tan\beta$ because it has a 
more obvious relationship to observables.  The running top Yukawa is 
defined at the high scale as one of the mSUGRA model parameters, 
which after renormalization group running predicts low-scale masses 
like the top mass, the superpartner masses, or the
light Higgs mass~\cite{m_h}. This approach should be taken for all
Standard Model parameters~\cite{ben}, couplings and masses, but their
errors are small compared to the error on the top mass (LHC target:
1~GeV; LC target: 0.12~GeV).\bigskip

\begin{figure}[t]
 \begin{minipage}{6cm}
    \begin{tabular}{l|rrrrrr}
     $\chi^2$&$m_0$ &$m_{1/2}$ &$\tan\beta$&$A_0$&$\mu$&$m_t$ \\ \hline
     0.09  &102.0 & 254.0 & 11.5 & -95.2  & $+$ & 172.4 \\
     1.50  &104.8 & 242.1 & 12.9 &-174.4  & $-$ & 172.3 \\
     73.2  &108.1 & 266.4 & 14.6 & 742.4  & $+$ & 173.7 \\
    139.5  &112.1 & 261.0 & 18.0 & 632.6  & $-$ & 173.0 \\
          \dots
     \end{tabular}
 \end{minipage} 
 \hspace*{2cm}
 \begin{minipage}{6cm}
 \includegraphics[width=7cm]{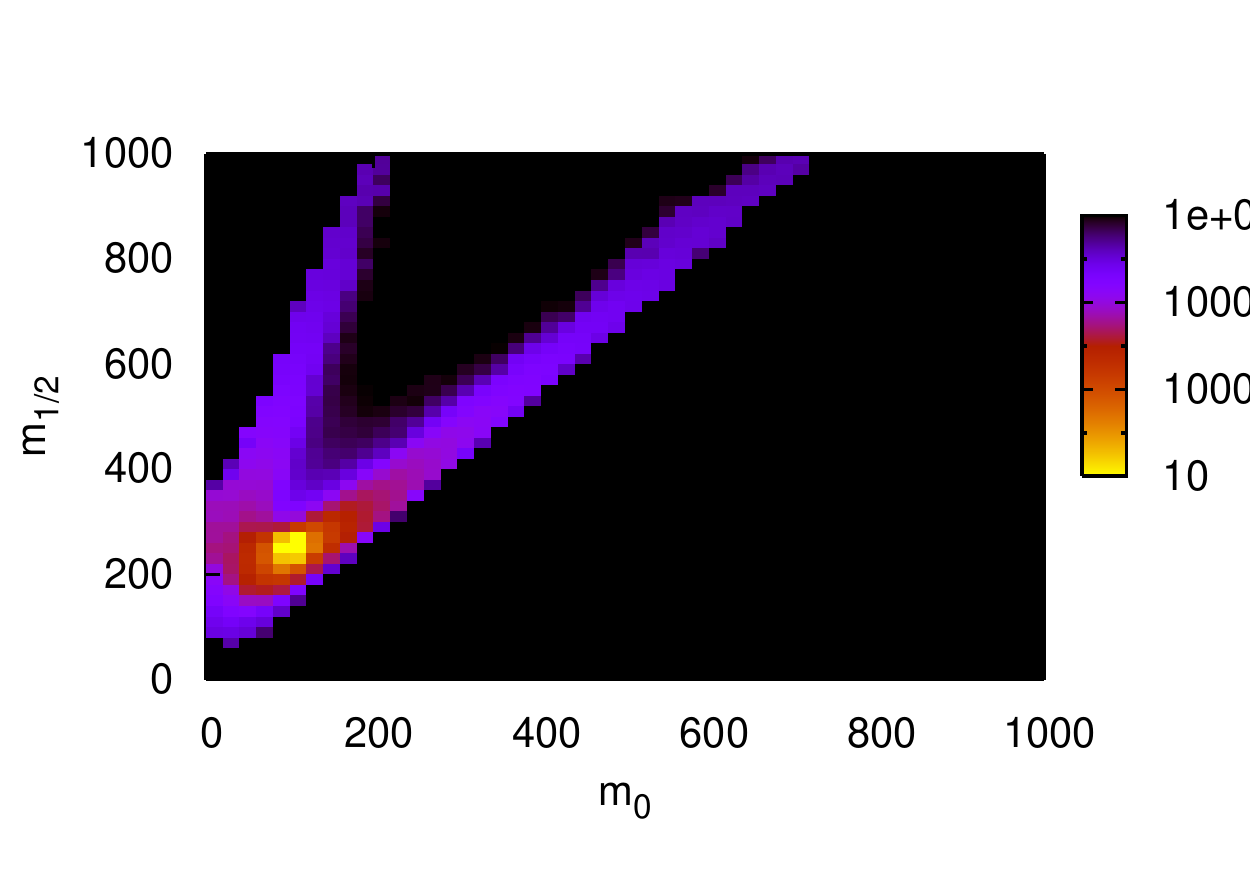}
 \end{minipage} \\
  \includegraphics[width=6cm]{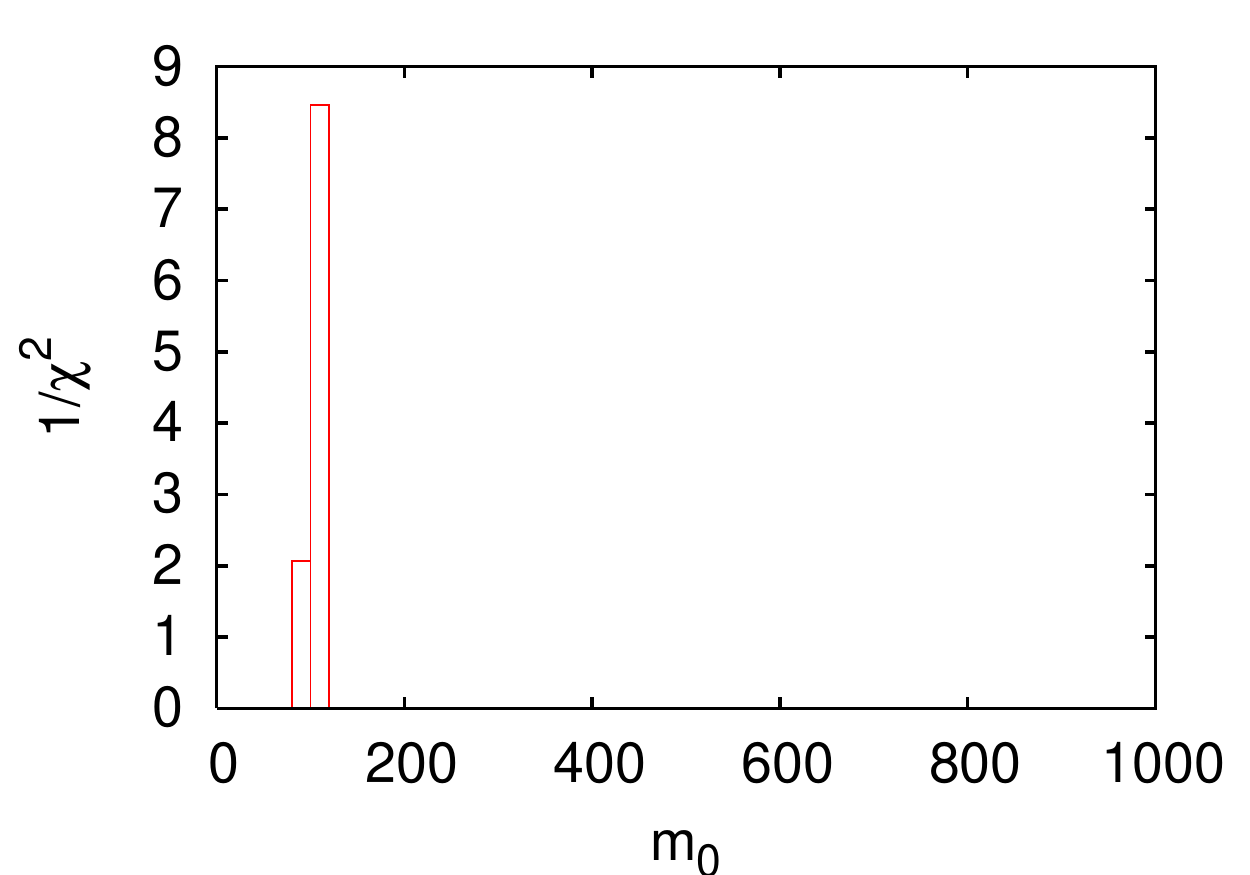} \hspace*{2cm}
  \includegraphics[width=6cm]{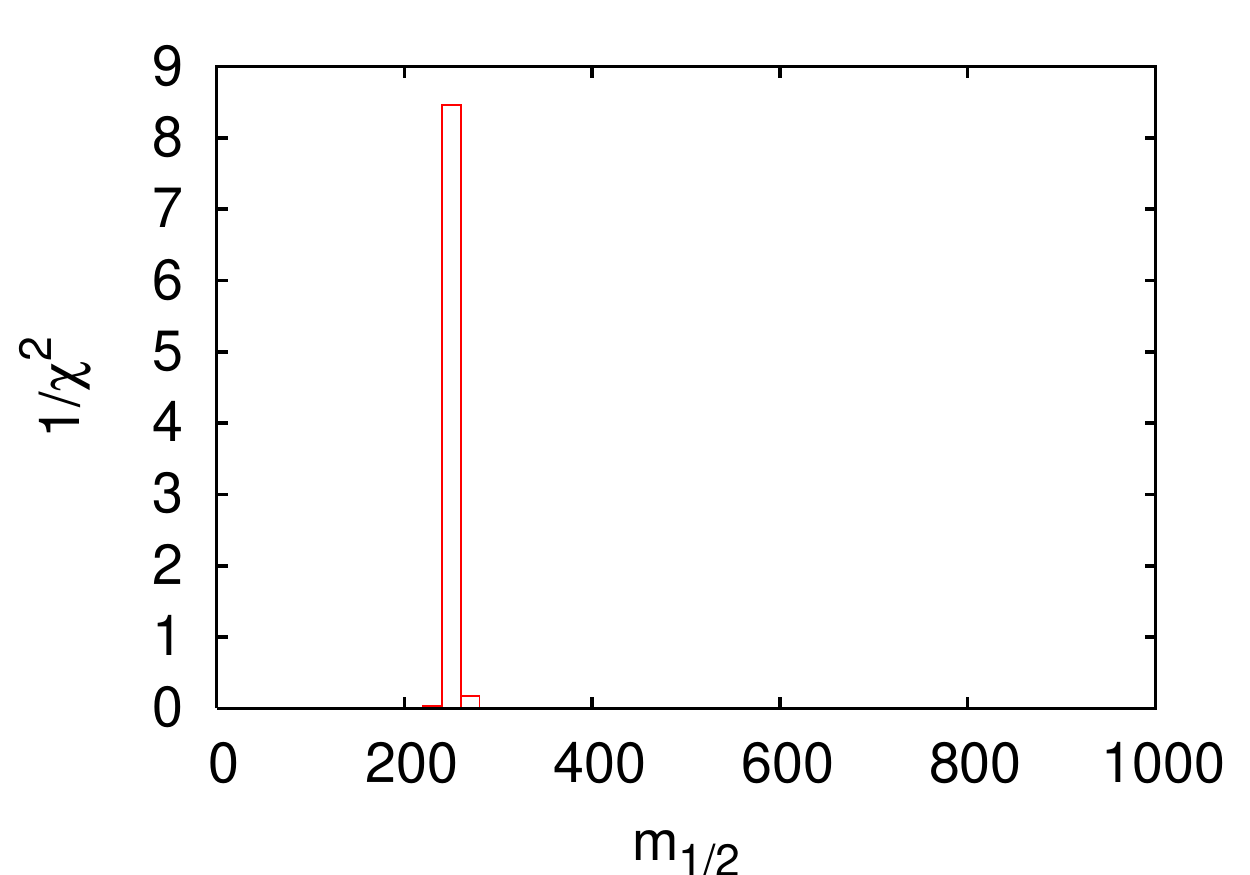}
\caption[]{Profile likelihoods for mSUGRA.  Upper left: list of the
  best log-likelihood values over the mSUGRA parameter space. Upper
  right: two-dimensional profile likelihood $\chi^2$ over the
  $m_0$-$m_{1/2}$ plane.  Lower: one-dimensional profile likelihoods
  $1/\chi^2$ for $m_0$ and $m_{1/2}$. All masses are given in
  GeV. Figure taken from Ref.~\cite{sfitter}.}
\label{fig:para_sugramap_f}
\end{figure}

Examining the projections for the SPS1a parameter point at the LHC, we
find that several parameters are heavily correlated while some are
only poorly constrained, and distinct log-likelihood maxima can differ
by $\om(1\!-\!10)$.  The first question is always if whether we can
unambiguously identify the correct parameters from a set of
observables and their errors, even in the simplified mSUGRA context.
In other words, which parameter point has the largest likelihood value
$p(d|m)$, evaluated as a function over the model space $m$ for given
data $d$.  Discrete model assumptions --- like mSUGRA vs extra
dimensions --- are not included in this question.

  In Fig.~\ref{fig:para_sugramap_f} we show
the best-fit points in the mSUGRA parameter space,
as obtained from the 5-dimensional likelihood map. A general pattern 
of four distinct maxima emerges as a result of two degeneracies:
for the first, the trilinear coupling can assume the correct
value of around $A_0\simeq -100$~GeV, but it can also become large 
and positive. This is correlated with a slight shift in the top mass and
is due to the light Higgs mass, so it depends on the dominant theory
error in $m_h$~\cite{m_h}. 
Secondly, the sign of $\mu$ and a slight shift in $\tan\beta$
compensate each other in the neutralino-chargino sector. Such a
degeneracy is worsened at the LHC where only one of the two heavy
neutralinos are observed. Including more precise measurements from
the linear collider might remove this degeneracy~\cite{Berger:2007yu}.

  In Fig.~\ref{fig:para_sugramap_f} we see how the significantly
different $\chi^2$ values would allow us to distinguish these four
candidate mSUGRA solutions. Amusingly, changing the theory errors from the
correct flat distribution to a possibly approximate gaussian shape affects 
the ranking of the maxima for our smeared measurements: for gaussian
theory errors $\chi^2$ values of 4.35, 26.1, 10.5, 22.6 appear in the
order shown in Fig.~\ref{fig:para_sugramap_f}.  Given that just 
smearing the measurements can shift the ordering of the best local 
maxima, it is important to study carefully more than just the best
solution and its local vicinity.
\bigskip

  For a complete analysis we also want to compute probability
distributions or likelihoods over subspaces of the model space. In
other words, we would like to eliminate dimensions of the parameter 
space until only one- or two-dimensional distributions remain. 
The likelihood cannot just be integrated unless we define an integration
measure in the model space. This measure introduces a prior and leads
us to a Bayesian probability.  To avoid this prior, we first define the
profile likelihood: for each binned parameter point in the
$(n-1)$--dimensional space we explore the $n$th direction which is to
be removed $\mathcal{L}(x_{1,...,n-1},x_n)$. Along this direction we
pick the best value of $\mathcal{L}^{\text{max}(n)}$ and identify its
value with the lower-dimensional parameter point
$\mathcal{L}(x_{1,...,n-1}) \equiv \mathcal{L}^{\text{max}(n)}
(x_{1,...,n-1},x_n)$. This projection most notably
guarantees that the best-fit points survive best to the final
representation. On the other hand, profile likelihoods are not
probability distributions because they do not protect the
normalization of the original likelihood map.

  An example of such a correlation is the profile likelihood in the
$m_0$--$m_{1/2}$ plane, after projecting away the $A_0$, $B\mu$,
$\sign(\mu)$ and $m_t$ directions. The maximum starts from
the true values $m_0=100$~GeV and $m_{1/2}=250$~GeV and continues along
two branches. They reflect the fact that extracting masses
from kinematic endpoints involves quadratic equations. Ignoring such
correlations, we can project the two-dimensional profile likelihood
onto each of the remaining directions. Both distributions show sharp 
maxima in the correct places because the resolution is not sufficient 
to resolve the four distinct solutions for 
$A_0$ and $\text{sign}(\mu)$.\bigskip

\begin{figure}[t]
 \begin{minipage}{6cm}
    \begin{tabular}{l|rrrrrr}
     $\chi^2$&$m_0$ &$m_{1/2}$ &$\tan\beta$&$A_0$&$\mu$&$m_t$ \\ \hline
     0.09  &102.0 & 254.0 & 11.5 & -95.2  & $+$ & 172.4 \\
     1.50  &104.8 & 242.1 & 12.9 &-174.4  & $-$ & 172.3 \\
     73.2  &108.1 & 266.4 & 14.6 & 742.4  & $+$ & 173.7 \\
    139.5  &112.1 & 261.0 & 18.0 & 632.6  & $-$ & 173.0 \\
          \dots
     \end{tabular}
 \end{minipage} 
 \hspace*{2cm}
 \begin{minipage}{6cm}
 \includegraphics[width=7cm]{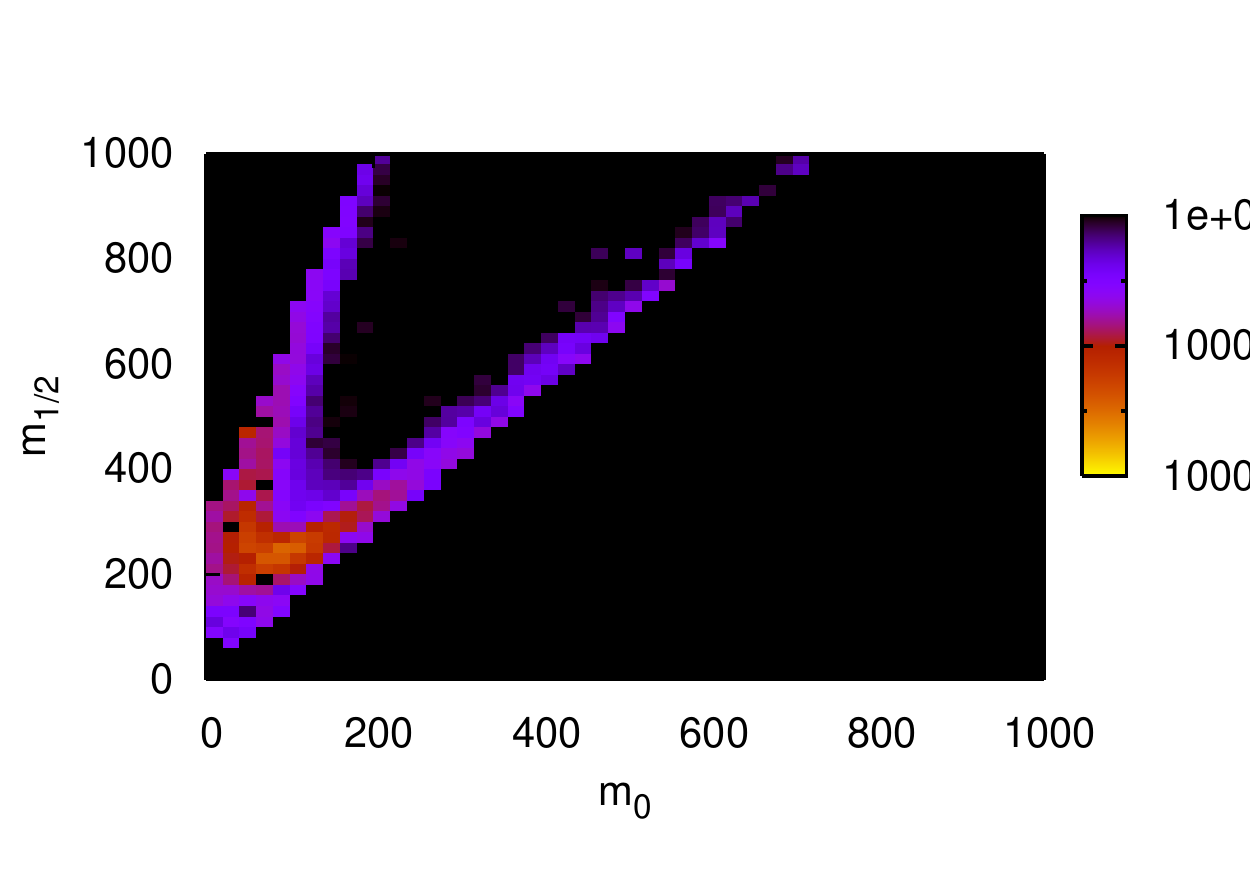}
 \end{minipage} \\
  \includegraphics[width=6cm]{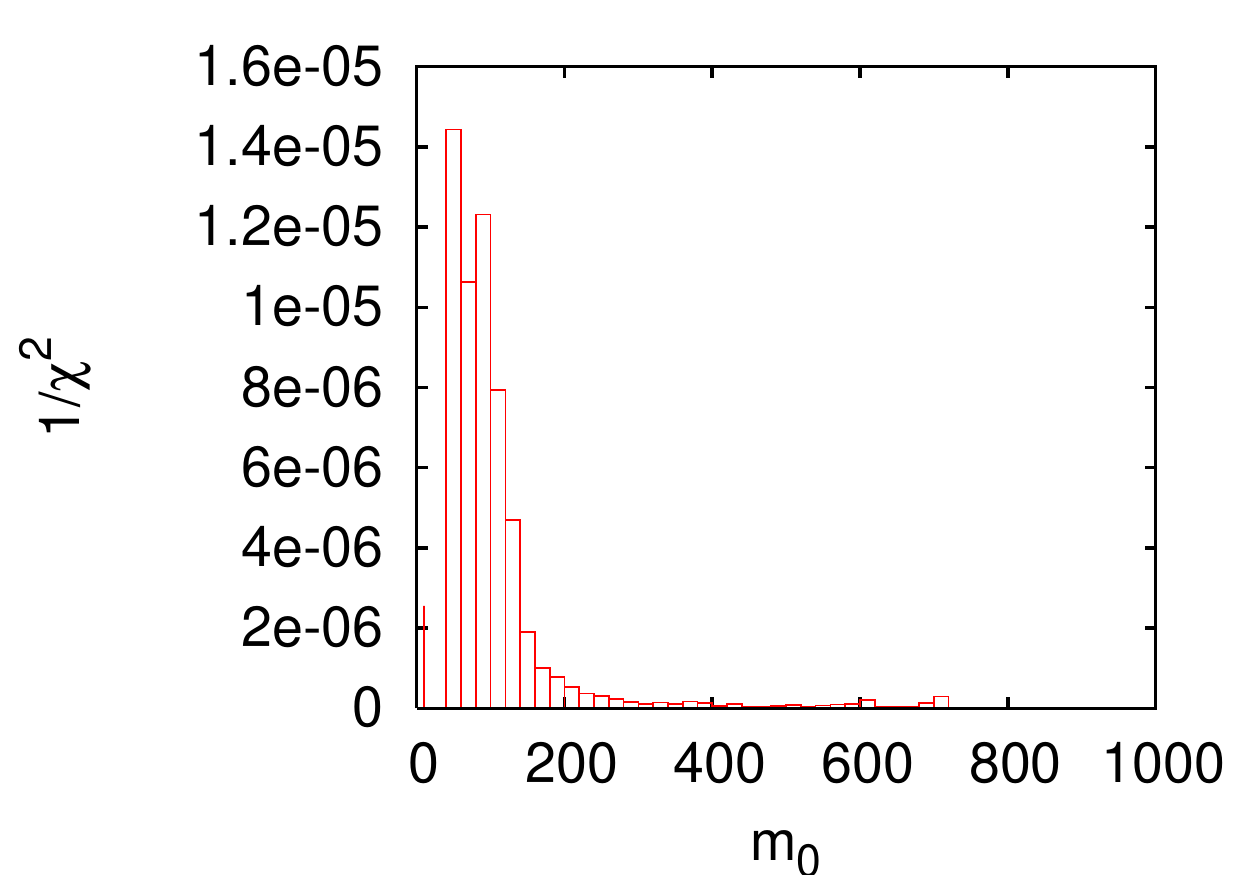} \hspace*{2cm}
  \includegraphics[width=6cm]{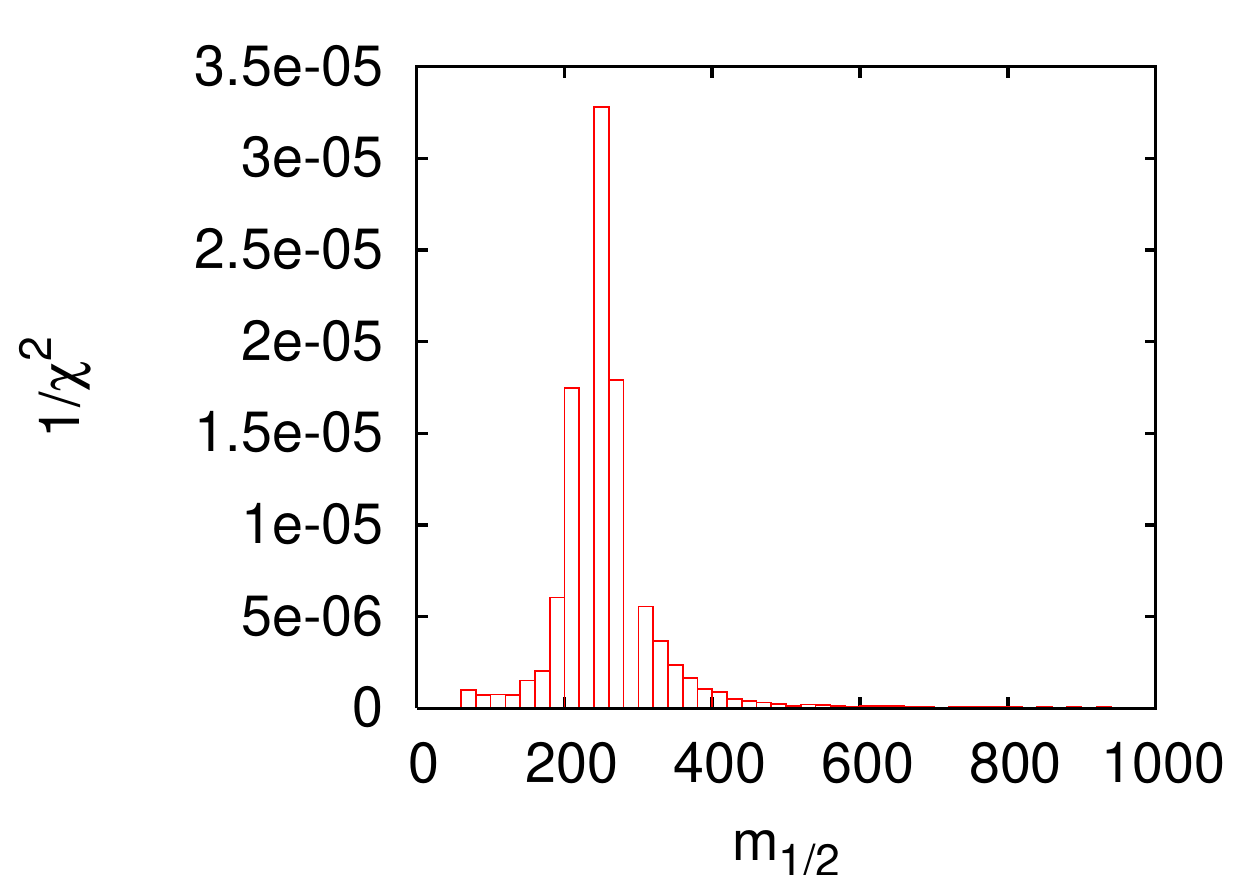}
\caption[]{Bayesian probabilities for mSUGRA.  Upper left: list of the
  largest log-likelihood values over the mSUGRA parameter space. Upper
  right: two-dimensional Bayesian probability $\chi^2$ over the
  $m_0$-$m_{1/2}$ plane marginalized over all other parameters.
  Lower: one-dimensional Bayesian probabilities $1/\chi^2$ for $m_0$
  and $m_{1/2}$.  All masses are given in GeV.  Figure taken from
  Ref.~\cite{sfitter}.}
\label{fig:para_sugramap_b}
\end{figure}

A likelihood analysis is unfortunately not designed to produce
probability distributions and to answer questions such as
which sign of $\mu$ is preferred in mSUGRA, in light of precision
electroweak bounds and dark matter constraints~\cite{dark_side}.
Note that this is not the same question as 
what the relative difference in the likelihood for the two best
points on each side of $\mu$ is. To answer the first question the
likelihood over each half of the parameter space needs to be
integrated. For such an integration we introduce an integration
measure or Bayesian prior. This has its advantages, but it can also
lead to unexpected effects~\cite{ben}. One might argue that such
questions are irrelevant because our goal is to find the correct
(\ie the most likely) parameter point. On the other hand, asking for a
reduced-dimensionality probability density could well be a very
typical situation in the LHC era. Questions such as what is the most
likely mechanism for dark matter annihilation deserve a well-defined
answer.\bigskip

Shifting from a frequentist to a Bayesian approach does not affect the
actual likelihood map, aside from possible differences in the
treatment and combination of errors or nuisance parameters.
Therefore, the top-likelihood points from
Fig.~\ref{fig:para_sugramap_f} also appear in the Bayesian-based
Fig.~\ref{fig:para_sugramap_b}. The second panel in
Fig.~\ref{fig:para_sugramap_b} now shows a two-dimensional
representation of the Bayesian probability over the mSUGRA parameter
space. All parameters except for $m_0$ and $m_{1/2}$ are marginalized
using flat priors.
We observe the same two-branch structure as for the profile
likelihood. However, there are two differences.  First, 
the area around the true parameter point is less pronounced in the Bayesian
probability. The integration over parameter space collects noise from
large regions with a finite but insignificant likelihood. This washes
out the peaks, while the profile likelihood by construction ignores
poor-fit areas.  This effect also considerably smears the
one-dimensional Bayesian probabilities.
Secondly, the branch structure is more pronounced in
Fig.~\ref{fig:para_sugramap_b}.  While in the profile likelihood the
compact area between the branches is filled by single good parameter
points, the Bayesian marginalization provides `typical' likelihood
values in this region which in general does not fit the data as
well.\bigskip

\begin{figure}[t]
 \includegraphics[width=7cm]{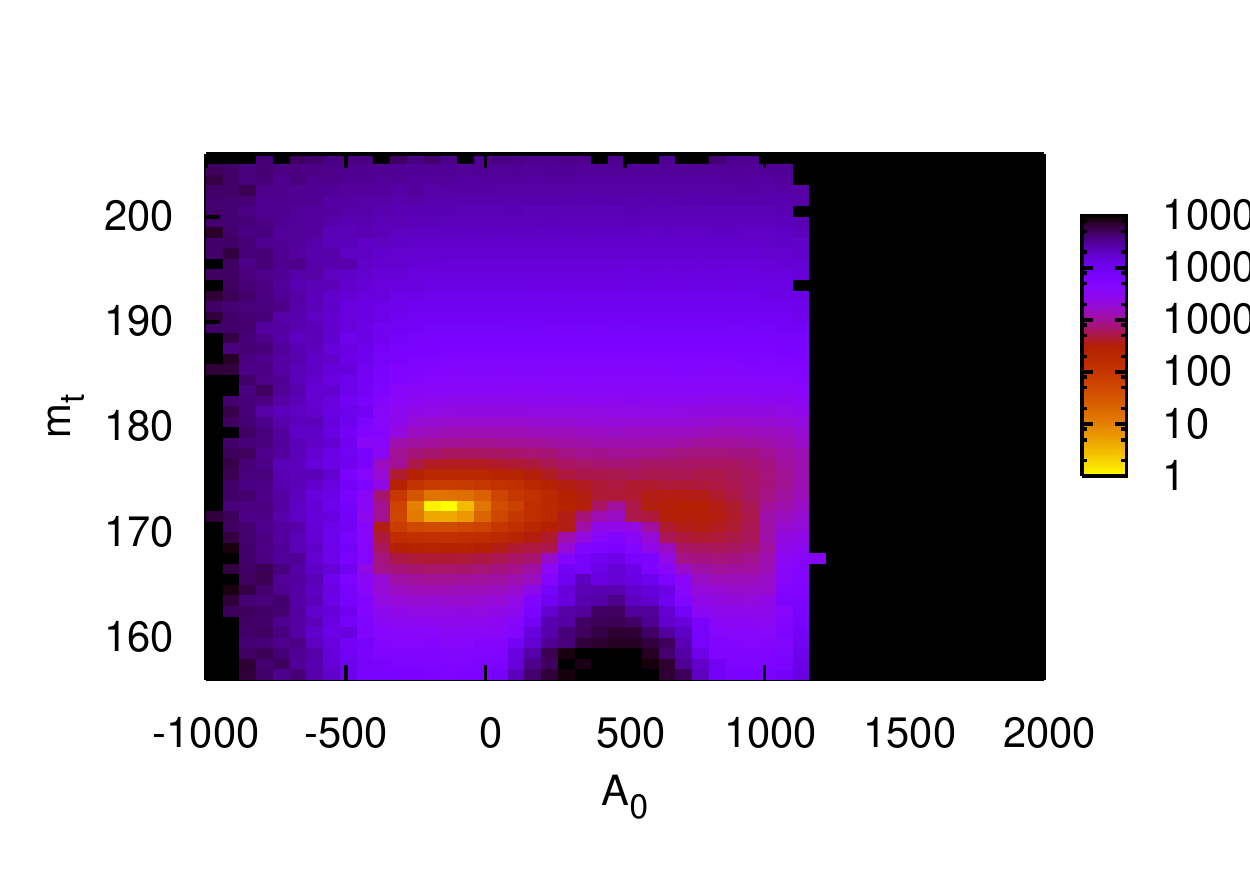} \hspace*{2cm}
 \includegraphics[width=7cm]{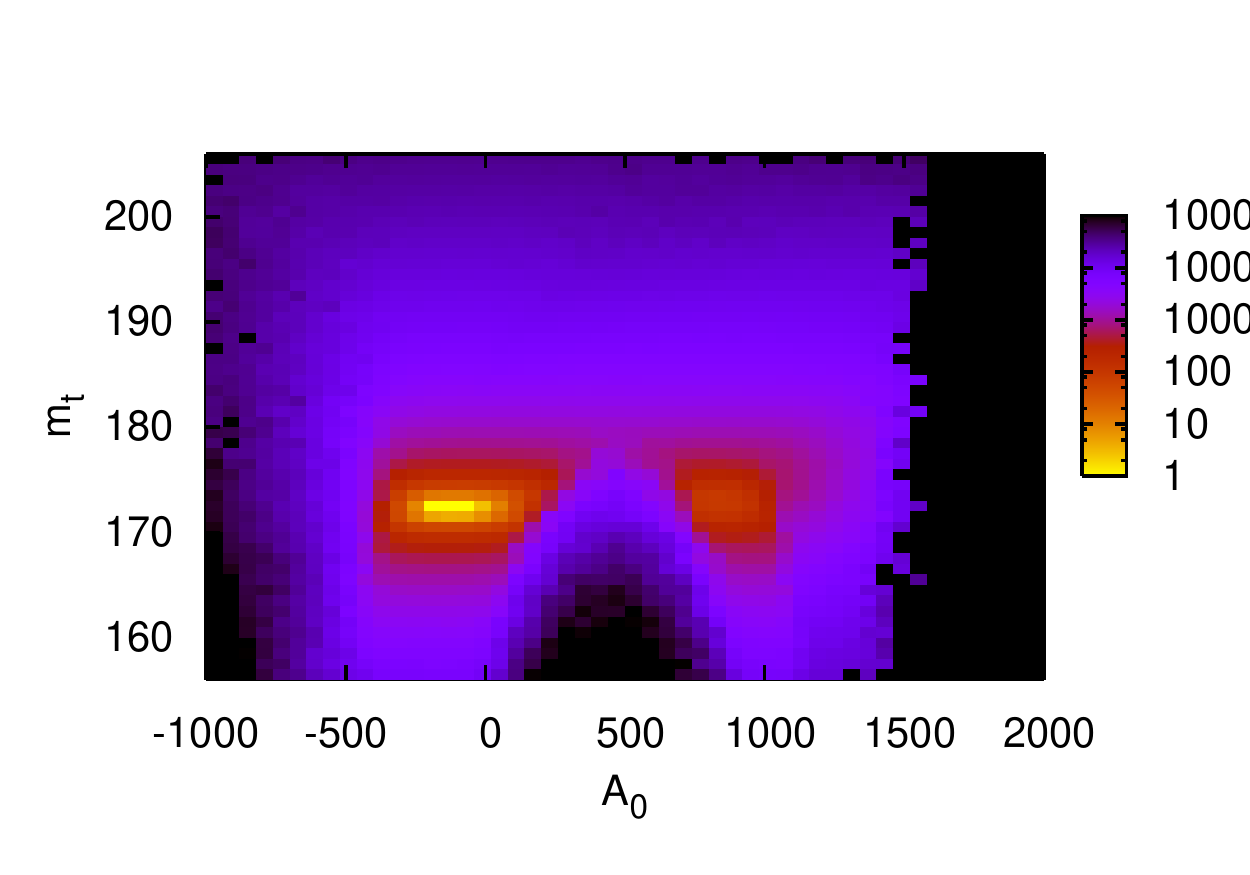} \\[-5mm]
 \includegraphics[width=7cm]{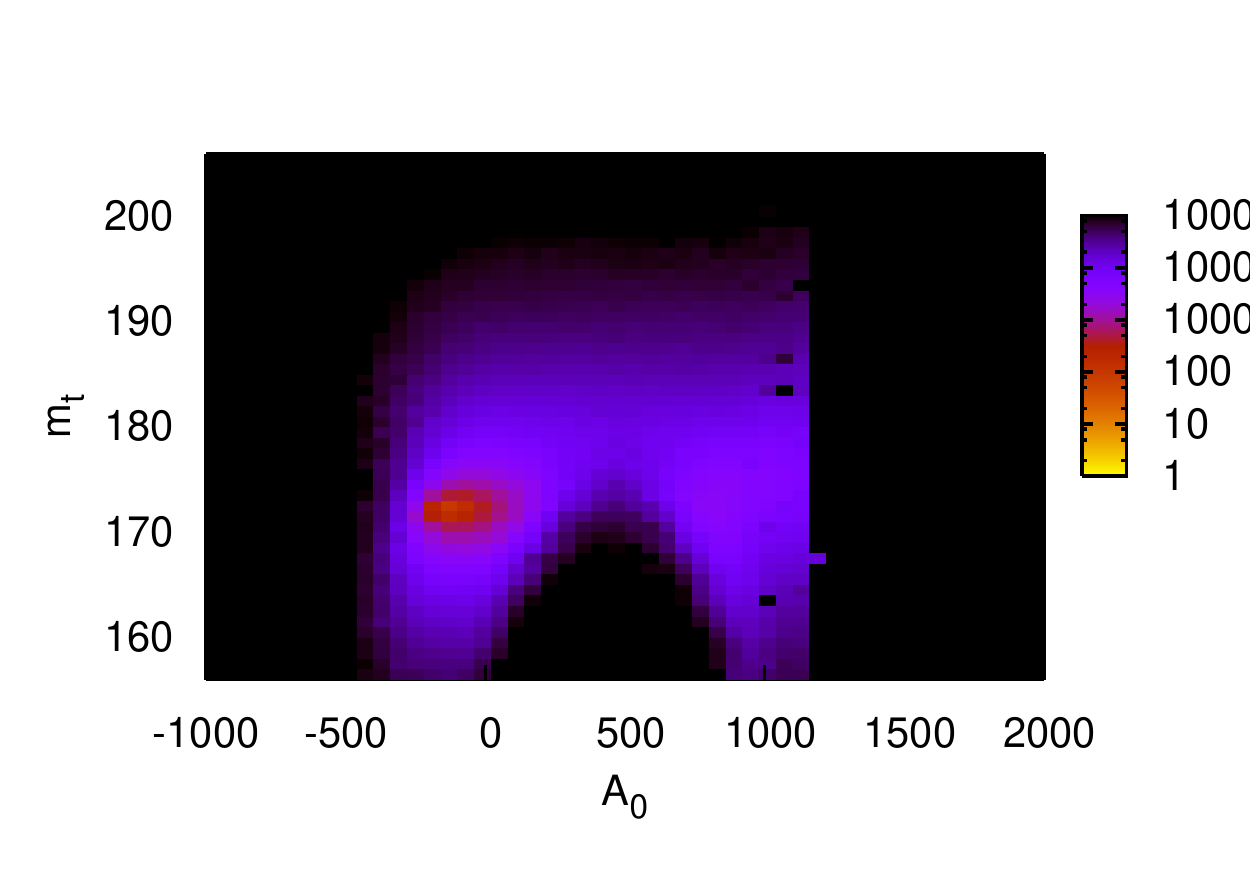} \hspace*{2cm}
 \includegraphics[width=7cm]{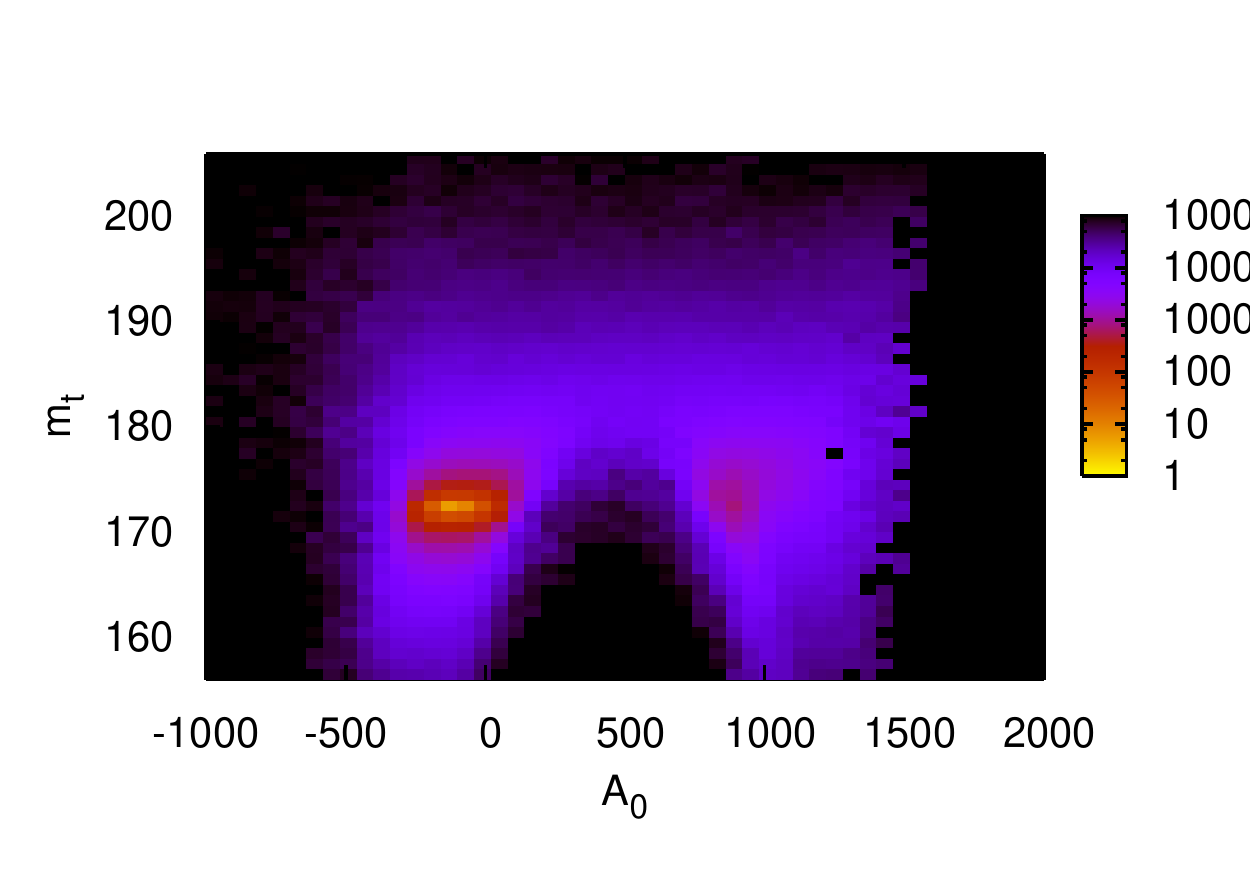} 
\caption[]{Profile likelihoods (upper) and Bayesian probabilities
  (lower) for $A_0$ and $y_t$. The two columns of one- and
  two-dimensional profile likelihoods correspond to $\mu<0$ (left) and
  $\mu>0$ (right). Figure taken from Ref.~\cite{sfitter}.}
\label{fig:para_sugramap_2}
\end{figure}

Eliminating a dimension in any parameter space means loss of
information. Fig.~\ref{fig:para_sugramap_2} shows the two-dimensional
profile likelihoods in the $m_t$--$A_0$ subspace for each sign of
$\mu$.  Locally, the two-dimensional profile likelihoods around the
maxima show little correlation between $m_t$ and $A_0$. The correct
value is preferred, but the alternative solution around $A_0\sim 700\,\gev$
is clearly visible. On top of this double-maximum structure, 
for both signs of $\mu$ there is a parabola-shaped correlation 
between $m_t$ and $A_0$ with an apex roughly 5~GeV above the best 
fits in $m_t$. This correlation becomes invisible once we project away 
one of the two parameter directions.  As usual, we compare the profile 
likelihood and the Bayesian probability for $A_0$ and see that volume 
effects in the latter significantly enhance the relative weight of the secondary
maximum.  Moreover, comparing the likelihood scales for $\mu<0$ and
(the correct) $\mu>0$ values, the relative enhancement of the Bayesian
probability is by almost an order of magnitude, while the profile
likelihoods for the two best-fit points differ by only a factor
five.\bigskip

\begin{table}[t]
\begin{small}
\begin{tabular}{|l|r|c|ccc|}
\hline
            & SPS1a  & $\Delta^\text{theo-exp}_\text{zero}$
                     & $\Delta^\text{expNoCorr}_\text{zero}$ 
                     & $\Delta^\text{theo-exp}_\text{zero}$ 
                     & $\Delta^\text{theo-exp}_\text{flat}$ \\
\hline
            &        & masses 
                     & \multicolumn{3}{c|}{endpoints} \\
\hline
$m_0$       & 100    & 4.11 & 1.08 & 0.50 & 2.17 \\
$m_{1/2}$   & 250    & 1.81 & 0.98 & 0.73 & 2.64 \\
$\tan\beta$ & 10     & 1.69 & 0.87 & 0.65 & 2.45 \\
$A_0$       & -100   & 36.2 & 23.3 & 21.2 & 49.6 \\
$m_t$       & 171.4  & 0.94 & 0.79 & 0.26 & 0.97 \\
\hline
\end{tabular}
\end{small}
\caption[]{Best-fit results for mSUGRA at the LHC derived from masses
  and endpoint measurements with absolute errors in GeV. The subscript
  represents neglected and flat theory errors. The experimental error
  includes correlations unless indicated otherwise in the
  superscript. The top mass is quoted in the on-shell scheme. Table
  taken from Ref.~\cite{sfitter}}
\label{tab:para_mass_edge}
\end{table}

  To determine model parameters, two different measurement sets are
available at the LHC: kinematic endpoints, or particle masses from a
fit to the endpoints without model assumptions~\cite{edges,sps1a_per,mgl_per}.
Because the extraction of masses from endpoints is highly correlated,
both approaches are only equivalent if the complete correlation matrix
of masses is taken into account. For the mSUGRA toy model we can
compare the endpoint results with an ansatz where we use the masses
while forgetting about their correlations. Again, this question is
completely independent of the supersymmetric framework --- the
question is what kind of measurements we should use to extract mass
spectra from cascade decays.

  Typical errors on the mSUGRA parameters for different assumptions are
shown in Table~\ref{tab:para_mass_edge}.  Switching from mass
measurements to endpoints for gaussian experimental errors and no
correlations improves the precision by more than a factor two. This
improvement arises from the absence of the correlation matrix between
the mass observables. If this matrix were known, the two results would
be similar. Taking into account the correlation of the systematic
energy-scale errors gives us another improvement by a
factor two for $m_0$.  Including theory uncertainties increases the
combined error bars on the experimentally well-measured scalar and gaugino
masses considerably.  For $\tan\beta$ the theory error on the light 
Higgs mass of about $2\,\gev$ is the dominant source, coming from the 
uncertainty in the top Yukawa as well as unknown higher orders~\cite{m_h}.  
The top-pole mass measurement at the LHC is expected to give an
experimental error of about $1\,\gev$, and as long as the experimental
error stays in this range, the theory error on the top mass from the
unknown renormalization scheme of $m_t$ at a hadron
collider~\cite{theo_mt} should be small, due to $\lambda_\text{QCD} \ll$~GeV.
This simple example shows how important
a proper treatment of errors can be, in particular in analyses where
the experimental measurements are precise and complex.

\subsubsection{Once high scale always high scale}
\label{sec:para_highscale}

\begin{figure}[t]
 \includegraphics[width=6cm]{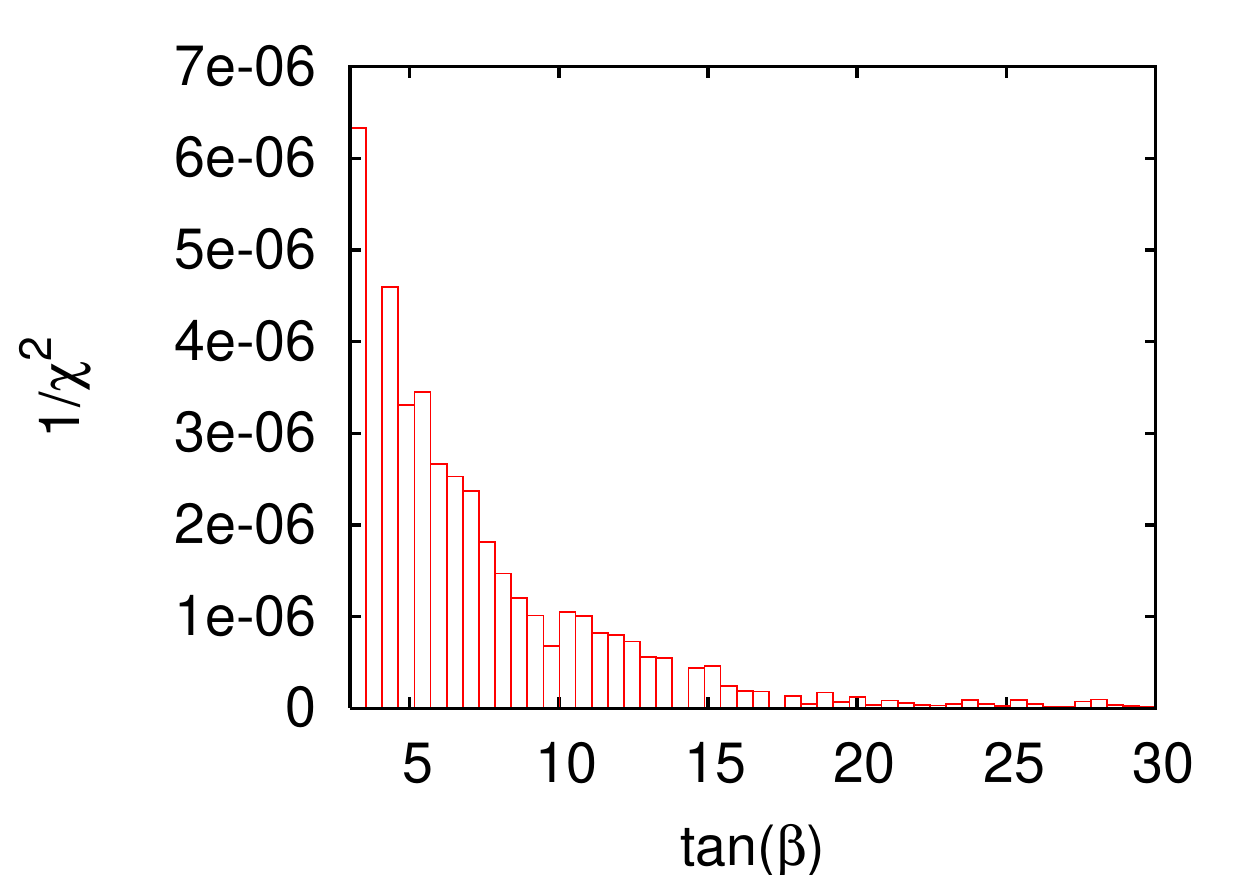} 
  \hspace*{10mm}
 \includegraphics[width=6cm]{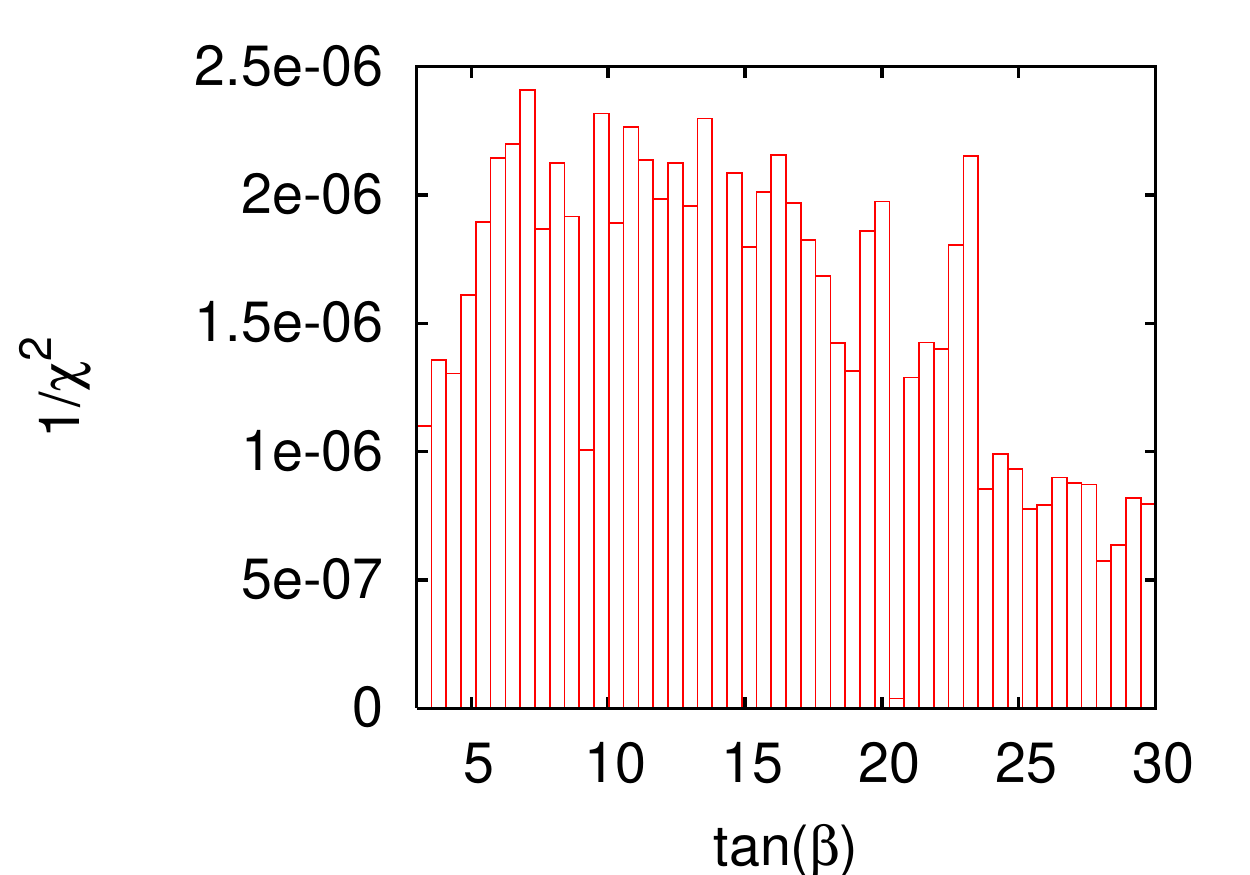} 
\caption[]{Bayesian probabilities for $\tan\beta$ with a flat prior in
  $B$ (left) and a flat prior in $\tan\beta$ (right). Figure taken
  from Ref.~\cite{sfitter}.}
\label{fig:para_highscale}
\end{figure}

A Bayesian analysis with its priors can depend on the way we define
our models in detail.  The usual set of mSUGRA model parameters
contains the high-scale mass parameters $m_0, m_{1/2}, A_0$, and on
the other hand the TeV-scale ratio of vacuum expectation values
$\tan\beta$, which assumes radiative electroweak symmetry
breaking.  Minimizing the potential in the directions of both VEVs
gives the two conditions~\cite{drees_martin}:
\begin{alignat}{5}
\mu^2 &=  \frac{m^2_{H_u} \sin^2 \beta - m^2_{H_u} \cos^2 \beta}
               {\cos 2 \beta}
        - \frac{1}{2} m_Z^2 \notag \\
2 B \mu &= \tan 2 \beta \; \left( m^2_{H_d} - m^2_{H_u} 
                           \right)
           + m_Z^2 \sin 2 \beta
\end{alignat}
The masses $m_{H_{u,d}}$ are the soft masses of the two Higgs doublets.
These are fixed in mSUGRA by the value of $m_0$ at the high input scale,
so with this and by specifying $\tan\beta$ we can eliminate $B\mu$
and $|\mu|$.

  Another approach is to replace $\tan\beta$ with $B\mu$ at
the high scale together with $m_0$ and compute $\tan\beta$ and
$|\mu|$ assuming electroweak symmetry breaking and the $Z$ mass. Now
all input parameters are high-scale mass parameters. This does not
make a difference for frequentist profile-likelihood map, but in a
Bayesian approach it does matter through volume effects. \bigskip
To illustrate the effects of flat priors either in $B\mu$ or in
$\tan\beta$ we show one-dimensional Bayesian $\tan\beta$ probabilities
in Fig.~\ref{fig:para_highscale}. From the best-fit points in
Fig.~\ref{fig:para_sugramap_f} even after including theory errors the
correct value for $\tan\beta$ can be determined.  However,
Fig.~\ref{fig:para_highscale} clearly shows that with a flat prior in
$B$ the one-dimensional Bayesian probability is largely dominated by
noise and by a bias towards as small as possible $\tan\beta$. This
bias is simply an effect of the prior. Switching to a flat prior in
$\tan\beta$, noise effects are still dominant, but the maximum of the
one-dimensional Bayesian probability is in the correct place. As
expected, the profile likelihood picks the correct central value of
$\tan\beta \sim 12$ for the smeared parameter point.

\subsubsection{TeV-scale MSSM}
\label{sec:para_mssm}

If any kind of new physics should be observed at the TeV scale, 
we need to determine as many of its model parameters as possible 
to understand the underlying structure.  High-scale models can then 
be inferred from TeV-scale data~\cite{sfitter,fittino,inverse,
Kane:2006yi,Kane:2006hd}.  On the other hand,
even at multi-purpose experiments like the LHC we can ignore some
new-physics parameters because no information about them is
expected. Such parameters will likely include
CP-violating phases already constrained by searches for electric dipole
moments~\cite{with_toby}
or similarly constrained non-symmetric flavor
structures~\cite{susy_flavor}. 
It could also include parameters which in a chiral model are proportional to
light Yukawa couplings, like the light-generation trilinear 
couplings $A_{l,u,d}$ in supersymmetry.  Third-generation $A_{\tau, b}$ can 
in principle play a role as off-diagonal entries in mass matrices,
and their effects could be easier to observe, but they are usually
subleading compared to $A_t$.
Applying these arguments leads to an effective 19--dimensional
TeV-scale MSSM parameter space listed for example in
Table~\ref{tab:para_secondary}. Obviously, such an assumption of
parameters being irrelevant for the TeV-scale likelihood map can 
and must be tested.\bigskip

\begin{table}
\begin{tabular}{|l|rrrr|rrrr|}
\hline
                     &\multicolumn{4}{c|}{$\mu<0$}&\multicolumn{4}{c|}{$\mu>0$}\\
\hline
                     &      &      &      &      &SPS1a&      &      &       \\
\hline \hline
$M_1$                &  96.6& 175.1& 103.5& 365.8&  98.3& 176.4& 105.9& 365.3\\
$M_2$                & 181.2&  98.4& 350.0& 130.9& 187.5& 103.9& 348.4& 137.8\\
$\mu$                &-354.1&-357.6&-177.7&-159.9& 347.8& 352.6& 178.0& 161.5\\
$\tan\beta$          &  14.6&  14.5&  29.1&  32.1&  15.0&  14.8&  29.2&  32.1\\
\hline
$M_3$                & 583.2& 583.3& 583.3& 583.5& 583.1& 583.1& 583.3& 583.4\\
$M_{\tilde{\tau}_L}$ & 114.9&2704.3& 128.3&4794.2& 128.0& 229.9&3269.3& 118.6\\
$M_{\tilde{\tau}_R}$ & 348.8& 129.9&1292.7& 130.1&2266.5& 138.5& 129.9& 255.1\\
$M_{\tilde{\mu}_L}$  & 192.7& 192.7& 192.7& 192.9& 192.6& 192.6& 192.7& 192.8\\
$M_{\tilde{\mu}_R}$  & 131.1& 131.1& 131.1& 131.3& 131.0& 131.0& 131.1& 131.2\\
$M_{\tilde{e}_L}$    & 186.3& 186.4& 186.4& 186.5& 186.2& 186.2& 186.4& 186.4\\
$M_{\tilde{e}_R}$    & 131.5& 131.5& 131.6& 131.7& 131.4& 131.4& 131.5& 131.6\\
$M_{\tilde{q}3_L}$   & 497.1& 497.2& 494.1& 494.0& 495.6& 495.6& 495.8& 495.0\\
$M_{\tilde{t}_R}$    &1073.9& 920.3& 547.9& 950.8& 547.9& 460.5& 978.2& 520.0\\
$M_{\tilde{b}_R}$    & 497.3& 497.3& 500.4& 500.9& 498.5& 498.5& 498.7& 499.6\\
$M_{\tilde{q}_L}$    & 525.1& 525.2& 525.3& 525.5& 525.0& 525.0& 525.2& 525.3\\
$M_{\tilde{q}_R}$    & 511.3& 511.3& 511.4& 511.5& 511.2& 511.2& 511.4& 511.5\\
$A_t$ $(-)$          &-252.3&-348.4&-477.1&-259.0&-470.0&-484.3&-243.4&-465.7\\
$A_t$ $(+)$          & 384.9& 481.8& 641.5& 432.5& 739.2& 774.7& 440.5& 656.9\\
$m_A$                & 350.3& 725.8& 263.1&1020.0& 171.6& 156.5& 897.6& 256.1\\
$m_t$                & 171.4& 171.4& 171.4& 171.4& 171.4& 171.4& 171.4& 171.4\\
\hline
\end{tabular}
\caption[]{List of the eight best-fitting points in the MSSM
  likelihood map with two alternative solutions for $A_t$.  All masses
  are given in GeV. The $\chi^2$ value for all points is approximately
  the same, so the ordering of the table is arbitrary.  The parameter
  point closest to the correct point is labeled as SPS1a.  Table taken
  from Ref.~\cite{sfitter}}
\label{tab:para_secondary}
\end{table}

  In Table~\ref{tab:para_secondary} we show the set of secondary local 
maxima in a likelihood map to the SPS1a parameter under the assumption
of a general 19--parameter $\tev$ scale MSSM.  The most interesting
feature of the different best-fitting points is the structure of the
neutralino sector. For a fixed sign of $\mu$, four equally good
solutions can be classified by the ordering of the mass parameters,
even though $M_1 < M_2 < |\mu|$ is the correct solution.  The reverse 
ordering of the two gaugino masses $M_2 < M_1 < |\mu|$ is equally likely. 
In both cases the missing neutralino will be a higgsino. Apart from these 
two light-gaugino scenarios, the second-lightest neutralino can be mostly a
higgsino, which corresponds to $M_1 < |\mu| < M_2$ and $M_2 < |\mu| <
M_1$. Note that given a set of LHC measurements the two gaugino
masses can always be switched as long as there are no chargino
constraints, even though we will see in Section~\ref{sec:para_dm} that
the gaugino and higgsino fraction usually decide if we are looking at
a model with a viable dark matter candidate or not. The one neutralino
which cannot be a higgsino is the LSP, because in that case $\mu$
would also affect the second neutralino mass and would have to be
heavily tuned, leading to a considerably larger smaller likelihood.
The shift in $\tan\beta$ for the correct SPS1a parameter point is an
effect of the smeared data set combined with the rather poor
constraints on this parameter and is within the errors.
\bigskip

Going back to Table~\ref{tab:para_edges} we see an important feature:
there are 22 measurements which should naively completely constrain
our 19--dimensional parameter space. However, all measurements are
constructed from only 15 underlying masses.  Five model parameters
turn out to be poorly constrained.  For small $\tan \beta$ one problem
is the heavy Higgs mass $m_A$. Other hardly determined parameters are
$M_{\tilde{t}_R}$ and $A_t$, which occur in the stop sector but not in
any of the edge measurements.  Number four is $\tan\beta$.  Looking at
the neutralino and sfermion mixing matrices, any effect in changing
$\tan\beta$ can always be accommodated by a change elsewhere. This is
particularly obvious in the stau sector. There, only the lighter of
$M_{\tilde{\tau}_L}$ or $M_{\tilde{\tau}_R}$ is determined from the
$m_{\tau\tau}$ edge. The heavier mass parameter and $\tan\beta$ can
compensate each other's effects freely. 

  A fairly typical situation is that there is exactly one measurement 
which strongly links all these five otherwise unconstrained parameters: 
the lightest Higgs mass. It defines a four-dimensional surface with 
a constant likelihood that fixes the range of all parameters involved.  
To illustrate the effect of the minimum surface we quote two values 
for $A_t$ in the table of minima. The
significant shift in $|A_t|$ shows the sizeable correlations with the
other parameters. Its origin is the stop contribution to the lightest
Higgs mass which contains sub-leading terms linear in $A_t$.  
If this minimum surface can be constrained by further measurements,
this degeneracy will vanish and correlations will require the other
parameters to shift, in order to accommodate two distinct point-like
minima. However, until we find a way to distinguish these alternative
minima, all secondary best-fit points in the TeV-scale MSSM have
exactly the same (perfect) likelihood, which is the most important
qualitative difference to a highly constrained model
like mSUGRA.\bigskip

\begin{table}[t]
\begin{tabular}{|l|rrr||l|rrr|}
\hline
                     &   SPS1a  & correct    & inverted       &    
                     &   SPS1a  & correct    & inverted       \\ \hline \hline
$M_1$                &    103.1 &     102.1 & 101.6          &
$M_2$                &    192.9 &     193.6 & 191.0          \\
$M_3$                &    577.9 &     582.0 & 582.1          &
$\tan\beta$          &     10.0 &       7.2 &   7.8          \\
$m_A$                &    394.9 &     394.0 & 299.3          &
$\mu$                &    353.7 &     347.7 & 369.3          \\ \hline
$M_{\tilde{e}_L}$    &    194.4 &     192.3 & 192.3          &
$M_{\tilde{e}_R}$    &    135.8 &     134.8 & 134.8          \\
$M_{\tilde{\mu}_L}$  &    194.4 &     191.0 & 191.0          &
$M_{\tilde{\mu}_R}$  &    135.8 &     134.7 & 134.7          \\
$M_{\tilde{\tau}_L}$ &    193.6 &     192.9 & 185.7          &
$M_{\tilde{\tau}_R}$ &    133.4 &     128.1 & 129.9          \\
$M_{\tilde{q}_L}$    &    526.6 &     527.0 & 527.1          &
$M_{\tilde{q}_R}$    &    508.1 &     514.8 & 514.9          \\
$M_{\tilde{q}3_L}$   &    480.8 &     477.9 & 478.5          &
$M_{\tilde{t}_R}$    &    408.3 &     423.6 & 187.6          \\
                     &          &\multicolumn{2}{c||}{}      & 
$M_{\tilde{b}_R}$    &    502.9 &     513.7 & 513.2          \\ \hline
$A_{l1,2}$           &   -251.1 &\multicolumn{2}{c||}{fixed 0}&
$A_\tau$             &   -249.4 &\multicolumn{2}{c|}{fixed 0}\\
$A_{d1,2}$           &   -821.8 &\multicolumn{2}{c||}{fixed 0}&
$A_b$                &   -763.4 &\multicolumn{2}{c|}{fixed 0}\\
$A_{u1,2}$           &   -657.2 &\multicolumn{2}{c||}{fixed 0}&
$A_t$                &   -490.9 &    -487.7 & -484.9         \\ \hline
$m_t$                &    171.4 &     172.2 &  172.2         &
                     &          &\multicolumn{2}{c|}{}        \\ \hline
\end{tabular}
\caption[]{Result for the MSSM parameter determination using the LHC
  endpoint measurements assuming either the third or fourth neutralino
  to be missing. The likelihood for both points is almost
  identical. All masses are given in GeV. Table taken from
  Ref.~\cite{sfitter}}
\label{tab:para_swapped}
\end{table}

According to the measurements listed in
Table~\ref{tab:para_mass_errors}, the LHC will only identify three out
of four neutralinos --- the third-heaviest neutralino we will miss due
to its higgsino nature. This situation might be typical at the LHC,
where there is no guarantee that we will be able to see all the particles
of the weakly interacting new-physics sector.  The question is what
happens if the fourth-heaviest neutralino is wrongly labeled as
third-heaviest. Obviously, following the above discussion we always
find a best-fitting parameter point. Two fits with the correct and
swapped neutralino mass assignments are shown in
Table~\ref{tab:para_swapped}. There are small shifts in all parameters
entering the neutralino mass matrix, but none of them appear
significant.  The shift in $\tan\beta$ is an effect of the smeared
parameter point. The relatively large shift in the heavy Higgs mass
between the two scenarios will turn out to be well within the wide
error bands.  In contrast, the lightest Higgs mass in both points is
identical, which means that for the typical LHC precision the
parameter point SPS1a is in the decoupling limit of the heavy MSSM
Higgs states. 

The key to distinguishing alternative minima is predicting more
properties of the `wrong' hypotheses, and searching for corresponding
signatures at the LHC or elsewhere. For example, it should be possible
to search for higgsinos in cascade decays involving gauge bosons. Such
a measurement or non-measurement would remove this particular
degeneracy.  The same would be true for chargino masses which
unfortunately are not part of the standard SPS1a sample~\cite{mihoko}.
Relying on such non-measurements poses a big problem for any kind
of automized approach.
What can and what cannot be seen at the LHC is
largely determined by Standard Model backgrounds and detector effects.
Either a detailed analysis or serious experimental intuition will be
needed, making an automated answering algorithm for such questions
unrealistic.\bigskip

\begin{figure}[t] 
 \includegraphics[width=7cm]{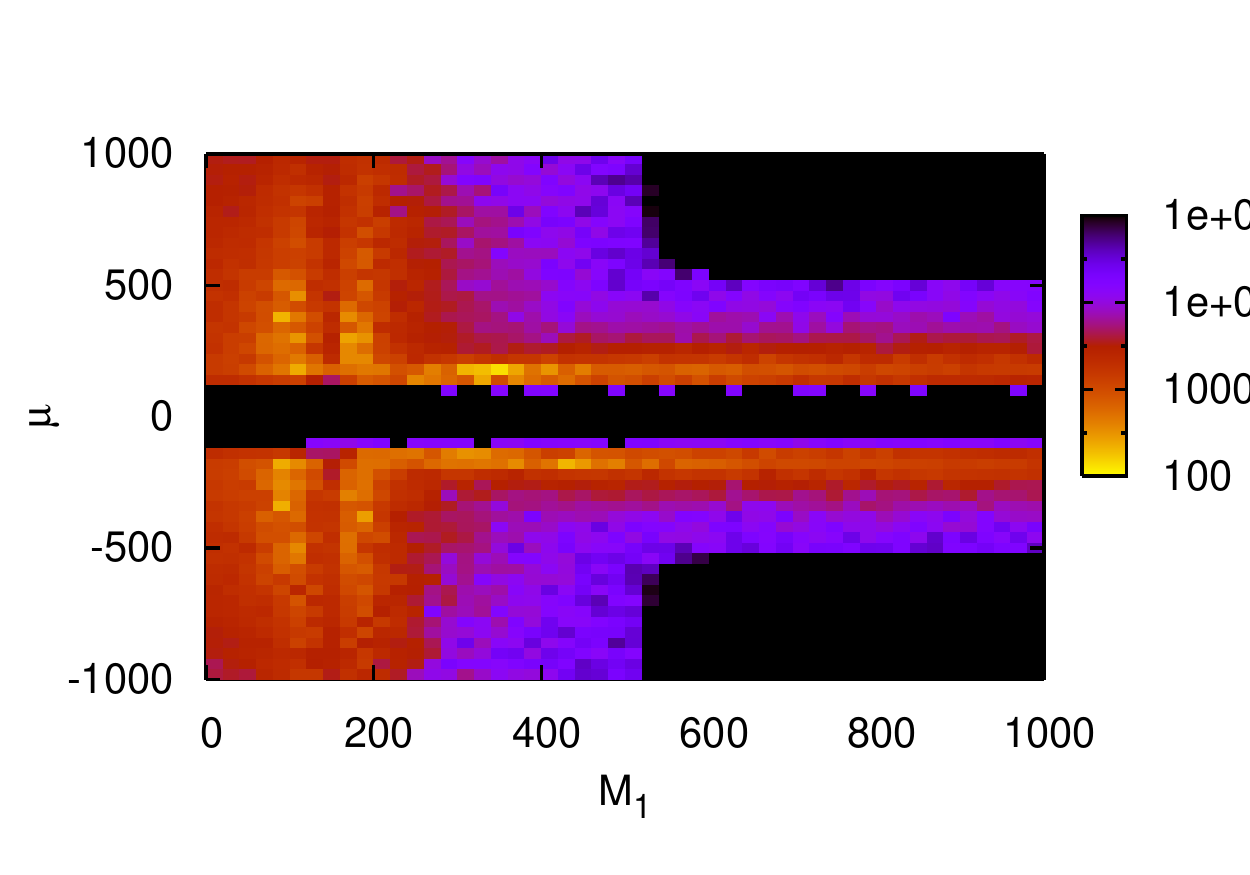} \hspace*{2cm}
 \includegraphics[width=7cm]{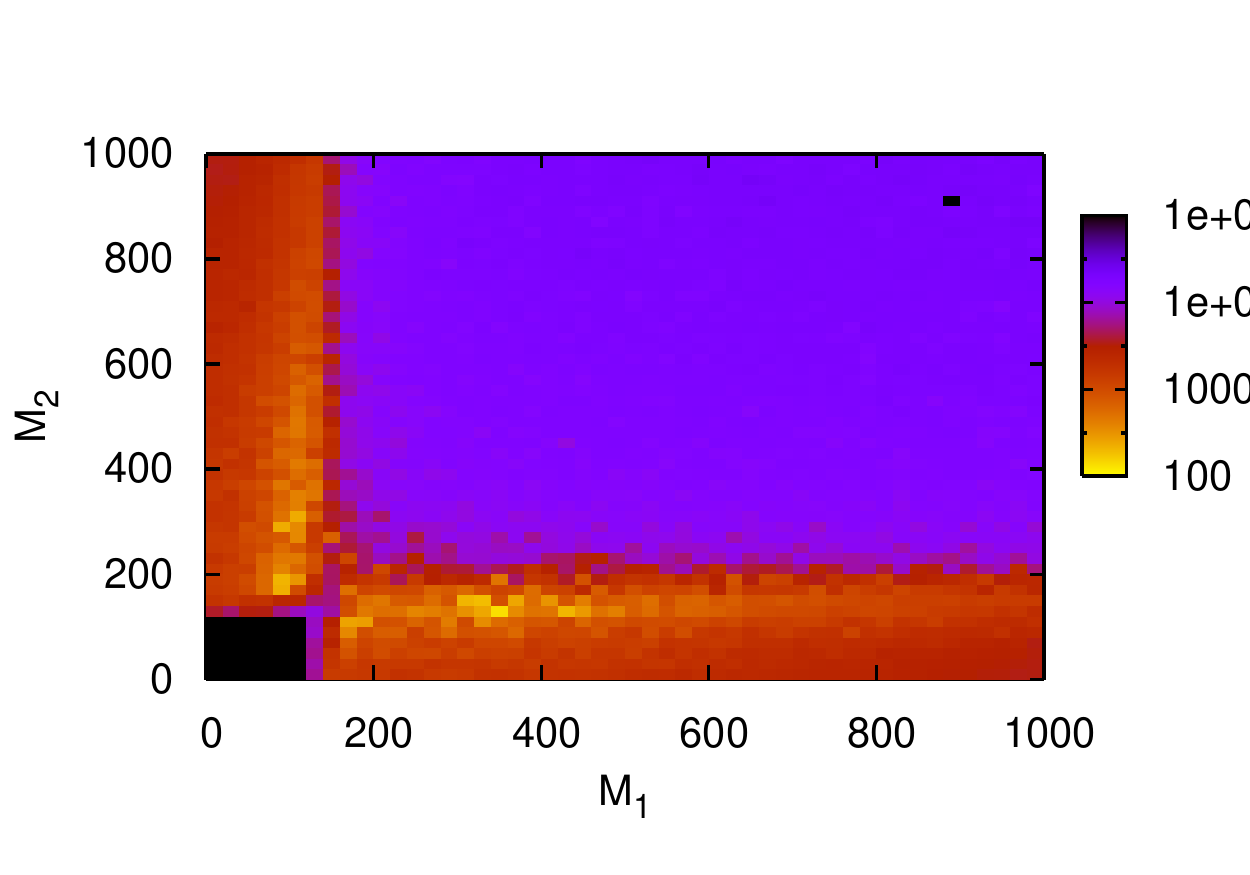} \\[-5mm]
 \includegraphics[width=7cm]{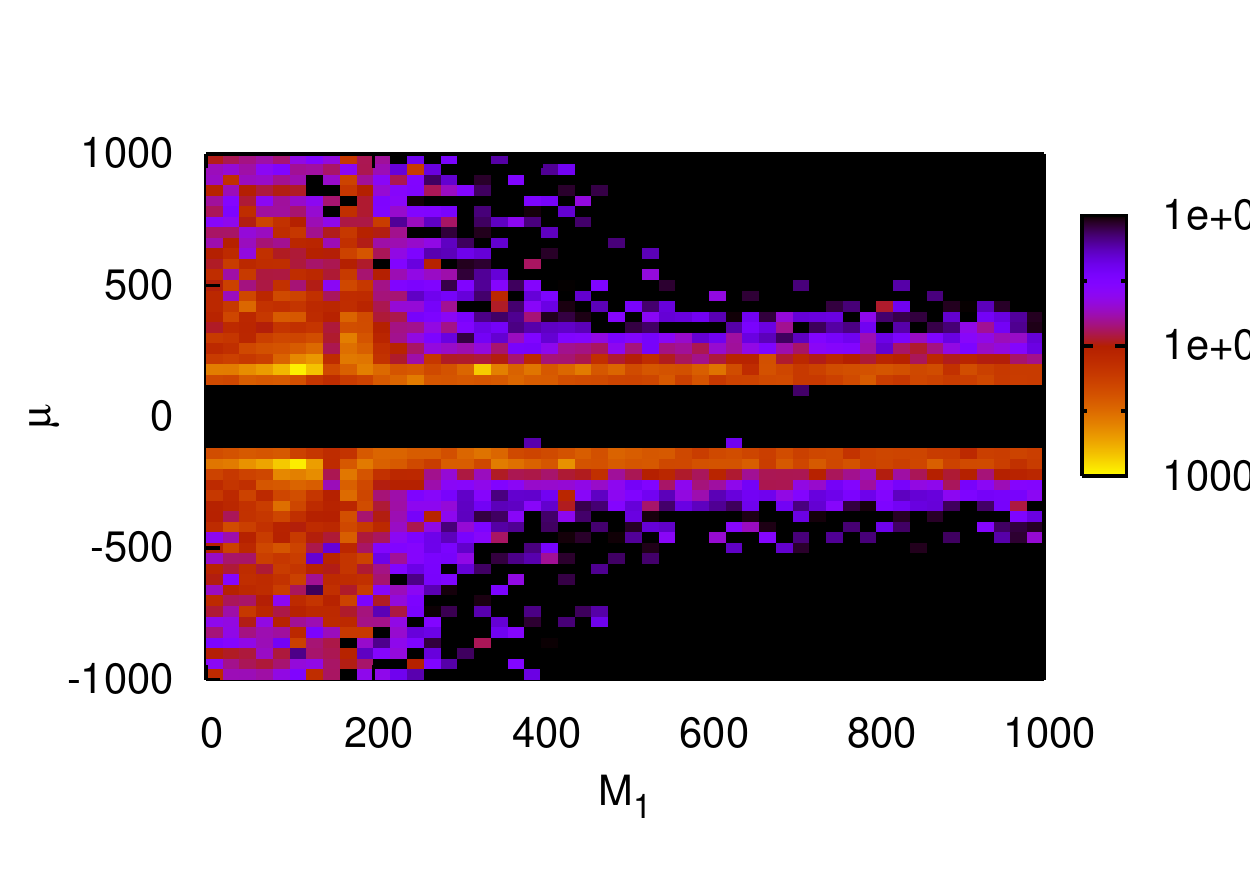} \hspace*{2cm}
 \includegraphics[width=7cm]{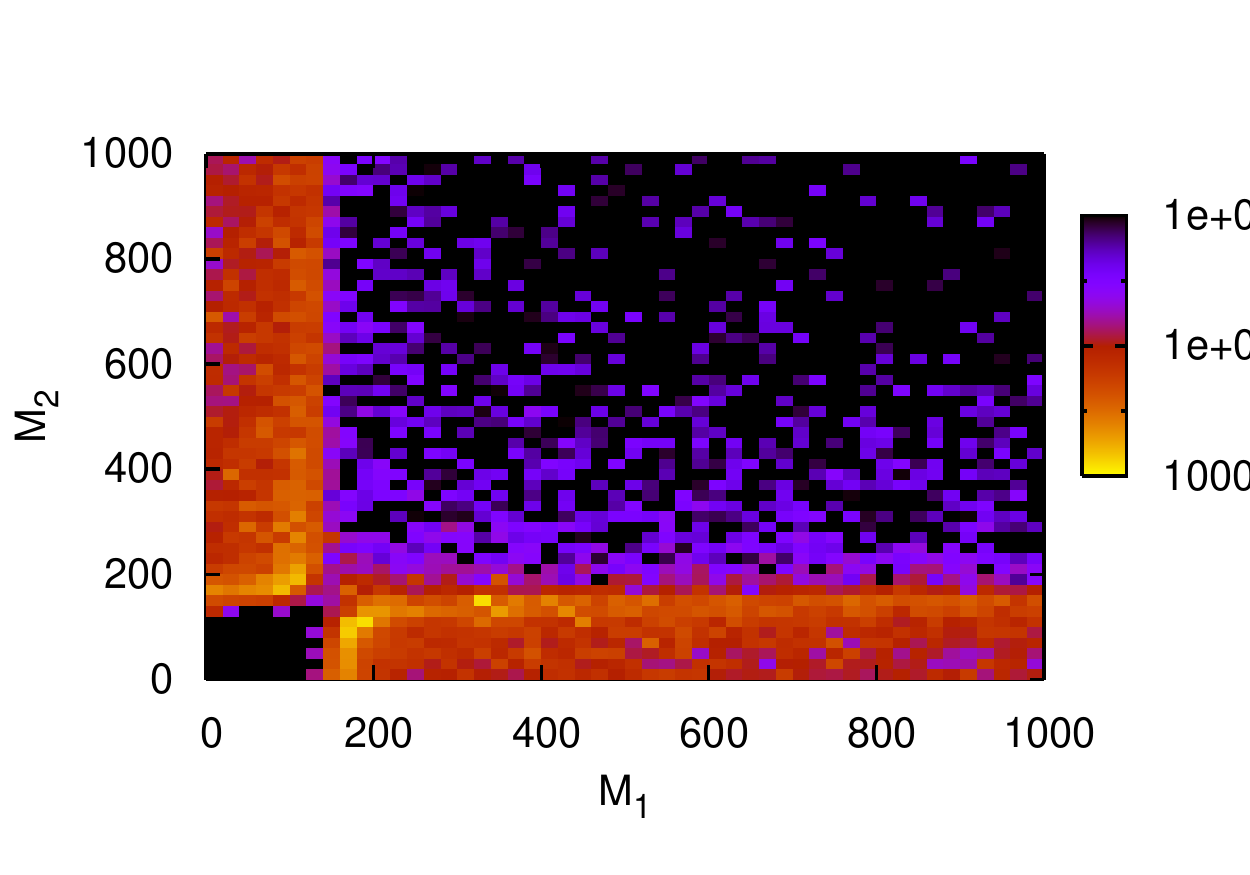} \\
 \includegraphics[width=5cm]{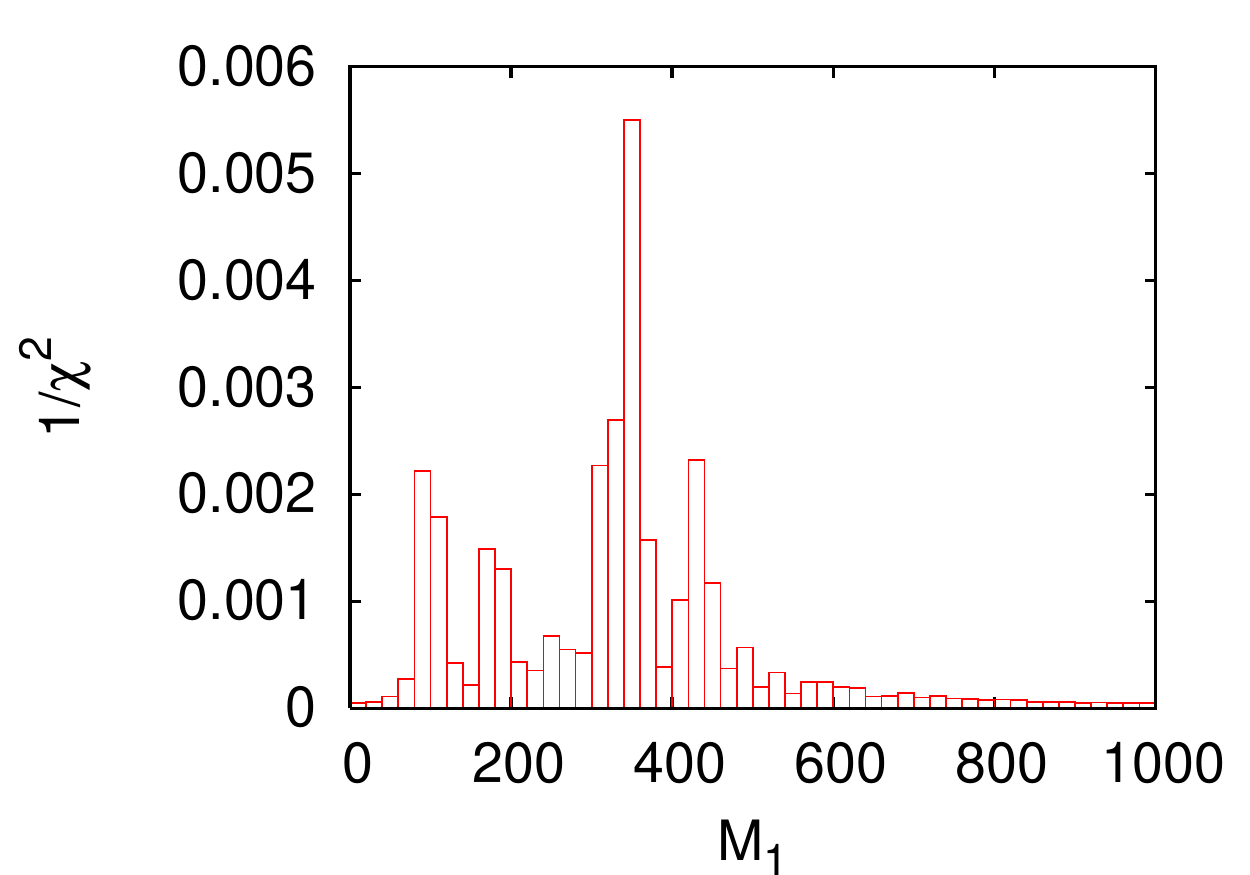} 
 \includegraphics[width=5cm]{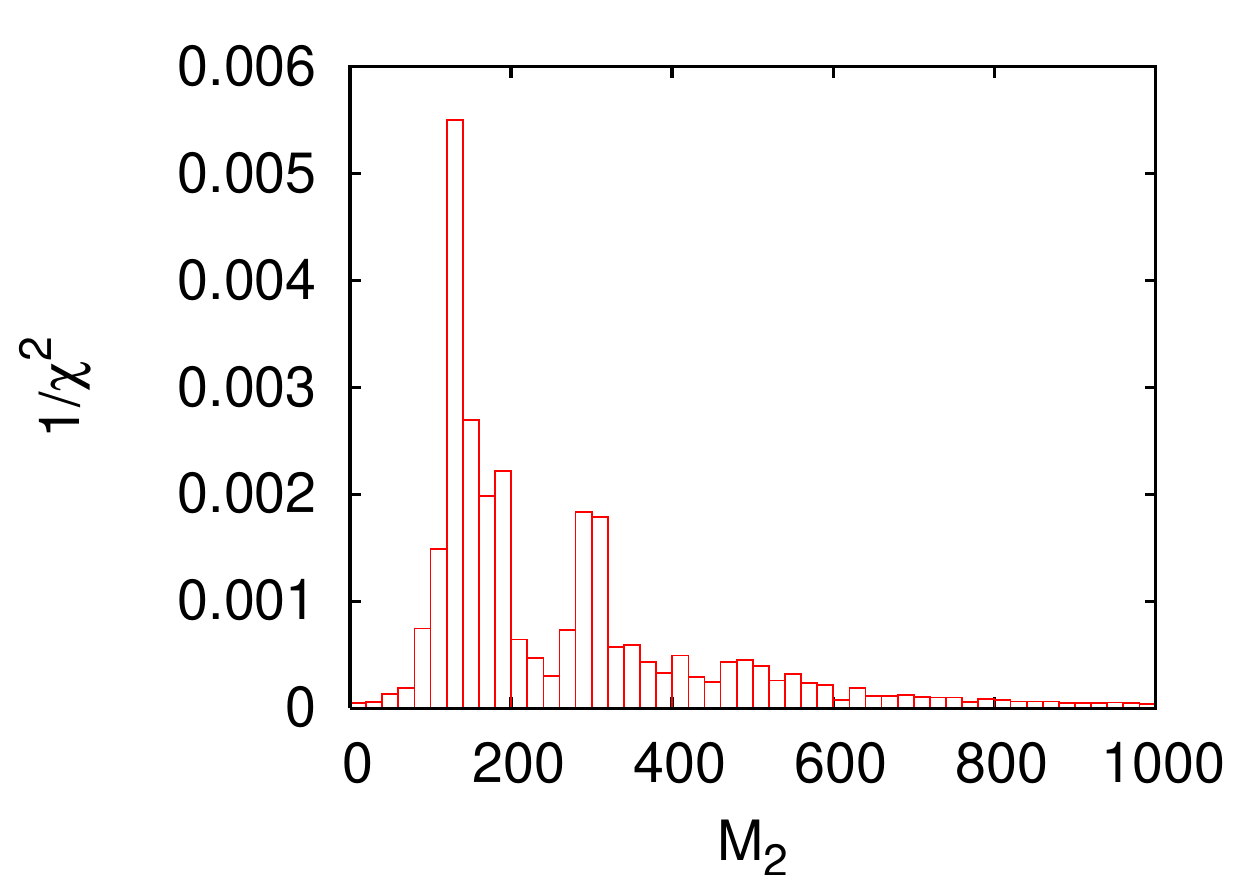} 
 \includegraphics[width=5cm]{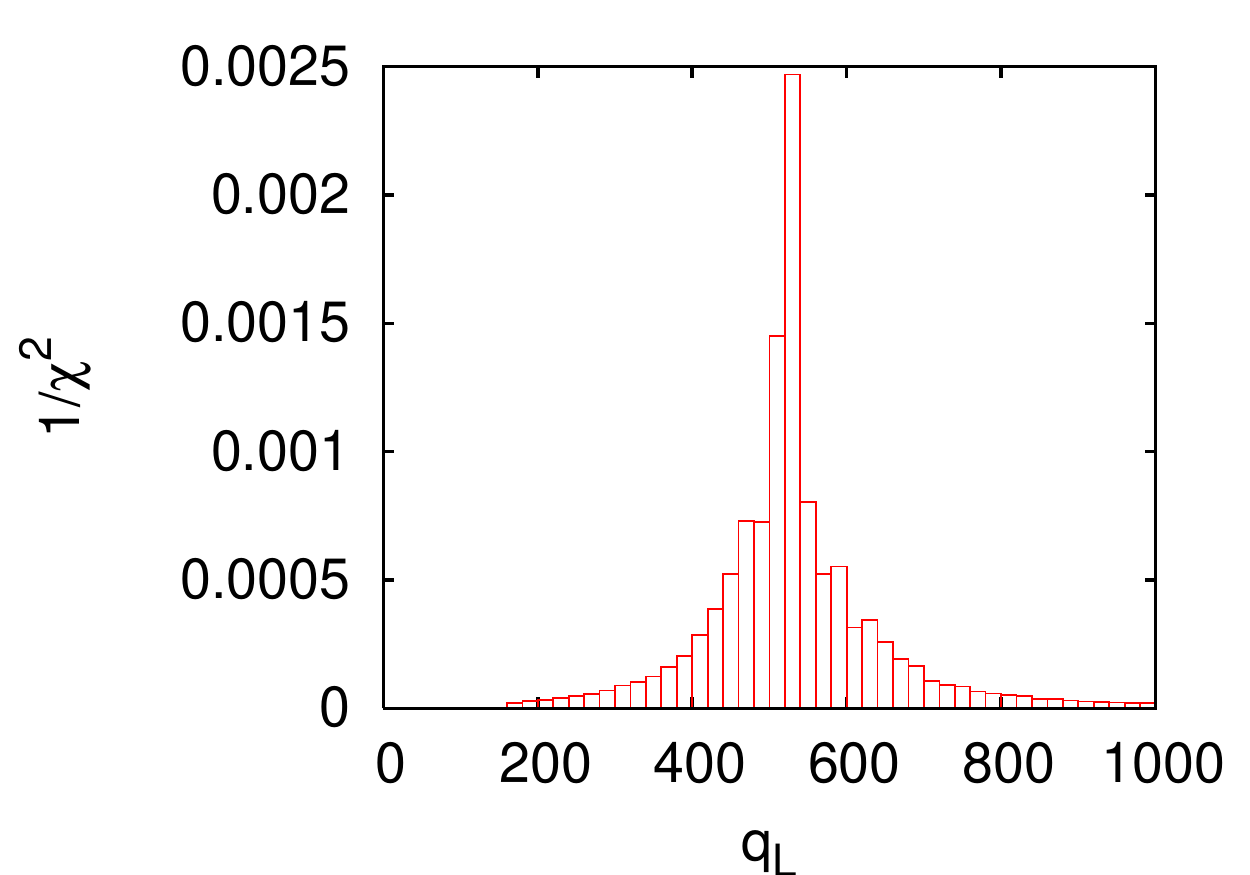}  \\
 \includegraphics[width=5cm]{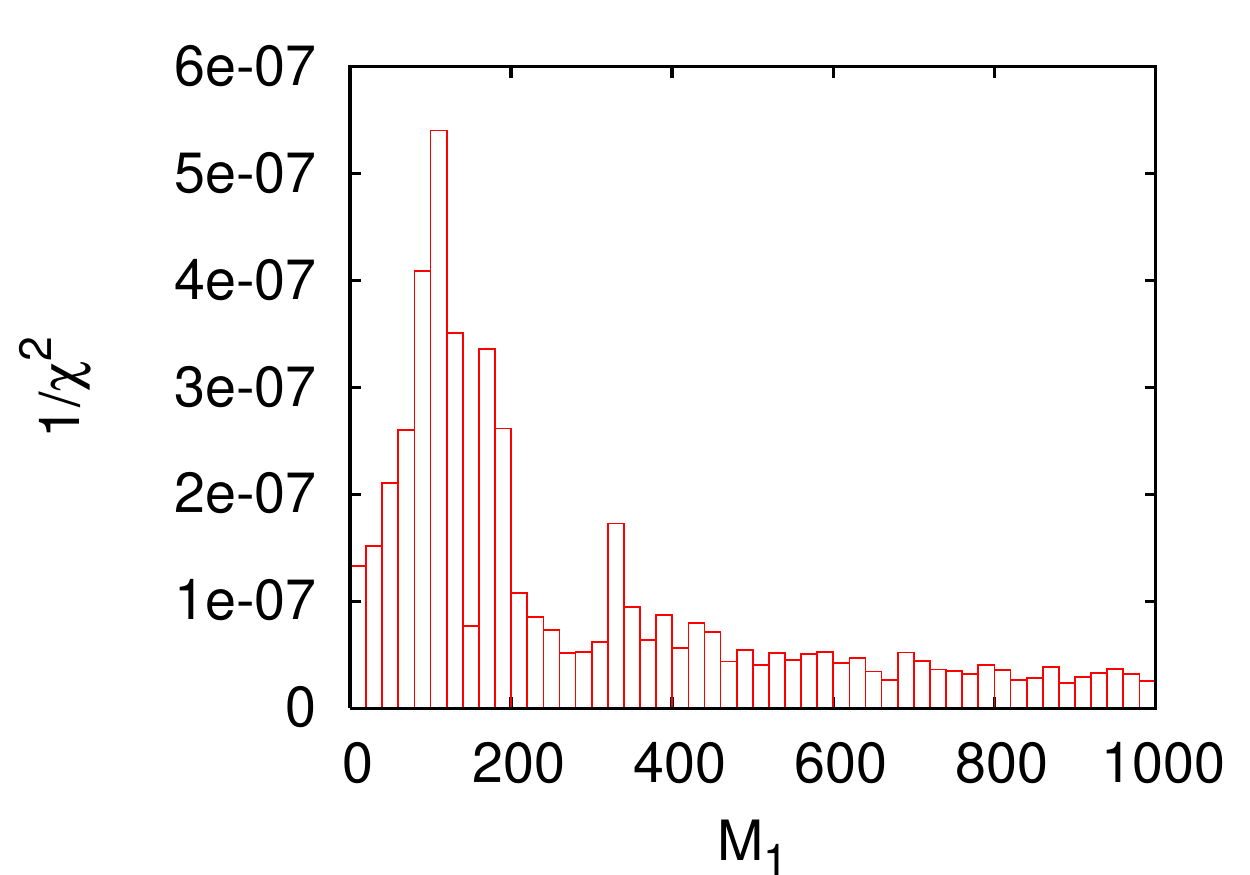} 
 \includegraphics[width=5cm]{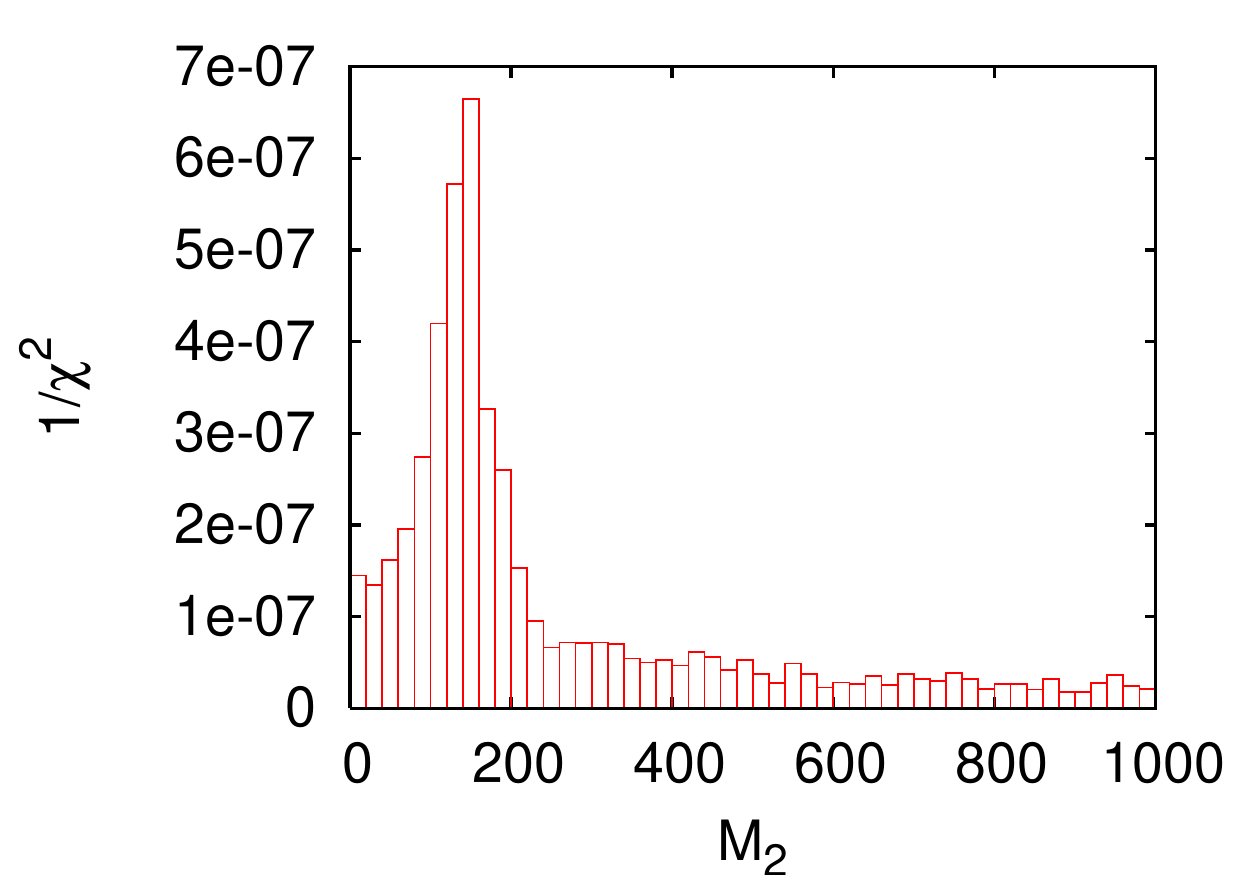} 
 \includegraphics[width=5cm]{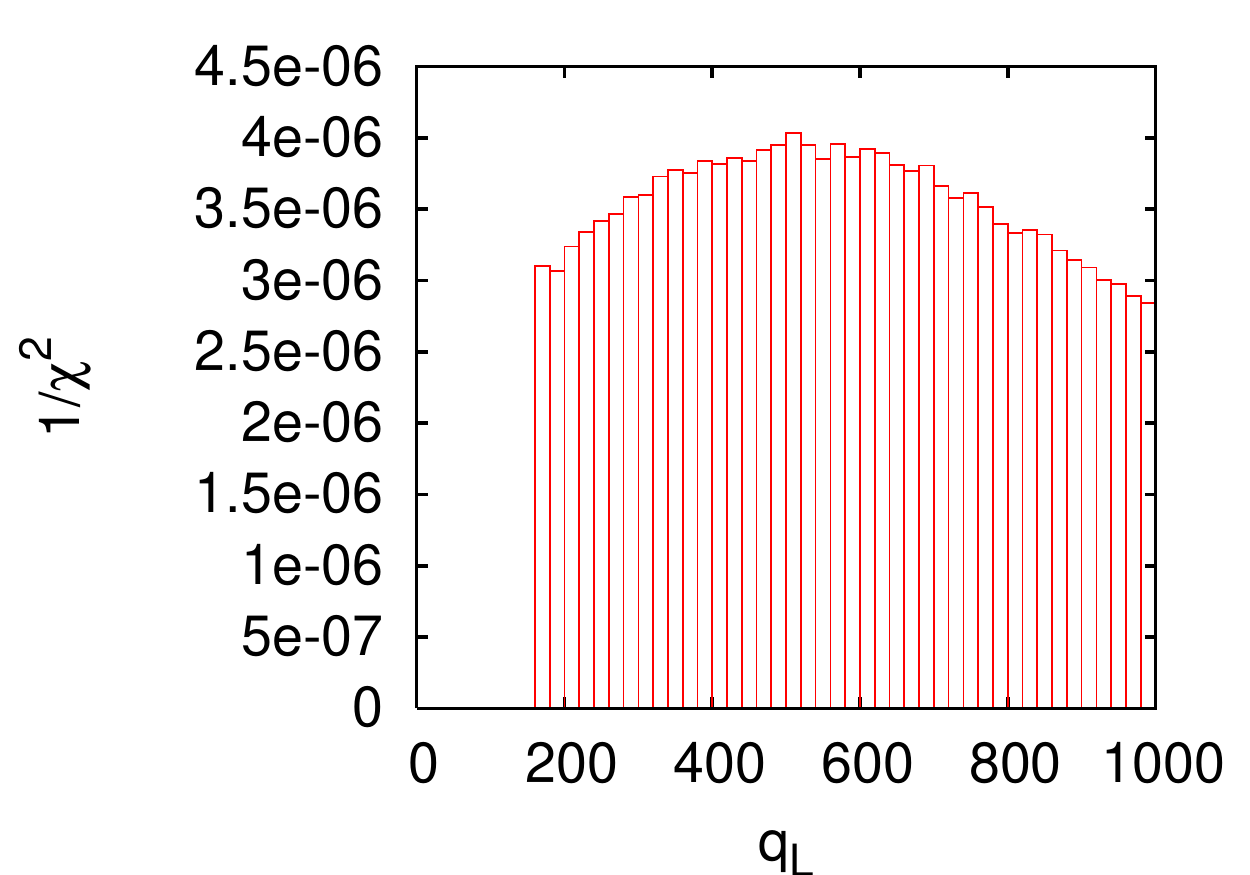} 
\caption[]{Profile likelihoods (respective upper panels) and Bayesian
  probabilities (respective lower panels) for the MSSM.  Figure taken
  from Ref.~\cite{sfitter}.}
\label{fig:para_mssm}
\end{figure}

In Fig.~\ref{fig:para_mssm} we show the profile likelihoods 
for $M_{1,2}$ and $\mu$. The
$M_1$--$\mu$ and $M_1$--$M_2$ planes reflect the different hierarchies
in the neutralino mass parameters: one of the two gaugino masses
corresponds to the measured LSP mass while the other gaugino mass can
in principle decouple. In the $M_1$--$\mu$ plane the three neutralino
masses can be identified in the $M_1$ directions. For light $M_{1,2}$
the higgsino mass parameter $|\mu|$ can be large, while for one heavy
gaugino $|\mu|$ needs to be small.  The one-dimensional profile
likelihood for example for $M_1$ again shows these three scenarios
with peaks around 100, 200 and 350~GeV, corresponding to the three
measured neutralino masses. The peak above 400~GeV is an alternative
likelihood maximum which does not correspond to a measured neutralino
mass. For $M_2$ there is again the 100~GeV peak, where the LSP is a
wino. The correct solution around 200~GeV is merged with the first
maximum, while the third peak around 300~GeV corresponds to at least
one light higgsino. 

The sfermion and neutralino masses appearing in long cascades are
correlated, reflecting the result from Section~\ref{sec:sig_cascade}
that mass differences are better constrained by decay kinematics than
absolute mass values. However, it turns out that for example the
correlation of the squark and slepton masses is numerically 
insignificant compared to noise effects and alternative maxima in the
profile likelihood. The one-dimensional profile likelihood for the
squark mass parameter, however, is clearly peaked around the correct
value.\bigskip

In addition to the profile likelihoods in Fig.~\ref{fig:para_mssm} we
also show Bayesian probability distributions. While the basic
structures of the results are very similar, the one-dimensional
histograms show two significant differences.  First, the Bayesian
probability for $M_{1,2}$ shows the same three physical solutions as
the corresponding profile likelihood, namely one peak around 100~GeV,
another one around 200~GeV, separated only by one bin from the edge of
the 100~GeV peak, and a heavy-neutralino peak above 300~GeV (more
visible for $M_1$). However, the peaks in the Bayesian probability are
much wider, as expected from the discussion of volume effects.  The
two lower peaks in $M_2$ appear as one, with a maximum around
150~GeV. The second difference is that the Bayesian probability
answers the question of which neutralino is the most likely to be
bino-like.  While for $M_1$ the best profile-likelihood entry lies
around 350~GeV, the Bayesian probability shows a clear maximum for the
input value around 100~GeV.

  In both approaches the difference between the two signs of $\mu$ is
small, but for the Bayesian probabilities $|\mu|$ is clearly driven to
small values by volume effects. This arises because of the decoupling
of one of the two gaugino masses for a light higgsino, while for two
light gauginos the higgsino mass is still determined by the fourth
neutralino.  The squark mass again shows the expected behavior: the
profile likelihood is much more strongly peaked than the Bayesian
probability.\bigskip

\begin{table}[t]
\begin{tabular}{|l|r@{$\pm$}r|r||l|r@{$\pm$}r|r|}
\hline
                     & \multicolumn{2}{c}{LHC}     & SPS1a &
                     & \multicolumn{2}{c}{LHC}     & SPS1a \\
\hline
$\tan\beta$          &      10.0 & 4.5             &     10.0 & 
$M_{\tilde{\tau}_L}$ &     227.8 & $\om(10^3)$       &    193.6 \\
$M_1$                &     102.1 & 7.8             &    103.1 &
$M_{\tilde{\tau}_R}$ &     164.1 & $\om(10^3)$       &    133.4 \\
$M_2$                &     193.3 & 7.8             &    192.9 &
$M_{\tilde{\mu}_L}$  &     193.2 & 8.8               &     194.4 \\
$M_3$                &     577.2 & 14.5            &    577.9 &
$M_{\tilde{\mu}_R}$  &     135.0 & 8.3             &    135.8 \\
&&&&
$M_{\tilde{e}_L}$    &     193.3 & 8.8              &    194.4 \\
$A_{l1,2}$           &\multicolumn{2}{c|}{fixed 0}  &   -251.1 &
$M_{\tilde{e}_R}$    &     135.0 & 8.3             &    135.8 \\
$A_\tau$             &\multicolumn{2}{c|}{fixed 0}  &   -249.4 &
$M_{\tilde{q}3_L}$   &     481.4 & 22.0            &    480.8 \\
$A_{u1,2}$           &\multicolumn{2}{c|}{fixed 0}  &   -657.2 &
$M_{\tilde{t}_R}$    &     415.8 & $\om(10^2)$     &    408.3 \\
$A_t$                &    -509.1 & 86.7            &   -490.9 &
$M_{\tilde{b}_R}$    &     501.7 & 17.9            &    502.9 \\
$A_b$                &\multicolumn{2}{c|}{fixed 0}  &   -763.4 &
$M_{\tilde{q}_L}$    &     524.6 & 14.5            &    526.6 \\
$A_{d1,2}$           &\multicolumn{2}{c|}{fixed 0}  &  -821.8 &
$M_{\tilde{q}_R}$    &     507.3 & 17.5            &    508.1 \\
$m_A$                &     406.3 & $\om(10^3)$     &    394.9 &
$\mu$                &     350.5 & 14.5            &   353.7 \\
$m_t$                &     171.4 & 1.0             &    171.4 &&&& \\
\hline
\end{tabular}
\caption[]{Result for the general MSSM parameter determination in
  SPS1a assuming flat theory errors. The kinematic endpoint
  measurements are given in Tab.~\ref{tab:para_edges}.  Shown are the
  nominal parameter values and the result after the LHC fit. All
  masses are given in GeV. Table taken from Ref.~\cite{sfitter}}
\label{tab:para_lhc}
\end{table}

A crucial result of any parameter extraction are the errors on the
extracted model parameters. The flat theory errors are now only
TeV-scale uncertainties, for example due to the translation of mass
parameters into physical masses or due to higher-order effects in the
observables.  For one best-fit parameter point (the correct one), we
show the results for the error determination in
Table~\ref{tab:para_lhc}. As discussed above, the LHC is insensitive
to several parameters.  Some of them, namely the trilinear mixing
terms $A_i$, are already fixed in the fit. Others, like the heavier
stau-mass and stop-mass parameters or the pseudoscalar Higgs mass, 
turn out to be unconstrained. Since the heavy Higgses are often
effectively decoupled as far as the relevant LHC signatures are concerned, 
the parameters in the Higgs
sector are $\tan\beta$ and the lightest stop mass which we cannot
determine well just from $m_h$.  Theory errors already degrade
the precision in the Higgs mass by a factor of about two. 
In particular the $m_{\ell\ell}$ theory error dominates over
the corresponding experimental error. On the other hand, we do see
that particularly in the strongly interacting new-physics sector the
LHC shows an impressive coverage of parameters determined largely
from cascade decays. A higher experimental precision on
these masses would require us to deal with the question of what kind of
parameter we actually measure, \ie what kind if renormalization scheme
do we use for cascade decays~\cite{def_masses}.

\subsubsection{Testing unification}
\label{sec:para_uni}

Once we have determined the parameters of a TeV-scale Lagrangian, the
next step is to extrapolate these parameters to higher energy scales. 
This way we can attempt to study the underlying fundamental 
interactions and symmetries.
If the fundamental theory of particles is renormalizable and
perturbative, this extrapolation is crucial. Inspired by the apparent
gauge coupling unification~\cite{coupling_uni} in the MSSM,
which itself can be tested at the LHC~\cite{Rainwater:2007qa},
it is sensible to ask whether any other model parameters unify 
as well~\cite{Altunkaynak:2009tg,blair}. For two reasons, the prime candidates for
unification are the gaugino masses.  First, in contrast to the scalar
masses, the three gaugino masses can well be argued to belong to a
unifying gauge sector.  Second, interactions between the
hidden SUSY-breaking sector or other sources of new physics
and the MSSM particle content can disturb
the unification pattern of the scalars, but have a much smaller effect
on the gauginos~\cite{ArkaniHamed:1998kj}.  In that case, scalar mass 
unification might have to be replaced by much less obvious sum rules 
for scalar masses at some high scale~\cite{Kane:2006hd,martin}.\bigskip

Technically, upwards running is considerably more
complicated~\cite{suspect,Kneur:2008ur} than starting from a
unification-scale and testing the unification hypothesis by comparing
to the TeV-scale particle spectrum. It is by no means guaranteed that
the renormalization group running will converge for TeV-scale input
values far away from the top-down prediction. In
Figure~\ref{fig:para_uni} we show the extrapolation of the central
values of the gaugino and scalar mass parameters using SuSpect. As
expected in SPS1a, the mass parameters unify at the GUT scale provided
we pick the correct solution~\cite{sfitter_uni}.

\begin{figure}[t]
 \includegraphics[width=8.1cm]{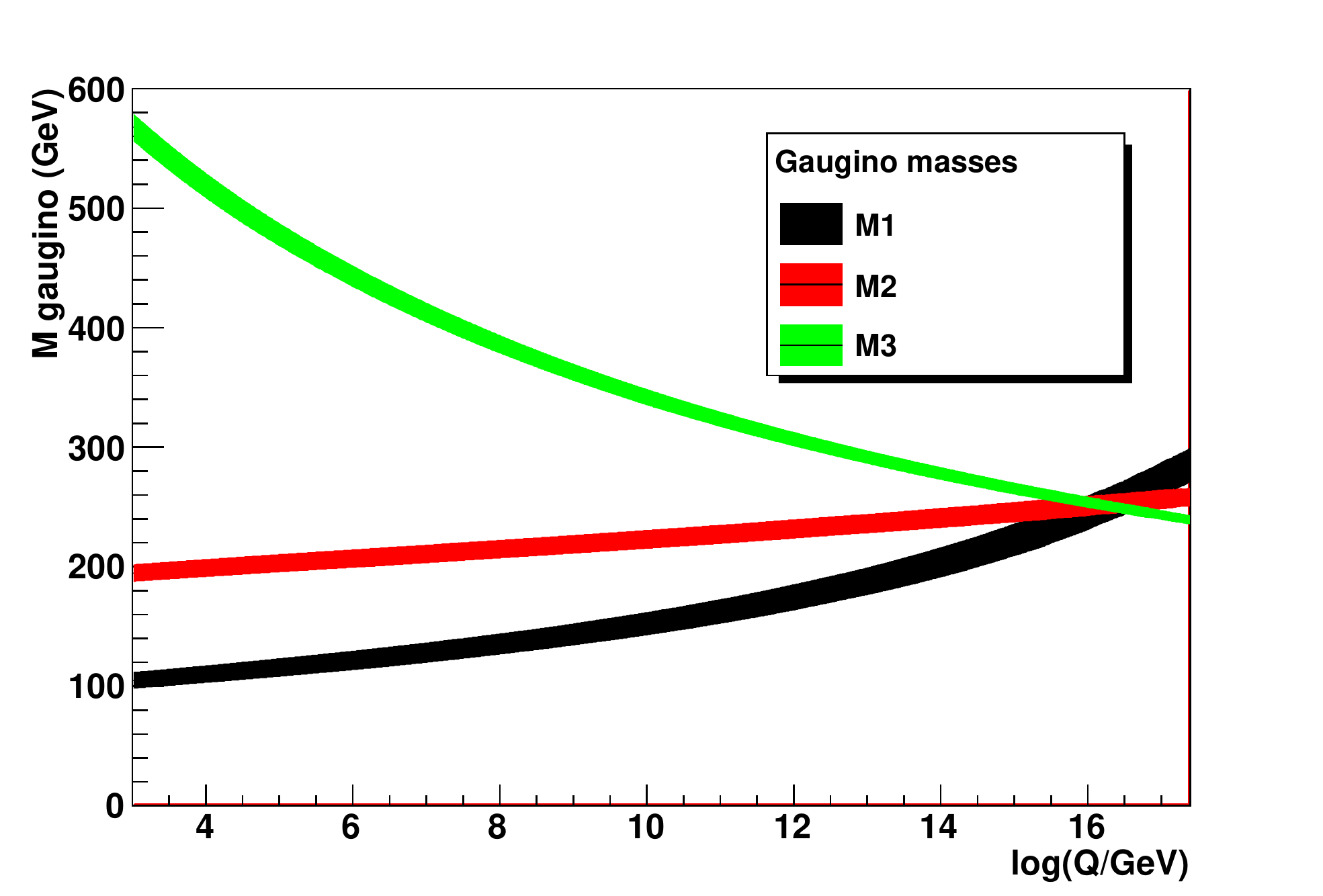}
 \hspace*{0mm}
 \includegraphics[width=8.1cm]{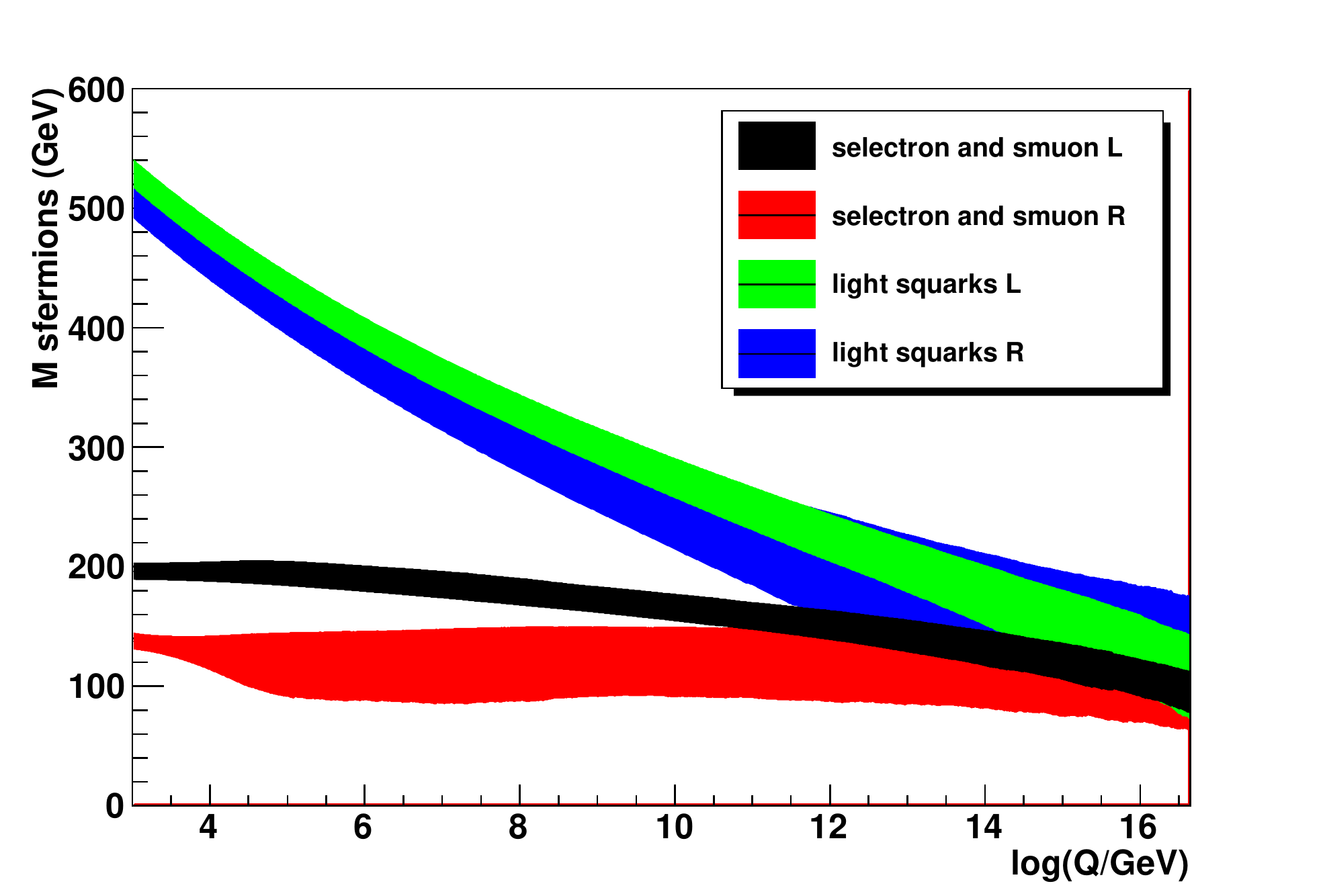}
\caption[]{Upward renormalization group running of the three gaugino
  masses (left) and the scalar masses (right) assuming only LHC
  measurements in the parameter point SPS1a. Figures taken from
  Ref.~\cite{sfitter_uni}}
\label{fig:para_uni}
\end{figure}

\subsection{Combination with $(g\!-\!2)_\mu$}
\label{sec:para_gm2}

  As discussed in Section~\ref{sec:para_mssm}, the full new physics
spectrum might not be directly observable at the LHC.
A further probe that we can combine with LHC measurements is the
anomalous magnetic moment of the muon, which currently provides a
suggestive hint for new physics beyond the Standard Model.
This parameter has been measured experimentally~\cite{gm2_ex} and 
computed in the Standard Model~\cite{gm2_th} to an extremely high precision:
\begin{alignat}{5}
a_\mu^\mathrm{(exp)} \equiv (g\!-\!2)_\mu/2 &= 116 592 080(63) \times 10^{-11}
\notag \\
a_\mu^\mathrm{(SM)} &= 116 591 785(61) \times 10^{-11}
\label{eq:gm2SM}
\end{alignat}
We use $e^+e^-$ data instead of the alternative tau
data~\cite{gm2_tau}. This difference between experiment and the
Standard model prediction might well be a new-physics effect.\bigskip

  Supersymmetry provides a particularly attractive explanation of this
discrepancy.  New loop corrections arise in supersymmetry that 
involve a neutralino (chargino) and a smuon ($\mu$-sneutrino),
and that depend on $\mu$, $\tan\beta$, and the
slepton masses.  For somewhat large values of $\tan\beta$ and
sleptons on the order of a few hundred~GeV, these corrections
can bring the theoretically predicted value of $a_{\mu}$ in line
with the observed value~\cite{gm2_bsm}.  This mass range
could well be accessible at the LHC.  Taking this apparent discrepancy
seriously and combining the measurement with other data will 
allow us to further constrain the allowed model parameters.
Indeed, for supersymmetry the leading correction to ($g\!-\!2$) 
is proportional to $\tan\beta$ and to the sign of $\mu$ and provides 
exactly the kind of information on the MSSM parameter space 
which is likely to be difficult to obtain from LHC data.

  If $\tan\beta$ is determined at the LHC, the value of $(g\!-\!2)_{\mu}$
will instead provide an important cross-check of the structure of the model.
For large enough values of $\tan\beta$, it can potentially be measured at 
the LHC through the production and decays of the heavy Higgs 
bosons~\cite{sasha_spirix,heavy_higgs}. This measurement is
sensitive to the re-scaled bottom Yukawa coupling $y_b \tan\beta$ with
its universal renormalization corrections $\Delta_b$~\cite{delta_b}.
Combining all errors, 
for Higgs masses below $\sim 500$~GeV and $\tan\beta \gtrsim 20$
 we arrive at a total error of $12\%$ to $16 \%$
on the $bbA/bbH$ Yukawa coupling, which is proportional to
$\tan\beta$~\cite{sasha_spirix}.  As discussed in
Section~\ref{sec:sig_qcd}, such a ($\sigma \times \br$) measurement
suffers from sizeable and hard to estimate QCD uncertainties.\bigskip

  To illustrate how to combine the $(g\!-\!2)_{\mu}$ measurement with LHC
data, we again turn to the SPS1a parameter point~\cite{Allanach:2002nj}. 
Its theory prediction is
$a_\mu^\mathrm{(SPS1a)} = a_\mu^\mathrm{(SM)} + 282 \times 10^{-11}$,
a deviation of $\Delta a_\mu = -13 \times 10^{-11}$ from the
experimentally observed value and well inside the experimental error
bounds. Therefore, we can safely use the current experimental value as
a hypothetical measurement, and estimate its impact not only on
the best-fitting parameter point but more importantly on the errors on
the MSSM parameters.

\begin{table}[t]
\begin{center} \begin{small}
\begin{tabular}{|l|r@{$\pm$}rr@{$\pm$}r|r@{$\pm$}rr@{$\pm$}r||r|}
\hline
                     & \multicolumn{4}{c|}{only experimental errors} & \multicolumn{4}{c||}{including flat theory errors}&  SPS1a\\
\hline
                     & \multicolumn{2}{c}{LHC}     &\multicolumn{2}{c|}{LHC $\otimes (g-2)$}&  \multicolumn{2}{c}{LHC}    &\multicolumn{2}{c||}{LHC $\otimes (g-2)$}&  \\
\hline                                                                           
$\tan\beta$          &       9.8 & 2.3             &       9.7 & 2.0             &      10.0 & 4.5             &      10.3 & 2.0             &     10.0 \\
$M_1$                &     101.5 & 4.6             &     101.1 & 3.6             &     102.1 & 7.8             &     102.7 & 5.9             &    103.1 \\
$M_2$                &     191.7 & 4.8             &     191.4 & 3.5             &     193.3 & 7.8             &     193.2 & 5.8             &    192.9 \\
$M_3$                &     575.7 & 7.7             &     575.4 & 7.3             &     577.2 & 14.5            &     578.2 & 12.1            &    577.9 \\
$M_{\tilde{\mu}_L}$  &     192.6 & 5.3             &     192.3 & 4.5             &     193.2 & 8.8             &     194.0 & 6.8             &    194.4 \\
$M_{\tilde{\mu}_R}$  &     134.0 & 4.8             &     133.6 & 3.9             &     135.0 & 8.3             &     135.6 & 6.3             &    135.8 \\
$m_A$                &     446.1 & $\mathcal{O}(10^3)$     &     473.9 & $\mathcal{O}(10^2)$     &     406.3 & $\mathcal{O}(10^3)$     &     411.1 & $\mathcal{O}(10^2)$     &    394.9 \\
$\mu$                &     350.9 & 7.3             &     350.2 & 6.5             &     350.5 & 14.5            &     352.5 & 10.8            &    353.7 \\
$m_t$                &     171.4 & 1.0             &     171.4 & 1.0             &     171.4 & 1.0             &     171.4 & 0.90            &    171.4 \\
\hline                 
\end{tabular}          
\end{small} \end{center}
\caption[]{Result for the general MSSM parameter determination at the
  LHC in SPS1a. The left part neglects all theory errors, the right
  one assumes flat theory errors. In the third and fifth column we
  include the current measurement of ($g\!-\!2$).  All masses are given in
  GeV. Table taken from Ref.~\cite{sfitter_lh}}
\label{tab:para_gm2}
\end{table}

Table~\ref{tab:para_gm2} shows the result of a SPS1a analysis. For
comparison we also include the result without $(g\!-\!2)_{\mu}$ and 
results with experimental errors only. The effect of the additional 
information on the accuracy of the parameter determination is impressive, 
particularly for $\tan\beta$.  This parameter it is not well determined
by the kinematic endpoints at all, and $(g\!-\!2)_{\mu}$ shrinks its error bar
by more than a factor of two. This improvement impacts all parameters
which are re-rotated when $\tan\beta$ is changed to reproduce the
same physical observables shown in
Table~\ref{tab:para_edges}. Correlations and loop corrections
propagate this improvement over almost the complete parameter space.

\subsection{Combination with flavor}
\label{sec:para_flavor}

 Following the last section, a third way of determining $\tan \beta$
are flavor observables such as the rare decay $B_s \to \mu^+ \mu^-$.
Flavor physics has not produced any conclusive evidence
for new physics and thus places extremely strong limits on the flavor
structure of any new physics beyond the Standard Model.
These will provide us with one of the most carefully studied set of 
measurements which we can combine with high-$p_T$ LHC measurements. 
This is certainly true for supersymmetric models, but also for extended 
Higgs sectors, little-Higgs models or models with large or warped 
extra dimensions.

  Independent of the underlying model, FCNC processes involving
down-type quarks are highly suppressed in the Standard Model and
therefore sensitive to modifications involving for example the
top-quark sector. Beyond their loop suppression, the unitarity and
hierarchical structure of the CKM matrix and the hierarchy $m_{t,W}
\gg m_{u,c}$ ensure GIM cancellations between the light-flavor loop
contributions, as experimentally required for example by the strong
kaon sector constraints. Turning this argument around, FCNCs have for
a long time been sensitive probes of new physics which might or might
not respect these symmetries and are a stringent constraint on many models. 
In supersymmetric models they provide a handle on the
stop and chargino masses, in particular when the supersymmetric flavor
structure is (close to) minimally
flavor-violating~\cite{mfv}.\bigskip

Computing the decay
widths $B_{s,d} \to \ell^+ \ell^-$ we observe the steepest $\tan\beta$
dependence of any known observable~\cite{fcnc_higgs,tgb_flavor}:
\begin{equation}
{\cal A}(B_q \to \ell^+ \ell^-) \propto \frac{y_b\, y_\ell}{\cos\beta\,
    m_{A}^2} \stackrel{\tan\beta \gg 1}{\propto} \frac{m_b m_\ell \tan^3\beta}{m_A^2} .
\end{equation}
This dependence makes it a prime suspect to extract $\tan \beta$
from~\cite{gerhard}.\bigskip 

  The good news for $B$ physics observables is that most studies of
cascade decays include the bottom sector. The reason is that sbottoms 
decay to bottom jets, which can be tagged and are less likely to be
confused with the overwhelming QCD radiation accompanying heavy
particles at the LHC~\cite{heavy_skands,heavy_mg,Plehn:2008ae}. 
A technical problem is that for $\tan\beta$ values below 10 or 15 
we are unlikely to observe any of the heavy Higgs bosons and fix their 
mass scales.  For that reason we
modify the usual parameter point and consider $\tan \beta = 30,40$,
where we can assume $m_A$ to be measured.

In addition, we need to include some very basic information about the
chargino-stop sector which governs the loop-mediated effective
$bs\phi$ couplings.  For simplicity, we assume that the set of SPS1a
cascade observables is not altered by this change, including the
sbottom mass determination.  While this assumption is
quantitatively naive, it should serve the purpose of estimating the
odds of combining flavor and high-$p_T$ information on $\tan\beta$.
In addition, we assume measurements from $\tilde{g} \rightarrow
\tilde{t}_1 t \rightarrow t \bar{t} \tilde{\chi}^0_1$, $\tilde{g}
\rightarrow \tilde{t}_1 t \rightarrow t \bar{b} \tilde{\chi}^+_1$, $
\tilde{q}_L \rightarrow \tilde{\chi}^0_3 q \rightarrow
\tilde{\chi}^0_2 Z q \rightarrow \tilde{\tau}_R \tau Z q \rightarrow
\tilde{\chi}^1_0 \tau \tau q Z$, $\tilde{q_L} \rightarrow
\tilde{\chi}^+_2 q \rightarrow \tilde{\chi}^+_1 Z q \rightarrow q W
\tilde{\chi}^0_1 Z q \rightarrow q q' q'' \tilde{\chi}^0_1 Z$, where
the $\tilde{\chi}^+_2$ cascade is strictly speaking not necessary for
the analysis~\cite{mihoko,sdecay}.  For a first study
we assume the observables listed in Table~\ref{tab:flavor_data}.
The three neutralino masses gives us indirect information on the
chargino mass parameters $M_2$ and $\mu$, and the left-handed stop
mass is linked to the left-handed sbottom mass via $SU(2)$.  The new
measurements we include are $\text{BR}(B_s \to \mu \mu)$ and the edge
of $m_{tb}$~\cite{mihoko}. For $B_s\to \mu \mu$, LHCb alone
expects about 100 events at the Standard Model rate after 5 years of
running. For our study we assume an integrated luminosity of 
$10~\text{fb}^{-1} $ for the $B_s$ sample. The Higgs-mediated contribution
always increases the corresponding events number.\bigskip

\begin{table}[t]
\begin{center}
\begin{small}
\begin{tabular}{|c|cc|cc|}
\hline
$\tan\beta$ & \multicolumn{2}{c|}{30} &
                \multicolumn{2}{c|}{40} \\
\hline
  & value & error & value & error \\
\hline
$m_h$            &  112.6   & 4.0  & 112.6 &  4.0 \\
$m_t$            & 174.5   & 2.0  & 174.5 &  2.0 \\
$m_{H^\pm}$       & 354.2   & 10.0 & 307.2 & 10.0 \\
$m_{\chi^0_1}$    & 98.4    & 4.8  & 98.7  & 4.8  \\
$m_{\chi^0_2}$    & 183.1   & 4.7  & 183.5 & 4.7  \\
$m_{\chi^0_3}$    & 353.0   & 5.1  & 350.7 & 5.1  \\
$m_{\chi^\pm_1}$  & 182.8  & 50.0 & 183.1 & 50.0   \\
$m_{\tilde{g}}$   & 607.7   & 8.0  & 607.6 &  8.0 \\
$m_{tb}$         & 404.2   & 5.0  & 404.2 & 5.0  \\
$\br(B_s\to \mu \mu)$ & 7.3 $\cdot 10^{-9}$   & $\sqrt{N} \otimes 15\%$ & 3.2$\cdot 10^{-8}$ & $\sqrt{N} \otimes 15\%$ \\
\hline
\end{tabular}
\end{small}
\end{center}
\caption{Set of toy measurements. The simple combined (absolute)
  errors are SPS1a-inspired. Table taken from
  Ref.~\cite{sfitter_lh}}
\label{tab:flavor_data}
\end{table}
 
  Instead of defining observables in which QCD effects cancel, we test
the possibility of using the actual $B_s$ branching ratio. In that
case, the theory error from the QCD prediction of the decay constant
$f_{B_s}$ becomes a serious issue. Because the Higgs exchange always
increases the decay rate $B_s \to \mu^+ \mu^-$, the statistical
experimental error on the decay width ranges around $10 \%$.  As a
consequence, the theory error, which at present is around $30\%$, will
soon dominate the total uncertainty, unless it can be reduced. This
might be expected from lattice QCD (see Ref.~\cite{dellamorte}
for a recent review). For our study we optimistically assume a reduction of
the error on $f_{B_s}$ to $7\%$, about half its present value and
commonly believed to be realistic over the next five years.\bigskip

\begin{table}[t]
\begin{center}
\begin{small}
\begin{tabular}{|c||ccc|cc||ccc|cc|}
\hline
& \multicolumn{3}{c|}{no theory error} & 
  \multicolumn{2}{c||}{$\Delta \text{BR}/\text{BR}=15\%$} & 
  \multicolumn{3}{c|}{no theory error} &  
  \multicolumn{2}{c|}{$\Delta \text{BR}/\text{BR}=15\%$} \\
\hline
                         & true  &  best   & error  & best   & error   & true     & best     & error  & best     & error \\
\hline\hline
$\tan \beta$  & 30      & 29.5    & 3.4    & 29.5    & 6.5       & 40        & 39.2     & 4.4   & 39.2   & 5.8    \\
$M_A$          & 344.3 & 344.4 & 33.8  & 344.3  & 31.2    & 295.5  & 304.4   & 35.4   & 295.6 & 33.9  \\
$M_1$          & 101.7 & 100.9 & 16.3  & 100.9  & 16.4    & 101.9  & 101.0   & 16.3   & 101.0 & 16.3 \\
$M_2$          & 192.0 & 200.3 & 18.9  & 200.3  & 18.8    & 192.3   & 200.3   & 20.0   & 200.7 & 18.9 \\
$\mu$           & 345.8 & 325.6 & 20.6  & 325.6  & 20.6    & 343.5   & 322.9   & 20.7   & 323.3 & 20.6  \\
$M_3$          & 586.4 & 575.8 & 28.8  & 575.8  & 28.7    & 586.9   & 576.0   & 28.7     & 575.8 & 29.0 \\
$M_{\tilde{Q}_L}$ &  494.4  & 494.4 &  78.1   &  494.3  &  78.0   & 487.1    &  487.6 & 79.4 & 487.5 & 78.9 \\
$M_{\tilde{t}_R}$ & 430.0   & 400.4  & 79.5   & 399.8    & 79.5    & 431.5    & 399.2  & 86.7 & 399.1 & 82.6 \\
\hline
\end{tabular}
\end{small}
\end{center}
\caption[]{The modified SPS1a point and the errors from the parameter
  fit for the two values of $\tan\beta=30,40$.  Dimensionful
  quantities are in units of GeV. For the measurement of $\text{BR}(B_s
  \to \mu \mu)$ we assume either no theory error or an expected
  improvement to $15\%$, as compared to the current status. Table
  taken from Ref.~\cite{sfitter_lh}}
\label{tab:flavor_fit}
\end{table}

 In Table~\ref{tab:flavor_fit} we see that, without taking into
account the theory error, $\tan\beta$ will be determined to roughly
$10\%$ from the combined toy data sample. Including a realistic theory
error increases this number to $15$--$20\%$.  The errors on the
remaining parameters remain largely unchanged. Slight shifts in either
direction are within the uncertainty on the determination of the error
bars, and the central fit value for the top mass is consistently lower
than the input value (by roughly half a standard deviation). Comparing
our error estimates with the results from
Section~\ref{sec:para_mssm}, we expect the situation to improve for all
model parameters once we include a more extensive set of measurements
and properly correlated errors.\bigskip

  The detailed results of this study should not be used at face value.
First of all, it is not clear if such a stop mass measurement can be
achieved in the SPS1 parameters point or with an increased value of
$\tan\beta$. Secondly, for the heavy/charged Higgs mass we only use a toy
measurement. And last but not least, we do not (yet) take into account
error correlations at this stage. None of these omissions we expect to
move the result of a complete analysis into a definite direction, but
there is certainly room for the final error bars to move.

This result shows, however, that the parameter $\tan\beta$ can indeed
be extracted from a combined cascade and flavor data sample. Moreover,
it shows that the combination of cascade-decay and flavor observables
will crucially depend on the quality of QCD predictions in all
sectors. Even our rough analysis shows that electroweak and sometimes
even QCD fixed-order corrections are not the primary issues in these
analyses. 
However, following the last two sections, the extraction of
$\tan \beta$ might turn into the most fascinating jigsaw in 
the LHC era.

\subsection{Combination with dark matter}
\label{sec:para_dm}

Before we combine collider data with dark matter observations we have
to start with an assumption concerning the nature of the dark matter
candidate.  We need to assume that the apparently stable
particle which carries away energy in the LHC experiments is in fact the
dark matter which fills the Universe~\cite{Jungman:1995df}. 
This means that we 
extrapolate from stable particles at the LHC and lifetimes of 
$\mathcal{O}(100~\ns)$ to a lifetime of order the age of the universe. 
On the other hand, there are many models where the lightest weakly interacting
particle marks the end of cascade decays at colliders, but
subsequently decays into the actual source of 
cosmological dark matter~\cite{Feng:2003xh}.  Given the range of
such very weak couplings, the lifetime of the apparent dark matter
agent produced in ATLAS or CMS can have an enormous range of values.
\bigskip

  If we indeed observe a dark matter candidate
at the LHC, the primary cosmological measurement is the relic
density $\Omega_\text{DM} h^2 = 0.1131\pm 0.0034$.
To apply this as a constraint on the BSM physics,
we must assume the dark matter was produced 
by thermal freeze-out, and that the cosmological evolution during and 
after freeze-out was of the standard form.   Neither of these 
assumptions are obviously correct,
and the comparison with the relic density may turn out to be a
more of a probe of cosmology at temperatures of order $\sim 10~\gev$
than a constraint on the parameters of a particle physics model.
In the case of a standard thermal relic, the density of dark matter
is controlled by pair annihilation of dark matter into SM particles
or co-annihilation with an almost mass degenerate
second-lightest new particle~\cite{Jungman:1995df,Griest:1990kh}. 
In the latter case the two lowest masses should agree to within 
a few percent. \bigskip

The link between dark matter and other TeV-scale observations 
is hazy, and detailed models may seem to indicate correlations
which are not representative. A good example is the common lore of a strong link
between dark matter, flavor observables and Tevatron signals for
supersymmetric dark matter: dark matter annihilation is most effective
in the presence of an $s$-channel (light) scalar or (heavy)
pseudo-scalar Higgs. This $s$-channel state needs to be on-shell, so
the annihilation strongly correlates Higgs masses with the LSP
mass. Flavor physics, as discussed in Section~\ref{sec:para_flavor},
is sensitive to the pseudoscalar Higgs mass and the stop mass. Last
but not least, Tevatron searches for tri-leptons rely heavily on one
slepton sitting between the second-lightest and the lightest
neutralino in mass. Unfortunately, the scalar masses of Higgs bosons,
squarks and sleptons are related by the unknown SUSY breaking
mechanism. In a TeV-scale analysis of
supersymmetry these three sets of observables could not be less
correlated. The lesson to be learned is that when combining very
distinct experiments we have to be even more careful what model we
really test~\cite{trileptons_dm}. 

  More generally, this example shows that when we consider 
weakly-interacting dark matter the relevant cosmological observables 
cannot be computed from the properties of the dark matter agent
alone. Assuming a light neutralino LSP in the mass range around
$\mathcal{O}(500)$~GeV there are four main scenarios which
determine what kind of measurements we need to link LHC measurements
with cosmological measurements~\cite{ilc_dm}. They can be organized by the
dark-matter annihilation process which predicts the relic density
after thermal production and freeze-out.  First, the annihilation of
dark matter can be dominated by fermionic final states, where in
supersymmetry the annihilation process is $t$-channel sfermion
exchange. These channels will be dominated by the lightest $t$-channel
particle, which could for example be the lightest stau --- based on
the experimental observation that sleptons can be lighter than squarks
and that for sizeable $\tan\beta$ the lightest stau can turn into the
lightest slepton mass eigenstate. For this bulk region of the
dark-matter parameter space the electroweak neutralino-slepton-lepton
couplings are crucial. In Ref.~\cite{ilc_dm} the parameter point LCC1 is
an example for such a scenario, in many ways similar to the 
SPS1a parameter point~\cite{Allanach:2002nj}. The updated SPS1a$^\prime$ 
improves the prediction of the relic density by shifting into the stau
coannihilation region. Secondly, the weakly interacting dark matter
agent can decay into pairs of $W$ bosons, which is typical for
Kaluza-Klein partners and for supersymmetric bino-wino mixed states
(LCC2). For such a scenario we do not require any detailed knowledge
of the new-physics spectrum, because the $s$-channel and
$t$-channel annihilation processes are described by the dark-matter
sector alone. A third channel for dark-matter annihilation is
co-annihilation with the second-lightest particle which has the same
new-physics parity at the LHC. This particle has to be within
$\mathcal{O}(10\%)$ of the LSP mass and can have electroweak as well
as strong couplings. Typically, it will not be strongly interacting,
because most of the candidates have strict mass limits from the
Tevatron which make the mechanism harder to realize in a given
model. However, in supersymmetry the co-annihilation
partner can for example be a stop, a stau 
or a chargino. And
last, but not least, LSP pairs can annihilate through an effective $2
\to 1$ process, where this on-shell state has a Standard-Model dark
matter parity. A good candidate is a Higgs scalar,
since heavy
Higgs scalars have large widths which we can use to adjust the size of
the annihilation rate. For such a rapid annihilation through a Higgs
boson we have to understand the couplings of the LSP to a non-minimal
Higgs sector. Since new physics at the TeV-scale typically does not
obtain masses through the weak Higgs mechanism, this can be
hard. Moreover, understanding the last two scenarios requires us
to carefully study not one but two new states, the LSP and its
annihilation partner or the funnel state. Unless the co-annihilation
partner is a strongly interacting stop, such a requirement makes the
LHC task of predicting the weakly interacting dark matter observables
at least twice as hard.\bigskip

In the spirit of proper bottom-up parameter extraction we start from
the TeV-scale supersymmetric Lagrangian. For any meaningful statement
we also ensure that the particle leaving the LHC detectors unseen is
indeed the dark matter agent. And finally, we assume that it is weakly
interacting. The first observable we reconstruct from LHC data and 
compare with cosmological experiments is
the neutralino relic density $\Omega_\text{DM}h^2$ in the SPS1a parameter
point~\cite{lhc_dm}.

  Problems with $\tan\beta$ measurements were discussed in
sections~\ref{sec:para_gm2} and \ref{sec:para_flavor}. The study
presented in Ref.\cite{lhc_dm} keeps it fixed and
conservatively treats the variation of the relic density prediction
with $\tan\beta$ as a systematic uncertainty. For simplicity we also
assume $\mu>0$ and that the heavier neutralino is indeed the
$\tilde{\chi}^0_4$. From Section~\ref{sec:para_mssm} we know that these
assumptions can be tested. Assuming the MSSM structure of the
neutralino/chargino mass matrices we can now extract the LSP's gaugino
and higgsino fractions. If the higgsino fraction is small we know that
the relic density will be determined by annihilation via the lightest
$t$-channel slepton, the light stau. To double check we can make sure
that the Higgs masses are not in the Higgs-funnel region $m_A \sim 2
m_{\tilde{\chi}_1^0}$.

To compute the dark matter annihilation cross section we need to know
all three entries of the stau mass matrix. The stau mixing angle fixes
the coupling strength to the lightest neutralino. One way of avoiding
nasty QCD errors is adding a ratio of measurements, for example a
ratio of branching ratios
$\br(\tilde\chi^0_2\rightarrow\tilde{\ell_R}\ell)/
\br(\tilde\chi^0_2\rightarrow\tilde{\tau}_1\tau)$ with a 10\%
error~\cite{sdecay}. Aside from $\tan\beta$
this ratio depends on the incoming and outgoing particle masses and on
the stau mixing angle which we can measure with an uncertainty of
roughly $35\%$. Now, we can predict the neutralino relic density for
fixed values of $\tan\beta$, $m_A$ and $m_{\tilde{\tau}_2}$ with an
error around $20\%$, dominated by a conservative 5~GeV error bar of
the $m_{\tau\tau}$ edge in Table~\ref{tab:para_edges}. Varying the
three input parameters within the experimental constraints gives us an
additional cumulative uncertainty of $(+2\%, -12\%)$.  The study of
less favorable model points shows, however, that in general additional
information from a linear collider is needed to discriminate different MSSM
parameter points giving very different predictions for the dark matter
relic density.\bigskip

  Going one step beyond this simple relic density estimate, 
we can also predict the direct and indirect detection cross sections. 
This combination has been successfully applied to an alternative
 supersymmetric dark matter candidate, the sneutrino: 
a sneutrino-neutrino-gaugino coupling
strength large enough to explain the measured relic density means that
a sneutrino should have been observed in direct detection
experiments~\cite{Falk:1994es}. Since this was not the case we can 
limit our analyses to neutralino LSPs 
or non-standard sneutrinos~\cite{Gopalakrishna:2006kr}. 

\begin{figure}[t]
 \includegraphics[width=6cm]{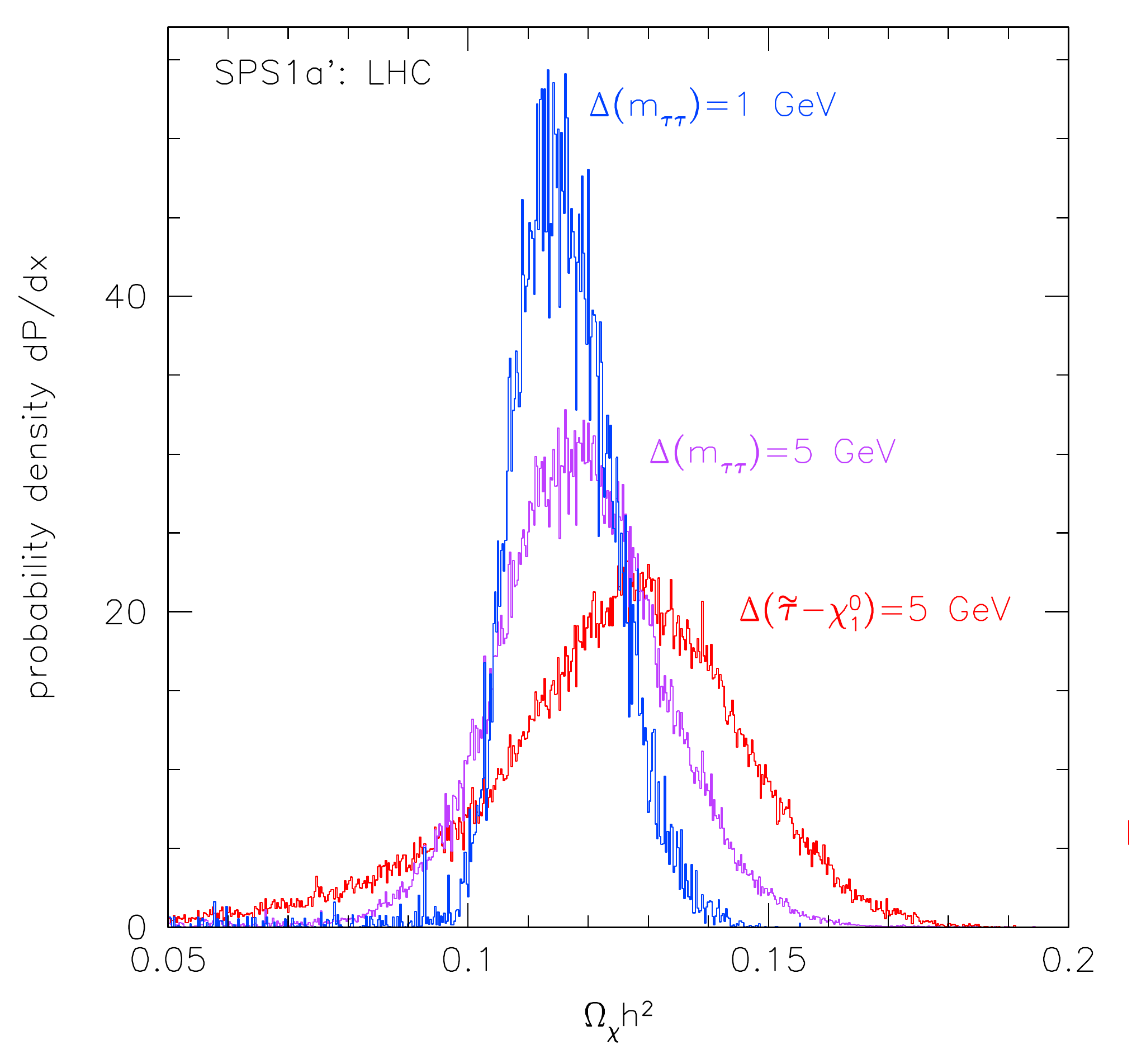}
 \hspace*{3cm}
 \includegraphics[width=6cm]{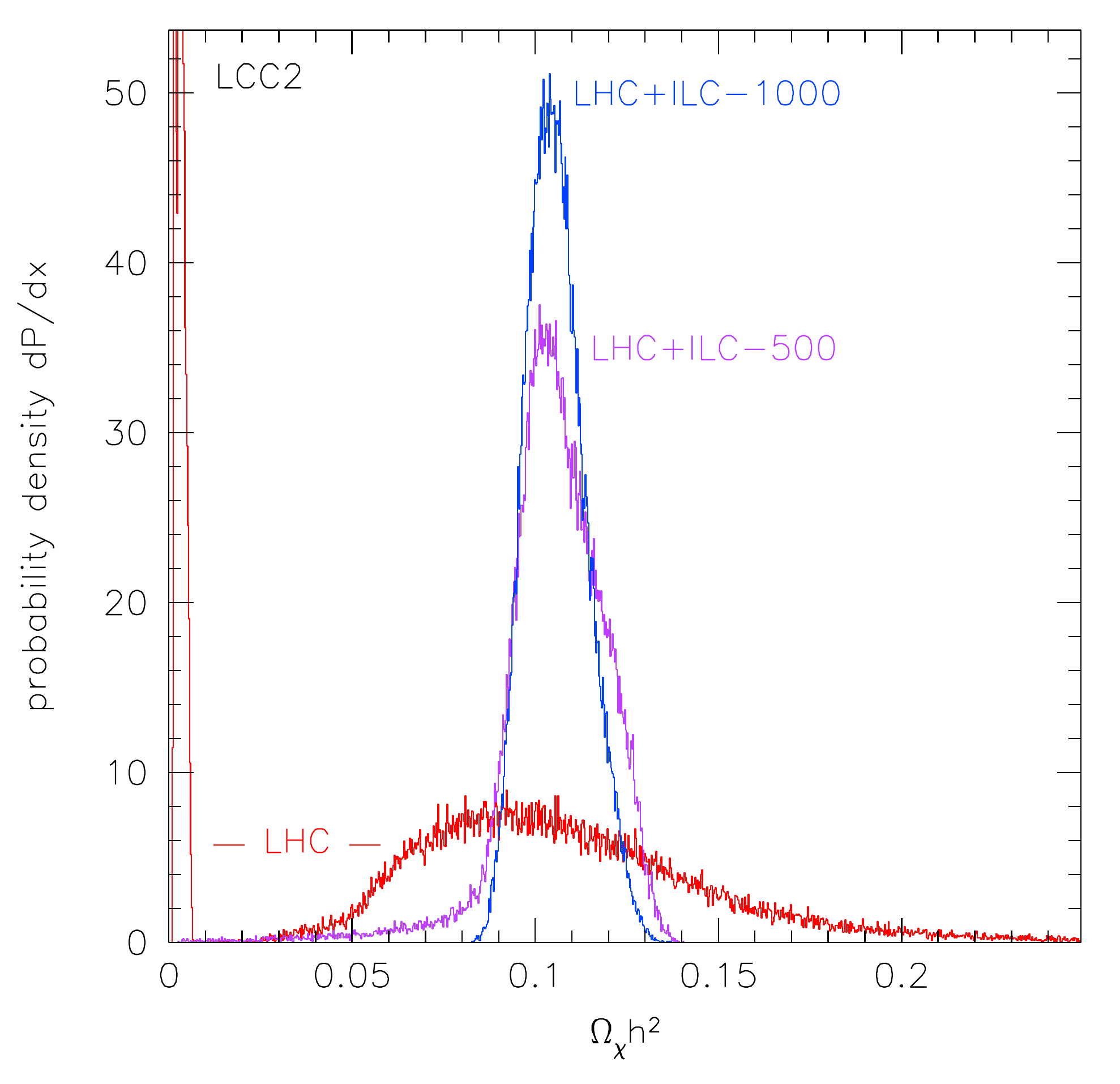} 
\caption[]{Dark matter density predictions for the parameter points
  SPS1a' (left) and LCC2 (right). For SPS1a' we show the LHC results
  with a 5~GeV error on the neutralino--stau mass difference and with
  a 5~GeV and 1~GeV error on $m_{\tau \tau}$\cite{spa}. For LCC2 we
  assume LHC and LC measurements following
  Ref.~\cite{measure_fp}. Figure taken from Ref.~\cite{ilc_dm}}
\label{fig:para_oh2}
\end{figure}

Technically, the best suited approach for the prediction of dark
matter properties is a Markov chain. As a matter of fact, such a
prediction is much closer to the original purpose of such Markov
chains than a simple likelihood map over parameter space; first, we
use some kind of measure like the likelihood given LHC mass
measurements.  Then, we compute a set of observables for these
parameter points and extract a probability distribution for these
observables. Two predictions of the relic density in the parameter
points SPS1a' and LCC2 we show in Fig.~\ref{fig:para_oh2}. Obviously,
the results we can hope for at the LHC are considerably less precise
than a later linear collider analysis, which can be easily understood from
Section~\ref{sec:para_ilc}. Moreover, since in SPS1a' the
co-annihilation 
helps us achieve the
measured relic density, its results are critically dependent on the
stau mass measurement and the mass difference between the NLSP-stau
and the LSP. Note that the $m_{\tau \tau}$ edge is a function of the
mass squared difference of these two states, so it has an increased
sensitivity on the stau--neutralino mass difference. From the
distribution we see that the tolerance of the measured relic density
on this mass difference is considerably smaller than the LHC error,
even with an optimistic 1~GeV error on the edge position. For LCC2 we
see two peaks in the relic density probability: one for a bino-like
LSP at the correct value, and another one for either a wino or a
higgsino solution to the LHC measurements. Such states annihilate much
more efficiently and predict a much smaller relic density.

As another example for predicting observables based on LHC (and LC)
data we show the distributions of spin-independent neutralino--proton
and neutralino--neutron cross sections in Fig.~\ref{fig:para_dd}. In
LCC1 the problem is that the main channel for neutralino--proton
scattering is the $t$-channel exchange of a heavy Higgs boson. Heavy
Higgs bosons, however, are only observable for large values of $\tan
\beta$, considerably larger than the $\tan \beta =10$ of LCC1, LCC2
and SPS1a. The peak at small cross sections observed in the upper left
panel of Fig.~\ref{fig:para_dd} corresponds to a light Higgs in the
$t$-channel. This poor result only changes, if for example at a
high-energy linear collider we observe the heavy Higgs states. For the LCC2 case
the $t$-channel exchange of a light Higgs gives the correct peak of
the neutralino-proton cross section, with additional peaks at larger
values.\bigskip

Such an analysis is by no means limited to supersymmetry and can be
extended to other dark matter models, such as UED.  In the 5d UED
case, the annihilation it usually predominantly into charged leptons,
and once again coannihilation, \eg with KK leptons may be
important~\cite{Servant:2002aq}.  In that case, precise measurements
of the mass difference between the LKP and the NLKP are crucial.  The
fact that the LKP in such theories is a massive vector, without
spin-suppressed interactions favors larger masses than is typical for
a neutralino, which itself makes the 5d UED theory with a thermal
relic harder to reconstruct at the LHC, because the rates are
correspondingly smaller.

\begin{figure}[t]
 \includegraphics[width=6cm]{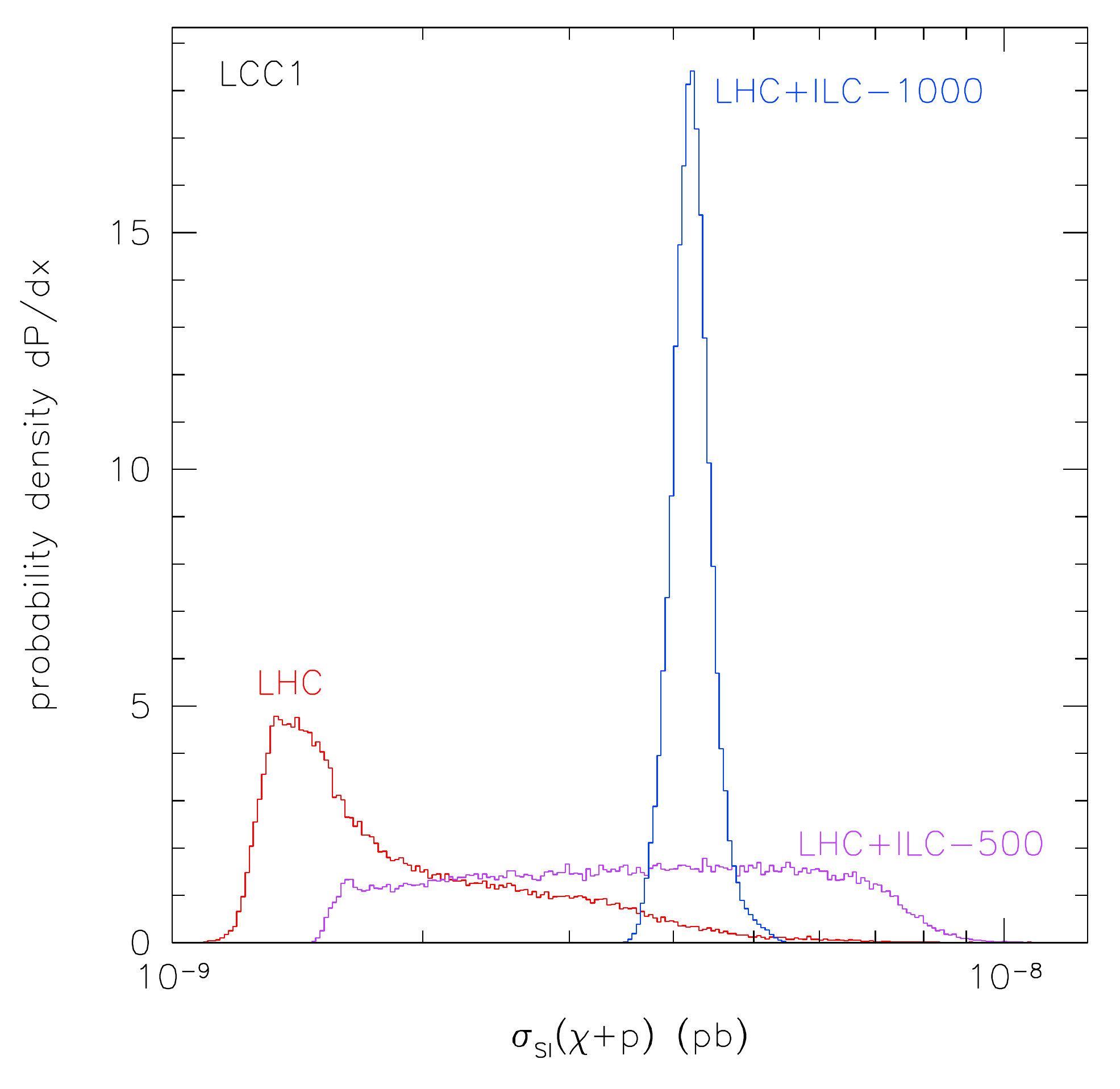}
 \hspace*{3cm}
 \includegraphics[width=6cm]{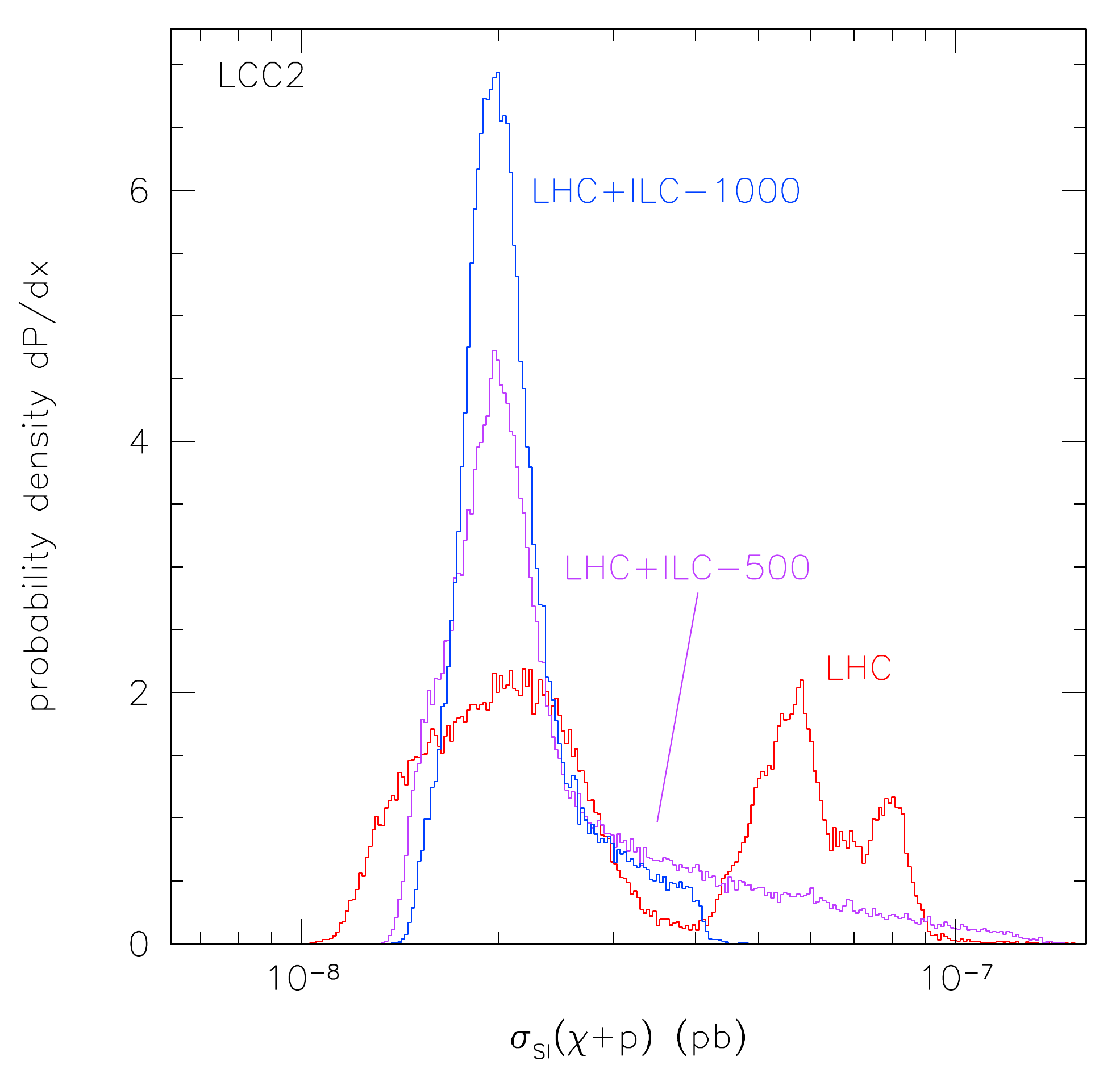} 
\caption[]{Expected spin-independent neutralino--proton cross
  sections. We show results for LCC1 and LCC2. Figure taken from
  Ref.~\cite{ilc_dm}}
\label{fig:para_dd}
\end{figure}

\subsection{Combination with a linear collider}
\label{sec:para_ilc}

\begin{table}[t]
\begin{tabular}{|l|r@{$\pm$}rr@{$\pm$}rr@{$\pm$}rr|}
\hline
                     & \multicolumn{2}{c}{LHC}    & \multicolumn{2}{c}{LC}     & \multicolumn{2}{c}{LHC+LC} & SPS1a \\
\hline
$\tan\beta$          &      10.0 & 4.5             &      12.1 & 7.0             &      12.6 & 6.2             &     10.0 \\
$M_1$                &     102.1 & 7.8             &     103.3 & 1.1             &     103.2 & 0.95            &    103.1 \\
$M_2$                &     193.3 & 7.8             &     194.1 & 3.3             &     193.3 & 2.6             &    192.9 \\
$M_3$                &     577.2 & 14.5            &\multicolumn{2}{c}{fixed 500}&     581.0 & 15.1            &    577.9 \\
$M_{\tilde{\tau}_L}$ &     227.8 & $\om(10^3)$     &     190.7 & 9.1             &     190.3 & 9.8             &    193.6 \\
$M_{\tilde{\tau}_R}$ &     164.1 & $\om(10^3)$     &     136.1 & 10.3            &     136.5 & 11.1            &    133.4 \\
$M_{\tilde{\mu}_L}$  &     193.2 & 8.8             &     194.5 & 1.3             &     194.5 & 1.2             &    194.4 \\
$M_{\tilde{\mu}_R}$  &     135.0 & 8.3             &     135.9 & 0.87            &     136.0 & 0.79            &    135.8 \\
$M_{\tilde{e}_L}$    &     193.3 & 8.8             &     194.4 & 0.91            &     194.4 & 0.84            &    194.4 \\
$M_{\tilde{e}_R}$    &     135.0 & 8.3             &     135.8 & 0.82            &     135.9 & 0.73            &    135.8 \\
$M_{\tilde{q}3_L}$   &     481.4 & 22.0            &     499.4 &$\om(10^2)$      &     493.1 & 23.2            &    480.8 \\
$M_{\tilde{t}_R}$    &     415.8 & $\om(10^2)$     &     434.7 &$\om(4\cdot10^2)$&     412.7 & 63.2            &    408.3 \\
$M_{\tilde{b}_R}$    &     501.7 & 17.9            &\multicolumn{2}{c}{fixed 500}&     502.4 & 23.8            &    502.9 \\
$M_{\tilde{q}_L}$    &     524.6 & 14.5            &\multicolumn{2}{c}{fixed 500}&     526.1 & 7.2             &    526.6 \\
$M_{\tilde{q}_R}$    &     507.3 & 17.5            &\multicolumn{2}{c}{fixed 500}&     509.0 & 19.2            &    508.1 \\
$A_\tau$             &\multicolumn{2}{c}{fixed 0}  &     613.4 & $\om(10^4)$     &     764.7 & $\om(10^4)$     &   -249.4 \\
$A_t$                &    -509.1 & 86.7            &    -524.1 & $\om(10^3)$     &    -493.1 & 262.9           &   -490.9 \\
$A_b$                &\multicolumn{2}{c}{fixed 0}  &\multicolumn{2}{c}{fixed 0}  &     199.6 & $\om(10^4)$     &   -763.4 \\
$A_{l1,2}$           &\multicolumn{2}{c}{fixed 0}  &\multicolumn{2}{c}{fixed 0}  &\multicolumn{2}{c}{fixed 0}  &   -251.1 \\
$A_{u1,2}$           &\multicolumn{2}{c}{fixed 0}  &\multicolumn{2}{c}{fixed 0}  &\multicolumn{2}{c}{fixed 0}  &   -657.2 \\
$A_{d1,2}$           &\multicolumn{2}{c}{fixed 0}  &\multicolumn{2}{c}{fixed 0}  &\multicolumn{2}{c}{fixed 0}  &   -821.8 \\
$m_A$                &     406.3 & $\om(10^3)$     &     393.8 & 1.6             &     393.7 & 1.6             &    394.9 \\
$\mu$                &     350.5 & 14.5            &     354.8 & 3.1             &     354.7 & 3.0             &    353.7 \\
$m_t$                &     171.4 & 1.0             &     171.4 & 0.12            &     171.4 & 0.12            &    171.4 \\
\hline
\end{tabular}
\caption[]{Result for the general MSSM parameter determination. Shown
  are the nominal parameter values and the result after fits to the
  different data sets. The LHC results correspond to
  Tab.~\ref{tab:para_lhc}. All masses are given in GeV. Table taken
  from Ref.~\cite{sfitter}}
\label{tab:para_ilc}
\end{table}

As should be apparent from the discussion in this section, a future
linear collider would be the perfect experiment to complement LHC and
other measurements, because the main challenge at the LHC is the
precision on the weakly interacting sector of new physics.  The one
alternative to a linear collider would be replacing precise
measurements by as precise biases, like for example postulating
supersymmetric GUT models to extract the neutralino/chargino sector
parameters from the gluino mass or the squark masses. Therefore, we
need to study how well the LC and the LHC can be combined to extract
the strongly interacting sector of new physics as well as the weakly
interacting sector.\bigskip

As usual, we have to rely on MSSM analyses, starting from the only
complete set of experimental studies available~\cite{sfitter,fittino}.
In Table~\ref{tab:para_mass_errors} we already list the precision of
LC mass measurements. The best of these measurements come from
threshold scans, where it is not clear for how many new particles such
scans can be performed. In principle, we can use this method for all
new particles as long as they have an electroweak charge, \ie all MSSM
states apart from the gluino. This method characterizes the kind of
linear collider which we assume in this section; it needs enough
energy to pair-produce all new weakly interacting particles, while at
the same time the energy should not be too large, to avoid a
spread-out beam energy distribution~\cite{clic}. 

For threshold scans
it is crucial that we precisely know the kinematics of the incoming
state.  These threshold scans can be combined with the kinematic
endpoints listed in Table~\ref{tab:para_edges}. For the best-fit
parameter point in Section~\ref{sec:para_mssm} we show the resulting
errors in Table~\ref{tab:para_ilc}. The general feature is that the
LHC is not sensitive to several of the TeV-scale model parameters, as
is the linear collider alone. The strength of the LHC is clearly visible in the
sector of heavy new particles with color charge. Whenever a parameter
is accessible at both colliders the linear collider dominates the precision.  For
example in the slepton and neutralino/chargino sectors, the LC
improves the precision by an order of magnitude.  While at the LHC and
LC separately not all parameters can be determined, the combination
of the two machines will allow us to measure essentially all
parameters, with the exception of the first and second generation
trilinear couplings. Given a sufficient linear collider energy this statement is
fairly general --- strongly interacting particles are hard to hide at
the LHC, at least as long as they decay to anything but light-flavor
jets. Weakly interacting particles will be pair-produced and measured
at a linear collider.

It is instructive to compare the effect of theory errors on the
parameter determination.  While the LC loses a factor five in
precision, going from a per-mille determination to half a percent, the
LHC loses less than a factor two. As discussed in
Section~\ref{sec:para_mssm} the latter is due to the one best-measured
LHC observable, the triangular $m_{\ell \ell}$ edge. For some
parameters the light supersymmetric Higgs mass and its theory error
seriously affect the LHC results. In that case we need to keep in mind
that for the calculation of the Higgs masses we not only have to
believe the perturbative error estimate but also the minimal structure
of the supersymmetric Higgs sector.\bigskip

\begin{figure}[t]
 \includegraphics[width=6.6cm]{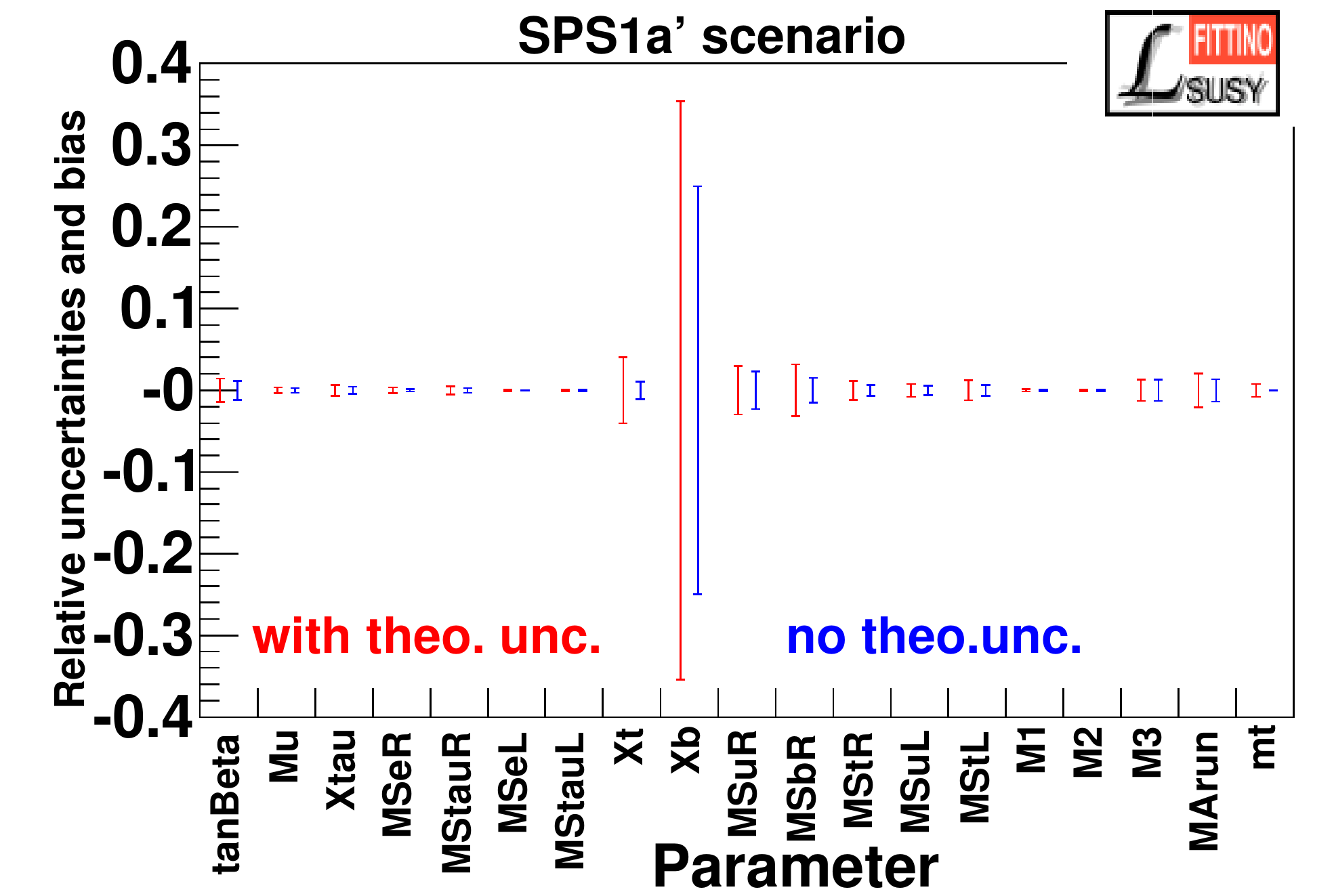} 
 \includegraphics[width=4.6cm]{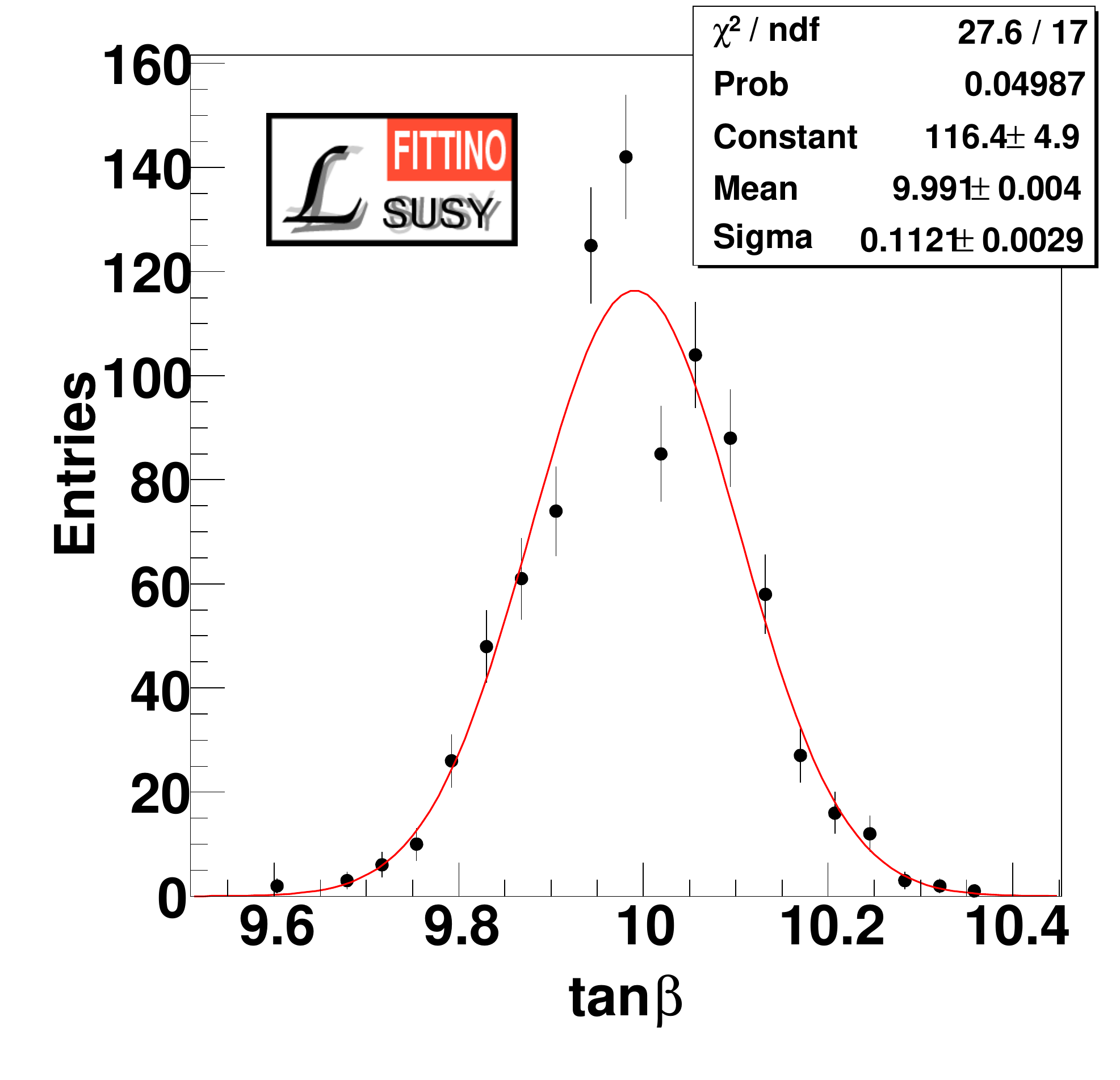}
 \includegraphics[width=4.6cm]{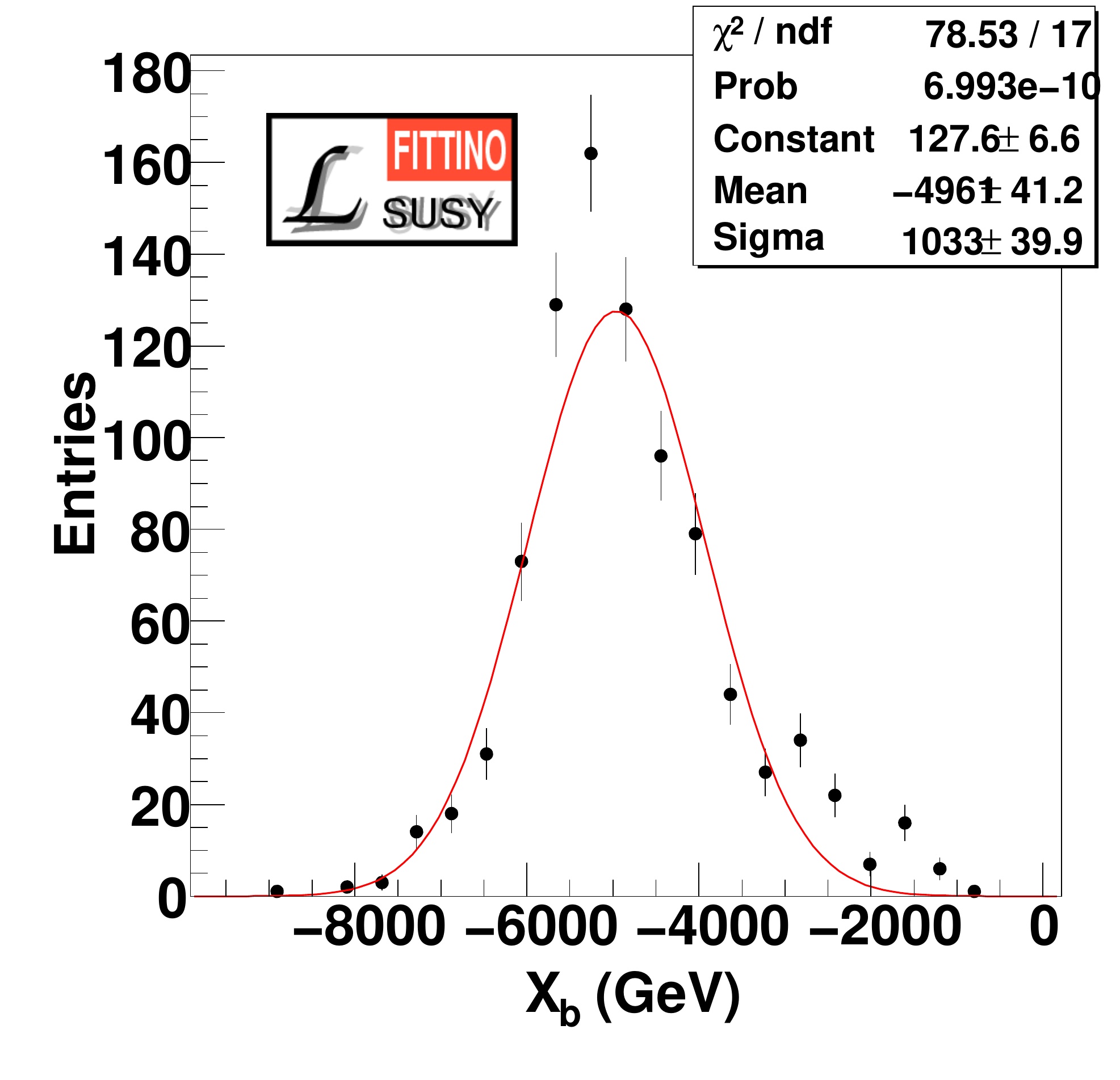} 
\caption[]{MSSM fit to combined LHC and LC data in the parameter
  point SPS1a'. The left panel shows the precision neglecting and
  including the theory error. The two right panels show the
  distributions of 1000 toy experiments smeared according to the
  measurement errors. Figure taken from Ref.~\cite{fittino}}
\label{fig:para_fittino}
\end{figure}

A similar Fittino analysis of the MSSM parameter space adds all
production cross sections times (observable) branching ratios to the
linear collider data set. In the clean and QCD-free environment we can compute
cross sections with a reasonable theory error and ensure that we see
all of the dominant decay channels. On the other hand, this analysis
assumes gaussian theory errors, which makes it hard to compare the
results. In Fig.~\ref{fig:para_fittino} we show the resulting errors
on some of the MSSM model parameters from a combined LHC-LC fit
including LC production rates. Again, we see that neglecting theory
errors will underestimate the errors on the model parameters by
typically a factor two. The best way of estimating the one-dimensional
error bars of each model parameter is a numerical error propagation:
running fits to 1000 smeared toy measurements and measuring the widths
of the resulting distributions. In Fig.~\ref{fig:para_fittino} we see
the results for the infamous $\tan \beta$ and 
$X_b = (A_b - \mu \tan \beta)$. 
While the extraction of $\tan \beta$ rests on the assumption
of a minimal Higgs sector and the corresponding neutralino and
chargino mixing matrices, the linear collider has the potential of measuring this
parameter with an error of a few percent. The sbottom mixing angle, or
$X_b$, also shows a clear peak in the likelihood, but with an error
around a TeV. Note that this result does not mean it will be
impossible to measure some model parameters, we might just be missing
the one good idea to do it.\bigskip

There are some parameters which we have not yet discussed yet ---
strictly speaking we assume a minimal Higgs sector and a common
supersymmetric mass scale when we construct the off-diagonal entries
in the neutralino and chargino mass matrices. The bino-wino mixing
terms are the Yukawa couplings of the mass eigenstates. Their
parameterization in terms of $m_{W,Z}$ and $\tan \beta$ will change
through renormalization group effects for example when we allow for
several stages of supersymmetry breaking. By studying all neutralinos
and charginos, such effects can be extracted at a linear 
collider~\cite{Kilian:2004uj,Bernal:2007uv,sfitter_lh}. Similarly, CP-violating 
phases can alter the structure of the neutralino and chargino 
mixing matrices. While all
phases relevant to neutralinos and charginos are strongly constrained
by the combination of electron, proton and mercury electric dipole
moments and even the linear collider has little hope to tell us
more~\cite{with_toby},
a large phase of $A_t$ might be visible as an
inconsistency in the fit to the minimal supersymmetric Higgs
sector~\cite{cp_higgs}. These caveats reflect the basic feature that
the tested hypothesis determines the appropriate observables as well
as the (theory) errors, which needs to be checked at every stage of
this analysis.\bigskip

In this section we have see that if we properly take into account all
theory errors in the combined analysis of TeV-scale models in the LHC
era we should be able to make definitive statements about their
underling structure.  If we do manage to control all potential
pitfalls, the combination of LHC and LC measurements and the links to
dedicated experiments like $(g-2)_\mu$, flavor physics or dark matter
observables~\cite{masters,leszek,ellis_olive,ilc_dm} can pave
a way to determining the nature of TeV-scale physics over the coming
years.

\newpage

\section{Outlook}

Of course, writing a review on new physics specifically for the LHC at
this point in time means that in the best of all worlds only one of
the models described in Section~\ref{sec:models} --- and hence a very
small fraction of this document --- will be known to be correct a few
years from now. As a matter of fact, there is a fair chance that none
of the models described in this review will turn out to describe the
TeV scale.  To put it bluntly, this review might turn out to be two
hundred pages of `wrong physics'.  So why write
or read it?
\bigskip

  Well, it is very likely that at least some of the signatures
analyzed in Section~\ref{sec:sig} will actually be seen at the
LHC.  The LHC is too complicated a machine and QCD too
complicated a theory for us to expect new physics effects to appear and
be understood automatically.  New physics at the LHC is something we
have to go and search for, and searching requires inspiration and guidelines.
This was our primary goal. 
Reading this review, it should be obvious that it would have been 
impossible to write the signatures section without having written 
the models section, and this is even more true for the section on 
deducing the underlying UV completion of the Standard Model.  
At the end of the LHC era, it is precisely in these latter stages of
analysis that we hope to be.  
\bigskip

  In this sense, even if ten years from now we will be forced to sit 
down and write a new Section~\ref{sec:models} and heavily modify
Section~\ref{sec:sig}, we hope that at least some readers will find
that the first edition of this review on new physics at the LHC 
came at exactly at the right time, and helped them to dig out 
the new and improved standard model of physics at the TeV scale.

\acknowledgments

We are grateful to (in alphabetical order) 
Uli Baur,
Ed Berger, 
Thomas Binoth, 
Kyle Cranmer, 
Sally Dawson, 
Tao Han, 
Gordy Kane, 
Mihoko Nojiri, 
Aaron Pierce, 
Giacomo Polesello, 
Dave Rainwater,
Carlos Wagner, 
James Wells, 
C.--P. Yuan, 
Dieter Zeppenfeld, 
Dirk and Peter Zerwas, 
and about a million people who taught us about new physics,
LHC physics, and QCD and all that in the past ten years.
It goes without saying that any shortcomings of this review are
solely due to the authors.  

We would also like to thank the Aspen Center for Physics --
where else would an idea like writing this kind of report be born and
immediately started on an innocent summer evening. 

Most of all, however, two of the authors would like to thank the
organizers and the participants of TASI~97. It is well possible that
those exciting four weeks are what got us hooked on new physics at
hadron colliders long before everybody knew that this is really the
thing to do. 
The youthful 
third author had a pretty great time at TASI~04 as well.

This work supported in part by the U.S. Department of Energy, Division of High Energy Physics, under Contract DE-AC02-06CH11357.

\end{document}